# Thermal Emission Spectroscopy for Single Nanoparticle Temperature Measurement: Optical System Design and Calibration

BRYAN A. LONG, DANIEL J. RODRIGUEZ, CHRIS Y. LAU, AND SCOTT L. ANDERSON*

*Department of Chemistry, University of Utah, 315 S. 1400 E., Salt Lake City, UT, 84112 USA*
**anderson@chem.utah.edu*

**Abstract:** We discuss the design of an optical system that allows measurement of 600 nm to 1650 nm emission spectra for individual nanoparticles (NPs), laser heated in an electrodynamic trap in controlled atmospheres. An approach to calibration of absolute intensity vs. wavelength for very low emission intensities is discussed, and examples of NP graphite and carbon black spectra are used to illustrate the methodology.

## 1. Introduction

We have developed an experiment to measure high temperature reaction kinetics for single nanoparticles (NPs) confined in a three dimensional quadrupole trap, with the goal of studying effects on NP surface chemistry and optical properties, of NP heterogeneity, i.e., variations in NP size and structure.[1-3] In essence, a series of single NPs are trapped and laser heated, allowing them to be detected optically via their thermal ("blackbody") emission. Changes in the NP mass vs. time are tracked to obtain the kinetics for mass-changing processes such as sublimation or oxidation. We are interested in NPs in the 5 to 50 nm range.

To use NP mass spectrometry as a kinetics tool, the NP temperature ($T_{NP}$) needs to be determined by fitting emission spectra measured during the kinetics experiments, with acquisition time scales of ~10 to 60 seconds. The wavelength ($\lambda$) range of interest spans the visible and near infrared (nIR) spectral regions, and we use a pair of spectrographs capable of covering the ~400 to 1650 nm range. The critical problem is to calibrate the detection sensitivity vs. $\lambda$ for the very low thermal emission intensities observed from single 5 to 50 nm NPs. $T_{NP}$ determination requires relative intensity vs. $\lambda$ calibration, but if absolute intensities can be extracted, absolute NP emissivities can be measured as a function of $\lambda$, $T_{NP}$, and NP size and composition – information that is largely unavailable.

Several groups have reported methodologies for emission spectroscopy from single trapped particles. For example, Bieske and co-workers reported dispersed emission spectra in the visible range for individual ~1 μm diameter dye-doped NPs in a trap similar to ours.[4-7] Our instrument must work with lower emission intensities, extend the spectra into the nIR where detectors are less sensitive, and must have well calibrated spectral intensities.

To our knowledge, the experiment most conceptually similar to ours was developed by Sarofim and co-workers.[8, 9] Individual particles were suspended in an AC trap and laser heated as they reacted in an oxidizing atmosphere. Particle temperature was measured by two-color (sometimes three-color) thermometry, where emission was measured with two or three broadband detectors, sensitive in different visible or IR wavelength ranges. From an optical perspective, the most important difference between their experiment and ours, is that their particles were ~1000 times larger than ours. Taking surface area and low emissivity for sub-$\lambda$ nanoparticles[10] into account, their intensities would have been $>10^6$ times higher. Furthermore, we felt that for our experiments – the first to measure dispersed emission for individual NPs as a function of $\lambda$, $T_{NP}$, NP size, and composition – it was essential to have dispersed spectra to extract emissivity vs. λ, and to look for any structured emission that might occur. To enable these experiments, a dual spectrograph optical system was developed, along with a method for calibrating absolute sensitivity vs. wavelength over the visible and near-IR ranges, at the very low intensities relevant to emission from sub-30 nm particles.

## 2. Experimental Methods

### *2.1 Optical system design*

Figure 1 shows the optical system layout. The instrument vacuum system, indicated schematically by the dashed box, has been described previously,[1] along with methods for NP mass and charge determination. Briefly, NPs are introduced into the vacuum system by electrospray ionization, and guided to the trap entrance by a combination of hexapole and quadrupole guides. NPs are injected into a split-ring-electrode quadrupole trap, based on a design by Gerlich, who also has discussed the motion of the trapped NP in detail.[11] During trapping, the center of the trap is irradiated by a 10 W cw $CO_2$ laser (10.6 μm) focused diagonally through the trap, or by a cw solid state laser focused through the trap perpendicular to the plane of the figure. Argon buffer gas is added at 1 to 15 mTorr pressure so that NPs undergo collisions as they pass through the trap, resulting in some probability that an NP will become trapped. Optical emission from the trap is monitored continuously, and injection is stopped as soon as an increase in emission indicates that a particle has become trapped.

Motion of trapped NPs is harmonic with well-defined frequencies for axial and radial motion, proportional to the charge to mass ratio, Q/M. To measure Q/M, a weak swept-frequency AC potential is applied across the trap, which causes a dip in collected emission intensity when the drive frequency is resonant with the NP axial frequency, $f_z$. Because each NP has distinct Q/M, the presence of more than one NP in the trap is obvious as multiple $f_z$ resonances, and when that happens, the trap is dumped and the injection process is repeated. All experiments described below were done with a single NP in the trap. To determine Q, and thus M, a vacuum ultraviolet lamp is used to induce photoemission from the NP, thus driving a series of single electron charge steps, which cause corresponding $\Delta f_z$ steps in $f_z$. Q is determined by fitting the steps: $Q = e \cdot f_z / \Delta f_z$. Once Q is determined, the lamp is turned off. To study NP sublimation or reactions, the NP mass is monitored continuously as the NP is laser heated in either inert or reactive atmospheres in the few mTorr range. The background pressure in the instrument is ~$10^{-8}$ Torr. For a given laser intensity, the NP temperature is determined by the balance between laser heating, and cooling processes which include thermal radiation, collisions with room temperature argon, and evaporative cooling. As discussed elsewhere,[3] for the conditions relevant here, thermal radiation and collisional cooling are dominant, and because both are temperature dependent the relationship between laser intensity and NP temperature is complex. For the spectra presented here, the NP was heated by a cw diode-pumped solid state laser operating at 532 nm, with power adjustable to 500 mW. For black carbon NPs, we have seen no indications that emission spectra or reactivity depend on excitation wavelength.

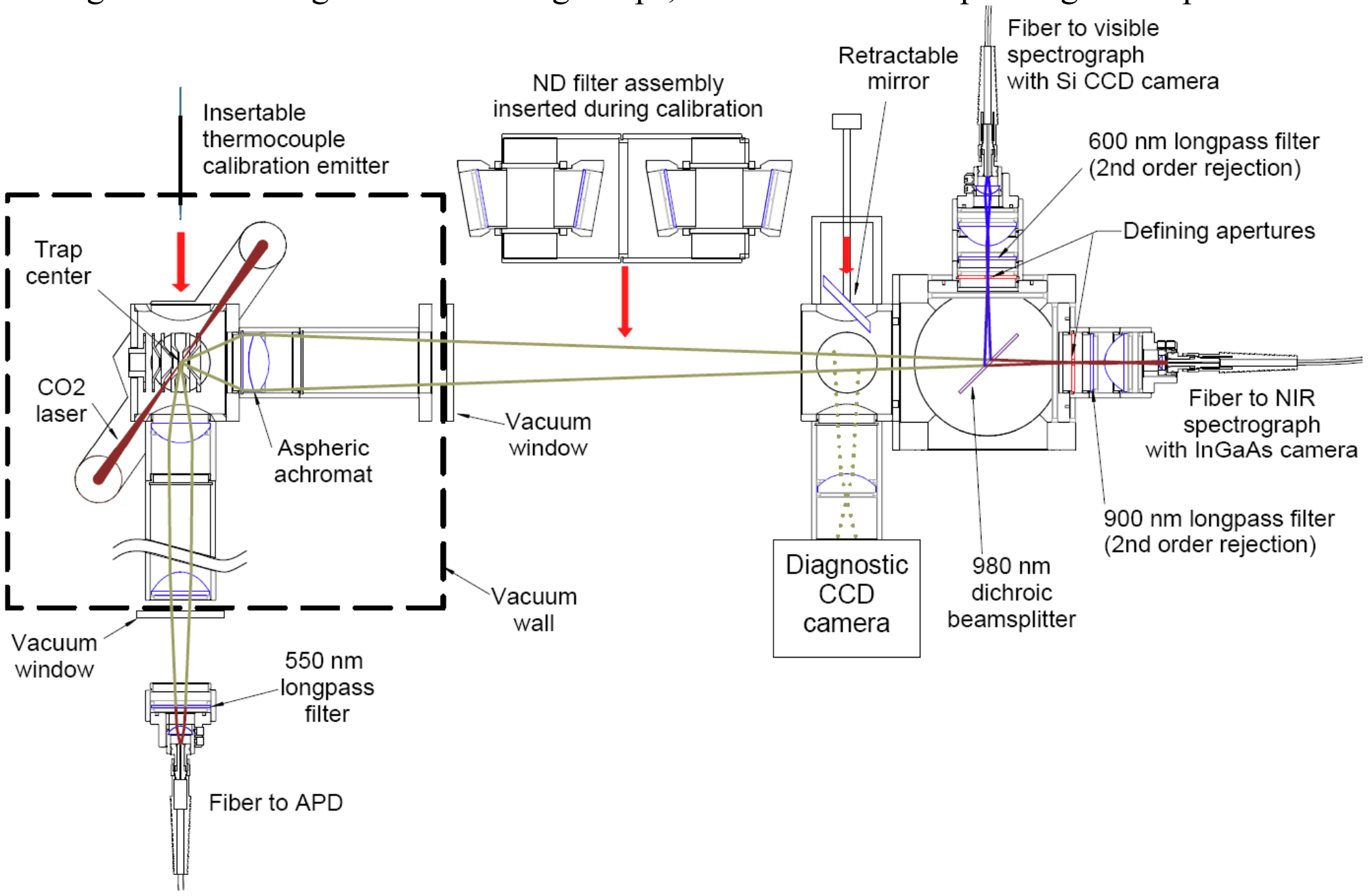

Fig. 1. Optical layout as described in text.

Light emitted by the trapped NP is collected along two directions. Light collected in the radial direction (solid angle ~0.4 steradians (sr)) is sent to a silicon avalanche photodiode (APD), monitoring intensity vs. time for NP M and Q determination, as described above. M is typically between 0.3 and 20 MDa (MDa = $10^6$ atomic mass units), and Q is typically 10 to 50 e. The trapped NP undergoes thermal motion with ~100 µm amplitude, thus from the optics perspective, the effective diameter of the NP emission source is ~200 µm.

Light emitted along the trap axis is collected and formed into a slowly converging beam by a 25 mm diameter, 30 mm effective focal length (efl) aspherized achromat (Edmund Optics). The beam can be diverted by a retractable mirror into a CCD camera for diagnostic purposes, but normally propagates to a dichroic beam splitter (Semrock), which reflects photons with $\lambda < 980$ nm by 90°, and passes $\lambda > 980$ nm. For simplicity, we will refer to the < 980 and > 980 nm spectral ranges as "visible" and "nIR", respectively.

The visible beam is focused into an optical fiber by a combination of a 25 mm diameter, 18.75 mm efl, 0.66 NA aspheric lens (Edmund Optics) and an 8.0 mm focal length, 0.50 NA, fixed focal length collimator (Thorlabs, 600-1050 nm AR coating). A 1000 µm single core, 0.50 NA, low OH fiber optic patch cable (Thorlabs) is used to guide the light to a small Czerny-Turner spectrograph (Andor Shamrock 163, f/3.6) equipped with Ag-coated mirrors and an Ag-coated 300 line/mm grating blazed at 760 nm. Diffracted light is detected by a back-illuminated, deep depletion 2000 x 256 pixel Si CCD camera, thermoelectrically cooled to -65°C (Andor, DU416A-LDC-DD). The camera-grating combination gives a detection bandwidth of ~569 nm, set for these experiments to cover the range from 437 to 1006 nm. For the $T_{NP}$ range of interest for materials like carbon, the spectral range below 600 nm is uninteresting, and to filter out scattered 532 nm laser light, or any light that might appear in 2nd order in the spectrum, the visible beam is passed through a 600 nm long-pass filter.

The nIR portion of the beam passes through a 900 nm long-pass filter, then is focused into another 1000 µm single core fiber optic cable using a 25 mm diameter, 18.75 efl aspheric lens, and a 4.64 mm focal length, 0.53 NA fixed focus collimator (Thorlabs, anti-reflection coating 1050-1620 nm). Another Czerny-Turner spectrograph (Andor Shamrock 163) equipped with Ag-coated mirrors and an Ag-coated 85 line/mm grating blazed at 1350 nm disperses the emission onto a thermoelectrically cooled (-80 °C) 512 x 1 pixel InGaAs photodiode camera (Andor, DU490A-1.7). To reduce the image height on the 500 µm tall camera array, the spectrograph includes a cylindrical lens.

Our interest is principally in analyzing thermal emission to obtain $T_{NP}$, and to improve sensitivity we operate the spectrographs with slits wide open, such that the resolution is controlled by the 1000 µm diameter optical fibers. The resolution and wavelength calibration were determined by measuring spectra for a mercury lamp. For the nIR spectrograph, the resolution is found to be ~60 nm fwhm, while for the visible spectrograph it is ~20 nm. When necessary, the slits can be closed to increase resolution at the cost of sensitivity.

Acquisition time for small NPs at relatively low temperatures is typically chosen to be 60 seconds, because that happens to match the time scale for NP mass determination. For larger

NPs or higher temperatures, where emission is strong, shorter acquisition times can be used to avoid saturation. Raw signal levels are kept below 10% of the manufacturer's stated saturation level, and under those conditions, we have never seen significant effects of acquisition time on the NP spectral shape or extracted NP temperature.

### *2.2 Background subtraction*

To allow background subtraction, spectra are measured under conditions identical to those used in the NP spectra, but with no NP in the trap. Therefore, the background spectra include all sources of background signal that are present in the NP spectra, except light scattering from the NP, which would be blocked by filters in the optical train. Typical raw NP spectra are shown in Fig. 2, along with the background spectra, for an NP pumped at 532 nm. The NP in this case was graphite with M = 18.0 MDa, corresponding to ~30 nm diameter if spherical with the bulk graphite density. The NP was held at relatively low $T_{NP}$ so that its mass loss rate from sublimation was slow, and 60 second acquisition time was used. As shown in Fig. 2 (left), when the 532 nm laser is used for particle heating, the visible background has a feature at ~700 nm, with a broader underlying feature extending from ~650 to 750 nm. The origin of this background signal is unclear (it is not seen when pumping with the $CO_2$ laser), but it is removed quantitatively during background subtraction.

The background signal for the nIR spectrograph (right frame, Fig. 2) mostly originates from the spectrograph, itself. Thermal radiation from room temperature surfaces inside the spectrograph impinges on the InGaAs array, which detects the small fraction with $\lambda$ < 1650 nm, producing a background "dark" signal. Because this dark signal originates in the spectrograph, it is unaffected by filters in the optical system. For long integration times (e.g. 60 sec, as in Fig. 2), which are necessary for small NPs at low $T_{NP}$, the dark signal can be significantly larger than the actual NP emission signal, as is the case here. Furthermore, while the room temperature dark signal is weak, it is strongly temperature dependent, such that a 0.1 K temperature change results in a nearly 1% change in the detected dark signal. Stabilizing the spectrograph temperature to much better than 0.1 K would be difficult, therefore we developed an approach in which the background is scaled to account for spectrograph temperature drift during the experiments. We take advantage of the fact that there is no real signal below 950 nm, due to a combination of the 980 nm dichroic beamsplitter and low grating and quantum efficiencies. The background spectrum is scaled so that the integrated signal in the <950 nm region matches that in the NP spectrum prior to subtraction. The >1650 nm range, where the real signal is also zero due to zero quantum efficiency for cold InGaAs, serves as a check on this procedure. This approach was tested by gently heating the spectrograph to verify that scaling correctly reproduces the temperature effects.

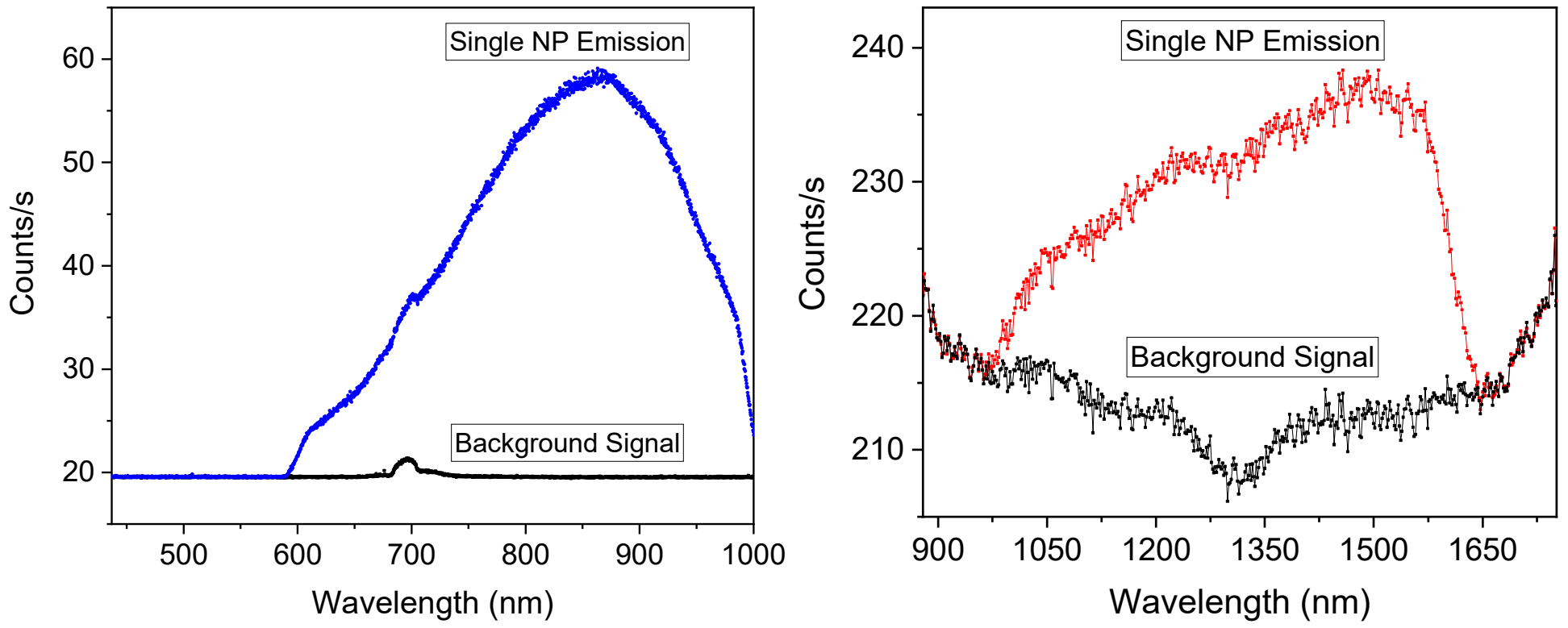

Fig. 2. Left: Raw NP and background spectra in the visible range for a ~18 MDa (~30 nm diameter) graphite NP. Right: Raw NP and background spectra in the nIR range. See **Data File 1** for underlying values.

Note that much of the scatter in both signal and background nIR spectra is not noise, but rather results from pixel-to-pixel variation in the camera sensitivity. As can be seen by close examination of the spectra in the <950 and > 1650 nm spectral regions, the pattern of scatter is quite reproducible, and mostly cancels when the background is subtracted. There is additional cancellation in the intensity calibration process described below.

### *2.3. Calibration of the optical system sensitivity*

To extract $T_{NP}$ from the spectra, it is necessary to correct the spectra quantitatively for the wavelength dependence of the optical detection sensitivity, S(λ), including collection efficiency, effects of focusing optics, vacuum windows, spatial filtering, spectrograph transmission, and detector quantum efficiency. The calibration process was developed with the following considerations: 1. Sensitivity is strongly affected by optical alignment with respect to the emitter position, i.e., the trap center. This requires that calibration be performed on the system as a whole, with the calibration source at the trap center. The calibration light source must, therefore, fit into the trap center through the 2 mm gap between the electrodes. 2. To avoid detector non-linearity effects, calibration should be done at intensities similar to that from single NP thermal emission. 3. The size of the calibration emitter should match the ~200 µm diameter of NP thermal motion, to avoid artifacts due to different source sizes during calibration and NP spectroscopy. 4. To enable frequent calibration, it should be possible to do the calibration without venting the vacuum system.

To provide a vacuum-compatible calibration emitter small enough to fit into the trap center, we use a type C (W-5% Re/W-26% Re) thermocouple (TC), fabricated from 78 µm diameter type C wire (Concept Alloys). The TC junction was fabricated by twisting the wire, then tightly folding and spot welding repeatedly to create a compact, roughly spherical junction volume. The TC is mounted to a precision XYZ vacuum manipulator via a ~1 mm diameter, dual bore ceramic tube, allowing the TC junction to be positioned at the trap center for calibration, then parked in a position well outside the trap for NP spectroscopy. When positioned at the trap center, the TC can be heated using the $CO_2$ laser to ~2600 K. The laser is focused through the trap by a pair of home-made off-axis paraboloids, resulting in a ~600 µm diameter beam waist[3] – roughly twice the junction diameter. To help heat the TC junction uniformly, the beam is retro-reflected after passing through the trap, so that the junction is irradiated from both sides.

The advantage of the large TC junction is that emission from the junction surface is much stronger than emission from the wires leading to the junction, which are not at a well-defined temperature. The concern is that a twisted/folded/welded junction might not have a uniform emitting surface temperature, but imaging of the junction shows uniform color temperature across the central ~300 µm region of the junction, presumably because a large number of high compression spot welds was used to form a compact junction that is heated uniformly by the weakly focused, retro-reflected laser. Because the absolute emissivity vs. wavelength and temperature, $\epsilon(\lambda,T)$, for W and Re are well known,[12-15] and the junction temperature can be read electrically, the TC provides an emitter with well-known intensity vs. λ, I(λ, T).

Sarofim *et al.*[8, 9] also used a laser-heated TC as an emitter to calibrate their two or three-color thermometry method. Because the particles studied in their experiments were in the tens-of-micron diameter range, their emission intensity was similar to that from their TC emitter. In our experiments, the single NP emission is roughly $10^8$ times weaker than that from the hot TC. Clearly the detectors cannot be expected to be linear over such a large intensity range, therefore the intensity from the TC must be attenuated by a factor of ~$10^8$, without affecting the optical path significantly. For this purpose, we use a set of four OD 2 reflective neutral density (ND) filters (Thorlabs, nickel-coated, 1 mm thick UV fused silica substrates) mounted in the holder shown in Fig. 1. The filters are tilted to prevent multiple reflections, and the angles alternate to eliminate lateral beam walk when the assembly is inserted during calibration. The tilt planes of the two pairs of filters are rotated 90° with respect to each other (not shown) to avoid introduction of polarization effects. Refraction in the filter substrates is calculated to increase the convergence distance of the light beam by ~0.3%, which has a negligible effect on the spatial filtering used to define the emitter.

We were unable to obtain reliable transmission spectra for the individual OD 2 filters using two different commercial spectrometers, therefore the transmission was measured using the optical setup in Fig. 1, with the laser-heated TC as the light source. A pair of angle-mounted OD 3 filters was used as a "pre-attenuator" to drop the intensity by a factor of roughly $10^6$, and both visible and nIR spectra were recorded. Spectra were then recorded after inserting one of the four OD 2 filters into the optical path, mounted perpendicular to the path to avoid beam walk, and far enough from the pre-attenuator to prevent multiple reflections between the filters. The ratio of the spectra, with and without the OD 2 filter, gives the transmission of that filter under the conditions used in the TC calibration. Repeated measurements were found to be identical within the signal/noise, and the transmission spectrum of each filter was obtained by averaging five measurements. The net transmission of the four OD 2 filters, ND(λ), was calculated as the product of the averaged transmission spectra for the individual filters. To check for possible intensity effects on ND(λ), transmission measurements were repeated using different ND filters in the pre-attenuator, varying the pre-attenuator transmission from ~$10^{-6}$ to $10^{-8}$. No significant intensity effects were observed, giving us confidence the measured ND(λ), and also demonstrating that the spectrograph cameras are quite linear over the intensity range of interest for our experiments.

As noted, we want the calibration emitter size to be similar to the diameter of the NP thermal motion, and this was achieved using defining apertures to control light from the trap. As shown in Fig. 1, the aspheric achromatic collection lens slowly focuses light collected from the trap center, forming images in both the visible and nIR branches of the optical train. Defining apertures are inserted at both focal planes to define the collection area, also efficiently suppressing scattered light and other background sources. The collection lens is only designed to be achromatic to 675 nm, however, the indices of refraction of the two glasses used varies by less than 2% between 675 and 1600 nm, thus we expect the magnifications at the visible and nIR aperture positions to be quite similar. The magnification for the visible image was measured by photographing the hot TC image projected on a paper screen at the aperture plane. 16 independent measurements of 78 ± 1 μm diameter TC wire leads were averaged to obtain a magnification factor of 12.3 ± 0.3. The junction, itself was found to be ~330 μm across. Given the 2.39 mm defining aperture diameter, this means that light is collected only if it is emitted from the central 194 ± 8 μm diameter region of the ~330 μm junction, and therefore that the effective TC emitter area is well matched to the ~200 μm diameter of NP thermal motion. A similar measurement was made on the nIR image of the TC, photographing the image through an IR viewer with λ sensitivity similar to that of the InGaAs spectrograph (Newport IRV1-1700). The nIR was not as sharp, resulting in greater uncertainty in the magnification, which was found to be ~12.9 ± 1.0. As expected from the lens properties, the magnifications, hence collection areas are identical within the uncertainty for the visible and nIR optics. For absolute calibration purposes we estimate that the emitter area is known to ±10%.

Finally, it is critical that the optical system be aligned to the NP position at the trap center, and that the TC junction be positioned at the trap center during calibration. A single NP is trapped and heated using a loose laser focus so that $T_{NP}$, hence emission intensity, is insensitive to slight laser misalignment. The optics, including defining apertures, are then carefully adjusted to maximize the NP signal, ensuring that the optical system is aligned to the trap center. For calibration, the TC is inserted and heated, and its position is adjusted to center the magnified junction image on the defining apertures.

### *2.4. Correcting NP spectral intensities.*

Using background-subtracted TC emission spectra, measured as described above, calibration of the optical system sensitivity vs. wavelength is straightforward. The intensity measured for the laser-heated TC ($I_{TC}^{meas}$) at temperature T is given by:

$$I_{TC}^{meas}(\lambda) = I_{BB}(\lambda, T) \cdot \epsilon_{TC}(\lambda, T) \cdot ND(\lambda) \cdot S(\lambda), \quad (1)$$

where $I_{BB}(\lambda, T)$ is the theoretical (Planck's law) spectrum for an ideal blackbody at temperature T, $\epsilon_{TC}(\lambda, T)$ is the emissivity of the W-Re thermocouple, ND(λ) is the transmission of the set of four OD 2 ND filters, and S(λ) is the desired optical system sensitivity vs. wavelength. Therefore:

$$S(\lambda) = \frac{I_{TC}^{meas}(\lambda)}{I_{BB}(\lambda,T)\cdot\epsilon_{TC}(\lambda,T)\cdot ND(\lambda)} \quad (2)$$

Once S(λ) is known, then background-subtracted NP spectra, $I_{NP}(\lambda,T)$, can be corrected by simply dividing by S(λ):

$$I_{NP}^{corr}(\lambda, T) = \frac{I_{NP}(\lambda,T)}{S(\lambda)} \quad (3)$$

To test this calibration procedure a TC calibration experiment with junction temperature of 2507 K was used to calculate S(λ). TC spectra were then electronically measured at three other junction temperatures ranging from 2136 to 2361 K, and these spectra were corrected using the S(λ) function determined from the 2507 K calibration:

$$I_{TC}^{corr}(\lambda, T_i) = \frac{I_{TC}^{meas}(\lambda,T_i)}{S(\lambda)^{2507K}}. \quad (4)$$

Finally, the corrected TC spectra at the three test temperatures were fit to the product of $I_{BB}(\lambda, T_i) \cdot \epsilon_{TC}(\lambda, T_i)$, to extract a spectral temperature, which was compared to the temperature read electronically from the TC during each measurement. For measurements at temperatures within 200 K of the 2507 K calibration temperature, the difference between the electronically and spectroscopically measured temperatures was less than 0.25%. At 2136 K, the fit temperature was 2129 K, – 378 K below the calibration temperature. This results in a spectroscopically determined temperature that is within 0.3% of the electrically measured temperature, i.e., well within the temperature error limits provided by the TC wire manufacturer (±1% for T > 425°C).

## 3. Results and Discussion

Figure 3 illustrates the process, and gives the data needed to calculate the wavelength-dependent sensitivity, S(λ). Frame A shows the background-subtracted visible and nIR spectra of the TC heated to 2454 K, with the signal attenuated by four OD 2 filters. The signal level is comparable to that for typical single NPs, and here it has been put on a *per* $m^2$ basis by dividing by the aperture-defined 2.93 x $10^{-8}$ emitter area. Regions where the signal is weak or strongly affected by camera quantum efficiency, filters, or the 980 nm dichroic beamsplitter are indicated with solid vertical lines. As in the NP spectra in Fig. 2, much of the data scatter results from pixel-to-pixel variations in camera sensitivity, which ultimately cancels out in the intensity-corrected NP spectra.

Frame B shows the Planck function for an ideal blackbody at 2454 K, in units of photons/second/sr/$m^2$ of surface area, assuming 1 nm bandwidth. The actual bandwidth subtended by each pixel varies from 0.296 to 0.268 nm for the visible spectrograph, and from 1.736 nm to 1.673 nm for the nIR spectrograph, however, this factor cancels exactly in the calibration process, so has not been included in Planck function calculation.

Frame C shows the absolute emissivity of tungsten, ϵ(λ), calculated at 2454 K using the empirical ϵ(λ,T) expression of Pon et al.,[12] which is based on experimental data of De Vos[13] and Larrabee.[14] The ϵ(λ,T) function for a type C thermocouple junction (average composition ~85% W, 15% Re) has not been reported to our knowledge, however, the ϵ(λ,T) data for W[13] and Re[15] are very similar for the λ range of interest, the main difference being that the absolute emissivity is ~3 % lower for Re. Therefore $\epsilon(\lambda,T)_{TC}$ estimated as a composition-weighted average of the W and Re emissivities differs from the W value by only ~0.5%, which is negligible compared to other sources of uncertainty in the absolute intensity calibration (e.g. aperture-defined emitter area). Therefore we use the W emissivity, for which an analytic $\epsilon(\lambda,T)_W$ expression is available.[12]

Frame D shows the net transmission, ND(λ), of the four OD 2 filters used to attenuate the TC signal. The transmission ranges between ~3.5 and ~5.5 x $10^{-8}$, and has significant noise resulting from the low signal level in the transmission measurements. Given the low resolution of the spectrographs, the high frequency noise cannot be real, and to avoid propagating it to the NP spectra, we used Savitsky-Golay smoothing to generate the solid black curve passing through the points. Note that it is important not to over-smooth ND(λ), because there is real structure, and if it is removed by over-smoothing, the result is spurious structure in the corrected NP spectra.

As shown in frame A, the signal in the region centered around ~1020 nm is low and strongly λ-dependent due to camera quantum efficiencies and grating efficiencies, and even the smoothed ND(λ) is noisy with spurious structure in this spectral region. To approximately interpolate ND(λ) across the gap, we fit the ND(λ) data in the 870 to 1271 nm range spanning the gap with a 3rd order polynomial. The interpolated values between 870 and 1100 nm (open purple symbols, frame D) are used only in the spectral range where the smoothed data are

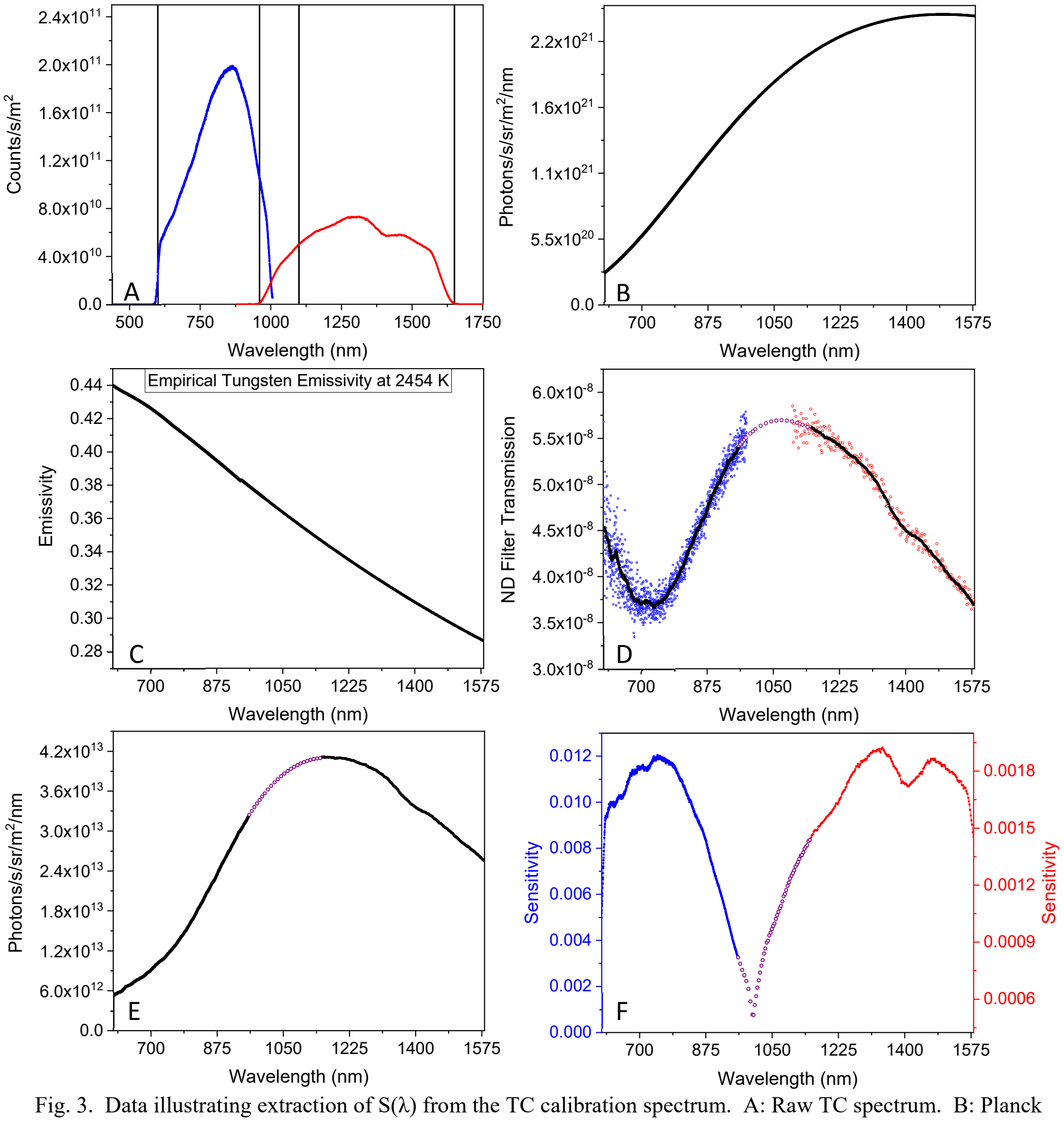

Fig. 3. Data illustrating extraction of S(λ) from the TC calibration spectrum. A: Raw TC spectrum. B: Planck function. C: Empirical W emissivity. D: ND filter transmission. E: Product of B through D. F: Final S(λ) function. See Data File 2 for underlying values.

unreliable, and only to check how the visible and nIR spectra connect, *not for $T_{NP}$ fitting.*

Frame A is the numerator of equation 2 above, and frame E (the product of frames B-D), is the denominator. Frame F is the resulting quotient, i.e., the desired sensitivity vs. wavelength, S(λ), in units of counts·nm·sr/photon. Note that the open symbols in frames E and F indicate points that are based on measured spectral intensities, but using interpolated ND(λ) values. The scatter for S(λ) has some contribution from pixel-to-pixel variations in the camera sensitivity, which cancels when the NP spectra are corrected using S(λ).

Comparing the spectra in frames A and E of Fig. 3 provides an opportunity to assess our understanding of the optical system, and its efficiency. Frame E is the predicted absolute emission intensity including transmission through the ND filters, in photons/second/sr/nm/$m^2$ of emitter area. Frame A is the measured number of counts/second/pixel/$m^2$ emitter area. To compare these, it is necessary to take into account the acceptance solid angle of the optical system, reflection losses from all the uncoated surfaces in the optical train, reflectivity of the Ag mirrors, the grating efficiencies, the actual bandwidth impinging on each pixel of the cameras, the quantum efficiency of the Si and InGaAs sensors, and the readout "sensitivity" of each camera, reported by the manufacturer (5.28 $e^-$/count for the Si CCD, 91.59 $e^-$/count for the InGaAs camera). Note that it is the much higher readout sensitivity used in the Si CCD camera that makes the raw visible counts so much higher than the nIR counts (Frame A), even though the emission intensity and bandwidth/pixel are both higher in the nIR.

Taking these factors into account, we predict that at 760 nm, near the peak of the visible spectrograph quantum and grating efficiencies, the signal should be $1.62 \times 10^{11}$ counts/pixel/second/$m^2$, compared to the measured $1.48 \times 10^{11}$ counts/pixel/second/$m^2$, i.e., there is ~8% unexplained signal loss. At 1350 nm, near the peak in the nIR region, we predict that there should be $1.12 \times 10^{11}$ counts/pixel/second/$m^2$, compared to the measured $6.88 \times 10^{10}$ counts/pixel/second/$m^2$, i.e., the nIR optics have a ~38 % unexplained loss. The origin of these inefficiencies is unclear, however, it is likely that much of the nIR loss results from overfilling the 500 μm tall camera array. Regardless, it is important to note that all losses are included in calculating S(λ), and therefore are accounted for in calculating intensity-corrected NP spectra using the procedure described above.

Fig. 4 shows the result of applying the S(λ) correction to the spectrum of the 18.0 MDa graphite NP that was used to illustrate the raw spectra in Fig. 2. The acquisition time was 60 seconds in this case, and repeated measurements are quite reproducible. The left frame shows the background-subtracted spectra prior to correction, i.e., the difference between the NP and background spectra given in Fig. 2. The right frame shows the corrected spectrum, obtained by dividing the left frame by the S(λ) function from Fig. 3F.

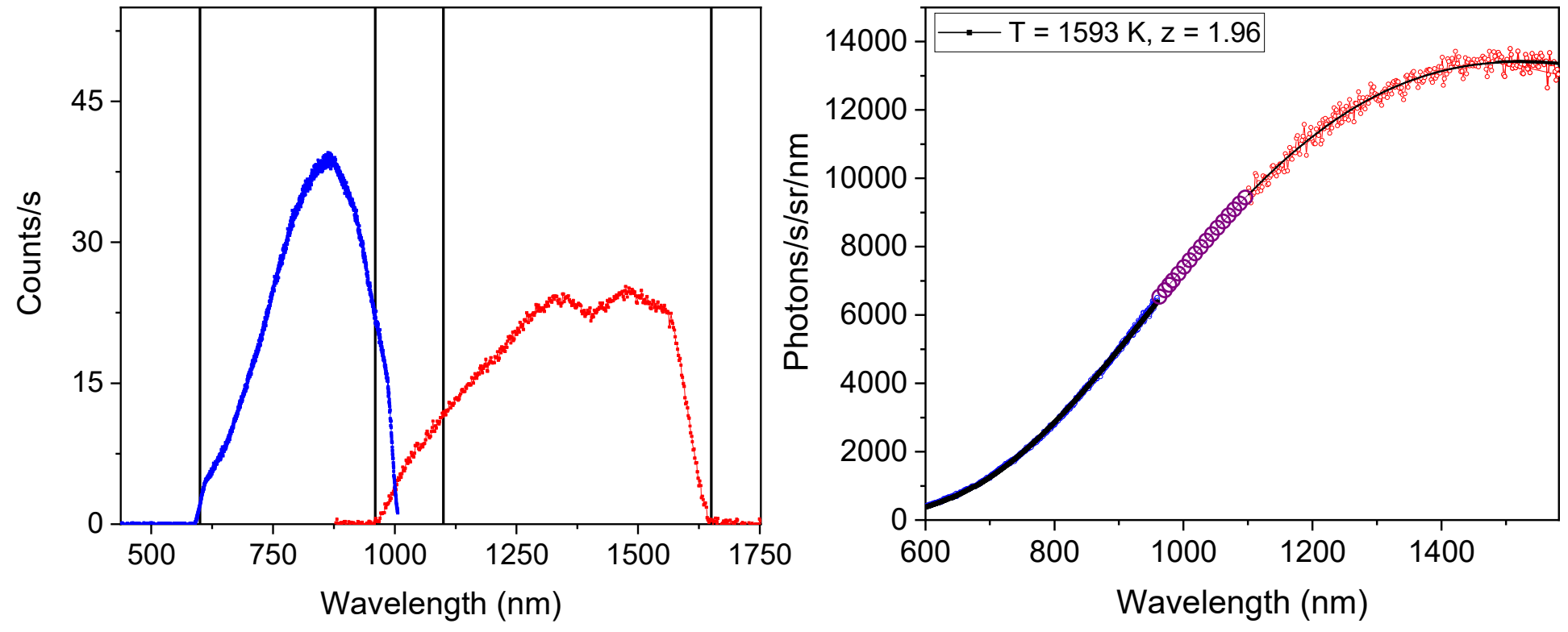

Fig. 4. (A) Raw emission and (B) corrected emission spectrum from a single ~18 MDa (~ 30 nm diameter) graphite NP. See **Data File 3** for underlying data.

The NP temperature can be estimated by fitting the corrected spectrum. The section of the fit that overlays data that were corrected with interpolated ND(λ) values is indicated by open circles. These data between 960 to 1100 nm were not included in the $T_{NP}$ fit, but are shown along with the fit to demonstrate that the ND(λ) interpolation process did not introduce significant artifacts into the corrected NP spectra. The function used to fit the data is obtained by multiplying Planck's law by a nanoparticle emissivity function, $\epsilon(\lambda)$. Emissivity for nanoparticles much smaller than the wavelength can be calculated from Mie theory,[16] however, for hot NPs this requires knowing the optical properties of the material at the temperature of interest over the visible/nIR wavelength range. There may also be effects on emissivity from surface or other localized states, which could vary from NP to NP. More detailed discussion of emissivity for individual carbon and non-carbon NPs will be the subject of a future publication. Here, for the purposes of illustration, we adopted the common procedure of approximating $\epsilon(\lambda)$ with a simple power law expression, $\epsilon(\lambda) \propto \lambda^{-z}$, z being a fitting parameter.[17-20] In this case, the fitted temperature is 1593 K, and the z parameter is 1.96. There are several sources of uncertainty, including ~±20 K from the ~1% uncertainty in the TC calibration temperature, and ~±15 K fitting uncertainty due to spectral noise.

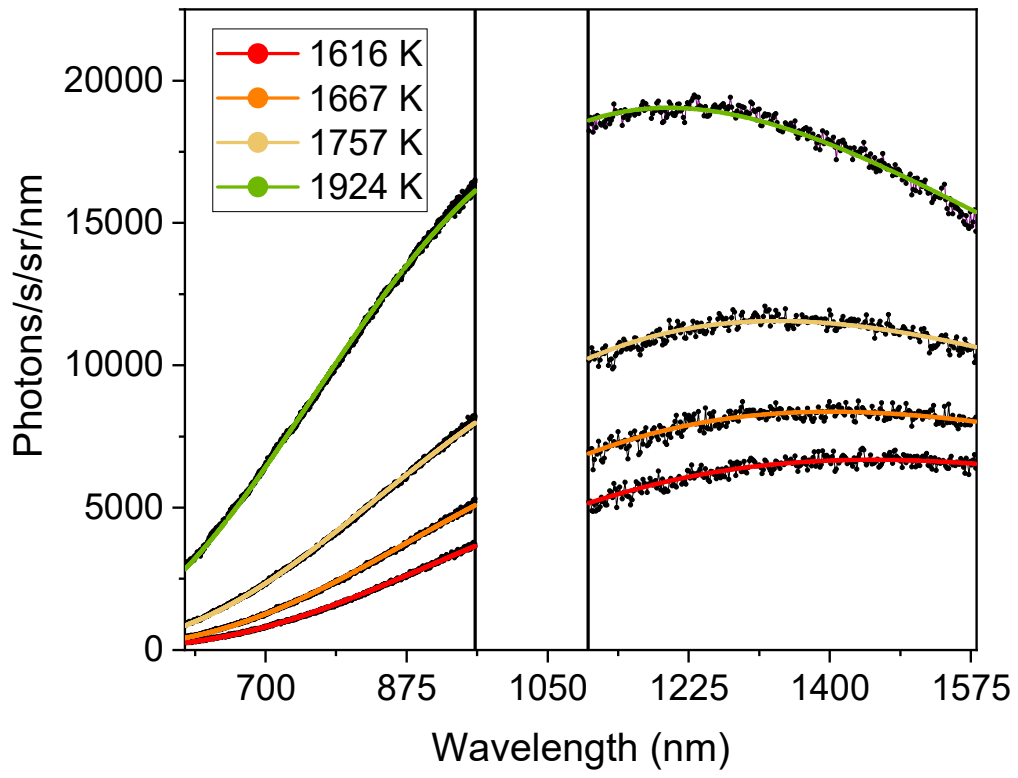

Fig. 5. Spectra for a single 5.8 MDa (~20.3 nm diameter) carbon black NP at different temperatures. See **Data File 4** for underlying values.

Fig. 5 shows a set of spectra for a smaller, 5.8 MDa (~20.3 nm) carbon black NP at four different 532 nm laser heating intensities, all recorded with 60 second acquisition time. The extracted $T_{NP}$ values vary from 1616 K to 1924 K.

## 4. Conclusions

We have discussed the design of an optical system and an intensity calibration method that allows thermal emission spectra for individual NPs to be measured over the 600 to 1600 nm range, on a time scale appropriate for measuring $T_{NP}$ to accompany NP sublimation and reaction kinetics experiments. The relative intensity vs. wavelength, i.e., the shape of the spectra, is critical in fitting the spectra for $T_{NP}$ determination, and we estimate that this is correct to within a few percent over the 600 – 1600 nm wavelength range. The approach used should also give accurate absolute intensity calibration. The main uncertainty is the area on the TC from which emission is collected, defined by the apertures placed at the visible and nIR image planes. We estimate that this area is known to within ~±10%.

The spectral accuracy and signal to noise is high enough that the main uncertainty in extracting $T_{NP}$ is modeling the NP emissivity. Careful examination of Fig. 5 shows that the simple power law emissivity model used fails to adequately fit the spectra, particularly at high $T_{NP}$. The effects of size, composition and temperature on NP emissivity is a complex topic, and will be the subject of a future publication.

### 5.1 Funding

The nIR spectrograph was purchased with funds from the Albaugh Scientific Equipment Endowment of the College of Science, University of Utah. This material is based upon work supported by the U.S. Department of Energy, Office of Science, Office of Basic Energy Sciences, Gas Phase Chemical Physics program under Award Number DE-SC- 0018049.

### 5.2 Acknowledgments

We thank Prof. Joel Harris (University of Utah, Chemistry Department) and Prof. Jordan Gerton (University of Utah, Physics Department) for many helpful discussions about optics and light detection.

### 5.3 Disclosures

The authors declare that there are no conflicts of interest related to this article.

Proposed TOC thumbnail:

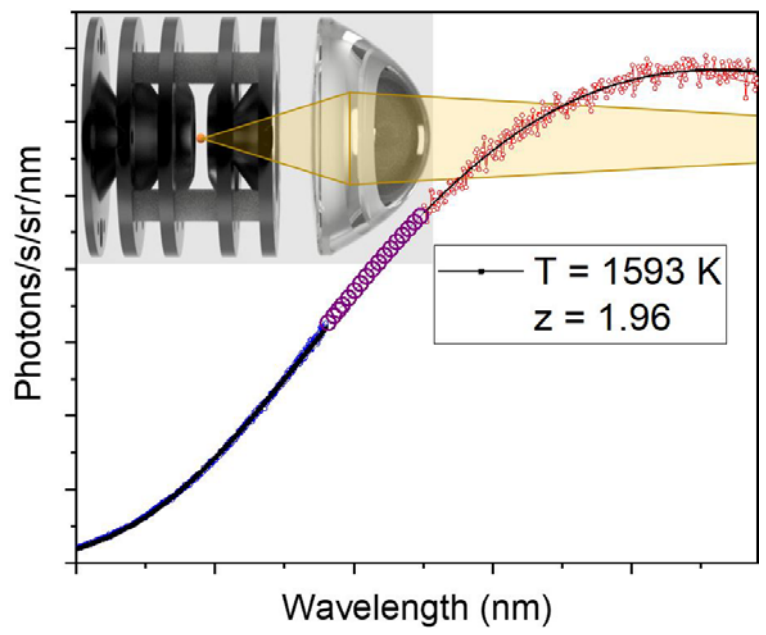

Data for figure two in the paper

| Wavelengtl | backgrounc | Visible Sign | Wavelengtl | backgrounc | nIR Signal |
|---|---|---|---|---|---|
| 437.661 | 19.55 | 19.6 | 879.476 | 221.5091 | 221.65 |
| 437.9568 | 19.5 | 19.51667 | 881.2142 | 222.618 | 223.1333 |
| 438.2526 | 19.55 | 19.63333 | 882.9523 | 222.0194 | 221.9278 |
| 438.5484 | 19.51667 | 19.56667 | 884.6903 | 221.8981 | 221.9806 |
| 438.8441 | 19.58333 | 19.56667 | 886.4282 | 221.9477 | 221.9972 |
| 439.1399 | 19.55 | 19.51667 | 888.166 | 220.3258 | 220.4028 |
| 439.4357 | 19.5 | 19.6 | 889.9037 | 218.85 | 219.0056 |
| 439.7314 | 19.55 | 19.61667 | 891.6413 | 219.7438 | 219.6056 |
| 440.0272 | 19.51667 | 19.58333 | 893.3788 | 220.7809 | 220.25 |
| 440.3229 | 19.58333 | 19.68333 | 895.1162 | 219.5341 | 219.7139 |
| 440.6187 | 19.51667 | 19.56667 | 896.8535 | 219.0266 | 219.3111 |
| 440.9144 | 19.55 | 19.61667 | 898.5908 | 218.3563 | 218.6361 |
| 441.2101 | 19.53333 | 19.66667 | 900.3279 | 218.2184 | 218.1583 |
| 441.5058 | 19.58333 | 19.6 | 902.0649 | 218.6708 | 218.5056 |
| 441.8015 | 19.55 | 19.53333 | 903.8019 | 218.0611 | 217.8694 |
| 442.0972 | 19.61667 | 19.56667 | 905.5387 | 216.8971 | 217.1389 |
| 442.3929 | 19.56667 | 19.53333 | 907.2754 | 217.2116 | 217.1694 |
| 442.6886 | 19.58333 | 19.56667 | 909.0121 | 218.4556 | 218.2639 |
| 442.9843 | 19.58333 | 19.55 | 910.7486 | 218.1632 | 218.3472 |
| 443.28 | 19.56667 | 19.61667 | 912.4851 | 217.3936 | 217.1444 |
| 443.5757 | 19.61667 | 19.56667 | 914.2214 | 218.3425 | 218.4833 |
| 443.8713 | 19.55 | 19.56667 | 915.9577 | 217.7936 | 217.9194 |
| 444.167 | 19.6 | 19.58333 | 917.6938 | 216.624 | 216.6806 |
| 444.4626 | 19.56667 | 19.55 | 919.4299 | 217.1536 | 216.8778 |
| 444.7583 | 19.55 | 19.6 | 921.1659 | 217.0488 | 216.8028 |
| 445.0539 | 19.56667 | 19.56667 | 922.9017 | 217.6943 | 217.625 |
| 445.3495 | 19.51667 | 19.55 | 924.6375 | 216.9826 | 217.2444 |
| 445.6451 | 19.53333 | 19.56667 | 926.3732 | 216.8199 | 217.0083 |
| 445.9408 | 19.53333 | 19.56667 | 928.1087 | 217.915 | 217.9194 |
| 446.2364 | 19.55 | 19.56667 | 929.8442 | 218.5687 | 218.4889 |
| 446.532 | 19.51667 | 19.56667 | 931.5796 | 217.915 | 217.55 |
| 446.8276 | 19.53333 | 19.58333 | 933.3149 | 216.6351 | 216.8278 |
| 447.1231 | 19.61667 | 19.6 | 935.05 | 217.1923 | 217.0639 |
| 447.4187 | 19.6 | 19.51667 | 936.7851 | 217.697 | 216.7333 |
| 447.7143 | 19.53333 | 19.56667 | 938.5201 | 217.4957 | 217.45 |
| 448.0099 | 19.51667 | 19.53333 | 940.255 | 217.1316 | 217.1 |
| 448.3054 | 19.55 | 19.53333 | 941.9898 | 216.1744 | 216.15 |
| 448.601 | 19.6 | 19.63333 | 943.7244 | 215.3469 | 215.4361 |
| 448.8965 | 19.48333 | 19.6 | 945.459 | 215.1483 | 215.2806 |
| 449.1921 | 19.58333 | 19.56667 | 947.1935 | 215.8489 | 216.1389 |
| 449.4876 | 19.51667 | 19.51667 | 948.9279 | 216.2599 | 216.2694 |
| 449.7831 | 19.53333 | 19.56667 | 950.6622 | 216.8613 | 217.0694 |
| 450.0786 | 19.58333 | 19.53333 | 952.3964 | 217.4902 | 217.4972 |
| 450.3742 | 19.56667 | 19.55 | 954.1304 | 217.093 | 216.825 |
| 450.6697 | 19.56667 | 19.61667 | 955.8644 | 216.0089 | 215.8611 |
| 450.9652 | 19.6 | 19.66667 | 957.5983 | 216.4916 | 216.5944 |
| 451.2606 | 19.51667 | 19.58333 | 959.3321 | 216.8171 | 217.1111 |
| 451.5561 | 19.6 | 19.55 | 961.0658 | 215.9427 | 216.55 |
| 451.8516 | 19.56667 | 19.55 | 962.7993 | 214.9221 | 215.2556 |
| 452.1471 | 19.6 | 19.61667 | 964.5328 | 215.3193 | 215.6417 |
| 452.4425 | 19.58333 | 19.53333 | 966.2662 | 216.2572 | 216.5417 |
| 452.738 | 19.53333 | 19.58333 | 967.9995 | 215.2917 | 215.575 |
| 453.0334 | 19.56667 | 19.58333 | 969.7327 | 216.5165 | 217.1389 |
| 453.3289 | 19.51667 | 19.58333 | 971.4657 | 215.06 | 216.2222 |
| 453.6243 | 19.56667 | 19.51667 | 973.1987 | 214.9469 | 216.0583 |
| 453.9197 | 19.51667 | 19.61667 | 974.9316 | 215.8434 | 217 |
| 454.2152 | 19.58333 | 19.5 | 976.6643 | 215.0159 | 216.225 |
| 454.5106 | 19.53333 | 19.51667 | 978.397 | 215.4986 | 216.7083 |
| 454.806 | 19.56667 | 19.55 | 980.1296 | 215.1952 | 217.3278 |
| 455.1014 | 19.55 | 19.55 | 981.862 | 214.7097 | 216.45 |
| 455.3968 | 19.6 | 19.58333 | 983.5944 | 214.8394 | 216.9083 |
| 455.6922 | 19.58333 | 19.6 | 985.3267 | 215.9648 | 218.0917 |
| 455.9875 | 19.58333 | 19.55 | 987.0588 | 216.0448 | 218.3 |
| 456.2829 | 19.55 | 19.56667 | 988.7909 | 216.6323 | 219.625 |
| 456.5783 | 19.53333 | 19.53333 | 990.5228 | 216.4116 | 218.9333 |
| 456.8736 | 19.51667 | 19.51667 | 992.2547 | 215.5207 | 218.2556 |
| 457.169 | 19.58333 | 19.58333 | 993.9864 | 216.0751 | 219.35 |
| 457.4643 | 19.56667 | 19.55 | 995.7181 | 214.3511 | 217.5556 |
| 457.7597 | 19.63333 | 19.56667 | 997.4496 | 214.6932 | 218.1861 |
| 458.055 | 19.53333 | 19.58333 | 999.1811 | 215.8213 | 219.6167 |
| 458.3503 | 19.53333 | 19.53333 | 1000.912 | 214.9 | 218.5472 |
| 458.6456 | 19.56667 | 19.51667 | 1002.644 | 214.9194 | 218.5944 |
| 458.9409 | 19.53333 | 19.58333 | 1004.375 | 215.9813 | 220.1722 |
| 459.2362 | 19.61667 | 19.63333 | 1006.106 | 215.9124 | 220.5667 |
| 459.5315 | 19.56667 | 19.61667 | 1007.837 | 216.3951 | 221.325 |
| 459.8268 | 19.58333 | 19.53333 | 1009.568 | 216.8557 | 221.6083 |
| 460.1221 | 19.5 | 19.56667 | 1011.298 | 216.031 | 221.35 |
| 460.4174 | 19.53333 | 19.56667 | 1013.029 | 216.5992 | 221.5833 |
| 460.7126 | 19.53333 | 19.55 | 1014.759 | 215.2476 | 220.6167 |
| 461.0079 | 19.53333 | 19.58333 | 1016.49 | 215.9013 | 221.9722 |
| 461.3031 | 19.55 | 19.56667 | 1018.22 | 216.7675 | 222.4972 |
| 461.5984 | 19.63333 | 19.63333 | 1019.95 | 214.7759 | 220.4639 |
| 461.8936 | 19.55 | 19.6 | 1021.68 | 216.784 | 222.7333 |
| 462.1888 | 19.53333 | 19.55 | 1023.41 | 215.7717 | 222.5694 |
| 462.4841 | 19.58333 | 19.58333 | 1025.14 | 214.2711 | 220.9417 |
| 462.7793 | 19.58333 | 19.6 | 1026.87 | 216.9302 | 223.3917 |
| 463.0745 | 19.5 | 19.51667 | 1028.6 | 216.0117 | 223.0306 |
| 463.3697 | 19.53333 | 19.55 | 1030.329 | 215.2642 | 222.0417 |
| 463.6649 | 19.51667 | 19.65 | 1032.059 | 216.5137 | 223.1389 |
| 463.96 | 19.55 | 19.55 | 1033.788 | 216.2489 | 223.875 |
| 464.2552 | 19.56667 | 19.55 | 1035.517 | 215.0076 | 222.3194 |
| 464.5504 | 19.46667 | 19.61667 | 1037.246 | 216.282 | 223.5639 |
| 464.8456 | 19.51667 | 19.58333 | 1038.975 | 216.3841 | 223.55 |
| 465.1407 | 19.58333 | 19.6 | 1040.704 | 215.6531 | 223.2056 |
| 465.4359 | 19.56667 | 19.55 | 1042.433 | 216.5137 | 224.8694 |
| 465.731 | 19.51667 | 19.51667 | 1044.162 | 216.3675 | 224.6917 |
| 466.0261 | 19.63333 | 19.51667 | 1045.89 | 216.0724 | 224.6611 |
| 466.3213 | 19.56667 | 19.58333 | 1047.619 | 216.4503 | 224.9861 |
| 466.6164 | 19.53333 | 19.6 | 1049.347 | 216.6902 | 224.7194 |
| 466.9115 | 19.53333 | 19.56667 | 1051.076 | 216.5882 | 224.5056 |
| 467.2066 | 19.58333 | 19.56667 | 1052.804 | 215.9096 | 224.0361 |
| 467.5017 | 19.6 | 19.56667 | 1054.532 | 216.2379 | 224.6889 |
| 467.7968 | 19.51667 | 19.58333 | 1056.26 | 213.8022 | 222.1667 |
| 468.0919 | 19.51667 | 19.48333 | 1057.988 | 213.8243 | 222.0583 |
| 468.3869 | 19.55 | 19.6 | 1059.716 | 216.4613 | 225.0111 |
| 468.682 | 19.58333 | 19.56667 | 1061.443 | 215.5841 | 224.175 |
| 468.9771 | 19.51667 | 19.56667 | 1063.171 | 215.3414 | 224.3028 |
| 469.2721 | 19.55 | 19.55 | 1064.898 | 215.4765 | 225.0917 |
| 469.5672 | 19.53333 | 19.58333 | 1066.625 | 215.2586 | 225.1417 |
| 469.8622 | 19.61667 | 19.58333 | 1068.353 | 215.2724 | 224.8056 |
| 470.1572 | 19.53333 | 19.61667 | 1070.08 | 215.0821 | 224.4139 |
| 470.4522 | 19.5 | 19.53333 | 1071.807 | 215.3497 | 224.5361 |
| 470.7473 | 19.6 | 19.58333 | 1073.534 | 215.0104 | 224.3556 |
| 471.0423 | 19.51667 | 19.51667 | 1075.26 | 214.6849 | 224.5333 |
| 471.3373 | 19.56667 | 19.51667 | 1076.987 | 215.0931 | 225.3694 |
| 471.6323 | 19.6 | 19.56667 | 1078.714 | 214.4146 | 225.375 |
| 471.9272 | 19.56667 | 19.63333 | 1080.44 | 215.2476 | 225.1389 |
| 472.2222 | 19.56667 | 19.5 | 1082.166 | 215.8793 | 225.7389 |
| 472.5172 | 19.53333 | 19.6 | 1083.893 | 215.5621 | 226.0222 |
| 472.8122 | 19.56667 | 19.61667 | 1085.619 | 215.5124 | 225.75 |
| 473.1071 | 19.56667 | 19.61667 | 1087.345 | 215.6365 | 226.6 |
| 473.4021 | 19.53333 | 19.58333 | 1089.071 | 215.3635 | 226.25 |
| 473.697 | 19.55 | 19.53333 | 1090.797 | 215.631 | 226.3139 |
| 473.9919 | 19.51667 | 19.58333 | 1092.522 | 214.9111 | 225.8111 |
| 474.2869 | 19.55 | 19.51667 | 1094.248 | 213.5733 | 224.9056 |
| 474.5818 | 19.58333 | 19.51667 | 1095.973 | 214.6104 | 225.925 |
| 474.8767 | 19.55 | 19.55 | 1097.699 | 213.4022 | 225.2222 |
| 475.1716 | 19.5 | 19.6 | 1099.424 | 214.3042 | 225.8056 |
| 475.4665 | 19.5 | 19.58333 | 1101.149 | 213.9153 | 225.8778 |
| 475.7614 | 19.55 | 19.6 | 1102.874 | 212.6216 | 224.5278 |
| 476.0562 | 19.58333 | 19.55 | 1104.599 | 215.0876 | 226.7056 |
| 476.3511 | 19.58333 | 19.56667 | 1106.324 | 213.6422 | 225.7389 |
| 476.646 | 19.55 | | 1108.049 | 213.3278 | 225.325 |
| 476.9408 | 19.5 | | 1109.773 | 214.9166 | 227.0722 |
| 477.2357 | 19.58333 | 19.56667 | 1111.498 | 213.3747 | 226.3278 |
| 477.5305 | 19.53333 | 19.56667 | 1113.222 | 211.2673 | 224.1306 |
| 477.8254 | 19.5 | 19.61667 | 1114.946 | 213.6836 | 226.0444 |
| 478.1202 | | 19.48333 | 1116.67 | 213.4243 | 226.2444 |
| 478.415 | | 19.58333 | 1118.394 | 213.1209 | 225.8917 |
| 478.7098 | 19.55 | 19.48333 | 1120.118 | 214.4808 | 227.1333 |
| 479.0046 | 19.55 | 19.55 | 1121.842 | 213.4547 | 226.3306 |
| 479.2994 | 19.55 | 19.6 | 1123.566 | 212.5416 | 225.8806 |
| 479.5942 | 19.56667 | 19.48333 | 1125.29 | 214.1498 | 227.45 |
| 479.889 | 19.58333 | 19.55 | 1127.013 | 213.9429 | 227.3889 |
| 480.1838 | 19.5 | 19.53333 | 1128.736 | 212.2851 | 225.6944 |
| 480.4785 | 19.56667 | 19.55 | 1130.46 | 213.0905 | 226.2944 |
| 480.7733 | 19.61667 | 19.51667 | 1132.183 | 213.0492 | 226.3833 |
| 481.068 | 19.56667 | 19.63333 | 1133.906 | 213.3912 | 227.0694 |
| 481.3628 | 19.58333 | 19.58333 | 1135.629 | 212.7761 | 226.6361 |
| 481.6575 | 19.56667 | 19.58333 | 1137.351 | 213.3526 | 227.6 |
| 481.9522 | 19.51667 | 19.53333 | 1139.074 | 213.5677 | 227.9083 |
| 482.247 | 19.53333 | 19.56667 | 1140.797 | 212.7761 | 226.8833 |
| 482.5417 | 19.56667 | 19.58333 | 1142.519 | 212.834 | 226.9528 |
| 482.8364 | 19.61667 | 19.55 | 1144.242 | 213.4822 | 227.5944 |
| 483.1311 | 19.51667 | 19.58333 | 1145.964 | 212.6768 | 227.2639 |
| 483.4257 | 19.56667 | 19.56667 | 1147.686 | 212.8588 | 227.3111 |
| 483.7204 | 19.55 | 19.55 | 1149.408 | 212.4534 | 227.2528 |
| 484.0151 | 19.56667 | 19.61667 | 1151.13 | 211.6065 | 226.4944 |
| 484.3098 | 19.58333 | 19.56667 | 1152.852 | 213.3885 | 228.3194 |
| 484.6044 | 19.58333 | 19.61667 | 1154.573 | 213.7994 | 228.6194 |
| 484.8991 | 19.51667 | 19.46667 | 1156.295 | 213.8105 | 228.6083 |
| 485.1937 | 19.53333 | 19.58333 | 1158.016 | 214.1884 | 229.7833 |
| 485.4883 | 19.65 | 19.56667 | 1159.738 | 214.1387 | 229.6917 |
| 485.783 | 19.56667 | 19.6 | 1161.459 | 213.1181 | 228.1361 |
| 486.0776 | 19.48333 | 19.56667 | 1163.18 | 212.3485 | 227.8444 |
| 486.3722 | 19.56667 | 19.56667 | 1164.901 | 212.9581 | 228.95 |
| 486.6668 | 19.53333 | 19.55 | 1166.622 | 211.5762 | 227.2028 |
| 486.9614 | 19.6 | 19.56667 | 1168.343 | 211.4907 | 227.0278 |
| 487.256 | 19.55 | 19.56667 | 1170.063 | 212.5223 | 228.3667 |
| 487.5506 | 19.55 | 19.65 | 1171.784 | 212.4837 | 228.5667 |
| 487.8451 | 19.53333 | 19.56667 | 1173.504 | 213.2285 | 229.2 |
| 488.1397 | 19.6 | 19.61667 | 1175.225 | 213.7884 | 230.3861 |
| 488.4342 | 19.48333 | 19.55 | 1176.945 | 212.5306 | 229.3694 |
| 488.7288 | 19.53333 | 19.53333 | 1178.665 | 212.6354 | 228.9556 |
| 489.0233 | 19.58333 | 19.53333 | 1180.385 | 212.9361 | 229.0361 |
| 489.3179 | 19.58333 | 19.6 | 1182.105 | 212.6051 | 229.1722 |
| 489.6124 | 19.58333 | 19.58333 | 1183.824 | 212.5665 | 229.3611 |
| 489.9069 | 19.51667 | 19.58333 | 1185.544 | 211.67 | 228.6528 |
| 490.2014 | 19.53333 | 19.53333 | 1187.263 | 212.4561 | 229.9889 |
| 490.4959 | 19.58333 | 19.55 | 1188.983 | 212.8644 | 229.6972 |
| 490.7904 | 19.56667 | 19.53333 | 1190.702 | 213.1154 | 229.3889 |
| 491.0849 | 19.6 | 19.56667 | 1192.421 | 212.4037 | 229.425 |
| 491.3793 | 19.61667 | 19.56667 | 1194.14 | 212.2741 | 229.4111 |
| 491.6738 | 19.61667 | 19.6 | 1195.859 | 212.3899 | 229.7417 |
| 491.9683 | 19.56667 | 19.56667 | 1197.578 | 212.8588 | 230.5722 |
| 492.2627 | 19.51667 | 19.55 | 1199.297 | 213.1567 | 230.3944 |
| 492.5572 | 19.53333 | 19.58333 | 1201.015 | 213.6946 | 230.8472 |
| 492.8516 | 19.51667 | 19.58333 | 1202.734 | 213.4795 | 230.6583 |
| 493.146 | 19.51667 | 19.58333 | 1204.452 | 212.9085 | 230.075 |
| 493.4404 | 19.5 | 19.65 | 1206.17 | 212.4616 | 230.2778 |
| 493.7348 | 19.6 | 19.55 | 1207.888 | 212.503 | 230.2417 |
| 494.0293 | 19.48333 | 19.51667 | 1209.606 | 213.6974 | 231.4139 |
| 494.3236 | 19.56667 | 19.65 | 1211.324 | 211.9513 | 230.3 |
| 494.618 | 19.6 | 19.6 | 1213.042 | 212.4506 | 229.9083 |
| 494.9124 | 19.55 | 19.55 | 1214.759 | 212.4285 | 230.0722 |
| 495.2068 | 19.53333 | 19.55 | 1216.477 | 212.3789 | 231.0028 |
| 495.5011 | 19.51667 | 19.5 | 1218.194 | 213.2202 | 231.4806 |
| 495.7955 | 19.53333 | 19.58333 | 1219.912 | 213.0823 | 231.9333 |
| 496.0898 | 19.55 | 19.53333 | 1221.629 | 213.8767 | 232.5306 |
| 496.3842 | 19.58333 | 19.58333 | 1223.346 | 212.9444 | 231.4389 |
| 496.6785 | 19.5 | 19.6 | 1225.063 | 211.1293 | 229.9389 |
| 496.9728 | 19.51667 | 19.56667 | 1226.779 | 211.7196 | 230.1 |
| 497.2671 | 19.58333 | 19.56667 | 1228.496 | 212.8478 | 232.0222 |
| 497.5614 | 19.58333 | 19.55 | 1230.213 | 212.1748 | 230.9833 |
| 497.8557 | 19.55 | 19.56667 | 1231.929 | 211.1431 | 230.6167 |
| 498.15 | 19.53333 | 19.55 | 1233.645 | 211.5983 | 231.5278 |
| 498.4443 | 19.58333 | 19.61667 | 1235.362 | 212.183 | 231.8806 |
| 498.7386 | 19.58333 | 19.6 | 1237.078 | 212.1472 | 232.1417 |
| 499.0329 | 19.51667 | 19.58333 | 1238.794 | 212.3265 | 231.6139 |
| 499.3271 | 19.58333 | 19.58333 | 1240.51 | 211.2645 | 231.1 |
| 499.6214 | 19.55 | 19.53333 | 1242.225 | 211.2148 | 231.6333 |
| 499.9156 | 19.48333 | 19.51667 | 1243.941 | 211.4576 | 230.9861 |
| 500.2098 | 19.55 | 19.55 | 1245.656 | 211.3886 | 231.0944 |
| 500.5041 | 19.55 | 19.53333 | 1247.372 | 211.3886 | 231.8917 |
| 500.7983 | 19.55 | 19.58333 | 1249.087 | 211.7224 | 232.0472 |
| 501.0925 | 19.53333 | 19.63333 | 1250.802 | 210.6577 | 230.8611 |
| 501.3867 | 19.51667 | 19.55 | 1252.517 | 209.9294 | 230.7667 |
| 501.6809 | 19.58333 | 19.63333 | 1254.232 | 211.5321 | 232.1306 |
| 501.975 | 19.53333 | 19.6 | 1255.947 | 210.9666 | 231.4111 |
| 502.2692 | 19.55 | 19.53333 | 1257.661 | 209.9736 | 231.4444 |
| 502.5634 | 19.5 | 19.55 | 1259.376 | 210.7625 | 231.7389 |
| 502.8575 | 19.48333 | 19.55 | 1261.09 | 210.0398 | 231.1361 |
| 503.1517 | 19.63333 | 19.53333 | 1262.805 | 209.5129 | 230.7667 |
| 503.4458 | 19.58333 | 19.6 | 1264.519 | 209.935 | 230.6528 |
| 503.74 | 19.55 | 19.58333 | 1266.233 | 209.5405 | 231.2833 |
| 504.0341 | 19.6 | 19.58333 | 1267.947 | 209.4164 | 231.1528 |
| 504.3282 | 19.56667 | 19.6 | 1269.661 | 210.1805 | 231.7694 |
| 504.6223 | 19.51667 | 19.58333 | 1271.374 | 209.375 | 231.8639 |
| 504.9164 | 19.58333 | 19.53333 | 1273.088 | 210.3266 | 232.5639 |
| 505.2105 | 19.56667 | 19.55 | 1274.801 | 211.0273 | 232.475 |
| 505.5046 | 19.51667 | 19.6 | 1276.514 | 209.0688 | 231.1972 |
| 505.7987 | 19.48333 | 19.55 | 1278.228 | 208.5558 | 230.8028 |
| 506.0927 | 19.55 | 19.5 | 1279.941 | 208.2937 | 230 |
| 506.3868 | 19.53333 | 19.58333 | 1281.654 | 208.713 | 230.5278 |
| 506.6808 | 19.56667 | 19.56667 | 1283.367 | 208.3185 | 230.6528 |
| 506.9749 | 19.66667 | 19.56667 | 1285.079 | 208.7571 | 231.2333 |
| 507.2689 | 19.53333 | 19.51667 | 1286.792 | 209.2647 | 231.5083 |
| 507.5629 | 19.51667 | 19.83333 | 1288.504 | 208.7406 | 231.2611 |
| 507.8569 | 19.51667 | 19.6 | 1290.217 | 208.5751 | 231.9417 |
| 508.151 | 19.53333 | 19.55 | 1291.929 | 209.0137 | 232.0667 |
| 508.445 | 19.6 | 19.55 | 1293.641 | 208.1006 | 230.7917 |
| 508.7389 | 19.58333 | 19.58333 | 1295.353 | 208.7544 | 232.0528 |
| 509.0329 | 19.78333 | 19.51667 | 1297.065 | 208.2358 | 231.2639 |
| 509.3269 | 19.58333 | 19.55 | 1298.776 | 206.1642 | 228.8389 |
| 509.6209 | 19.55 | 19.58333 | 1300.488 | 207.4414 | 230.4417 |
| 509.9148 | 19.5 | 19.56667 | 1302.199 | 207.5434 | 230.3417 |
| 510.2088 | 19.58333 | 19.61667 | 1303.911 | 207.7807 | 230.6417 |
| 510.5027 | 19.55 | 19.55 | 1305.622 | 208.8702 | 231.9028 |
| 510.7966 | 19.53333 | 19.58333 | 1307.333 | 209.375 | 232.0222 |
| 511.0906 | 19.58333 | 19.51667 | 1309.044 | 208.5585 | 231.8472 |
| 511.3845 | 19.56667 | 19.51667 | 1310.755 | 208.8895 | 232.0583 |
| 511.6784 | 19.56667 | 19.56667 | 1312.466 | 207.6703 | 231.2972 |
| 511.9723 | 19.56667 | 19.55 | 1314.176 | 206.6966 | 230.1583 |
| 512.2662 | 19.53333 | 19.55 | 1315.887 | 207.5655 | 230.7778 |
| 512.56 | 19.55 | 19.51667 | 1317.597 | 207.5186 | 231.5361 |
| 512.8539 | 19.53333 | 19.5 | 1319.307 | 207.6758 | 231.4889 |
| 513.1478 | 19.53333 | 19.58333 | 1321.017 | 207.72 | 231.4583 |
| 513.4416 | 19.55 | 19.53333 | 1322.727 | 208.3434 | 231.9778 |
| 513.7355 | 19.56667 | 19.55 | 1324.437 | 208.8757 | 233.0611 |
| 514.0293 | 19.51667 | 19.58333 | 1326.147 | 207.7669 | 232.05 |
| 514.3232 | 19.61667 | 19.53333 | 1327.857 | 208.0786 | 232.0139 |
| 514.617 | 19.61667 | 19.6 | 1329.566 | 208.9033 | 233.2194 |
| 514.9108 | 19.53333 | 19.55 | 1331.275 | 207.731 | 231.9444 |
| 515.2046 | 19.55 | 19.5 | 1332.985 | 208.9475 | 233.1778 |
| 515.4984 | 19.58333 | 19.63333 | 1334.694 | 207.9875 | 231.7611 |
| 515.7922 | 19.6 | 19.55 | 1336.403 | 207.2676 | 230.7722 |
| 516.0859 | 19.55 | 19.6 | 1338.112 | 209.1323 | 233.05 |
| 516.3797 | 19.56667 | 19.58333 | 1339.82 | 207.9296 | 231.4528 |
| 516.6735 | 19.51667 | 19.56667 | 1341.529 | 208.6909 | 231.9083 |
| 516.9672 | 19.51667 | 19.55 | 1343.237 | 209.2867 | 232.825 |
| 517.261 | 19.58333 | 19.55 | 1344.946 | 209.4026 | 233.9028 |
| 517.5547 | 19.56667 | 19.55 | 1346.654 | 209.3998 | 233.8778 |
| 517.8484 | 19.56667 | 19.61667 | 1348.362 | 208.2055 | 231.8861 |
| 518.1421 | 19.63333 | 19.63333 | 1350.07 | 208.6164 | 232.6667 |
| 518.4359 | 19.6 | 19.55 | 1351.778 | 210.2356 | 234.3278 |
| 518.7296 | 19.58333 | 19.53333 | 1353.486 | 209.4302 | 233.6222 |
| 519.0232 | 19.56667 | 19.63333 | 1355.193 | 209.1157 | 233.3667 |
| 519.3169 | 19.5 | 19.56667 | 1356.901 | 210.1639 | 234.4806 |
| 519.6106 | 19.53333 | 19.58333 | 1358.608 | 209.8439 | 233.725 |
| 519.9043 | 19.55 | 19.56667 | 1360.315 | 210.0012 | 233.6778 |
| 520.1979 | 19.58333 | 19.63333 | 1362.022 | 210.5611 | 234.1889 |
| 520.4916 | 19.68333 | 19.55 | 1363.729 | 211.5017 | 234.7667 |
| 520.7852 | 19.51667 | 19.5 | 1365.436 | 210.5501 | 233.9806 |
| 521.0788 | 19.5 | 19.53333 | 1367.143 | 211.2535 | 235.0556 |
| 521.3725 | 19.55 | 19.51667 | 1368.849 | 211.3969 | 234.7694 |
| 521.6661 | 19.58333 | 19.6 | 1370.556 | 209.7557 | 232.825 |
| 521.9597 | 19.55 | 19.65 | 1372.262 | 210.7073 | 233.7389 |
| 522.2533 | 19.56667 | 19.55 | 1373.968 | 210.9666 | 234.1111 |
| 522.5469 | 19.51667 | 19.51667 | 1375.674 | 210.5942 | 233.9222 |
| 522.8404 | 19.55 | 19.58333 | 1377.38 | 210.8342 | 234.3306 |
| 523.134 | 19.51667 | 19.56667 | 1379.086 | 211.4686 | 234.8528 |
| 523.4276 | 19.6 | 19.55 | 1380.792 | 210.6273 | 234.0028 |
| 523.7211 | 19.55 | 19.5 | 1382.497 | 211.7665 | 235.0056 |
| 524.0147 | 19.51667 | 19.6 | 1384.203 | 212.6437 | 235.3028 |
| 524.3082 | 19.51667 | 19.61667 | 1385.908 | 211.6893 | 234.0806 |
| 524.6017 | 19.55 | 19.58333 | 1387.613 | 211.7251 | 234.3028 |
| 524.8952 | 19.6 | 19.61667 | 1389.318 | 210.6797 | 233.4083 |
| 525.1887 | 19.55 | 19.55 | 1391.023 | 210.2715 | 232.4083 |
| 525.4822 | 19.51667 | 19.56667 | 1392.728 | 210.6659 | 232.9917 |
| 525.7757 | 19.51667 | 19.56667 | 1394.432 | 211.7803 | 234.2444 |
| 526.0692 | 19.5 | 19.55 | 1396.137 | 212.6023 | 235.3694 |
| 526.3627 | 19.55 | 19.51667 | 1397.841 | 212.2437 | 234.6694 |
| 526.6561 | 19.58333 | 19.53333 | 1399.545 | 211.7196 | 234.35 |
| 526.9496 | 19.5 | 19.6 | 1401.25 | 210.5528 | 233.5778 |
| 527.243 | 19.55 | 19.58333 | 1402.954 | 212.2134 | 233.8389 |
| 527.5365 | 19.58333 | 19.56667 | 1404.657 | 211.2176 | 233.2 |
| 527.8299 | 19.6 | 19.5 | 1406.361 | 210.6383 | 232.9333 |
| 528.1233 | 19.55 | 19.51667 | 1408.065 | 211.3224 | 233.375 |
| 528.4167 | 19.55 | 19.61667 | 1409.768 | 211.2259 | 233.9917 |
| 528.7101 | 19.51667 | 19.53333 | 1411.471 | 212.183 | 234.8639 |
| 529.0035 | 19.55 | 19.53333 | 1413.175 | 212.7871 | 235.2639 |
| 529.2969 | 19.53333 | 19.53333 | 1414.878 | 213.1374 | 235.7444 |
| 529.5903 | 19.53333 | 19.58333 | 1416.581 | 212.1086 | 234.9889 |
| 529.8836 | 19.55 | 19.53333 | 1418.284 | 211.8438 | 234.7 |
| 530.177 | 19.51667 | 19.58333 | 1419.986 | 212.9857 | 236.05 |
| 530.4703 | 19.55 | 19.56667 | 1421.689 | 212.6685 | 236.1944 |
| 530.7637 | 19.5 | 19.55 | 1423.391 | 211.27 | 234.2194 |
| 531.057 | 19.53333 | 19.63333 | 1425.093 | 211.9541 | 235.1833 |
| 531.3503 | 19.51667 | 19.55 | 1426.796 | 211.8658 | 235.375 |
| 531.6436 | 19.55 | 19.63333 | 1428.498 | 212.0865 | 235.0667 |
| 531.9369 | 19.51667 | 19.55 | 1430.2 | 211.8989 | 234.7667 |
| 532.2302 | 19.61667 | 19.56667 | 1431.901 | 212.1858 | 235.1778 |
| 532.5235 | 19.51667 | 19.51667 | 1433.603 | 214.5194 | 237.2333 |
| 532.8168 | 19.55 | 19.58333 | 1435.304 | 213.3498 | 236.6417 |
| 533.1101 | 19.53333 | 19.55 | 1437.006 | 212.2934 | 236.0861 |
| 533.4033 | 19.53333 | 19.56667 | 1438.707 | 213.4464 | 236.5111 |
| 533.6966 | 19.55 | 19.56667 | 1440.408 | 211.7748 | 235.45 |
| 533.9898 | 19.5 | 19.6 | 1442.109 | 212.1941 | 236.1278 |
| 534.283 | 19.56667 | 19.56667 | 1443.81 | 212.5692 | 236.2333 |
| 534.5762 | 19.55 | 19.55 | 1445.511 | 210.3873 | 234.15 |
| 534.8695 | 19.55 | 19.55 | 1447.211 | 211.83 | 235.5972 |
| 535.1627 | 19.51667 | 19.51667 | 1448.912 | 212.0782 | 235.8806 |
| 535.4558 | 19.55 | 19.56667 | 1450.612 | 211.5955 | 235.8083 |
| 535.749 | 19.55 | 19.53333 | 1452.312 | 214.1111 | 238.1 |
| 536.0422 | 19.56667 | 19.63333 | 1454.012 | 213.6422 | 237.6833 |
| 536.3354 | 19.55 | 19.63333 | 1455.712 | 212.4782 | 236.8583 |
| 536.6285 | 19.58333 | 19.58333 | 1457.412 | 214.2187 | 238.3139 |
| 536.9217 | 19.61667 | 19.58333 | 1459.111 | 212.1775 | 236.5611 |
| 537.2148 | 19.5 | 19.56667 | 1460.811 | 211.1597 | 235.7583 |
| 537.5079 | 19.51667 | 19.63333 | 1462.51 | 211.3914 | 235.8083 |
| 537.8011 | 19.51667 | 19.53333 | 1464.209 | 209.8632 | 234.4667 |
| 538.0942 | 19.53333 | 19.56667 | 1465.909 | 211.3362 | 235.8972 |
| 538.3873 | 19.55 | 19.58333 | 1467.607 | 212.3485 | 237.1056 |
| 538.6804 | 19.53333 | 19.58333 | 1469.306 | 212.7733 | 236.9722 |
| 538.9735 | 19.61667 | 19.55 | 1471.005 | 212.241 | 236.125 |
| 539.2665 | 19.58333 | 19.53333 | 1472.704 | 211.681 | 236.5139 |
| 539.5596 | 19.55 | 19.6 | 1474.402 | 212.3844 | 237.6667 |
| 539.8526 | 19.6 | 19.56667 | 1476.1 | 212.2548 | 236.8056 |
| 540.1457 | 19.55 | 19.56667 | 1477.798 | 213.063 | 237.2361 |
| 540.4387 | 19.53333 | 19.53333 | 1479.496 | 213.0271 | 237.7472 |
| 540.7318 | 19.5 | 19.53333 | 1481.194 | 211.008 | 236.1194 |
| 541.0248 | 19.51667 | 19.53333 | 1482.892 | 210.928 | 235.6556 |
| 541.3178 | 19.5 | 19.58333 | 1484.59 | 212.5996 | 237.15 |
| 541.6108 | 19.61667 | 19.58333 | 1486.287 | 213.0685 | 237.7083 |
| 541.9038 | 19.53333 | 19.56667 | 1487.984 | 212.3292 | 237.325 |
| 542.1968 | 19.56667 | 19.56667 | 1489.682 | 212.6851 | 237.4833 |
| 542.4897 | 19.53333 | 19.55 | 1491.379 | 213.6864 | 238.2389 |
| 542.7827 | 19.53333 | 19.55 | 1493.076 | 212.8975 | 237.5972 |
| 543.0756 | 19.55 | 19.53333 | 1494.772 | 213.5181 | 237.8417 |
| 543.3686 | 19.53333 | 19.58333 | 1496.469 | 213.0492 | 236.6861 |
| 543.6615 | 19.51667 | 19.6 | 1498.166 | 212.7016 | 236.2111 |
| 543.9544 | 19.53333 | 19.5 | 1499.862 | 211.9596 | 236.1389 |
| 544.2474 | 19.53333 | 19.55 | 1501.558 | 211.6396 | 235.95 |
| 544.5403 | 19.58333 | 19.55 | 1503.254 | 212.7623 | 236.8806 |
| 544.8332 | 19.51667 | 19.51667 | 1504.95 | 213.5567 | 237.6167 |
| 545.1261 | 19.56667 | 19.56667 | 1506.646 | 213.4547 | 238.3278 |
| 545.4189 | 19.55 | 19.53333 | 1508.342 | 211.7941 | 236.1694 |
| 545.7118 | 19.53333 | 19.55 | 1510.038 | 211.5183 | 235.0861 |
| 546.0047 | 19.51667 | 19.56667 | 1511.733 | 212.3513 | 236.1056 |
| 546.2975 | 19.51667 | 19.56667 | 1513.428 | 212.7209 | 236.6222 |
| 546.5904 | 19.5 | 19.6 | 1515.123 | 213.256 | 236.7361 |
| 546.8832 | 19.48333 | 19.61667 | 1516.818 | 213.0712 | 236.6417 |
| 547.176 | 19.55 | 19.55 | 1518.513 | 211.681 | 235.9 |
| 547.4688 | 19.56667 | 19.61667 | 1520.208 | 212.241 | 235.9056 |
| 547.7616 | 19.51667 | 19.5 | 1521.903 | 213.5926 | 237.0444 |
| 548.0544 | 19.51667 | 19.63333 | 1523.597 | 212.9609 | 236.5472 |
| 548.3472 | 19.53333 | 19.5 | 1525.291 | 212.3651 | 235.4 |
| 548.64 | 19.51667 | 19.5 | 1526.986 | 212.5223 | 235.4611 |
| 548.9328 | 19.55 | 19.53333 | 1528.68 | 211.7941 | 235.1861 |
| 549.2255 | 19.51667 | 19.6 | 1530.374 | 211.5072 | 235.3139 |
| 549.5183 | 19.53333 | 19.56667 | 1532.067 | 212.7016 | 236.3833 |
| 549.811 | 19.53333 | 19.55 | 1533.761 | 213.3471 | 236.3917 |
| 550.1037 | 19.55 | 19.56667 | 1535.455 | 211.7776 | 235.1333 |
| 550.3964 | 19.56667 | 19.56667 | 1537.148 | 210.7073 | 233.4 |
| 550.6891 | 19.5 | 19.55 | 1538.841 | 211.3886 | 234.3889 |
| 550.9818 | 19.51667 | 19.58333 | 1540.534 | 212.4258 | 236.0667 |
| 551.2745 | 19.51667 | 19.55 | 1542.227 | 213.0299 | 236.1139 |
| 551.5672 | 19.55 | 19.5 | 1543.92 | 213.0492 | 236.1472 |
| 551.8599 | 19.56667 | 19.58333 | 1545.613 | 213.7581 | 236.7417 |
| 552.1526 | 19.56667 | 19.53333 | 1547.305 | 213.9981 | 237.2861 |
| 552.4452 | 19.56667 | 19.65 | 1548.998 | 213.3995 | 236.1722 |
| 552.7378 | 19.51667 | 19.51667 | 1550.69 | 212.9609 | 235.55 |
| 553.0305 | 19.51667 | 19.53333 | 1552.382 | 212.7044 | 235.7611 |
| 553.3231 | 19.51667 | 19.55 | 1554.074 | 211.6838 | 234.9639 |
| 553.6157 | 19.51667 | 19.56667 | 1555.766 | 210.9942 | 233.8611 |
| 553.9083 | 19.53333 | 19.56667 | 1557.457 | 211.1514 | 233.7917 |
| 554.2009 | 19.56667 | 19.68333 | 1559.149 | 212.7209 | 235.225 |
| 554.4935 | 19.53333 | 19.56667 | 1560.84 | 212.6961 | 235.3222 |
| 554.7861 | 19.53333 | 19.51667 | 1562.532 | 212.8781 | 235.3278 |
| 555.0786 | 19.56667 | 19.53333 | 1564.223 | 213.9622 | 234.9444 |
| 555.3712 | 19.53333 | 19.61667 | 1565.914 | 213.7277 | 235.3139 |
| 555.6637 | 19.6 | 19.53333 | 1567.605 | 212.8726 | 235.3194 |
| 555.9563 | 19.48333 | 19.48333 | 1569.295 | 212.9195 | 234.5417 |
| 556.2488 | 19.56667 | 19.53333 | 1570.986 | 214.5635 | 235.9333 |
| 556.5413 | 19.51667 | 19.5 | 1572.676 | 212.9499 | 233.8611 |
| 556.8338 | 19.51667 | 19.51667 | 1574.367 | 211.5872 | 232.2389 |
| 557.1263 | 19.56667 | 19.53333 | 1576.057 | 213.6146 | 233.5583 |
| 557.4188 | 19.55 | 19.61667 | 1577.747 | 213.7415 | 232.9722 |
| 557.7113 | 19.51667 | 19.53333 | 1579.437 | 213.0878 | 232.3056 |
| 558.0038 | 19.51667 | 19.56667 | 1581.126 | 213.9346 | 232.5472 |
| 558.2962 | 19.51667 | 19.56667 | 1582.816 | 214.2491 | 232.1944 |
| 558.5887 | 19.55 | 19.58333 | 1584.505 | 214.0698 | 231.4972 |
| 558.8811 | 19.53333 | 19.55 | 1586.195 | 213.8215 | 231.4361 |
| 559.1736 | 19.53333 | 19.58333 | 1587.884 | 213.8988 | 230.8167 |
| 559.466 | 19.53333 | 19.53333 | 1589.573 | 213.6008 | 229.4389 |
| 559.7584 | 19.55 | 19.58333 | 1591.262 | 213.5126 | 229.1167 |
| 560.0508 | 19.56667 | 19.61667 | 1592.95 | 213.9374 | 228.9 |
| 560.3432 | 19.55 | 19.55 | 1594.639 | 213.7746 | 228.4139 |
| 560.6356 | 19.48333 | 19.5 | 1596.327 | 214.0725 | 227.5222 |
| 560.9279 | 19.5 | 19.55 | 1598.016 | 213.554 | 226.3694 |
| 561.2203 | 19.6 | 19.55 | 1599.704 | 213.4353 | 226.3917 |
| 561.5127 | 19.61667 | 19.55 | 1601.392 | 215.4876 | 227.8667 |
| 561.805 | 19.55 | 19.55 | 1603.08 | 214.5525 | 226.0889 |
| 562.0973 | 19.55 | 19.56667 | 1604.767 | 213.4988 | 224.775 |
| 562.3897 | 19.56667 | 19.55 | 1606.455 | 214.5111 | 224.95 |
| 562.682 | 19.51667 | 19.6 | 1608.142 | 213.3581 | 223.0639 |
| 562.9743 | 19.55 | 19.58333 | 1609.83 | 213.7443 | 223.0861 |
| 563.2666 | 19.58333 | 19.55 | 1611.517 | 214.0339 | 222.7972 |
| 563.5589 | 19.55 | 19.55 | 1613.204 | 213.3774 | 221.275 |
| 563.8511 | 19.56667 | 19.56667 | 1614.891 | 212.8119 | 220.8 |
| 564.1434 | 19.56667 | 19.5 | 1616.577 | 214.7538 | 221.7667 |
| 564.4357 | 19.46667 | 19.55 | 1618.264 | 214.878 | 221.3889 |
| 564.7279 | 19.51667 | 19.53333 | 1619.95 | 214.4311 | 220.9639 |
| 565.0201 | 19.58333 | 19.6 | 1621.637 | 214.8642 | 220.4389 |
| 565.3124 | 19.55 | 19.5 | 1623.323 | 214.2353 | 219.1028 |
| 565.6046 | 19.53333 | 19.63333 | 1625.009 | 214.9966 | 219.0722 |
| 565.8968 | 19.53333 | 19.61667 | 1626.695 | 214.1194 | 218.7083 |
| 566.189 | 19.53333 | 19.53333 | 1628.38 | 213.3057 | 217.4028 |
| 566.4812 | 19.56667 | 19.53333 | 1630.066 | 214.0367 | 216.8361 |
| 566.7733 | 19.61667 | 19.56667 | 1631.751 | 213.9732 | 216.9444 |
| 567.0655 | 19.63333 | 19.51667 | 1633.437 | 214.1332 | 216.7528 |
| 567.3577 | 19.6 | 19.5 | 1635.122 | 214.2022 | 216.2528 |
| 567.6498 | 19.6 | 19.61667 | 1636.807 | 213.7195 | 215.925 |
| 567.9419 | 19.51667 | 19.5 | 1638.492 | 213.9346 | 216.1722 |
| 568.2341 | 19.51667 | 19.6 | 1640.176 | 214.5249 | 216.3083 |
| 568.5262 | 19.51667 | 19.55 | 1641.861 | 214.1277 | 214.8 |
| 568.8183 | 19.51667 | 19.53333 | 1643.545 | 214.6214 | 215.0667 |
| 569.1104 | 19.58333 | 19.51667 | 1645.23 | 213.6698 | 213.8528 |
| 569.4025 | 19.5 | 19.56667 | 1646.914 | 212.6933 | 213.0028 |
| 569.6946 | 19.5 | 19.56667 | 1648.598 | 213.3747 | 213.9028 |
| 569.9866 | 19.56667 | 19.53333 | 1650.282 | 213.9346 | 214.3111 |
| 570.2787 | 19.53333 | 19.55 | 1651.965 | 214.4808 | 214.9111 |
| 570.5707 | 19.51667 | 19.53333 | 1653.649 | 214.387 | 214.0417 |
| 570.8628 | 19.61667 | 19.58333 | 1655.332 | 215.1979 | 215.4806 |
| 571.1548 | 19.53333 | 19.55 | 1657.015 | 215.3772 | 215.6806 |
| 571.4468 | 19.53333 | 19.63333 | 1658.698 | 214.0201 | 214.6611 |
| 571.7388 | 19.51667 | 19.56667 | 1660.381 | 214.649 | 214.9306 |
| 572.0308 | 19.56667 | 19.53333 | 1662.064 | 214.9662 | 214.8667 |
| 572.3228 | 19.56667 | 19.58333 | 1663.747 | 213.325 | 213.9111 |
| 572.6148 | 19.55 | 19.56667 | 1665.429 | 214.307 | 214.0111 |
| 572.9067 | 19.65 | 19.51667 | 1667.112 | 214.6325 | 214.1056 |
| 573.1987 | 19.55 | 19.51667 | 1668.794 | 215.0986 | 214.75 |
| 573.4906 | 19.51667 | 19.63333 | 1670.476 | 214.4587 | 214.3389 |
| 573.7826 | 19.51667 | 19.55 | 1672.158 | 213.9732 | 213.9972 |
| 574.0745 | 19.56667 | 19.53333 | 1673.84 | 215.6007 | 215.6917 |
| 574.3664 | 19.61667 | 19.51667 | 1675.521 | 215.1428 | 215.4722 |
| 574.6583 | 19.51667 | 19.55 | 1677.203 | 215.4214 | 215.375 |
| 574.9502 | 19.56667 | 19.53333 | 1678.884 | 214.5911 | 214.7111 |
| 575.2421 | 19.56667 | 19.6 | 1680.565 | 214.3511 | 214.75 |
| 575.534 | 19.51667 | 19.58333 | 1682.246 | 214.6242 | 214.6861 |
| 575.8259 | 19.51667 | 19.58333 | 1683.927 | 214.3897 | 214.7222 |
| 576.1177 | 19.51667 | 19.55 | 1685.608 | 214.9 | 214.7944 |
| 576.4096 | 19.56667 | 19.58333 | 1687.289 | 216.3482 | 216.3056 |
| 576.7014 | 19.55 | 19.63333 | 1688.969 | 216.7151 | 217.1361 |
| 576.9932 | 19.58333 | 19.53333 | 1690.649 | 216.8144 | 216.3306 |
| 577.285 | 19.53333 | 19.58333 | 1692.329 | 216.6847 | 216.5139 |
| 577.5768 | 19.55 | 19.58333 | 1694.009 | 216.8971 | 217.0472 |
| 577.8686 | 19.51667 | 19.51667 | 1695.689 | 217.0047 | 217 |
| 578.1604 | 19.56667 | 19.51667 | 1697.369 | 217.2667 | 217.4611 |
| 578.4522 | 19.5 | 19.56667 | 1699.049 | 218.2046 | 218.1944 |
| 578.744 | 19.56667 | 19.58333 | 1700.728 | 217.1895 | 217.4639 |
| 579.0357 | 19.51667 | 19.58333 | 1702.407 | 218.3342 | 218.8611 |
| 579.3275 | 19.55 | 19.58333 | 1704.086 | 217.9563 | 218.3889 |
| 579.6192 | 19.5 | 19.56667 | 1705.765 | 217.3164 | 217.0056 |
| 579.9109 | 19.56667 | 19.56667 | 1707.444 | 218.3673 | 218.0278 |
| 580.2026 | 19.51667 | 19.56667 | 1709.123 | 218.7728 | 218.6083 |
| 580.4943 | 19.5 | 19.55 | 1710.801 | 219.501 | 219.0639 |
| 580.786 | 19.56667 | 19.55 | 1712.48 | 218.4584 | 218.4972 |
| 581.0777 | 19.55 | 19.6 | 1714.158 | 218.2901 | 218.45 |
| 581.3694 | 19.53333 | 19.61667 | 1715.836 | 218.0611 | 218.1028 |
| 581.661 | 19.51667 | 19.55 | 1717.514 | 217.7164 | 217.7389 |
| 581.9527 | 19.53333 | 19.55 | 1719.191 | 218.1163 | 218.2028 |
| 582.2443 | 19.58333 | 19.58333 | 1720.869 | 218.668 | 218.4472 |
| 582.536 | 19.53333 | 19.51667 | 1722.547 | 220.061 | 219.6556 |
| 582.8276 | 19.51667 | 19.51667 | 1724.224 | 220.1934 | 219.9 |
| 583.1192 | 19.56667 | 19.56667 | 1725.901 | 219.5617 | 219.3 |
| 583.4108 | 19.56667 | 19.53333 | 1727.578 | 220.6044 | 220.7444 |
| 583.7024 | 19.48333 | 19.6 | 1729.255 | 220.0279 | 219.8667 |
| 583.994 | 19.55 | 19.58333 | 1730.932 | 220.392 | 220.3278 |
| 584.2855 | 19.56667 | 19.55 | 1732.608 | 221.1588 | 221.1111 |
| 584.5771 | 19.51667 | 19.53333 | 1734.285 | 220.643 | 220.7194 |
| 584.8686 | 19.55 | 19.56667 | 1735.961 | 220.7037 | 220.5333 |
| 585.1602 | 19.56667 | 19.58333 | 1737.637 | 221.3216 | 220.7611 |
| 585.4517 | 19.55 | 19.55 | 1739.313 | 220.8995 | 220.8194 |
| 585.7432 | 19.56667 | 19.55 | 1740.989 | 220.654 | 220.7222 |
| 586.0347 | 19.51667 | 19.6 | 1742.664 | 222.7201 | 222.6694 |
| 586.3262 | 19.48333 | 19.6 | 1744.34 | 223.2635 | 223.1722 |
| 586.6177 | 19.5 | 19.58333 | 1746.015 | 221.6553 | 221.6417 |
| 586.9092 | 19.51667 | 19.58333 | 1747.69 | 223.4303 | 223.3667 |
| 587.2007 | 19.55 | 19.56667 | 1749.365 | 225.9943 | 226.5333 |
| 587.4921 | 19.51667 | 19.55 | 1751.04 | 220.7478 | 221.1167 |
| 587.7836 | 19.51667 | 19.65 | 1752.715 | 226.0439 | 225.85 |
| 588.075 | 19.55 | 19.58333 | | | |
| 588.3664 | 19.56667 | 19.58333 | | | |
| 588.6578 | 19.51667 | 19.65 | | | |
| 588.9492 | 19.5 | 19.58333 | | | |
| 589.2406 | 19.61667 | 19.58333 | | | |
| 589.532 | 19.46667 | 19.61667 | | | |
| 589.8234 | 19.55 | 19.6 | | | |
| 590.1147 | 19.55 | 19.68333 | | | |
| 590.4061 | 19.55 | 19.73333 | | | |
| 590.6974 | 19.51667 | 19.71667 | | | |
| 590.9888 | 19.5 | 19.8 | | | |
| 591.2801 | 19.56667 | 19.83333 | | | |
| 591.5714 | 19.51667 | 19.85 | | | |
| 591.8627 | 19.51667 | 19.91667 | | | |
| 592.154 | 19.51667 | 19.9 | | | |
| 592.4453 | 19.51667 | 20.13333 | | | |
| 592.7365 | 19.56667 | 20.03333 | | | |
| 593.0278 | 19.58333 | 20.21667 | | | |
| 593.319 | 19.58333 | 20.11667 | | | |
| 593.6103 | 19.55 | 20.18333 | | | |
| 593.9015 | 19.61667 | 20.3 | | | |
| 594.1927 | 19.51667 | 20.58333 | | | |
| 594.4839 | 19.53333 | 20.36667 | | | |
| 594.7751 | 19.61667 | 20.45 | | | |
| 595.0663 | 19.56667 | 20.28333 | | | |
| 595.3574 | 19.51667 | 20.53333 | | | |
| 595.6486 | 19.66667 | 20.4 | | | |
| 595.9398 | 19.55 | 20.75 | | | |
| 596.2309 | 19.51667 | 20.85 | | | |
| 596.522 | 19.61667 | 20.76667 | | | |
| 596.8131 | 19.55 | 20.78333 | | | |
| 597.1043 | 19.46667 | 20.88333 | | | |
| 597.3954 | 19.56667 | 21.05 | | | |
| 597.6864 | 19.55 | 21.06667 | | | |
| 597.9775 | 19.53333 | 21.15 | | | |
| 598.2686 | 19.5 | 21.23333 | | | |
| 598.5596 | 19.5 | 21.21667 | | | |
| 598.8507 | 19.53333 | 21.33333 | | | |
| 599.1417 | 19.55 | 21.5 | | | |
| 599.4327 | 19.55 | 21.51667 | | | |
| 599.7237 | 19.58333 | 21.53333 | | | |
| 600.0147 | 19.53333 | 21.66667 | | | |
| 600.3057 | 19.5 | 21.56667 | | | |
| 600.5967 | 19.55 | 21.71667 | | | |
| 600.8877 | 19.58333 | 21.95 | | | |
| 601.1786 | 19.56667 | 21.88333 | | | |
| 601.4696 | 19.51667 | 22 | | | |
| 601.7605 | 19.51667 | 21.98333 | | | |
| 602.0515 | 19.56667 | 22.16667 | | | |

| | | |
|---|---|---|
| 602.3424 | 19.58333 | 22.3 |
| 602.6333 | 19.58333 | 22.23333 |
| 602.9242 | 19.63333 | 22.28333 |
| 603.215 | 19.48333 | 22.6 |
| 603.5059 | 19.53333 | 22.58333 |
| 603.7968 | 19.56667 | 22.68333 |
| 604.0876 | 19.56667 | 22.6 |
| 604.3785 | 19.51667 | 22.7 |
| 604.6693 | 19.61667 | 22.81667 |
| 604.9601 | 19.55 | 23.01667 |
| 605.2509 | 19.58333 | 22.91667 |
| 605.5417 | 19.53333 | 23.18333 |
| 605.8325 | 19.55 | 23.16667 |
| 606.1233 | 19.56667 | 23.03333 |
| 606.414 | 19.5 | 23.05 |
| 606.7048 | 19.6 | 23.43333 |
| 606.9955 | 19.55 | 23.41667 |
| 607.2863 | 19.51667 | 23.26667 |
| 607.577 | 19.51667 | 23.4 |
| 607.8677 | 19.58333 | 23.31667 |
| 608.1584 | 19.53333 | 23.8 |
| 608.4491 | 19.55 | 23.63333 |
| 608.7398 | 19.51667 | 23.63333 |
| 609.0304 | 19.55 | 23.81667 |
| 609.3211 | 19.6 | 23.73333 |
| 609.6117 | 19.51667 | 23.76667 |
| 609.9023 | 19.58333 | 23.93333 |
| 610.193 | 19.53333 | 24 |
| 610.4836 | 19.6 | 23.91667 |
| 610.7742 | 19.58333 | 24.05 |
| 611.0648 | 19.55 | 24.03333 |
| 611.3553 | 19.53333 | 24.06667 |
| 611.6459 | 19.53333 | 24.3 |
| 611.9365 | 19.55 | 24.23333 |
| 612.227 | 19.56667 | 24.1 |
| 612.5175 | 19.53333 | 24.31667 |
| 612.8081 | 19.55 | 24.18333 |
| 613.0986 | 19.53333 | 24.11667 |
| 613.3891 | 19.56667 | 24.35 |
| 613.6796 | 19.51667 | 24.18333 |
| 613.97 | 19.6 | 24.3 |
| 614.2605 | 19.53333 | 24.43333 |
| 614.551 | 19.55 | 24.23333 |
| 614.8414 | 19.51667 | 24.33333 |
| 615.1318 | 19.51667 | 24.3 |
| 615.4223 | 19.56667 | 24.58333 |
| 615.7127 | 19.55 | 24.41667 |
| 616.0031 | 19.53333 | 24.4 |
| 616.2935 | 19.53333 | 24.65 |
| 616.5838 | 19.53333 | 24.3 |
| 616.8742 | 19.56667 | 24.53333 |
| 617.1646 | 19.6 | 24.41667 |
| 617.4549 | 19.55 | 24.78333 |
| 617.7452 | 19.56667 | 24.73333 |
| 618.0356 | 19.48333 | 24.51667 |
| 618.3259 | 19.6 | 24.61667 |
| 618.6162 | 19.55 | 24.66667 |
| 618.9065 | 19.58333 | 24.85 |
| 619.1967 | 19.56667 | 24.75 |
| 619.487 | 19.56667 | 24.81667 |
| 619.7773 | 19.53333 | 24.93333 |
| 620.0675 | 19.48333 | 24.8 |
| 620.3577 | 19.51667 | 24.61667 |
| 620.648 | 19.58333 | 25.15 |
| 620.9382 | 19.53333 | 24.81667 |
| 621.2284 | 19.58333 | 24.68333 |
| 621.5186 | 19.53333 | 24.9 |
| 621.8087 | 19.6 | 24.75 |
| 622.0989 | 19.55 | 24.86667 |
| 622.389 | 19.53333 | 24.68333 |
| 622.6792 | 19.51667 | 25.15 |
| 622.9693 | 19.58333 | 24.88333 |
| 623.2594 | 19.51667 | 25.05 |
| 623.5495 | 19.58333 | 25.38333 |
| 623.8396 | 19.56667 | 25.06667 |
| 624.1297 | 19.55 | 25.43333 |
| 624.4198 | 19.53333 | 24.95 |
| 624.7099 | 19.53333 | 25.15 |
| 624.9999 | 19.53333 | 25.23333 |
| 625.29 | 19.56667 | 25.25 |
| 625.58 | 19.55 | 25.16667 |
| 625.87 | 19.61667 | 25.28333 |
| 626.16 | 19.51667 | 25.2 |
| 626.45 | 19.53333 | 25.18333 |
| 626.74 | 19.56667 | 25.46667 |
| 627.03 | 19.51667 | 25.5 |
| 627.3199 | 19.51667 | 25.31667 |
| 627.6099 | 19.6 | 25.61667 |
| 627.8998 | 19.53333 | 25.38333 |
| 628.1897 | 19.55 | 25.98333 |
| 628.4797 | 19.51667 | 25.71667 |
| 628.7696 | 19.58333 | 25.65 |
| 629.0595 | 19.5 | 25.43333 |
| 629.3493 | 19.46667 | 25.53333 |
| 629.6392 | 19.51667 | 25.83333 |
| 629.9291 | 19.55 | 25.61667 |
| 630.2189 | 19.55 | 25.63333 |
| 630.5087 | 19.56667 | 25.68333 |
| 630.7986 | 19.51667 | 25.51667 |
| 631.0884 | 19.56667 | 25.68333 |
| 631.3782 | 19.53333 | 25.78333 |
| 631.668 | 19.55 | 26.18333 |
| 631.9577 | 19.61667 | 26.05 |
| 632.2475 | 19.56667 | 25.98333 |
| 632.5373 | 19.51667 | 25.98333 |
| 632.827 | 19.6 | 25.93333 |
| 633.1167 | 19.55 | 26.1 |
| 633.4065 | 19.6 | 26.1 |
| 633.6962 | 19.55 | 26.05 |
| 633.9859 | 19.51667 | 25.98333 |
| 634.2756 | 19.58333 | 26.11667 |
| 634.5652 | 19.56667 | 25.98333 |
| 634.8549 | 19.55 | 26.33333 |
| 635.1445 | 19.51667 | 25.95 |
| 635.4342 | 19.51667 | 26.31667 |
| 635.7238 | 19.51667 | 26.38333 |
| 636.0134 | 19.56667 | 26.3 |
| 636.303 | 19.56667 | 26.35 |
| 636.5926 | 19.53333 | 26.58333 |
| 636.8822 | 19.51667 | 26.56667 |
| 637.1718 | 19.58333 | 26.31667 |
| 637.4613 | 19.58333 | 26.45 |
| 637.7509 | 19.55 | 26.66667 |
| 638.0404 | 19.53333 | 26.55 |
| 638.3299 | 19.63333 | |
| 638.6194 | 19.6 | 26.41667 |
| 638.9089 | 19.55 | 26.66667 |
| 639.1984 | 19.56667 | 26.61667 |
| 639.4879 | 19.56667 | 26.61667 |
| 639.7774 | | 26.7 |
| 640.0668 | 19.65 | 26.93333 |
| 640.3563 | 19.58333 | 26.63333 |
| 640.6457 | 19.56667 | 26.98333 |
| 640.9351 | 19.51667 | 26.61667 |
| 641.2245 | 19.63333 | 26.53333 |
| 641.5139 | 19.58333 | 27 |
| 641.8033 | 19.56667 | 26.71667 |
| 642.0927 | 19.58333 | 26.43333 |
| 642.382 | 19.6 | 26.95 |
| 642.6714 | 19.56667 | 27.05 |
| 642.9607 | 19.53333 | 26.58333 |
| 643.25 | 19.58333 | 26.85 |
| 643.5393 | 19.56667 | 27.05 |
| 643.8286 | 19.51667 | 26.98333 |
| 644.1179 | 19.56667 | 27.43333 |
| 644.4072 | 19.56667 | 27.23333 |
| 644.6965 | 19.48333 | 27.06667 |
| 644.9857 | 19.51667 | 27.33333 |
| 645.2749 | 19.55 | 27.21667 |
| 645.5642 | 19.51667 | 27 |
| 645.8534 | 19.58333 | 26.98333 |
| 646.1426 | 19.53333 | 27.81667 |
| 646.4318 | 19.58333 | 27.26667 |
| 646.721 | 19.56667 | 27.25 |
| 647.0101 | 19.58333 | 27.1 |
| 647.2993 | 19.55 | 26.81667 |
| 647.5884 | 19.6 | 27.48333 |
| 647.8776 | 19.51667 | 27.5 |
| 648.1667 | 19.51667 | 27.55 |
| 648.4558 | 19.55 | 27.35 |
| 648.7449 | 19.55 | 27.56667 |
| 649.034 | 19.58333 | 27.7 |
| 649.3231 | 19.51667 | 27.3 |
| 649.6121 | 19.53333 | 27.58333 |
| 649.9012 | 19.56667 | 27.45 |
| 650.1902 | 19.51667 | 27.66667 |
| 650.4792 | 19.58333 | 27.81667 |
| 650.7683 | 19.58333 | 27.68333 |
| 651.0573 | 19.6 | 27.5 |
| 651.3462 | 19.58333 | 27.58333 |
| 651.6352 | 19.55 | 27.61667 |
| 651.9242 | 19.56667 | 27.81667 |
| 652.2131 | 19.56667 | 28.18333 |
| 652.5021 | 19.55 | 27.9 |
| 652.791 | 19.5 | 27.61667 |
| 653.0799 | 19.55 | 28.18333 |
| 653.3688 | 19.53333 | 28.18333 |
| 653.6577 | 19.5 | 27.76667 |
| 653.9466 | 19.55 | 28.31667 |
| 654.2355 | 19.61667 | 27.91667 |
| 654.5243 | 19.63333 | 28.21667 |
| 654.8132 | 19.56667 | 28.25 |
| 655.102 | 19.63333 | 28.3 |
| 655.3908 | 19.65 | 28.15 |
| 655.6797 | 19.58333 | 28.16667 |
| 655.9685 | 19.51667 | 28.41667 |
| 656.2572 | 19.58333 | 28.55 |
| 656.546 | 19.58333 | 28.23333 |
| 656.8348 | 19.61667 | 28.51667 |
| 657.1235 | 19.6 | 28.4 |
| 657.4123 | 19.6 | 28.43333 |
| 657.701 | 19.56667 | 28.43333 |
| 657.9897 | 19.6 | 28.73333 |
| 658.2784 | 19.56667 | 28.73333 |
| 658.5671 | 19.6 | 28.83333 |
| 658.8558 | 19.6 | 28.38333 |
| 659.1444 | 19.61667 | 29.05 |
| 659.4331 | 19.55 | 28.98333 |
| 659.7217 | 19.6 | 28.86667 |
| 660.0103 | 19.58333 | 28.83333 |
| 660.299 | 19.6 | 28.71667 |
| 660.5876 | 19.6 | 29.03333 |
| 660.8762 | 19.58333 | 29.05 |
| 661.1647 | 19.55 | 28.7 |
| 661.4533 | 19.6 | 29.38333 |
| 661.7419 | 19.55 | 29.23333 |
| 662.0304 | 19.61667 | 29.03333 |
| 662.3189 | 19.55 | 29.56667 |
| 662.6074 | 19.68333 | 29.4 |
| 662.8959 | 19.63333 | 29.38333 |
| 663.1844 | 19.65 | 29.25 |
| 663.4729 | 19.66667 | 29.45 |
| 663.7614 | 19.6 | 29.3 |
| 664.0498 | 19.61667 | 29.65 |
| 664.3383 | 19.63333 | 29.55 |
| 664.6267 | 19.63333 | 29.55 |
| 664.9151 | 19.68333 | 29.53333 |
| 665.2035 | 19.68333 | 29.85 |
| 665.4919 | 19.63333 | 29.61667 |
| 665.7803 | 19.61667 | 29.51667 |
| 666.0687 | 19.63333 | 29.96667 |
| 666.357 | 19.66667 | 29.5 |
| 666.6454 | 19.61667 | 29.93333 |
| 666.9337 | 19.68333 | 29.81667 |
| 667.222 | 19.65 | 30 |
| 667.5104 | 19.61667 | 29.96667 |
| 667.7986 | 19.6 | 29.93333 |
| 668.0869 | 19.68333 | 30.31667 |
| 668.3752 | 19.61667 | 29.98333 |
| 668.6635 | 19.66667 | 30.13333 |
| 668.9517 | 19.7 | 30.5 |
| 669.2399 | 19.68333 | 30.58333 |
| 669.5282 | 19.7 | 30.53333 |
| 669.8164 | 19.71667 | 30.61667 |
| 670.1046 | 19.7 | 30.5 |
| 670.3927 | 19.86667 | 30.91667 |
| 670.6809 | 19.71667 | 30.71667 |
| 670.9691 | 19.61667 | 30.56667 |
| 671.2572 | 19.68333 | 30.88333 |
| 671.5454 | 19.65 | 30.7 |
| 671.8335 | 19.73333 | 30.83333 |
| 672.1216 | 19.65 | 31.08333 |
| 672.4097 | 19.65 | 31.26667 |
| 672.6978 | 19.71667 | 30.8 |
| 672.9858 | 19.65 | 31.03333 |
| 673.2739 | 19.68333 | 31 |
| 673.5619 | 19.81667 | 30.96667 |
| 673.85 | 19.78333 | 31.18333 |
| 674.138 | 19.76667 | 31 |
| 674.426 | 19.73333 | 30.9 |
| 674.714 | 19.7 | 31.36667 |
| 675.002 | 19.73333 | 31.31667 |
| 675.29 | 19.75 | 31.4 |
| 675.5779 | 19.71667 | 31.48333 |
| 675.8659 | 19.66667 | |
| 676.1538 | 19.7 | 31.68333 |
| 676.4417 | 19.66667 | 31.6 |
| 676.7296 | 19.68333 | 31.58333 |
| 677.0175 | 19.71667 | 31.65 |
| 677.3054 | | 31.75 |
| 677.5933 | 20.18333 | 31.48333 |
| 677.8811 | 19.81667 | 31.5 |
| 678.169 | 19.7 | 31.83333 |
| 678.4568 | 19.7 | 31.5 |
| 678.7446 | 19.71667 | 32.15 |
| 679.0324 | 19.73333 | 31.86667 |
| 679.3202 | 19.71667 | 31.88333 |
| 679.608 | 19.76667 | 32.2 |
| 679.8958 | 19.65 | 31.86667 |
| 680.1835 | 19.71667 | 32.06667 |
| 680.4713 | 19.71667 | 31.93333 |
| 680.759 | 19.61667 | 32.38333 |
| 681.0467 | 19.73333 | 32.36667 |
| 681.3344 | 19.7 | 32.56667 |
| 681.6221 | 19.75 | 32.73333 |
| 681.9098 | 19.73333 | 32.83333 |
| 682.1975 | 19.68333 | 32.15 |
| 682.4851 | 19.7 | 32.81667 |
| 682.7728 | 19.65 | 32.41667 |
| 683.0604 | 19.8 | 32.7 |
| 683.348 | 19.75 | 32.9 |
| 683.6356 | 19.73333 | 32.98333 |
| 683.9232 | 19.73333 | 33.1 |
| 684.2108 | 19.86667 | 33.15 |
| 684.4983 | 19.85 | 33.1 |
| 684.7859 | 19.91667 | 33.7 |
| 685.0734 | 19.96667 | 33.91667 |
| 685.361 | 20 | 33.55 |
| 685.6485 | 20.2 | 33.66667 |
| 685.936 | 20.21667 | 33.71667 |
| 686.2235 | 20.25 | 33.65 |
| 686.5109 | 20.38333 | 33.81667 |
| 686.7984 | 20.38333 | 33.75 |
| 687.0858 | 20.41667 | 33.86667 |
| 687.3733 | 20.53333 | 34.1 |
| 687.6607 | 20.46667 | 34.4 |
| 687.9481 | 20.61667 | 34.38333 |
| 688.2355 | 20.66667 | 34.6 |
| 688.5229 | 20.6 | 34.38333 |
| 688.8103 | 20.55 | 34.58333 |
| 689.0976 | 20.76667 | 34.88333 |
| 689.385 | 20.78333 | 34.83333 |
| 689.6723 | 20.71667 | 34.6 |
| 689.9596 | 20.8 | 34.98333 |
| 690.2469 | 20.83333 | 34.91667 |
| 690.5342 | 20.98333 | 34.78333 |
| 690.8215 | 20.9 | 34.95 |
| 691.1088 | 20.95 | 34.85 |
| 691.396 | 20.85 | 35.35 |
| 691.6833 | 20.98333 | 35.43333 |
| 691.9705 | 20.91667 | 35.3 |
| 692.2577 | 20.81667 | 35.16667 |
| 692.5449 | 21 | 35.86667 |
| 692.8321 | 21.15 | 35.3 |
| 693.1193 | 21.05 | 35.28333 |
| 693.4065 | 21.18333 | 35.36667 |
| 693.6936 | 20.96667 | 35.58333 |
| 693.9808 | 21.11667 | 35.31667 |
| 694.2679 | 21.1 | 35.83333 |
| 694.555 | 21.15 | 35.85 |
| 694.8421 | 21.1 | 36.01667 |
| 695.1292 | 21.26667 | 36.11667 |
| 695.4163 | 21.13333 | 35.7 |
| 695.7033 | 21.08333 | 36.05 |
| 695.9904 | 21.18333 | 35.88333 |
| 696.2774 | 21.15 | 36.05 |
| 696.5644 | 21.25 | 36.01667 |
| 696.8514 | 21.25 | 36.63333 |
| 697.1384 | 21.36667 | 35.95 |
| 697.4254 | 21.28333 | 36.33333 |
| 697.7124 | 21.15 | 36.51667 |
| 697.9993 | 21.16667 | 36.75 |
| 698.2863 | 21.26667 | 36.56667 |
| 698.5732 | 21.36667 | 37.01667 |
| 698.8601 | 21.23333 | 36.36667 |
| 699.147 | 21.23333 | 37 |
| 699.4339 | 21.3 | 36.86667 |
| 699.7208 | 21.11667 | 37.13333 |
| 700.0077 | 21.25 | 36.91667 |
| 700.2945 | 21.11667 | 36.71667 |
| 700.5814 | 21.26667 | 36.86667 |
| 700.8682 | 21.18333 | 37.28333 |
| 701.155 | 21.25 | 37.05 |
| 701.4418 | 21.15 | 37.11667 |
| 701.7286 | 21.18333 | 37.05 |
| 702.0153 | 21.15 | 37.11667 |
| 702.3021 | 21.18333 | 36.83333 |
| 702.5888 | 21.01667 | 36.65 |
| 702.8756 | 20.96667 | 37.18333 |
| 703.1623 | 21.05 | 37.03333 |
| 703.449 | 20.93333 | 36.91667 |
| 703.7357 | 20.98333 | 37.15 |
| 704.0224 | 20.96667 | 37.15 |
| 704.309 | 20.85 | 36.85 |
| 704.5957 | 20.83333 | 36.73333 |
| 704.8823 | 20.76667 | 36.81667 |
| 705.169 | 20.76667 | 36.43333 |
| 705.4556 | 20.65 | 36.26667 |
| 705.7422 | 20.5 | 36.73333 |
| 706.0288 | 20.4 | 37.08333 |
| 706.3153 | 20.31667 | 37.08333 |
| 706.6019 | 20.33333 | 36.75 |
| 706.8884 | 20.31667 | 37.18333 |
| 707.175 | 20.2 | 37.23333 |
| 707.4615 | 20.21667 | 37.01667 |
| 707.748 | 20.15 | 37.35 |
| 708.0345 | 20.21667 | 37.11667 |
| 708.321 | 20.16667 | 36.75 |
| 708.6074 | 20.25 | 37.5 |
| 708.8939 | 20.11667 | 37.43333 |
| 709.1803 | 20.03333 | 37.23333 |
| 709.4668 | 20.1 | 37.45 |
| 709.7532 | 20.03333 | 37.6 |
| 710.0396 | 20.06667 | 37.03333 |
| 710.3259 | 20.06667 | 37.45 |
| 710.6123 | 20.11667 | 37.71667 |
| 710.8987 | 20.2 | 37.48333 |
| 711.185 | 20.18333 | 37.9 |
| 711.4714 | 20.15 | 37.26667 |
| 711.7577 | 20.11667 | 37.65 |
| 712.044 | 20.08333 | 37.53333 |
| 712.3303 | 20.16667 | 37.98333 |
| 712.6165 | 20.05 | 37.65 |
| 712.9028 | 20.1 | 37.75 |
| 713.1891 | 20.11667 | 37.55 |
| 713.4753 | 20.06667 | 38.11667 |
| 713.7615 | 20.13333 | 37.88333 |
| 714.0477 | 20.18333 | 37.45 |
| 714.3339 | 20.21667 | 37.81667 |
| 714.6201 | 20.15 | 38.11667 |
| 714.9063 | 20.06667 | 38.58333 |
| 715.1924 | 20.13333 | 37.88333 |
| 715.4786 | 20.05 | 38.28333 |
| 715.7647 | 19.95 | 38.21667 |
| 716.0508 | 20.18333 | 37.93333 |
| 716.3369 | 20.2 | 38.11667 |
| 716.623 | 20.08333 | 38.3 |
| 716.9091 | 20.15 | 38.4 |
| 717.1952 | 20 | 38.68333 |
| 717.4812 | 20.06667 | 38.26667 |
| 717.7672 | 20.13333 | 38.65 |
| 718.0533 | 20.03333 | 38.8 |
| 718.3393 | 20.08333 | 38.53333 |
| 718.6253 | 20.06667 | 38.68333 |
| 718.9112 | 20.11667 | 38.93333 |
| 719.1972 | 20.01667 | 38.3 |
| 719.4832 | 19.98333 | 39.1 |
| 719.7691 | 20.06667 | 38.68333 |
| 720.055 | 20.06667 | 38.71667 |
| 720.3409 | 20.06667 | 38.61667 |
| 720.6268 | 19.98333 | 39.56667 |
| 720.9127 | 19.98333 | 39.1 |
| 721.1986 | 20.06667 | 38.8 |
| 721.4845 | 20.01667 | 39.13333 |
| 721.7703 | 20.06667 | 39.01667 |
| 722.0561 | 20.03333 | 38.91667 |
| 722.3419 | 20.11667 | 38.93333 |
| 722.6278 | 20 | 39.48333 |
| 722.9135 | 20.08333 | 39.06667 |
| 723.1993 | 19.98333 | 39.61667 |
| 723.4851 | 19.95 | 39.35 |
| 723.7708 | 19.95 | 39.73333 |
| 724.0566 | 19.95 | 39.58333 |
| 724.3423 | 19.91667 | 39.01667 |
| 724.628 | 19.96667 | 39.93333 |
| 724.9137 | 19.93333 | 40.05 |
| 725.1994 | 19.95 | 39.6 |
| 725.485 | 19.9 | 39.46667 |
| 725.7707 | 19.93333 | 39.73333 |
| 726.0563 | 19.85 | 39.43333 |
| 726.3419 | 19.93333 | 39.8 |
| 726.6275 | 19.95 | 39.76667 |
| 726.9131 | 19.85 | 39.96667 |
| 727.1987 | 19.86667 | 40 |
| 727.4843 | 19.9 | 40.45 |
| 727.7698 | 19.9 | 40.28333 |
| 728.0554 | 19.93333 | 39.51667 |
| 728.3409 | 19.85 | 40.2 |
| 728.6264 | 19.8 | 39.86667 |
| 728.9119 | 19.83333 | 40.35 |
| 729.1974 | 19.8 | 40.66667 |
| 729.4829 | 19.81667 | 40.05 |
| 729.7683 | 19.8 | 40.63333 |
| 730.0538 | 19.71667 | 40.41667 |
| 730.3392 | 19.86667 | 41 |
| 730.6246 | 19.76667 | 40.5 |
| 730.91 | 19.75 | 41.21667 |
| 731.1954 | 19.76667 | 40.75 |
| 731.4808 | 19.8 | 40.61667 |
| 731.7662 | 19.76667 | 40.51667 |
| 732.0515 | 19.78333 | 41.2 |
| 732.3368 | 19.73333 | 40.95 |
| 732.6222 | 19.8 | 40.5 |
| 732.9075 | 19.78333 | 41.16667 |
| 733.1928 | 19.81667 | 41.23333 |
| 733.478 | 19.73333 | 40.91667 |
| 733.7633 | 19.76667 | 41.65 |
| 734.0485 | 19.71667 | 41.45 |
| 734.3338 | 19.66667 | 41.26667 |
| 734.619 | 19.73333 | 41.28333 |
| 734.9042 | 19.76667 | 41.18333 |
| 735.1894 | 19.68333 | 40.86667 |
| 735.4746 | 19.75 | 41.63333 |
| 735.7597 | 19.7 | 41.48333 |
| 736.0449 | 19.76667 | 41.5 |
| 736.33 | 19.7 | 42.1 |
| 736.6152 | 19.73333 | 41.5 |
| 736.9003 | 19.73333 | 41.95 |
| 737.1854 | 19.71667 | 41.86667 |
| 737.4704 | 19.68333 | 42.35 |
| 737.7555 | 19.63333 | 41.65 |
| 738.0406 | 19.68333 | 41.93333 |
| 738.3256 | 19.71667 | 42.11667 |
| 738.6106 | 19.7 | 42.23333 |
| 738.8956 | 19.61667 | 42.56667 |
| 739.1806 | 19.71667 | 41.71667 |
| 739.4656 | 19.63333 | 42.08333 |
| 739.7506 | 19.66667 | 42.5 |
| 740.0355 | 19.7 | 42.45 |
| 740.3205 | 19.68333 | 42.73333 |
| 740.6054 | 19.53333 | 42.18333 |
| 740.8903 | 19.68333 | 42.8 |
| 741.1752 | 19.61667 | 42.81667 |
| 741.4601 | 19.65 | 42.51667 |
| 741.745 | 19.68333 | 42.65 |
| 742.0298 | 19.61667 | 42.56667 |
| 742.3147 | 19.63333 | 42.66667 |
| 742.5995 | 19.71667 | 43.33333 |
| 742.8843 | 19.6 | 42.9 |
| 743.1691 | 19.65 | 42.83333 |
| 743.4539 | 19.61667 | 43.41667 |
| 743.7387 | 19.68333 | 43.31667 |
| 744.0234 | 19.63333 | 42.91667 |
| 744.3082 | 19.66667 | 43.11667 |
| 744.5929 | 19.6 | 43.23333 |
| 744.8776 | 19.63333 | 43.8 |
| 745.1623 | 19.63333 | 43.48333 |
| 745.447 | 19.65 | 43.9 |
| 745.7317 | 19.65 | 43.38333 |
| 746.0163 | 19.58333 | 43.85 |
| 746.301 | 19.63333 | 44.08333 |
| 746.5856 | 19.63333 | 44.11667 |
| 746.8702 | 19.6 | 43.61667 |
| 747.1548 | 19.65 | 43.96667 |
| 747.4394 | 19.65 | 44.25 |
| 747.7239 | 19.61667 | 44.08333 |
| 748.0085 | 19.61667 | 43.96667 |
| 748.293 | 19.6 | 44.15 |
| 748.5776 | 19.63333 | 44.16667 |
| 748.8621 | 19.6 | 44.1 |
| 749.1466 | 19.58333 | 44.71667 |
| 749.4311 | 19.65 | 44.2 |
| 749.7155 | 19.58333 | 44.41667 |
| 750 | 19.68333 | 44.48333 |
| 750.2844 | 19.61667 | 44.8 |
| 750.5689 | 19.68333 | 44.9 |
| 750.8533 | 19.61667 | 44.7 |
| 751.1377 | 19.58333 | 44.66667 |
| 751.4221 | 19.6 | 44.88333 |
| 751.7064 | 19.63333 | 44.58333 |
| 751.9908 | 19.63333 | 45.01667 |
| 752.2751 | 19.66667 | 44.43333 |
| 752.5595 | 19.65 | 45.33333 |
| 752.8438 | 19.66667 | 45.33333 |
| 753.1281 | 19.63333 | 45.36667 |
| 753.4124 | 19.58333 | 45.88333 |
| 753.6966 | 19.61667 | 45.03333 |
| 753.9809 | 19.58333 | 45.26667 |
| 754.2651 | 19.61667 | 45.61667 |
| 754.5493 | 19.58333 | 45.3 |
| 754.8336 | 19.53333 | 45.41667 |
| 755.1178 | 19.61667 | 45.75 |
| 755.4019 | 19.58333 | 45.31667 |
| 755.6861 | 19.61667 | 45.36667 |
| 755.9703 | 19.61667 | 45.81667 |
| 756.2544 | 19.53333 | 45.41667 |
| 756.5385 | 19.63333 | 45.26667 |
| 756.8226 | 19.65 | 45.61667 |
| 757.1067 | 19.66667 | 46.28333 |
| 757.3908 | 19.56667 | 45.91667 |
| 757.6749 | 19.56667 | 45.78333 |
| 757.9589 | 19.61667 | 46.15 |
| 758.243 | 19.58333 | 45.71667 |
| 758.527 | 19.65 | 45.7 |
| 758.811 | 19.58333 | 45.5 |
| 759.095 | 19.63333 | 45.51667 |
| 759.379 | 19.63333 | 46.3 |
| 759.6629 | 19.58333 | 46.4 |
| 759.9469 | 19.61667 | 46.73333 |
| 760.2308 | 19.6 | 46.53333 |
| 760.5147 | 19.56667 | 46.56667 |
| 760.7986 | 19.56667 | 46.73333 |
| 761.0825 | 19.65 | 46.48333 |
| 761.3664 | 19.63333 | 45.81667 |
| 761.6503 | 19.66667 | 47.08333 |
| 761.9341 | 19.63333 | 46.6 |
| 762.218 | 19.56667 | 47.01667 |
| 762.5018 | 19.61667 | 46.75 |
| 762.7856 | 19.6 | 47.1 |
| 763.0694 | 19.58333 | 46.7 |
| 763.3531 | 19.63333 | 46.76667 |
| 763.6369 | 19.58333 | 47.38333 |
| 763.9207 | 19.58333 | 47.51667 |
| 764.2044 | 19.6 | 47.1 |
| 764.4881 | 19.53333 | 47.3 |
| 764.7718 | 19.56667 | 47.73333 |
| 765.0555 | 19.6 | 47.8 |
| 765.3392 | 19.66667 | 47.05 |
| 765.6228 | 19.58333 | 48.11667 |
| 765.9065 | 19.65 | 47.15 |
| 766.1901 | 19.55 | 47.55 |
| 766.4737 | 19.56667 | 47.68333 |
| 766.7573 | 19.58333 | 47.91667 |
| 767.0409 | 19.6 | 48.13333 |
| 767.3244 | 19.61667 | 48.11667 |
| 767.608 | 19.61667 | 47.4 |
| 767.8915 | 19.63333 | 47.25 |
| 768.1751 | 19.6 | 47.95 |
| 768.4586 | 19.58333 | 47.73333 |
| 768.7421 | 19.61667 | 47.63333 |
| 769.0256 | 19.58333 | 47.8 |
| 769.309 | 19.56667 | 47.95 |
| 769.5925 | 19.6 | 47.38333 |
| 769.8759 | 19.6 | 48 |
| 770.1593 | 19.51667 | 48.78333 |
| 770.4427 | 19.56667 | 47.88333 |
| 770.7261 | 19.61667 | 48.85 |
| 771.0095 | 19.61667 | 48.36667 |
| 771.2929 | 19.6 | 48.2 |
| 771.5762 | 19.58333 | 47.95 |
| 771.8595 | 19.58333 | 48.68333 |
| 772.1429 | 19.53333 | 48.33333 |
| 772.4262 | 19.7 | 49.28333 |
| 772.7094 | 19.6 | 48.95 |
| 772.9927 | 19.6 | 48.83333 |
| 773.276 | 19.56667 | 48.66667 |
| 773.5592 | 19.58333 | 49.08333 |
| 773.8424 | 19.63333 | 48.75 |
| 774.1257 | 19.6 | 49.26667 |
| 774.4089 | 19.61667 | 49.28333 |
| 774.692 | 19.63333 | 48.93333 |
| 774.9752 | 19.6 | 48.88333 |
| 775.2584 | 19.65 | 48.6 |
| 775.5415 | 19.6 | 48.76667 |
| 775.8246 | 19.58333 | 49.53333 |
| 776.1077 | 19.58333 | 48.81667 |
| 776.3908 | 19.55 | 49.21667 |
| 776.6739 | 19.66667 | 49.43333 |
| 776.957 | 19.6 | 48.95 |
| 777.24 | 19.66667 | 49.38333 |
| 777.5231 | 19.56667 | 49.35 |
| 777.8061 | 19.6 | 49.78333 |
| 778.0891 | 19.56667 | 49.73333 |
| 778.3721 | 19.6 | 49.73333 |
| 778.655 | 19.53333 | 49.63333 |
| 778.938 | 19.58333 | 49.5 |
| 779.2209 | 19.56667 | 49.96667 |
| 779.5039 | 19.63333 | 49.35 |
| 779.7868 | 19.61667 | 50.15 |
| 780.0697 | 19.65 | 50 |
| 780.3526 | 19.58333 | 51 |
| 780.6354 | 19.6 | 50.15 |
| 780.9183 | 19.53333 | 50.36667 |
| 781.2011 | 19.53333 | 50.55 |
| 781.484 | 19.56667 | 49.8 |
| 781.7668 | 19.58333 | 50.58333 |
| 782.0496 | 19.6 | 50.68333 |
| 782.3323 | 19.6 | 50.63333 |
| 782.6151 | 19.55 | 50.33333 |
| 782.8978 | 19.56667 | 50.8 |
| 783.1806 | 19.63333 | 50.68333 |
| 783.4633 | 19.6 | 50.85 |
| 783.746 | 19.6 | 50.35 |
| 784.0287 | 19.55 | 51.03333 |
| 784.3114 | 19.56667 | 50.3 |
| 784.594 | 19.6 | 51.45 |
| 784.8767 | 19.58333 | 51.11667 |
| 785.1593 | 19.55 | 50.91667 |
| 785.4419 | 19.56667 | 50.61667 |
| 785.7245 | 19.56667 | 51.23333 |
| 786.0071 | 19.6 | 51.85 |
| 786.2896 | 19.56667 | 50.93333 |
| 786.5722 | 19.58333 | 51.13333 |
| 786.8547 | 19.61667 | 51.6 |
| 787.1372 | 19.58333 | 51.7 |
| 787.4197 | 19.6 | 52.15 |
| 787.7022 | 19.6 | 51.41667 |
| 787.9847 | 19.58333 | 50.93333 |
| 788.2672 | 19.55 | 51.58333 |
| 788.5496 | 19.61667 | 51.23333 |
| 788.832 | 19.55 | 51.58333 |
| 789.1145 | 19.53333 | 52.06667 |
| 789.3969 | 19.55 | 51.53333 |
| 789.6792 | 19.61667 | 51.73333 |
| 789.9616 | 19.56667 | 52.86667 |
| 790.244 | 19.51667 | 52.53333 |
| 790.5263 | 19.58333 | 51.65 |
| 790.8086 | 19.6 | 52.11667 |
| 791.0909 | 19.6 | 52.01667 |
| 791.3732 | 19.48333 | 52.33333 |
| 791.6555 | 19.58333 | 51.26667 |
| 791.9378 | 19.6 | 52.63333 |
| 792.22 | 19.51667 | 52.5 |
| 792.5022 | 19.61667 | 52.08333 |
| 792.7845 | 19.56667 | 52.13333 |
| 793.0667 | 19.5 | 52.31667 |
| 793.3488 | 19.53333 | 52.71667 |
| 793.631 | 19.58333 | 52.15 |
| 793.9132 | 19.68333 | 52.38333 |
| 794.1953 | 19.58333 | 52.25 |
| 794.4774 | 19.55 | 52.48333 |
| 794.7595 | 19.53333 | 52.85 |
| 795.0416 | 19.58333 | 52.46667 |
| 795.3237 | 19.6 | 53.06667 |
| 795.6058 | 19.58333 | 52.26667 |
| 795.8878 | 19.53333 | 52.31667 |
| 796.1698 | 19.53333 | 52.78333 |
| 796.4519 | 19.58333 | 53.26667 |
| 796.7339 | 19.6 | 52.71667 |
| 797.0158 | 19.6 | 52.71667 |
| 797.2978 | 19.48333 | 52.5 |
| 797.5798 | 19.58333 | 52.48333 |
| 797.8617 | 19.53333 | 53.18333 |
| 798.1436 | 19.55 | 52.6 |
| 798.4255 | 19.61667 | 52.96667 |
| 798.7074 | 19.5 | 53.35 |
| 798.9893 | 19.56667 | 53.08333 |
| 799.2712 | 19.53333 | 53.21667 |
| 799.553 | 19.56667 | 53.3 |
| 799.8348 | 19.55 | 52.66667 |
| 800.1167 | 19.55 | 53.43333 |
| 800.3985 | 19.61667 | 53.43333 |
| 800.6802 | 19.55 | 53.01667 |
| 800.962 | 19.55 | 53.61667 |
| 801.2438 | 19.56667 | 53.53333 |
| 801.5255 | 19.58333 | 53.61667 |
| 801.8072 | 19.55 | 53.3 |
| 802.0889 | 19.56667 | 53.23333 |
| 802.3706 | 19.61667 | 53.65 |
| 802.6523 | 19.55 | 53.61667 |
| 802.9339 | 19.6 | 53.46667 |
| 803.2156 | 19.63333 | 53.05 |
| 803.4972 | 19.56667 | 53.75 |
| 803.7788 | 19.56667 | 54.08333 |
| 804.0604 | 19.55 | 53.7 |
| 804.342 | 19.55 | 54.13333 |
| 804.6236 | 19.56667 | 54 |
| 804.9051 | 19.56667 | 54.3 |
| 805.1867 | 19.56667 | 54.25 |
| 805.4682 | 19.53333 | 54.43333 |
| 805.7497 | 19.5 | 53.65 |
| 806.0312 | 19.6 | 53.9 |
| 806.3126 | 19.58333 | 54 |
| 806.5941 | 19.56667 | 53.31667 |
| 806.8755 | 19.55 | 53.95 |
| 807.157 | 19.51667 | 53.66667 |
| 807.4384 | 19.55 | 53.85 |
| 807.7198 | 19.61667 | 54.48333 |
| 808.0011 | 19.53333 | 54.36667 |
| 808.2825 | 19.56667 | 54.23333 |
| 808.5639 | 19.51667 | 53.71667 |
| 808.8452 | 19.55 | 54.18333 |
| 809.1265 | 19.58333 | 54.56667 |
| 809.4078 | 19.56667 | 54.66667 |
| 809.6891 | 19.53333 | 54.01667 |
| 809.9704 | 19.51667 | 53.91667 |
| 810.2516 | 19.56667 | 54.41667 |
| 810.5328 | 19.55 | 55.16667 |
| 810.8141 | 19.53333 | 54.55 |
| 811.0953 | 19.6 | 54.16667 |
| 811.3765 | 19.56667 | 54.95 |
| 811.6576 | 19.58333 | 54.81667 |
| 811.9388 | 19.53333 | 54.58333 |
| 812.2199 | 19.55 | 54.61667 |
| 812.5011 | 19.51667 | 55.48333 |
| 812.7822 | 19.56667 | 54.71667 |
| 813.0633 | 19.56667 | 54.48333 |
| 813.3443 | 19.5 | 54.51667 |
| 813.6254 | 19.53333 | 55.51667 |
| 813.9065 | 19.53333 | 54.9 |
| 814.1875 | 19.55 | 54.85 |
| 814.4685 | 19.55 | 55.5 |
| 814.7495 | 19.51667 | 55.41667 |
| 815.0305 | 19.55 | 54.78333 |
| 815.3115 | 19.53333 | 54.58333 |
| 815.5924 | 19.53333 | 54.75 |
| 815.8733 | 19.53333 | 55.61667 |
| 816.1543 | 19.61667 | 54.56667 |
| 816.4352 | 19.61667 | 55.1 |
| 816.7161 | 19.53333 | 54.93333 |
| 816.9969 | 19.5 | 56.08333 |
| 817.2778 | 19.53333 | 55.3 |
| 817.5586 | 19.5 | 55.45 |
| 817.8395 | 19.56667 | 55.38333 |
| 818.1203 | 19.51667 | 55.18333 |
| 818.4011 | 19.56667 | 55.98333 |
| 818.6818 | 19.53333 | 55.75 |
| 818.9626 | 19.61667 | 55.1 |
| 819.2434 | 19.58333 | 55.31667 |
| 819.5241 | 19.53333 | 55.41667 |
| 819.8048 | 19.53333 | 55.81667 |
| 820.0855 | 19.51667 | 55.33333 |
| 820.3662 | 19.56667 | 56.21667 |
| 820.6468 | 19.53333 | 55.4 |
| 820.9275 | 19.53333 | 55.61667 |
| 821.2081 | 19.55 | 55.03333 |
| 821.4887 | 19.56667 | 55.16667 |
| 821.7693 | 19.51667 | 56.13333 |
| 822.0499 | 19.56667 | 56 |
| 822.3305 | 19.56667 | 55.96667 |
| 822.6111 | 19.58333 | 55.71667 |
| 822.8916 | 19.46667 | 55.78333 |
| 823.1721 | 19.53333 | 55.45 |
| 823.4526 | 19.55 | 56.2 |
| 823.7331 | 19.55 | 56.13333 |
| 824.0136 | 19.55 | 55.58333 |
| 824.2941 | 19.63333 | 55.73333 |
| 824.5745 | 19.56667 | 55.71667 |
| 824.8549 | 19.53333 | 55.66667 |
| 825.1353 | 19.6 | 55.51667 |
| 825.4157 | 19.56667 | 55.95 |
| 825.6961 | 19.51667 | 56.2 |
| 825.9765 | 19.5 | 56.08333 |
| 826.2568 | 19.48333 | 55.26667 |
| 826.5371 | 19.6 | 56.08333 |
| 826.8175 | 19.53333 | 55.91667 |
| 827.0978 | 19.46667 | 55.63333 |
| 827.378 | 19.5 | 56.53333 |
| 827.6583 | 19.58333 | 56.53333 |
| 827.9386 | 19.5 | 55.81667 |
| 828.2188 | 19.56667 | 56.43333 |
| 828.499 | 19.51667 | 56.23333 |
| 828.7792 | 19.55 | 55.66667 |
| 829.0594 | 19.51667 | 56.46667 |
| 829.3396 | 19.56667 | 56.28333 |
| 829.6197 | 19.51667 | 56.13333 |
| 829.8998 | 19.53333 | 56.85 |
| 830.18 | 19.55 | 57.21667 |
| 830.4601 | 19.53333 | 56.71667 |
| 830.7402 | 19.51667 | 56.93333 |
| 831.0202 | 19.55 | 56.4 |
| 831.3003 | 19.55 | 56.7 |
| 831.5803 | 19.53333 | 56.95 |
| 831.8603 | 19.45 | 56.96667 |
| 832.1404 | 19.55 | 56.55 |
| 832.4203 | 19.51667 | 56.43333 |
| 832.7003 | 19.51667 | 56.71667 |
| 832.9803 | 19.53333 | 56.66667 |
| 833.2602 | 19.53333 | 57.06667 |
| 833.5401 | 19.48333 | 56.53333 |
| 833.8201 | 19.51667 | 56.56667 |
| 834.0999 | 19.56667 | 56.28333 |
| 834.3798 | 19.58333 | 57.65 |
| 834.6597 | 19.51667 | 56.71667 |
| 834.9395 | 19.5 | 56.5 |
| 835.2194 | 19.56667 | 57.5 |
| 835.4992 | 19.53333 | 56.66667 |
| 835.779 | 19.58333 | 57.7 |
| 836.0587 | 19.55 | 57.03333 |
| 836.3385 | 19.55 | 56.71667 |
| 836.6183 | 19.45 | 56.8 |
| 836.898 | 19.51667 | 57.08333 |
| 837.1777 | 19.51667 | 56.61667 |
| 837.4574 | 19.51667 | 56.95 |
| 837.7371 | 19.56667 | 57.5 |
| 838.0167 | 19.48333 | 57.83333 |
| 838.2964 | 19.55 | 57.08333 |
| 838.576 | 19.51667 | 56.51667 |
| 838.8556 | 19.51667 | 57.26667 |
| 839.1352 | 19.51667 | 57.78333 |
| 839.4148 | 19.56667 | 57.1 |
| 839.6944 | 19.53333 | 57.08333 |
| 839.9739 | 19.46667 | 57.4 |
| 840.2535 | 19.58333 | 57.4 |
| 840.533 | 19.55 | 56.8 |
| 840.8125 | 19.58333 | 57.6 |
| 841.092 | 19.56667 | 57.43333 |
| 841.3715 | 19.41667 | 56.96667 |
| 841.6509 | 19.55 | 57.08333 |
| 841.9303 | 19.53333 | 56.78333 |
| 842.2098 | 19.6 | 57.48333 |
| 842.4892 | 19.63333 | 56.96667 |
| 842.7685 | 19.56667 | 58.23333 |
| 843.0479 | 19.5 | 57.21667 |
| 843.3273 | 19.5 | 57.71667 |
| 843.6066 | 19.5 | 57.25 |
| 843.8859 | 19.51667 | 58.15 |
| 844.1652 | 19.48333 | 57.65 |
| 844.4445 | 19.53333 | 57.61667 |
| 844.7238 | 19.6 | 57.08333 |
| 845.003 | 19.55 | 57.4 |
| 845.2823 | 19.51667 | 57.78333 |
| 845.5615 | 19.58333 | 57.46667 |
| 845.8407 | 19.53333 | 57.43333 |
| 846.1199 | 19.56667 | 57.33333 |
| 846.3991 | 19.55 | 57.08333 |
| 846.6782 | 19.5 | 57.73333 |
| 846.9574 | 19.51667 | 57.76667 |
| 847.2365 | 19.5 | 57.71667 |
| 847.5156 | 19.58333 | 58.11667 |
| 847.7947 | 19.48333 | 57.26667 |
| 848.0738 | 19.51667 | 57.46667 |
| 848.3528 | 19.53333 | 57.63333 |
| 848.6319 | 19.51667 | 57.11667 |
| 848.9109 | 19.51667 | 57.73333 |
| 849.1899 | 19.61667 | 57.36667 |
| 849.4689 | 19.53333 | 57.8 |
| 849.7478 | 19.55 | 57.65 |
| 850.0268 | 19.56667 | 58.18333 |
| 850.3057 | 19.51667 | 57.83333 |
| 850.5847 | 19.53333 | 57.05 |
| 850.8636 | 19.53333 | 57.8 |
| 851.1425 | 19.53333 | 57.46667 |
| 851.4213 | 19.51667 | 57.98333 |
| 851.7002 | 19.55 | 58.06667 |
| 851.979 | 19.55 | 57.15 |
| 852.2579 | 19.51667 | 57.5 |
| 852.5367 | 19.53333 | 58.05 |
| 852.8155 | 19.55 | 58.56667 |
| 853.0942 | 19.55 | 57.58333 |
| 853.373 | 19.51667 | 58.11667 |
| 853.6517 | 19.55 | 57.71667 |
| 853.9305 | 19.51667 | 58 |
| 854.2092 | 19.51667 | 57.63333 |
| 854.4878 | 19.53333 | 57.91667 |
| 854.7665 | 19.51667 | 57.43333 |
| 855.0452 | 19.48333 | 58.73333 |
| 855.3238 | 19.5 | 58.21667 |
| 855.6024 | 19.56667 | 58.06667 |
| 855.881 | 19.51667 | 58.18333 |
| 856.1596 | 19.55 | 57.71667 |
| 856.4382 | 19.53333 | 58.03333 |
| 856.7168 | 19.53333 | 58.03333 |
| 856.9953 | 19.53333 | 58.06667 |
| 857.2738 | 19.56667 | 57.75 |
| 857.5523 | 19.46667 | 57.38333 |
| 857.8308 | 19.56667 | 57.81667 |
| 858.1093 | 19.5 | 58.11667 |
| 858.3878 | 19.5 | 58.05 |
| 858.6662 | 19.51667 | 58.06667 |
| 858.9446 | 19.48333 | 58.83333 |
| 859.223 | 19.56667 | 58.16667 |
| 859.5014 | 19.53333 | 57.71667 |
| 859.7798 | 19.51667 | 58.48333 |
| 860.0581 | 19.6 | 58.58333 |
| 860.3365 | 19.51667 | 58.41667 |
| 860.6148 | 19.51667 | 58.01667 |
| 860.8931 | 19.55 | 57.86667 |
| 861.1714 | 19.46667 | 58.31667 |
| 861.4497 | 19.51667 | 58.25 |
| 861.7279 | 19.51667 | 58.63333 |
| 862.0061 | 19.56667 | 58.3 |
| 862.2844 | 19.6 | 58.25 |
| 862.5626 | 19.55 | 58.5 |
| 862.8408 | 19.55 | 57.58333 |
| 863.1189 | 19.53333 | 59.1 |
| 863.3971 | 19.51667 | 58.78333 |
| 863.6752 | 19.53333 | 57.66667 |
| 863.9533 | 19.53333 | 58.06667 |
| 864.2314 | 19.53333 | 58.01667 |
| 864.5095 | 19.55 | 58.16667 |
| 864.7876 | 19.56667 | 57.8 |
| 865.0656 | 19.6 | 57.93333 |
| 865.3437 | 19.5 | 58.18333 |
| 865.6217 | 19.56667 | 58.11667 |
| 865.8997 | 19.61667 | 58.95 |
| 866.1777 | 19.5 | 58.23333 |
| 866.4556 | 19.51667 | 58.43333 |
| 866.7336 | 19.51667 | 57.81667 |
| 867.0115 | 19.61667 | 58.43333 |
| 867.2894 | 19.48333 | 58.75 |
| 867.5673 | 19.5 | 58.95 |
| 867.8452 | 19.5 | 58.21667 |
| 868.1231 | 19.55 | 58.03333 |
| 868.4009 | 19.56667 | 58.11667 |
| 868.6788 | 19.53333 | 57.3 |
| 868.9566 | 19.5 | 58.23333 |
| 869.2344 | 19.55 | 58.86667 |
| 869.5122 | 19.5 | 58.18333 |
| 869.7899 | 19.5 | 58.16667 |
| 870.0677 | 19.48333 | 58.8 |
| 870.3454 | 19.53333 | 58.36667 |
| 870.6231 | 19.53333 | 58.63333 |
| 870.9008 | 19.53333 | 57.71667 |
| 871.1785 | 19.53333 | 58.51667 |
| 871.4561 | 19.56667 | 58.05 |
| 871.7338 | 19.58333 | 58.55 |
| 872.0114 | 19.55 | 57.98333 |
| 872.289 | 19.55 | 57.7 |
| 872.5666 | 19.5 | 58.08333 |
| 872.8442 | 19.58333 | 58.21667 |
| 873.1217 | 19.6 | 58.46667 |
| 873.3993 | 19.5 | 58.45 |
| 873.6768 | 19.55 | 58.3 |
| 873.9543 | 19.56667 | 58.41667 |
| 874.2318 | 19.51667 | 58.25 |
| 874.5093 | 19.56667 | 58.06667 |
| 874.7867 | 19.58333 | 57.85 |
| 875.0642 | 19.51667 | 58.16667 |
| 875.3416 | 19.53333 | 58.03333 |
| 875.619 | 19.51667 | 58.28333 |
| 875.8964 | 19.55 | 57.88333 |
| 876.1737 | 19.48333 | 58.4 |
| 876.4511 | 19.55 | 57.7 |
| 876.7284 | 19.56667 | 58.13333 |
| 877.0057 | 19.51667 | 58.46667 |
| 877.283 | 19.55 | 57.86667 |
| 877.5603 | 19.55 | 57.86667 |
| 877.8376 | 19.51667 | 57.9 |
| 878.1148 | 19.58333 | 58.31667 |
| 878.3921 | 19.45 | 57.46667 |
| 878.6693 | 19.48333 | 58.33333 |
| 878.9465 | 19.53333 | 57.36667 |
| 879.2237 | 19.53333 | 57.7 |
| 879.5008 | 19.53333 | 57.8 |
| 879.778 | 19.56667 | 57.51667 |
| 880.0551 | 19.56667 | 57.88333 |
| 880.3322 | 19.5 | 57.53333 |
| 880.6093 | 19.58333 | 57.76667 |
| 880.8864 | 19.5 | 57.18333 |
| 881.1634 | 19.53333 | 57.1 |
| 881.4405 | 19.61667 | 57.16667 |
| 881.7175 | 19.53333 | 57.43333 |
| 881.9945 | 19.51667 | 57.73333 |
| 882.2715 | 19.51667 | 56.96667 |
| 882.5485 | 19.51667 | 57.66667 |
| 882.8254 | 19.51667 | 57.18333 |
| 883.1024 | 19.5 | 57.03333 |
| 883.3793 | 19.53333 | 56.9 |
| 883.6562 | 19.48333 | 56.73333 |
| 883.9331 | 19.6 | 56.95 |
| 884.2099 | 19.56667 | 57.43333 |
| 884.4868 | 19.56667 | 56.86667 |
| 884.7636 | 19.51667 | 57.21667 |
| 885.0404 | 19.55 | 56.65 |
| 885.3172 | 19.48333 | 57.43333 |
| 885.594 | 19.51667 | 57.16667 |
| 885.8708 | 19.51667 | 56.98333 |
| 886.1475 | 19.51667 | 56.66667 |
| 886.4243 | 19.5 | 56.9 |
| 886.701 | 19.5 | 56.73333 |
| 886.9777 | 19.51667 | 56.4 |
| 887.2544 | 19.51667 | 56.85 |
| 887.531 | 19.58333 | 57.06667 |
| 887.8077 | 19.5 | 56.91667 |
| 888.0843 | 19.53333 | 56.45 |
| 888.3609 | 19.55 | 56.48333 |
| 888.6375 | 19.53333 | 56.25 |
| 888.9141 | 19.5 | 56.98333 |
| 889.1906 | 19.55 | 56.53333 |
| 889.4671 | 19.53333 | 56.01667 |
| 889.7437 | 19.51667 | 56.21667 |
| 890.0202 | 19.51667 | 56.5 |
| 890.2967 | 19.55 | 56.68333 |
| 890.5731 | 19.53333 | 56.06667 |
| 890.8496 | 19.51667 | 56.58333 |
| 891.126 | 19.5 | 56.16667 |
| 891.4024 | 19.55 | 56.66667 |
| 891.6788 | 19.51667 | 56.23333 |
| 891.9552 | 19.53333 | 55.71667 |
| 892.2316 | 19.53333 | 55.75 |
| 892.5079 | 19.5 | 56.26667 |
| 892.7842 | 19.53333 | 56.68333 |
| 893.0606 | 19.6 | 56.26667 |
| 893.3368 | 19.53333 | 55.76667 |
| 893.6131 | 19.46667 | 56.18333 |
| 893.8894 | 19.53333 | 55.65 |
| 894.1656 | 19.58333 | 55.85 |
| 894.4418 | 19.56667 | 55.4 |
| 894.718 | 19.51667 | 55.86667 |
| 894.9942 | 19.56667 | 56.05 |
| 895.2704 | 19.56667 | 56.35 |
| 895.5466 | 19.51667 | 55.88333 |
| 895.8227 | 19.56667 | 55.88333 |
| 896.0988 | 19.5 | 55.46667 |
| 896.3749 | 19.51667 | 55.61667 |
| 896.651 | 19.55 | 55.71667 |
| 896.9271 | 19.53333 | 56.11667 |
| 897.2031 | 19.5 | 54.76667 |
| 897.4791 | 19.5 | 55.71667 |
| 897.7551 | 19.55 | 55.35 |
| 898.0311 | 19.51667 | 55.76667 |
| 898.3071 | 19.55 | 55.35 |
| 898.5831 | 19.55 | 55.2 |
| 898.859 | 19.53333 | 55.13333 |
| 899.1349 | 19.5 | 55.28333 |
| 899.4108 | 19.55 | 55.23333 |
| 899.6867 | 19.5 | 55.26667 |
| 899.9626 | 19.58333 | 55.16667 |
| 900.2385 | 19.53333 | 55.13333 |
| 900.5143 | 19.56667 | 55.46667 |
| 900.7901 | 19.51667 | 55.83333 |
| 901.0659 | 19.48333 | 55.33333 |
| 901.3417 | 19.5 | 54.98333 |
| 901.6175 | 19.51667 | 54.45 |
| 901.8932 | 19.56667 | 54.7 |
| 902.1689 | 19.48333 | 55.01667 |
| 902.4446 | 19.46667 | 54.65 |
| 902.7203 | 19.53333 | 55.38333 |
| 902.996 | 19.51667 | 54.53333 |
| 903.2717 | 19.53333 | 55.28333 |
| 903.5473 | 19.55 | 55.26667 |
| 903.8229 | 19.53333 | 54.46667 |
| 904.0985 | 19.55 | 54.85 |
| 904.3741 | 19.55 | 55.01667 |
| 904.6497 | 19.51667 | 54.45 |
| 904.9252 | 19.5 | 54.88333 |
| 905.2008 | 19.6 | 54.5 |
| 905.4763 | 19.51667 | 54.23333 |
| 905.7518 | 19.51667 | 54.5 |
| 906.0273 | 19.53333 | 54.25 |
| 906.3027 | 19.53333 | 54.56667 |
| 906.5782 | 19.55 | 54.26667 |
| 906.8536 | 19.53333 | 54.41667 |
| 907.129 | 19.46667 | 54.58333 |
| 907.4044 | 19.53333 | 53.93333 |
| 907.6798 | 19.6 | 54.73333 |
| 907.9551 | 19.53333 | 54 |
| 908.2305 | 19.51667 | 54.41667 |
| 908.5058 | 19.51667 | 53.78333 |
| 908.7811 | 19.46667 | 54.28333 |
| 909.0564 | 19.5 | 54.33333 |
| 909.3316 | 19.48333 | 53.86667 |
| 909.6069 | 19.5 | 53.68333 |
| 909.8821 | 19.55 | 53.51667 |
| 910.1573 | 19.53333 | 54.1 |
| 910.4325 | 19.53333 | 53.73333 |
| 910.7077 | 19.51667 | 53.9 |
| 910.9829 | 19.56667 | 53.78333 |
| 911.258 | 19.5 | 53.91667 |
| 911.5331 | 19.5 | 53.61667 |
| 911.8082 | 19.5 | 53.41667 |
| 912.0833 | 19.48333 | 54.11667 |
| 912.3584 | 19.51667 | 53.53333 |
| 912.6334 | 19.51667 | 53.51667 |
| 912.9085 | 19.55 | 54.01667 |
| 913.1835 | 19.51667 | 53.43333 |
| 913.4585 | 19.51667 | 53.65 |
| 913.7335 | 19.51667 | 53.53333 |
| 914.0084 | 19.53333 | 53.58333 |
| 914.2834 | 19.55 | 53.58333 |
| 914.5583 | 19.6 | 53.46667 |
| 914.8332 | 19.6 | 53.26667 |
| 915.1081 | 19.53333 | 53.08333 |
| 915.383 | 19.46667 | 53.66667 |
| 915.6578 | 19.5 | 53.38333 |
| 915.9327 | 19.58333 | 52.93333 |
| 916.2075 | 19.46667 | 53.16667 |
| 916.4823 | 19.51667 | 53.23333 |
| 916.7571 | 19.55 | 52.95 |
| 917.0318 | 19.58333 | 52.91667 |
| 917.3066 | 19.5 | 53.11667 |
| 917.5813 | 19.55 | 52.48333 |
| 917.856 | 19.51667 | 52.5 |
| 918.1307 | 19.51667 | 53.18333 |
| 918.4054 | 19.51667 | 52.48333 |
| 918.68 | 19.51667 | 53.03333 |
| 918.9547 | 19.56667 | 52.21667 |
| 919.2293 | 19.63333 | 52 |
| 919.5039 | 19.51667 | 52.51667 |
| 919.7785 | 19.6 | 51.91667 |
| 920.053 | 19.53333 | 52.38333 |
| 920.3276 | 19.58333 | 52.23333 |
| 920.6021 | 19.65 | 52.7 |
| 920.8766 | 19.55 | 52.2 |
| 921.1511 | 19.6 | 51.95 |
| 921.4256 | 19.5 | 52.93333 |
| 921.7 | 19.46667 | 51.68333 |
| 921.9745 | 19.53333 | 52.3 |
| 922.2489 | 19.51667 | 52.18333 |
| 922.5233 | 19.53333 | 51.6 |
| 922.7977 | 19.53333 | 52.25 |
| 923.072 | 19.51667 | 51.88333 |
| 923.3464 | 19.55 | 51.66667 |
| 923.6207 | 19.5 | 51.46667 |
| 923.895 | 19.56667 | 51.48333 |
| 924.1693 | 19.51667 | 50.8 |
| 924.4436 | 19.53333 | 51.51667 |
| 924.7179 | 19.51667 | 51.55 |
| 924.9921 | 19.53333 | 50.86667 |
| 925.2663 | 19.53333 | 51.01667 |
| 925.5405 | 19.51667 | 51.36667 |
| 925.8147 | 19.55 | 50.61667 |
| 926.0889 | 19.55 | 50.8 |
| 926.363 | 19.53333 | 50.51667 |
| 926.6371 | 19.55 | 50.83333 |
| 926.9112 | 19.53333 | 49.98333 |
| 927.1853 | 19.55 | 51.28333 |
| 927.4594 | 19.53333 | 50.68333 |
| 927.7335 | 19.63333 | 50.81667 |
| 928.0075 | 19.58333 | 50.33333 |
| 928.2815 | 19.5 | 50.51667 |
| 928.5555 | 19.53333 | 50.11667 |
| 928.8295 | 19.56667 | 50.05 |
| 929.1035 | 19.53333 | 50.21667 |
| 929.3774 | 19.51667 | 49.83333 |
| 929.6514 | 19.6 | 50.26667 |
| 929.9253 | 19.56667 | 49.86667 |
| 930.1992 | 19.53333 | 49.83333 |
| 930.473 | 19.58333 | 49.25 |
| 930.7469 | 19.56667 | 49.83333 |
| 931.0207 | 19.55 | 49.91667 |
| 931.2945 | 19.56667 | 49.46667 |
| 931.5683 | 19.51667 | 49.98333 |
| 931.8421 | 19.53333 | 49.4 |
| 932.1159 | 19.6 | 49.48333 |
| 932.3896 | 19.61667 | 49.2 |
| 932.6634 | 19.65 | 49.41667 |
| 932.9371 | 19.51667 | 49.08333 |
| 933.2107 | 19.56667 | 49.36667 |
| 933.4844 | 19.51667 | 49.13333 |
| 933.7581 | 19.53333 | 49.3 |
| 934.0317 | 19.55 | 49.18333 |
| 934.3053 | 19.48333 | 49.56667 |
| 934.5789 | 19.53333 | 48.53333 |
| 934.8525 | 19.58333 | 49.38333 |
| 935.1261 | 19.58333 | 48.58333 |
| 935.3996 | 19.55 | 48.16667 |
| 935.6731 | 19.46667 | 48.81667 |
| 935.9466 | 19.56667 | 47.88333 |
| 936.2201 | 19.53333 | 47.91667 |
| 936.4936 | 19.5 | 48.31667 |
| 936.767 | 19.55 | 47.88333 |
| 937.0405 | 19.53333 | 48.31667 |
| 937.3139 | 19.53333 | 47.53333 |
| 937.5873 | 19.55 | 48.28333 |
| 937.8607 | 19.53333 | 47.83333 |
| 938.134 | 19.51667 | 47.93333 |
| 938.4074 | 19.51667 | 47.73333 |
| 938.6807 | 19.58333 | 47.61667 |
| 938.954 | 19.48333 | 48 |
| 939.2273 | 19.55 | 47.96667 |
| 939.5005 | 19.51667 | 47.4 |
| 939.7738 | 19.5 | 47.65 |
| 940.047 | 19.63333 | 47.08333 |
| 940.3202 | 19.53333 | 47.31667 |
| 940.5934 | 19.51667 | 46.8 |
| 940.8666 | 19.53333 | 47.31667 |
| 941.1397 | 19.55 | 46.83333 |
| 941.4129 | 19.53333 | 46.98333 |
| 941.686 | 19.5 | 47.11667 |
| 941.9591 | 19.46667 | 46.68333 |
| 942.2322 | 19.53333 | 46.21667 |
| 942.5052 | 19.53333 | 46.21667 |
| 942.7783 | 19.5 | 46.03333 |
| 943.0513 | 19.53333 | 46.05 |
| 943.3243 | 19.58333 | 45.93333 |
| 943.5973 | 19.56667 | 45.91667 |
| 943.8703 | 19.55 | 46.36667 |
| 944.1432 | 19.53333 | 45.51667 |
| 944.4162 | 19.48333 | 45.26667 |
| 944.6891 | 19.56667 | 45.21667 |
| 944.962 | 19.48333 | 45.96667 |
| 945.2348 | 19.53333 | 45.36667 |
| 945.5077 | 19.55 | 45.65 |
| 945.7805 | 19.56667 | 45 |
| 946.0534 | 19.53333 | 44.93333 |
| 946.3262 | 19.55 | 45.1 |
| 946.5989 | 19.56667 | 45.15 |
| 946.8717 | 19.53333 | 45 |
| 947.1445 | 19.51667 | 44.53333 |
| 947.4172 | 19.53333 | 44.96667 |
| 947.6899 | 19.58333 | 45.05 |
| 947.9626 | 19.5 | 45.06667 |
| 948.2352 | 19.53333 | 44.86667 |
| 948.5079 | 19.55 | 44.76667 |
| 948.7805 | 19.51667 | 45.13333 |
| 949.0531 | 19.58333 | 43.96667 |
| 949.3257 | 19.51667 | 44.16667 |
| 949.5983 | 19.53333 | 44.83333 |
| 949.8709 | 19.5 | 44.01667 |
| 950.1434 | 19.5 | 44.35 |
| 950.4159 | 19.5 | 44.45 |
| 950.6884 | 19.51667 | 44.15 |
| 950.9609 | 19.53333 | 44.01667 |
| 951.2334 | 19.58333 | 43.93333 |
| 951.5058 | 19.63333 | 43.31667 |
| 951.7783 | 19.55 | 43.78333 |
| 952.0507 | 19.53333 | 43.88333 |
| 952.3231 | 19.48333 | 43.7 |
| 952.5954 | 19.55 | 43.63333 |
| 952.8678 | 19.56667 | 43.18333 |
| 953.1401 | 19.55 | 43.33333 |
| 953.4124 | 19.48333 | 43.36667 |
| 953.6847 | 19.51667 | 43.4 |
| 953.957 | 19.55 | 42.96667 |
| 954.2293 | 19.5 | 43.18333 |
| 954.5015 | 19.55 | 42.78333 |
| 954.7737 | 19.53333 | 42.76667 |
| 955.0459 | 19.55 | 42.91667 |
| 955.3181 | 19.46667 | 42.88333 |
| 955.5903 | 19.5 | 43.05 |
| 955.8624 | 19.6 | 42.45 |
| 956.1346 | 19.46667 | 42.75 |
| 956.4067 | 19.55 | 42.7 |
| 956.6788 | 19.56667 | 42.06667 |
| 956.9508 | 19.56667 | 42.36667 |
| 957.2229 | 19.51667 | 42.63333 |
| 957.4949 | 19.58333 | 42.31667 |
| 957.7669 | 19.51667 | 42.35 |
| 958.0389 | 19.5 | 41.85 |
| 958.3109 | 19.56667 | 41.53333 |
| 958.5829 | 19.56667 | 41.96667 |
| 958.8548 | 19.6 | 41.51667 |
| 959.1267 | 19.51667 | 41.98333 |
| 959.3986 | 19.5 | 41.93333 |
| 959.6705 | 19.61667 | 41.96667 |
| 959.9424 | 19.55 | 41.6 |
| 960.2142 | 19.53333 | 40.93333 |
| 960.4861 | 19.51667 | 41.51667 |
| 960.7579 | 19.51667 | 41.38333 |
| 961.0296 | 19.51667 | 41.06667 |
| 961.3014 | 19.5 | 41 |
| 961.5732 | 19.53333 | 42.18333 |
| 961.8449 | 19.51667 | 41.08333 |
| 962.1166 | 19.48333 | 40.75 |
| 962.3883 | 19.51667 | 40.75 |
| 962.66 | 19.55 | 40.68333 |
| 962.9316 | 19.51667 | 40.58333 |
| 963.2033 | 19.58333 | 41.03333 |
| 963.4749 | 19.53333 | 41.03333 |
| 963.7465 | 19.51667 | 40.95 |
| 964.0181 | 19.48333 | 40.85 |
| 964.2896 | 19.55 | 40.98333 |
| 964.5612 | 19.55 | 40.53333 |
| 964.8327 | 19.51667 | 40.26667 |
| 965.1042 | 19.51667 | 40.83333 |
| 965.3757 | 19.53333 | 40.58333 |
| 965.6471 | 19.51667 | 40.25 |
| 965.9186 | 19.53333 | 40.85 |
| 966.19 | 19.56667 | 40.06667 |
| 966.4614 | 19.51667 | 40.23333 |
| 966.7328 | 19.51667 | 39.81667 |
| 967.0042 | 19.53333 | 39.73333 |
| 967.2755 | 19.56667 | 40.08333 |
| 967.5469 | 19.55 | 39.88333 |
| 967.8182 | 19.51667 | 38.9 |
| 968.0895 | 19.48333 | 39.65 |
| 968.3607 | 19.53333 | 39.33333 |
| 968.632 | 19.56667 | 39.68333 |
| 968.9032 | 19.51667 | 39.73333 |
| 969.1745 | 19.55 | 39.4 |
| 969.4457 | 19.53333 | 40.18333 |
| 969.7168 | 19.51667 | 39.61667 |
| 969.988 | 19.51667 | 39.36667 |
| 970.2591 | 19.63333 | 39.01667 |
| 970.5303 | 19.51667 | 38.8 |
| 970.8014 | 19.5 | 38.86667 |
| 971.0724 | 19.48333 | 38.55 |
| 971.3435 | 19.51667 | 38.61667 |
| 971.6146 | 19.53333 | 38.71667 |
| 971.8856 | 19.55 | 38.93333 |
| 972.1566 | 19.51667 | 38.08333 |
| 972.4276 | 19.53333 | 38.8 |
| 972.6985 | 19.63333 | 38.86667 |
| 972.9695 | 19.56667 | 38.35 |
| 973.2404 | 19.56667 | 38.46667 |
| 973.5113 | 19.5 | 38.48333 |
| 973.7822 | 19.53333 | 37.91667 |
| 974.0531 | 19.55 | 38.16667 |
| 974.324 | 19.48333 | 37.76667 |
| 974.5948 | 19.6 | 37.93333 |
| 974.8656 | 19.6 | 37.75 |
| 975.1364 | 19.58333 | 37.46667 |
| 975.4072 | 19.51667 | 37.5 |
| 975.6779 | 19.48333 | 37.65 |
| 975.9487 | 19.56667 | 37.96667 |
| 976.2194 | 19.51667 | 37.58333 |
| 976.4901 | 19.55 | 37.03333 |
| 976.7608 | 19.55 | 37.33333 |
| 977.0314 | 19.53333 | 37.11667 |
| 977.3021 | 19.5 | 37.28333 |
| 977.5727 | 19.51667 | 37.26667 |
| 977.8433 | 19.55 | 37.21667 |
| 978.1139 | 19.5 | 37.16667 |
| 978.3845 | 19.5 | 36.96667 |
| 978.655 | 19.53333 | 36.96667 |
| 978.9255 | 19.58333 | 36.8 |
| 979.196 | 19.56667 | 36.38333 |
| 979.4665 | 19.5 | 36.86667 |
| 979.737 | 19.58333 | 36.55 |
| 980.0075 | 19.56667 | 36.23333 |
| 980.2779 | 19.51667 | 36.25 |
| 980.5483 | 19.5 | 36.06667 |
| 980.8187 | 19.53333 | 36.26667 |
| 981.0891 | 19.51667 | 36.38333 |
| 981.3594 | 19.48333 | 35.6 |
| 981.6298 | 19.48333 | 36.11667 |
| 981.9001 | 19.53333 | 35.81667 |
| 982.1704 | 19.53333 | 35.56667 |
| 982.4406 | 19.51667 | 35.85 |
| 982.7109 | 19.58333 | 35.8 |
| 982.9811 | 19.53333 | 35.21667 |
| 983.2514 | 19.51667 | 35.51667 |
| 983.5215 | 19.51667 | 35.55 |
| 983.7917 | 19.58333 | 35.13333 |
| 984.0619 | 19.6 | 35.5 |
| 984.332 | 19.61667 | 35.11667 |
| 984.6022 | 19.58333 | 34.73333 |
| 984.8723 | 19.53333 | 35.13333 |
| 985.1423 | 19.53333 | 35.03333 |
| 985.4124 | 19.51667 | 35.01667 |
| 985.6824 | 19.53333 | 34 |
| 985.9525 | 19.48333 | 34.8 |
| 986.2225 | 19.56667 | 33.98333 |
| 986.4925 | 19.56667 | 34.2 |
| 986.7624 | 19.48333 | 33.61667 |
| 987.0324 | 19.53333 | 33.9 |
| 987.3023 | 19.6 | 33.2 |
| 987.5722 | 19.5 | 33.51667 |
| 987.8421 | 19.58333 | 33.28333 |
| 988.112 | 19.56667 | 33.41667 |
| 988.3818 | 19.56667 | 32.53333 |
| 988.6516 | 19.58333 | 32.5 |
| 988.9214 | 19.51667 | 32.83333 |
| 989.1912 | 19.51667 | 32.31667 |
| 989.461 | 19.48333 | 32.18333 |
| 989.7308 | 19.51667 | 32.03333 |
| 990.0005 | 19.5 | 31.48333 |
| 990.2702 | 19.55 | 31.65 |
| 990.5399 | 19.51667 | 31.71667 |
| 990.8096 | 19.48333 | 30.8 |
| 991.0792 | 19.5 | 30.66667 |
| 991.3489 | 19.56667 | 30.6 |
| 991.6185 | 19.55 | 30.48333 |
| 991.8881 | 19.55 | 30.01667 |
| 992.1577 | 19.56667 | 30.46667 |
| 992.4272 | 19.55 | 29.9 |
| 992.6967 | 19.48333 | 29.8 |
| 992.9663 | 19.51667 | 29.26667 |
| 993.2358 | 19.55 | 29.18333 |
| 993.5052 | 19.51667 | 28.91667 |
| 993.7747 | 19.58333 | 28.61667 |
| 994.0441 | 19.53333 | 28.31667 |
| 994.3136 | 19.58333 | 28.18333 |
| 994.583 | 19.53333 | 28.23333 |
| 994.8523 | 19.53333 | 27.71667 |
| 995.1217 | 19.53333 | 27.41667 |
| 995.391 | 19.51667 | 27.31667 |
| 995.6604 | 19.5 | 27.13333 |
| 995.9297 | 19.55 | 26.86667 |
| 996.199 | 19.53333 | 26.53333 |
| 996.4682 | 19.56667 | 26.46667 |
| 996.7375 | 19.5 | 26.55 |
| 997.0067 | 19.68333 | 26.01667 |
| 997.2759 | 19.53333 | 25.8 |
| 997.5451 | 19.53333 | 25.61667 |
| 997.8142 | 19.51667 | 25.51667 |
| 998.0834 | 19.56667 | 25.11667 |
| 998.3525 | 19.5 | 24.61667 |
| 998.6216 | 19.55 | 24.68333 |
| 998.8907 | 19.51667 | 24.53333 |
| 999.1598 | 19.53333 | 24.45 |
| 999.4288 | 19.51667 | 24.05 |
| 999.6978 | 19.48333 | 24.06667 |
| 999.9668 | 19.5 | 23.9 |
| 1000.236 | 19.51667 | 23.6 |
| 1000.505 | 19.63333 | 23.48333 |
| 1000.774 | 19.51667 | 23.31667 |
| 1001.043 | 19.61667 | 23.11667 |
| 1001.312 | 19.51667 | 22.75 |
| 1001.58 | 19.53333 | 22.86667 |
| 1001.849 | 19.6 | 22.63333 |
| 1002.118 | 19.46667 | 22.53333 |
| 1002.387 | 19.51667 | 22.46667 |
| 1002.656 | 19.5 | 22.11667 |
| 1002.925 | 19.51667 | 22.18333 |
| 1003.193 | 19.55 | 21.88333 |
| 1003.462 | 19.48333 | 21.85 |
| 1003.731 | 19.53333 | 21.68333 |
| 1004 | 19.5 | 21.4 |
| 1004.268 | 19.56667 | 21.61667 |
| 1004.537 | 19.56667 | 21.33333 |
| 1004.806 | 19.55 | 21.23333 |
| 1005.074 | 19.53333 | 21.16667 |
| 1005.343 | 19.48333 | 21.06667 |
| 1005.611 | 19.56667 | 20.93333 |
| 1005.88 | 19.56667 | 20.96667 |
| 1006.149 | 19.53333 | 20.75 |

Data for Fig. 3

| Figure 3.A | Figure 3.A | Figure 3.A | Figure 3.A | Figure 3.B | Figure 3.B | Figure 3.C | Figure 3.C |
|---|---|---|---|---|---|---|---|
| Wavelengt | (Visible) Cc | Wavelengt | (nIR) Counts/s/m^2 T( | wavelengtl | P/s/sr/m^2/nm at T = | Wavelengt | Empirical Emissivity at |
| 437.661 | 2.73E+07 | 879.476 | -1.98E+08 | 600.015 | 2.64E+20 | 600.015 | 0.44002 |
| 437.9568 | 5.45E+07 | 881.2142 | -2.04E+08 | 600.306 | 2.65E+20 | 600.306 | 0.43998 |
| 438.2526 | 2.73E+07 | 882.9523 | -1.25E+08 | 600.597 | 2.66E+20 | 600.597 | 0.43993 |
| 438.5484 | 3.41E+07 | 884.6903 | -1.06E+08 | 600.888 | 2.66E+20 | 600.888 | 0.43988 |
| 438.8442 | 4.77E+07 | 886.4282 | -2.45E+08 | 601.179 | 2.67E+20 | 601.179 | 0.43983 |
| 439.1399 | 5.45E+07 | 888.166 | -1.53E+08 | 601.47 | 2.68E+20 | 601.47 | 0.43979 |
| 439.4357 | 2.04E+07 | 889.9037 | -1.05E+08 | 601.761 | 2.69E+20 | 601.761 | 0.43974 |
| 439.7314 | 6.13E+07 | 891.6413 | -1.56E+08 | 602.051 | 2.69E+20 | 602.051 | 0.4397 |
| 440.0272 | 6.82E+07 | 893.3788 | -1.79E+08 | 602.342 | 2.70E+20 | 602.342 | 0.43965 |
| 440.3229 | 4.09E+07 | 895.1162 | -7.27E+07 | 602.633 | 2.71E+20 | 602.633 | 0.4396 |
| 440.6187 | 5.45E+07 | 896.8535 | -6.59E+07 | 602.924 | 2.72E+20 | 602.924 | 0.43956 |
| 440.9144 | 6.82E+07 | 898.5908 | -1.93E+07 | 603.215 | 2.72E+20 | 603.215 | 0.43951 |
| 441.2101 | 5.45E+07 | 900.3279 | -4.77E+07 | 603.506 | 2.73E+20 | 603.506 | 0.43946 |
| 441.5058 | 6.82E+07 | 902.0649 | -2.85E+08 | 603.797 | 2.74E+20 | 603.797 | 0.43942 |
| 441.8015 | 2.73E+07 | 903.8019 | -1.41E+08 | 604.088 | 2.75E+20 | 604.088 | 0.43937 |
| 442.0972 | 5.45E+07 | 905.5387 | -9.09E+07 | 604.378 | 2.75E+20 | 604.378 | 0.43933 |
| 442.3929 | 4.77E+07 | 907.2754 | -8.52E+07 | 604.669 | 2.76E+20 | 604.669 | 0.43928 |
| 442.6886 | 2.73E+07 | 909.0121 | -5.11E+07 | 604.96 | 2.77E+20 | 604.96 | 0.43924 |
| 442.9843 | 6.82E+07 | 910.7486 | -2.27E+06 | 605.251 | 2.78E+20 | 605.251 | 0.43919 |
| 443.28 | 4.77E+07 | 912.4851 | -1.92E+08 | 605.542 | 2.78E+20 | 605.542 | 0.43915 |
| 443.5757 | 6.13E+07 | 914.2214 | 3.29E+07 | 605.833 | 2.79E+20 | 605.833 | 0.4391 |
| 443.8713 | 8.86E+07 | 915.9577 | 9.09E+07 | 606.123 | 2.80E+20 | 606.123 | 0.43906 |
| 444.167 | 5.45E+07 | 917.6939 | -8.63E+07 | 606.414 | 2.81E+20 | 606.414 | 0.43901 |
| 444.4626 | 5.45E+07 | 919.4299 | 5.00E+07 | 606.705 | 2.81E+20 | 606.705 | 0.43897 |
| 444.7583 | 4.09E+07 | 921.1659 | -3.75E+07 | 606.996 | 2.82E+20 | 606.996 | 0.43892 |
| 445.0539 | 6.13E+07 | 922.9017 | 1.11E+08 | 607.286 | 2.83E+20 | 607.286 | 0.43888 |
| 445.3495 | 5.45E+07 | 924.6375 | 1.33E+08 | 607.577 | 2.84E+20 | 607.577 | 0.43883 |
| 445.6451 | 4.09E+07 | 926.3732 | 2.16E+07 | 607.868 | 2.84E+20 | 607.868 | 0.43879 |
| 445.9408 | 8.18E+07 | 928.1088 | 3.41E+06 | 608.158 | 2.85E+20 | 608.158 | 0.43875 |
| 446.2364 | 5.45E+07 | 929.8442 | 1.32E+08 | 608.449 | 2.86E+20 | 608.449 | 0.4387 |
| 446.532 | 6.82E+07 | 931.5796 | 2.04E+07 | 608.74 | 2.87E+20 | 608.74 | 0.43866 |
| 446.8276 | 8.18E+07 | 933.3149 | -1.18E+08 | 609.03 | 2.88E+20 | 609.03 | 0.43862 |
| 447.1232 | 4.77E+07 | 935.05 | 4.09E+07 | 609.321 | 2.88E+20 | 609.321 | 0.43857 |
| 447.4187 | 4.09E+07 | 936.7851 | 1.36E+07 | 609.612 | 2.89E+20 | 609.612 | 0.43853 |
| 447.7143 | 5.45E+07 | 938.5201 | -2.50E+07 | 609.902 | 2.90E+20 | 609.902 | 0.43849 |
| 448.0099 | 6.82E+07 | 940.255 | 2.95E+07 | 610.193 | 2.91E+20 | 610.193 | 0.43844 |
| 448.3054 | 6.82E+07 | 941.9898 | -1.14E+07 | 610.484 | 2.91E+20 | 610.484 | 0.4384 |
| 448.601 | 6.82E+07 | 943.7244 | 6.47E+07 | 610.774 | 2.92E+20 | 610.774 | 0.43836 |
| 448.8965 | 6.13E+07 | 945.459 | 1.37E+08 | 611.065 | 2.93E+20 | 611.065 | 0.43832 |
| 449.1921 | 8.18E+07 | 947.1935 | 1.01E+08 | 611.355 | 2.94E+20 | 611.355 | 0.43827 |
| 449.4876 | 5.45E+07 | 948.9279 | 1.19E+08 | 611.646 | 2.95E+20 | 611.646 | 0.43823 |
| 449.7831 | 6.82E+07 | 950.6622 | 5.57E+07 | 611.936 | 2.95E+20 | 611.936 | 0.43819 |
| 450.0786 | 6.82E+07 | 952.3964 | 1.01E+08 | 612.227 | 2.96E+20 | 612.227 | 0.43815 |
| 450.3742 | 4.77E+07 | 954.1304 | 2.66E+08 | 612.518 | 2.97E+20 | 612.518 | 0.4381 |
| 450.6697 | 6.13E+07 | 955.8644 | 3.95E+08 | 612.808 | 2.98E+20 | 612.808 | 0.43806 |
| 450.9652 | 4.77E+07 | 957.5983 | 6.00E+08 | 613.099 | 2.98E+20 | 613.099 | 0.43802 |
| 451.2606 | 6.13E+07 | 959.3321 | 9.20E+08 | 613.389 | 2.99E+20 | 613.389 | 0.43798 |
| 451.5561 | 6.13E+07 | 961.0658 | 1.07E+09 | 613.68 | 3.00E+20 | 613.68 | 0.43794 |
| 451.8516 | 6.82E+07 | 962.7993 | 1.51E+09 | 613.97 | 3.01E+20 | 613.97 | 0.43789 |
| 452.1471 | 1.36E+07 | 964.5328 | 1.90E+09 | 614.261 | 3.02E+20 | 614.261 | 0.43785 |
| 452.4425 | 4.09E+07 | 966.2662 | 2.54E+09 | 614.551 | 3.02E+20 | 614.551 | 0.43781 |
| 452.738 | 4.77E+07 | 967.9995 | 3.12E+09 | 614.841 | 3.03E+20 | 614.841 | 0.43777 |
| 453.0334 | 4.09E+07 | 969.7327 | 3.83E+09 | 615.132 | 3.04E+20 | 615.132 | 0.43773 |
| 453.3289 | 6.82E+07 | 971.4657 | 4.44E+09 | 615.422 | 3.05E+20 | 615.422 | 0.43769 |
| 453.6243 | 1.02E+08 | 973.1987 | 5.08E+09 | 615.713 | 3.06E+20 | 615.713 | 0.43765 |
| 453.9197 | 6.82E+07 | 974.9316 | 5.90E+09 | 616.003 | 3.06E+20 | 616.003 | 0.43761 |
| 454.2152 | 6.13E+07 | 976.6643 | 6.65E+09 | 616.293 | 3.07E+20 | 616.293 | 0.43757 |
| 454.5106 | 6.13E+07 | 978.397 | 7.28E+09 | 616.584 | 3.08E+20 | 616.584 | 0.43753 |
| 454.806 | 4.77E+07 | 980.1296 | 8.20E+09 | 616.874 | 3.09E+20 | 616.874 | 0.43749 |
| 455.1014 | 6.82E+07 | 981.862 | 8.93E+09 | 617.165 | 3.10E+20 | 617.165 | 0.43745 |
| 455.3968 | 7.50E+07 | 983.5944 | 9.65E+09 | 617.455 | 3.10E+20 | 617.455 | 0.43741 |
| 455.6922 | 4.09E+07 | 985.3267 | 1.04E+10 | 617.745 | 3.11E+20 | 617.745 | 0.43737 |
| 455.9875 | 6.82E+07 | 987.0588 | 1.15E+10 | 618.036 | 3.12E+20 | 618.036 | 0.43733 |
| 456.2829 | 6.82E+07 | 988.7909 | 1.22E+10 | 618.326 | 3.13E+20 | 618.326 | 0.43729 |
| 456.5783 | 7.50E+07 | 990.5228 | 1.31E+10 | 618.616 | 3.14E+20 | 618.616 | 0.43725 |
| 456.8736 | 6.13E+07 | 992.2547 | 1.41E+10 | 618.906 | 3.14E+20 | 618.906 | 0.43721 |
| 457.169 | 6.82E+07 | 993.9864 | 1.48E+10 | 619.197 | 3.15E+20 | 619.197 | 0.43717 |
| 457.4643 | 7.50E+07 | 995.7181 | 1.56E+10 | 619.487 | 3.16E+20 | 619.487 | 0.43713 |
| 457.7597 | 3.41E+07 | 997.4496 | 1.68E+10 | 619.777 | 3.17E+20 | 619.777 | 0.43709 |
| 458.055 | 3.41E+07 | 999.1811 | 1.75E+10 | 620.068 | 3.18E+20 | 620.068 | 0.43705 |
| 458.3503 | 3.41E+07 | 1000.912 | 1.84E+10 | 620.358 | 3.18E+20 | 620.358 | 0.43701 |
| 458.6456 | 6.82E+07 | 1002.644 | 1.92E+10 | 620.648 | 3.19E+20 | 620.648 | 0.43698 |
| 458.9409 | 4.09E+07 | 1004.375 | 2.04E+10 | 620.938 | 3.20E+20 | 620.938 | 0.43694 |
| 459.2362 | 4.77E+07 | 1006.106 | 2.14E+10 | 621.228 | 3.21E+20 | 621.228 | 0.4369 |
| 459.5315 | 4.77E+07 | 1007.837 | 2.22E+10 | 621.519 | 3.22E+20 | 621.519 | 0.43686 |
| 459.8268 | 8.18E+07 | 1009.568 | 2.31E+10 | 621.809 | 3.23E+20 | 621.809 | 0.43682 |
| 460.1221 | 4.09E+07 | 1011.298 | 2.40E+10 | 622.099 | 3.23E+20 | 622.099 | 0.43678 |
| 460.4174 | 6.82E+07 | 1013.029 | 2.47E+10 | 622.389 | 3.24E+20 | 622.389 | 0.43675 |
| 460.7126 | 7.50E+07 | 1014.759 | 2.55E+10 | 622.679 | 3.25E+20 | 622.679 | 0.43671 |
| 461.0079 | 6.13E+07 | 1016.49 | 2.63E+10 | 622.969 | 3.26E+20 | 622.969 | 0.43667 |
| 461.3031 | 5.45E+07 | 1018.22 | 2.72E+10 | 623.259 | 3.27E+20 | 623.259 | 0.43663 |
| 461.5984 | 9.54E+07 | 1019.95 | 2.78E+10 | 623.55 | 3.27E+20 | 623.55 | 0.4366 |
| 461.8936 | 4.09E+07 | 1021.68 | 2.87E+10 | 623.84 | 3.28E+20 | 623.84 | 0.43656 |
| 462.1888 | 3.41E+07 | 1023.41 | 2.95E+10 | 624.13 | 3.29E+20 | 624.13 | 0.43652 |
| 462.4841 | 5.45E+07 | 1025.14 | 2.99E+10 | 624.42 | 3.30E+20 | 624.42 | 0.43649 |
| 462.7793 | 5.45E+07 | 1026.87 | 3.08E+10 | 624.71 | 3.31E+20 | 624.71 | 0.43645 |
| 463.0745 | 4.77E+07 | 1028.6 | 3.14E+10 | 625 | 3.32E+20 | 625 | 0.43641 |
| 463.3697 | 4.77E+07 | 1030.329 | 3.19E+10 | 625.29 | 3.32E+20 | 625.29 | 0.43638 |
| 463.6649 | 3.41E+07 | 1032.059 | 3.25E+10 | 625.58 | 3.33E+20 | 625.58 | 0.43634 |
| 463.9601 | 4.77E+07 | 1033.788 | 3.32E+10 | 625.87 | 3.34E+20 | 625.87 | 0.4363 |
| 464.2552 | 7.50E+07 | 1035.517 | 3.39E+10 | 626.16 | 3.35E+20 | 626.16 | 0.43627 |
| 464.5504 | 7.50E+07 | 1037.246 | 3.40E+10 | 626.45 | 3.36E+20 | 626.45 | 0.43623 |
| 464.8456 | 6.82E+07 | 1038.975 | 3.44E+10 | 626.74 | 3.37E+20 | 626.74 | 0.43619 |
| 465.1407 | 6.13E+07 | 1040.704 | 3.50E+10 | 627.03 | 3.37E+20 | 627.03 | 0.43616 |
| 465.4359 | 7.50E+07 | 1042.433 | 3.56E+10 | 627.32 | 3.38E+20 | 627.32 | 0.43612 |
| 465.731 | 6.82E+07 | 1044.162 | 3.60E+10 | 627.61 | 3.39E+20 | 627.61 | 0.43609 |
| 466.0261 | 8.18E+07 | 1045.89 | 3.64E+10 | 627.9 | 3.40E+20 | 627.9 | 0.43605 |
| 466.3213 | 5.45E+07 | 1047.619 | 3.68E+10 | 628.19 | 3.41E+20 | 628.19 | 0.43601 |

| Figure 3.D | Figure 3.D | Figure 3.D | Figure 3.D | Figure 3.D | Figure 3.D | Figure 3.D | Figure 3.D | Figure 3.D | Figure 3.D | Figure 3.D | Figure 3.D |
|---|---|---|---|---|---|---|---|---|---|---|---|
| Wavelengt | Savitsky-G( | Wavelengt | Savitsky-G( | Wavelengt | Raw ND Fil | Wavelengt | Raw ND Fil | Wavelengt | Interpolate | Wavelengt | Interpolated Data (nIR |
| 600.015 | 4.54E-08 | 1151.13 | 5.62E-08 | 600.015 | 4.20E-08 | 1101.15 | 5.85E-08 | 958.039 | 5.41E-08 | 988.791 | 5.55E-08 |
| 600.306 | 4.54E-08 | 1152.85 | 5.62E-08 | 600.306 | 4.82E-08 | 1102.87 | 5.46E-08 | 958.311 | 5.41E-08 | 990.523 | 5.56E-08 |
| 600.597 | 4.54E-08 | 1154.57 | 5.61E-08 | 600.597 | 4.29E-08 | 1104.6 | 5.66E-08 | 958.583 | 5.41E-08 | 992.255 | 5.56E-08 |
| 600.888 | 4.53E-08 | 1156.29 | 5.60E-08 | 600.888 | 5.68E-08 | 1106.32 | 5.80E-08 | 958.855 | 5.41E-08 | 993.986 | 5.57E-08 |
| 601.179 | 4.53E-08 | 1158.02 | 5.61E-08 | 601.179 | 4.97E-08 | 1108.05 | 5.58E-08 | 959.127 | 5.41E-08 | 995.718 | 5.58E-08 |
| 601.47 | 4.53E-08 | 1159.74 | 5.59E-08 | 601.47 | 4.15E-08 | 1109.77 | 5.74E-08 | 959.399 | 5.42E-08 | 997.45 | 5.58E-08 |
| 601.761 | 4.52E-08 | 1161.46 | 5.58E-08 | 601.761 | 4.18E-08 | 1111.5 | 5.51E-08 | 959.671 | 5.42E-08 | 999.181 | 5.59E-08 |
| 602.051 | 4.52E-08 | 1163.18 | 5.58E-08 | 602.051 | 4.86E-08 | 1113.22 | 5.57E-08 | 959.942 | 5.42E-08 | 1000.91 | 5.59E-08 |
| 602.342 | 4.52E-08 | 1164.9 | 5.58E-08 | 602.342 | 4.49E-08 | 1114.95 | 5.68E-08 | 960.214 | 5.42E-08 | 1002.64 | 5.60E-08 |
| 602.633 | 4.52E-08 | 1166.62 | 5.58E-08 | 602.633 | 4.18E-08 | 1116.67 | 5.40E-08 | 960.486 | 5.42E-08 | 1004.37 | 5.60E-08 |
| 602.924 | 4.51E-08 | 1168.34 | 5.57E-08 | 602.924 | 4.70E-08 | 1118.39 | 5.64E-08 | 960.758 | 5.42E-08 | 1006.11 | 5.61E-08 |
| 603.215 | 4.51E-08 | 1170.06 | 5.56E-08 | 603.215 | 4.82E-08 | 1120.12 | 5.68E-08 | 961.03 | 5.43E-08 | 1007.84 | 5.61E-08 |
| 603.506 | 4.51E-08 | 1171.78 | 5.57E-08 | 603.506 | 3.79E-08 | 1121.84 | 5.53E-08 | 961.301 | 5.43E-08 | 1009.57 | 5.62E-08 |
| 603.797 | 4.51E-08 | 1173.5 | 5.56E-08 | 603.797 | 4.38E-08 | 1123.57 | 5.77E-08 | 961.573 | 5.43E-08 | 1011.3 | 5.62E-08 |
| 604.088 | 4.50E-08 | 1175.22 | 5.56E-08 | 604.088 | 4.42E-08 | 1125.29 | 5.54E-08 | 961.845 | 5.43E-08 | 1013.03 | 5.63E-08 |
| 604.378 | 4.50E-08 | 1176.94 | 5.55E-08 | 604.378 | 5.14E-08 | 1127.01 | 5.64E-08 | 962.117 | 5.43E-08 | 1014.76 | 5.63E-08 |
| 604.669 | 4.50E-08 | 1178.66 | 5.54E-08 | 604.669 | 4.53E-08 | 1128.74 | 5.72E-08 | 962.388 | 5.43E-08 | 1016.49 | 5.64E-08 |
| 604.96 | 4.49E-08 | 1180.38 | 5.53E-08 | 604.96 | 4.13E-08 | 1130.46 | 5.45E-08 | 962.66 | 5.43E-08 | 1018.22 | 5.64E-08 |
| 605.251 | 4.49E-08 | 1182.1 | 5.54E-08 | 605.251 | 4.16E-08 | 1132.18 | 5.58E-08 | 962.932 | 5.44E-08 | 1019.95 | 5.64E-08 |
| 605.542 | 4.49E-08 | 1183.82 | 5.53E-08 | 605.542 | 4.68E-08 | 1133.91 | 5.55E-08 | 963.203 | 5.44E-08 | 1021.68 | 5.65E-08 |
| 605.833 | 4.49E-08 | 1185.54 | 5.52E-08 | 605.833 | 4.38E-08 | 1135.63 | 5.64E-08 | 963.475 | 5.44E-08 | 1023.41 | 5.65E-08 |
| 606.123 | 4.49E-08 | 1187.26 | 5.50E-08 | 606.123 | 4.29E-08 | 1137.35 | 5.71E-08 | 963.746 | 5.44E-08 | 1025.14 | 5.65E-08 |
| 606.414 | 4.48E-08 | 1188.98 | 5.50E-08 | 606.414 | 4.29E-08 | 1139.07 | 5.45E-08 | 964.018 | 5.44E-08 | 1026.87 | 5.66E-08 |
| 606.705 | 4.48E-08 | 1190.7 | 5.51E-08 | 606.705 | 4.36E-08 | 1140.8 | 5.65E-08 | 964.29 | 5.44E-08 | 1028.6 | 5.66E-08 |
| 606.996 | 4.45E-08 | 1192.42 | 5.50E-08 | 606.996 | 4.25E-08 | 1142.52 | 5.60E-08 | 964.561 | 5.44E-08 | 1030.33 | 5.66E-08 |
| 607.286 | 4.43E-08 | 1194.14 | 5.50E-08 | 607.286 | 4.06E-08 | 1144.24 | 5.55E-08 | 964.833 | 5.45E-08 | 1032.06 | 5.67E-08 |
| 607.577 | 4.43E-08 | 1195.86 | 5.50E-08 | 607.577 | 4.53E-08 | 1145.96 | 5.60E-08 | 965.104 | 5.45E-08 | 1033.79 | 5.67E-08 |
| 607.868 | 4.44E-08 | 1197.58 | 5.50E-08 | 607.868 | 4.04E-08 | 1147.69 | 5.63E-08 | 965.376 | 5.45E-08 | 1035.52 | 5.67E-08 |
| 608.158 | 4.43E-08 | 1199.3 | 5.50E-08 | 608.158 | 4.97E-08 | 1149.41 | 5.82E-08 | 965.647 | 5.45E-08 | 1037.25 | 5.67E-08 |
| 608.449 | 4.43E-08 | 1201.02 | 5.49E-08 | 608.449 | 3.77E-08 | 1151.13 | 5.59E-08 | 965.919 | 5.45E-08 | 1038.98 | 5.68E-08 |
| 608.74 | 4.43E-08 | 1202.73 | 5.48E-08 | 608.74 | 4.63E-08 | 1152.85 | 5.59E-08 | 966.19 | 5.45E-08 | 1040.7 | 5.68E-08 |
| 609.03 | 4.43E-08 | 1204.45 | 5.48E-08 | 609.03 | 4.59E-08 | 1154.57 | 5.67E-08 | 966.461 | 5.45E-08 | 1042.43 | 5.68E-08 |
| 609.321 | 4.41E-08 | 1206.17 | 5.48E-08 | 609.321 | 5.09E-08 | 1156.29 | 5.55E-08 | 966.733 | 5.45E-08 | 1044.16 | 5.68E-08 |
| 609.612 | 4.42E-08 | 1207.89 | 5.47E-08 | 609.612 | 4.85E-08 | 1158.02 | 5.56E-08 | 967.004 | 5.46E-08 | 1045.89 | 5.68E-08 |
| 609.902 | 4.42E-08 | 1209.61 | 5.46E-08 | 609.902 | 3.87E-08 | 1159.74 | 5.76E-08 | 967.276 | 5.46E-08 | 1047.62 | 5.69E-08 |
| 610.193 | 4.42E-08 | 1211.32 | 5.46E-08 | 610.193 | 4.76E-08 | 1161.46 | 5.67E-08 | 967.547 | 5.46E-08 | 1049.35 | 5.69E-08 |
| 610.484 | 4.41E-08 | 1213.04 | 5.46E-08 | 610.484 | 4.83E-08 | 1163.18 | 5.53E-08 | 967.818 | 5.46E-08 | 1051.08 | 5.69E-08 |
| 610.774 | 4.40E-08 | 1214.76 | 5.46E-08 | 610.774 | 4.47E-08 | 1164.9 | 5.57E-08 | 968.089 | 5.46E-08 | 1052.8 | 5.69E-08 |
| 611.065 | 4.41E-08 | 1216.48 | 5.45E-08 | 611.065 | 4.99E-08 | 1166.62 | 5.62E-08 | 968.361 | 5.46E-08 | 1054.53 | 5.69E-08 |
| 611.355 | 4.40E-08 | 1218.19 | 5.44E-08 | 611.355 | 4.32E-08 | 1168.34 | 5.68E-08 | 968.632 | 5.46E-08 | 1056.26 | 5.69E-08 |
| 611.646 | 4.39E-08 | 1219.91 | 5.43E-08 | 611.646 | 4.00E-08 | 1170.06 | 5.53E-08 | 968.903 | 5.47E-08 | 1057.99 | 5.69E-08 |
| 611.936 | 4.39E-08 | 1221.63 | 5.43E-08 | 611.936 | 4.32E-08 | 1171.78 | 5.42E-08 | 969.174 | 5.47E-08 | 1059.72 | 5.69E-08 |
| 612.227 | 4.39E-08 | 1223.35 | 5.43E-08 | 612.227 | 4.47E-08 | 1173.5 | 5.57E-08 | 969.446 | 5.47E-08 | 1061.44 | 5.69E-08 |
| 612.518 | 4.38E-08 | 1225.06 | 5.42E-08 | 612.518 | 4.28E-08 | 1175.22 | 5.62E-08 | 969.717 | 5.47E-08 | 1063.17 | 5.70E-08 |
| 612.808 | 4.39E-08 | 1226.78 | 5.42E-08 | 612.808 | 4.36E-08 | 1176.94 | 5.37E-08 | 969.988 | 5.47E-08 | 1064.9 | 5.70E-08 |
| 613.099 | 4.39E-08 | 1228.5 | 5.41E-08 | 613.099 | 4.26E-08 | 1178.66 | 5.34E-08 | 970.259 | 5.47E-08 | 1066.63 | 5.70E-08 |
| 613.389 | 4.38E-08 | 1230.21 | 5.40E-08 | 613.389 | 4.24E-08 | 1180.38 | 5.58E-08 | 970.53 | 5.47E-08 | 1068.35 | 5.70E-08 |
| 613.68 | 4.39E-08 | 1231.93 | 5.40E-08 | 613.68 | 4.33E-08 | 1182.1 | 5.52E-08 | 970.801 | 5.47E-08 | 1070.08 | 5.70E-08 |
| 613.97 | 4.39E-08 | 1233.65 | 5.40E-08 | 613.97 | 4.80E-08 | 1183.82 | 5.63E-08 | 971.072 | 5.48E-08 | 1071.81 | 5.70E-08 |
| 614.261 | 4.37E-08 | 1235.36 | 5.40E-08 | 614.261 | 4.16E-08 | 1185.54 | 5.63E-08 | 971.344 | 5.48E-08 | 1073.53 | 5.70E-08 |
| 614.551 | 4.38E-08 | 1237.08 | 5.39E-08 | 614.551 | 4.46E-08 | 1187.26 | 5.45E-08 | 971.615 | 5.48E-08 | 1075.26 | 5.70E-08 |
| 614.841 | 4.37E-08 | 1238.79 | 5.38E-08 | 614.841 | 4.37E-08 | 1188.98 | 5.61E-08 | 971.886 | 5.48E-08 | 1076.99 | 5.70E-08 |
| 615.132 | 4.35E-08 | 1240.51 | 5.37E-08 | 615.132 | 4.32E-08 | 1190.7 | 5.52E-08 | 972.157 | 5.48E-08 | 1078.71 | 5.70E-08 |
| 615.422 | 4.34E-08 | 1242.23 | 5.36E-08 | 615.422 | 4.22E-08 | 1192.42 | 5.51E-08 | 972.428 | 5.48E-08 | 1080.44 | 5.70E-08 |
| 615.713 | 4.33E-08 | 1243.94 | 5.35E-08 | 615.713 | 4.11E-08 | 1194.14 | 5.47E-08 | 972.699 | 5.48E-08 | 1082.17 | 5.69E-08 |
| 616.003 | 4.33E-08 | 1245.66 | 5.35E-08 | 616.003 | 4.73E-08 | 1195.86 | 5.41E-08 | 972.969 | 5.49E-08 | 1083.89 | 5.69E-08 |
| 616.293 | 4.32E-08 | 1247.37 | 5.34E-08 | 616.293 | 4.37E-08 | 1197.58 | 5.63E-08 | 973.24 | 5.49E-08 | 1085.62 | 5.69E-08 |
| 616.584 | 4.30E-08 | 1249.09 | 5.34E-08 | 616.584 | 4.23E-08 | 1199.3 | 5.59E-08 | 973.511 | 5.49E-08 | 1087.34 | 5.69E-08 |
| 616.874 | 4.29E-08 | 1250.8 | 5.33E-08 | 616.874 | 4.45E-08 | 1201.02 | 5.38E-08 | 973.782 | 5.49E-08 | 1089.07 | 5.69E-08 |
| 617.165 | 4.27E-08 | 1252.52 | 5.32E-08 | 617.165 | 4.10E-08 | 1202.73 | 5.40E-08 | 974.053 | 5.49E-08 | 1090.8 | 5.69E-08 |
| 617.455 | 4.27E-08 | 1254.23 | 5.32E-08 | 617.455 | 4.00E-08 | 1204.45 | 5.37E-08 | 974.324 | 5.49E-08 | 1092.52 | 5.69E-08 |
| 617.745 | 4.27E-08 | 1255.95 | 5.31E-08 | 617.745 | 4.62E-08 | 1206.17 | 5.51E-08 | 974.595 | 5.49E-08 | 1094.25 | 5.69E-08 |
| 618.036 | 4.27E-08 | 1257.66 | 5.31E-08 | 618.036 | 4.20E-08 | 1207.89 | 5.55E-08 | 974.866 | 5.49E-08 | 1095.97 | 5.69E-08 |
| 618.326 | 4.27E-08 | 1259.38 | 5.30E-08 | 618.326 | 3.93E-08 | 1209.61 | 5.48E-08 | 975.136 | 5.49E-08 | 1097.7 | 5.68E-08 |
| 618.616 | 4.26E-08 | 1261.09 | 5.30E-08 | 618.616 | 4.71E-08 | 1211.32 | 5.50E-08 | 975.407 | 5.50E-08 | 1099.42 | 5.68E-08 |
| 618.906 | 4.26E-08 | 1262.8 | 5.29E-08 | 618.906 | 4.09E-08 | 1213.04 | 5.37E-08 | 975.678 | 5.50E-08 | 1101.15 | 5.68E-08 |
| 619.197 | 4.26E-08 | 1264.52 | 5.29E-08 | 619.197 | 3.92E-08 | 1214.76 | 5.41E-08 | 975.949 | 5.50E-08 | 1102.87 | 5.68E-08 |
| 619.487 | 4.26E-08 | 1266.23 | 5.29E-08 | 619.487 | 4.88E-08 | 1216.48 | 5.48E-08 | 976.219 | 5.50E-08 | 1104.6 | 5.68E-08 |
| 619.777 | 4.25E-08 | 1267.95 | 5.29E-08 | 619.777 | 3.90E-08 | 1218.19 | 5.41E-08 | 976.49 | 5.50E-08 | 1106.32 | 5.68E-08 |
| 620.068 | 4.22E-08 | 1269.66 | 5.27E-08 | 620.068 | 4.02E-08 | 1219.91 | 5.49E-08 | 976.761 | 5.50E-08 | 1108.05 | 5.67E-08 |
| 620.358 | 4.22E-08 | 1271.37 | 5.26E-08 | 620.358 | 4.39E-08 | 1221.63 | 5.52E-08 | 977.031 | 5.50E-08 | 1109.77 | 5.67E-08 |
| 620.648 | 4.20E-08 | 1273.09 | 5.25E-08 | 620.648 | 4.17E-08 | 1223.35 | 5.45E-08 | 977.302 | 5.50E-08 | 1111.5 | 5.67E-08 |
| 620.938 | 4.20E-08 | 1274.8 | 5.25E-08 | 620.938 | 3.95E-08 | 1225.06 | 5.40E-08 | 977.573 | 5.51E-08 | 1113.22 | 5.67E-08 |
| 621.228 | 4.19E-08 | 1276.51 | 5.24E-08 | 621.228 | 4.43E-08 | 1226.78 | 5.43E-08 | 977.843 | 5.51E-08 | 1114.95 | 5.67E-08 |
| 621.519 | 4.19E-08 | 1278.23 | 5.23E-08 | 621.519 | 4.39E-08 | 1228.5 | 5.45E-08 | 978.114 | 5.51E-08 | 1116.67 | 5.66E-08 |
| 621.809 | 4.19E-08 | 1279.94 | 5.23E-08 | 621.809 | 4.13E-08 | 1230.21 | 5.38E-08 | 978.384 | 5.51E-08 | 1118.39 | 5.66E-08 |
| 622.099 | 4.19E-08 | 1281.65 | 5.22E-08 | 622.099 | 4.29E-08 | 1231.93 | 5.39E-08 | 978.655 | 5.51E-08 | 1120.12 | 5.66E-08 |
| 622.389 | 4.18E-08 | 1283.37 | 5.21E-08 | 622.389 | 3.97E-08 | 1233.65 | 5.45E-08 | 978.926 | 5.51E-08 | 1121.84 | 5.66E-08 |
| 622.679 | 4.19E-08 | 1285.08 | 5.21E-08 | 622.679 | 4.11E-08 | 1235.36 | 5.36E-08 | 979.196 | 5.51E-08 | 1123.57 | 5.65E-08 |
| 622.969 | 4.18E-08 | 1286.79 | 5.20E-08 | 622.969 | 4.15E-08 | 1237.08 | 5.29E-08 | 979.467 | 5.51E-08 | 1125.29 | 5.65E-08 |
| 623.259 | 4.18E-08 | 1288.5 | 5.20E-08 | 623.259 | 4.07E-08 | 1238.79 | 5.39E-08 | 979.737 | 5.52E-08 | 1127.01 | 5.65E-08 |
| 623.55 | 4.19E-08 | 1290.22 | 5.19E-08 | 623.55 | 4.29E-08 | 1240.51 | 5.39E-08 | 980.13 | 5.52E-08 | 1128.74 | 5.64E-08 |
| 623.84 | 4.19E-08 | 1291.93 | 5.19E-08 | 623.84 | 4.33E-08 | 1242.23 | 5.32E-08 | 981.862 | 5.52E-08 | 1130.46 | 5.64E-08 |
| 624.13 | 4.19E-08 | 1293.64 | 5.18E-08 | 624.13 | 4.26E-08 | 1243.94 | 5.38E-08 | 983.594 | 5.53E-08 | 1132.18 | 5.64E-08 |
| 624.42 | 4.19E-08 | 1295.35 | 5.18E-08 | 624.42 | 4.05E-08 | 1245.66 | 5.34E-08 | 985.327 | 5.54E-08 | 1133.91 | 5.64E-08 |
| 624.71 | 4.19E-08 | 1297.06 | 5.17E-08 | 624.71 | 4.23E-08 | 1247.37 | 5.29E-08 | 987.059 | 5.54E-08 | 1135.63 | 5.63E-08 |
| 625 | 4.20E-08 | 1298.78 | 5.16E-08 | 625 | 4.44E-08 | 1249.09 | 5.30E-08 | 988.791 | 5.55E-08 | 1137.35 | 5.63E-08 |
| 625.29 | 4.21E-08 | 1300.49 | 5.14E-08 | 625.29 | 4.04E-08 | 1250.8 | 5.45E-08 | 990.523 | 5.56E-08 | 1139.07 | 5.63E-08 |
| 625.58 | 4.20E-08 | 1302.2 | 5.13E-08 | 625.58 | 3.86E-08 | 1252.52 | 5.43E-08 | 992.255 | 5.56E-08 | 1140.8 | 5.62E-08 |
| 625.87 | 4.20E-08 | 1303.91 | 5.12E-08 | 625.87 | 3.79E-08 | 1254.23 | 5.21E-08 | 993.986 | 5.57E-08 | 1142.52 | 5.62E-08 |
| 626.16 | 4.19E-08 | 1305.62 | 5.12E-08 | 626.16 | 4.17E-08 | 1255.95 | 5.29E-08 | 995.718 | 5.58E-08 | 1144.24 | 5.62E-08 |
| 626.45 | 4.20E-08 | 1307.33 | 5.11E-08 | 626.45 | 3.70E-08 | 1257.66 | 5.40E-08 | 997.45 | 5.58E-08 | 1145.96 | 5.61E-08 |
| 626.74 | 4.20E-08 | 1309.04 | 5.11E-08 | 626.74 | 4.20E-08 | 1259.38 | 5.24E-08 | 999.181 | 5.59E-08 | 1147.69 | 5.61E-08 |
| 627.03 | 4.22E-08 | 1310.75 | 5.10E-08 | 627.03 | 4.02E-08 | 1261.09 | 5.24E-08 | | | 1149.41 | 5.61E-08 |
| 627.32 | 4.22E-08 | 1312.47 | 5.09E-08 | 627.32 | 4.02E-08 | 1262.8 | 5.31E-08 | | | 1151.13 | 5.60E-08 |
| 627.61 | 4.21E-08 | 1314.18 | 5.08E-08 | 627.61 | 4.31E-08 | 1264.52 | 5.28E-08 | | | | |
| 627.9 | 4.20E-08 | 1315.89 | 5.06E-08 | 627.9 | 4.46E-08 | 1266.23 | 5.32E-08 | | | | |
| 628.19 | 4.21E-08 | 1317.6 | 5.05E-08 | 628.19 | 4.12E-08 | 1267.95 | 5.24E-08 | | | | |

| Figure 3.E | Figure 3.E | Figure 3.E | Figure 3.E | Figure 3.E | Figure 3.E | Figure 3.F | Figure 3.F | Figure 3.F | Figure 3.F | Figure 3.F | Figure 3.F | Figure 3.F | Figure 3.F |
|---|---|---|---|---|---|---|---|---|---|---|---|---|---|
| Wavelengt | Photons/s/ | Wavelengt | Interpolati | Wavelengt | Interpolation nIR | Wavelengt | Sensitivity | Wavelengt | Sensitivity | Wavelengt | Interpolati | Wavelengt | Interpolatic |
| 600.015 | 5.28E+12 | 960.214 | 3.24E+13 | 1000.91 | 3.57E+13 | 600.015 | 0.00499 | 1151.13 | 0.00144 | 960.214 | 0.00326 | 1000.91 | 5.16E-04 |
| 600.306 | 5.29E+12 | 960.486 | 3.25E+13 | 1002.64 | 3.58E+13 | 600.306 | 0.00518 | 1152.85 | 0.00145 | 960.486 | 0.00322 | 1002.64 | 5.38E-04 |
| 600.597 | 5.30E+12 | 960.758 | 3.25E+13 | 1004.37 | 3.59E+13 | 600.597 | 0.00529 | 1154.57 | 0.00146 | 960.758 | 0.00322 | 1004.37 | 5.69E-04 |
| 600.888 | 5.31E+12 | 961.03 | 3.25E+13 | 1006.11 | 3.60E+13 | 600.888 | 0.00549 | 1156.29 | 0.00146 | 961.03 | 0.00321 | 1006.11 | 5.94E-04 |
| 601.179 | 5.32E+12 | 961.301 | 3.25E+13 | 1007.84 | 3.61E+13 | 601.179 | 0.00558 | 1158.02 | 0.00147 | 961.301 | 0.0032 | 1007.84 | 6.15E-04 |
| 601.47 | 5.33E+12 | 961.573 | 3.26E+13 | 1009.57 | 3.63E+13 | 601.47 | 0.0057 | 1159.74 | 0.00148 | 961.573 | 0.00319 | 1009.57 | 6.38E-04 |
| 601.761 | 5.34E+12 | 961.845 | 3.26E+13 | 1011.3 | 3.64E+13 | 601.761 | 0.00593 | 1161.46 | 0.00149 | 961.845 | 0.00316 | 1011.3 | 6.61E-04 |
| 602.051 | 5.35E+12 | 962.117 | 3.26E+13 | 1013.03 | 3.65E+13 | 602.051 | 0.00604 | 1163.18 | 0.00149 | 962.117 | 0.00315 | 1013.03 | 6.78E-04 |
| 602.342 | 5.37E+12 | 962.388 | 3.26E+13 | 1014.76 | 3.66E+13 | 602.342 | 0.00628 | 1164.9 | 0.00149 | 962.388 | 0.00313 | 1014.76 | 6.99E-04 |
| 602.633 | 5.38E+12 | 962.66 | 3.27E+13 | 1016.49 | 3.67E+13 | 602.633 | 0.00647 | 1166.62 | 0.00149 | 962.66 | 0.00312 | 1016.49 | 7.19E-04 |
| 602.924 | 5.39E+12 | 962.932 | 3.27E+13 | 1018.22 | 3.68E+13 | 602.924 | 0.00647 | 1168.34 | 0.00149 | 962.932 | 0.0031 | 1018.22 | 7.40E-04 |
| 603.215 | 5.40E+12 | 963.203 | 3.27E+13 | 1019.95 | 3.69E+13 | 603.215 | 0.00665 | 1170.06 | 0.00149 | 963.203 | 0.00313 | 1019.95 | 7.54E-04 |
| 603.506 | 5.41E+12 | 963.475 | 3.27E+13 | 1021.68 | 3.70E+13 | 603.506 | 0.00672 | 1171.78 | 0.00151 | 963.475 | 0.00312 | 1021.68 | 7.76E-04 |
| 603.797 | 5.42E+12 | 963.746 | 3.28E+13 | 1023.41 | 3.71E+13 | 603.797 | 0.00692 | 1173.5 | 0.00151 | 963.746 | 0.00309 | 1023.41 | 7.95E-04 |
| 604.088 | 5.43E+12 | 964.018 | 3.28E+13 | 1025.14 | 3.72E+13 | 604.088 | 0.00714 | 1175.22 | 0.00151 | 964.018 | 0.00309 | 1025.14 | 8.05E-04 |
| 604.378 | 5.44E+12 | 964.29 | 3.28E+13 | 1026.87 | 3.73E+13 | 604.378 | 0.00721 | 1176.94 | 0.00152 | 964.29 | 0.00307 | 1026.87 | 8.27E-04 |
| 604.669 | 5.45E+12 | 964.561 | 3.28E+13 | 1028.6 | 3.74E+13 | 604.669 | 0.00735 | 1178.66 | 0.00152 | 964.561 | 0.00306 | 1028.6 | 8.40E-04 |
| 604.96 | 5.46E+12 | 964.833 | 3.28E+13 | 1030.33 | 3.75E+13 | 604.96 | 0.00748 | 1180.38 | 0.00153 | 964.833 | 0.00303 | 1030.33 | 8.50E-04 |
| 605.251 | 5.48E+12 | 965.104 | 3.29E+13 | 1032.06 | 3.76E+13 | 605.251 | 0.00759 | 1182.1 | 0.00153 | 965.104 | 0.00301 | 1032.06 | 8.67E-04 |
| 605.542 | 5.49E+12 | 965.376 | 3.29E+13 | 1033.79 | 3.77E+13 | 605.542 | 0.00773 | 1183.82 | 0.00153 | 965.376 | 0.00301 | 1033.79 | 8.82E-04 |
| 605.833 | 5.50E+12 | 965.647 | 3.29E+13 | 1035.52 | 3.78E+13 | 605.833 | 0.00788 | 1185.54 | 0.00154 | 965.647 | 0.00298 | 1035.52 | 8.98E-04 |
| 606.123 | 5.52E+12 | 965.919 | 3.29E+13 | 1037.25 | 3.79E+13 | 606.123 | 0.00804 | 1187.26 | 0.00154 | 965.919 | 0.00296 | 1037.25 | 9.00E-04 |
| 606.414 | 5.52E+12 | 966.19 | 3.30E+13 | 1038.98 | 3.80E+13 | 606.414 | 0.00818 | 1188.98 | 0.00155 | 966.19 | 0.00294 | 1038.98 | 9.08E-04 |
| 606.705 | 5.53E+12 | 966.461 | 3.30E+13 | 1040.7 | 3.81E+13 | 606.705 | 0.00816 | 1190.7 | 0.00155 | 966.461 | 0.00296 | 1040.7 | 9.20E-04 |
| 606.996 | 5.51E+12 | 966.733 | 3.30E+13 | 1042.43 | 3.81E+13 | 606.996 | 0.00845 | 1192.42 | 0.00156 | 966.733 | 0.00293 | 1042.43 | 9.35E-04 |
| 607.286 | 5.50E+12 | 967.004 | 3.30E+13 | 1044.16 | 3.82E+13 | 607.286 | 0.00859 | 1194.14 | 0.00156 | 967.004 | 0.00292 | 1044.16 | 9.42E-04 |
| 607.577 | 5.52E+12 | 967.276 | 3.31E+13 | 1045.89 | 3.83E+13 | 607.577 | 0.00869 | 1195.86 | 0.00156 | 967.276 | 0.0029 | 1045.89 | 9.50E-04 |
| 607.868 | 5.54E+12 | 967.547 | 3.31E+13 | 1047.62 | 3.84E+13 | 607.868 | 0.00877 | 1197.58 | 0.00157 | 967.547 | 0.00288 | 1047.62 | 9.60E-04 |
| 608.158 | 5.55E+12 | 967.818 | 3.31E+13 | 1049.35 | 3.85E+13 | 608.158 | 0.0089 | 1199.3 | 0.00157 | 967.818 | 0.00288 | 1049.35 | 9.69E-04 |
| 608.449 | 5.55E+12 | 968.089 | 3.31E+13 | 1051.08 | 3.86E+13 | 608.449 | 0.00893 | 1201.02 | 0.00157 | 968.089 | 0.00286 | 1051.08 | 9.78E-04 |
| 608.74 | 5.58E+12 | 968.361 | 3.31E+13 | 1052.8 | 3.86E+13 | 608.74 | 0.00903 | 1202.73 | 0.00158 | 968.361 | 0.00286 | 1052.8 | 9.86E-04 |
| 609.03 | 5.58E+12 | 968.632 | 3.32E+13 | 1054.53 | 3.87E+13 | 609.03 | 0.00908 | 1204.45 | 0.00158 | 968.632 | 0.00285 | 1054.53 | 9.97E-04 |
| 609.321 | 5.58E+12 | 968.903 | 3.32E+13 | 1056.26 | 3.88E+13 | 609.321 | 0.0092 | 1206.17 | 0.00158 | 968.903 | 0.00283 | 1056.26 | 9.99E-04 |
| 609.612 | 5.61E+12 | 969.174 | 3.32E+13 | 1057.99 | 3.89E+13 | 609.612 | 0.00925 | 1207.89 | 0.00159 | 969.174 | 0.00281 | 1057.99 | 0.00101 |
| 609.902 | 5.61E+12 | 969.446 | 3.32E+13 | 1059.72 | 3.89E+13 | 609.902 | 0.00932 | 1209.61 | 0.0016 | 969.446 | 0.00281 | 1059.72 | 0.00102 |
| 610.193 | 5.64E+12 | 969.717 | 3.33E+13 | 1061.44 | 3.90E+13 | 610.193 | 0.00932 | 1211.32 | 0.0016 | 969.717 | 0.00278 | 1061.44 | 0.00103 |
| 610.484 | 5.63E+12 | 969.988 | 3.33E+13 | 1063.17 | 3.91E+13 | 610.484 | 0.0094 | 1213.04 | 0.0016 | 969.988 | 0.00277 | 1063.17 | 0.00104 |
| 610.774 | 5.63E+12 | 970.259 | 3.33E+13 | 1064.9 | 3.91E+13 | 610.774 | 0.00933 | 1214.76 | 0.00161 | 970.259 | 0.00276 | 1064.9 | 0.00105 |
| 611.065 | 5.66E+12 | 970.53 | 3.33E+13 | 1066.63 | 3.92E+13 | 611.065 | 0.00928 | 1216.48 | 0.00162 | 970.53 | 0.00272 | 1066.63 | 0.00106 |
| 611.355 | 5.67E+12 | 970.801 | 3.34E+13 | 1068.35 | 3.93E+13 | 611.355 | 0.00934 | 1218.19 | 0.00163 | 970.801 | 0.00273 | 1068.35 | 0.00107 |
| 611.646 | 5.66E+12 | 971.072 | 3.34E+13 | 1070.08 | 3.93E+13 | 611.646 | 0.0094 | 1219.91 | 0.00163 | 971.072 | 0.00272 | 1070.08 | 0.00108 |
| 611.936 | 5.69E+12 | 971.344 | 3.34E+13 | 1071.81 | 3.94E+13 | 611.936 | 0.00936 | 1221.63 | 0.00163 | 971.344 | 0.00269 | 1071.81 | 0.00108 |
| 612.227 | 5.70E+12 | 971.615 | 3.34E+13 | 1073.53 | 3.95E+13 | 612.227 | 0.00931 | 1223.35 | 0.00163 | 971.615 | 0.00268 | 1073.53 | 0.00109 |
| 612.518 | 5.70E+12 | 971.886 | 3.34E+13 | 1075.26 | 3.95E+13 | 612.518 | 0.00939 | 1225.06 | 0.00163 | 971.886 | 0.00269 | 1075.26 | 0.0011 |
| 612.808 | 5.73E+12 | 972.157 | 3.35E+13 | 1076.99 | 3.96E+13 | 612.808 | 0.0094 | 1226.78 | 0.00165 | 972.157 | 0.00268 | 1076.99 | 0.00111 |
| 613.099 | 5.74E+12 | 972.428 | 3.35E+13 | 1078.71 | 3.97E+13 | 613.099 | 0.00939 | 1228.5 | 0.00165 | 972.428 | 0.00267 | 1078.71 | 0.00113 |
| 613.389 | 5.75E+12 | 972.699 | 3.35E+13 | 1080.44 | 3.97E+13 | 613.389 | 0.00943 | 1230.21 | 0.00166 | 972.699 | 0.00266 | 1080.44 | 0.00114 |
| 613.68 | 5.77E+12 | 972.969 | 3.35E+13 | 1082.17 | 3.98E+13 | 613.68 | 0.00938 | 1231.93 | 0.00166 | 972.969 | 0.00263 | 1082.17 | 0.00115 |
| 613.97 | 5.78E+12 | 973.24 | 3.36E+13 | 1083.89 | 3.98E+13 | 613.97 | 0.00952 | 1233.65 | 0.00168 | 973.24 | 0.00264 | 1083.89 | 0.00115 |
| 614.261 | 5.77E+12 | 973.511 | 3.36E+13 | 1085.62 | 3.99E+13 | 614.261 | 0.0094 | 1235.36 | 0.00168 | 973.511 | 0.00261 | 1085.62 | 0.00117 |
| 614.551 | 5.80E+12 | 973.782 | 3.36E+13 | 1087.34 | 3.99E+13 | 614.551 | 0.00938 | 1237.08 | 0.00169 | 973.782 | 0.00261 | 1087.34 | 0.00117 |
| 614.841 | 5.80E+12 | 974.053 | 3.36E+13 | 1089.07 | 4.00E+13 | 614.841 | 0.00939 | 1238.79 | 0.0017 | 974.053 | 0.00257 | 1089.07 | 0.00118 |
| 615.132 | 5.79E+12 | 974.324 | 3.36E+13 | 1090.8 | 4.00E+13 | 615.132 | 0.00936 | 1240.51 | 0.0017 | 974.324 | 0.00258 | 1090.8 | 0.0012 |
| 615.422 | 5.79E+12 | 974.595 | 3.37E+13 | 1092.52 | 4.01E+13 | 615.422 | 0.00947 | 1242.23 | 0.0017 | 974.595 | 0.00257 | 1092.52 | 0.0012 |
| 615.713 | 5.79E+12 | 974.866 | 3.37E+13 | 1094.25 | 4.01E+13 | 615.713 | 0.00948 | 1243.94 | 0.00171 | 974.866 | 0.00255 | 1094.25 | 0.00121 |
| 616.003 | 5.81E+12 | 975.136 | 3.37E+13 | 1095.97 | 4.02E+13 | 616.003 | 0.00953 | 1245.66 | 0.00172 | 975.136 | 0.00254 | 1095.97 | 0.00123 |
| 616.293 | 5.81E+12 | 975.407 | 3.37E+13 | 1097.7 | 4.02E+13 | 616.293 | 0.00953 | 1247.37 | 0.00173 | 975.407 | 0.00254 | 1097.7 | 0.00123 |
| 616.584 | 5.80E+12 | 975.678 | 3.38E+13 | 1099.42 | 4.02E+13 | 616.584 | 0.00956 | 1249.09 | 0.00174 | 975.678 | 0.00251 | 1099.42 | 0.00124 |
| 616.874 | 5.80E+12 | 975.949 | 3.38E+13 | 1101.15 | 4.03E+13 | 616.874 | 0.00967 | 1250.8 | 0.00174 | 975.949 | 0.00252 | 1101.15 | 0.00125 |
| 617.165 | 5.79E+12 | 976.219 | 3.38E+13 | 1102.87 | 4.03E+13 | 617.165 | 0.00969 | 1252.52 | 0.00175 | 976.219 | 0.0025 | 1102.87 | 0.00125 |
| 617.455 | 5.79E+12 | 976.49 | 3.38E+13 | 1104.6 | 4.04E+13 | 617.455 | 0.00967 | 1254.23 | 0.00176 | 976.49 | 0.00248 | 1104.6 | 0.00126 |
| 617.745 | 5.82E+12 | 976.761 | 3.38E+13 | 1106.32 | 4.04E+13 | 617.745 | 0.00964 | 1255.95 | 0.00177 | 976.761 | 0.00245 | 1106.32 | 0.00127 |
| 618.036 | 5.83E+12 | 977.031 | 3.39E+13 | 1108.05 | 4.04E+13 | 618.036 | 0.0097 | 1257.66 | 0.00177 | 977.031 | 0.00245 | 1108.05 | 0.00128 |
| 618.326 | 5.84E+12 | 977.302 | 3.39E+13 | 1109.77 | 4.05E+13 | 618.326 | 0.00976 | 1259.38 | 0.00178 | 977.302 | 0.00245 | 1109.77 | 0.00129 |
| 618.616 | 5.85E+12 | 977.573 | 3.39E+13 | 1111.5 | 4.05E+13 | 618.616 | 0.00965 | 1261.09 | 0.00178 | 977.573 | 0.00243 | 1111.5 | 0.0013 |
| 618.906 | 5.86E+12 | 977.843 | 3.39E+13 | 1113.22 | 4.05E+13 | 618.906 | 0.00966 | 1262.8 | 0.00179 | 977.843 | 0.00242 | 1113.22 | 0.00129 |
| 619.197 | 5.88E+12 | 978.114 | 3.40E+13 | 1114.95 | 4.06E+13 | 619.197 | 0.00991 | 1264.52 | 0.00179 | 978.114 | 0.00243 | 1114.95 | 0.00131 |
| 619.487 | 5.88E+12 | 978.384 | 3.40E+13 | 1116.67 | 4.06E+13 | 619.487 | 0.00972 | 1266.23 | 0.0018 | 978.384 | 0.00239 | 1116.67 | 0.00132 |
| 619.777 | 5.88E+12 | 978.655 | 3.40E+13 | 1118.39 | 4.06E+13 | 619.777 | 0.00977 | 1267.95 | 0.0018 | 978.655 | 0.0024 | 1118.39 | 0.00132 |
| 620.068 | 5.86E+12 | 978.926 | 3.40E+13 | 1120.12 | 4.07E+13 | 620.068 | 0.00979 | 1269.66 | 0.00181 | 978.926 | 0.00237 | 1120.12 | 0.00133 |
| 620.358 | 5.88E+12 | 979.196 | 3.40E+13 | 1121.84 | 4.07E+13 | 620.358 | 0.00972 | 1271.37 | 0.00182 | 979.196 | 0.00236 | 1121.84 | 0.00134 |
| 620.648 | 5.87E+12 | 979.467 | 3.41E+13 | 1123.57 | 4.07E+13 | 620.648 | 0.00982 | 1273.09 | 0.00182 | 979.467 | 0.00236 | 1123.57 | 0.00134 |
| 620.938 | 5.87E+12 | 979.737 | 3.41E+13 | 1125.29 | 4.07E+13 | 620.938 | 0.00985 | 1274.8 | 0.00183 | 979.737 | 0.00232 | 1125.29 | 0.00135 |
| 621.228 | 5.88E+12 | 980.007 | 3.41E+13 | 1127.01 | 4.08E+13 | 621.228 | 0.00993 | 1276.51 | 0.00184 | 980.007 | 0.00232 | 1127.01 | 0.00136 |
| 621.519 | 5.89E+12 | 980.278 | 3.41E+13 | 1128.74 | 4.08E+13 | 621.519 | 0.00994 | 1278.23 | 0.00182 | 980.278 | 0.00231 | 1128.74 | 0.00136 |
| 621.809 | 5.91E+12 | 980.548 | 3.42E+13 | 1130.46 | 4.08E+13 | 621.809 | 0.00987 | 1279.94 | 0.00182 | 980.548 | 0.00228 | 1130.46 | 0.00137 |
| 622.099 | 5.91E+12 | 980.819 | 3.42E+13 | 1132.18 | 4.08E+13 | 622.099 | 0.00989 | 1281.65 | 0.00183 | 980.819 | 0.00229 | 1132.18 | 0.00137 |
| 622.389 | 5.92E+12 | 981.089 | 3.42E+13 | 1133.91 | 4.09E+13 | 622.389 | 0.00981 | 1283.37 | 0.00184 | 981.089 | 0.00226 | 1133.91 | 0.00138 |
| 622.679 | 5.95E+12 | 981.359 | 3.42E+13 | 1135.63 | 4.09E+13 | 622.679 | 0.00979 | 1285.08 | 0.00185 | 981.359 | 0.00226 | 1135.63 | 0.00139 |
| 622.969 | 5.95E+12 | 981.63 | 3.42E+13 | 1137.35 | 4.09E+13 | 622.969 | 0.0099 | 1286.79 | 0.00185 | 981.63 | 0.00224 | 1137.35 | 0.0014 |
| 623.259 | 5.96E+12 | 981.9 | 3.43E+13 | 1139.07 | 4.09E+13 | 623.259 | 0.01001 | 1288.5 | 0.00185 | 981.9 | 0.00223 | 1139.07 | 0.0014 |
| 623.55 | 5.99E+12 | 982.17 | 3.43E+13 | 1140.8 | 4.09E+13 | 623.55 | 0.00993 | 1290.22 | 0.00186 | 982.17 | 0.00221 | 1140.8 | 0.00141 |
| 623.84 | 6.00E+12 | 982.441 | 3.43E+13 | 1142.52 | 4.09E+13 | 623.84 | 0.00993 | 1291.93 | 0.00186 | 982.441 | 0.00219 | 1142.52 | 0.00141 |
| 624.13 | 6.02E+12 | 982.711 | 3.43E+13 | 1144.24 | 4.09E+13 | 624.13 | 0.01001 | 1293.64 | 0.00185 | 982.711 | 0.00218 | 1144.24 | 0.00142 |
| 624.42 | 6.04E+12 | 982.981 | 3.43E+13 | 1145.96 | 4.10E+13 | 624.42 | 0.00997 | 1295.35 | 0.00186 | 982.981 | 0.00217 | 1145.96 | 0.00143 |
| 624.71 | 6.05E+12 | 983.251 | 3.44E+13 | 1147.69 | 4.10E+13 | 624.71 | 0.00987 | 1297.06 | 0.00186 | 983.251 | 0.00214 | 1147.69 | 0.00144 |
| 625 | 6.07E+12 | 983.522 | 3.44E+13 | 1149.41 | 4.10E+13 | 625 | 0.00988 | 1298.78 | 0.00186 | 983.522 | 0.00214 | 1149.41 | 0.00144 |
| 625.29 | 6.11E+12 | 983.792 | 3.44E+13 | 1151.13 | 4.11E+13 | 625.29 | 0.00991 | 1300.49 | 0.00187 | 983.792 | 0.00213 | | |
| 625.58 | 6.10E+12 | 984.062 | 3.44E+13 | | | 625.58 | 0.00999 | 1302.2 | 0.00188 | 984.062 | 0.00213 | | |
| 625.87 | 6.12E+12 | 984.332 | 3.44E+13 | | | 625.87 | 0.00993 | 1303.91 | 0.00188 | 984.332 | 0.0021 | | |
| 626.16 | 6.13E+12 | 984.602 | 3.45E+13 | | | 626.16 | 0.00996 | 1305.62 | 0.00188 | 984.602 | 0.00208 | | |
| 626.45 | 6.15E+12 | 984.872 | 3.45E+13 | | | 626.45 | 0.0099 | 1307.33 | 0.00189 | 984.872 | 0.00205 | | |
| 626.74 | 6.17E+12 | 985.142 | 3.45E+13 | | | 626.74 | 0.01 | 1309.04 | 0.00189 | 985.142 | 0.00205 | | |
| 627.03 | 6.21E+12 | 985.412 | 3.45E+13 | | | 627.03 | 0.00995 | 1310.75 | 0.0019 | 985.412 | 0.00203 | | |
| 627.32 | 6.22E+12 | 985.682 | 3.46E+13 | | | 627.32 | 0.00993 | 1312.47 | 0.0019 | 985.682 | 0.00201 | | |
| 627.61 | 6.23E+12 | 985.952 | 3.46E+13 | | | 627.61 | 0.00999 | 1314.18 | 0.0019 | 985.952 | 0.00199 | | |
| 627.9 | 6.23E+12 | 986.222 | 3.46E+13 | | | 627.9 | 0.00998 | 1315.89 | 0.00191 | 986.222 | 0.00198 | | |
| 628.19 | 6.25E+12 | 986.492 | 3.46E+13 | | | 628.19 | 0.00992 | 1317.6 | 0.0019 | 986.492 | 0.00195 | | |

| | | | | | | | |
|---|---|---|---|---|---|---|---|
| 466.6164 | 3.41E+07 | 1049.347 | 3.72E+10 | 628.48 | 3.42E+20 | 628.48 | 0.43598 |
| 466.9115 | 6.82E+07 | 1051.076 | 3.77E+10 | 628.77 | 3.42E+20 | 628.77 | 0.43595 |
| 467.2066 | 5.45E+07 | 1052.804 | 3.80E+10 | 629.059 | 3.43E+20 | 629.059 | 0.43591 |
| 467.5017 | 4.09E+07 | 1054.532 | 3.86E+10 | 629.349 | 3.44E+20 | 629.349 | 0.43588 |
| 467.7968 | 7.50E+07 | 1056.26 | 3.87E+10 | 629.639 | 3.45E+20 | 629.639 | 0.43584 |
| 468.0919 | 5.45E+07 | 1057.988 | 3.91E+10 | 629.929 | 3.46E+20 | 629.929 | 0.43581 |
| 468.3869 | 6.82E+07 | 1059.716 | 3.98E+10 | 630.219 | 3.47E+20 | 630.219 | 0.43577 |
| 468.682 | 8.18E+07 | 1061.443 | 4.01E+10 | 630.509 | 3.47E+20 | 630.509 | 0.43574 |
| 468.9771 | 8.86E+07 | 1063.171 | 4.07E+10 | 630.799 | 3.48E+20 | 630.799 | 0.4357 |
| 469.2721 | 8.18E+07 | 1064.898 | 4.10E+10 | 631.088 | 3.49E+20 | 631.088 | 0.43567 |
| 469.5672 | 2.73E+07 | 1066.625 | 4.15E+10 | 631.378 | 3.50E+20 | 631.378 | 0.43563 |
| 469.8622 | 5.45E+07 | 1068.353 | 4.21E+10 | 631.668 | 3.51E+20 | 631.668 | 0.4356 |
| 470.1572 | 5.45E+07 | 1070.08 | 4.24E+10 | 631.958 | 3.52E+20 | 631.958 | 0.43557 |
| 470.4523 | 5.45E+07 | 1071.807 | 4.27E+10 | 632.248 | 3.53E+20 | 632.248 | 0.43553 |
| 470.7473 | 8.18E+07 | 1073.534 | 4.30E+10 | 632.537 | 3.53E+20 | 632.537 | 0.4355 |
| 471.0423 | 5.45E+07 | 1075.26 | 4.35E+10 | 632.827 | 3.54E+20 | 632.827 | 0.43546 |
| 471.3373 | 4.09E+07 | 1076.987 | 4.41E+10 | 633.117 | 3.55E+20 | 633.117 | 0.43543 |
| 471.6323 | 5.45E+07 | 1078.714 | 4.46E+10 | 633.406 | 3.56E+20 | 633.406 | 0.4354 |
| 471.9273 | 5.45E+07 | 1080.44 | 4.53E+10 | 633.696 | 3.57E+20 | 633.696 | 0.43537 |
| 472.2222 | 6.82E+07 | 1082.166 | 4.55E+10 | 633.986 | 3.58E+20 | 633.986 | 0.43533 |
| 472.5172 | 4.77E+07 | 1083.893 | 4.59E+10 | 634.276 | 3.59E+20 | 634.276 | 0.4353 |
| 472.8122 | 4.77E+07 | 1085.619 | 4.66E+10 | 634.565 | 3.59E+20 | 634.565 | 0.43527 |
| 473.1071 | 6.13E+07 | 1087.345 | 4.68E+10 | 634.855 | 3.60E+20 | 634.855 | 0.43523 |
| 473.4021 | 7.50E+07 | 1089.071 | 4.72E+10 | 635.145 | 3.61E+20 | 635.145 | 0.4352 |
| 473.697 | 3.41E+07 | 1090.797 | 4.78E+10 | 635.434 | 3.62E+20 | 635.434 | 0.43517 |
| 473.9919 | 4.77E+07 | 1092.522 | 4.80E+10 | 635.724 | 3.63E+20 | 635.724 | 0.43514 |
| 474.2869 | 5.45E+07 | 1094.248 | 4.87E+10 | 636.013 | 3.64E+20 | 636.013 | 0.43511 |
| 474.5818 | 4.09E+07 | 1095.973 | 4.92E+10 | 636.303 | 3.65E+20 | 636.303 | 0.43507 |
| 474.8767 | 4.77E+07 | 1097.699 | 4.96E+10 | 636.593 | 3.65E+20 | 636.593 | 0.43504 |
| 475.1716 | 6.13E+07 | 1099.424 | 5.00E+10 | 636.882 | 3.66E+20 | 636.882 | 0.43501 |
| 475.4665 | 5.45E+07 | 1101.149 | 5.03E+10 | 637.172 | 3.67E+20 | 637.172 | 0.43498 |
| 475.7614 | 7.50E+07 | 1102.874 | 5.06E+10 | 637.461 | 3.68E+20 | 637.461 | 0.43495 |
| 476.0563 | 6.82E+07 | 1104.599 | 5.09E+10 | 637.751 | 3.69E+20 | 637.751 | 0.43492 |
| 476.3511 | 3.41E+07 | 1106.324 | 5.15E+10 | 638.04 | 3.70E+20 | 638.04 | 0.43488 |
| 476.646 | 3.41E+07 | 1108.049 | 5.16E+10 | 638.33 | 3.71E+20 | 638.33 | 0.43485 |
| 476.9408 | 7.50E+07 | 1109.773 | 5.21E+10 | 638.619 | 3.72E+20 | 638.619 | 0.43482 |
| 477.2357 | 4.09E+07 | 1111.498 | 5.25E+10 | 638.909 | 3.72E+20 | 638.909 | 0.43479 |
| 477.5305 | 6.13E+07 | 1113.222 | 5.24E+10 | 639.198 | 3.73E+20 | 639.198 | 0.43476 |
| 477.8254 | 5.45E+07 | 1114.946 | 5.31E+10 | 639.488 | 3.74E+20 | 639.488 | 0.43473 |
| 478.1202 | 5.45E+07 | 1116.67 | 5.35E+10 | 639.777 | 3.75E+20 | 639.777 | 0.4347 |
| 478.415 | 4.09E+07 | 1118.394 | 5.37E+10 | 640.067 | 3.76E+20 | 640.067 | 0.43455 |
| 478.7098 | 6.13E+07 | 1120.118 | 5.40E+10 | 640.356 | 3.77E+20 | 640.356 | 0.43452 |
| 479.0046 | 6.13E+07 | 1121.842 | 5.44E+10 | 640.646 | 3.78E+20 | 640.646 | 0.43448 |
| 479.2994 | 4.77E+07 | 1123.566 | 5.47E+10 | 640.935 | 3.79E+20 | 640.935 | 0.43445 |
| 479.5942 | 7.50E+07 | 1125.29 | 5.51E+10 | 641.225 | 3.79E+20 | 641.225 | 0.43442 |
| 479.889 | 5.45E+07 | 1127.013 | 5.54E+10 | 641.514 | 3.80E+20 | 641.514 | 0.43438 |
| 480.1838 | 3.41E+07 | 1128.736 | 5.54E+10 | 641.803 | 3.81E+20 | 641.803 | 0.43434 |
| 480.4785 | 6.13E+07 | 1130.46 | 5.58E+10 | 642.093 | 3.82E+20 | 642.093 | 0.43431 |
| 480.7733 | 8.18E+07 | 1132.183 | 5.60E+10 | 642.382 | 3.83E+20 | 642.382 | 0.43428 |
| 481.068 | 3.41E+07 | 1133.906 | 5.62E+10 | 642.671 | 3.84E+20 | 642.671 | 0.43424 |
| 481.3628 | 6.13E+07 | 1135.629 | 5.68E+10 | 642.961 | 3.85E+20 | 642.961 | 0.43421 |
| 481.6575 | 7.50E+07 | 1137.352 | 5.70E+10 | 643.25 | 3.86E+20 | 643.25 | 0.43417 |
| 481.9522 | 4.09E+07 | 1139.074 | 5.73E+10 | 643.539 | 3.87E+20 | 643.539 | 0.43414 |
| 482.247 | 6.13E+07 | 1140.797 | 5.76E+10 | 643.829 | 3.87E+20 | 643.829 | 0.4341 |
| 482.5417 | 7.50E+07 | 1142.519 | 5.77E+10 | 644.118 | 3.88E+20 | 644.118 | 0.43406 |
| 482.8364 | 4.77E+07 | 1144.242 | 5.80E+10 | 644.407 | 3.89E+20 | 644.407 | 0.43403 |
| 483.1311 | 4.09E+07 | 1145.964 | 5.86E+10 | 644.696 | 3.90E+20 | 644.696 | 0.434 |
| 483.4258 | 7.50E+07 | 1147.686 | 5.88E+10 | 644.986 | 3.91E+20 | 644.986 | 0.43396 |
| 483.7204 | 6.13E+07 | 1149.408 | 5.89E+10 | 645.275 | 3.92E+20 | 645.275 | 0.43392 |
| 484.0151 | 6.13E+07 | 1151.13 | 5.92E+10 | 645.564 | 3.93E+20 | 645.564 | 0.43389 |
| 484.3098 | 3.41E+07 | 1152.852 | 5.96E+10 | 645.853 | 3.94E+20 | 645.853 | 0.43385 |
| 484.6044 | 5.45E+07 | 1154.573 | 6.00E+10 | 646.143 | 3.95E+20 | 646.143 | 0.43382 |
| 484.8991 | 5.45E+07 | 1156.295 | 6.01E+10 | 646.432 | 3.95E+20 | 646.432 | 0.43378 |
| 485.1937 | 6.82E+07 | 1158.016 | 6.04E+10 | 646.721 | 3.96E+20 | 646.721 | 0.43375 |
| 485.4883 | 6.82E+07 | 1159.738 | 6.08E+10 | 647.01 | 3.97E+20 | 647.01 | 0.43371 |
| 485.783 | 6.82E+07 | 1161.459 | 6.10E+10 | 647.299 | 3.98E+20 | 647.299 | 0.43367 |
| 486.0776 | 6.82E+07 | 1163.18 | 6.12E+10 | 647.588 | 3.99E+20 | 647.588 | 0.43364 |
| 486.3722 | 2.73E+07 | 1164.901 | 6.12E+10 | 647.878 | 4.00E+20 | 647.878 | 0.4336 |
| 486.6668 | 5.45E+07 | 1166.622 | 6.12E+10 | 648.167 | 4.01E+20 | 648.167 | 0.43357 |
| 486.9614 | 7.50E+07 | 1168.343 | 6.11E+10 | 648.456 | 4.02E+20 | 648.456 | 0.43353 |
| 487.256 | 6.13E+07 | 1170.063 | 6.13E+10 | 648.745 | 4.03E+20 | 648.745 | 0.4335 |
| 487.5506 | 6.13E+07 | 1171.784 | 6.19E+10 | 649.034 | 4.04E+20 | 649.034 | 0.43346 |
| 487.8451 | 5.45E+07 | 1173.504 | 6.20E+10 | 649.323 | 4.04E+20 | 649.323 | 0.43342 |
| 488.1397 | 8.86E+07 | 1175.225 | 6.21E+10 | 649.612 | 4.05E+20 | 649.612 | 0.43339 |
| 488.4342 | 3.41E+07 | 1176.945 | 6.23E+10 | 649.901 | 4.06E+20 | 649.901 | 0.43335 |
| 488.7288 | 5.45E+07 | 1178.665 | 6.24E+10 | 650.19 | 4.07E+20 | 650.19 | 0.43332 |
| 489.0233 | 6.13E+07 | 1180.385 | 6.26E+10 | 650.479 | 4.08E+20 | 650.479 | 0.43328 |
| 489.3179 | 5.45E+07 | 1182.105 | 6.27E+10 | 650.768 | 4.09E+20 | 650.768 | 0.43324 |
| 489.6124 | 5.45E+07 | 1183.824 | 6.29E+10 | 651.057 | 4.10E+20 | 651.057 | 0.43321 |
| 489.9069 | 6.13E+07 | 1185.544 | 6.29E+10 | 651.346 | 4.11E+20 | 651.346 | 0.43317 |
| 490.2014 | 5.45E+07 | 1187.263 | 6.29E+10 | 651.635 | 4.12E+20 | 651.635 | 0.43313 |
| 490.4959 | 6.82E+07 | 1188.983 | 6.32E+10 | 651.924 | 4.13E+20 | 651.924 | 0.4331 |
| 490.7904 | 6.82E+07 | 1190.702 | 6.34E+10 | 652.213 | 4.14E+20 | 652.213 | 0.43306 |
| 491.0849 | 6.82E+07 | 1192.421 | 6.37E+10 | 652.502 | 4.14E+20 | 652.502 | 0.43302 |
| 491.3793 | 6.82E+07 | 1194.14 | 6.39E+10 | 652.791 | 4.15E+20 | 652.791 | 0.43299 |
| 491.6738 | 3.41E+07 | 1195.859 | 6.39E+10 | 653.08 | 4.16E+20 | 653.08 | 0.43295 |
| 491.9683 | 7.50E+07 | 1197.578 | 6.42E+10 | 653.369 | 4.17E+20 | 653.369 | 0.43291 |
| 492.2627 | 6.13E+07 | 1199.297 | 6.44E+10 | 653.658 | 4.18E+20 | 653.658 | 0.43288 |
| 492.5572 | 6.82E+07 | 1201.015 | 6.44E+10 | 653.947 | 4.19E+20 | 653.947 | 0.43284 |
| 492.8516 | 7.50E+07 | 1202.734 | 6.46E+10 | 654.235 | 4.20E+20 | 654.235 | 0.4328 |
| 493.146 | 8.86E+07 | 1204.452 | 6.48E+10 | 654.524 | 4.21E+20 | 654.524 | 0.43276 |
| 493.4404 | 7.50E+07 | 1206.17 | 6.47E+10 | 654.813 | 4.22E+20 | 654.813 | 0.43273 |
| 493.7349 | 6.82E+07 | 1207.888 | 6.49E+10 | 655.102 | 4.23E+20 | 655.102 | 0.43269 |
| 494.0293 | 5.45E+07 | 1209.606 | 6.52E+10 | 655.391 | 4.24E+20 | 655.391 | 0.43265 |
| 494.3237 | 7.50E+07 | 1211.324 | 6.53E+10 | 655.68 | 4.25E+20 | 655.68 | 0.43262 |
| 494.618 | 4.77E+07 | 1213.042 | 6.56E+10 | 655.968 | 4.26E+20 | 655.968 | 0.43258 |
| 494.9124 | 4.77E+07 | 1214.759 | 6.58E+10 | 656.257 | 4.26E+20 | 656.257 | 0.43254 |
| 495.2068 | 6.82E+07 | 1216.477 | 6.60E+10 | 656.546 | 4.27E+20 | 656.546 | 0.4325 |
| 495.5011 | 4.77E+07 | 1218.194 | 6.63E+10 | 656.835 | 4.28E+20 | 656.835 | 0.43247 |
| 495.7955 | 7.50E+07 | 1219.912 | 6.64E+10 | 657.124 | 4.29E+20 | 657.124 | 0.43243 |
| 496.0898 | 8.18E+07 | 1221.629 | 6.64E+10 | 657.412 | 4.30E+20 | 657.412 | 0.43239 |
| 628.48 | 4.21E-08 | 1319.31 | 5.05E-08 | 628.48 | 4.71E-08 | 1269.66 | 5.23E-08 |
| 628.77 | 4.22E-08 | 1321.02 | 5.04E-08 | 628.77 | 4.13E-08 | 1271.37 | 5.30E-08 |
| 629.059 | 4.23E-08 | 1322.73 | 5.03E-08 | 629.059 | 3.95E-08 | 1273.09 | 5.24E-08 |
| 629.349 | 4.23E-08 | 1324.44 | 5.02E-08 | 629.349 | 4.41E-08 | 1274.8 | 5.24E-08 |
| 629.639 | 4.22E-08 | 1326.15 | 5.01E-08 | 629.639 | 4.56E-08 | 1276.51 | 5.28E-08 |
| 629.929 | 4.22E-08 | 1327.86 | 5.00E-08 | 629.929 | 4.20E-08 | 1278.23 | 5.27E-08 |
| 630.219 | 4.21E-08 | 1329.57 | 4.99E-08 | 630.219 | 4.15E-08 | 1279.94 | 5.26E-08 |
| 630.509 | 4.21E-08 | 1331.28 | 4.97E-08 | 630.509 | 4.58E-08 | 1281.65 | 5.33E-08 |
| 630.799 | 4.23E-08 | 1332.98 | 4.95E-08 | 630.799 | 4.35E-08 | 1283.37 | 5.30E-08 |
| 631.088 | 4.23E-08 | 1334.69 | 4.94E-08 | 631.088 | 4.47E-08 | 1285.08 | 5.18E-08 |
| 631.378 | 4.23E-08 | 1336.4 | 4.93E-08 | 631.378 | 4.26E-08 | 1286.79 | 5.18E-08 |
| 631.668 | 4.24E-08 | 1338.11 | 4.92E-08 | 631.668 | 4.05E-08 | 1288.5 | 5.14E-08 |
| 631.958 | 4.25E-08 | 1339.82 | 4.90E-08 | 631.958 | 3.83E-08 | 1290.22 | 5.08E-08 |
| 632.248 | 4.25E-08 | 1341.53 | 4.88E-08 | 632.248 | 4.66E-08 | 1291.93 | 5.15E-08 |
| 632.537 | 4.26E-08 | 1343.24 | 4.88E-08 | 632.537 | 4.31E-08 | 1293.64 | 5.13E-08 |
| 632.827 | 4.27E-08 | 1344.95 | 4.87E-08 | 632.827 | 4.59E-08 | 1295.35 | 5.14E-08 |
| 633.117 | 4.26E-08 | 1346.65 | 4.86E-08 | 633.117 | 4.28E-08 | 1297.06 | 5.27E-08 |
| 633.406 | 4.28E-08 | 1348.36 | 4.85E-08 | 633.406 | 4.34E-08 | 1298.78 | 5.10E-08 |
| 633.696 | 4.27E-08 | 1350.07 | 4.83E-08 | 633.696 | 3.69E-08 | 1300.49 | 5.13E-08 |
| 633.986 | 4.27E-08 | 1351.78 | 4.82E-08 | 633.986 | 4.51E-08 | 1302.2 | 5.21E-08 |
| 634.276 | 4.26E-08 | 1353.49 | 4.81E-08 | 634.276 | 4.08E-08 | 1303.91 | 5.12E-08 |
| 634.565 | 4.25E-08 | 1355.19 | 4.79E-08 | 634.565 | 4.64E-08 | 1305.62 | 5.13E-08 |
| 634.855 | 4.25E-08 | 1356.9 | 4.76E-08 | 634.855 | 4.35E-08 | 1307.33 | 5.06E-08 |
| 635.145 | 4.26E-08 | 1358.61 | 4.75E-08 | 635.145 | 3.90E-08 | 1309.04 | 5.21E-08 |
| 635.434 | 4.25E-08 | 1360.32 | 4.74E-08 | 635.434 | 4.06E-08 | 1310.75 | 5.18E-08 |
| 635.724 | 4.23E-08 | 1362.02 | 4.72E-08 | 635.724 | 4.24E-08 | 1312.47 | 5.13E-08 |
| 636.013 | 4.23E-08 | 1363.73 | 4.71E-08 | 636.013 | 3.98E-08 | 1314.18 | 5.13E-08 |
| 636.303 | 4.22E-08 | 1365.44 | 4.70E-08 | 636.303 | 4.15E-08 | 1315.89 | 4.95E-08 |
| 636.593 | 4.21E-08 | 1367.14 | 4.69E-08 | 636.593 | 4.75E-08 | 1317.6 | 5.11E-08 |
| 636.882 | 4.19E-08 | 1368.85 | 4.68E-08 | 636.882 | 4.41E-08 | 1319.31 | 5.07E-08 |
| 637.172 | 4.19E-08 | 1370.56 | 4.67E-08 | 637.172 | 4.36E-08 | 1321.02 | 4.95E-08 |
| 637.461 | 4.18E-08 | 1372.26 | 4.66E-08 | 637.461 | 4.08E-08 | 1322.73 | 5.10E-08 |
| 637.751 | 4.18E-08 | 1373.97 | 4.64E-08 | 637.751 | 4.30E-08 | 1324.44 | 4.96E-08 |
| 638.04 | 4.19E-08 | 1375.67 | 4.64E-08 | 638.04 | 4.15E-08 | 1326.15 | 5.00E-08 |
| 638.33 | 4.18E-08 | 1377.38 | 4.63E-08 | 638.33 | 4.28E-08 | 1327.86 | 4.96E-08 |
| 638.619 | 4.17E-08 | 1379.09 | 4.62E-08 | 638.619 | 4.42E-08 | 1329.57 | 4.94E-08 |
| 638.909 | 4.16E-08 | 1380.79 | 4.61E-08 | 638.909 | 3.77E-08 | 1331.28 | 4.93E-08 |
| 639.198 | 4.17E-08 | 1382.5 | 4.60E-08 | 639.198 | 4.85E-08 | 1332.98 | 4.86E-08 |
| 639.488 | 4.15E-08 | 1384.2 | 4.58E-08 | 639.488 | 3.80E-08 | 1334.69 | 4.94E-08 |
| 639.777 | 4.16E-08 | 1385.91 | 4.57E-08 | 639.777 | 4.21E-08 | 1336.4 | 5.05E-08 |
| 640.067 | 4.14E-08 | 1387.61 | 4.56E-08 | 640.067 | 3.98E-08 | 1338.11 | 5.05E-08 |
| 640.356 | 4.14E-08 | 1389.32 | 4.56E-08 | 640.356 | 4.34E-08 | 1339.82 | 4.91E-08 |
| 640.646 | 4.13E-08 | 1391.02 | 4.55E-08 | 640.646 | 3.87E-08 | 1341.53 | 5.00E-08 |
| 640.935 | 4.13E-08 | 1392.73 | 4.54E-08 | 640.935 | 4.28E-08 | 1343.24 | 4.86E-08 |
| 641.225 | 4.13E-08 | 1394.43 | 4.54E-08 | 641.225 | 4.03E-08 | 1344.95 | 4.86E-08 |
| 641.514 | 4.13E-08 | 1396.14 | 4.52E-08 | 641.514 | 3.71E-08 | 1346.65 | 4.92E-08 |
| 641.803 | 4.12E-08 | 1397.84 | 4.52E-08 | 641.803 | 4.27E-08 | 1348.36 | 4.76E-08 |
| 642.093 | 4.12E-08 | 1399.55 | 4.51E-08 | 642.093 | 4.00E-08 | 1350.07 | 4.76E-08 |
| 642.382 | 4.11E-08 | 1401.25 | 4.50E-08 | 642.382 | 3.84E-08 | 1351.78 | 4.79E-08 |
| 642.671 | 4.09E-08 | 1402.95 | 4.49E-08 | 642.671 | 3.77E-08 | 1353.49 | 4.81E-08 |
| 642.961 | 4.09E-08 | 1404.66 | 4.49E-08 | 642.961 | 4.38E-08 | 1355.19 | 4.83E-08 |
| 643.25 | 4.09E-08 | 1406.36 | 4.48E-08 | 643.25 | 4.02E-08 | 1356.9 | 4.68E-08 |
| 643.539 | 4.09E-08 | 1408.06 | 4.47E-08 | 643.539 | 3.94E-08 | 1358.61 | 4.67E-08 |
| 643.829 | 4.08E-08 | 1409.77 | 4.47E-08 | 643.829 | 4.08E-08 | 1360.32 | 4.86E-08 |
| 644.118 | 4.07E-08 | 1411.47 | 4.47E-08 | 644.118 | 4.26E-08 | 1362.02 | 4.73E-08 |
| 644.407 | 4.07E-08 | 1413.17 | 4.46E-08 | 644.407 | 4.03E-08 | 1363.73 | 4.73E-08 |
| 644.696 | 4.07E-08 | 1414.88 | 4.46E-08 | 644.696 | 4.28E-08 | 1365.44 | 4.75E-08 |
| 644.986 | 4.07E-08 | 1416.58 | 4.45E-08 | 644.986 | 4.40E-08 | 1367.14 | 4.64E-08 |
| 645.275 | 4.06E-08 | 1418.28 | 4.44E-08 | 645.275 | 3.66E-08 | 1368.85 | 4.64E-08 |
| 645.564 | 4.06E-08 | 1419.99 | 4.44E-08 | 645.564 | 3.91E-08 | 1370.56 | 4.63E-08 |
| 645.853 | 4.05E-08 | 1421.69 | 4.44E-08 | 645.853 | 4.05E-08 | 1372.26 | 4.63E-08 |
| 646.143 | 4.06E-08 | 1423.39 | 4.44E-08 | 646.143 | 4.05E-08 | 1373.97 | 4.52E-08 |
| 646.432 | 4.05E-08 | 1425.09 | 4.44E-08 | 646.432 | 4.28E-08 | 1375.67 | 4.63E-08 |
| 646.721 | 4.05E-08 | 1426.8 | 4.44E-08 | 646.721 | 3.99E-08 | 1377.38 | 4.74E-08 |
| 647.01 | 4.04E-08 | 1428.5 | 4.43E-08 | 647.01 | 4.16E-08 | 1379.09 | 4.58E-08 |
| 647.299 | 4.04E-08 | 1430.2 | 4.42E-08 | 647.299 | 3.82E-08 | 1380.79 | 4.62E-08 |
| 647.588 | 4.04E-08 | 1431.9 | 4.41E-08 | 647.588 | 3.83E-08 | 1382.5 | 4.70E-08 |
| 647.878 | 4.03E-08 | 1433.6 | 4.41E-08 | 647.878 | 4.16E-08 | 1384.2 | 4.61E-08 |
| 648.167 | 4.04E-08 | 1435.3 | 4.41E-08 | 648.167 | 3.82E-08 | 1385.91 | 4.45E-08 |
| 648.456 | 4.04E-08 | 1437.01 | 4.40E-08 | 648.456 | 3.84E-08 | 1387.61 | 4.55E-08 |
| 648.745 | 4.04E-08 | 1438.71 | 4.40E-08 | 648.745 | 4.46E-08 | 1389.32 | 4.69E-08 |
| 649.034 | 4.03E-08 | 1440.41 | 4.39E-08 | 649.034 | 4.30E-08 | 1391.02 | 4.48E-08 |
| 649.323 | 4.02E-08 | 1442.11 | 4.38E-08 | 649.323 | 4.23E-08 | 1392.73 | 4.52E-08 |
| 649.612 | 4.02E-08 | 1443.81 | 4.38E-08 | 649.612 | 3.81E-08 | 1394.43 | 4.61E-08 |
| 649.901 | 4.03E-08 | 1445.51 | 4.37E-08 | 649.901 | 3.78E-08 | 1396.14 | 4.52E-08 |
| 650.19 | 4.02E-08 | 1447.21 | 4.36E-08 | 650.19 | 4.03E-08 | 1397.84 | 4.54E-08 |
| 650.479 | 4.02E-08 | 1448.91 | 4.35E-08 | 650.479 | 4.35E-08 | 1399.55 | 4.50E-08 |
| 650.768 | 4.00E-08 | 1450.61 | 4.34E-08 | 650.768 | 4.10E-08 | 1401.25 | 4.40E-08 |
| 651.057 | 3.99E-08 | 1452.31 | 4.33E-08 | 651.057 | 4.16E-08 | 1402.95 | 4.38E-08 |
| 651.346 | 3.99E-08 | 1454.01 | 4.33E-08 | 651.346 | 3.87E-08 | 1404.66 | 4.43E-08 |
| 651.635 | 4.01E-08 | 1455.71 | 4.32E-08 | 651.635 | 3.89E-08 | 1406.36 | 4.55E-08 |
| 651.924 | 4.00E-08 | 1457.41 | 4.31E-08 | 651.924 | 4.35E-08 | 1408.06 | 4.45E-08 |
| 652.213 | 4.01E-08 | 1459.11 | 4.30E-08 | 652.213 | 3.90E-08 | 1409.77 | 4.47E-08 |
| 652.502 | 4.01E-08 | 1460.81 | 4.29E-08 | 652.502 | 3.79E-08 | 1411.47 | 4.46E-08 |
| 652.791 | 4.01E-08 | 1462.51 | 4.27E-08 | 652.791 | 3.81E-08 | 1413.17 | 4.47E-08 |
| 653.08 | 4.01E-08 | 1464.21 | 4.27E-08 | 653.08 | 4.11E-08 | 1414.88 | 4.47E-08 |
| 653.369 | 4.01E-08 | 1465.91 | 4.27E-08 | 653.369 | 4.02E-08 | 1416.58 | 4.48E-08 |
| 653.658 | 4.02E-08 | 1467.61 | 4.27E-08 | 653.658 | 3.78E-08 | 1418.28 | 4.49E-08 |
| 653.947 | 4.01E-08 | 1469.31 | 4.25E-08 | 653.947 | 4.10E-08 | 1419.99 | 4.36E-08 |
| 654.235 | 4.02E-08 | 1471 | 4.24E-08 | 654.235 | 3.90E-08 | 1421.69 | 4.44E-08 |
| 654.524 | 4.02E-08 | 1472.7 | 4.23E-08 | 654.524 | 3.82E-08 | 1423.39 | 4.44E-08 |
| 654.813 | 4.00E-08 | 1474.4 | 4.22E-08 | 654.813 | 3.96E-08 | 1425.09 | 4.40E-08 |
| 655.102 | 4.00E-08 | 1476.1 | 4.22E-08 | 655.102 | 3.71E-08 | 1426.8 | 4.47E-08 |
| 655.391 | 3.99E-08 | 1477.8 | 4.21E-08 | 655.391 | 4.00E-08 | 1428.5 | 4.51E-08 |
| 655.68 | 4.00E-08 | 1479.5 | 4.20E-08 | 655.68 | 4.27E-08 | 1430.2 | 4.49E-08 |
| 655.968 | 4.00E-08 | 1481.19 | 4.20E-08 | 655.968 | 3.84E-08 | 1431.9 | 4.48E-08 |
| 656.257 | 3.99E-08 | 1482.89 | 4.19E-08 | 656.257 | 4.11E-08 | 1433.6 | 4.29E-08 |
| 656.546 | 3.98E-08 | 1484.59 | 4.18E-08 | 656.546 | 3.67E-08 | 1435.3 | 4.32E-08 |
| 656.835 | 3.97E-08 | 1486.29 | 4.17E-08 | 656.835 | 3.72E-08 | 1437.01 | 4.39E-08 |
| 657.124 | 3.96E-08 | 1487.98 | 4.17E-08 | 657.124 | 3.90E-08 | 1438.71 | 4.45E-08 |
| 657.412 | 3.96E-08 | 1489.68 | 4.16E-08 | 657.412 | 4.72E-08 | 1440.41 | 4.40E-08 |

| | | | | | | | | | |
|---|---|---|---|---|---|---|---|---|---|
| 628.48 | 6.27E+12 | 986.762 | 3.46E+13 | 628.48 | 0.00987 | 1319.31 | 0.0019 | 986.762 | 0.00194 |
| 628.77 | 6.31E+12 | 987.032 | 3.47E+13 | 628.77 | 0.00998 | 1321.02 | 0.0019 | 987.032 | 0.00191 |
| 629.059 | 6.33E+12 | 987.302 | 3.47E+13 | 629.059 | 0.01 | 1322.73 | 0.00191 | 987.302 | 0.00189 |
| 629.349 | 6.34E+12 | 987.572 | 3.47E+13 | 629.349 | 0.00994 | 1324.44 | 0.0019 | 987.572 | 0.00187 |
| 629.639 | 6.34E+12 | 987.842 | 3.47E+13 | 629.639 | 0.00996 | 1326.15 | 0.00191 | 987.842 | 0.00185 |
| 629.929 | 6.36E+12 | 988.112 | 3.47E+13 | 629.929 | 0.0099 | 1327.86 | 0.00191 | 988.112 | 0.00183 |
| 630.219 | 6.36E+12 | 988.382 | 3.48E+13 | 630.219 | 0.00986 | 1329.57 | 0.00191 | 988.382 | 0.00179 |
| 630.509 | 6.38E+12 | 988.652 | 3.48E+13 | 630.509 | 0.01 | 1331.28 | 0.00191 | 988.652 | 0.00177 |
| 630.799 | 6.41E+12 | 988.921 | 3.48E+13 | 630.799 | 0.00997 | 1332.98 | 0.00192 | 988.921 | 0.00175 |
| 631.088 | 6.43E+12 | 989.191 | 3.48E+13 | 631.088 | 0.00983 | 1334.69 | 0.0019 | 989.191 | 0.00171 |
| 631.378 | 6.45E+12 | 989.461 | 3.48E+13 | 631.378 | 0.01002 | 1336.4 | 0.0019 | 989.461 | 0.00169 |
| 631.668 | 6.48E+12 | 989.731 | 3.49E+13 | 631.668 | 0.00992 | 1338.11 | 0.00192 | 989.731 | 0.00166 |
| 631.958 | 6.51E+12 | 990 | 3.49E+13 | 631.958 | 0.00989 | 1339.82 | 0.00192 | 990 | 0.00164 |
| 632.248 | 6.53E+12 | 990.27 | 3.49E+13 | 632.248 | 0.00978 | 1341.53 | 0.00192 | 990.27 | 0.0016 |
| 632.537 | 6.56E+12 | 990.54 | 3.49E+13 | 632.537 | 0.00989 | 1343.24 | 0.00192 | 990.54 | 0.00158 |
| 632.827 | 6.59E+12 | 990.81 | 3.49E+13 | 632.827 | 0.00983 | 1344.95 | 0.00191 | 990.81 | 0.00155 |
| 633.117 | 6.59E+12 | 991.079 | 3.50E+13 | 633.117 | 0.0098 | 1346.65 | 0.00191 | 991.079 | 0.00153 |
| 633.406 | 6.64E+12 | 991.349 | 3.50E+13 | 633.406 | 0.00981 | 1348.36 | 0.00189 | 991.349 | 0.00149 |
| 633.696 | 6.64E+12 | 991.618 | 3.50E+13 | 633.696 | 0.00988 | 1350.07 | 0.00189 | 991.618 | 0.00146 |
| 633.986 | 6.64E+12 | 991.888 | 3.50E+13 | 633.986 | 0.00986 | 1351.78 | 0.00189 | 991.888 | 0.00143 |
| 634.276 | 6.65E+12 | 992.158 | 3.50E+13 | 634.276 | 0.00986 | 1353.49 | 0.00188 | 992.158 | 0.00139 |
| 634.565 | 6.65E+12 | 992.427 | 3.51E+13 | 634.565 | 0.00988 | 1355.19 | 0.00188 | 992.427 | 0.00137 |
| 634.855 | 6.66E+12 | 992.697 | 3.51E+13 | 634.855 | 0.00982 | 1356.9 | 0.00189 | 992.697 | 0.00135 |
| 635.145 | 6.69E+12 | 992.966 | 3.51E+13 | 635.145 | 0.00983 | 1358.61 | 0.00188 | 992.966 | 0.00133 |
| 635.434 | 6.69E+12 | 993.236 | 3.51E+13 | 635.434 | 0.00987 | 1360.32 | 0.00188 | 993.236 | 0.0013 |
| 635.724 | 6.67E+12 | 993.505 | 3.51E+13 | 635.724 | 0.00993 | 1362.02 | 0.00187 | 993.505 | 0.00126 |
| 636.013 | 6.69E+12 | 993.775 | 3.52E+13 | 636.013 | 0.00985 | 1363.73 | 0.00186 | 993.775 | 0.00123 |
| 636.303 | 6.70E+12 | 994.044 | 3.52E+13 | 636.303 | 0.00994 | 1365.44 | 0.00185 | 994.044 | 0.0012 |
| 636.593 | 6.69E+12 | 994.314 | 3.52E+13 | 636.593 | 0.00991 | 1367.14 | 0.00185 | 994.314 | 0.00116 |
| 636.882 | 6.68E+12 | 994.583 | 3.52E+13 | 636.882 | 0.01 | 1368.85 | 0.00184 | 994.583 | 0.00114 |
| 637.172 | 6.69E+12 | 994.852 | 3.52E+13 | 637.172 | 0.01001 | 1370.56 | 0.00183 | 994.852 | 0.0011 |
| 637.461 | 6.70E+12 | 995.122 | 3.53E+13 | 637.461 | 0.00992 | 1372.26 | 0.00183 | 995.122 | 0.00107 |
| 637.751 | 6.71E+12 | 995.391 | 3.53E+13 | 637.751 | 0.01001 | 1373.97 | 0.00182 | 995.391 | 0.00104 |
| 638.04 | 6.73E+12 | 995.66 | 3.53E+13 | 638.04 | 0.00997 | 1375.67 | 0.00181 | 995.66 | 0.00102 |
| 638.33 | 6.73E+12 | 995.93 | 3.53E+13 | 638.33 | 0.00999 | 1377.38 | 0.00181 | 995.93 | 0.00101 |
| 638.619 | 6.74E+12 | 996.199 | 3.53E+13 | 638.619 | 0.01 | 1379.09 | 0.00181 | 996.199 | 9.52E-04 |
| 638.909 | 6.74E+12 | 996.468 | 3.54E+13 | 638.909 | 0.00998 | 1380.79 | 0.0018 | 996.468 | 9.38E-04 |
| 639.198 | 6.76E+12 | 996.737 | 3.54E+13 | 639.198 | 0.01002 | 1382.5 | 0.00179 | 996.737 | 9.03E-04 |
| 639.488 | 6.75E+12 | 997.007 | 3.54E+13 | 639.488 | 0.01006 | 1384.2 | 0.00179 | 997.007 | 8.87E-04 |
| 639.777 | 6.77E+12 | 997.276 | 3.54E+13 | 639.777 | 0.01005 | 1385.91 | 0.00177 | 997.276 | 8.54E-04 |
| 640.067 | 6.77E+12 | 997.545 | 3.54E+13 | 640.067 | 0.01012 | 1387.61 | 0.00178 | 997.545 | 8.24E-04 |
| 640.356 | 6.78E+12 | 997.814 | 3.54E+13 | 640.356 | 0.01001 | 1389.32 | 0.00176 | 997.814 | 7.85E-04 |
| 640.646 | 6.79E+12 | 998.083 | 3.55E+13 | 640.646 | 0.01003 | 1391.02 | 0.00174 | 998.083 | 7.74E-04 |
| 640.935 | 6.79E+12 | 998.353 | 3.55E+13 | 640.935 | 0.01014 | 1392.73 | 0.00174 | 998.353 | 7.44E-04 |
| 641.225 | 6.81E+12 | 998.622 | 3.55E+13 | 641.225 | 0.01003 | 1394.43 | 0.00175 | 998.622 | 7.14E-04 |
| 641.514 | 6.82E+12 | 998.891 | 3.55E+13 | 641.514 | 0.01011 | 1396.14 | 0.00174 | 998.891 | 6.82E-04 |
| 641.803 | 6.82E+12 | 999.16 | 3.55E+13 | 641.803 | 0.01019 | 1397.84 | 0.00174 | 999.16 | 6.55E-04 |
| 642.093 | 6.84E+12 | 999.429 | 3.56E+13 | 642.093 | 0.01011 | 1399.55 | 0.00174 | 999.429 | 6.33E-04 |
| 642.382 | 6.84E+12 | 999.698 | 3.56E+13 | 642.382 | 0.01011 | 1401.25 | 0.00173 | 999.698 | 6.11E-04 |
| 642.671 | 6.82E+12 | 999.967 | 3.56E+13 | 642.671 | 0.01021 | 1402.95 | 0.00173 | 999.967 | 5.85E-04 |
| 642.961 | 6.84E+12 | | | 642.961 | 0.01019 | 1404.66 | 0.00173 | | |
| 643.25 | 6.85E+12 | | | 643.25 | 0.01015 | 1406.36 | 0.00172 | | |
| 643.539 | 6.87E+12 | | | 643.539 | 0.01014 | 1408.06 | 0.00172 | | |
| 643.829 | 6.87E+12 | | | 643.829 | 0.01021 | 1409.77 | 0.00172 | | |
| 644.118 | 6.87E+12 | | | 644.118 | 0.01021 | 1411.47 | 0.00172 | | |
| 644.407 | 6.87E+12 | | | 644.407 | 0.01018 | 1413.17 | 0.00173 | | |
| 644.696 | 6.88E+12 | | | 644.696 | 0.0102 | 1414.88 | 0.00173 | | |
| 644.986 | 6.91E+12 | | | 644.986 | 0.01022 | 1416.58 | 0.00174 | | |
| 645.275 | 6.90E+12 | | | 645.275 | 0.01017 | 1418.28 | 0.00174 | | |
| 645.564 | 6.92E+12 | | | 645.564 | 0.01024 | 1419.99 | 0.00174 | | |
| 645.853 | 6.92E+12 | | | 645.853 | 0.0103 | 1421.69 | 0.00174 | | |
| 646.143 | 6.95E+12 | | | 646.143 | 0.01017 | 1423.39 | 0.00175 | | |
| 646.432 | 6.95E+12 | | | 646.432 | 0.0102 | 1425.09 | 0.00175 | | |
| 646.721 | 6.96E+12 | | | 646.721 | 0.0102 | 1426.8 | 0.00175 | | |
| 647.01 | 6.95E+12 | | | 647.01 | 0.01018 | 1428.5 | 0.00176 | | |
| 647.299 | 6.97E+12 | | | 647.299 | 0.01023 | 1430.2 | 0.00177 | | |
| 647.588 | 7.00E+12 | | | 647.588 | 0.01018 | 1431.9 | 0.00177 | | |
| 647.878 | 6.99E+12 | | | 647.878 | 0.0102 | 1433.6 | 0.00177 | | |
| 648.167 | 7.01E+12 | | | 648.167 | 0.01019 | 1435.3 | 0.00177 | | |
| 648.456 | 7.03E+12 | | | 648.456 | 0.01023 | 1437.01 | 0.00177 | | |
| 648.745 | 7.05E+12 | | | 648.745 | 0.01019 | 1438.71 | 0.00179 | | |
| 649.034 | 7.05E+12 | | | 649.034 | 0.0102 | 1440.41 | 0.00179 | | |
| 649.323 | 7.05E+12 | | | 649.323 | 0.01031 | 1442.11 | 0.00179 | | |
| 649.612 | 7.07E+12 | | | 649.612 | 0.01021 | 1443.81 | 0.0018 | | |
| 649.901 | 7.09E+12 | | | 649.901 | 0.01022 | 1445.51 | 0.00179 | | |
| 650.19 | 7.09E+12 | | | 650.19 | 0.01027 | 1447.21 | 0.0018 | | |
| 650.479 | 7.10E+12 | | | 650.479 | 0.01026 | 1448.91 | 0.00181 | | |
| 650.768 | 7.09E+12 | | | 650.768 | 0.0103 | 1450.61 | 0.00182 | | |
| 651.057 | 7.08E+12 | | | 651.057 | 0.01028 | 1452.31 | 0.00182 | | |
| 651.346 | 7.10E+12 | | | 651.346 | 0.01025 | 1454.01 | 0.00183 | | |
| 651.635 | 7.15E+12 | | | 651.635 | 0.0102 | 1455.71 | 0.00184 | | |
| 651.924 | 7.15E+12 | | | 651.924 | 0.01014 | 1457.41 | 0.00184 | | |
| 652.213 | 7.18E+12 | | | 652.213 | 0.01022 | 1459.11 | 0.00185 | | |
| 652.502 | 7.19E+12 | | | 652.502 | 0.01022 | 1460.81 | 0.00186 | | |
| 652.791 | 7.21E+12 | | | 652.791 | 0.01018 | 1462.51 | 0.00185 | | |
| 653.08 | 7.22E+12 | | | 653.08 | 0.01022 | 1464.21 | 0.00184 | | |
| 653.369 | 7.25E+12 | | | 653.369 | 0.01018 | 1465.91 | 0.00185 | | |
| 653.658 | 7.27E+12 | | | 653.658 | 0.01021 | 1467.61 | 0.00185 | | |
| 653.947 | 7.28E+12 | | | 653.947 | 0.0102 | 1469.31 | 0.00186 | | |
| 654.235 | 7.30E+12 | | | 654.235 | 0.01021 | 1471 | 0.00186 | | |
| 654.524 | 7.32E+12 | | | 654.524 | 0.01017 | 1472.7 | 0.00186 | | |
| 654.813 | 7.30E+12 | | | 654.813 | 0.01022 | 1474.4 | 0.00187 | | |
| 655.102 | 7.31E+12 | | | 655.102 | 0.01029 | 1476.1 | 0.00186 | | |
| 655.391 | 7.31E+12 | | | 655.391 | 0.0102 | 1477.8 | 0.00186 | | |
| 655.68 | 7.34E+12 | | | 655.68 | 0.01029 | 1479.5 | 0.00187 | | |
| 655.968 | 7.36E+12 | | | 655.968 | 0.01025 | 1481.19 | 0.00186 | | |
| 656.257 | 7.37E+12 | | | 656.257 | 0.01026 | 1482.89 | 0.00185 | | |
| 656.546 | 7.37E+12 | | | 656.546 | 0.01021 | 1484.59 | 0.00185 | | |
| 656.835 | 7.36E+12 | | | 656.835 | 0.01027 | 1486.29 | 0.00185 | | |
| 657.124 | 7.36E+12 | | | 657.124 | 0.01034 | 1487.98 | 0.00185 | | |
| 657.412 | 7.37E+12 | | | 657.412 | 0.01036 | 1489.68 | 0.00185 | | |

| | | | |
|---|---|---|---|
| 496.3842 | 3.41E+07 | 1223.346 | 6.63E+10 |
| 496.6785 | 4.77E+07 | 1225.063 | 6.65E+10 |
| 496.9728 | 5.45E+07 | 1226.779 | 6.70E+10 |
| 497.2671 | 5.45E+07 | 1228.496 | 6.72E+10 |
| 497.5615 | 4.09E+07 | 1230.213 | 6.73E+10 |
| 497.8558 | 4.09E+07 | 1231.929 | 6.75E+10 |
| 498.15 | 6.82E+07 | 1233.645 | 6.81E+10 |
| 498.4443 | 8.86E+07 | 1235.362 | 6.84E+10 |
| 498.7386 | 6.82E+07 | 1237.078 | 6.85E+10 |
| 499.0329 | 2.04E+07 | 1238.794 | 6.88E+10 |
| 499.3271 | 8.18E+07 | 1240.51 | 6.87E+10 |
| 499.6214 | 6.82E+07 | 1242.225 | 6.88E+10 |
| 499.9156 | 6.82E+07 | 1243.941 | 6.91E+10 |
| 500.2098 | 6.13E+07 | 1245.656 | 6.93E+10 |
| 500.5041 | 6.82E+07 | 1247.372 | 6.97E+10 |
| 500.7983 | 4.77E+07 | 1249.087 | 7.01E+10 |
| 501.0925 | 5.45E+07 | 1250.802 | 7.00E+10 |
| 501.3867 | 5.45E+07 | 1252.517 | 7.02E+10 |
| 501.6809 | 5.45E+07 | 1254.232 | 7.05E+10 |
| 501.9751 | 8.18E+07 | 1255.947 | 7.08E+10 |
| 502.2692 | 6.13E+07 | 1257.661 | 7.09E+10 |
| 502.5634 | 4.09E+07 | 1259.376 | 7.12E+10 |
| 502.8575 | 6.13E+07 | 1261.09 | 7.13E+10 |
| 503.1517 | 6.82E+07 | 1262.805 | 7.16E+10 |
| 503.4458 | 7.50E+07 | 1264.519 | 7.17E+10 |
| 503.74 | 7.50E+07 | 1266.233 | 7.18E+10 |
| 504.0341 | 6.82E+07 | 1267.947 | 7.20E+10 |
| 504.3282 | 6.13E+07 | 1269.661 | 7.21E+10 |
| 504.6223 | 8.18E+07 | 1271.374 | 7.25E+10 |
| 504.9164 | 6.82E+07 | 1273.088 | 7.24E+10 |
| 505.2105 | 8.18E+07 | 1274.801 | 7.25E+10 |
| 505.5046 | 6.82E+07 | 1276.515 | 7.28E+10 |
| 505.7987 | 4.77E+07 | 1278.228 | 7.21E+10 |
| 506.0927 | 7.50E+07 | 1279.941 | 7.20E+10 |
| 506.3868 | 4.77E+07 | 1281.654 | 7.23E+10 |
| 506.6808 | 4.77E+07 | 1283.367 | 7.26E+10 |
| 506.9749 | 3.41E+07 | 1285.079 | 7.30E+10 |
| 507.2689 | 4.77E+07 | 1286.792 | 7.29E+10 |
| 507.5629 | 5.45E+07 | 1288.504 | 7.26E+10 |
| 507.857 | 5.45E+07 | 1290.217 | 7.28E+10 |
| 508.151 | 5.45E+07 | 1291.929 | 7.28E+10 |
| 508.445 | 8.18E+07 | 1293.641 | 7.27E+10 |
| 508.7389 | 8.18E+07 | 1295.353 | 7.29E+10 |
| 509.0329 | 6.13E+07 | 1297.065 | 7.28E+10 |
| 509.3269 | 6.82E+07 | 1298.776 | 7.25E+10 |
| 509.6209 | 7.50E+07 | 1300.488 | 7.26E+10 |
| 509.9148 | 6.82E+07 | 1302.199 | 7.29E+10 |
| 510.2088 | 4.77E+07 | 1303.911 | 7.28E+10 |
| 510.5027 | 3.41E+07 | 1305.622 | 7.27E+10 |
| 510.7966 | 6.82E+07 | 1307.333 | 7.29E+10 |
| 511.0906 | 7.50E+07 | 1309.044 | 7.30E+10 |
| 511.3845 | 4.09E+07 | 1310.755 | 7.30E+10 |
| 511.6784 | 6.13E+07 | 1312.466 | 7.29E+10 |
| 511.9723 | 8.18E+07 | 1314.176 | 7.28E+10 |
| 512.2662 | 7.50E+07 | 1315.887 | 7.29E+10 |
| 512.5601 | 6.82E+07 | 1317.597 | 7.26E+10 |
| 512.8539 | 6.82E+07 | 1319.307 | 7.25E+10 |
| 513.1478 | 6.13E+07 | 1321.017 | 7.25E+10 |
| 513.4416 | 4.09E+07 | 1322.727 | 7.24E+10 |
| 513.7355 | 8.86E+07 | 1324.437 | 7.22E+10 |
| 514.0293 | 2.73E+07 | 1326.147 | 7.22E+10 |
| 514.3232 | 8.18E+07 | 1327.857 | 7.22E+10 |
| 514.617 | 7.50E+07 | 1329.566 | 7.20E+10 |
| 514.9108 | 6.13E+07 | 1331.275 | 7.16E+10 |
| 515.2046 | 6.13E+07 | 1332.985 | 7.16E+10 |
| 515.4984 | 6.82E+07 | 1334.694 | 7.09E+10 |
| 515.7922 | 4.09E+07 | 1336.403 | 7.05E+10 |
| 516.086 | 8.86E+07 | 1338.112 | 7.10E+10 |
| 516.3797 | 6.13E+07 | 1339.82 | 7.10E+10 |
| 516.6735 | 1.02E+08 | 1341.529 | 7.08E+10 |
| 516.9672 | 6.13E+07 | 1343.237 | 7.05E+10 |
| 517.261 | 8.18E+07 | 1344.946 | 7.00E+10 |
| 517.5547 | 6.82E+07 | 1346.654 | 6.97E+10 |
| 517.8484 | 4.77E+07 | 1348.362 | 6.91E+10 |
| 518.1422 | 4.77E+07 | 1350.07 | 6.88E+10 |
| 518.4359 | 4.77E+07 | 1351.778 | 6.85E+10 |
| 518.7296 | 6.13E+07 | 1353.486 | 6.80E+10 |
| 519.0232 | 5.45E+07 | 1355.193 | 6.77E+10 |
| 519.3169 | 6.82E+07 | 1356.901 | 6.75E+10 |
| 519.6106 | 6.82E+07 | 1358.608 | 6.70E+10 |
| 519.9043 | 4.09E+07 | 1360.315 | 6.68E+10 |
| 520.1979 | 4.77E+07 | 1362.022 | 6.64E+10 |
| 520.4916 | 8.18E+07 | 1363.729 | 6.59E+10 |
| 520.7852 | 4.77E+07 | 1365.436 | 6.54E+10 |
| 521.0788 | 4.09E+07 | 1367.143 | 6.51E+10 |
| 521.3725 | 8.18E+07 | 1368.849 | 6.45E+10 |
| 521.6661 | 5.45E+07 | 1370.556 | 6.39E+10 |
| 521.9597 | 8.18E+07 | 1372.262 | 6.38E+10 |
| 522.2533 | 6.13E+07 | 1373.968 | 6.34E+10 |
| 522.5469 | 6.13E+07 | 1375.674 | 6.30E+10 |
| 522.8404 | 8.18E+07 | 1377.38 | 6.29E+10 |
| 523.134 | 6.82E+07 | 1379.086 | 6.25E+10 |
| 523.4276 | 8.86E+07 | 1380.792 | 6.20E+10 |
| 523.7211 | 6.13E+07 | 1382.497 | 6.15E+10 |
| 524.0147 | 8.18E+07 | 1384.203 | 6.12E+10 |
| 524.3082 | 7.50E+07 | 1385.908 | 6.06E+10 |
| 524.6017 | 4.77E+07 | 1387.613 | 6.05E+10 |
| 524.8952 | 7.50E+07 | 1389.318 | 5.97E+10 |
| 525.1887 | 8.18E+07 | 1391.023 | 5.89E+10 |
| 525.4822 | 5.45E+07 | 1392.728 | 5.91E+10 |
| 525.7757 | 9.54E+07 | 1394.432 | 5.91E+10 |
| 657.701 | 4.31E+20 | 657.701 | 0.43235 |
| 657.99 | 4.32E+20 | 657.99 | 0.43232 |
| 658.278 | 4.33E+20 | 658.278 | 0.43228 |
| 658.567 | 4.34E+20 | 658.567 | 0.43224 |
| 658.856 | 4.35E+20 | 658.856 | 0.4322 |
| 659.144 | 4.36E+20 | 659.144 | 0.43216 |
| 659.433 | 4.37E+20 | 659.433 | 0.43213 |
| 659.722 | 4.38E+20 | 659.722 | 0.43209 |
| 660.01 | 4.39E+20 | 660.01 | 0.43205 |
| 660.299 | 4.39E+20 | 660.299 | 0.43201 |
| 660.588 | 4.40E+20 | 660.588 | 0.43197 |
| 660.876 | 4.41E+20 | 660.876 | 0.43194 |
| 661.165 | 4.42E+20 | 661.165 | 0.4319 |
| 661.453 | 4.43E+20 | 661.453 | 0.43186 |
| 661.742 | 4.44E+20 | 661.742 | 0.43182 |
| 662.03 | 4.45E+20 | 662.03 | 0.43178 |
| 662.319 | 4.46E+20 | 662.319 | 0.43174 |
| 662.607 | 4.47E+20 | 662.607 | 0.43171 |
| 662.896 | 4.48E+20 | 662.896 | 0.43167 |
| 663.184 | 4.49E+20 | 663.184 | 0.43163 |
| 663.473 | 4.50E+20 | 663.473 | 0.43159 |
| 663.761 | 4.51E+20 | 663.761 | 0.43155 |
| 664.05 | 4.52E+20 | 664.05 | 0.43151 |
| 664.338 | 4.53E+20 | 664.338 | 0.43147 |
| 664.627 | 4.54E+20 | 664.627 | 0.43143 |
| 664.915 | 4.55E+20 | 664.915 | 0.43139 |
| 665.204 | 4.56E+20 | 665.204 | 0.43135 |
| 665.492 | 4.57E+20 | 665.492 | 0.43132 |
| 665.78 | 4.57E+20 | 665.78 | 0.43128 |
| 666.069 | 4.58E+20 | 666.069 | 0.43124 |
| 666.357 | 4.59E+20 | 666.357 | 0.4312 |
| 666.645 | 4.60E+20 | 666.645 | 0.43116 |
| 666.934 | 4.61E+20 | 666.934 | 0.43112 |
| 667.222 | 4.62E+20 | 667.222 | 0.43108 |
| 667.51 | 4.63E+20 | 667.51 | 0.43104 |
| 667.799 | 4.64E+20 | 667.799 | 0.431 |
| 668.087 | 4.65E+20 | 668.087 | 0.43096 |
| 668.375 | 4.66E+20 | 668.375 | 0.43092 |
| 668.663 | 4.67E+20 | 668.663 | 0.43088 |
| 668.952 | 4.68E+20 | 668.952 | 0.43084 |
| 669.24 | 4.69E+20 | 669.24 | 0.4308 |
| 669.528 | 4.70E+20 | 669.528 | 0.43076 |
| 669.816 | 4.71E+20 | 669.816 | 0.43072 |
| 670.105 | 4.72E+20 | 670.105 | 0.43068 |
| 670.393 | 4.73E+20 | 670.393 | 0.43064 |
| 670.681 | 4.74E+20 | 670.681 | 0.4306 |
| 670.969 | 4.75E+20 | 670.969 | 0.43056 |
| 671.257 | 4.76E+20 | 671.257 | 0.43052 |
| 671.545 | 4.77E+20 | 671.545 | 0.43048 |
| 671.833 | 4.78E+20 | 671.833 | 0.43044 |
| 672.122 | 4.79E+20 | 672.122 | 0.4304 |
| 672.41 | 4.80E+20 | 672.41 | 0.43036 |
| 672.698 | 4.81E+20 | 672.698 | 0.43032 |
| 672.986 | 4.82E+20 | 672.986 | 0.43028 |
| 673.274 | 4.82E+20 | 673.274 | 0.43024 |
| 673.562 | 4.83E+20 | 673.562 | 0.4302 |
| 673.85 | 4.84E+20 | 673.85 | 0.43016 |
| 674.138 | 4.85E+20 | 674.138 | 0.43012 |
| 674.426 | 4.86E+20 | 674.426 | 0.43008 |
| 674.714 | 4.87E+20 | 674.714 | 0.43004 |
| 675.002 | 4.88E+20 | 675.002 | 0.43 |
| 675.29 | 4.89E+20 | 675.29 | 0.42996 |
| 675.578 | 4.90E+20 | 675.578 | 0.42992 |
| 675.866 | 4.91E+20 | 675.866 | 0.42988 |
| 676.154 | 4.92E+20 | 676.154 | 0.42983 |
| 676.442 | 4.93E+20 | 676.442 | 0.42979 |
| 676.73 | 4.94E+20 | 676.73 | 0.42975 |
| 677.018 | 4.95E+20 | 677.018 | 0.42971 |
| 677.305 | 4.96E+20 | 677.305 | 0.42967 |
| 677.593 | 4.97E+20 | 677.593 | 0.42963 |
| 677.881 | 4.98E+20 | 677.881 | 0.42959 |
| 678.169 | 4.99E+20 | 678.169 | 0.42955 |
| 678.457 | 5.00E+20 | 678.457 | 0.4295 |
| 678.745 | 5.01E+20 | 678.745 | 0.42946 |
| 679.032 | 5.02E+20 | 679.032 | 0.42942 |
| 679.32 | 5.03E+20 | 679.32 | 0.42938 |
| 679.608 | 5.04E+20 | 679.608 | 0.42934 |
| 679.896 | 5.05E+20 | 679.896 | 0.4293 |
| 680.184 | 5.06E+20 | 680.184 | 0.42925 |
| 680.471 | 5.07E+20 | 680.471 | 0.42921 |
| 680.759 | 5.08E+20 | 680.759 | 0.42917 |
| 681.047 | 5.09E+20 | 681.047 | 0.42913 |
| 681.334 | 5.10E+20 | 681.334 | 0.42909 |
| 681.622 | 5.11E+20 | 681.622 | 0.42905 |
| 681.91 | 5.12E+20 | 681.91 | 0.429 |
| 682.197 | 5.13E+20 | 682.197 | 0.42896 |
| 682.485 | 5.14E+20 | 682.485 | 0.42892 |
| 682.773 | 5.15E+20 | 682.773 | 0.42888 |
| 683.06 | 5.16E+20 | 683.06 | 0.42883 |
| 683.348 | 5.17E+20 | 683.348 | 0.42879 |
| 683.636 | 5.18E+20 | 683.636 | 0.42875 |
| 683.923 | 5.19E+20 | 683.923 | 0.42871 |
| 684.211 | 5.20E+20 | 684.211 | 0.42866 |
| 684.498 | 5.21E+20 | 684.498 | 0.42862 |
| 684.786 | 5.22E+20 | 684.786 | 0.42858 |
| 685.073 | 5.23E+20 | 685.073 | 0.42854 |
| 685.361 | 5.24E+20 | 685.361 | 0.42849 |
| 685.648 | 5.25E+20 | 685.648 | 0.42845 |
| 685.936 | 5.26E+20 | 685.936 | 0.42841 |
| 686.223 | 5.27E+20 | 686.223 | 0.42837 |
| 686.511 | 5.28E+20 | 686.511 | 0.42832 |

| | | | | | | | |
|---|---|---|---|---|---|---|---|
| 657.701 | 3.96E-08 | 1491.38 | 4.15E-08 | 657.701 | 3.67E-08 | 1442.11 | 4.44E-08 |
| 657.99 | 3.96E-08 | 1493.08 | 4.14E-08 | 657.99 | 4.40E-08 | 1443.81 | 4.46E-08 |
| 658.278 | 3.96E-08 | 1494.77 | 4.13E-08 | 658.278 | 4.06E-08 | 1445.51 | 4.26E-08 |
| 658.567 | 3.96E-08 | 1496.47 | 4.12E-08 | 658.567 | 4.12E-08 | 1447.21 | 4.30E-08 |
| 658.856 | 3.97E-08 | 1498.17 | 4.12E-08 | 658.856 | 4.11E-08 | 1448.91 | 4.38E-08 |
| 659.144 | 3.96E-08 | 1499.86 | 4.10E-08 | 659.144 | 4.02E-08 | 1450.61 | 4.45E-08 |
| 659.433 | 3.95E-08 | 1501.56 | 4.09E-08 | 659.433 | 3.98E-08 | 1452.31 | 4.39E-08 |
| 659.722 | 3.96E-08 | 1503.25 | 4.08E-08 | 659.722 | 4.11E-08 | 1454.01 | 4.29E-08 |
| 660.01 | 3.95E-08 | 1504.95 | 4.07E-08 | 660.01 | 3.89E-08 | 1455.71 | 4.29E-08 |
| 660.299 | 3.95E-08 | 1506.65 | 4.06E-08 | 660.299 | 3.96E-08 | 1457.41 | 4.29E-08 |
| 660.588 | 3.96E-08 | 1508.34 | 4.05E-08 | 660.588 | 3.72E-08 | 1459.11 | 4.31E-08 |
| 660.876 | 3.96E-08 | 1510.04 | 4.05E-08 | 660.876 | 4.17E-08 | 1460.81 | 4.32E-08 |
| 661.165 | 3.96E-08 | 1511.73 | 4.04E-08 | 661.165 | 3.87E-08 | 1462.51 | 4.25E-08 |
| 661.453 | 3.96E-08 | 1513.43 | 4.03E-08 | 661.453 | 4.15E-08 | 1464.21 | 4.23E-08 |
| 661.742 | 3.95E-08 | 1515.12 | 4.01E-08 | 661.742 | 3.85E-08 | 1465.91 | 4.27E-08 |
| 662.03 | 3.94E-08 | 1516.82 | 4.00E-08 | 662.03 | 3.78E-08 | 1467.61 | 4.28E-08 |
| 662.319 | 3.94E-08 | 1518.51 | 4.00E-08 | 662.319 | 4.00E-08 | 1469.31 | 4.23E-08 |
| 662.607 | 3.95E-08 | 1520.21 | 3.99E-08 | 662.607 | 3.69E-08 | 1471 | 4.24E-08 |
| 662.896 | 3.95E-08 | 1521.9 | 3.98E-08 | 662.896 | 3.75E-08 | 1472.7 | 4.22E-08 |
| 663.184 | 3.95E-08 | 1523.6 | 3.97E-08 | 663.184 | 3.87E-08 | 1474.4 | 4.16E-08 |
| 663.473 | 3.93E-08 | 1525.29 | 3.96E-08 | 663.473 | 3.88E-08 | 1476.1 | 4.22E-08 |
| 663.761 | 3.94E-08 | 1526.99 | 3.96E-08 | 663.761 | 4.18E-08 | 1477.8 | 4.22E-08 |
| 664.05 | 3.93E-08 | 1528.68 | 3.95E-08 | 664.05 | 4.04E-08 | 1479.5 | 4.10E-08 |
| 664.338 | 3.92E-08 | 1530.37 | 3.94E-08 | 664.338 | 3.85E-08 | 1481.19 | 4.27E-08 |
| 664.627 | 3.92E-08 | 1532.07 | 3.94E-08 | 664.627 | 4.11E-08 | 1482.89 | 4.36E-08 |
| 664.915 | 3.91E-08 | 1533.76 | 3.93E-08 | 664.915 | 3.67E-08 | 1484.59 | 4.24E-08 |
| 665.204 | 3.90E-08 | 1535.45 | 3.92E-08 | 665.204 | 3.63E-08 | 1486.29 | 4.17E-08 |
| 665.492 | 3.90E-08 | 1537.15 | 3.92E-08 | 665.492 | 4.10E-08 | 1487.98 | 4.18E-08 |
| 665.78 | 3.89E-08 | 1538.84 | 3.91E-08 | 665.78 | 3.83E-08 | 1489.68 | 4.10E-08 |
| 666.069 | 3.88E-08 | 1540.53 | 3.91E-08 | 666.069 | 3.74E-08 | 1491.38 | 4.06E-08 |
| 666.357 | 3.88E-08 | 1542.23 | 3.90E-08 | 666.357 | 4.26E-08 | 1493.08 | 4.17E-08 |
| 666.645 | 3.88E-08 | 1543.92 | 3.89E-08 | 666.645 | 3.80E-08 | 1494.77 | 4.13E-08 |
| 666.934 | 3.87E-08 | 1545.61 | 3.88E-08 | 666.934 | 3.84E-08 | 1496.47 | 4.19E-08 |
| 667.222 | 3.86E-08 | 1547.31 | 3.88E-08 | 667.222 | 3.90E-08 | 1498.17 | 4.17E-08 |
| 667.51 | 3.86E-08 | 1549 | 3.86E-08 | 667.51 | 3.84E-08 | 1499.86 | 4.08E-08 |
| 667.799 | 3.86E-08 | 1550.69 | 3.86E-08 | 667.799 | 3.75E-08 | 1501.56 | 4.12E-08 |
| 668.087 | 3.86E-08 | 1552.38 | 3.85E-08 | 668.087 | 3.84E-08 | 1503.25 | 4.05E-08 |
| 668.375 | 3.86E-08 | 1554.07 | 3.85E-08 | 668.375 | 3.99E-08 | 1504.95 | 4.06E-08 |
| 668.663 | 3.86E-08 | 1555.77 | 3.84E-08 | 668.663 | 3.80E-08 | 1506.65 | 4.10E-08 |
| 668.952 | 3.87E-08 | 1557.46 | 3.83E-08 | 668.952 | 4.16E-08 | 1508.34 | 4.03E-08 |
| 669.24 | 3.86E-08 | 1559.15 | 3.82E-08 | 669.24 | 3.92E-08 | 1510.04 | 3.99E-08 |
| 669.528 | 3.87E-08 | 1560.84 | 3.81E-08 | 669.528 | 3.97E-08 | 1511.73 | 4.03E-08 |
| 669.816 | 3.86E-08 | 1562.53 | 3.80E-08 | 669.816 | 3.85E-08 | 1513.43 | 4.00E-08 |
| 670.105 | 3.86E-08 | 1564.22 | 3.79E-08 | 670.105 | 3.67E-08 | 1515.12 | 4.02E-08 |
| 670.393 | 3.86E-08 | 1565.91 | 3.78E-08 | 670.393 | 4.07E-08 | 1516.82 | 3.97E-08 |
| 670.681 | 3.84E-08 | 1567.6 | 3.77E-08 | 670.681 | 3.64E-08 | 1518.51 | 4.00E-08 |
| 670.969 | 3.85E-08 | 1569.3 | 3.76E-08 | 670.969 | 3.85E-08 | 1520.21 | 4.02E-08 |
| 671.257 | 3.85E-08 | 1570.99 | 3.75E-08 | 671.257 | 3.76E-08 | 1521.9 | 3.96E-08 |
| 671.545 | 3.84E-08 | 1572.68 | 3.74E-08 | 671.545 | 3.78E-08 | 1523.6 | 3.99E-08 |
| 671.833 | 3.84E-08 | 1574.37 | 3.73E-08 | 671.833 | 3.60E-08 | 1525.29 | 3.97E-08 |
| 672.122 | 3.84E-08 | 1576.06 | 3.72E-08 | 672.122 | 3.93E-08 | 1526.99 | 3.97E-08 |
| 672.41 | 3.83E-08 | 1577.75 | 3.71E-08 | 672.41 | 3.61E-08 | 1528.68 | 3.90E-08 |
| 672.698 | 3.83E-08 | 1579.44 | 3.70E-08 | 672.698 | 3.65E-08 | 1530.37 | 3.95E-08 |
| 672.986 | 3.83E-08 | 1581.13 | 3.69E-08 | 672.986 | 3.78E-08 | 1532.07 | 3.95E-08 |
| 673.274 | 3.83E-08 | | | 673.274 | 4.08E-08 | 1533.76 | 3.91E-08 |
| 673.562 | 3.83E-08 | | | 673.562 | 3.69E-08 | 1535.45 | 3.95E-08 |
| 673.85 | 3.83E-08 | | | 673.85 | 3.98E-08 | 1537.15 | 3.88E-08 |
| 674.138 | 3.82E-08 | | | 674.138 | 3.75E-08 | 1538.84 | 3.86E-08 |
| 674.426 | 3.81E-08 | | | 674.426 | 3.70E-08 | 1540.53 | 3.92E-08 |
| 674.714 | 3.80E-08 | | | 674.714 | 4.15E-08 | 1542.23 | 3.99E-08 |
| 675.002 | 3.79E-08 | | | 675.002 | 3.74E-08 | 1543.92 | 3.87E-08 |
| 675.29 | 3.79E-08 | | | 675.29 | 4.07E-08 | 1545.61 | 3.85E-08 |
| 675.578 | 3.78E-08 | | | 675.578 | 3.87E-08 | 1547.31 | 3.90E-08 |
| 675.866 | 3.78E-08 | | | 675.866 | 3.91E-08 | 1549 | 3.82E-08 |
| 676.154 | 3.79E-08 | | | 676.154 | 3.83E-08 | 1550.69 | 3.88E-08 |
| 676.442 | 3.78E-08 | | | 676.442 | 3.54E-08 | 1552.38 | 3.86E-08 |
| 676.73 | 3.77E-08 | | | 676.73 | 3.78E-08 | 1554.07 | 3.85E-08 |
| 677.018 | 3.77E-08 | | | 677.018 | 3.85E-08 | 1555.77 | 3.94E-08 |
| 677.305 | 3.77E-08 | | | 677.305 | 3.65E-08 | 1557.46 | 3.89E-08 |
| 677.593 | 3.77E-08 | | | 677.593 | 3.81E-08 | 1559.15 | 3.88E-08 |
| 677.881 | 3.77E-08 | | | 677.881 | 3.78E-08 | 1560.84 | 3.76E-08 |
| 678.169 | 3.77E-08 | | | 678.169 | 3.73E-08 | 1562.53 | 3.70E-08 |
| 678.457 | 3.77E-08 | | | 678.457 | 3.99E-08 | 1564.22 | 3.78E-08 |
| 678.745 | 3.77E-08 | | | 678.745 | 3.85E-08 | 1565.91 | 3.71E-08 |
| 679.032 | 3.77E-08 | | | 679.032 | 3.97E-08 | 1567.6 | 3.86E-08 |
| 679.32 | 3.76E-08 | | | 679.32 | 3.72E-08 | 1569.3 | 3.80E-08 |
| 679.608 | 3.76E-08 | | | 679.608 | 3.64E-08 | 1570.99 | 3.76E-08 |
| 679.896 | 3.76E-08 | | | 679.896 | 3.34E-08 | 1572.68 | 3.78E-08 |
| 680.184 | 3.75E-08 | | | 680.184 | 3.78E-08 | 1574.37 | 3.65E-08 |
| 680.471 | 3.75E-08 | | | 680.471 | 3.39E-08 | 1576.06 | 3.70E-08 |
| 680.759 | 3.75E-08 | | | 680.759 | 3.78E-08 | 1577.75 | 3.73E-08 |
| 681.047 | 3.75E-08 | | | 681.047 | 3.86E-08 | 1579.44 | 3.75E-08 |
| 681.334 | 3.74E-08 | | | 681.334 | 3.75E-08 | 1581.13 | 3.68E-08 |
| 681.622 | 3.74E-08 | | | 681.622 | 3.85E-08 | | |
| 681.91 | 3.74E-08 | | | 681.91 | 3.78E-08 | | |
| 682.197 | 3.74E-08 | | | 682.197 | 3.60E-08 | | |
| 682.485 | 3.74E-08 | | | 682.485 | 3.51E-08 | | |
| 682.773 | 3.74E-08 | | | 682.773 | 3.87E-08 | | |
| 683.06 | 3.75E-08 | | | 683.06 | 3.53E-08 | | |
| 683.348 | 3.75E-08 | | | 683.348 | 3.84E-08 | | |
| 683.636 | 3.75E-08 | | | 683.636 | 3.75E-08 | | |
| 683.923 | 3.75E-08 | | | 683.923 | 3.76E-08 | | |
| 684.211 | 3.75E-08 | | | 684.211 | 3.72E-08 | | |
| 684.498 | 3.74E-08 | | | 684.498 | 3.69E-08 | | |
| 684.786 | 3.74E-08 | | | 684.786 | 3.58E-08 | | |
| 685.073 | 3.75E-08 | | | 685.073 | 3.82E-08 | | |
| 685.361 | 3.74E-08 | | | 685.361 | 3.68E-08 | | |
| 685.648 | 3.74E-08 | | | 685.648 | 3.86E-08 | | |
| 685.936 | 3.75E-08 | | | 685.936 | 3.64E-08 | | |
| 686.223 | 3.75E-08 | | | 686.223 | 3.69E-08 | | |
| 686.511 | 3.76E-08 | | | 686.511 | 3.88E-08 | | |

| | | | | | |
|---|---|---|---|---|---|
| 657.701 | 7.39E+12 | 657.701 | 0.0103 | 1491.38 | 0.00184 |
| 657.99 | 7.40E+12 | 657.99 | 0.01034 | 1493.08 | 0.00184 |
| 658.278 | 7.42E+12 | 658.278 | 0.01038 | 1494.77 | 0.00184 |
| 658.567 | 7.44E+12 | 658.567 | 0.01033 | 1496.47 | 0.00183 |
| 658.856 | 7.46E+12 | 658.856 | 0.01031 | 1498.17 | 0.00182 |
| 659.144 | 7.46E+12 | 659.144 | 0.01036 | 1499.86 | 0.00183 |
| 659.433 | 7.46E+12 | 659.433 | 0.01046 | 1501.56 | 0.00183 |
| 659.722 | 7.49E+12 | 659.722 | 0.0104 | 1503.25 | 0.00183 |
| 660.01 | 7.49E+12 | 660.01 | 0.0104 | 1504.95 | 0.00183 |
| 660.299 | 7.50E+12 | 660.299 | 0.01044 | 1506.65 | 0.00182 |
| 660.588 | 7.53E+12 | 660.588 | 0.01035 | 1508.34 | 0.00181 |
| 660.876 | 7.54E+12 | 660.876 | 0.01042 | 1510.04 | 0.00181 |
| 661.165 | 7.56E+12 | 661.165 | 0.01037 | 1511.73 | 0.00181 |
| 661.453 | 7.57E+12 | 661.453 | 0.01044 | 1513.43 | 0.00181 |
| 661.742 | 7.57E+12 | 661.742 | 0.01041 | 1515.12 | 0.00181 |
| 662.03 | 7.58E+12 | 662.03 | 0.0104 | 1516.82 | 0.0018 |
| 662.319 | 7.58E+12 | 662.319 | 0.01051 | 1518.51 | 0.00179 |
| 662.607 | 7.61E+12 | 662.607 | 0.01045 | 1520.21 | 0.00179 |
| 662.896 | 7.63E+12 | 662.896 | 0.01052 | 1521.9 | 0.0018 |
| 663.184 | 7.66E+12 | 663.184 | 0.01048 | 1523.6 | 0.0018 |
| 663.473 | 7.64E+12 | 663.473 | 0.01055 | 1525.29 | 0.0018 |
| 663.761 | 7.67E+12 | 663.761 | 0.01051 | 1526.99 | 0.00179 |
| 664.05 | 7.66E+12 | 664.05 | 0.01052 | 1528.68 | 0.00178 |
| 664.338 | 7.65E+12 | 664.338 | 0.01059 | 1530.37 | 0.00178 |
| 664.627 | 7.67E+12 | 664.627 | 0.01059 | 1532.07 | 0.00178 |
| 664.915 | 7.66E+12 | 664.915 | 0.01069 | 1533.76 | 0.00178 |
| 665.204 | 7.67E+12 | 665.204 | 0.01061 | 1535.45 | 0.00177 |
| 665.492 | 7.67E+12 | 665.492 | 0.01068 | 1537.15 | 0.00177 |
| 665.78 | 7.67E+12 | 665.78 | 0.01067 | 1538.84 | 0.00175 |
| 666.069 | 7.67E+12 | 666.069 | 0.01066 | 1540.53 | 0.00175 |
| 666.357 | 7.69E+12 | 666.357 | 0.01067 | 1542.23 | 0.00175 |
| 666.645 | 7.70E+12 | 666.645 | 0.0107 | 1543.92 | 0.00174 |
| 666.934 | 7.69E+12 | 666.934 | 0.01075 | 1545.61 | 0.00175 |
| 667.222 | 7.70E+12 | 667.222 | 0.01086 | 1547.31 | 0.00175 |
| 667.51 | 7.71E+12 | 667.51 | 0.01075 | 1549 | 0.00174 |
| 667.799 | 7.72E+12 | 667.799 | 0.01088 | 1550.69 | 0.00173 |
| 668.087 | 7.74E+12 | 668.087 | 0.01088 | 1552.38 | 0.00172 |
| 668.375 | 7.74E+12 | 668.375 | 0.01088 | 1554.07 | 0.00173 |
| 668.663 | 7.76E+12 | 668.663 | 0.01077 | 1555.77 | 0.00171 |
| 668.952 | 7.80E+12 | 668.952 | 0.01087 | 1557.46 | 0.00171 |
| 669.24 | 7.80E+12 | 669.24 | 0.01082 | 1559.15 | 0.00171 |
| 669.528 | 7.83E+12 | 669.528 | 0.01093 | 1560.84 | 0.00171 |
| 669.816 | 7.83E+12 | 669.816 | 0.0108 | 1562.53 | 0.0017 |
| 670.105 | 7.84E+12 | 670.105 | 0.0109 | 1564.22 | 0.00168 |
| 670.393 | 7.85E+12 | 670.393 | 0.01086 | 1565.91 | 0.00168 |
| 670.681 | 7.84E+12 | 670.681 | 0.01101 | 1567.6 | 0.00166 |
| 670.969 | 7.86E+12 | 670.969 | 0.01086 | 1569.3 | 0.00164 |
| 671.257 | 7.89E+12 | 671.257 | 0.01101 | 1570.99 | 0.00162 |
| 671.545 | 7.88E+12 | 671.545 | 0.01094 | 1572.68 | 0.0016 |
| 671.833 | 7.89E+12 | 671.833 | 0.01099 | 1574.37 | 0.00157 |
| 672.122 | 7.91E+12 | 672.122 | 0.01102 | 1576.06 | 0.00154 |
| 672.41 | 7.90E+12 | 672.41 | 0.01097 | 1577.75 | 0.00152 |
| 672.698 | 7.92E+12 | 672.698 | 0.01099 | 1579.44 | 0.00148 |
| 672.986 | 7.94E+12 | 672.986 | 0.01098 | 1581.13 | 0.00145 |
| 673.274 | 7.96E+12 | 673.274 | 0.01104 | | |
| 673.562 | 7.97E+12 | 673.562 | 0.01099 | | |
| 673.85 | 7.98E+12 | 673.85 | 0.01105 | | |
| 674.138 | 7.97E+12 | 674.138 | 0.0111 | | |
| 674.426 | 7.97E+12 | 674.426 | 0.0111 | | |
| 674.714 | 7.97E+12 | 674.714 | 0.01116 | | |
| 675.002 | 7.96E+12 | 675.002 | 0.01114 | | |
| 675.29 | 7.97E+12 | 675.29 | 0.01126 | | |
| 675.578 | 7.98E+12 | 675.578 | 0.01113 | | |
| 675.866 | 7.99E+12 | 675.866 | 0.01121 | | |
| 676.154 | 8.01E+12 | 676.154 | 0.01113 | | |
| 676.442 | 8.00E+12 | 676.442 | 0.01116 | | |
| 676.73 | 8.01E+12 | 676.73 | 0.01126 | | |
| 677.018 | 8.03E+12 | 677.018 | 0.01115 | | |
| 677.305 | 8.03E+12 | 677.305 | 0.01123 | | |
| 677.593 | 8.05E+12 | 677.593 | 0.01124 | | |
| 677.881 | 8.07E+12 | 677.881 | 0.0112 | | |
| 678.169 | 8.08E+12 | 678.169 | 0.0112 | | |
| 678.457 | 8.10E+12 | 678.457 | 0.01126 | | |
| 678.745 | 8.12E+12 | 678.745 | 0.01124 | | |
| 679.032 | 8.12E+12 | 679.032 | 0.01114 | | |
| 679.32 | 8.12E+12 | 679.32 | 0.01126 | | |
| 679.608 | 8.14E+12 | 679.608 | 0.01122 | | |
| 679.896 | 8.15E+12 | 679.896 | 0.01125 | | |
| 680.184 | 8.16E+12 | 680.184 | 0.01134 | | |
| 680.471 | 8.17E+12 | 680.471 | 0.01124 | | |
| 680.759 | 8.17E+12 | 680.759 | 0.01135 | | |
| 681.047 | 8.19E+12 | 681.047 | 0.01133 | | |
| 681.334 | 8.19E+12 | 681.334 | 0.01136 | | |
| 681.622 | 8.20E+12 | 681.622 | 0.01141 | | |
| 681.91 | 8.21E+12 | 681.91 | 0.01131 | | |
| 682.197 | 8.23E+12 | 682.197 | 0.01129 | | |
| 682.485 | 8.25E+12 | 682.485 | 0.01127 | | |
| 682.773 | 8.26E+12 | 682.773 | 0.0113 | | |
| 683.06 | 8.29E+12 | 683.06 | 0.01121 | | |
| 683.348 | 8.31E+12 | 683.348 | 0.0113 | | |
| 683.636 | 8.34E+12 | 683.636 | 0.01121 | | |
| 683.923 | 8.35E+12 | 683.923 | 0.01129 | | |
| 684.211 | 8.36E+12 | 684.211 | 0.01121 | | |
| 684.498 | 8.36E+12 | 684.498 | 0.01133 | | |
| 684.786 | 8.37E+12 | 684.786 | 0.01134 | | |
| 685.073 | 8.39E+12 | 685.073 | 0.01126 | | |
| 685.361 | 8.41E+12 | 685.361 | 0.01136 | | |
| 685.648 | 8.42E+12 | 685.648 | 0.0113 | | |
| 685.936 | 8.45E+12 | 685.936 | 0.01128 | | |
| 686.223 | 8.45E+12 | 686.223 | 0.01137 | | |
| 686.511 | 8.50E+12 | 686.511 | 0.0113 | | |

| | | | | | | | | | | | | | | | |
|---|---|---|---|---|---|---|---|---|---|---|---|---|---|---|---|
| 526.0692 | 7.50E+07 | 1396.137 | 5.86E+10 | 686.798 | 5.29E+20 | 686.798 | 0.42828 | 686.798 | 3.76E-08 | 686.798 | 3.74E-08 | 686.798 | 8.51E+12 | 686.798 | 0.0112 |
| 526.3627 | 1.02E+08 | 1397.841 | 5.84E+10 | 687.086 | 5.30E+20 | 687.086 | 0.42824 | 687.086 | 3.75E-08 | 687.086 | 3.88E-08 | 687.086 | 8.51E+12 | 687.086 | 0.01131 |
| 526.6561 | 7.50E+07 | 1399.545 | 5.84E+10 | 687.373 | 5.31E+20 | 687.373 | 0.42819 | 687.373 | 3.75E-08 | 687.373 | 3.71E-08 | 687.373 | 8.52E+12 | 687.373 | 0.01135 |
| 526.9496 | 8.18E+07 | 1401.25 | 5.80E+10 | 687.661 | 5.32E+20 | 687.661 | 0.42815 | 687.661 | 3.74E-08 | 687.661 | 3.81E-08 | 687.661 | 8.53E+12 | 687.661 | 0.01132 |
| 527.243 | 9.54E+07 | 1402.954 | 5.79E+10 | 687.948 | 5.33E+20 | 687.948 | 0.42811 | 687.948 | 3.74E-08 | 687.948 | 3.94E-08 | 687.948 | 8.53E+12 | 687.948 | 0.01142 |
| 527.5365 | 9.54E+07 | 1404.657 | 5.79E+10 | 688.236 | 5.34E+20 | 688.236 | 0.42806 | 688.236 | 3.74E-08 | 688.236 | 3.75E-08 | 688.236 | 8.55E+12 | 688.236 | 0.01135 |
| 527.8299 | 6.13E+07 | 1406.361 | 5.75E+10 | 688.523 | 5.35E+20 | 688.523 | 0.42802 | 688.523 | 3.75E-08 | 688.523 | 3.65E-08 | 688.523 | 8.58E+12 | 688.523 | 0.01135 |
| 528.1233 | 5.45E+07 | 1408.065 | 5.71E+10 | 688.81 | 5.36E+20 | 688.81 | 0.42798 | 688.81 | 3.74E-08 | 688.81 | 4.15E-08 | 688.81 | 8.59E+12 | 688.81 | 0.01129 |
| 528.4167 | 6.82E+07 | 1409.768 | 5.71E+10 | 689.098 | 5.37E+20 | 689.098 | 0.42793 | 689.098 | 3.75E-08 | 689.098 | 3.65E-08 | 689.098 | 8.61E+12 | 689.098 | 0.01133 |
| 528.7101 | 8.86E+07 | 1411.472 | 5.71E+10 | 689.385 | 5.38E+20 | 689.385 | 0.42789 | 689.385 | 3.75E-08 | 689.385 | 4.03E-08 | 689.385 | 8.62E+12 | 689.385 | 0.01126 |
| 529.0035 | 5.45E+07 | 1413.175 | 5.72E+10 | 689.672 | 5.39E+20 | 689.672 | 0.42785 | 689.672 | 3.75E-08 | 689.672 | 3.70E-08 | 689.672 | 8.64E+12 | 689.672 | 0.01132 |
| 529.2969 | 5.45E+07 | 1414.878 | 5.72E+10 | 689.96 | 5.40E+20 | 689.96 | 0.4278 | 689.96 | 3.74E-08 | 689.96 | 3.68E-08 | 689.96 | 8.65E+12 | 689.96 | 0.01126 |
| 529.5903 | 6.13E+07 | 1416.581 | 5.72E+10 | 690.247 | 5.41E+20 | 690.247 | 0.42776 | 690.247 | 3.75E-08 | 690.247 | 3.72E-08 | 690.247 | 8.67E+12 | 690.247 | 0.01135 |
| 529.8836 | 6.13E+07 | 1418.284 | 5.73E+10 | 690.534 | 5.42E+20 | 690.534 | 0.42772 | 690.534 | 3.75E-08 | 690.534 | 3.77E-08 | 690.534 | 8.68E+12 | 690.534 | 0.01134 |
| 530.177 | 8.86E+07 | 1419.986 | 5.72E+10 | 690.822 | 5.43E+20 | 690.822 | 0.42767 | 690.822 | 3.75E-08 | 690.822 | 4.07E-08 | 690.822 | 8.70E+12 | 690.822 | 0.0113 |
| 530.4703 | 6.13E+07 | 1421.689 | 5.74E+10 | 691.109 | 5.44E+20 | 691.109 | 0.42763 | 691.109 | 3.74E-08 | 691.109 | 3.70E-08 | 691.109 | 8.71E+12 | 691.109 | 0.01128 |
| 530.7637 | 8.18E+07 | 1423.391 | 5.75E+10 | 691.396 | 5.45E+20 | 691.396 | 0.42758 | 691.396 | 3.74E-08 | 691.396 | 3.57E-08 | 691.396 | 8.73E+12 | 691.396 | 0.01135 |
| 531.057 | 6.82E+07 | 1425.093 | 5.75E+10 | 691.683 | 5.46E+20 | 691.683 | 0.42754 | 691.683 | 3.74E-08 | 691.683 | 3.60E-08 | 691.683 | 8.73E+12 | 691.683 | 0.01142 |
| 531.3503 | 5.45E+07 | 1426.796 | 5.74E+10 | 691.971 | 5.47E+20 | 691.971 | 0.4275 | 691.971 | 3.74E-08 | 691.971 | 3.62E-08 | 691.971 | 8.74E+12 | 691.971 | 0.01128 |
| 531.6436 | 8.86E+07 | 1428.498 | 5.75E+10 | 692.258 | 5.48E+20 | 692.258 | 0.42745 | 692.258 | 3.74E-08 | 692.258 | 3.95E-08 | 692.258 | 8.77E+12 | 692.258 | 0.01125 |
| 531.9369 | 6.82E+07 | 1430.2 | 5.76E+10 | 692.545 | 5.49E+20 | 692.545 | 0.42741 | 692.545 | 3.74E-08 | 692.545 | 3.63E-08 | 692.545 | 8.77E+12 | 692.545 | 0.01143 |
| 532.2302 | 8.86E+07 | 1431.901 | 5.75E+10 | 692.832 | 5.50E+20 | 692.832 | 0.42737 | 692.832 | 3.74E-08 | 692.832 | 3.60E-08 | 692.832 | 8.79E+12 | 692.832 | 0.01135 |
| 532.5235 | 6.82E+07 | 1433.603 | 5.77E+10 | 693.119 | 5.51E+20 | 693.119 | 0.42732 | 693.119 | 3.73E-08 | 693.119 | 3.76E-08 | 693.119 | 8.79E+12 | 693.119 | 0.01128 |
| 532.8168 | 8.18E+07 | 1435.304 | 5.75E+10 | 693.406 | 5.52E+20 | 693.406 | 0.42728 | 693.406 | 3.73E-08 | 693.406 | 3.61E-08 | 693.406 | 8.81E+12 | 693.406 | 0.01134 |
| 533.1101 | 7.50E+07 | 1437.006 | 5.74E+10 | 693.694 | 5.53E+20 | 693.694 | 0.42723 | 693.694 | 3.73E-08 | 693.694 | 3.59E-08 | 693.694 | 8.81E+12 | 693.694 | 0.01142 |
| 533.4033 | 5.45E+07 | 1438.707 | 5.78E+10 | 693.981 | 5.54E+20 | 693.981 | 0.42719 | 693.981 | 3.72E-08 | 693.981 | 3.62E-08 | 693.981 | 8.81E+12 | 693.981 | 0.01141 |
| 533.6966 | 8.86E+07 | 1440.408 | 5.77E+10 | 694.268 | 5.55E+20 | 694.268 | 0.42715 | 694.268 | 3.72E-08 | 694.268 | 3.77E-08 | 694.268 | 8.82E+12 | 694.268 | 0.01155 |
| 533.9898 | 6.82E+07 | 1442.109 | 5.78E+10 | 694.555 | 5.56E+20 | 694.555 | 0.4271 | 694.555 | 3.72E-08 | 694.555 | 3.81E-08 | 694.555 | 8.84E+12 | 694.555 | 0.01141 |
| 534.283 | 5.45E+07 | 1443.81 | 5.78E+10 | 694.842 | 5.57E+20 | 694.842 | 0.42706 | 694.842 | 3.71E-08 | 694.842 | 3.64E-08 | 694.842 | 8.83E+12 | 694.842 | 0.01147 |
| 534.5762 | 8.86E+07 | 1445.511 | 5.73E+10 | 695.129 | 5.58E+20 | 695.129 | 0.42701 | 695.129 | 3.72E-08 | 695.129 | 3.79E-08 | 695.129 | 8.86E+12 | 695.129 | 0.01144 |
| 534.8695 | 5.45E+07 | 1447.211 | 5.75E+10 | 695.416 | 5.59E+20 | 695.416 | 0.42697 | 695.416 | 3.70E-08 | 695.416 | 3.74E-08 | 695.416 | 8.85E+12 | 695.416 | 0.01145 |
| 535.1627 | 7.50E+07 | 1448.912 | 5.78E+10 | 695.703 | 5.60E+20 | 695.703 | 0.42692 | 695.703 | 3.71E-08 | 695.703 | 3.66E-08 | 695.703 | 8.88E+12 | 695.703 | 0.01153 |
| 535.4559 | 6.13E+07 | 1450.612 | 5.78E+10 | 695.99 | 5.61E+20 | 695.99 | 0.42688 | 695.99 | 3.71E-08 | 695.99 | 3.77E-08 | 695.99 | 8.90E+12 | 695.99 | 0.01145 |
| 535.749 | 8.86E+07 | 1452.312 | 5.79E+10 | 696.277 | 5.62E+20 | 696.277 | 0.42683 | 696.277 | 3.71E-08 | 696.277 | 3.69E-08 | 696.277 | 8.91E+12 | 696.277 | 0.01147 |
| 536.0422 | 6.82E+07 | 1454.012 | 5.80E+10 | 696.564 | 5.63E+20 | 696.564 | 0.42679 | 696.564 | 3.71E-08 | 696.564 | 3.69E-08 | 696.564 | 8.93E+12 | 696.564 | 0.01151 |
| 536.3354 | 7.50E+07 | 1455.712 | 5.81E+10 | 696.851 | 5.64E+20 | 696.851 | 0.42674 | 696.851 | 3.70E-08 | 696.851 | 3.68E-08 | 696.851 | 8.92E+12 | 696.851 | 0.01151 |
| 536.6285 | 8.18E+07 | 1457.412 | 5.81E+10 | 697.138 | 5.66E+20 | 697.138 | 0.4267 | 697.138 | 3.70E-08 | 697.138 | 3.66E-08 | 697.138 | 8.94E+12 | 697.138 | 0.01158 |
| 536.9217 | 8.86E+07 | 1459.111 | 5.80E+10 | 697.425 | 5.67E+20 | 697.425 | 0.42665 | 697.425 | 3.71E-08 | 697.425 | 3.63E-08 | 697.425 | 8.97E+12 | 697.425 | 0.01153 |
| 537.2148 | 6.13E+07 | 1460.811 | 5.81E+10 | 697.712 | 5.68E+20 | 697.712 | 0.42661 | 697.712 | 3.71E-08 | 697.712 | 3.61E-08 | 697.712 | 8.99E+12 | 697.712 | 0.01145 |
| 537.5079 | 9.54E+07 | 1462.51 | 5.76E+10 | 697.999 | 5.69E+20 | 697.999 | 0.42656 | 697.999 | 3.71E-08 | 697.999 | 3.99E-08 | 697.999 | 9.00E+12 | 697.999 | 0.01151 |
| 537.8011 | 8.18E+07 | 1464.209 | 5.74E+10 | 698.286 | 5.70E+20 | 698.286 | 0.42652 | 698.286 | 3.70E-08 | 698.286 | 3.53E-08 | 698.286 | 9.00E+12 | 698.286 | 0.0115 |
| 538.0942 | 6.13E+07 | 1465.909 | 5.76E+10 | 698.573 | 5.71E+20 | 698.573 | 0.42647 | 698.573 | 3.71E-08 | 698.573 | 3.74E-08 | 698.573 | 9.02E+12 | 698.573 | 0.0115 |
| 538.3873 | 6.13E+07 | 1467.607 | 5.77E+10 | 698.86 | 5.72E+20 | 698.86 | 0.42643 | 698.86 | 3.71E-08 | 698.86 | 3.71E-08 | 698.86 | 9.04E+12 | 698.86 | 0.01149 |
| 538.6804 | 9.54E+07 | 1469.306 | 5.77E+10 | 699.147 | 5.73E+20 | 699.147 | 0.42638 | 699.147 | 3.71E-08 | 699.147 | 3.73E-08 | 699.147 | 9.07E+12 | 699.147 | 0.01155 |
| 538.9735 | 7.50E+07 | 1471.005 | 5.75E+10 | 699.434 | 5.74E+20 | 699.434 | 0.42634 | 699.434 | 3.71E-08 | 699.434 | 3.64E-08 | 699.434 | 9.08E+12 | 699.434 | 0.01147 |
| 539.2665 | 8.18E+07 | 1472.704 | 5.73E+10 | 699.721 | 5.75E+20 | 699.721 | 0.42629 | 699.721 | 3.71E-08 | 699.721 | 3.58E-08 | 699.721 | 9.10E+12 | 699.721 | 0.01152 |
| 539.5596 | 6.82E+07 | 1474.402 | 5.73E+10 | 700.008 | 5.76E+20 | 700.008 | 0.42625 | 700.008 | 3.72E-08 | 700.008 | 3.66E-08 | 700.008 | 9.13E+12 | 700.008 | 0.0114 |
| 539.8526 | 6.82E+07 | 1476.1 | 5.70E+10 | 700.295 | 5.77E+20 | 700.295 | 0.4262 | 700.295 | 3.71E-08 | 700.295 | 3.78E-08 | 700.295 | 9.13E+12 | 700.295 | 0.01151 |
| 540.1457 | 8.86E+07 | 1477.798 | 5.67E+10 | 700.581 | 5.78E+20 | 700.581 | 0.42616 | 700.581 | 3.71E-08 | 700.581 | 3.78E-08 | 700.581 | 9.13E+12 | 700.581 | 0.01156 |
| 540.4387 | 1.09E+08 | 1479.496 | 5.69E+10 | 700.868 | 5.79E+20 | 700.868 | 0.42611 | 700.868 | 3.72E-08 | 700.868 | 3.86E-08 | 700.868 | 9.17E+12 | 700.868 | 0.01149 |
| 540.7318 | 7.50E+07 | 1481.194 | 5.65E+10 | 701.155 | 5.80E+20 | 701.155 | 0.42607 | 701.155 | 3.71E-08 | 701.155 | 3.54E-08 | 701.155 | 9.16E+12 | 701.155 | 0.01146 |
| 541.0248 | 4.77E+07 | 1482.892 | 5.61E+10 | 701.442 | 5.81E+20 | 701.442 | 0.42602 | 701.442 | 3.70E-08 | 701.442 | 3.92E-08 | 701.442 | 9.17E+12 | 701.442 | 0.01153 |
| 541.3178 | 6.13E+07 | 1484.59 | 5.60E+10 | 701.729 | 5.82E+20 | 701.729 | 0.42597 | 701.729 | 3.71E-08 | 701.729 | 3.87E-08 | 701.729 | 9.19E+12 | 701.729 | 0.01149 |
| 541.6108 | 8.18E+07 | 1486.287 | 5.58E+10 | 702.015 | 5.83E+20 | 702.015 | 0.42593 | 702.015 | 3.71E-08 | 702.015 | 3.67E-08 | 702.015 | 9.22E+12 | 702.015 | 0.01148 |
| 541.9038 | 6.82E+07 | 1487.984 | 5.57E+10 | 702.302 | 5.84E+20 | 702.302 | 0.42588 | 702.302 | 3.71E-08 | 702.302 | 3.71E-08 | 702.302 | 9.23E+12 | 702.302 | 0.01145 |
| 542.1968 | 3.41E+07 | 1489.682 | 5.55E+10 | 702.589 | 5.85E+20 | 702.589 | 0.42584 | 702.589 | 3.71E-08 | 702.589 | 3.68E-08 | 702.589 | 9.26E+12 | 702.589 | 0.01151 |
| 542.4897 | 7.50E+07 | 1491.379 | 5.51E+10 | 702.876 | 5.86E+20 | 702.876 | 0.42579 | 702.876 | 3.72E-08 | 702.876 | 3.77E-08 | 702.876 | 9.28E+12 | 702.876 | 0.01143 |
| 542.7827 | 6.13E+07 | 1493.076 | 5.50E+10 | 703.162 | 5.87E+20 | 703.162 | 0.42575 | 703.162 | 3.72E-08 | 703.162 | 3.86E-08 | 703.162 | 9.29E+12 | 703.162 | 0.0115 |
| 543.0756 | 7.50E+07 | 1494.772 | 5.49E+10 | 703.449 | 5.88E+20 | 703.449 | 0.4257 | 703.449 | 3.72E-08 | 703.449 | 3.70E-08 | 703.449 | 9.32E+12 | 703.449 | 0.01144 |
| 543.3686 | 6.13E+07 | 1496.469 | 5.44E+10 | 703.736 | 5.89E+20 | 703.736 | 0.42565 | 703.736 | 3.72E-08 | 703.736 | 3.50E-08 | 703.736 | 9.33E+12 | 703.736 | 0.01149 |
| 543.6615 | 1.16E+08 | 1498.166 | 5.41E+10 | 704.022 | 5.90E+20 | 704.022 | 0.42561 | 704.022 | 3.71E-08 | 704.022 | 3.62E-08 | 704.022 | 9.33E+12 | 704.022 | 0.01154 |
| 543.9545 | 6.82E+07 | 1499.862 | 5.40E+10 | 704.309 | 5.91E+20 | 704.309 | 0.42556 | 704.309 | 3.72E-08 | 704.309 | 3.78E-08 | 704.309 | 9.37E+12 | 704.309 | 0.01149 |
| 544.2474 | 8.86E+07 | 1501.558 | 5.38E+10 | 704.596 | 5.92E+20 | 704.596 | 0.42552 | 704.596 | 3.72E-08 | 704.596 | 3.67E-08 | 704.596 | 9.39E+12 | 704.596 | 0.01153 |
| 544.5403 | 8.18E+07 | 1503.254 | 5.36E+10 | 704.882 | 5.93E+20 | 704.882 | 0.42547 | 704.882 | 3.73E-08 | 704.882 | 3.93E-08 | 704.882 | 9.41E+12 | 704.882 | 0.01148 |
| 544.8332 | 8.86E+07 | 1504.95 | 5.34E+10 | 705.169 | 5.94E+20 | 705.169 | 0.42542 | 705.169 | 3.72E-08 | 705.169 | 3.60E-08 | 705.169 | 9.42E+12 | 705.169 | 0.01149 |
| 545.1261 | 8.86E+07 | 1506.646 | 5.31E+10 | 705.456 | 5.96E+20 | 705.456 | 0.42538 | 705.456 | 3.73E-08 | 705.456 | 3.65E-08 | 705.456 | 9.44E+12 | 705.456 | 0.01155 |
| 545.4189 | 9.54E+07 | 1508.342 | 5.26E+10 | 705.742 | 5.97E+20 | 705.742 | 0.42533 | 705.742 | 3.73E-08 | 705.742 | 3.87E-08 | 705.742 | 9.46E+12 | 705.742 | 0.01153 |
| 545.7118 | 6.82E+07 | 1510.038 | 5.25E+10 | 706.029 | 5.98E+20 | 706.029 | 0.42528 | 706.029 | 3.73E-08 | 706.029 | 3.47E-08 | 706.029 | 9.48E+12 | 706.029 | 0.01149 |
| 546.0047 | 1.09E+08 | 1511.733 | 5.23E+10 | 706.315 | 5.99E+20 | 706.315 | 0.42524 | 706.315 | 3.72E-08 | 706.315 | 3.65E-08 | 706.315 | 9.48E+12 | 706.315 | 0.01153 |
| 546.2975 | 8.86E+07 | 1513.428 | 5.19E+10 | 706.602 | 6.00E+20 | 706.602 | 0.42519 | 706.602 | 3.72E-08 | 706.602 | 3.95E-08 | 706.602 | 9.50E+12 | 706.602 | 0.01152 |
| 546.5904 | 9.54E+07 | 1515.123 | 5.18E+10 | 706.888 | 6.01E+20 | 706.888 | 0.42515 | 706.888 | 3.72E-08 | 706.888 | 3.50E-08 | 706.888 | 9.50E+12 | 706.888 | 0.01156 |
| 546.8832 | 4.77E+07 | 1516.818 | 5.15E+10 | 707.175 | 6.02E+20 | 707.175 | 0.4251 | 707.175 | 3.72E-08 | 707.175 | 3.53E-08 | 707.175 | 9.53E+12 | 707.175 | 0.01158 |
| 547.176 | 6.82E+07 | 1518.513 | 5.11E+10 | 707.461 | 6.03E+20 | 707.461 | 0.42505 | 707.461 | 3.73E-08 | 707.461 | 3.88E-08 | 707.461 | 9.55E+12 | 707.461 | 0.01154 |
| 547.4688 | 4.77E+07 | 1520.208 | 5.09E+10 | 707.748 | 6.04E+20 | 707.748 | 0.42501 | 707.748 | 3.72E-08 | 707.748 | 3.85E-08 | 707.748 | 9.55E+12 | 707.748 | 0.01157 |
| 547.7616 | 8.18E+07 | 1521.903 | 5.08E+10 | 708.034 | 6.05E+20 | 708.034 | 0.42496 | 708.034 | 3.72E-08 | 708.034 | 3.71E-08 | 708.034 | 9.57E+12 | 708.034 | 0.0115 |
| 548.0544 | 9.54E+07 | 1523.597 | 5.07E+10 | 708.321 | 6.06E+20 | 708.321 | 0.42491 | 708.321 | 3.72E-08 | 708.321 | 3.84E-08 | 708.321 | 9.58E+12 | 708.321 | 0.01163 |
| 548.3472 | 7.50E+07 | 1525.291 | 5.06E+10 | 708.607 | 6.07E+20 | 708.607 | 0.42486 | 708.607 | 3.72E-08 | 708.607 | 3.80E-08 | 708.607 | 9.60E+12 | 708.607 | 0.01157 |
| 548.64 | 1.09E+08 | 1526.986 | 5.04E+10 | 708.894 | 6.08E+20 | 708.894 | 0.42482 | 708.894 | 3.72E-08 | 708.894 | 3.65E-08 | 708.894 | 9.60E+12 | 708.894 | 0.01154 |
| 548.9328 | 6.82E+07 | 1528.68 | 5.00E+10 | 709.18 | 6.09E+20 | 709.18 | 0.42477 | 709.18 | 3.71E-08 | 709.18 | 3.87E-08 | 709.18 | 9.60E+12 | 709.18 | 0.01159 |
| 549.2255 | 1.02E+08 | 1530.374 | 4.99E+10 | 709.467 | 6.10E+20 | 709.467 | 0.42472 | 709.467 | 3.72E-08 | 709.467 | 3.50E-08 | 709.467 | 9.63E+12 | 709.467 | 0.01159 |
| 549.5183 | 8.86E+07 | 1532.067 | 4.97E+10 | 709.753 | 6.11E+20 | 709.753 | 0.42468 | 709.753 | 3.72E-08 | 709.753 | 3.72E-08 | 709.753 | 9.67E+12 | 709.753 | 0.01148 |
| 549.811 | 1.02E+08 | 1533.761 | 4.94E+10 | 710.04 | 6.12E+20 | 710.04 | 0.42463 | 710.04 | 3.73E-08 | 710.04 | 3.84E-08 | 710.04 | 9.69E+12 | 710.04 | 0.01151 |
| 550.1037 | 6.82E+07 | 1535.455 | 4.92E+10 | 710.326 | 6.13E+20 | 710.326 | 0.42458 | 710.326 | 3.73E-08 | 710.326 | 3.81E-08 | 710.326 | 9.71E+12 | 710.326 | 0.01147 |
| 550.3964 | 7.50E+07 | 1537.148 | 4.89E+10 | 710.612 | 6.14E+20 | 710.612 | 0.42454 | 710.612 | 3.73E-08 | 710.612 | 3.88E-08 | 710.612 | 9.73E+12 | 710.612 | 0.01145 |
| 550.6892 | 6.13E+07 | 1538.841 | 4.84E+10 | 710.899 | 6.15E+20 | 710.899 | 0.42449 | 710.899 | 3.73E-08 | 710.899 | 3.55E-08 | 710.899 | 9.74E+12 | 710.899 | 0.01142 |
| 550.9819 | 8.18E+07 | 1540.534 | 4.82E+10 | 711.185 | 6.17E+20 | 711.185 | 0.42444 | 711.185 | 3.73E-08 | 711.185 | 3.78E-08 | 711.185 | 9.76E+12 | 711.185 | 0.01147 |
| 551.2745 | 8.86E+07 | 1542.227 | 4.81E+10 | 711.471 | 6.18E+20 | 711.471 | 0.42439 | 711.471 | 3.73E-08 | 711.471 | 3.60E-08 | 711.471 | 9.78E+12 | 711.471 | 0.01152 |
| 551.5672 | 8.18E+07 | 1543.92 | 4.77E+10 | 711.758 | 6.19E+20 | 711.758 | 0.42435 | 711.758 | 3.73E-08 | 711.758 | 3.70E-08 | 711.758 | 9.79E+12 | 711.758 | 0.01146 |
| 551.8599 | 7.50E+07 | 1545.613 | 4.77E+10 | 712.044 | 6.20E+20 | 712.044 | 0.4243 | 712.044 | 3.74E-08 | 712.044 | 3.63E-08 | 712.044 | 9.83E+12 | 712.044 | 0.0115 |
| 552.1526 | 1.02E+08 | 1547.305 | 4.76E+10 | 712.33 | 6.21E+20 | 712.33 | 0.42425 | 712.33 | 3.74E-08 | 712.33 | 3.78E-08 | 712.33 | 9.85E+12 | 712.33 | 0.01152 |
| 552.4452 | 8.18E+07 | 1548.998 | 4.73E+10 | 712.617 | 6.22E+20 | 712.617 | 0.4242 | 712.617 | 3.74E-08 | 712.617 | 3.71E-08 | 712.617 | 9.86E+12 | 712.617 | 0.01136 |
| 552.7379 | 6.82E+07 | 1550.69 | 4.68E+10 | 712.903 | 6.23E+20 | 712.903 | 0.42416 | 712.903 | 3.74E-08 | 712.903 | 3.66E-08 | 712.903 | 9.88E+12 | 712.903 | 0.01151 |
| 553.0305 | 7.50E+07 | 1552.382 | 4.66E+10 | 713.189 | 6.24E+20 | 713.189 | 0.42411 | 713.189 | 3.75E-08 | 713.189 | 3.96E-08 | 713.189 | 9.91E+12 | 713.189 | 0.01144 |
| 553.3231 | 7.50E+07 | 1554.074 | 4.65E+10 | 713.475 | 6.25E+20 | 713.475 | 0.42406 | 713.475 | 3.74E-08 | 713.475 | 3.67E-08 | 713.475 | 9.92E+12 | 713.475 | 0.01141 |
| 553.6157 | 1.02E+08 | 1555.766 | 4.59E+10 | 713.762 | 6.26E+20 | 713.762 | 0.42401 | 713.762 | 3.74E-08 | 713.762 | 3.68E-08 | 713.762 | 9.92E+12 | 713.762 | 0.01142 |
| 553.9083 | 1.02E+08 | 1557.457 | 4.59E+10 | 714.048 | 6.27E+20 | 714.048 | 0.42396 | 714.048 | 3.74E-08 | 714.048 | 3.73E-08 | 714.048 | 9.93E+12 | 714.048 | 0.01149 |
| 554.2009 | 1.09E+08 | 1559.149 | 4.58E+10 | 714.334 | 6.28E+20 | 714.334 | 0.42392 | 714.334 | 3.73E-08 | 714.334 | 3.69E-08 | 714.334 | 9.93E+12 | 714.334 | 0.01147 |
| 554.4935 | 8.18E+07 | 1560.84 | 4.55E+10 | 714.62 | 6.29E+20 | 714.62 | 0.42387 | 714.62 | 3.72E-08 | 714.62 | 3.54E-08 | 714.62 | 9.93E+12 | 714.62 | 0.01145 |
| 554.7861 | 9.54E+07 | 1562.532 | 4.50E+10 | 714.906 | 6.30E+20 | 714.906 | 0.42382 | 714.906 | 3.73E-08 | 714.906 | 3.64E-08 | 714.906 | 9.95E+12 | 714.906 | 0.0114 |
| 555.0787 | 4.09E+07 | 1564.223 | 4.45E+10 | 715.192 | 6.31E+20 | 715.192 | 0.42377 | 715.192 | 3.72E-08 | 715.192 | 3.95E-08 | 715.192 | 9.96E+12 | 715.192 | 0.01145 |
| 555.3712 | 6.82E+07 | 1565.914 | 4.43E+10 | 715.479 | 6.32E+20 | 715.479 | 0.42372 | 715.479 | 3.73E-08 | 715.479 | 3.82E-08 | 715.479 | 9.99E+12 | 715.479 | 0.01145 |
| 555.6638 | 6.82E+07 | 1567.605 | 4.35E+10 | 715.765 | 6.33E+20 | 715.765 | 0.42368 | 715.765 | 3.73E-08 | 715.765 | 3.73E-08 | 715.765 | 1.00E+13 | 715.765 | 0.01144 |
| 555.9563 | 8.86E+07 | 1569.295 | 4.29E+10 | 716.051 | 6.35E+20 | 716.051 | 0.42363 | 716.051 | 3.72E-08 | 716.051 | 3.82E-08 | 716.051 | 1.00E+13 | 716.051 | 0.01143 |
| 556.2488 | 6.13E+07 | 1570.986 | 4.23E+10 | 716.337 | 6.36E+20 | 716.337 | 0.42358 | 716.337 | 3.72E-08 | 716.337 | 3.84E-08 | 716.337 | 1.00E+13 | 716.337 | 0.01152 |
| 556.5413 | 7.50E+07 | 1572.676 | 4.14E+10 | 716.623 | 6.37E+20 | 716.623 | 0.42353 | 716.623 | 3.72E-08 | 716.623 | 3.73E-08 | 716.623 | 1.00E+13 | 716.623 | 0.01146 |
| 556.8338 | 8.86E+07 | 1574.367 | 4.05E+10 | 716.909 | 6.38E+20 | 716.909 | 0.42348 | 716.909 | 3.72E-08 | 716.909 | 3.68E-08 | 716.909 | 1.01E+13 | 716.909 | 0.01147 |
| 557.1263 | 8.86E+07 | 1576.057 | 3.97E+10 | 717.195 | 6.39E+20 | 717.195 | 0.42343 | 717.195 | 3.72E-08 | 717.195 | 3.81E-08 | 717.195 | 1.01E+13 | 717.195 | 0.01146 |
| 557.4188 | 4.77E+07 | 1577.747 | 3.90E+10 | 717.481 | 6.40E+20 | 717.481 | 0.42339 | 717.481 | 3.72E-08 | 717.481 | 3.73E-08 | 717.481 | 1.01E+13 | 717.481 | 0.01143 |
| 557.7113 | 8.86E+07 | 1579.437 | 3.78E+10 | 717.767 | 6.41E+20 | 717.767 | 0.42334 | 717.767 | 3.72E-08 | 717.767 | 3.88E-08 | 717.767 | 1.01E+13 | 717.767 | 0.01147 |
| 558.0038 | 6.82E+07 | 1581.126 | 3.71E+10 | 718.053 | 6.42E+20 | 718.053 | 0.42329 | 718.053 | 3.73E-08 | 718.053 | 3.67E-08 | 718.053 | 1.01E+13 | 718.053 | 0.01144 |
| 558.2962 | 6.13E+07 | 1582.816 | 3.61E+10 | 718.339 | 6.43E+20 | 718.339 | 0.42324 | 718.339 | 3.73E-08 | 718.339 | 3.82E-08 | 718.339 | 1.01E+13 | 718.339 | 0.01143 |
| 558.5887 | 8.86E+07 | 1584.505 | 3.50E+10 | 718.625 | 6.44E+20 | 718.625 | 0.42319 | 718.625 | 3.73E-08 | 718.625 | 3.66E-08 | 718.625 | 1.02E+13 | 718.625 | 0.01142 |
| 558.8811 | 1.02E+08 | 1586.195 | 3.38E+10 | 718.911 | 6.45E+20 | 718.911 | 0.42314 | 718.911 | 3.73E-08 | 718.911 | 3.72E-08 | 718.911 | 1.02E+13 | 718.911 | 0.01142 |
| 559.1736 | 8.18E+07 | 1587.884 | 3.25E+10 | 719.197 | 6.46E+20 | 719.197 | 0.42309 | 719.197 | 3.72E-08 | 719.197 | 3.74E-08 | 719.197 | 1.02E+13 | 719.197 | 0.01136 |
| 559.466 | 8.18E+07 | 1589.573 | 3.13E+10 | 719.483 | 6.47E+20 | 719.483 | 0.42305 | 719.483 | 3.72E-08 | 719.483 | 3.67E-08 | 719.483 | 1.02E+13 | 719.483 | 0.01148 |
| 559.7584 | 6.13E+07 | 1591.262 | 3.01E+10 | 719.769 | 6.48E+20 | 719.769 | 0.423 | 719.769 | 3.72E-08 | 719.769 | 3.67E-08 | 719.769 | 1.02E+13 | 719.769 | 0.01141 |
| 560.0508 | 8.18E+07 | 1592.95 | 2.89E+10 | 720.055 | 6.49E+20 | 720.055 | 0.42295 | 720.055 | 3.72E-08 | 720.055 | 3.58E-08 | 720.055 | 1.02E+13 | 720.055 | 0.01145 |
| 560.3432 | 7.50E+07 | 1594.639 | 2.78E+10 | 720.341 | 6.50E+20 | 720.341 | 0.4229 | 720.341 | 3.72E-08 | 720.341 | 3.56E-08 | 720.341 | 1.02E+13 | 720.341 | 0.01141 |
| 560.6356 | 6.13E+07 | 1596.327 | 2.66E+10 | 720.627 | 6.52E+20 | 720.627 | 0.42285 | 720.627 | 3.73E-08 | 720.627 | 3.72E-08 | 720.627 | 1.03E+13 | 720.627 | 0.01144 |
| 560.9279 | 5.45E+07 | 1598.016 | 2.57E+10 | 720.913 | 6.53E+20 | 720.913 | 0.4228 | 720.913 | 3.73E-08 | 720.913 | 3.74E-08 | 720.913 | 1.03E+13 | 720.913 | 0.01137 |
| 561.2203 | 6.82E+07 | 1599.704 | 2.43E+10 | 721.199 | 6.54E+20 | 721.199 | 0.42275 | 721.199 | 3.73E-08 | 721.199 | 3.70E-08 | 721.199 | 1.03E+13 | 721.199 | 0.0115 |
| 561.5127 | 7.50E+07 | 1601.392 | 2.32E+10 | 721.484 | 6.55E+20 | 721.484 | 0.4227 | 721.484 | 3.72E-08 | 721.484 | 3.77E-08 | 721.484 | 1.03E+13 | 721.484 | 0.01143 |
| 561.805 | 1.02E+08 | 1603.08 | 2.22E+10 | 721.77 | 6.56E+20 | 721.77 | 0.42265 | 721.77 | 3.72E-08 | 721.77 | 3.65E-08 | 721.77 | 1.03E+13 | 721.77 | 0.01155 |
| 562.0973 | 1.36E+08 | 1604.767 | 2.10E+10 | 722.056 | 6.57E+20 | 722.056 | 0.4226 | 722.056 | 3.72E-08 | 722.056 | 3.68E-08 | 722.056 | 1.03E+13 | 722.056 | 0.0115 |
| 562.3897 | 7.50E+07 | 1606.455 | 1.99E+10 | 722.342 | 6.58E+20 | 722.342 | 0.42255 | 722.342 | 3.72E-08 | 722.342 | 3.77E-08 | 722.342 | 1.03E+13 | 722.342 | 0.01152 |
| 562.682 | 8.18E+07 | 1608.142 | 1.88E+10 | 722.628 | 6.59E+20 | 722.628 | 0.42251 | 722.628 | 3.72E-08 | 722.628 | 3.74E-08 | 722.628 | 1.04E+13 | 722.628 | 0.01141 |
| 562.9743 | 8.86E+07 | 1609.83 | 1.74E+10 | 722.914 | 6.60E+20 | 722.914 | 0.42246 | 722.914 | 3.72E-08 | 722.914 | 3.83E-08 | 722.914 | 1.04E+13 | 722.914 | 0.01141 |
| 563.2666 | 9.54E+07 | 1611.517 | 1.64E+10 | 723.199 | 6.61E+20 | 723.199 | 0.42241 | 723.199 | 3.72E-08 | 723.199 | 3.58E-08 | 723.199 | 1.04E+13 | 723.199 | 0.01148 |
| 563.5589 | 1.16E+08 | 1613.204 | 1.54E+10 | 723.485 | 6.62E+20 | 723.485 | 0.42236 | 723.485 | 3.72E-08 | 723.485 | 3.71E-08 | 723.485 | 1.04E+13 | 723.485 | 0.0115 |
| 563.8511 | 9.54E+07 | 1614.891 | 1.42E+10 | 723.771 | 6.63E+20 | 723.771 | 0.42231 | 723.771 | 3.71E-08 | 723.771 | 3.74E-08 | 723.771 | 1.04E+13 | 723.771 | 0.01156 |
| 564.1434 | 1.02E+08 | 1616.577 | 1.33E+10 | 724.057 | 6.64E+20 | 724.057 | 0.42226 | 724.057 | 3.71E-08 | 724.057 | 3.80E-08 | 724.057 | 1.04E+13 | 724.057 | 0.01151 |
| 564.4357 | 8.86E+07 | 1618.264 | 1.23E+10 | 724.342 | 6.65E+20 | 724.342 | 0.42221 | 724.342 | 3.71E-08 | 724.342 | 3.82E-08 | 724.342 | 1.04E+13 | 724.342 | 0.01157 |
| 564.7279 | 8.18E+07 | 1619.95 | 1.11E+10 | 724.628 | 6.67E+20 | 724.628 | 0.42216 | 724.628 | 3.71E-08 | 724.628 | 3.64E-08 | 724.628 | 1.04E+13 | 724.628 | 0.01149 |
| 565.0201 | 8.18E+07 | 1621.637 | 1.03E+10 | 724.914 | 6.68E+20 | 724.914 | 0.42211 | 724.914 | 3.70E-08 | 724.914 | 3.50E-08 | 724.914 | 1.04E+13 | 724.914 | 0.01149 |
| 565.3124 | 1.09E+08 | 1623.323 | 9.41E+09 | 725.199 | 6.69E+20 | 725.199 | 0.42206 | 725.199 | 3.70E-08 | 725.199 | 3.92E-08 | 725.199 | 1.04E+13 | 725.199 | 0.01153 |
| 565.6046 | 8.18E+07 | 1625.009 | 8.27E+09 | 725.485 | 6.70E+20 | 725.485 | 0.42201 | 725.485 | 3.70E-08 | 725.485 | 3.59E-08 | 725.485 | 1.05E+13 | 725.485 | 0.0116 |
| 565.8968 | 1.16E+08 | 1626.695 | 7.48E+09 | 725.771 | 6.71E+20 | 725.771 | 0.42196 | 725.771 | 3.70E-08 | 725.771 | 3.73E-08 | 725.771 | 1.05E+13 | 725.771 | 0.0116 |
| 566.189 | 9.54E+07 | 1628.38 | 6.75E+09 | 726.056 | 6.72E+20 | 726.056 | 0.42191 | 726.056 | 3.71E-08 | 726.056 | 3.77E-08 | 726.056 | 1.05E+13 | 726.056 | 0.01147 |
| 566.4812 | 7.50E+07 | 1630.066 | 5.94E+09 | 726.342 | 6.73E+20 | 726.342 | 0.42186 | 726.342 | 3.71E-08 | 726.342 | 3.75E-08 | 726.342 | 1.05E+13 | 726.342 | 0.01149 |
| 566.7733 | 9.54E+07 | 1631.751 | 5.28E+09 | 726.628 | 6.74E+20 | 726.628 | 0.42181 | 726.628 | 3.70E-08 | 726.628 | 3.59E-08 | 726.628 | 1.05E+13 | 726.628 | 0.0116 |
| 567.0655 | 1.29E+08 | 1633.437 | 4.48E+09 | 726.913 | 6.75E+20 | 726.913 | 0.42176 | 726.913 | 3.70E-08 | 726.913 | 3.91E-08 | 726.913 | 1.05E+13 | 726.913 | 0.01153 |
| 567.3577 | 1.02E+08 | 1635.122 | 3.82E+09 | 727.199 | 6.76E+20 | 727.199 | 0.42171 | 727.199 | 3.70E-08 | 727.199 | 3.55E-08 | 727.199 | 1.06E+13 | 727.199 | 0.01154 |
| 567.6498 | 1.02E+08 | 1636.807 | 3.32E+09 | 727.484 | 6.77E+20 | 727.484 | 0.42166 | 727.484 | 3.70E-08 | 727.484 | 3.71E-08 | 727.484 | 1.06E+13 | 727.484 | 0.01158 |
| 567.9419 | 1.23E+08 | 1638.492 | 2.83E+09 | 727.77 | 6.78E+20 | 727.77 | 0.42161 | 727.77 | 3.69E-08 | 727.77 | 3.90E-08 | 727.77 | 1.06E+13 | 727.77 | 0.01159 |
| 568.2341 | 1.02E+08 | 1640.176 | 2.32E+09 | 728.055 | 6.80E+20 | 728.055 | 0.42156 | 728.055 | 3.70E-08 | 728.055 | 3.73E-08 | 728.055 | 1.06E+13 | 728.055 | 0.01167 |
| 568.5262 | 1.23E+08 | 1641.861 | 2.01E+09 | 728.341 | 6.81E+20 | 728.341 | 0.42151 | 728.341 | 3.69E-08 | 728.341 | 3.72E-08 | 728.341 | 1.06E+13 | 728.341 | 0.0116 |
| 568.8183 | 1.09E+08 | 1643.545 | 1.57E+09 | 728.626 | 6.82E+20 | 728.626 | 0.42146 | 728.626 | 3.69E-08 | 728.626 | 3.73E-08 | 728.626 | 1.06E+13 | 728.626 | 0.0116 |
| 569.1104 | 1.09E+08 | 1645.23 | 1.36E+09 | 728.912 | 6.83E+20 | 728.912 | 0.4214 | 728.912 | 3.69E-08 | 728.912 | 3.73E-08 | 728.912 | 1.06E+13 | 728.912 | 0.01162 |
| 569.4025 | 1.57E+08 | 1646.914 | 1.03E+09 | 729.197 | 6.84E+20 | 729.197 | 0.42135 | 729.197 | 3.69E-08 | 729.197 | 3.73E-08 | 729.197 | 1.06E+13 | 729.197 | 0.01158 |
| 569.6946 | 1.02E+08 | 1648.598 | 7.96E+08 | 729.483 | 6.85E+20 | 729.483 | 0.4213 | 729.483 | 3.69E-08 | 729.483 | 3.74E-08 | 729.483 | 1.07E+13 | 729.483 | 0.01163 |
| 569.9866 | 1.23E+08 | 1650.282 | 7.61E+08 | 729.768 | 6.86E+20 | 729.768 | 0.42125 | 729.768 | 3.69E-08 | 729.768 | 3.61E-08 | 729.768 | 1.07E+13 | 729.768 | 0.0117 |
| 570.2787 | 1.29E+08 | 1651.965 | 5.91E+08 | 730.054 | 6.87E+20 | 730.054 | 0.4212 | 730.054 | 3.69E-08 | 730.054 | 3.56E-08 | 730.054 | 1.07E+13 | 730.054 | 0.01168 |
| 570.5707 | 1.43E+08 | 1653.649 | 4.86E+08 | 730.339 | 6.88E+20 | 730.339 | 0.42115 | 730.339 | 3.69E-08 | 730.339 | 3.71E-08 | 730.339 | 1.07E+13 | 730.339 | 0.01167 |
| 570.8628 | 1.50E+08 | 1655.332 | 3.77E+08 | 730.625 | 6.89E+20 | 730.625 | 0.4211 | 730.625 | 3.68E-08 | 730.625 | 3.57E-08 | 730.625 | 1.07E+13 | 730.625 | 0.01171 |
| 571.1548 | 1.36E+08 | 1657.015 | 2.58E+08 | 730.91 | 6.90E+20 | 730.91 | 0.42105 | 730.91 | 3.69E-08 | 730.91 | 3.68E-08 | 730.91 | 1.07E+13 | 730.91 | 0.01173 |
| 571.4468 | 1.23E+08 | 1658.698 | 2.14E+08 | 731.195 | 6.91E+20 | 731.195 | 0.421 | 731.195 | 3.68E-08 | 731.195 | 3.64E-08 | 731.195 | 1.07E+13 | 731.195 | 0.01168 |
| 571.7388 | 1.57E+08 | 1660.381 | 2.64E+08 | 731.481 | 6.93E+20 | 731.481 | 0.42095 | 731.481 | 3.68E-08 | 731.481 | 3.70E-08 | 731.481 | 1.07E+13 | 731.481 | 0.01173 |
| 572.0308 | 1.23E+08 | 1662.064 | 2.47E+08 | 731.766 | 6.94E+20 | 731.766 | 0.4209 | 731.766 | 3.68E-08 | 731.766 | 3.76E-08 | 731.766 | 1.07E+13 | 731.766 | 0.01174 |
| 572.3228 | 1.02E+08 | 1663.747 | 1.17E+08 | 732.052 | 6.95E+20 | 732.052 | 0.42085 | 732.052 | 3.67E-08 | 732.052 | 3.65E-08 | 732.052 | 1.07E+13 | 732.052 | 0.0117 |
| 572.6148 | 1.09E+08 | 1665.429 | 3.29E+07 | 732.337 | 6.96E+20 | 732.337 | 0.42079 | 732.337 | 3.67E-08 | 732.337 | 3.54E-08 | 732.337 | 1.08E+13 | 732.337 | 0.01172 |
| 572.9067 | 1.36E+08 | 1667.112 | -1.93E+07 | 732.622 | 6.97E+20 | 732.622 | 0.42074 | 732.622 | 3.68E-08 | 732.622 | 3.58E-08 | 732.622 | 1.08E+13 | 732.622 | 0.01171 |
| 573.1987 | 1.50E+08 | 1668.794 | -7.61E+07 | 732.907 | 6.98E+20 | 732.907 | 0.42069 | 732.907 | 3.67E-08 | 732.907 | 3.67E-08 | 732.907 | 1.08E+13 | 732.907 | 0.01175 |
| 573.4906 | 1.09E+08 | 1670.476 | -1.16E+08 | 733.193 | 6.99E+20 | 733.193 | 0.42064 | 733.193 | 3.68E-08 | 733.193 | 3.58E-08 | 733.193 | 1.08E+13 | 733.193 | 0.01177 |
| 573.7826 | 1.16E+08 | 1672.158 | -6.70E+07 | 733.478 | 7.00E+20 | 733.478 | 0.42059 | 733.478 | 3.68E-08 | 733.478 | 3.65E-08 | 733.478 | 1.08E+13 | 733.478 | 0.0118 |
| 574.0745 | 1.36E+08 | 1673.84 | -3.52E+07 | 733.763 | 7.01E+20 | 733.763 | 0.42054 | 733.763 | 3.68E-08 | 733.763 | 3.74E-08 | 733.763 | 1.08E+13 | 733.763 | 0.01172 |
| 574.3664 | 1.91E+08 | 1675.521 | -5.91E+07 | 734.049 | 7.02E+20 | 734.049 | 0.42049 | 734.049 | 3.68E-08 | 734.049 | 3.71E-08 | 734.049 | 1.09E+13 | 734.049 | 0.01178 |
| 574.6583 | 1.16E+08 | 1677.203 | 4.54E+07 | 734.334 | 7.03E+20 | 734.334 | 0.42044 | 734.334 | 3.67E-08 | 734.334 | 3.60E-08 | 734.334 | 1.09E+13 | 734.334 | 0.01182 |
| 574.9502 | 1.16E+08 | 1678.884 | 3.75E+07 | 734.619 | 7.04E+20 | 734.619 | 0.42039 | 734.619 | 3.67E-08 | 734.619 | 3.62E-08 | 734.619 | 1.09E+13 | 734.619 | 0.01178 |
| 575.2421 | 1.23E+08 | 1680.565 | -8.86E+07 | 734.904 | 7.06E+20 | 734.904 | 0.42033 | 734.904 | 3.68E-08 | 734.904 | 3.71E-08 | 734.904 | 1.09E+13 | 734.904 | 0.01186 |
| 575.534 | 1.50E+08 | 1682.246 | -1.26E+08 | 735.189 | 7.07E+20 | 735.189 | 0.42028 | 735.189 | 3.67E-08 | 735.189 | 3.84E-08 | 735.189 | 1.09E+13 | 735.189 | 0.01186 |
| 575.8259 | 1.16E+08 | 1683.927 | -1.73E+08 | 735.475 | 7.08E+20 | 735.475 | 0.42023 | 735.475 | 3.67E-08 | 735.475 | 3.65E-08 | 735.475 | 1.09E+13 | 735.475 | 0.01179 |
| 576.1177 | 1.16E+08 | 1685.608 | -2.75E+08 | 735.76 | 7.09E+20 | 735.76 | 0.42018 | 735.76 | 3.67E-08 | 735.76 | 3.74E-08 | 735.76 | 1.09E+13 | 735.76 | 0.01183 |
| 576.4096 | 1.57E+08 | 1687.289 | -4.66E+07 | 736.045 | 7.10E+20 | 736.045 | 0.42013 | 736.045 | 3.67E-08 | 736.045 | 3.71E-08 | 736.045 | 1.10E+13 | 736.045 | 0.0118 |
| 576.7014 | 1.64E+08 | 1688.969 | 2.39E+07 | 736.33 | 7.11E+20 | 736.33 | 0.42007 | 736.33 | 3.67E-08 | 736.33 | 3.53E-08 | 736.33 | 1.10E+13 | 736.33 | 0.01193 |
| 576.9932 | 1.64E+08 | 1690.649 | -1.47E+08 | 736.615 | 7.12E+20 | 736.615 | 0.42002 | 736.615 | 3.68E-08 | 736.615 | 3.61E-08 | 736.615 | 1.10E+13 | 736.615 | 0.01196 |
| 577.285 | 1.57E+08 | 1692.329 | -2.61E+07 | 736.9 | 7.13E+20 | 736.9 | 0.41997 | 736.9 | 3.67E-08 | 736.9 | 3.55E-08 | 736.9 | 1.10E+13 | 736.9 | 0.01184 |
| 577.5768 | 1.57E+08 | 1694.009 | -1.92E+08 | 737.185 | 7.14E+20 | 737.185 | 0.41992 | 737.185 | 3.67E-08 | 737.185 | 3.71E-08 | 737.185 | 1.10E+13 | 737.185 | 0.01191 |
| 577.8686 | 1.50E+08 | 1695.689 | -2.12E+08 | 737.47 | 7.15E+20 | 737.47 | 0.41987 | 737.47 | 3.67E-08 | 737.47 | 3.55E-08 | 737.47 | 1.10E+13 | 737.47 | 0.01184 |
| 578.1604 | 1.77E+08 | 1697.369 | -2.10E+08 | 737.756 | 7.16E+20 | 737.756 | 0.41981 | 737.756 | 3.67E-08 | 737.756 | 3.69E-08 | 737.756 | 1.10E+13 | 737.756 | 0.0119 |
| 578.4522 | 1.50E+08 | 1699.049 | -2.47E+08 | 738.041 | 7.18E+20 | 738.041 | 0.41976 | 738.041 | 3.67E-08 | 738.041 | 3.75E-08 | 738.041 | 1.11E+13 | 738.041 | 0.01193 |
| 578.744 | 1.64E+08 | 1700.728 | -1.11E+08 | 738.326 | 7.19E+20 | 738.326 | 0.41971 | 738.326 | 3.67E-08 | 738.326 | 3.64E-08 | 738.326 | 1.11E+13 | 738.326 | 0.01199 |
| 579.0357 | 1.77E+08 | 1702.407 | -8.29E+07 | 738.611 | 7.20E+20 | 738.611 | 0.41966 | 738.611 | 3.68E-08 | 738.611 | 3.75E-08 | 738.611 | 1.11E+13 | 738.611 | 0.01193 |
| 579.3275 | 1.98E+08 | 1704.086 | -1.60E+08 | 738.896 | 7.21E+20 | 738.896 | 0.41961 | 738.896 | 3.68E-08 | 738.896 | 3.86E-08 | 738.896 | 1.11E+13 | 738.896 | 0.01186 |
| 579.6192 | 2.04E+08 | 1705.765 | -2.76E+08 | 739.181 | 7.22E+20 | 739.181 | 0.41955 | 739.181 | 3.68E-08 | 739.181 | 3.71E-08 | 739.181 | 1.12E+13 | 739.181 | 0.0119 |
| 579.9109 | 2.04E+08 | 1707.444 | -2.68E+08 | 739.466 | 7.23E+20 | 739.466 | 0.4195 | 739.466 | 3.69E-08 | 739.466 | 3.81E-08 | 739.466 | 1.12E+13 | 739.466 | 0.01183 |
| 580.2026 | 1.91E+08 | 1709.123 | -2.06E+08 | 739.751 | 7.24E+20 | 739.751 | 0.41945 | 739.751 | 3.69E-08 | 739.751 | 3.73E-08 | 739.751 | 1.12E+13 | 739.751 | 0.01181 |
| 580.4943 | 2.25E+08 | 1710.801 | -1.91E+08 | 740.036 | 7.25E+20 | 740.036 | 0.4194 | 740.036 | 3.69E-08 | 740.036 | 3.55E-08 | 740.036 | 1.12E+13 | 740.036 | 0.01192 |
| 580.786 | 2.39E+08 | 1712.48 | -2.45E+08 | 740.32 | 7.26E+20 | 740.32 | 0.41934 | 740.32 | 3.69E-08 | 740.32 | 3.73E-08 | 740.32 | 1.13E+13 | 740.32 | 0.01188 |
| 581.0777 | 2.25E+08 | 1714.158 | -3.98E+07 | 740.605 | 7.27E+20 | 740.605 | 0.41929 | 740.605 | 3.69E-08 | 740.605 | 3.85E-08 | 740.605 | 1.13E+13 | 740.605 | 0.01186 |
| 581.3694 | 1.98E+08 | 1715.836 | -1.14E+07 | 740.89 | 7.29E+20 | 740.89 | 0.41924 | 740.89 | 3.69E-08 | 740.89 | 3.60E-08 | 740.89 | 1.13E+13 | 740.89 | 0.01186 |
| 581.661 | 2.32E+08 | 1717.514 | -2.34E+08 | 741.175 | 7.30E+20 | 741.175 | 0.41919 | 741.175 | 3.69E-08 | 741.175 | 3.73E-08 | 741.175 | 1.13E+13 | 741.175 | 0.01189 |
| 581.9527 | 1.98E+08 | 1719.191 | -1.94E+08 | 741.46 | 7.31E+20 | 741.46 | 0.41913 | 741.46 | 3.70E-08 | 741.46 | 3.54E-08 | 741.46 | 1.13E+13 | 741.46 | 0.01189 |
| 582.2443 | 1.77E+08 | 1720.869 | -1.43E+08 | 741.745 | 7.32E+20 | 741.745 | 0.41908 | 741.745 | 3.70E-08 | 741.745 | 3.63E-08 | 741.745 | 1.13E+13 | 741.745 | 0.01185 |
| 582.536 | 2.39E+08 | 1722.547 | -1.12E+08 | 742.03 | 7.33E+20 | 742.03 | 0.41903 | 742.03 | 3.69E-08 | 742.03 | 3.66E-08 | 742.03 | 1.13E+13 | 742.03 | 0.01186 |
| 582.8276 | 2.52E+08 | 1724.224 | -9.09E+07 | 742.315 | 7.34E+20 | 742.315 | 0.41897 | 742.315 | 3.70E-08 | 742.315 | 3.78E-08 | 742.315 | 1.14E+13 | 742.315 | 0.01194 |
| 583.1192 | 2.25E+08 | 1725.901 | -1.65E+08 | 742.599 | 7.35E+20 | 742.599 | 0.41892 | 742.599 | 3.71E-08 | 742.599 | 3.63E-08 | 742.599 | 1.14E+13 | 742.599 | 0.01194 |
| 583.4108 | 2.73E+08 | 1727.578 | -1.74E+08 | 742.884 | 7.36E+20 | 742.884 | 0.41887 | 742.884 | 3.71E-08 | 742.884 | 3.54E-08 | 742.884 | 1.15E+13 | 742.884 | 0.01183 |
| 583.7024 | 2.59E+08 | 1729.255 | -2.37E+08 | 743.169 | 7.37E+20 | 743.169 | 0.41882 | 743.169 | 3.72E-08 | 743.169 | 3.72E-08 | 743.169 | 1.15E+13 | 743.169 | 0.0119 |
| 583.994 | 2.32E+08 | 1730.932 | -1.68E+08 | 743.454 | 7.38E+20 | 743.454 | 0.41876 | 743.454 | 3.72E-08 | 743.454 | 3.74E-08 | 743.454 | 1.15E+13 | 743.454 | 0.01181 |
| 584.2855 | 2.73E+08 | 1732.608 | -8.29E+07 | 743.739 | 7.40E+20 | 743.739 | 0.41871 | 743.739 | 3.72E-08 | 743.739 | 3.62E-08 | 743.739 | 1.15E+13 | 743.739 | 0.01186 |
| 584.5771 | 2.66E+08 | 1734.285 | -2.25E+08 | 744.023 | 7.41E+20 | 744.023 | 0.41866 | 744.023 | 3.71E-08 | 744.023 | 3.71E-08 | 744.023 | 1.15E+13 | 744.023 | 0.01187 |
| 584.8686 | 3.20E+08 | 1735.961 | -2.10E+08 | 744.308 | 7.42E+20 | 744.308 | 0.4186 | 744.308 | 3.72E-08 | 744.308 | 3.84E-08 | 744.308 | 1.15E+13 | 744.308 | 0.01186 |
| 585.1602 | 3.14E+08 | 1737.637 | -1.03E+08 | 744.593 | 7.43E+20 | 744.593 | 0.41855 | 744.593 | 3.72E-08 | 744.593 | 3.71E-08 | 744.593 | 1.16E+13 | 744.593 | 0.01187 |
| 585.4517 | 2.79E+08 | 1739.313 | -2.22E+08 | 744.878 | 7.44E+20 | 744.878 | 0.4185 | 744.878 | 3.71E-08 | 744.878 | 3.67E-08 | 744.878 | 1.16E+13 | 744.878 | 0.01192 |
| 585.7432 | 3.27E+08 | 1740.989 | -7.95E+06 | 745.162 | 7.45E+20 | 745.162 | 0.41844 | 745.162 | 3.71E-08 | 745.162 | 3.74E-08 | 745.162 | 1.16E+13 | 745.162 | 0.01186 |
| 586.0347 | 2.66E+08 | 1742.664 | 6.02E+07 | 745.447 | 7.46E+20 | 745.447 | 0.41839 | 745.447 | 3.71E-08 | 745.447 | 3.72E-08 | 745.447 | 1.16E+13 | 745.447 | 0.01197 |
| 586.3262 | 2.66E+08 | 1744.34 | -2.45E+08 | 745.732 | 7.47E+20 | 745.732 | 0.41834 | 745.732 | 3.71E-08 | 745.732 | 3.82E-08 | 745.732 | 1.16E+13 | 745.732 | 0.01191 |
| 586.6177 | 2.79E+08 | 1746.015 | -1.53E+08 | 746.016 | 7.48E+20 | 746.016 | 0.41828 | 746.016 | 3.71E-08 | 746.016 | 3.83E-08 | 746.016 | 1.16E+13 | 746.016 | 0.01192 |
| 586.9092 | 2.86E+08 | 1747.69 | -1.81E+08 | 746.301 | 7.49E+20 | 746.301 | 0.41823 | 746.301 | 3.71E-08 | 746.301 | 3.60E-08 | 746.301 | 1.16E+13 | 746.301 | 0.01192 |
| 587.2007 | 3.41E+08 | 1749.365 | -2.45E+08 | 746.586 | 7.51E+20 | 746.586 | 0.41818 | 746.586 | 3.70E-08 | 746.586 | 3.74E-08 | 746.586 | 1.16E+13 | 746.586 | 0.01191 |
| 587.4921 | 4.16E+08 | 1751.04 | -2.11E+08 | 746.87 | 7.52E+20 | 746.87 | 0.41812 | 746.87 | 3.70E-08 | 746.87 | 3.62E-08 | 746.87 | 1.16E+13 | 746.87 | 0.01191 |
| 587.7836 | 4.23E+08 | 1752.715 | -2.93E+08 | 747.155 | 7.53E+20 | 747.155 | 0.41807 | 747.155 | 3.71E-08 | 747.155 | 3.90E-08 | 747.155 | 1.17E+13 | 747.155 | 0.01203 |
| 588.075 | 4.63E+08 | | | 747.439 | 7.54E+20 | 747.439 | 0.41801 | 747.439 | 3.71E-08 | 747.439 | 3.74E-08 | 747.439 | 1.17E+13 | 747.439 | 0.01188 |
| 588.3664 | 5.38E+08 | | | 747.724 | 7.55E+20 | 747.724 | 0.41796 | 747.724 | 3.71E-08 | 747.724 | 3.67E-08 | 747.724 | 1.17E+13 | 747.724 | 0.01205 |
| 588.6578 | 7.09E+08 | | | 748.009 | 7.56E+20 | 748.009 | 0.41791 | 748.009 | 3.71E-08 | 748.009 | 3.66E-08 | 748.009 | 1.17E+13 | 748.009 | 0.01196 |
| 588.9492 | 9.54E+08 | | | 748.293 | 7.57E+20 | 748.293 | 0.41785 | 748.293 | 3.71E-08 | 748.293 | 3.96E-08 | 748.293 | 1.18E+13 | 748.293 | 0.01191 |
| 589.2406 | 1.22E+09 | | | 748.578 | 7.58E+20 | 748.578 | 0.4178 | 748.578 | 3.72E-08 | 748.578 | 3.84E-08 | 748.578 | 1.18E+13 | 748.578 | 0.01193 |
| 589.532 | 1.55E+09 | | | 748.862 | 7.59E+20 | 748.862 | 0.41774 | 748.862 | 3.72E-08 | 748.862 | 3.91E-08 | 748.862 | 1.18E+13 | 748.862 | 0.01193 |
| 589.8234 | 1.92E+09 | | | 749.147 | 7.60E+20 | 749.147 | 0.41769 | 749.147 | 3.73E-08 | 749.147 | 3.67E-08 | 749.147 | 1.18E+13 | 749.147 | 0.01194 |
| 590.1147 | 2.32E+09 | | | 749.431 | 7.62E+20 | 749.431 | 0.41764 | 749.431 | 3.72E-08 | 749.431 | 3.50E-08 | 749.431 | 1.18E+13 | 749.431 | 0.0119 |
| 590.4061 | 2.68E+09 | | | 749.716 | 7.63E+20 | 749.716 | 0.41758 | 749.716 | 3.73E-08 | 749.716 | 3.59E-08 | 749.716 | 1.19E+13 | 749.716 | 0.01196 |
| 590.6974 | 3.22E+09 | | | 750 | 7.64E+20 | 750 | 0.41753 | 750 | 3.73E-08 | 750 | 3.84E-08 | 750 | 1.19E+13 | 750 | 0.01201 |
| 590.9888 | 3.65E+09 | | | 750.284 | 7.65E+20 | 750.284 | 0.41747 | 750.284 | 3.72E-08 | 750.284 | 3.70E-08 | 750.284 | 1.19E+13 | 750.284 | 0.01193 |
| 591.2801 | 4.15E+09 | | | 750.569 | 7.66E+20 | 750.569 | 0.41742 | 750.569 | 3.72E-08 | 750.569 | 3.74E-08 | 750.569 | 1.19E+13 | 750.569 | 0.01197 |
| 591.5714 | 4.73E+09 | | | 750.853 | 7.67E+20 | 750.853 | 0.41736 | 750.853 | 3.72E-08 | 750.853 | 3.72E-08 | 750.853 | 1.19E+13 | 750.853 | 0.01187 |
| 591.8627 | 5.28E+09 | | | 751.138 | 7.68E+20 | 751.138 | 0.41731 | 751.138 | 3.73E-08 | 751.138 | 3.77E-08 | 751.138 | 1.20E+13 | 751.138 | 0.01186 |
| 592.154 | 5.83E+09 | | | 751.422 | 7.69E+20 | 751.422 | 0.41726 | 751.422 | 3.73E-08 | 751.422 | 3.57E-08 | 751.422 | 1.20E+13 | 751.422 | 0.01194 |
| 592.4453 | 6.55E+09 | | | 751.706 | 7.70E+20 | 751.706 | 0.4172 | 751.706 | 3.73E-08 | 751.706 | 3.56E-08 | 751.706 | 1.20E+13 | 751.706 | 0.01191 |
| 592.7365 | 7.12E+09 | | | 751.991 | 7.72E+20 | 751.991 | 0.41715 | 751.991 | 3.72E-08 | 751.991 | 3.61E-08 | 751.991 | 1.20E+13 | 751.991 | 0.01195 |
| 593.0278 | 7.68E+09 | | | 752.275 | 7.73E+20 | 752.275 | 0.41709 | 752.275 | 3.73E-08 | 752.275 | 3.65E-08 | 752.275 | 1.20E+13 | 752.275 | 0.01194 |
| 593.319 | 8.11E+09 | | | 752.559 | 7.74E+20 | 752.559 | 0.41704 | 752.559 | 3.73E-08 | 752.559 | 3.73E-08 | 752.559 | 1.20E+13 | 752.559 | 0.01198 |
| 593.6103 | 8.96E+09 | | | 752.844 | 7.75E+20 | 752.844 | 0.41698 | 752.844 | 3.73E-08 | 752.844 | 3.81E-08 | 752.844 | 1.21E+13 | 752.844 | 0.01188 |
| 593.9015 | 9.61E+09 | | | 753.128 | 7.76E+20 | 753.128 | 0.41693 | 753.128 | 3.73E-08 | 753.128 | 3.68E-08 | 753.128 | 1.21E+13 | 753.128 | 0.012 |
| 594.1927 | 1.04E+10 | | | 753.412 | 7.77E+20 | 753.412 | 0.41687 | 753.412 | 3.73E-08 | 753.412 | 3.75E-08 | 753.412 | 1.21E+13 | 753.412 | 0.0119 |
| 594.4839 | 1.12E+10 | | | 753.697 | 7.78E+20 | 753.697 | 0.41682 | 753.697 | 3.73E-08 | 753.697 | 3.74E-08 | 753.697 | 1.21E+13 | 753.697 | 0.0119 |
| 594.7751 | 1.20E+10 | | | 753.981 | 7.79E+20 | 753.981 | 0.41676 | 753.981 | 3.74E-08 | 753.981 | 3.78E-08 | 753.981 | 1.21E+13 | 753.981 | 0.0119 |
| 595.0663 | 1.29E+10 | | | 754.265 | 7.80E+20 | 754.265 | 0.41671 | 754.265 | 3.73E-08 | 754.265 | 3.79E-08 | 754.265 | 1.21E+13 | 754.265 | 0.01193 |
| 595.3575 | 1.31E+10 | | | 754.549 | 7.82E+20 | 754.549 | 0.41665 | 754.549 | 3.73E-08 | 754.549 | 3.76E-08 | 754.549 | 1.21E+13 | 754.549 | 0.01201 |
| 595.6486 | 1.41E+10 | | | 754.834 | 7.83E+20 | 754.834 | 0.4166 | 754.834 | 3.72E-08 | 754.834 | 3.84E-08 | 754.834 | 1.21E+13 | 754.834 | 0.01193 |
| 595.9398 | 1.47E+10 | | | 755.118 | 7.84E+20 | 755.118 | 0.41654 | 755.118 | 3.72E-08 | 755.118 | 3.60E-08 | 755.118 | 1.21E+13 | 755.118 | 0.01197 |
| 596.2309 | 1.55E+10 | | | 755.402 | 7.85E+20 | 755.402 | 0.41649 | 755.402 | 3.72E-08 | 755.402 | 3.75E-08 | 755.402 | 1.22E+13 | 755.402 | 0.01194 |
| 596.522 | 1.65E+10 | | | 755.686 | 7.86E+20 | 755.686 | 0.41643 | 755.686 | 3.73E-08 | 755.686 | 3.65E-08 | 755.686 | 1.22E+13 | 755.686 | 0.01194 |
| 596.8132 | 1.73E+10 | | | 755.97 | 7.87E+20 | 755.97 | 0.41638 | 755.97 | 3.73E-08 | 755.97 | 3.68E-08 | 755.97 | 1.22E+13 | 755.97 | 0.01195 |
| 597.1043 | 1.80E+10 | | | 756.254 | 7.88E+20 | 756.254 | 0.41632 | 756.254 | 3.73E-08 | 756.254 | 3.76E-08 | 756.254 | 1.22E+13 | 756.254 | 0.01198 |
| 597.3954 | 1.89E+10 | | | 756.539 | 7.89E+20 | 756.539 | 0.41627 | 756.539 | 3.73E-08 | 756.539 | 3.73E-08 | 756.539 | 1.23E+13 | 756.539 | 0.01196 |
| 597.6864 | 1.98E+10 | | | 756.823 | 7.90E+20 | 756.823 | 0.41621 | 756.823 | 3.74E-08 | 756.823 | 3.93E-08 | 756.823 | 1.23E+13 | 756.823 | 0.01191 |
| 597.9775 | 2.06E+10 | | | 757.107 | 7.92E+20 | 757.107 | 0.41616 | 757.107 | 3.74E-08 | 757.107 | 3.80E-08 | 757.107 | 1.23E+13 | 757.107 | 0.01192 |
| 598.2686 | 2.15E+10 | | | 757.391 | 7.93E+20 | 757.391 | 0.4161 | 757.391 | 3.74E-08 | 757.391 | 3.70E-08 | 757.391 | 1.23E+13 | 757.391 | 0.01186 |
| 598.5596 | 2.20E+10 | | | 757.675 | 7.94E+20 | 757.675 | 0.41605 | 757.675 | 3.74E-08 | 757.675 | 3.65E-08 | 757.675 | 1.23E+13 | 757.675 | 0.01185 |
| 598.8507 | 2.29E+10 | | | 757.959 | 7.95E+20 | 757.959 | 0.41599 | 757.959 | 3.75E-08 | 757.959 | 3.79E-08 | 757.959 | 1.24E+13 | 757.959 | 0.01189 |
| 599.1417 | 2.36E+10 | | | 758.243 | 7.96E+20 | 758.243 | 0.41593 | 758.243 | 3.75E-08 | 758.243 | 3.96E-08 | 758.243 | 1.24E+13 | 758.243 | 0.01192 |
| 599.4327 | 2.44E+10 | | | 758.527 | 7.97E+20 | 758.527 | 0.41588 | 758.527 | 3.75E-08 | 758.527 | 3.54E-08 | 758.527 | 1.24E+13 | 758.527 | 0.01185 |
| 599.7238 | 2.58E+10 | | | 758.811 | 7.98E+20 | 758.811 | 0.41582 | 758.811 | 3.75E-08 | 758.811 | 3.84E-08 | 758.811 | 1.24E+13 | 758.811 | 0.0119 |
| 600.0148 | 2.64E+10 | | | 759.095 | 7.99E+20 | 759.095 | 0.41577 | 759.095 | 3.75E-08 | 759.095 | 3.80E-08 | 759.095 | 1.25E+13 | 759.095 | 0.01194 |
| 600.3057 | 2.75E+10 | | | 759.379 | 8.00E+20 | 759.379 | 0.41571 | 759.379 | 3.75E-08 | 759.379 | 3.76E-08 | 759.379 | 1.25E+13 | 759.379 | 0.01185 |
| 600.5967 | 2.81E+10 | | | 759.663 | 8.02E+20 | 759.663 | 0.41566 | 759.663 | 3.76E-08 | 759.663 | 3.71E-08 | 759.663 | 1.25E+13 | 759.663 | 0.01184 |
| 600.8877 | 2.92E+10 | | | 759.947 | 8.03E+20 | 759.947 | 0.4156 | 759.947 | 3.76E-08 | 759.947 | 3.78E-08 | 759.947 | 1.25E+13 | 759.947 | 0.01195 |
| 601.1786 | 2.98E+10 | | | 760.231 | 8.04E+20 | 760.231 | 0.4156 | 760.231 | 3.76E-08 | 760.231 | 3.63E-08 | 760.231 | 1.26E+13 | 760.231 | 0.01181 |
| 601.4696 | 3.05E+10 | | | 760.515 | 8.05E+20 | 760.515 | 0.41555 | 760.515 | 3.76E-08 | 760.515 | 3.59E-08 | 760.515 | 1.26E+13 | 760.515 | 0.01181 |
| 601.7605 | 3.18E+10 | | | 760.799 | 8.06E+20 | 760.799 | 0.4155 | 760.799 | 3.76E-08 | 760.799 | 3.76E-08 | 760.799 | 1.26E+13 | 760.799 | 0.01192 |
| 602.0515 | 3.24E+10 | | | 761.083 | 8.07E+20 | 761.083 | 0.41545 | 761.083 | 3.76E-08 | 761.083 | 3.71E-08 | 761.083 | 1.26E+13 | 761.083 | 0.01188 |
| 602.3424 | 3.37E+10 | | | 761.366 | 8.08E+20 | 761.366 | 0.4154 | 761.366 | 3.76E-08 | 761.366 | 3.84E-08 | 761.366 | 1.26E+13 | 761.366 | 0.01185 |
| 602.6333 | 3.48E+10 | | | 761.65 | 8.09E+20 | 761.65 | 0.41535 | 761.65 | 3.76E-08 | 761.65 | 3.74E-08 | 761.65 | 1.26E+13 | 761.65 | 0.01187 |
| 602.9242 | 3.49E+10 | | | 761.934 | 8.10E+20 | 761.934 | 0.4153 | 761.934 | 3.77E-08 | 761.934 | 3.75E-08 | 761.934 | 1.27E+13 | 761.934 | 0.01186 |
| 603.2151 | 3.60E+10 | | | 762.218 | 8.12E+20 | 762.218 | 0.41525 | 762.218 | 3.76E-08 | 762.218 | 3.80E-08 | 762.218 | 1.27E+13 | 762.218 | 0.01194 |
| 603.5059 | 3.64E+10 | | | 762.502 | 8.13E+20 | 762.502 | 0.4152 | 762.502 | 3.76E-08 | 762.502 | 3.87E-08 | 762.502 | 1.27E+13 | 762.502 | 0.01191 |
| 603.7968 | 3.76E+10 | | | 762.786 | 8.14E+20 | 762.786 | 0.41515 | 762.786 | 3.76E-08 | 762.786 | 3.79E-08 | 762.786 | 1.27E+13 | 762.786 | 0.01182 |
| 604.0876 | 3.88E+10 | | | 763.069 | 8.15E+20 | 763.069 | 0.4151 | 763.069 | 3.76E-08 | 763.069 | 3.71E-08 | 763.069 | 1.27E+13 | 763.069 | 0.01189 |
| 604.3785 | 3.93E+10 | | | 763.353 | 8.16E+20 | 763.353 | 0.41506 | 763.353 | 3.76E-08 | 763.353 | 3.55E-08 | 763.353 | 1.27E+13 | 763.353 | 0.0119 |
| 604.6693 | 4.01E+10 | | | 763.637 | 8.17E+20 | 763.637 | 0.41501 | 763.637 | 3.77E-08 | 763.637 | 3.86E-08 | 763.637 | 1.28E+13 | 763.637 | 0.01178 |
| 604.9601 | 4.09E+10 | | | 763.921 | 8.18E+20 | 763.921 | 0.41496 | 763.921 | 3.77E-08 | 763.921 | 3.88E-08 | 763.921 | 1.28E+13 | 763.921 | 0.01187 |
| 605.2509 | 4.16E+10 | | | 764.204 | 8.19E+20 | 764.204 | 0.41491 | 764.204 | 3.76E-08 | 764.204 | 3.60E-08 | 764.204 | 1.28E+13 | 764.204 | 0.01189 |
| 605.5417 | 4.25E+10 | | | 764.488 | 8.21E+20 | 764.488 | 0.41486 | 764.488 | 3.77E-08 | 764.488 | 3.76E-08 | 764.488 | 1.28E+13 | 764.488 | 0.01183 |
| 605.8325 | 4.34E+10 | | | 764.772 | 8.22E+20 | 764.772 | 0.41481 | 764.772 | 3.77E-08 | 764.772 | 3.75E-08 | 764.772 | 1.29E+13 | 764.772 | 0.0118 |
| 606.1233 | 4.44E+10 | | | 765.055 | 8.23E+20 | 765.055 | 0.41476 | 765.055 | 3.77E-08 | 765.055 | 3.98E-08 | 765.055 | 1.29E+13 | 765.055 | 0.01187 |
| 606.414 | 4.52E+10 | | | 765.339 | 8.24E+20 | 765.339 | 0.41471 | 765.339 | 3.77E-08 | 765.339 | 3.81E-08 | 765.339 | 1.29E+13 | 765.339 | 0.01187 |
| 606.7048 | 4.52E+10 | | | 765.623 | 8.25E+20 | 765.623 | 0.41466 | 765.623 | 3.77E-08 | 765.623 | 3.82E-08 | 765.623 | 1.29E+13 | 765.623 | 0.01192 |
| 606.9955 | 4.66E+10 | | | 765.906 | 8.26E+20 | 765.906 | 0.41461 | 765.906 | 3.77E-08 | 765.906 | 3.91E-08 | 765.906 | 1.29E+13 | 765.906 | 0.01184 |
| 607.2863 | 4.73E+10 | | | 766.19 | 8.27E+20 | 766.19 | 0.41456 | 766.19 | 3.77E-08 | 766.19 | 3.79E-08 | 766.19 | 1.29E+13 | 766.19 | 0.01189 |
| 607.577 | 4.80E+10 | | | 766.474 | 8.28E+20 | 766.474 | 0.41451 | 766.474 | 3.78E-08 | 766.474 | 3.74E-08 | 766.474 | 1.30E+13 | 766.474 | 0.0118 |
| 607.8677 | 4.86E+10 | | | 766.757 | 8.29E+20 | 766.757 | 0.41446 | 766.757 | 3.78E-08 | 766.757 | 3.70E-08 | 766.757 | 1.30E+13 | 766.757 | 0.01181 |
| 608.1584 | 4.95E+10 | | | 767.041 | 8.31E+20 | 767.041 | 0.41441 | 767.041 | 3.78E-08 | 767.041 | 3.73E-08 | 767.041 | 1.30E+13 | 767.041 | 0.01184 |
| 608.4491 | 4.97E+10 | | | 767.324 | 8.32E+20 | 767.324 | 0.41436 | 767.324 | 3.78E-08 | 767.324 | 3.75E-08 | 767.324 | 1.30E+13 | 767.324 | 0.01184 |
| 608.7398 | 5.04E+10 | | | 767.608 | 8.33E+20 | 767.608 | 0.41431 | 767.608 | 3.79E-08 | 767.608 | 3.82E-08 | 767.608 | 1.31E+13 | 767.608 | 0.01188 |
| 609.0304 | 5.07E+10 | | | 767.892 | 8.34E+20 | 767.892 | 0.41426 | 767.892 | 3.79E-08 | 767.892 | 3.73E-08 | 767.892 | 1.31E+13 | 767.892 | 0.01177 |
| 609.3211 | 5.14E+10 | | | 768.175 | 8.35E+20 | 768.175 | 0.41422 | 768.175 | 3.79E-08 | 768.175 | 3.67E-08 | 768.175 | 1.31E+13 | 768.175 | 0.01186 |
| 609.6117 | 5.20E+10 | | | 768.459 | 8.36E+20 | 768.459 | 0.41417 | 768.459 | 3.79E-08 | 768.459 | 3.73E-08 | 768.459 | 1.31E+13 | 768.459 | 0.01175 |
| 609.9023 | 5.24E+10 | | | 768.742 | 8.37E+20 | 768.742 | 0.41412 | 768.742 | 3.79E-08 | 768.742 | 3.84E-08 | 768.742 | 1.31E+13 | 768.742 | 0.01182 |
| 610.193 | 5.27E+10 | | | 769.026 | 8.38E+20 | 769.026 | 0.41407 | 769.026 | 3.79E-08 | 769.026 | 3.79E-08 | 769.026 | 1.32E+13 | 769.026 | 0.01187 |
| 610.4836 | 5.30E+10 | | | 769.309 | 8.40E+20 | 769.309 | 0.41402 | 769.309 | 3.80E-08 | 769.309 | 3.86E-08 | 769.309 | 1.32E+13 | 769.309 | 0.01169 |
| 610.7742 | 5.27E+10 | | | 769.592 | 8.41E+20 | 769.592 | 0.41397 | 769.592 | 3.80E-08 | 769.592 | 3.80E-08 | 769.592 | 1.32E+13 | 769.592 | 0.01185 |
| 611.0648 | 5.26E+10 | | | 769.876 | 8.42E+20 | 769.876 | 0.41392 | 769.876 | 3.80E-08 | 769.876 | 3.82E-08 | 769.876 | 1.32E+13 | 769.876 | 0.01185 |
| 611.3553 | 5.31E+10 | | | 770.159 | 8.43E+20 | 770.159 | 0.41387 | 770.159 | 3.81E-08 | 770.159 | 3.82E-08 | 770.159 | 1.33E+13 | 770.159 | 0.01178 |
| 611.6459 | 5.33E+10 | | | 770.443 | 8.44E+20 | 770.443 | 0.41382 | 770.443 | 3.81E-08 | 770.443 | 3.87E-08 | 770.443 | 1.33E+13 | 770.443 | 0.01175 |
| 611.9365 | 5.33E+10 | | | 770.726 | 8.45E+20 | 770.726 | 0.41377 | 770.726 | 3.82E-08 | 770.726 | 3.68E-08 | 770.726 | 1.33E+13 | 770.726 | 0.01175 |
| 612.227 | 5.31E+10 | | | 771.009 | 8.46E+20 | 771.009 | 0.41372 | 771.009 | 3.81E-08 | 771.009 | 3.79E-08 | 771.009 | 1.34E+13 | 771.009 | 0.01174 |
| 612.5175 | 5.36E+10 | | | 771.293 | 8.47E+20 | 771.293 | 0.41367 | 771.293 | 3.81E-08 | 771.293 | 3.76E-08 | 771.293 | 1.34E+13 | 771.293 | 0.01173 |
| 612.8081 | 5.39E+10 | | | 771.576 | 8.49E+20 | 771.576 | 0.41362 | 771.576 | 3.81E-08 | 771.576 | 3.73E-08 | 771.576 | 1.34E+13 | 771.576 | 0.01182 |
| 613.0986 | 5.40E+10 | | | 771.86 | 8.50E+20 | 771.86 | 0.41357 | 771.86 | 3.81E-08 | 771.86 | 3.85E-08 | 771.86 | 1.34E+13 | 771.86 | 0.01181 |
| 613.3891 | 5.43E+10 | | | 772.143 | 8.51E+20 | 772.143 | 0.41352 | 772.143 | 3.82E-08 | 772.143 | 3.80E-08 | 772.143 | 1.34E+13 | 772.143 | 0.0118 |
| 613.6796 | 5.42E+10 | | | 772.426 | 8.52E+20 | 772.426 | 0.41347 | 772.426 | 3.82E-08 | 772.426 | 3.85E-08 | 772.426 | 1.35E+13 | 772.426 | 0.01181 |
| 613.97 | 5.51E+10 | | | 772.709 | 8.53E+20 | 772.709 | 0.41342 | 772.709 | 3.83E-08 | 772.709 | 3.86E-08 | 772.709 | 1.35E+13 | 772.709 | 0.01176 |
| 614.2605 | 5.43E+10 | | | 772.993 | 8.54E+20 | 772.993 | 0.41337 | 772.993 | 3.83E-08 | 772.993 | 3.82E-08 | 772.993 | 1.35E+13 | 772.993 | 0.01174 |

| | | | | | | | | | | | | | |
|---|---|---|---|---|---|---|---|---|---|---|---|---|---|
| 614.551 | 5.45E+10 | 773.276 | 8.55E+20 | 773.276 | 0.41332 | 773.276 | 3.83E-08 | 773.276 | 3.85E-08 | 773.276 | 1.36E+13 | 773.276 | 0.01177 |
| 614.8414 | 5.46E+10 | 773.559 | 8.56E+20 | 773.559 | 0.41327 | 773.559 | 3.83E-08 | 773.559 | 3.81E-08 | 773.559 | 1.36E+13 | 773.559 | 0.01177 |
| 615.1318 | 5.43E+10 | 773.842 | 8.58E+20 | 773.842 | 0.41322 | 773.842 | 3.84E-08 | 773.842 | 3.84E-08 | 773.842 | 1.36E+13 | 773.842 | 0.01176 |
| 615.4223 | 5.49E+10 | 774.126 | 8.59E+20 | 774.126 | 0.41318 | 774.126 | 3.84E-08 | 774.126 | 3.72E-08 | 774.126 | 1.36E+13 | 774.126 | 0.01183 |
| 615.7127 | 5.49E+10 | 774.409 | 8.60E+20 | 774.409 | 0.41313 | 774.409 | 3.84E-08 | 774.409 | 3.90E-08 | 774.409 | 1.37E+13 | 774.409 | 0.01178 |
| 616.0031 | 5.55E+10 | 774.692 | 8.61E+20 | 774.692 | 0.41308 | 774.692 | 3.84E-08 | 774.692 | 3.87E-08 | 774.692 | 1.37E+13 | 774.692 | 0.01179 |
| 616.2935 | 5.54E+10 | 774.975 | 8.62E+20 | 774.975 | 0.41303 | 774.975 | 3.84E-08 | 774.975 | 3.75E-08 | 774.975 | 1.37E+13 | 774.975 | 0.01176 |
| 616.5838 | 5.55E+10 | 775.258 | 8.63E+20 | 775.258 | 0.41298 | 775.258 | 3.85E-08 | 775.258 | 3.95E-08 | 775.258 | 1.37E+13 | 775.258 | 0.01177 |
| 616.8742 | 5.62E+10 | 775.542 | 8.64E+20 | 775.542 | 0.41293 | 775.542 | 3.85E-08 | 775.542 | 3.89E-08 | 775.542 | 1.37E+13 | 775.542 | 0.01176 |
| 617.1646 | 5.62E+10 | 775.825 | 8.65E+20 | 775.825 | 0.41288 | 775.825 | 3.85E-08 | 775.825 | 3.92E-08 | 775.825 | 1.37E+13 | 775.825 | 0.01177 |
| 617.4549 | 5.61E+10 | 776.108 | 8.67E+20 | 776.108 | 0.41283 | 776.108 | 3.85E-08 | 776.108 | 3.89E-08 | 776.108 | 1.38E+13 | 776.108 | 0.01171 |
| 617.7452 | 5.61E+10 | 776.391 | 8.68E+20 | 776.391 | 0.41278 | 776.391 | 3.85E-08 | 776.391 | 3.97E-08 | 776.391 | 1.38E+13 | 776.391 | 0.01175 |
| 618.0356 | 5.66E+10 | 776.674 | 8.69E+20 | 776.674 | 0.41273 | 776.674 | 3.85E-08 | 776.674 | 3.91E-08 | 776.674 | 1.38E+13 | 776.674 | 0.01174 |
| 618.3259 | 5.71E+10 | 776.957 | 8.70E+20 | 776.957 | 0.41268 | 776.957 | 3.85E-08 | 776.957 | 3.84E-08 | 776.957 | 1.38E+13 | 776.957 | 0.01173 |
| 618.6162 | 5.65E+10 | 777.24 | 8.71E+20 | 777.24 | 0.41263 | 777.24 | 3.85E-08 | 777.24 | 3.76E-08 | 777.24 | 1.38E+13 | 777.24 | 0.01179 |
| 618.9065 | 5.67E+10 | 777.523 | 8.72E+20 | 777.523 | 0.41258 | 777.523 | 3.86E-08 | 777.523 | 3.92E-08 | 777.523 | 1.39E+13 | 777.523 | 0.01164 |
| 619.1967 | 5.83E+10 | 777.806 | 8.73E+20 | 777.806 | 0.41253 | 777.806 | 3.86E-08 | 777.806 | 3.99E-08 | 777.806 | 1.39E+13 | 777.806 | 0.01173 |
| 619.487 | 5.73E+10 | 778.089 | 8.74E+20 | 778.089 | 0.41248 | 778.089 | 3.87E-08 | 778.089 | 3.96E-08 | 778.089 | 1.39E+13 | 778.089 | 0.01174 |
| 619.7773 | 5.76E+10 | 778.372 | 8.76E+20 | 778.372 | 0.41243 | 778.372 | 3.87E-08 | 778.372 | 3.85E-08 | 778.372 | 1.40E+13 | 778.372 | 0.01178 |
| 620.0675 | 5.75E+10 | 778.655 | 8.77E+20 | 778.655 | 0.41238 | 778.655 | 3.87E-08 | 778.655 | 3.90E-08 | 778.655 | 1.40E+13 | 778.655 | 0.01174 |
| 620.3577 | 5.72E+10 | 778.938 | 8.78E+20 | 778.938 | 0.41233 | 778.938 | 3.88E-08 | 778.938 | 3.81E-08 | 778.938 | 1.40E+13 | 778.938 | 0.01165 |
| 620.648 | 5.77E+10 | 779.221 | 8.79E+20 | 779.221 | 0.41228 | 779.221 | 3.88E-08 | 779.221 | 3.68E-08 | 779.221 | 1.41E+13 | 779.221 | 0.01167 |
| 620.9382 | 5.80E+10 | 779.504 | 8.80E+20 | 779.504 | 0.41223 | 779.504 | 3.88E-08 | 779.504 | 4.03E-08 | 779.504 | 1.41E+13 | 779.504 | 0.01171 |
| 621.2284 | 5.85E+10 | 779.787 | 8.81E+20 | 779.787 | 0.41218 | 779.787 | 3.88E-08 | 779.787 | 3.76E-08 | 779.787 | 1.41E+13 | 779.787 | 0.01171 |
| 621.5186 | 5.86E+10 | 780.07 | 8.82E+20 | 780.07 | 0.41213 | 780.07 | 3.89E-08 | 780.07 | 3.89E-08 | 780.07 | 1.41E+13 | 780.07 | 0.01171 |
| 621.8087 | 5.84E+10 | 780.353 | 8.83E+20 | 780.353 | 0.41208 | 780.353 | 3.88E-08 | 780.353 | 3.73E-08 | 780.353 | 1.41E+13 | 780.353 | 0.01172 |
| 622.0989 | 5.85E+10 | 780.635 | 8.85E+20 | 780.635 | 0.41203 | 780.635 | 3.89E-08 | 780.635 | 3.87E-08 | 780.635 | 1.42E+13 | 780.635 | 0.01164 |
| 622.3891 | 5.82E+10 | 780.918 | 8.86E+20 | 780.918 | 0.41198 | 780.918 | 3.89E-08 | 780.918 | 4.03E-08 | 780.918 | 1.42E+13 | 780.918 | 0.01167 |
| 622.6792 | 5.83E+10 | 781.201 | 8.87E+20 | 781.201 | 0.41193 | 781.201 | 3.89E-08 | 781.201 | 3.86E-08 | 781.201 | 1.42E+13 | 781.201 | 0.01163 |
| 622.9693 | 5.90E+10 | 781.484 | 8.88E+20 | 781.484 | 0.41188 | 781.484 | 3.89E-08 | 781.484 | 3.81E-08 | 781.484 | 1.42E+13 | 781.484 | 0.01171 |
| 623.2594 | 5.98E+10 | 781.767 | 8.89E+20 | 781.767 | 0.41183 | 781.767 | 3.89E-08 | 781.767 | 3.93E-08 | 781.767 | 1.42E+13 | 781.767 | 0.01172 |
| 623.5496 | 5.96E+10 | 782.05 | 8.90E+20 | 782.05 | 0.41178 | 782.05 | 3.89E-08 | 782.05 | 3.82E-08 | 782.05 | 1.43E+13 | 782.05 | 0.01167 |
| 623.8397 | 5.97E+10 | 782.332 | 8.91E+20 | 782.332 | 0.41173 | 782.332 | 3.89E-08 | 782.332 | 3.72E-08 | 782.332 | 1.43E+13 | 782.332 | 0.01171 |
| 624.1297 | 6.03E+10 | 782.615 | 8.92E+20 | 782.615 | 0.41168 | 782.615 | 3.89E-08 | 782.615 | 3.97E-08 | 782.615 | 1.43E+13 | 782.615 | 0.01166 |
| 624.4198 | 6.03E+10 | 782.898 | 8.94E+20 | 782.898 | 0.41163 | 782.898 | 3.89E-08 | 782.898 | 3.69E-08 | 782.898 | 1.43E+13 | 782.898 | 0.01166 |
| 624.7099 | 5.98E+10 | 783.181 | 8.95E+20 | 783.181 | 0.41158 | 783.181 | 3.89E-08 | 783.181 | 4.00E-08 | 783.181 | 1.43E+13 | 783.181 | 0.01165 |
| 624.9999 | 6.01E+10 | 783.463 | 8.96E+20 | 783.463 | 0.41153 | 783.463 | 3.89E-08 | 783.463 | 3.86E-08 | 783.463 | 1.43E+13 | 783.463 | 0.01168 |
| 625.29 | 6.06E+10 | 783.746 | 8.97E+20 | 783.746 | 0.41148 | 783.746 | 3.89E-08 | 783.746 | 4.03E-08 | 783.746 | 1.43E+13 | 783.746 | 0.01172 |
| 625.58 | 6.10E+10 | 784.029 | 8.98E+20 | 784.029 | 0.41143 | 784.029 | 3.88E-08 | 784.029 | 3.90E-08 | 784.029 | 1.44E+13 | 784.029 | 0.01172 |
| 625.87 | 6.09E+10 | 784.311 | 8.99E+20 | 784.311 | 0.41138 | 784.311 | 3.89E-08 | 784.311 | 3.97E-08 | 784.311 | 1.44E+13 | 784.311 | 0.01167 |
| 626.16 | 6.11E+10 | 784.594 | 9.00E+20 | 784.594 | 0.41133 | 784.594 | 3.89E-08 | 784.594 | 4.13E-08 | 784.594 | 1.44E+13 | 784.594 | 0.01164 |
| 626.45 | 6.10E+10 | 784.877 | 9.01E+20 | 784.877 | 0.41128 | 784.877 | 3.89E-08 | 784.877 | 3.91E-08 | 784.877 | 1.44E+13 | 784.877 | 0.0117 |
| 626.74 | 6.19E+10 | 785.159 | 9.03E+20 | 785.159 | 0.41123 | 785.159 | 3.90E-08 | 785.159 | 4.01E-08 | 785.159 | 1.45E+13 | 785.159 | 0.0117 |
| 627.03 | 6.19E+10 | 785.442 | 9.04E+20 | 785.442 | 0.41118 | 785.442 | 3.90E-08 | 785.442 | 3.88E-08 | 785.442 | 1.45E+13 | 785.442 | 0.0117 |
| 627.3199 | 6.19E+10 | 785.724 | 9.05E+20 | 785.724 | 0.41113 | 785.724 | 3.91E-08 | 785.724 | 3.87E-08 | 785.724 | 1.45E+13 | 785.724 | 0.01165 |
| 627.6099 | 6.24E+10 | 786.007 | 9.06E+20 | 786.007 | 0.41108 | 786.007 | 3.90E-08 | 786.007 | 3.74E-08 | 786.007 | 1.45E+13 | 786.007 | 0.01167 |
| 627.8998 | 6.23E+10 | 786.29 | 9.07E+20 | 786.29 | 0.41103 | 786.29 | 3.91E-08 | 786.29 | 3.93E-08 | 786.29 | 1.46E+13 | 786.29 | 0.01161 |
| 628.1897 | 6.21E+10 | 786.572 | 9.08E+20 | 786.572 | 0.41098 | 786.572 | 3.91E-08 | 786.572 | 3.89E-08 | 786.572 | 1.46E+13 | 786.572 | 0.01169 |
| 628.4797 | 6.20E+10 | 786.855 | 9.09E+20 | 786.855 | 0.41093 | 786.855 | 3.92E-08 | 786.855 | 3.98E-08 | 786.855 | 1.46E+13 | 786.855 | 0.01164 |
| 628.7696 | 6.31E+10 | 787.137 | 9.10E+20 | 787.137 | 0.41089 | 787.137 | 3.92E-08 | 787.137 | 3.92E-08 | 787.137 | 1.47E+13 | 787.137 | 0.01158 |
| 629.0595 | 6.34E+10 | 787.42 | 9.12E+20 | 787.42 | 0.41084 | 787.42 | 3.92E-08 | 787.42 | 3.88E-08 | 787.42 | 1.47E+13 | 787.42 | 0.01156 |
| 629.3493 | 6.31E+10 | 787.702 | 9.13E+20 | 787.702 | 0.41079 | 787.702 | 3.92E-08 | 787.702 | 4.02E-08 | 787.702 | 1.47E+13 | 787.702 | 0.01164 |
| 629.6392 | 6.33E+10 | 787.985 | 9.14E+20 | 787.985 | 0.41074 | 787.985 | 3.92E-08 | 787.985 | 3.76E-08 | 787.985 | 1.47E+13 | 787.985 | 0.01158 |
| 629.9291 | 6.30E+10 | 788.267 | 9.15E+20 | 788.267 | 0.41069 | 788.267 | 3.93E-08 | 788.267 | 3.89E-08 | 788.267 | 1.48E+13 | 788.267 | 0.01158 |
| 630.2189 | 6.28E+10 | 788.55 | 9.16E+20 | 788.55 | 0.41063 | 788.55 | 3.93E-08 | 788.55 | 3.79E-08 | 788.55 | 1.48E+13 | 788.55 | 0.01161 |
| 630.5087 | 6.39E+10 | 788.832 | 9.17E+20 | 788.832 | 0.41058 | 788.832 | 3.93E-08 | 788.832 | 3.93E-08 | 788.832 | 1.48E+13 | 788.832 | 0.01159 |
| 630.7986 | 6.40E+10 | 789.114 | 9.18E+20 | 789.114 | 0.41053 | 789.114 | 3.93E-08 | 789.114 | 3.91E-08 | 789.114 | 1.48E+13 | 789.114 | 0.01165 |
| 631.0884 | 6.33E+10 | 789.397 | 9.19E+20 | 789.397 | 0.41048 | 789.397 | 3.94E-08 | 789.397 | 3.85E-08 | 789.397 | 1.49E+13 | 789.397 | 0.01153 |
| 631.3782 | 6.48E+10 | 789.679 | 9.21E+20 | 789.679 | 0.41043 | 789.679 | 3.94E-08 | 789.679 | 3.85E-08 | 789.679 | 1.49E+13 | 789.679 | 0.0115 |
| 631.668 | 6.44E+10 | 789.962 | 9.22E+20 | 789.962 | 0.41039 | 789.962 | 3.94E-08 | 789.962 | 3.94E-08 | 789.962 | 1.49E+13 | 789.962 | 0.01159 |
| 631.9578 | 6.45E+10 | 790.244 | 9.23E+20 | 790.244 | 0.41034 | 790.244 | 3.94E-08 | 790.244 | 3.95E-08 | 790.244 | 1.49E+13 | 790.244 | 0.01154 |
| 632.2475 | 6.40E+10 | 790.526 | 9.24E+20 | 790.526 | 0.41029 | 790.526 | 3.94E-08 | 790.526 | 4.07E-08 | 790.526 | 1.49E+13 | 790.526 | 0.01162 |
| 632.5373 | 6.50E+10 | 790.809 | 9.25E+20 | 790.809 | 0.41024 | 790.809 | 3.94E-08 | 790.809 | 3.95E-08 | 790.809 | 1.49E+13 | 790.809 | 0.01157 |
| 632.827 | 6.49E+10 | 791.091 | 9.26E+20 | 791.091 | 0.41019 | 791.091 | 3.94E-08 | 791.091 | 4.04E-08 | 791.091 | 1.50E+13 | 791.091 | 0.01152 |
| 633.1167 | 6.47E+10 | 791.373 | 9.27E+20 | 791.373 | 0.41013 | 791.373 | 3.94E-08 | 791.373 | 3.98E-08 | 791.373 | 1.50E+13 | 791.373 | 0.01154 |
| 633.4065 | 6.52E+10 | 791.656 | 9.29E+20 | 791.656 | 0.41008 | 791.656 | 3.94E-08 | 791.656 | 3.76E-08 | 791.656 | 1.50E+13 | 791.656 | 0.01156 |
| 633.6962 | 6.56E+10 | 791.938 | 9.30E+20 | 791.938 | 0.41003 | 791.938 | 3.94E-08 | 791.938 | 3.94E-08 | 791.938 | 1.50E+13 | 791.938 | 0.01158 |
| 633.9859 | 6.56E+10 | 792.22 | 9.31E+20 | 792.22 | 0.40998 | 792.22 | 3.94E-08 | 792.22 | 4.10E-08 | 792.22 | 1.50E+13 | 792.22 | 0.01152 |
| 634.2756 | 6.57E+10 | 792.502 | 9.32E+20 | 792.502 | 0.40993 | 792.502 | 3.94E-08 | 792.502 | 4.06E-08 | 792.502 | 1.51E+13 | 792.502 | 0.0115 |
| 634.5652 | 6.58E+10 | 792.784 | 9.33E+20 | 792.784 | 0.40988 | 792.784 | 3.95E-08 | 792.784 | 3.91E-08 | 792.784 | 1.51E+13 | 792.784 | 0.01154 |
| 634.8549 | 6.55E+10 | 793.067 | 9.34E+20 | 793.067 | 0.40983 | 793.067 | 3.95E-08 | 793.067 | 3.93E-08 | 793.067 | 1.51E+13 | 793.067 | 0.01155 |
| 635.1445 | 6.58E+10 | 793.349 | 9.35E+20 | 793.349 | 0.40978 | 793.349 | 3.95E-08 | 793.349 | 3.89E-08 | 793.349 | 1.51E+13 | 793.349 | 0.01146 |
| 635.4342 | 6.61E+10 | 793.631 | 9.36E+20 | 793.631 | 0.40973 | 793.631 | 3.95E-08 | 793.631 | 3.98E-08 | 793.631 | 1.52E+13 | 793.631 | 0.01151 |
| 635.7238 | 6.64E+10 | 793.913 | 9.38E+20 | 793.913 | 0.40968 | 793.913 | 3.96E-08 | 793.913 | 3.90E-08 | 793.913 | 1.52E+13 | 793.913 | 0.01149 |
| 636.0134 | 6.60E+10 | 794.195 | 9.39E+20 | 794.195 | 0.40963 | 794.195 | 3.96E-08 | 794.195 | 3.97E-08 | 794.195 | 1.52E+13 | 794.195 | 0.01148 |
| 636.303 | 6.67E+10 | 794.477 | 9.40E+20 | 794.477 | 0.40958 | 794.477 | 3.96E-08 | 794.477 | 4.00E-08 | 794.477 | 1.53E+13 | 794.477 | 0.0115 |
| 636.5926 | 6.64E+10 | 794.76 | 9.41E+20 | 794.76 | 0.40953 | 794.76 | 3.96E-08 | 794.76 | 4.03E-08 | 794.76 | 1.53E+13 | 794.76 | 0.01144 |
| 636.8822 | 6.69E+10 | 795.042 | 9.42E+20 | 795.042 | 0.40948 | 795.042 | 3.97E-08 | 795.042 | 4.01E-08 | 795.042 | 1.53E+13 | 795.042 | 0.01138 |
| 637.1718 | 6.71E+10 | 795.324 | 9.43E+20 | 795.324 | 0.40943 | 795.324 | 3.97E-08 | 795.324 | 4.06E-08 | 795.324 | 1.53E+13 | 795.324 | 0.01149 |
| 637.4613 | 6.65E+10 | 795.606 | 9.44E+20 | 795.606 | 0.40938 | 795.606 | 3.98E-08 | 795.606 | 3.79E-08 | 795.606 | 1.54E+13 | 795.606 | 0.01143 |
| 637.7509 | 6.72E+10 | 795.888 | 9.45E+20 | 795.888 | 0.40933 | 795.888 | 3.98E-08 | 795.888 | 3.98E-08 | 795.888 | 1.54E+13 | 795.888 | 0.01139 |
| 638.0404 | 6.72E+10 | 796.17 | 9.47E+20 | 796.17 | 0.40928 | 796.17 | 3.98E-08 | 796.17 | 4.15E-08 | 796.17 | 1.54E+13 | 796.17 | 0.01141 |
| 638.3299 | 6.73E+10 | 796.452 | 9.48E+20 | 796.452 | 0.40923 | 796.452 | 3.98E-08 | 796.452 | 3.92E-08 | 796.452 | 1.54E+13 | 796.452 | 0.01137 |
| 638.6194 | 6.75E+10 | 796.734 | 9.49E+20 | 796.734 | 0.40918 | 796.734 | 3.98E-08 | 796.734 | 3.92E-08 | 796.734 | 1.54E+13 | 796.734 | 0.01141 |
| 638.9089 | 6.74E+10 | 797.016 | 9.50E+20 | 797.016 | 0.40913 | 797.016 | 3.98E-08 | 797.016 | 4.01E-08 | 797.016 | 1.55E+13 | 797.016 | 0.0114 |
| 639.1984 | 6.79E+10 | 797.298 | 9.51E+20 | 797.298 | 0.40908 | 797.298 | 3.98E-08 | 797.298 | 3.89E-08 | 797.298 | 1.55E+13 | 797.298 | 0.01143 |
| 639.4879 | 6.80E+10 | 797.58 | 9.52E+20 | 797.58 | 0.40903 | 797.58 | 3.99E-08 | 797.58 | 3.92E-08 | 797.58 | 1.55E+13 | 797.58 | 0.0113 |
| 639.7774 | 6.82E+10 | 797.862 | 9.53E+20 | 797.862 | 0.40898 | 797.862 | 4.00E-08 | 797.862 | 3.90E-08 | 797.862 | 1.56E+13 | 797.862 | 0.01129 |
| 640.0668 | 6.86E+10 | 798.144 | 9.54E+20 | 798.144 | 0.40893 | 798.144 | 4.00E-08 | 798.144 | 3.87E-08 | 798.144 | 1.56E+13 | 798.144 | 0.01134 |
| 640.3563 | 6.80E+10 | 798.426 | 9.56E+20 | 798.426 | 0.40888 | 798.426 | 4.00E-08 | 798.426 | 4.13E-08 | 798.426 | 1.56E+13 | 798.426 | 0.0113 |
| 640.6457 | 6.81E+10 | 798.707 | 9.57E+20 | 798.707 | 0.40883 | 798.707 | 4.00E-08 | 798.707 | 4.02E-08 | 798.707 | 1.56E+13 | 798.707 | 0.01127 |
| 640.9351 | 6.89E+10 | 798.989 | 9.58E+20 | 798.989 | 0.40878 | 798.989 | 4.00E-08 | 798.989 | 4.00E-08 | 798.989 | 1.57E+13 | 798.989 | 0.01133 |
| 641.2245 | 6.84E+10 | 799.271 | 9.59E+20 | 799.271 | 0.40873 | 799.271 | 4.01E-08 | 799.271 | 4.05E-08 | 799.271 | 1.57E+13 | 799.271 | 0.01127 |
| 641.5139 | 6.90E+10 | 799.553 | 9.60E+20 | 799.553 | 0.40868 | 799.553 | 4.01E-08 | 799.553 | 4.02E-08 | 799.553 | 1.57E+13 | 799.553 | 0.01125 |
| 641.8033 | 6.95E+10 | 799.835 | 9.61E+20 | 799.835 | 0.40863 | 799.835 | 4.01E-08 | 799.835 | 3.96E-08 | 799.835 | 1.58E+13 | 799.835 | 0.01125 |
| 642.0927 | 6.92E+10 | 800.117 | 9.62E+20 | 800.117 | 0.40858 | 800.117 | 4.02E-08 | 800.117 | 3.98E-08 | 800.117 | 1.58E+13 | 800.117 | 0.01125 |
| 642.382 | 6.93E+10 | 800.398 | 9.64E+20 | 800.398 | 0.40853 | 800.398 | 4.02E-08 | 800.398 | 3.96E-08 | 800.398 | 1.58E+13 | 800.398 | 0.01129 |
| 642.6714 | 6.97E+10 | 800.68 | 9.65E+20 | 800.68 | 0.40848 | 800.68 | 4.02E-08 | 800.68 | 3.94E-08 | 800.68 | 1.58E+13 | 800.68 | 0.01124 |
| 642.9607 | 6.98E+10 | 800.962 | 9.66E+20 | 800.962 | 0.40843 | 800.962 | 4.02E-08 | 800.962 | 4.10E-08 | 800.962 | 1.59E+13 | 800.962 | 0.01121 |
| 643.25 | 6.96E+10 | 801.244 | 9.67E+20 | 801.244 | 0.40838 | 801.244 | 4.02E-08 | 801.244 | 4.11E-08 | 801.244 | 1.59E+13 | 801.244 | 0.01125 |
| 643.5393 | 6.98E+10 | 801.525 | 9.68E+20 | 801.525 | 0.40833 | 801.525 | 4.02E-08 | 801.525 | 4.04E-08 | 801.525 | 1.59E+13 | 801.525 | 0.01122 |
| 643.8286 | 7.02E+10 | 801.807 | 9.69E+20 | 801.807 | 0.40828 | 801.807 | 4.02E-08 | 801.807 | 4.02E-08 | 801.807 | 1.59E+13 | 801.807 | 0.01116 |
| 644.1179 | 7.02E+10 | 802.089 | 9.70E+20 | 802.089 | 0.40823 | 802.089 | 4.02E-08 | 802.089 | 4.04E-08 | 802.089 | 1.59E+13 | 802.089 | 0.01121 |
| 644.4072 | 7.00E+10 | 802.371 | 9.71E+20 | 802.371 | 0.40818 | 802.371 | 4.03E-08 | 802.371 | 3.95E-08 | 802.371 | 1.60E+13 | 802.371 | 0.01124 |
| 644.6965 | 7.03E+10 | 802.652 | 9.73E+20 | 802.652 | 0.40813 | 802.652 | 4.03E-08 | 802.652 | 4.02E-08 | 802.652 | 1.60E+13 | 802.652 | 0.01115 |
| 644.9857 | 7.08E+10 | 802.934 | 9.74E+20 | 802.934 | 0.40808 | 802.934 | 4.04E-08 | 802.934 | 4.17E-08 | 802.934 | 1.60E+13 | 802.934 | 0.01117 |
| 645.275 | 7.02E+10 | 803.216 | 9.75E+20 | 803.216 | 0.40803 | 803.216 | 4.05E-08 | 803.216 | 4.06E-08 | 803.216 | 1.61E+13 | 803.216 | 0.01113 |
| 645.5642 | 7.09E+10 | 803.497 | 9.76E+20 | 803.497 | 0.40797 | 803.497 | 4.05E-08 | 803.497 | 4.21E-08 | 803.497 | 1.61E+13 | 803.497 | 0.01116 |
| 645.8534 | 7.14E+10 | 803.779 | 9.77E+20 | 803.779 | 0.40792 | 803.779 | 4.06E-08 | 803.779 | 4.02E-08 | 803.779 | 1.62E+13 | 803.779 | 0.01112 |
| 646.1426 | 7.08E+10 | 804.06 | 9.78E+20 | 804.06 | 0.40787 | 804.06 | 4.06E-08 | 804.06 | 4.11E-08 | 804.06 | 1.62E+13 | 804.06 | 0.01114 |
| 646.4318 | 7.10E+10 | 804.342 | 9.79E+20 | 804.342 | 0.40782 | 804.342 | 4.06E-08 | 804.342 | 3.95E-08 | 804.342 | 1.62E+13 | 804.342 | 0.01112 |
| 646.721 | 7.11E+10 | 804.624 | 9.80E+20 | 804.624 | 0.40777 | 804.624 | 4.06E-08 | 804.624 | 4.05E-08 | 804.624 | 1.62E+13 | 804.624 | 0.01111 |
| 647.0101 | 7.09E+10 | 804.905 | 9.82E+20 | 804.905 | 0.40772 | 804.905 | 4.07E-08 | 804.905 | 4.19E-08 | 804.905 | 1.63E+13 | 804.905 | 0.01104 |
| 647.2993 | 7.14E+10 | 805.187 | 9.83E+20 | 805.187 | 0.40767 | 805.187 | 4.07E-08 | 805.187 | 3.99E-08 | 805.187 | 1.63E+13 | 805.187 | 0.01104 |
| 647.5884 | 7.14E+10 | 805.468 | 9.84E+20 | 805.468 | 0.40762 | 805.468 | 4.07E-08 | 805.468 | 4.06E-08 | 805.468 | 1.63E+13 | 805.468 | 0.01111 |
| 647.8776 | 7.15E+10 | 805.75 | 9.85E+20 | 805.75 | 0.40757 | 805.75 | 4.07E-08 | 805.75 | 4.17E-08 | 805.75 | 1.64E+13 | 805.75 | 0.01105 |
| 648.1667 | 7.16E+10 | 806.031 | 9.86E+20 | 806.031 | 0.40752 | 806.031 | 4.08E-08 | 806.031 | 4.07E-08 | 806.031 | 1.64E+13 | 806.031 | 0.011 |
| 648.4558 | 7.21E+10 | 806.313 | 9.87E+20 | 806.313 | 0.40747 | 806.313 | 4.08E-08 | 806.313 | 3.98E-08 | 806.313 | 1.64E+13 | 806.313 | 0.01107 |
| 648.7449 | 7.19E+10 | 806.594 | 9.88E+20 | 806.594 | 0.40742 | 806.594 | 4.08E-08 | 806.594 | 4.05E-08 | 806.594 | 1.64E+13 | 806.594 | 0.01098 |
| 649.034 | 7.20E+10 | 806.876 | 9.89E+20 | 806.876 | 0.40737 | 806.876 | 4.08E-08 | 806.876 | 4.02E-08 | 806.876 | 1.65E+13 | 806.876 | 0.01101 |
| 649.3231 | 7.28E+10 | 807.157 | 9.91E+20 | 807.157 | 0.40732 | 807.157 | 4.08E-08 | 807.157 | 4.00E-08 | 807.157 | 1.65E+13 | 807.157 | 0.01098 |
| 649.6121 | 7.22E+10 | 807.438 | 9.92E+20 | 807.438 | 0.40727 | 807.438 | 4.09E-08 | 807.438 | 3.94E-08 | 807.438 | 1.65E+13 | 807.438 | 0.01101 |
| 649.9012 | 7.26E+10 | 807.72 | 9.93E+20 | 807.72 | 0.40722 | 807.72 | 4.09E-08 | 807.72 | 4.11E-08 | 807.72 | 1.65E+13 | 807.72 | 0.01103 |
| 650.1902 | 7.29E+10 | 808.001 | 9.94E+20 | 808.001 | 0.40717 | 808.001 | 4.09E-08 | 808.001 | 4.20E-08 | 808.001 | 1.66E+13 | 808.001 | 0.01096 |
| 650.4792 | 7.30E+10 | 808.283 | 9.95E+20 | 808.283 | 0.40712 | 808.283 | 4.09E-08 | 808.283 | 4.10E-08 | 808.283 | 1.66E+13 | 808.283 | 0.01092 |
| 650.7683 | 7.32E+10 | 808.564 | 9.96E+20 | 808.564 | 0.40707 | 808.564 | 4.10E-08 | 808.564 | 4.15E-08 | 808.564 | 1.66E+13 | 808.564 | 0.01094 |
| 651.0573 | 7.29E+10 | 808.845 | 9.97E+20 | 808.845 | 0.40702 | 808.845 | 4.10E-08 | 808.845 | 4.32E-08 | 808.845 | 1.66E+13 | 808.845 | 0.01099 |
| 651.3463 | 7.29E+10 | 809.127 | 9.99E+20 | 809.127 | 0.40697 | 809.127 | 4.10E-08 | 809.127 | 4.16E-08 | 809.127 | 1.66E+13 | 809.127 | 0.01091 |
| 651.6352 | 7.31E+10 | 809.408 | 1.00E+21 | 809.408 | 0.40692 | 809.408 | 4.09E-08 | 809.408 | 4.07E-08 | 809.408 | 1.66E+13 | 809.408 | 0.01095 |
| 651.9242 | 7.27E+10 | 809.689 | 1.00E+21 | 809.689 | 0.40686 | 809.689 | 4.10E-08 | 809.689 | 4.15E-08 | 809.689 | 1.67E+13 | 809.689 | 0.01095 |
| 652.2132 | 7.36E+10 | 809.97 | 1.00E+21 | 809.97 | 0.40681 | 809.97 | 4.10E-08 | 809.97 | 3.97E-08 | 809.97 | 1.67E+13 | 809.97 | 0.01097 |
| 652.5021 | 7.36E+10 | 810.252 | 1.00E+21 | 810.252 | 0.40676 | 810.252 | 4.10E-08 | 810.252 | 4.15E-08 | 810.252 | 1.67E+13 | 810.252 | 0.01096 |
| 652.791 | 7.35E+10 | 810.533 | 1.00E+21 | 810.533 | 0.40671 | 810.533 | 4.10E-08 | 810.533 | 4.18E-08 | 810.533 | 1.67E+13 | 810.533 | 0.01095 |
| 653.0799 | 7.39E+10 | 810.814 | 1.01E+21 | 810.814 | 0.40666 | 810.814 | 4.10E-08 | 810.814 | 4.24E-08 | 810.814 | 1.67E+13 | 810.814 | 0.011 |
| 653.3688 | 7.39E+10 | 811.095 | 1.01E+21 | 811.095 | 0.40661 | 811.095 | 4.10E-08 | 811.095 | 3.99E-08 | 811.095 | 1.68E+13 | 811.095 | 0.01091 |
| 653.6577 | 7.43E+10 | 811.376 | 1.01E+21 | 811.376 | 0.40656 | 811.376 | 4.10E-08 | 811.376 | 4.04E-08 | 811.376 | 1.68E+13 | 811.376 | 0.01096 |
| 653.9466 | 7.44E+10 | 811.658 | 1.01E+21 | 811.658 | 0.40651 | 811.658 | 4.10E-08 | 811.658 | 4.10E-08 | 811.658 | 1.68E+13 | 811.658 | 0.01089 |
| 654.2355 | 7.46E+10 | 811.939 | 1.01E+21 | 811.939 | 0.40646 | 811.939 | 4.10E-08 | 811.939 | 3.98E-08 | 811.939 | 1.68E+13 | 811.939 | 0.01089 |
| 654.5244 | 7.46E+10 | 812.22 | 1.01E+21 | 812.22 | 0.40641 | 812.22 | 4.11E-08 | 812.22 | 4.19E-08 | 812.22 | 1.69E+13 | 812.22 | 0.01093 |
| 654.8132 | 7.47E+10 | 812.501 | 1.01E+21 | 812.501 | 0.40636 | 812.501 | 4.11E-08 | 812.501 | 4.10E-08 | 812.501 | 1.69E+13 | 812.501 | 0.0109 |
| 655.102 | 7.53E+10 | 812.782 | 1.01E+21 | 812.782 | 0.40631 | 812.782 | 4.12E-08 | 812.782 | 4.14E-08 | 812.782 | 1.69E+13 | 812.782 | 0.01084 |
| 655.3909 | 7.47E+10 | 813.063 | 1.01E+21 | 813.063 | 0.40626 | 813.063 | 4.12E-08 | 813.063 | 4.11E-08 | 813.063 | 1.70E+13 | 813.063 | 0.01091 |
| 655.6797 | 7.57E+10 | 813.344 | 1.02E+21 | 813.344 | 0.40621 | 813.344 | 4.13E-08 | 813.344 | 4.05E-08 | 813.344 | 1.70E+13 | 813.344 | 0.01082 |
| 655.9685 | 7.56E+10 | 813.625 | 1.02E+21 | 813.625 | 0.40616 | 813.625 | 4.13E-08 | 813.625 | 4.26E-08 | 813.625 | 1.70E+13 | 813.625 | 0.01089 |
| 656.2572 | 7.57E+10 | 813.906 | 1.02E+21 | 813.906 | 0.4061 | 813.906 | 4.13E-08 | 813.906 | 4.05E-08 | 813.906 | 1.71E+13 | 813.906 | 0.01085 |
| 656.546 | 7.53E+10 | 814.187 | 1.02E+21 | 814.187 | 0.40606 | 814.187 | 4.14E-08 | 814.187 | 4.20E-08 | 814.187 | 1.71E+13 | 814.187 | 0.01075 |
| 656.8348 | 7.57E+10 | 814.469 | 1.02E+21 | 814.469 | 0.406 | 814.469 | 4.14E-08 | 814.469 | 4.16E-08 | 814.469 | 1.71E+13 | 814.469 | 0.01077 |
| 657.1235 | 7.62E+10 | 814.75 | 1.02E+21 | 814.75 | 0.40595 | 814.75 | 4.14E-08 | 814.75 | 4.02E-08 | 814.75 | 1.72E+13 | 814.75 | 0.01078 |
| 657.4123 | 7.65E+10 | 815.03 | 1.02E+21 | 815.03 | 0.4059 | 815.03 | 4.14E-08 | 815.03 | 4.02E-08 | 815.03 | 1.72E+13 | 815.03 | 0.01074 |
| 657.701 | 7.62E+10 | 815.311 | 1.02E+21 | 815.311 | 0.40585 | 815.311 | 4.14E-08 | 815.311 | 4.24E-08 | 815.311 | 1.72E+13 | 815.311 | 0.01075 |
| 657.9897 | 7.66E+10 | 815.592 | 1.02E+21 | 815.592 | 0.4058 | 815.592 | 4.15E-08 | 815.592 | 4.18E-08 | 815.592 | 1.72E+13 | 815.592 | 0.01071 |
| 658.2784 | 7.71E+10 | 815.873 | 1.03E+21 | 815.873 | 0.40575 | 815.873 | 4.16E-08 | 815.873 | 4.07E-08 | 815.873 | 1.73E+13 | 815.873 | 0.01074 |
| 658.5671 | 7.70E+10 | 816.154 | 1.03E+21 | 816.154 | 0.4057 | 816.154 | 4.16E-08 | 816.154 | 3.97E-08 | 816.154 | 1.73E+13 | 816.154 | 0.01067 |
| 658.8558 | 7.71E+10 | 816.435 | 1.03E+21 | 816.435 | 0.40565 | 816.435 | 4.16E-08 | 816.435 | 4.04E-08 | 816.435 | 1.73E+13 | 816.435 | 0.01071 |
| 659.1444 | 7.74E+10 | 816.716 | 1.03E+21 | 816.716 | 0.4056 | 816.716 | 4.16E-08 | 816.716 | 4.15E-08 | 816.716 | 1.74E+13 | 816.716 | 0.01071 |
| 659.4331 | 7.81E+10 | 816.997 | 1.03E+21 | 816.997 | 0.40555 | 816.997 | 4.17E-08 | 816.997 | 4.00E-08 | 816.997 | 1.74E+13 | 816.997 | 0.01066 |
| 659.7217 | 7.80E+10 | 817.278 | 1.03E+21 | 817.278 | 0.4055 | 817.278 | 4.17E-08 | 817.278 | 4.22E-08 | 817.278 | 1.74E+13 | 817.278 | 0.01071 |
| 660.0103 | 7.80E+10 | 817.559 | 1.03E+21 | 817.559 | 0.40545 | 817.559 | 4.17E-08 | 817.559 | 4.14E-08 | 817.559 | 1.75E+13 | 817.559 | 0.01065 |
| 660.299 | 7.84E+10 | 817.839 | 1.03E+21 | 817.839 | 0.4054 | 817.839 | 4.18E-08 | 817.839 | 4.08E-08 | 817.839 | 1.75E+13 | 817.839 | 0.01067 |
| 660.5876 | 7.81E+10 | 818.12 | 1.03E+21 | 818.12 | 0.40535 | 818.12 | 4.18E-08 | 818.12 | 4.26E-08 | 818.12 | 1.75E+13 | 818.12 | 0.01069 |
| 660.8762 | 7.87E+10 | 818.401 | 1.04E+21 | 818.401 | 0.40529 | 818.401 | 4.18E-08 | 818.401 | 4.28E-08 | 818.401 | 1.76E+13 | 818.401 | 0.01072 |
| 661.1647 | 7.86E+10 | 818.682 | 1.04E+21 | 818.682 | 0.40524 | 818.682 | 4.18E-08 | 818.682 | 4.14E-08 | 818.682 | 1.76E+13 | 818.682 | 0.01068 |
| 661.4533 | 7.92E+10 | 818.963 | 1.04E+21 | 818.963 | 0.40519 | 818.963 | 4.19E-08 | 818.963 | 4.26E-08 | 818.963 | 1.76E+13 | 818.963 | 0.01062 |
| 661.7419 | 7.89E+10 | 819.243 | 1.04E+21 | 819.243 | 0.40514 | 819.243 | 4.20E-08 | 819.243 | 4.15E-08 | 819.243 | 1.77E+13 | 819.243 | 0.01058 |
| 662.0304 | 7.90E+10 | 819.524 | 1.04E+21 | 819.524 | 0.40509 | 819.524 | 4.20E-08 | 819.524 | 4.17E-08 | 819.524 | 1.77E+13 | 819.524 | 0.01058 |
| 662.3189 | 7.98E+10 | 819.805 | 1.04E+21 | 819.805 | 0.40504 | 819.805 | 4.20E-08 | 819.805 | 4.40E-08 | 819.805 | 1.77E+13 | 819.805 | 0.01059 |
| 662.6074 | 7.97E+10 | 820.085 | 1.04E+21 | 820.085 | 0.40499 | 820.085 | 4.20E-08 | 820.085 | 4.32E-08 | 820.085 | 1.77E+13 | 820.085 | 0.01059 |
| 662.8959 | 8.05E+10 | 820.366 | 1.04E+21 | 820.366 | 0.40494 | 820.366 | 4.20E-08 | 820.366 | 4.24E-08 | 820.366 | 1.78E+13 | 820.366 | 0.0106 |
| 663.1844 | 8.04E+10 | 820.647 | 1.04E+21 | 820.647 | 0.40489 | 820.647 | 4.21E-08 | 820.647 | 4.24E-08 | 820.647 | 1.78E+13 | 820.647 | 0.01054 |
| 663.4729 | 8.07E+10 | 820.927 | 1.05E+21 | 820.927 | 0.40484 | 820.927 | 4.22E-08 | 820.927 | 4.21E-08 | 820.927 | 1.79E+13 | 820.927 | 0.01052 |
| 663.7614 | 8.07E+10 | 821.208 | 1.05E+21 | 821.208 | 0.40479 | 821.208 | 4.22E-08 | 821.208 | 4.39E-08 | 821.208 | 1.79E+13 | 821.208 | 0.01056 |
| 664.0498 | 8.07E+10 | 821.489 | 1.05E+21 | 821.489 | 0.40474 | 821.489 | 4.22E-08 | 821.489 | 4.28E-08 | 821.489 | 1.79E+13 | 821.489 | 0.01054 |
| 664.3383 | 8.12E+10 | 821.769 | 1.05E+21 | 821.769 | 0.40468 | 821.769 | 4.23E-08 | 821.769 | 4.14E-08 | 821.769 | 1.79E+13 | 821.769 | 0.0105 |
| 664.6267 | 8.13E+10 | 822.05 | 1.05E+21 | 822.05 | 0.40463 | 822.05 | 4.23E-08 | 822.05 | 4.17E-08 | 822.05 | 1.80E+13 | 822.05 | 0.01047 |
| 664.9151 | 8.20E+10 | 822.331 | 1.05E+21 | 822.331 | 0.40458 | 822.331 | 4.24E-08 | 822.331 | 4.35E-08 | 822.331 | 1.80E+13 | 822.331 | 0.01045 |
| 665.2035 | 8.14E+10 | 822.611 | 1.05E+21 | 822.611 | 0.40453 | 822.611 | 4.24E-08 | 822.611 | 4.27E-08 | 822.611 | 1.81E+13 | 822.611 | 0.01044 |
| 665.4919 | 8.21E+10 | 822.892 | 1.05E+21 | 822.892 | 0.40448 | 822.892 | 4.25E-08 | 822.892 | 4.27E-08 | 822.892 | 1.81E+13 | 822.892 | 0.01048 |
| 665.7803 | 8.20E+10 | 823.172 | 1.05E+21 | 823.172 | 0.40443 | 823.172 | 4.25E-08 | 823.172 | 4.17E-08 | 823.172 | 1.81E+13 | 823.172 | 0.0104 |
| 666.0687 | 8.19E+10 | 823.453 | 1.06E+21 | 823.453 | 0.40438 | 823.453 | 4.26E-08 | 823.453 | 4.20E-08 | 823.453 | 1.82E+13 | 823.453 | 0.01042 |
| 666.357 | 8.21E+10 | 823.733 | 1.06E+21 | 823.733 | 0.40433 | 823.733 | 4.26E-08 | 823.733 | 4.24E-08 | 823.733 | 1.82E+13 | 823.733 | 0.01032 |
| 666.6454 | 8.25E+10 | 824.014 | 1.06E+21 | 824.014 | 0.40428 | 824.014 | 4.27E-08 | 824.014 | 4.23E-08 | 824.014 | 1.83E+13 | 824.014 | 0.01038 |
| 666.9337 | 8.28E+10 | 824.294 | 1.06E+21 | 824.294 | 0.40423 | 824.294 | 4.26E-08 | 824.294 | 4.19E-08 | 824.294 | 1.83E+13 | 824.294 | 0.01038 |
| 667.222 | 8.37E+10 | 824.574 | 1.06E+21 | 824.574 | 0.40418 | 824.574 | 4.26E-08 | 824.574 | 4.27E-08 | 824.574 | 1.83E+13 | 824.574 | 0.01037 |
| 667.5104 | 8.30E+10 | 824.855 | 1.06E+21 | 824.855 | 0.40413 | 824.855 | 4.27E-08 | 824.855 | 4.41E-08 | 824.855 | 1.83E+13 | 824.855 | 0.01037 |
| 667.7986 | 8.41E+10 | 825.135 | 1.06E+21 | 825.135 | 0.40407 | 825.135 | 4.27E-08 | 825.135 | 4.24E-08 | 825.135 | 1.83E+13 | 825.135 | 0.01034 |
| 668.0869 | 8.44E+10 | 825.416 | 1.06E+21 | 825.416 | 0.40402 | 825.416 | 4.28E-08 | 825.416 | 4.18E-08 | 825.416 | 1.84E+13 | 825.416 | 0.01032 |
| 668.3752 | 8.44E+10 | 825.696 | 1.06E+21 | 825.696 | 0.40397 | 825.696 | 4.28E-08 | 825.696 | 4.32E-08 | 825.696 | 1.84E+13 | 825.696 | 0.01029 |
| 668.6635 | 8.37E+10 | 825.976 | 1.07E+21 | 825.976 | 0.40392 | 825.976 | 4.28E-08 | 825.976 | 4.26E-08 | 825.976 | 1.84E+13 | 825.976 | 0.01031 |
| 668.9517 | 8.49E+10 | 826.257 | 1.07E+21 | 826.257 | 0.40387 | 826.257 | 4.28E-08 | 826.257 | 4.31E-08 | 826.257 | 1.85E+13 | 826.257 | 0.0103 |
| 669.2399 | 8.46E+10 | 826.537 | 1.07E+21 | 826.537 | 0.40382 | 826.537 | 4.28E-08 | 826.537 | 4.25E-08 | 826.537 | 1.85E+13 | 826.537 | 0.01031 |
| 669.5282 | 8.57E+10 | 826.817 | 1.07E+21 | 826.817 | 0.40377 | 826.817 | 4.28E-08 | 826.817 | 4.38E-08 | 826.817 | 1.85E+13 | 826.817 | 0.01027 |
| 669.8164 | 8.47E+10 | 827.098 | 1.07E+21 | 827.098 | 0.40372 | 827.098 | 4.28E-08 | 827.098 | 4.09E-08 | 827.098 | 1.85E+13 | 827.098 | 0.01026 |
| 670.1046 | 8.56E+10 | 827.378 | 1.07E+21 | 827.378 | 0.40367 | 827.378 | 4.28E-08 | 827.378 | 4.35E-08 | 827.378 | 1.85E+13 | 827.378 | 0.0103 |
| 670.3927 | 8.54E+10 | 827.658 | 1.07E+21 | 827.658 | 0.40362 | 827.658 | 4.29E-08 | 827.658 | 4.19E-08 | 827.658 | 1.86E+13 | 827.658 | 0.01029 |
| 670.6809 | 8.64E+10 | 827.939 | 1.07E+21 | 827.939 | 0.40356 | 827.939 | 4.29E-08 | 827.939 | 4.37E-08 | 827.939 | 1.86E+13 | 827.939 | 0.01027 |
| 670.9691 | 8.55E+10 | 828.219 | 1.08E+21 | 828.219 | 0.40351 | 828.219 | 4.29E-08 | 828.219 | 4.37E-08 | 828.219 | 1.86E+13 | 828.219 | 0.01025 |
| 671.2572 | 8.70E+10 | 828.499 | 1.08E+21 | 828.499 | 0.40346 | 828.499 | 4.29E-08 | 828.499 | 4.31E-08 | 828.499 | 1.86E+13 | 828.499 | 0.01022 |
| 671.5454 | 8.64E+10 | 828.779 | 1.08E+21 | 828.779 | 0.40341 | 828.779 | 4.30E-08 | 828.779 | 4.19E-08 | 828.779 | 1.87E+13 | 828.779 | 0.01014 |
| 671.8335 | 8.69E+10 | 829.059 | 1.08E+21 | 829.059 | 0.40336 | 829.059 | 4.30E-08 | 829.059 | 4.41E-08 | 829.059 | 1.87E+13 | 829.059 | 0.0102 |
| 672.1216 | 8.73E+10 | 829.34 | 1.08E+21 | 829.34 | 0.40331 | 829.34 | 4.30E-08 | 829.34 | 4.38E-08 | 829.34 | 1.87E+13 | 829.34 | 0.01016 |
| 672.4097 | 8.68E+10 | 829.62 | 1.08E+21 | 829.62 | 0.40326 | 829.62 | 4.31E-08 | 829.62 | 4.31E-08 | 829.62 | 1.88E+13 | 829.62 | 0.0102 |
| 672.6978 | 8.72E+10 | 829.9 | 1.08E+21 | 829.9 | 0.40321 | 829.9 | 4.31E-08 | 829.9 | 4.19E-08 | 829.9 | 1.88E+13 | 829.9 | 0.01018 |
| 672.9858 | 8.73E+10 | 830.18 | 1.08E+21 | 830.18 | 0.40316 | 830.18 | 4.31E-08 | 830.18 | 4.17E-08 | 830.18 | 1.88E+13 | 830.18 | 0.01016 |
| 673.2739 | 8.80E+10 | 830.46 | 1.08E+21 | 830.46 | 0.4031 | 830.46 | 4.32E-08 | 830.46 | 4.30E-08 | 830.46 | 1.89E+13 | 830.46 | 0.01012 |
| 673.5619 | 8.77E+10 | 830.74 | 1.09E+21 | 830.74 | 0.40305 | 830.74 | 4.32E-08 | 830.74 | 4.36E-08 | 830.74 | 1.89E+13 | 830.74 | 0.0101 |
| 673.85 | 8.83E+10 | 831.02 | 1.09E+21 | 831.02 | 0.403 | 831.02 | 4.32E-08 | 831.02 | 4.41E-08 | 831.02 | 1.89E+13 | 831.02 | 0.01014 |
| 674.138 | 8.86E+10 | 831.3 | 1.09E+21 | 831.3 | 0.40295 | 831.3 | 4.33E-08 | 831.3 | 4.49E-08 | 831.3 | 1.90E+13 | 831.3 | 0.01001 |
| 674.426 | 8.86E+10 | 831.58 | 1.09E+21 | 831.58 | 0.4029 | 831.58 | 4.33E-08 | 831.58 | 4.32E-08 | 831.58 | 1.90E+13 | 831.58 | 0.01007 |
| 674.714 | 8.90E+10 | 831.86 | 1.09E+21 | 831.86 | 0.40285 | 831.86 | 4.33E-08 | 831.86 | 4.38E-08 | 831.86 | 1.90E+13 | 831.86 | 0.01008 |
| 675.002 | 8.89E+10 | 832.14 | 1.09E+21 | 832.14 | 0.4028 | 832.14 | 4.33E-08 | 832.14 | 4.33E-08 | 832.14 | 1.90E+13 | 832.14 | 0.01011 |
| 675.29 | 8.99E+10 | 832.42 | 1.09E+21 | 832.42 | 0.40275 | 832.42 | 4.33E-08 | 832.42 | 4.18E-08 | 832.42 | 1.90E+13 | 832.42 | 0.01008 |
| 675.5779 | 8.89E+10 | 832.7 | 1.09E+21 | 832.7 | 0.4027 | 832.7 | 4.33E-08 | 832.7 | 4.37E-08 | 832.7 | 1.91E+13 | 832.7 | 0.01013 |
| 675.8659 | 8.97E+10 | 832.98 | 1.09E+21 | 832.98 | 0.40264 | 832.98 | 4.34E-08 | 832.98 | 4.15E-08 | 832.98 | 1.91E+13 | 832.98 | 0.01008 |
| 676.1538 | 8.94E+10 | 833.26 | 1.10E+21 | 833.26 | 0.40259 | 833.26 | 4.34E-08 | 833.26 | 4.43E-08 | 833.26 | 1.91E+13 | 833.26 | 0.01006 |
| 676.4417 | 8.95E+10 | 833.54 | 1.10E+21 | 833.54 | 0.40254 | 833.54 | 4.35E-08 | 833.54 | 4.31E-08 | 833.54 | 1.92E+13 | 833.54 | 0.01001 |
| 676.7296 | 9.03E+10 | 833.82 | 1.10E+21 | 833.82 | 0.40249 | 833.82 | 4.34E-08 | 833.82 | 4.43E-08 | 833.82 | 1.92E+13 | 833.82 | 0.01002 |
| 677.0175 | 8.97E+10 | 834.1 | 1.10E+21 | 834.1 | 0.40244 | 834.1 | 4.34E-08 | 834.1 | 4.37E-08 | 834.1 | 1.92E+13 | 834.1 | 0.01006 |
| 677.3054 | 9.03E+10 | 834.38 | 1.10E+21 | 834.38 | 0.40239 | 834.38 | 4.34E-08 | 834.38 | 4.40E-08 | 834.38 | 1.92E+13 | 834.38 | 0.01004 |
| 677.5933 | 9.06E+10 | 834.66 | 1.10E+21 | 834.66 | 0.40234 | 834.66 | 4.35E-08 | 834.66 | 4.27E-08 | 834.66 | 1.92E+13 | 834.66 | 0.01007 |
| 677.8811 | 9.05E+10 | 834.94 | 1.10E+21 | 834.94 | 0.40229 | 834.94 | 4.35E-08 | 834.94 | 4.36E-08 | 834.94 | 1.93E+13 | 834.94 | 0.01007 |
| 678.169 | 9.07E+10 | 835.219 | 1.10E+21 | 835.219 | 0.40224 | 835.219 | 4.35E-08 | 835.219 | 4.35E-08 | 835.219 | 1.93E+13 | 835.219 | 0.01001 |
| 678.4568 | 9.14E+10 | 835.499 | 1.10E+21 | 835.499 | 0.40218 | 835.499 | 4.35E-08 | 835.499 | 4.25E-08 | 835.499 | 1.93E+13 | 835.499 | 0.01003 |
| 678.7446 | 9.14E+10 | 835.779 | 1.11E+21 | 835.779 | 0.40213 | 835.779 | 4.35E-08 | 835.779 | 4.36E-08 | 835.779 | 1.94E+13 | 835.779 | 0.01002 |
| 679.0324 | 9.06E+10 | 836.059 | 1.11E+21 | 836.059 | 0.40208 | 836.059 | 4.36E-08 | 836.059 | 4.47E-08 | 836.059 | 1.94E+13 | 836.059 | 0.00997 |
| 679.3202 | 9.16E+10 | 836.339 | 1.11E+21 | 836.339 | 0.40203 | 836.339 | 4.36E-08 | 836.339 | 4.35E-08 | 836.339 | 1.94E+13 | 836.339 | 0.00996 |
| 679.608 | 9.15E+10 | 836.618 | 1.11E+21 | 836.618 | 0.40198 | 836.618 | 4.36E-08 | 836.618 | 4.54E-08 | 836.618 | 1.94E+13 | 836.618 | 0.01001 |
| 679.8958 | 9.18E+10 | 836.898 | 1.11E+21 | 836.898 | 0.40193 | 836.898 | 4.37E-08 | 836.898 | 4.41E-08 | 836.898 | 1.95E+13 | 836.898 | 0.00991 |
| 680.1835 | 9.26E+10 | 837.178 | 1.11E+21 | 837.178 | 0.40188 | 837.178 | 4.37E-08 | 837.178 | 4.35E-08 | 837.178 | 1.95E+13 | 837.178 | 0.00994 |
| 680.4713 | 9.20E+10 | 837.457 | 1.11E+21 | 837.457 | 0.40183 | 837.457 | 4.37E-08 | 837.457 | 4.23E-08 | 837.457 | 1.95E+13 | 837.457 | 0.00994 |
| 680.759 | 9.29E+10 | 837.737 | 1.11E+21 | 837.737 | 0.40177 | 837.737 | 4.37E-08 | 837.737 | 4.26E-08 | 837.737 | 1.95E+13 | 837.737 | 0.00985 |
| 681.0467 | 9.29E+10 | 838.017 | 1.11E+21 | 838.017 | 0.40172 | 838.017 | 4.37E-08 | 838.017 | 4.32E-08 | 838.017 | 1.96E+13 | 838.017 | 0.00988 |
| 681.3344 | 9.32E+10 | 838.296 | 1.12E+21 | 838.296 | 0.40167 | 838.296 | 4.38E-08 | 838.296 | 4.48E-08 | 838.296 | 1.96E+13 | 838.296 | 0.00984 |
| 681.6221 | 9.37E+10 | 838.576 | 1.12E+21 | 838.576 | 0.40162 | 838.576 | 4.38E-08 | 838.576 | 4.36E-08 | 838.576 | 1.97E+13 | 838.576 | 0.00986 |
| 681.9098 | 9.30E+10 | 838.856 | 1.12E+21 | 838.856 | 0.40157 | 838.856 | 4.39E-08 | 838.856 | 4.47E-08 | 838.856 | 1.97E+13 | 838.856 | 0.00979 |
| 682.1975 | 9.30E+10 | 839.135 | 1.12E+21 | 839.135 | 0.40152 | 839.135 | 4.40E-08 | 839.135 | 4.39E-08 | 839.135 | 1.98E+13 | 839.135 | 0.00985 |
| 682.4851 | 9.32E+10 | 839.415 | 1.12E+21 | 839.415 | 0.40147 | 839.415 | 4.40E-08 | 839.415 | 4.31E-08 | 839.415 | 1.98E+13 | 839.415 | 0.00981 |
| 682.7728 | 9.35E+10 | 839.694 | 1.12E+21 | 839.694 | 0.40142 | 839.694 | 4.41E-08 | 839.694 | 4.34E-08 | 839.694 | 1.98E+13 | 839.694 | 0.0098 |
| 683.0604 | 9.32E+10 | 839.974 | 1.12E+21 | 839.974 | 0.40136 | 839.974 | 4.41E-08 | 839.974 | 4.37E-08 | 839.974 | 1.98E+13 | 839.974 | 0.00978 |
| 683.348 | 9.41E+10 | 840.253 | 1.12E+21 | 840.253 | 0.40131 | 840.253 | 4.41E-08 | 840.253 | 4.25E-08 | 840.253 | 1.99E+13 | 840.253 | 0.00974 |
| 683.6356 | 9.36E+10 | 840.533 | 1.12E+21 | 840.533 | 0.40126 | 840.533 | 4.42E-08 | 840.533 | 4.44E-08 | 840.533 | 1.99E+13 | 840.533 | 0.00979 |
| 683.9232 | 9.44E+10 | 840.812 | 1.13E+21 | 840.812 | 0.40121 | 840.812 | 4.42E-08 | 840.812 | 4.37E-08 | 840.812 | 1.99E+13 | 840.812 | 0.00975 |
| 684.2108 | 9.39E+10 | 841.092 | 1.13E+21 | 841.092 | 0.40116 | 841.092 | 4.42E-08 | 841.092 | 4.48E-08 | 841.092 | 2.00E+13 | 841.092 | 0.00972 |
| 684.4983 | 9.49E+10 | 841.371 | 1.13E+21 | 841.371 | 0.40111 | 841.371 | 4.43E-08 | 841.371 | 4.31E-08 | 841.371 | 2.00E+13 | 841.371 | 0.00972 |
| 684.7859 | 9.51E+10 | 841.651 | 1.13E+21 | 841.651 | 0.40106 | 841.651 | 4.43E-08 | 841.651 | 4.51E-08 | 841.651 | 2.01E+13 | 841.651 | 0.00971 |
| 685.0734 | 9.46E+10 | 841.93 | 1.13E+21 | 841.93 | 0.40101 | 841.93 | 4.44E-08 | 841.93 | 4.39E-08 | 841.93 | 2.01E+13 | 841.93 | 0.00973 |
| 685.361 | 9.56E+10 | 842.21 | 1.13E+21 | 842.21 | 0.40095 | 842.21 | 4.44E-08 | 842.21 | 4.30E-08 | 842.21 | 2.01E+13 | 842.21 | 0.00961 |
| 685.6485 | 9.53E+10 | 842.489 | 1.13E+21 | 842.489 | 0.4009 | 842.489 | 4.44E-08 | 842.489 | 4.56E-08 | 842.489 | 2.01E+13 | 842.489 | 0.00967 |
| 685.936 | 9.55E+10 | 842.769 | 1.13E+21 | 842.769 | 0.40085 | 842.769 | 4.44E-08 | 842.769 | 4.50E-08 | 842.769 | 2.02E+13 | 842.769 | 0.00969 |
| 686.2235 | 9.63E+10 | 843.048 | 1.13E+21 | 843.048 | 0.4008 | 843.048 | 4.44E-08 | 843.048 | 4.42E-08 | 843.048 | 2.02E+13 | 843.048 | 0.00967 |
| 686.5109 | 9.62E+10 | 843.327 | 1.14E+21 | 843.327 | 0.40075 | 843.327 | 4.44E-08 | 843.327 | 4.47E-08 | 843.327 | 2.02E+13 | 843.327 | 0.00964 |
| 686.7984 | 9.55E+10 | 843.607 | 1.14E+21 | 843.607 | 0.4007 | 843.607 | 4.45E-08 | 843.607 | 4.35E-08 | 843.607 | 2.03E+13 | 843.607 | 0.00963 |
| 687.0859 | 9.64E+10 | 843.886 | 1.14E+21 | 843.886 | 0.40064 | 843.886 | 4.45E-08 | 843.886 | 4.53E-08 | 843.886 | 2.03E+13 | 843.886 | 0.00961 |
| 687.3733 | 9.69E+10 | 844.165 | 1.14E+21 | 844.165 | 0.40059 | 844.165 | 4.45E-08 | 844.165 | 4.57E-08 | 844.165 | 2.03E+13 | 844.165 | 0.00957 |
| 687.6607 | 9.67E+10 | 844.445 | 1.14E+21 | 844.445 | 0.40054 | 844.445 | 4.46E-08 | 844.445 | 4.49E-08 | 844.445 | 2.04E+13 | 844.445 | 0.00958 |
| 687.9481 | 9.76E+10 | 844.724 | 1.14E+21 | 844.724 | 0.40049 | 844.724 | 4.46E-08 | 844.724 | 4.70E-08 | 844.724 | 2.04E+13 | 844.724 | 0.00957 |
| 688.2355 | 9.72E+10 | 845.003 | 1.14E+21 | 845.003 | 0.40044 | 845.003 | 4.47E-08 | 845.003 | 4.48E-08 | 845.003 | 2.04E+13 | 845.003 | 0.00956 |
| 688.5229 | 9.75E+10 | 845.282 | 1.14E+21 | 845.282 | 0.40039 | 845.282 | 4.47E-08 | 845.282 | 4.49E-08 | 845.282 | 2.05E+13 | 845.282 | 0.00953 |
| 688.8103 | 9.71E+10 | 845.562 | 1.14E+21 | 845.562 | 0.40034 | 845.562 | 4.47E-08 | 845.562 | 4.44E-08 | 845.562 | 2.05E+13 | 845.562 | 0.00953 |
| 689.0976 | 9.77E+10 | 845.841 | 1.15E+21 | 845.841 | 0.40029 | 845.841 | 4.48E-08 | 845.841 | 4.62E-08 | 845.841 | 2.05E+13 | 845.841 | 0.00946 |
| 689.385 | 9.73E+10 | 846.12 | 1.15E+21 | 846.12 | 0.40023 | 846.12 | 4.48E-08 | 846.12 | 4.41E-08 | 846.12 | 2.06E+13 | 846.12 | 0.0095 |
| 689.6723 | 9.80E+10 | 846.399 | 1.15E+21 | 846.399 | 0.40018 | 846.399 | 4.49E-08 | 846.399 | 4.41E-08 | 846.399 | 2.06E+13 | 846.399 | 0.00947 |
| 689.9596 | 9.75E+10 | 846.678 | 1.15E+21 | 846.678 | 0.40013 | 846.678 | 4.49E-08 | 846.678 | 4.49E-08 | 846.678 | 2.06E+13 | 846.678 | 0.00945 |
| 690.2469 | 9.85E+10 | 846.957 | 1.15E+21 | 846.957 | 0.40008 | 846.957 | 4.49E-08 | 846.957 | 4.50E-08 | 846.957 | 2.06E+13 | 846.957 | 0.00949 |
| 690.5342 | 9.86E+10 | 847.236 | 1.15E+21 | 847.236 | 0.40003 | 847.236 | 4.50E-08 | 847.236 | 4.57E-08 | 847.236 | 2.07E+13 | 847.236 | 0.0094 |
| 690.8215 | 9.85E+10 | 847.516 | 1.15E+21 | 847.516 | 0.39998 | 847.516 | 4.50E-08 | 847.516 | 4.73E-08 | 847.516 | 2.07E+13 | 847.516 | 0.0094 |
| 691.1088 | 9.85E+10 | 847.795 | 1.15E+21 | 847.795 | 0.39992 | 847.795 | 4.50E-08 | 847.795 | 4.53E-08 | 847.795 | 2.08E+13 | 847.795 | 0.00941 |
| 691.396 | 9.92E+10 | 848.074 | 1.15E+21 | 848.074 | 0.39987 | 848.074 | 4.51E-08 | 848.074 | 4.38E-08 | 848.074 | 2.08E+13 | 848.074 | 0.00938 |
| 691.6833 | 9.99E+10 | 848.353 | 1.16E+21 | 848.353 | 0.39982 | 848.353 | 4.50E-08 | 848.353 | 4.33E-08 | 848.353 | 2.08E+13 | 848.353 | 0.00943 |
| 691.9705 | 9.88E+10 | 848.632 | 1.16E+21 | 848.632 | 0.39977 | 848.632 | 4.50E-08 | 848.632 | 4.30E-08 | 848.632 | 2.08E+13 | 848.632 | 0.00945 |
| 692.2577 | 9.88E+10 | 848.911 | 1.16E+21 | 848.911 | 0.39972 | 848.911 | 4.50E-08 | 848.911 | 4.45E-08 | 848.911 | 2.08E+13 | 848.911 | 0.00937 |
| 692.5449 | 1.00E+11 | 849.19 | 1.16E+21 | 849.19 | 0.39967 | 849.19 | 4.50E-08 | 849.19 | 4.59E-08 | 849.19 | 2.09E+13 | 849.19 | 0.00942 |
| 692.8321 | 9.98E+10 | 849.469 | 1.16E+21 | 849.469 | 0.39962 | 849.469 | 4.51E-08 | 849.469 | 4.50E-08 | 849.469 | 2.09E+13 | 849.469 | 0.00934 |
| 693.1193 | 9.93E+10 | 849.748 | 1.16E+21 | 849.748 | 0.39956 | 849.748 | 4.51E-08 | 849.748 | 4.53E-08 | 849.748 | 2.09E+13 | 849.748 | 0.00937 |
| 693.4065 | 1.00E+11 | 850.027 | 1.16E+21 | 850.027 | 0.39951 | 850.027 | 4.51E-08 | 850.027 | 4.59E-08 | 850.027 | 2.09E+13 | 850.027 | 0.0094 |
| 693.6936 | 1.01E+11 | 850.306 | 1.16E+21 | 850.306 | 0.39946 | 850.306 | 4.51E-08 | 850.306 | 4.65E-08 | 850.306 | 2.09E+13 | 850.306 | 0.00934 |
| 693.9808 | 1.01E+11 | 850.585 | 1.16E+21 | 850.585 | 0.39941 | 850.585 | 4.50E-08 | 850.585 | 4.47E-08 | 850.585 | 2.09E+13 | 850.585 | 0.00941 |
| 694.2679 | 1.02E+11 | 850.864 | 1.17E+21 | 850.864 | 0.39936 | 850.864 | 4.51E-08 | 850.864 | 4.41E-08 | 850.864 | 2.10E+13 | 850.864 | 0.00936 |
| 694.555 | 1.01E+11 | 851.142 | 1.17E+21 | 851.142 | 0.39931 | 851.142 | 4.50E-08 | 851.142 | 4.54E-08 | 851.142 | 2.10E+13 | 851.142 | 0.00935 |
| 694.8421 | 1.01E+11 | 851.421 | 1.17E+21 | 851.421 | 0.39926 | 851.421 | 4.51E-08 | 851.421 | 4.52E-08 | 851.421 | 2.10E+13 | 851.421 | 0.0093 |
| 695.1292 | 1.02E+11 | 851.7 | 1.17E+21 | 851.7 | 0.3992 | 851.7 | 4.51E-08 | 851.7 | 4.52E-08 | 851.7 | 2.10E+13 | 851.7 | 0.00932 |
| 695.4163 | 1.02E+11 | 851.979 | 1.17E+21 | 851.979 | 0.39915 | 851.979 | 4.51E-08 | 851.979 | 4.58E-08 | 851.979 | 2.11E+13 | 851.979 | 0.00932 |
| 695.7033 | 1.02E+11 | 852.258 | 1.17E+21 | 852.258 | 0.3991 | 852.258 | 4.52E-08 | 852.258 | 4.42E-08 | 852.258 | 2.11E+13 | 852.258 | 0.0093 |
| 695.9904 | 1.02E+11 | 852.537 | 1.17E+21 | 852.537 | 0.39905 | 852.537 | 4.52E-08 | 852.537 | 4.50E-08 | 852.537 | 2.12E+13 | 852.537 | 0.00926 |
| 696.2774 | 1.02E+11 | 852.815 | 1.17E+21 | 852.815 | 0.399 | 852.815 | 4.52E-08 | 852.815 | 4.62E-08 | 852.815 | 2.12E+13 | 852.815 | 0.00925 |
| 696.5644 | 1.03E+11 | 853.094 | 1.17E+21 | 853.094 | 0.39894 | 853.094 | 4.52E-08 | 853.094 | 4.63E-08 | 853.094 | 2.12E+13 | 853.094 | 0.00922 |
| 696.8514 | 1.03E+11 | 853.373 | 1.18E+21 | 853.373 | 0.39889 | 853.373 | 4.52E-08 | 853.373 | 4.60E-08 | 853.373 | 2.12E+13 | 853.373 | 0.00929 |
| 697.1384 | 1.04E+11 | 853.652 | 1.18E+21 | 853.652 | 0.39884 | 853.652 | 4.52E-08 | 853.652 | 4.38E-08 | 853.652 | 2.12E+13 | 853.652 | 0.00922 |
| 697.4254 | 1.04E+11 | 853.93 | 1.18E+21 | 853.93 | 0.39879 | 853.93 | 4.52E-08 | 853.93 | 4.41E-08 | 853.93 | 2.12E+13 | 853.93 | 0.0092 |
| 697.7124 | 1.03E+11 | 854.209 | 1.18E+21 | 854.209 | 0.39874 | 854.209 | 4.54E-08 | 854.209 | 4.36E-08 | 854.209 | 2.13E+13 | 854.209 | 0.00918 |
| 697.9993 | 1.04E+11 | 854.488 | 1.18E+21 | 854.488 | 0.39869 | 854.488 | 4.54E-08 | 854.488 | 4.56E-08 | 854.488 | 2.14E+13 | 854.488 | 0.00916 |
| 698.2863 | 1.04E+11 | 854.767 | 1.18E+21 | 854.767 | 0.39864 | 854.767 | 4.55E-08 | 854.767 | 4.50E-08 | 854.767 | 2.14E+13 | 854.767 | 0.00919 |
| 698.5732 | 1.04E+11 | 855.045 | 1.18E+21 | 855.045 | 0.39858 | 855.045 | 4.54E-08 | 855.045 | 4.51E-08 | 855.045 | 2.14E+13 | 855.045 | 0.00922 |
| 698.8601 | 1.04E+11 | 855.324 | 1.18E+21 | 855.324 | 0.39853 | 855.324 | 4.54E-08 | 855.324 | 4.54E-08 | 855.324 | 2.14E+13 | 855.324 | 0.00912 |
| 699.147 | 1.05E+11 | 855.602 | 1.18E+21 | 855.602 | 0.39848 | 855.602 | 4.55E-08 | 855.602 | 4.51E-08 | 855.602 | 2.14E+13 | 855.602 | 0.00911 |
| 699.4339 | 1.04E+11 | 855.881 | 1.19E+21 | 855.881 | 0.39843 | 855.881 | 4.55E-08 | 855.881 | 4.54E-08 | 855.881 | 2.15E+13 | 855.881 | 0.00914 |
| 699.7208 | 1.05E+11 | 856.16 | 1.19E+21 | 856.16 | 0.39838 | 856.16 | 4.55E-08 | 856.16 | 4.55E-08 | 856.16 | 2.15E+13 | 856.16 | 0.00918 |
| 700.0077 | 1.04E+11 | 856.438 | 1.19E+21 | 856.438 | 0.39832 | 856.438 | 4.55E-08 | 856.438 | 4.54E-08 | 856.438 | 2.15E+13 | 856.438 | 0.00915 |
| 700.2945 | 1.05E+11 | 856.717 | 1.19E+21 | 856.717 | 0.39827 | 856.717 | 4.56E-08 | 856.717 | 4.42E-08 | 856.717 | 2.16E+13 | 856.717 | 0.00918 |
| 700.5814 | 1.06E+11 | 856.995 | 1.19E+21 | 856.995 | 0.39822 | 856.995 | 4.56E-08 | 856.995 | 4.77E-08 | 856.995 | 2.16E+13 | 856.995 | 0.00911 |
| 700.8682 | 1.06E+11 | 857.274 | 1.19E+21 | 857.274 | 0.39817 | 857.274 | 4.57E-08 | 857.274 | 4.54E-08 | 857.274 | 2.17E+13 | 857.274 | 0.0091 |
| 701.155 | 1.05E+11 | 857.552 | 1.19E+21 | 857.552 | 0.39812 | 857.552 | 4.57E-08 | 857.552 | 4.61E-08 | 857.552 | 2.17E+13 | 857.552 | 0.00911 |
| 701.4418 | 1.06E+11 | 857.831 | 1.19E+21 | 857.831 | 0.39807 | 857.831 | 4.57E-08 | 857.831 | 4.60E-08 | 857.831 | 2.17E+13 | 857.831 | 0.00912 |
| 701.7286 | 1.06E+11 | 858.109 | 1.19E+21 | 858.109 | 0.39802 | 858.109 | 4.57E-08 | 858.109 | 4.64E-08 | 858.109 | 2.17E+13 | 858.109 | 0.0091 |
| 702.0153 | 1.06E+11 | 858.388 | 1.20E+21 | 858.388 | 0.39796 | 858.388 | 4.58E-08 | 858.388 | 4.55E-08 | 858.388 | 2.18E+13 | 858.388 | 0.00902 |
| 702.3021 | 1.06E+11 | 858.666 | 1.20E+21 | 858.666 | 0.39791 | 858.666 | 4.58E-08 | 858.666 | 4.60E-08 | 858.666 | 2.18E+13 | 858.666 | 0.00902 |
| 702.5889 | 1.07E+11 | 858.945 | 1.20E+21 | 858.945 | 0.39786 | 858.945 | 4.58E-08 | 858.945 | 4.45E-08 | 858.945 | 2.18E+13 | 858.945 | 0.00902 |
| 702.8756 | 1.06E+11 | 859.223 | 1.20E+21 | 859.223 | 0.39781 | 859.223 | 4.58E-08 | 859.223 | 4.69E-08 | 859.223 | 2.18E+13 | 859.223 | 0.00901 |
| 703.1623 | 1.07E+11 | 859.501 | 1.20E+21 | 859.501 | 0.39776 | 859.501 | 4.59E-08 | 859.501 | 4.46E-08 | 859.501 | 2.19E+13 | 859.501 | 0.00903 |
| 703.449 | 1.07E+11 | 859.78 | 1.20E+21 | 859.78 | 0.3977 | 859.78 | 4.59E-08 | 859.78 | 4.80E-08 | 859.78 | 2.19E+13 | 859.78 | 0.00901 |
| 703.7357 | 1.07E+11 | 860.058 | 1.20E+21 | 860.058 | 0.39765 | 860.058 | 4.60E-08 | 860.058 | 4.52E-08 | 860.058 | 2.20E+13 | 860.058 | 0.00899 |
| 704.0224 | 1.08E+11 | 860.336 | 1.20E+21 | 860.336 | 0.3976 | 860.336 | 4.60E-08 | 860.336 | 4.62E-08 | 860.336 | 2.20E+13 | 860.336 | 0.00897 |
| 704.309 | 1.08E+11 | 860.615 | 1.20E+21 | 860.615 | 0.39755 | 860.615 | 4.61E-08 | 860.615 | 4.43E-08 | 860.615 | 2.20E+13 | 860.615 | 0.009 |
| 704.5957 | 1.08E+11 | 860.893 | 1.21E+21 | 860.893 | 0.3975 | 860.893 | 4.60E-08 | 860.893 | 4.49E-08 | 860.893 | 2.21E+13 | 860.893 | 0.00897 |
| 704.8823 | 1.08E+11 | 861.171 | 1.21E+21 | 861.171 | 0.39744 | 861.171 | 4.61E-08 | 861.171 | 4.70E-08 | 861.171 | 2.21E+13 | 861.171 | 0.00894 |
| 705.169 | 1.08E+11 | 861.45 | 1.21E+21 | 861.45 | 0.39739 | 861.45 | 4.61E-08 | 861.45 | 4.70E-08 | 861.45 | 2.21E+13 | 861.45 | 0.00894 |
| 705.4556 | 1.09E+11 | 861.728 | 1.21E+21 | 861.728 | 0.39734 | 861.728 | 4.61E-08 | 861.728 | 4.68E-08 | 861.728 | 2.21E+13 | 861.728 | 0.00896 |
| 705.7422 | 1.09E+11 | 862.006 | 1.21E+21 | 862.006 | 0.39729 | 862.006 | 4.61E-08 | 862.006 | 4.62E-08 | 862.006 | 2.22E+13 | 862.006 | 0.00893 |
| 706.0288 | 1.09E+11 | 862.284 | 1.21E+21 | 862.284 | 0.39724 | 862.284 | 4.61E-08 | 862.284 | 4.65E-08 | 862.284 | 2.22E+13 | 862.284 | 0.0089 |
| 706.3153 | 1.09E+11 | 862.563 | 1.21E+21 | 862.563 | 0.39719 | 862.563 | 4.62E-08 | 862.563 | 4.80E-08 | 862.563 | 2.22E+13 | 862.563 | 0.0089 |
| 706.6019 | 1.10E+11 | 862.841 | 1.21E+21 | 862.841 | 0.39713 | 862.841 | 4.61E-08 | 862.841 | 4.67E-08 | 862.841 | 2.22E+13 | 862.841 | 0.00889 |
| 706.8884 | 1.10E+11 | 863.119 | 1.21E+21 | 863.119 | 0.39708 | 863.119 | 4.61E-08 | 863.119 | 4.63E-08 | 863.119 | 2.22E+13 | 863.119 | 0.00883 |
| 707.175 | 1.11E+11 | 863.397 | 1.21E+21 | 863.397 | 0.39703 | 863.397 | 4.62E-08 | 863.397 | 4.60E-08 | 863.397 | 2.23E+13 | 863.397 | 0.00887 |
| 707.4615 | 1.10E+11 | 863.675 | 1.22E+21 | 863.675 | 0.39698 | 863.675 | 4.62E-08 | 863.675 | 4.55E-08 | 863.675 | 2.23E+13 | 863.675 | 0.00885 |
| 707.748 | 1.11E+11 | 863.953 | 1.22E+21 | 863.953 | 0.39693 | 863.953 | 4.62E-08 | 863.953 | 4.60E-08 | 863.953 | 2.23E+13 | 863.953 | 0.0089 |
| 708.0345 | 1.10E+11 | 864.231 | 1.22E+21 | 864.231 | 0.39687 | 864.231 | 4.62E-08 | 864.231 | 4.61E-08 | 864.231 | 2.23E+13 | 864.231 | 0.00885 |
| 708.321 | 1.12E+11 | 864.51 | 1.22E+21 | 864.51 | 0.39682 | 864.51 | 4.63E-08 | 864.51 | 4.69E-08 | 864.51 | 2.24E+13 | 864.51 | 0.0088 |
| 708.6074 | 1.11E+11 | 864.788 | 1.22E+21 | 864.788 | 0.39677 | 864.788 | 4.62E-08 | 864.788 | 4.77E-08 | 864.788 | 2.24E+13 | 864.788 | 0.00883 |
| 708.8939 | 1.11E+11 | 865.066 | 1.22E+21 | 865.066 | 0.39672 | 865.066 | 4.62E-08 | 865.066 | 4.56E-08 | 865.066 | 2.24E+13 | 865.066 | 0.00886 |
| 709.1803 | 1.11E+11 | 865.344 | 1.22E+21 | 865.344 | 0.39667 | 865.344 | 4.63E-08 | 865.344 | 4.63E-08 | 865.344 | 2.24E+13 | 865.344 | 0.00884 |
| 709.4668 | 1.12E+11 | 865.622 | 1.22E+21 | 865.622 | 0.39662 | 865.622 | 4.62E-08 | 865.622 | 4.64E-08 | 865.622 | 2.24E+13 | 865.622 | 0.00881 |
| 709.7532 | 1.11E+11 | 865.9 | 1.22E+21 | 865.9 | 0.39656 | 865.9 | 4.63E-08 | 865.9 | 4.73E-08 | 865.9 | 2.25E+13 | 865.9 | 0.00882 |
| 710.0396 | 1.12E+11 | 866.178 | 1.23E+21 | 866.178 | 0.39651 | 866.178 | 4.63E-08 | 866.178 | 4.62E-08 | 866.178 | 2.25E+13 | 866.178 | 0.00882 |
| 710.326 | 1.12E+11 | 866.456 | 1.23E+21 | 866.456 | 0.39646 | 866.456 | 4.63E-08 | 866.456 | 4.44E-08 | 866.456 | 2.25E+13 | 866.456 | 0.00874 |
| 710.6123 | 1.12E+11 | 866.734 | 1.23E+21 | 866.734 | 0.39641 | 866.734 | 4.63E-08 | 866.734 | 4.66E-08 | 866.734 | 2.26E+13 | 866.734 | 0.00875 |
| 710.8987 | 1.11E+11 | 867.012 | 1.23E+21 | 867.012 | 0.39636 | 867.012 | 4.64E-08 | 867.012 | 4.70E-08 | 867.012 | 2.26E+13 | 867.012 | 0.00874 |
| 711.185 | 1.12E+11 | 867.289 | 1.23E+21 | 867.289 | 0.3963 | 867.289 | 4.64E-08 | 867.289 | 4.47E-08 | 867.289 | 2.26E+13 | 867.289 | 0.00873 |
| 711.4714 | 1.13E+11 | 867.567 | 1.23E+21 | 867.567 | 0.39625 | 867.567 | 4.64E-08 | 867.567 | 4.58E-08 | 867.567 | 2.26E+13 | 867.567 | 0.00874 |
| 711.7577 | 1.12E+11 | 867.845 | 1.23E+21 | 867.845 | 0.3962 | 867.845 | 4.64E-08 | 867.845 | 4.61E-08 | 867.845 | 2.26E+13 | 867.845 | 0.00872 |
| 712.044 | 1.13E+11 | 868.123 | 1.23E+21 | 868.123 | 0.39615 | 868.123 | 4.64E-08 | 868.123 | 4.64E-08 | 868.123 | 2.27E+13 | 868.123 | 0.00871 |
| 712.3303 | 1.14E+11 | 868.401 | 1.23E+21 | 868.401 | 0.39609 | 868.401 | 4.64E-08 | 868.401 | 4.60E-08 | 868.401 | 2.27E+13 | 868.401 | 0.00869 |
| 712.6166 | 1.12E+11 | 868.679 | 1.24E+21 | 868.679 | 0.39604 | 868.679 | 4.64E-08 | 868.679 | 4.58E-08 | 868.679 | 2.27E+13 | 868.679 | 0.00867 |
| 712.9028 | 1.14E+11 | 868.957 | 1.24E+21 | 868.957 | 0.39599 | 868.957 | 4.65E-08 | 868.957 | 4.72E-08 | 868.957 | 2.28E+13 | 868.957 | 0.00862 |
| 713.1891 | 1.14E+11 | 869.234 | 1.24E+21 | 869.234 | 0.39594 | 869.234 | 4.65E-08 | 869.234 | 4.73E-08 | 869.234 | 2.28E+13 | 869.234 | 0.0086 |
| 713.4753 | 1.13E+11 | 869.512 | 1.24E+21 | 869.512 | 0.39589 | 869.512 | 4.65E-08 | 869.512 | 4.60E-08 | 869.512 | 2.28E+13 | 869.512 | 0.00863 |
| 713.7615 | 1.13E+11 | 869.79 | 1.24E+21 | 869.79 | 0.39583 | 869.79 | 4.66E-08 | 869.79 | 4.66E-08 | 869.79 | 2.29E+13 | 869.79 | 0.00861 |
| 714.0477 | 1.14E+11 | 870.068 | 1.24E+21 | 870.068 | 0.39578 | 870.068 | 4.66E-08 | 870.068 | 4.74E-08 | 870.068 | 2.29E+13 | 870.068 | 0.00862 |
| 714.3339 | 1.14E+11 | 870.345 | 1.24E+21 | 870.345 | 0.39573 | 870.345 | 4.66E-08 | 870.345 | 4.42E-08 | 870.345 | 2.29E+13 | 870.345 | 0.0086 |
| 714.6201 | 1.14E+11 | 870.623 | 1.24E+21 | 870.623 | 0.39568 | 870.623 | 4.66E-08 | 870.623 | 4.60E-08 | 870.623 | 2.29E+13 | 870.623 | 0.00855 |
| 714.9063 | 1.14E+11 | 870.901 | 1.24E+21 | 870.901 | 0.39563 | 870.901 | 4.66E-08 | 870.901 | 4.68E-08 | 870.901 | 2.30E+13 | 870.901 | 0.00855 |
| 715.1925 | 1.14E+11 | 871.178 | 1.25E+21 | 871.178 | 0.39558 | 871.178 | 4.67E-08 | 871.178 | 4.61E-08 | 871.178 | 2.30E+13 | 871.178 | 0.00855 |
| 715.4786 | 1.15E+11 | 871.456 | 1.25E+21 | 871.456 | 0.39552 | 871.456 | 4.67E-08 | 871.456 | 4.70E-08 | 871.456 | 2.30E+13 | 871.456 | 0.00852 |
| 715.7647 | 1.15E+11 | 871.734 | 1.25E+21 | 871.734 | 0.39547 | 871.734 | 4.68E-08 | 871.734 | 4.56E-08 | 871.734 | 2.31E+13 | 871.734 | 0.0085 |
| 716.0508 | 1.15E+11 | 872.011 | 1.25E+21 | 872.011 | 0.39542 | 872.011 | 4.68E-08 | 872.011 | 4.66E-08 | 872.011 | 2.31E+13 | 872.011 | 0.00848 |
| 716.3369 | 1.16E+11 | 872.289 | 1.25E+21 | 872.289 | 0.39537 | 872.289 | 4.69E-08 | 872.289 | 4.64E-08 | 872.289 | 2.32E+13 | 872.289 | 0.00848 |
| 716.623 | 1.15E+11 | 872.567 | 1.25E+21 | 872.567 | 0.39532 | 872.567 | 4.69E-08 | 872.567 | 4.73E-08 | 872.567 | 2.32E+13 | 872.567 | 0.00847 |
| 716.9091 | 1.16E+11 | 872.844 | 1.25E+21 | 872.844 | 0.39526 | 872.844 | 4.69E-08 | 872.844 | 4.68E-08 | 872.844 | 2.32E+13 | 872.844 | 0.00842 |
| 717.1952 | 1.16E+11 | 873.122 | 1.25E+21 | 873.122 | 0.39521 | 873.122 | 4.70E-08 | 873.122 | 4.70E-08 | 873.122 | 2.33E+13 | 873.122 | 0.00841 |
| 717.4812 | 1.15E+11 | 873.399 | 1.25E+21 | 873.399 | 0.39516 | 873.399 | 4.70E-08 | 873.399 | 4.69E-08 | 873.399 | 2.33E+13 | 873.399 | 0.00845 |
| 717.7673 | 1.16E+11 | 873.677 | 1.26E+21 | 873.677 | 0.39511 | 873.677 | 4.71E-08 | 873.677 | 4.70E-08 | 873.677 | 2.33E+13 | 873.677 | 0.0084 |
| 718.0533 | 1.16E+11 | 873.954 | 1.26E+21 | 873.954 | 0.39505 | 873.954 | 4.71E-08 | 873.954 | 4.82E-08 | 873.954 | 2.34E+13 | 873.954 | 0.00842 |
| 718.3393 | 1.16E+11 | 874.232 | 1.26E+21 | 874.232 | 0.395 | 874.232 | 4.71E-08 | 874.232 | 4.76E-08 | 874.232 | 2.34E+13 | 874.232 | 0.00838 |
| 718.6253 | 1.16E+11 | 874.509 | 1.26E+21 | 874.509 | 0.39495 | 874.509 | 4.72E-08 | 874.509 | 4.82E-08 | 874.509 | 2.34E+13 | 874.509 | 0.00839 |
| 718.9113 | 1.16E+11 | 874.787 | 1.26E+21 | 874.787 | 0.3949 | 874.787 | 4.72E-08 | 874.787 | 4.75E-08 | 874.787 | 2.35E+13 | 874.787 | 0.00838 |
| 719.1972 | 1.16E+11 | 875.064 | 1.26E+21 | 875.064 | 0.39485 | 875.064 | 4.72E-08 | 875.064 | 4.71E-08 | 875.064 | 2.35E+13 | 875.064 | 0.00832 |
| 719.4832 | 1.17E+11 | 875.342 | 1.26E+21 | 875.342 | 0.39479 | 875.342 | 4.72E-08 | 875.342 | 4.74E-08 | 875.342 | 2.35E+13 | 875.342 | 0.00834 |
| 719.7691 | 1.17E+11 | 875.619 | 1.26E+21 | 875.619 | 0.39474 | 875.619 | 4.72E-08 | 875.619 | 4.65E-08 | 875.619 | 2.35E+13 | 875.619 | 0.00833 |
| 720.055 | 1.17E+11 | 875.896 | 1.26E+21 | 875.896 | 0.39469 | 875.896 | 4.72E-08 | 875.896 | 4.67E-08 | 875.896 | 2.36E+13 | 875.896 | 0.0083 |
| 720.341 | 1.17E+11 | 876.174 | 1.26E+21 | 876.174 | 0.39464 | 876.174 | 4.73E-08 | 876.174 | 4.89E-08 | 876.174 | 2.36E+13 | 876.174 | 0.00832 |
| 720.6268 | 1.18E+11 | 876.451 | 1.27E+21 | 876.451 | 0.39459 | 876.451 | 4.74E-08 | 876.451 | 4.72E-08 | 876.451 | 2.37E+13 | 876.451 | 0.00828 |
| 720.9127 | 1.17E+11 | 876.728 | 1.27E+21 | 876.728 | 0.39453 | 876.728 | 4.74E-08 | 876.728 | 4.84E-08 | 876.728 | 2.37E+13 | 876.728 | 0.00826 |
| 721.1986 | 1.19E+11 | 877.006 | 1.27E+21 | 877.006 | 0.39448 | 877.006 | 4.75E-08 | 877.006 | 4.79E-08 | 877.006 | 2.38E+13 | 877.006 | 0.00819 |
| 721.4845 | 1.18E+11 | 877.283 | 1.27E+21 | 877.283 | 0.39443 | 877.283 | 4.76E-08 | 877.283 | 4.81E-08 | 877.283 | 2.38E+13 | 877.283 | 0.00816 |
| 721.7703 | 1.19E+11 | 877.56 | 1.27E+21 | 877.56 | 0.39438 | 877.56 | 4.76E-08 | 877.56 | 4.65E-08 | 877.56 | 2.39E+13 | 877.56 | 0.00817 |
| 722.0561 | 1.19E+11 | 877.838 | 1.27E+21 | 877.838 | 0.39432 | 877.838 | 4.77E-08 | 877.838 | 4.92E-08 | 877.838 | 2.39E+13 | 877.838 | 0.00817 |
| 722.342 | 1.19E+11 | 878.115 | 1.27E+21 | 878.115 | 0.39427 | 878.115 | 4.77E-08 | 878.115 | 4.69E-08 | 878.115 | 2.39E+13 | 878.115 | 0.00818 |
| 722.6278 | 1.18E+11 | 878.392 | 1.27E+21 | 878.392 | 0.39422 | 878.392 | 4.77E-08 | 878.392 | 4.76E-08 | 878.392 | 2.40E+13 | 878.392 | 0.00815 |
| 722.9135 | 1.19E+11 | 878.669 | 1.27E+21 | 878.669 | 0.39417 | 878.669 | 4.78E-08 | 878.669 | 4.87E-08 | 878.669 | 2.40E+13 | 878.669 | 0.0081 |
| 723.1993 | 1.19E+11 | 878.946 | 1.28E+21 | 878.946 | 0.39411 | 878.946 | 4.78E-08 | 878.946 | 4.73E-08 | 878.946 | 2.40E+13 | 878.946 | 0.00809 |
| 723.4851 | 1.20E+11 | 879.224 | 1.28E+21 | 879.224 | 0.39406 | 879.224 | 4.78E-08 | 879.224 | 4.79E-08 | 879.224 | 2.41E+13 | 879.224 | 0.00808 |
| 723.7708 | 1.21E+11 | 879.501 | 1.28E+21 | 879.501 | 0.39401 | 879.501 | 4.78E-08 | 879.501 | 4.68E-08 | 879.501 | 2.41E+13 | 879.501 | 0.00802 |
| 724.0566 | 1.20E+11 | 879.778 | 1.28E+21 | 879.778 | 0.39396 | 879.778 | 4.79E-08 | 879.778 | 4.75E-08 | 879.778 | 2.41E+13 | 879.778 | 0.00802 |
| 724.3423 | 1.21E+11 | 880.055 | 1.28E+21 | 880.055 | 0.39391 | 880.055 | 4.79E-08 | 880.055 | 4.77E-08 | 880.055 | 2.41E+13 | 880.055 | 0.008 |
| 724.628 | 1.20E+11 | 880.332 | 1.28E+21 | 880.332 | 0.39385 | 880.332 | 4.79E-08 | 880.332 | 4.76E-08 | 880.332 | 2.42E+13 | 880.332 | 0.008 |
| 724.9137 | 1.20E+11 | 880.609 | 1.28E+21 | 880.609 | 0.3938 | 880.609 | 4.79E-08 | 880.609 | 4.88E-08 | 880.609 | 2.42E+13 | 880.609 | 0.00801 |
| 725.1994 | 1.21E+11 | 880.886 | 1.28E+21 | 880.886 | 0.39375 | 880.886 | 4.79E-08 | 880.886 | 4.66E-08 | 880.886 | 2.42E+13 | 880.886 | 0.00792 |
| 725.485 | 1.22E+11 | 881.163 | 1.28E+21 | 881.163 | 0.3937 | 881.163 | 4.80E-08 | 881.163 | 4.65E-08 | 881.163 | 2.42E+13 | 881.163 | 0.00793 |
| 725.7707 | 1.22E+11 | 881.44 | 1.29E+21 | 881.44 | 0.39364 | 881.44 | 4.80E-08 | 881.44 | 4.80E-08 | 881.44 | 2.43E+13 | 881.44 | 0.00792 |
| 726.0563 | 1.21E+11 | 881.718 | 1.29E+21 | 881.718 | 0.39359 | 881.718 | 4.81E-08 | 881.718 | 4.79E-08 | 881.718 | 2.44E+13 | 881.718 | 0.00793 |
| 726.3419 | 1.21E+11 | 881.995 | 1.29E+21 | 881.995 | 0.39354 | 881.995 | 4.81E-08 | 881.995 | 4.76E-08 | 881.995 | 2.44E+13 | 881.995 | 0.00789 |
| 726.6276 | 1.22E+11 | 882.272 | 1.29E+21 | 882.272 | 0.39349 | 882.272 | 4.82E-08 | 882.272 | 4.92E-08 | 882.272 | 2.44E+13 | 882.272 | 0.00784 |
| 726.9131 | 1.22E+11 | 882.548 | 1.29E+21 | 882.548 | 0.39343 | 882.548 | 4.82E-08 | 882.548 | 4.97E-08 | 882.548 | 2.44E+13 | 882.548 | 0.00784 |
| 727.1987 | 1.22E+11 | 882.825 | 1.29E+21 | 882.825 | 0.39338 | 882.825 | 4.82E-08 | 882.825 | 4.85E-08 | 882.825 | 2.45E+13 | 882.825 | 0.00782 |
| 727.4843 | 1.22E+11 | 883.102 | 1.29E+21 | 883.102 | 0.39333 | 883.102 | 4.82E-08 | 883.102 | 4.89E-08 | 883.102 | 2.45E+13 | 883.102 | 0.00778 |
| 727.7699 | 1.23E+11 | 883.379 | 1.29E+21 | 883.379 | 0.39328 | 883.379 | 4.83E-08 | 883.379 | 4.85E-08 | 883.379 | 2.45E+13 | 883.379 | 0.00781 |
| 728.0554 | 1.24E+11 | 883.656 | 1.29E+21 | 883.656 | 0.39322 | 883.656 | 4.82E-08 | 883.656 | 4.78E-08 | 883.656 | 2.46E+13 | 883.656 | 0.00777 |
| 728.3409 | 1.23E+11 | 883.933 | 1.30E+21 | 883.933 | 0.39317 | 883.933 | 4.83E-08 | 883.933 | 4.88E-08 | 883.933 | 2.46E+13 | 883.933 | 0.0078 |
| 728.6264 | 1.23E+11 | 884.21 | 1.30E+21 | 884.21 | 0.39312 | 884.21 | 4.83E-08 | 884.21 | 4.81E-08 | 884.21 | 2.46E+13 | 884.21 | 0.00776 |
| 728.9119 | 1.23E+11 | 884.487 | 1.30E+21 | 884.487 | 0.39307 | 884.487 | 4.83E-08 | 884.487 | 4.79E-08 | 884.487 | 2.46E+13 | 884.487 | 0.00775 |
| 729.1974 | 1.23E+11 | 884.764 | 1.30E+21 | 884.764 | 0.39302 | 884.764 | 4.83E-08 | 884.764 | 4.77E-08 | 884.764 | 2.47E+13 | 884.764 | 0.00774 |
| 729.4829 | 1.24E+11 | 885.04 | 1.30E+21 | 885.04 | 0.39296 | 885.04 | 4.84E-08 | 885.04 | 4.82E-08 | 885.04 | 2.47E+13 | 885.04 | 0.00767 |
| 729.7683 | 1.25E+11 | 885.317 | 1.30E+21 | 885.317 | 0.39291 | 885.317 | 4.84E-08 | 885.317 | 4.83E-08 | 885.317 | 2.47E+13 | 885.317 | 0.00763 |
| 730.0538 | 1.25E+11 | 885.594 | 1.30E+21 | 885.594 | 0.39286 | 885.594 | 4.85E-08 | 885.594 | 4.86E-08 | 885.594 | 2.48E+13 | 885.594 | 0.00765 |
| 730.3392 | 1.25E+11 | 885.871 | 1.30E+21 | 885.871 | 0.39281 | 885.871 | 4.85E-08 | 885.871 | 4.76E-08 | 885.871 | 2.48E+13 | 885.871 | 0.00757 |
| 730.6246 | 1.25E+11 | 886.148 | 1.30E+21 | 886.148 | 0.39275 | 886.148 | 4.86E-08 | 886.148 | 4.80E-08 | 886.148 | 2.49E+13 | 886.148 | 0.0076 |
| 730.91 | 1.26E+11 | 886.424 | 1.30E+21 | 886.424 | 0.3927 | 886.424 | 4.86E-08 | 886.424 | 5.00E-08 | 886.424 | 2.49E+13 | 886.424 | 0.00758 |
| 731.1954 | 1.25E+11 | 886.701 | 1.31E+21 | 886.701 | 0.39265 | 886.701 | 4.87E-08 | 886.701 | 4.80E-08 | 886.701 | 2.49E+13 | 886.701 | 0.00756 |
| 731.4808 | 1.26E+11 | 886.978 | 1.31E+21 | 886.978 | 0.3926 | 886.978 | 4.87E-08 | 886.978 | 4.87E-08 | 886.978 | 2.50E+13 | 886.978 | 0.00753 |
| 731.7662 | 1.26E+11 | 887.254 | 1.31E+21 | 887.254 | 0.39254 | 887.254 | 4.87E-08 | 887.254 | 5.05E-08 | 887.254 | 2.50E+13 | 887.254 | 0.00755 |
| 732.0515 | 1.26E+11 | 887.531 | 1.31E+21 | 887.531 | 0.39249 | 887.531 | 4.87E-08 | 887.531 | 4.97E-08 | 887.531 | 2.50E+13 | 887.531 | 0.00752 |
| 732.3368 | 1.26E+11 | 887.808 | 1.31E+21 | 887.808 | 0.39244 | 887.808 | 4.88E-08 | 887.808 | 4.88E-08 | 887.808 | 2.51E+13 | 887.808 | 0.00745 |
| 732.6222 | 1.26E+11 | 888.084 | 1.31E+21 | 888.084 | 0.39239 | 888.084 | 4.88E-08 | 888.084 | 4.85E-08 | 888.084 | 2.51E+13 | 888.084 | 0.00746 |
| 732.9075 | 1.27E+11 | 888.361 | 1.31E+21 | 888.361 | 0.39233 | 888.361 | 4.88E-08 | 888.361 | 4.84E-08 | 888.361 | 2.51E+13 | 888.361 | 0.00744 |
| 733.1928 | 1.27E+11 | 888.637 | 1.31E+21 | 888.637 | 0.39228 | 888.637 | 4.88E-08 | 888.637 | 4.85E-08 | 888.637 | 2.51E+13 | 888.637 | 0.00747 |
| 733.478 | 1.28E+11 | 888.914 | 1.31E+21 | 888.914 | 0.39223 | 888.914 | 4.88E-08 | 888.914 | 4.97E-08 | 888.914 | 2.52E+13 | 888.914 | 0.00744 |
| 733.7633 | 1.27E+11 | 889.191 | 1.32E+21 | 889.191 | 0.39218 | 889.191 | 4.88E-08 | 889.191 | 4.85E-08 | 889.191 | 2.52E+13 | 889.191 | 0.00742 |
| 734.0486 | 1.28E+11 | 889.467 | 1.32E+21 | 889.467 | 0.39212 | 889.467 | 4.88E-08 | 889.467 | 4.80E-08 | 889.467 | 2.52E+13 | 889.467 | 0.00741 |
| 734.3338 | 1.29E+11 | 889.744 | 1.32E+21 | 889.744 | 0.39207 | 889.744 | 4.88E-08 | 889.744 | 4.88E-08 | 889.744 | 2.52E+13 | 889.744 | 0.00737 |
| 734.619 | 1.28E+11 | 890.02 | 1.32E+21 | 890.02 | 0.39202 | 890.02 | 4.88E-08 | 890.02 | 4.79E-08 | 890.02 | 2.52E+13 | 890.02 | 0.00739 |
| 734.9042 | 1.30E+11 | 890.297 | 1.32E+21 | 890.297 | 0.39197 | 890.297 | 4.89E-08 | 890.297 | 4.90E-08 | 890.297 | 2.53E+13 | 890.297 | 0.00736 |
| 735.1894 | 1.30E+11 | 890.573 | 1.32E+21 | 890.573 | 0.39191 | 890.573 | 4.89E-08 | 890.573 | 4.99E-08 | 890.573 | 2.53E+13 | 890.573 | 0.00737 |
| 735.4746 | 1.29E+11 | 890.85 | 1.32E+21 | 890.85 | 0.39186 | 890.85 | 4.89E-08 | 890.85 | 4.89E-08 | 890.85 | 2.53E+13 | 890.85 | 0.00732 |
| 735.7598 | 1.30E+11 | 891.126 | 1.32E+21 | 891.126 | 0.39181 | 891.126 | 4.90E-08 | 891.126 | 4.89E-08 | 891.126 | 2.54E+13 | 891.126 | 0.0073 |
| 736.0449 | 1.29E+11 | 891.402 | 1.32E+21 | 891.402 | 0.39176 | 891.402 | 4.90E-08 | 891.402 | 5.02E-08 | 891.402 | 2.54E+13 | 891.402 | 0.00729 |
| 736.33 | 1.31E+11 | 891.679 | 1.32E+21 | 891.679 | 0.3917 | 891.679 | 4.90E-08 | 891.679 | 4.96E-08 | 891.679 | 2.54E+13 | 891.679 | 0.00733 |
| 736.6152 | 1.32E+11 | 891.955 | 1.33E+21 | 891.955 | 0.39165 | 891.955 | 4.90E-08 | 891.955 | 5.05E-08 | 891.955 | 2.55E+13 | 891.955 | 0.00727 |
| 736.9003 | 1.30E+11 | 892.232 | 1.33E+21 | 892.232 | 0.3916 | 892.232 | 4.90E-08 | 892.232 | 4.84E-08 | 892.232 | 2.55E+13 | 892.232 | 0.00728 |
| 737.1854 | 1.31E+11 | 892.508 | 1.33E+21 | 892.508 | 0.39155 | 892.508 | 4.90E-08 | 892.508 | 4.77E-08 | 892.508 | 2.55E+13 | 892.508 | 0.00729 |
| 737.4705 | 1.31E+11 | 892.784 | 1.33E+21 | 892.784 | 0.39149 | 892.784 | 4.91E-08 | 892.784 | 4.88E-08 | 892.784 | 2.55E+13 | 892.784 | 0.00722 |
| 737.7555 | 1.32E+11 | 893.061 | 1.33E+21 | 893.061 | 0.39144 | 893.061 | 4.91E-08 | 893.061 | 4.96E-08 | 893.061 | 2.56E+13 | 893.061 | 0.00722 |
| 738.0406 | 1.32E+11 | 893.337 | 1.33E+21 | 893.337 | 0.39139 | 893.337 | 4.90E-08 | 893.337 | 5.03E-08 | 893.337 | 2.56E+13 | 893.337 | 0.00722 |
| 738.3256 | 1.33E+11 | 893.613 | 1.33E+21 | 893.613 | 0.39134 | 893.613 | 4.91E-08 | 893.613 | 4.93E-08 | 893.613 | 2.56E+13 | 893.613 | 0.00718 |
| 738.6106 | 1.33E+11 | 893.889 | 1.33E+21 | 893.889 | 0.39128 | 893.889 | 4.91E-08 | 893.889 | 4.74E-08 | 893.889 | 2.56E+13 | 893.889 | 0.00716 |
| 738.8956 | 1.32E+11 | 894.166 | 1.33E+21 | 894.166 | 0.39123 | 894.166 | 4.92E-08 | 894.166 | 5.06E-08 | 894.166 | 2.57E+13 | 894.166 | 0.00718 |
| 739.1806 | 1.33E+11 | 894.442 | 1.34E+21 | 894.442 | 0.39118 | 894.442 | 4.92E-08 | 894.442 | 4.83E-08 | 894.442 | 2.57E+13 | 894.442 | 0.00715 |
| 739.4656 | 1.33E+11 | 894.718 | 1.34E+21 | 894.718 | 0.39113 | 894.718 | 4.92E-08 | 894.718 | 4.74E-08 | 894.718 | 2.57E+13 | 894.718 | 0.0071 |
| 739.7506 | 1.32E+11 | 894.994 | 1.34E+21 | 894.994 | 0.39107 | 894.994 | 4.92E-08 | 894.994 | 5.00E-08 | 894.994 | 2.57E+13 | 894.994 | 0.00709 |
| 740.0355 | 1.34E+11 | 895.27 | 1.34E+21 | 895.27 | 0.39102 | 895.27 | 4.93E-08 | 895.27 | 4.83E-08 | 895.27 | 2.58E+13 | 895.27 | 0.00706 |
| 740.3205 | 1.34E+11 | 895.547 | 1.34E+21 | 895.547 | 0.39097 | 895.547 | 4.93E-08 | 895.547 | 4.91E-08 | 895.547 | 2.58E+13 | 895.547 | 0.00705 |
| 740.6054 | 1.34E+11 | 895.823 | 1.34E+21 | 895.823 | 0.39092 | 895.823 | 4.94E-08 | 895.823 | 4.89E-08 | 895.823 | 2.59E+13 | 895.823 | 0.00702 |
| 740.8903 | 1.34E+11 | 896.099 | 1.34E+21 | 896.099 | 0.39086 | 896.099 | 4.94E-08 | 896.099 | 4.96E-08 | 896.099 | 2.59E+13 | 896.099 | 0.007 |
| 741.1752 | 1.34E+11 | 896.375 | 1.34E+21 | 896.375 | 0.39081 | 896.375 | 4.94E-08 | 896.375 | 4.86E-08 | 896.375 | 2.59E+13 | 896.375 | 0.00698 |
| 741.4601 | 1.35E+11 | 896.651 | 1.34E+21 | 896.651 | 0.39076 | 896.651 | 4.94E-08 | 896.651 | 5.04E-08 | 896.651 | 2.60E+13 | 896.651 | 0.00697 |
| 741.745 | 1.34E+11 | 896.927 | 1.35E+21 | 896.927 | 0.39071 | 896.927 | 4.95E-08 | 896.927 | 4.80E-08 | 896.927 | 2.60E+13 | 896.927 | 0.00695 |
| 742.0298 | 1.35E+11 | 897.203 | 1.35E+21 | 897.203 | 0.39065 | 897.203 | 4.95E-08 | 897.203 | 4.97E-08 | 897.203 | 2.60E+13 | 897.203 | 0.00695 |
| 742.3147 | 1.36E+11 | 897.479 | 1.35E+21 | 897.479 | 0.3906 | 897.479 | 4.95E-08 | 897.479 | 4.91E-08 | 897.479 | 2.60E+13 | 897.479 | 0.00692 |
| 742.5995 | 1.37E+11 | 897.755 | 1.35E+21 | 897.755 | 0.39055 | 897.755 | 4.95E-08 | 897.755 | 4.78E-08 | 897.755 | 2.61E+13 | 897.755 | 0.00691 |
| 742.8843 | 1.36E+11 | 898.031 | 1.35E+21 | 898.031 | 0.39049 | 898.031 | 4.95E-08 | 898.031 | 4.99E-08 | 898.031 | 2.61E+13 | 898.031 | 0.0069 |
| 743.1691 | 1.37E+11 | 898.307 | 1.35E+21 | 898.307 | 0.39044 | 898.307 | 4.96E-08 | 898.307 | 5.09E-08 | 898.307 | 2.61E+13 | 898.307 | 0.00686 |
| 743.4539 | 1.36E+11 | 898.583 | 1.35E+21 | 898.583 | 0.39039 | 898.583 | 4.96E-08 | 898.583 | 5.11E-08 | 898.583 | 2.62E+13 | 898.583 | 0.0069 |
| 743.7387 | 1.37E+11 | 898.859 | 1.35E+21 | 898.859 | 0.39034 | 898.859 | 4.96E-08 | 898.859 | 4.78E-08 | 898.859 | 2.62E+13 | 898.859 | 0.00683 |
| 744.0234 | 1.37E+11 | 899.135 | 1.35E+21 | 899.135 | 0.39028 | 899.135 | 4.96E-08 | 899.135 | 4.92E-08 | 899.135 | 2.62E+13 | 899.135 | 0.00682 |
| 744.3082 | 1.37E+11 | 899.411 | 1.35E+21 | 899.411 | 0.39023 | 899.411 | 4.96E-08 | 899.411 | 5.15E-08 | 899.411 | 2.62E+13 | 899.411 | 0.00683 |
| 744.5929 | 1.37E+11 | 899.687 | 1.36E+21 | 899.687 | 0.39018 | 899.687 | 4.96E-08 | 899.687 | 4.95E-08 | 899.687 | 2.63E+13 | 899.687 | 0.00679 |
| 744.8776 | 1.38E+11 | 899.963 | 1.36E+21 | 899.963 | 0.39013 | 899.963 | 4.97E-08 | 899.963 | 5.13E-08 | 899.963 | 2.63E+13 | 899.963 | 0.0068 |
| 745.1623 | 1.37E+11 | 900.238 | 1.36E+21 | 900.238 | 0.39007 | 900.238 | 4.97E-08 | 900.238 | 4.93E-08 | 900.238 | 2.63E+13 | 900.238 | 0.0068 |
| 745.447 | 1.39E+11 | 900.514 | 1.36E+21 | 900.514 | 0.39002 | 900.514 | 4.98E-08 | 900.514 | 4.86E-08 | 900.514 | 2.64E+13 | 900.514 | 0.00676 |
| 745.7317 | 1.38E+11 | 900.79 | 1.36E+21 | 900.79 | 0.38997 | 900.79 | 4.97E-08 | 900.79 | 5.05E-08 | 900.79 | 2.64E+13 | 900.79 | 0.00676 |
| 746.0163 | 1.39E+11 | 901.066 | 1.36E+21 | 901.066 | 0.38991 | 901.066 | 4.98E-08 | 901.066 | 5.02E-08 | 901.066 | 2.64E+13 | 901.066 | 0.0068 |
| 746.301 | 1.39E+11 | 901.342 | 1.36E+21 | 901.342 | 0.38986 | 901.342 | 4.98E-08 | 901.342 | 4.97E-08 | 901.342 | 2.64E+13 | 901.342 | 0.00672 |
| 746.5856 | 1.39E+11 | 901.617 | 1.36E+21 | 901.617 | 0.38981 | 901.617 | 4.98E-08 | 901.617 | 5.02E-08 | 901.617 | 2.65E+13 | 901.617 | 0.00675 |
| 746.8702 | 1.39E+11 | 901.893 | 1.36E+21 | 901.893 | 0.38976 | 901.893 | 4.99E-08 | 901.893 | 5.07E-08 | 901.893 | 2.65E+13 | 901.893 | 0.00667 |
| 747.1548 | 1.41E+11 | 902.169 | 1.37E+21 | 902.169 | 0.3897 | 902.169 | 5.00E-08 | 902.169 | 4.91E-08 | 902.169 | 2.66E+13 | 902.169 | 0.0067 |
| 747.4394 | 1.39E+11 | 902.445 | 1.37E+21 | 902.445 | 0.38965 | 902.445 | 4.99E-08 | 902.445 | 5.15E-08 | 902.445 | 2.66E+13 | 902.445 | 0.00665 |
| 747.724 | 1.41E+11 | 902.72 | 1.37E+21 | 902.72 | 0.3896 | 902.72 | 5.00E-08 | 902.72 | 5.08E-08 | 902.72 | 2.66E+13 | 902.72 | 0.00666 |
| 748.0085 | 1.41E+11 | 902.996 | 1.37E+21 | 902.996 | 0.38954 | 902.996 | 5.00E-08 | 902.996 | 4.96E-08 | 902.996 | 2.66E+13 | 902.996 | 0.00664 |
| 748.2931 | 1.40E+11 | 903.272 | 1.37E+21 | 903.272 | 0.38949 | 903.272 | 5.00E-08 | 903.272 | 5.00E-08 | 903.272 | 2.67E+13 | 903.272 | 0.00665 |
| 748.5776 | 1.41E+11 | 903.547 | 1.37E+21 | 903.547 | 0.38944 | 903.547 | 5.01E-08 | 903.547 | 5.06E-08 | 903.547 | 2.67E+13 | 903.547 | 0.0066 |
| 748.8621 | 1.41E+11 | 903.823 | 1.37E+21 | 903.823 | 0.38939 | 903.823 | 5.01E-08 | 903.823 | 4.89E-08 | 903.823 | 2.67E+13 | 903.823 | 0.00662 |
| 749.1466 | 1.42E+11 | 904.099 | 1.37E+21 | 904.099 | 0.38933 | 904.099 | 5.01E-08 | 904.099 | 5.05E-08 | 904.099 | 2.68E+13 | 904.099 | 0.00657 |
| 749.4311 | 1.41E+11 | 904.374 | 1.37E+21 | 904.374 | 0.38928 | 904.374 | 5.01E-08 | 904.374 | 5.09E-08 | 904.374 | 2.68E+13 | 904.374 | 0.00657 |
| 749.7155 | 1.42E+11 | 904.65 | 1.37E+21 | 904.65 | 0.38923 | 904.65 | 5.01E-08 | 904.65 | 4.73E-08 | 904.65 | 2.68E+13 | 904.65 | 0.00656 |
| 750 | 1.43E+11 | 904.925 | 1.38E+21 | 904.925 | 0.38917 | 904.925 | 5.02E-08 | 904.925 | 4.95E-08 | 904.925 | 2.69E+13 | 904.925 | 0.00654 |
| 750.2844 | 1.42E+11 | 905.201 | 1.38E+21 | 905.201 | 0.38912 | 905.201 | 5.01E-08 | 905.201 | 5.07E-08 | 905.201 | 2.69E+13 | 905.201 | 0.00651 |
| 750.5689 | 1.43E+11 | 905.476 | 1.38E+21 | 905.476 | 0.38907 | 905.476 | 5.02E-08 | 905.476 | 5.09E-08 | 905.476 | 2.69E+13 | 905.476 | 0.00647 |
| 750.8533 | 1.42E+11 | 905.752 | 1.38E+21 | 905.752 | 0.38902 | 905.752 | 5.01E-08 | 905.752 | 4.83E-08 | 905.752 | 2.69E+13 | 905.752 | 0.00649 |
| 751.1377 | 1.42E+11 | 906.027 | 1.38E+21 | 906.027 | 0.38896 | 906.027 | 5.02E-08 | 906.027 | 5.15E-08 | 906.027 | 2.69E+13 | 906.027 | 0.00646 |
| 751.4221 | 1.43E+11 | 906.303 | 1.38E+21 | 906.303 | 0.38891 | 906.303 | 5.02E-08 | 906.303 | 4.94E-08 | 906.303 | 2.70E+13 | 906.303 | 0.00647 |
| 751.7064 | 1.43E+11 | 906.578 | 1.38E+21 | 906.578 | 0.38886 | 906.578 | 5.02E-08 | 906.578 | 4.94E-08 | 906.578 | 2.70E+13 | 906.578 | 0.00646 |
| 751.9908 | 1.43E+11 | 906.854 | 1.38E+21 | 906.854 | 0.38881 | 906.854 | 5.03E-08 | 906.854 | 4.94E-08 | 906.854 | 2.70E+13 | 906.854 | 0.00642 |
| 752.2751 | 1.44E+11 | 907.129 | 1.38E+21 | 907.129 | 0.38875 | 907.129 | 5.03E-08 | 907.129 | 5.17E-08 | 907.129 | 2.70E+13 | 907.129 | 0.00639 |
| 752.5595 | 1.45E+11 | 907.404 | 1.38E+21 | 907.404 | 0.3887 | 907.404 | 5.03E-08 | 907.404 | 5.22E-08 | 907.404 | 2.71E+13 | 907.404 | 0.0064 |
| 752.8438 | 1.44E+11 | 907.68 | 1.39E+21 | 907.68 | 0.38865 | 907.68 | 5.03E-08 | 907.68 | 5.04E-08 | 907.68 | 2.71E+13 | 907.68 | 0.00641 |
| 753.1281 | 1.45E+11 | 907.955 | 1.39E+21 | 907.955 | 0.38859 | 907.955 | 5.03E-08 | 907.955 | 4.96E-08 | 907.955 | 2.71E+13 | 907.955 | 0.00638 |
| 753.4124 | 1.44E+11 | 908.23 | 1.39E+21 | 908.23 | 0.38854 | 908.23 | 5.03E-08 | 908.23 | 5.01E-08 | 908.23 | 2.71E+13 | 908.23 | 0.00638 |
| 753.6966 | 1.44E+11 | 908.506 | 1.39E+21 | 908.506 | 0.38849 | 908.506 | 5.03E-08 | 908.506 | 5.03E-08 | 908.506 | 2.71E+13 | 908.506 | 0.00636 |
| 753.9809 | 1.45E+11 | 908.781 | 1.39E+21 | 908.781 | 0.38843 | 908.781 | 5.02E-08 | 908.781 | 4.91E-08 | 908.781 | 2.71E+13 | 908.781 | 0.00638 |
| 754.2651 | 1.45E+11 | 909.056 | 1.39E+21 | 909.056 | 0.38838 | 909.056 | 5.03E-08 | 909.056 | 5.15E-08 | 909.056 | 2.72E+13 | 909.056 | 0.00633 |
| 754.5493 | 1.46E+11 | 909.332 | 1.39E+21 | 909.332 | 0.38833 | 909.332 | 5.03E-08 | 909.332 | 5.03E-08 | 909.332 | 2.72E+13 | 909.332 | 0.00634 |
| 754.8336 | 1.45E+11 | 909.607 | 1.39E+21 | 909.607 | 0.38828 | 909.607 | 5.04E-08 | 909.607 | 5.13E-08 | 909.607 | 2.72E+13 | 909.607 | 0.00631 |
| 755.1178 | 1.46E+11 | 909.882 | 1.39E+21 | 909.882 | 0.38822 | 909.882 | 5.03E-08 | 909.882 | 5.05E-08 | 909.882 | 2.73E+13 | 909.882 | 0.00631 |
| 755.4019 | 1.46E+11 | 910.157 | 1.40E+21 | 910.157 | 0.38817 | 910.157 | 5.04E-08 | 910.157 | 5.00E-08 | 910.157 | 2.73E+13 | 910.157 | 0.00627 |
| 755.6861 | 1.46E+11 | 910.433 | 1.40E+21 | 910.433 | 0.38812 | 910.433 | 5.05E-08 | 910.433 | 5.08E-08 | 910.433 | 2.74E+13 | 910.433 | 0.00628 |
| 755.9703 | 1.46E+11 | 910.708 | 1.40E+21 | 910.708 | 0.38806 | 910.708 | 5.05E-08 | 910.708 | 4.93E-08 | 910.708 | 2.74E+13 | 910.708 | 0.00623 |
| 756.2544 | 1.47E+11 | 910.983 | 1.40E+21 | 910.983 | 0.38801 | 910.983 | 5.05E-08 | 910.983 | 5.11E-08 | 910.983 | 2.74E+13 | 910.983 | 0.00621 |
| 756.5385 | 1.47E+11 | 911.258 | 1.40E+21 | 911.258 | 0.38796 | 911.258 | 5.05E-08 | 911.258 | 5.00E-08 | 911.258 | 2.74E+13 | 911.258 | 0.00621 |
| 756.8226 | 1.47E+11 | 911.533 | 1.40E+21 | 911.533 | 0.38791 | 911.533 | 5.06E-08 | 911.533 | 5.14E-08 | 911.533 | 2.75E+13 | 911.533 | 0.00619 |
| 757.1067 | 1.47E+11 | 911.808 | 1.40E+21 | 911.808 | 0.38785 | 911.808 | 5.06E-08 | 911.808 | 5.03E-08 | 911.808 | 2.75E+13 | 911.808 | 0.00618 |
| 757.3908 | 1.47E+11 | 912.083 | 1.40E+21 | 912.083 | 0.3878 | 912.083 | 5.07E-08 | 912.083 | 5.11E-08 | 912.083 | 2.76E+13 | 912.083 | 0.00615 |
| 757.6749 | 1.47E+11 | 912.358 | 1.40E+21 | 912.358 | 0.38775 | 912.358 | 5.08E-08 | 912.358 | 5.04E-08 | 912.358 | 2.76E+13 | 912.358 | 0.00612 |
| 757.9589 | 1.47E+11 | 912.633 | 1.40E+21 | 912.633 | 0.38769 | 912.633 | 5.08E-08 | 912.633 | 5.08E-08 | 912.633 | 2.77E+13 | 912.633 | 0.00612 |
| 758.243 | 1.48E+11 | 912.908 | 1.41E+21 | 912.908 | 0.38764 | 912.908 | 5.08E-08 | 912.908 | 5.01E-08 | 912.908 | 2.77E+13 | 912.908 | 0.0061 |
| 758.527 | 1.47E+11 | 913.183 | 1.41E+21 | 913.183 | 0.38759 | 913.183 | 5.07E-08 | 913.183 | 5.09E-08 | 913.183 | 2.77E+13 | 913.183 | 0.0061 |
| 758.811 | 1.48E+11 | 913.458 | 1.41E+21 | 913.458 | 0.38753 | 913.458 | 5.08E-08 | 913.458 | 4.94E-08 | 913.458 | 2.77E+13 | 913.458 | 0.00608 |
| 759.095 | 1.49E+11 | 913.733 | 1.41E+21 | 913.733 | 0.38748 | 913.733 | 5.08E-08 | 913.733 | 5.02E-08 | 913.733 | 2.77E+13 | 913.733 | 0.00606 |
| 759.379 | 1.48E+11 | 914.008 | 1.41E+21 | 914.008 | 0.38743 | 914.008 | 5.08E-08 | 914.008 | 5.13E-08 | 914.008 | 2.77E+13 | 914.008 | 0.00607 |
| 759.6629 | 1.48E+11 | 914.283 | 1.41E+21 | 914.283 | 0.38738 | 914.283 | 5.08E-08 | 914.283 | 4.87E-08 | 914.283 | 2.78E+13 | 914.283 | 0.00607 |
| 759.9469 | 1.50E+11 | 914.558 | 1.41E+21 | 914.558 | 0.38732 | 914.558 | 5.09E-08 | 914.558 | 5.11E-08 | 914.558 | 2.78E+13 | 914.558 | 0.00602 |
| 760.2308 | 1.49E+11 | 914.833 | 1.41E+21 | 914.833 | 0.38727 | 914.833 | 5.09E-08 | 914.833 | 5.15E-08 | 914.833 | 2.79E+13 | 914.833 | 0.00602 |
| 760.5147 | 1.49E+11 | 915.108 | 1.41E+21 | 915.108 | 0.38721 | 915.108 | 5.10E-08 | 915.108 | 5.15E-08 | 915.108 | 2.79E+13 | 915.108 | 0.00599 |
| 760.7986 | 1.50E+11 | 915.383 | 1.41E+21 | 915.383 | 0.38716 | 915.383 | 5.10E-08 | 915.383 | 5.03E-08 | 915.383 | 2.79E+13 | 915.383 | 0.00597 |
| 761.0825 | 1.50E+11 | 915.658 | 1.42E+21 | 915.658 | 0.38711 | 915.658 | 5.09E-08 | 915.658 | 5.23E-08 | 915.658 | 2.79E+13 | 915.658 | 0.00594 |
| 761.3664 | 1.50E+11 | 915.933 | 1.42E+21 | 915.933 | 0.38706 | 915.933 | 5.10E-08 | 915.933 | 5.11E-08 | 915.933 | 2.80E+13 | 915.933 | 0.00594 |
| 761.6503 | 1.50E+11 | 916.207 | 1.42E+21 | 916.207 | 0.387 | 916.207 | 5.10E-08 | 916.207 | 5.22E-08 | 916.207 | 2.80E+13 | 916.207 | 0.00593 |
| 761.9341 | 1.51E+11 | 916.482 | 1.42E+21 | 916.482 | 0.38695 | 916.482 | 5.11E-08 | 916.482 | 4.89E-08 | 916.482 | 2.80E+13 | 916.482 | 0.00593 |
| 762.218 | 1.52E+11 | 916.757 | 1.42E+21 | 916.757 | 0.3869 | 916.757 | 5.11E-08 | 916.757 | 5.09E-08 | 916.757 | 2.81E+13 | 916.757 | 0.00589 |
| 762.5018 | 1.51E+11 | 917.032 | 1.42E+21 | 917.032 | 0.38684 | 917.032 | 5.12E-08 | 917.032 | 5.21E-08 | 917.032 | 2.81E+13 | 917.032 | 0.00588 |
| 762.7856 | 1.50E+11 | 917.307 | 1.42E+21 | 917.307 | 0.38679 | 917.307 | 5.11E-08 | 917.307 | 5.20E-08 | 917.307 | 2.81E+13 | 917.307 | 0.00587 |
| 763.0694 | 1.51E+11 | 917.581 | 1.42E+21 | 917.581 | 0.38674 | 917.581 | 5.12E-08 | 917.581 | 5.36E-08 | 917.581 | 2.82E+13 | 917.581 | 0.00584 |
| 763.3531 | 1.52E+11 | 917.856 | 1.42E+21 | 917.856 | 0.38668 | 917.856 | 5.12E-08 | 917.856 | 5.26E-08 | 917.856 | 2.82E+13 | 917.856 | 0.00586 |
| 763.6369 | 1.51E+11 | 918.131 | 1.43E+21 | 918.131 | 0.38663 | 918.131 | 5.12E-08 | 918.131 | 5.01E-08 | 918.131 | 2.82E+13 | 918.131 | 0.00582 |
| 763.9207 | 1.52E+11 | 918.405 | 1.43E+21 | 918.405 | 0.38658 | 918.405 | 5.12E-08 | 918.405 | 5.20E-08 | 918.405 | 2.82E+13 | 918.405 | 0.00579 |
| 764.2044 | 1.52E+11 | 918.68 | 1.43E+21 | 918.68 | 0.38653 | 918.68 | 5.12E-08 | 918.68 | 4.91E-08 | 918.68 | 2.82E+13 | 918.68 | 0.00579 |
| 764.4881 | 1.52E+11 | 918.955 | 1.43E+21 | 918.955 | 0.38647 | 918.955 | 5.12E-08 | 918.955 | 5.27E-08 | 918.955 | 2.83E+13 | 918.955 | 0.00577 |
| 764.7718 | 1.52E+11 | 919.229 | 1.43E+21 | 919.229 | 0.38642 | 919.229 | 5.13E-08 | 919.229 | 5.00E-08 | 919.229 | 2.83E+13 | 919.229 | 0.00576 |
| 765.0555 | 1.53E+11 | 919.504 | 1.43E+21 | 919.504 | 0.38637 | 919.504 | 5.13E-08 | 919.504 | 5.01E-08 | 919.504 | 2.83E+13 | 919.504 | 0.00574 |
| 765.3392 | 1.53E+11 | 919.778 | 1.43E+21 | 919.778 | 0.38631 | 919.778 | 5.13E-08 | 919.778 | 5.14E-08 | 919.778 | 2.84E+13 | 919.778 | 0.00573 |
| 765.6228 | 1.54E+11 | 920.053 | 1.43E+21 | 920.053 | 0.38626 | 920.053 | 5.14E-08 | 920.053 | 5.04E-08 | 920.053 | 2.84E+13 | 920.053 | 0.00569 |
| 765.9065 | 1.53E+11 | 920.328 | 1.43E+21 | 920.328 | 0.38621 | 920.328 | 5.14E-08 | 920.328 | 5.39E-08 | 920.328 | 2.84E+13 | 920.328 | 0.00568 |
| 766.1901 | 1.54E+11 | 920.602 | 1.43E+21 | 920.602 | 0.38615 | 920.602 | 5.14E-08 | 920.602 | 5.16E-08 | 920.602 | 2.85E+13 | 920.602 | 0.00569 |
| 766.4737 | 1.53E+11 | 920.877 | 1.44E+21 | 920.877 | 0.3861 | 920.877 | 5.14E-08 | 920.877 | 5.13E-08 | 920.877 | 2.85E+13 | 920.877 | 0.00568 |
| 766.7573 | 1.54E+11 | 921.151 | 1.44E+21 | 921.151 | 0.38605 | 921.151 | 5.15E-08 | 921.151 | 4.98E-08 | 921.151 | 2.85E+13 | 921.151 | 0.00563 |
| 767.0409 | 1.54E+11 | 921.426 | 1.44E+21 | 921.426 | 0.38599 | 921.426 | 5.14E-08 | 921.426 | 5.22E-08 | 921.426 | 2.85E+13 | 921.426 | 0.00562 |
| 767.3245 | 1.55E+11 | 921.7 | 1.44E+21 | 921.7 | 0.38594 | 921.7 | 5.15E-08 | 921.7 | 5.19E-08 | 921.7 | 2.86E+13 | 921.7 | 0.00559 |
| 767.608 | 1.55E+11 | 921.974 | 1.44E+21 | 921.974 | 0.38589 | 921.974 | 5.15E-08 | 921.974 | 5.07E-08 | 921.974 | 2.86E+13 | 921.974 | 0.00561 |
| 767.8915 | 1.54E+11 | 922.249 | 1.44E+21 | 922.249 | 0.38583 | 922.249 | 5.16E-08 | 922.249 | 5.24E-08 | 922.249 | 2.87E+13 | 922.249 | 0.00557 |
| 768.1751 | 1.56E+11 | 922.523 | 1.44E+21 | 922.523 | 0.38578 | 922.523 | 5.16E-08 | 922.523 | 5.32E-08 | 922.523 | 2.87E+13 | 922.523 | 0.00554 |
| 768.4586 | 1.54E+11 | 922.798 | 1.44E+21 | 922.798 | 0.38573 | 922.798 | 5.15E-08 | 922.798 | 4.96E-08 | 922.798 | 2.87E+13 | 922.798 | 0.00552 |
| 768.7421 | 1.56E+11 | 923.072 | 1.44E+21 | 923.072 | 0.38567 | 923.072 | 5.16E-08 | 923.072 | 5.30E-08 | 923.072 | 2.87E+13 | 923.072 | 0.00551 |
| 769.0256 | 1.56E+11 | 923.346 | 1.44E+21 | 923.346 | 0.38562 | 923.346 | 5.16E-08 | 923.346 | 5.02E-08 | 923.346 | 2.87E+13 | 923.346 | 0.00552 |
| 769.309 | 1.54E+11 | 923.621 | 1.45E+21 | 923.621 | 0.38557 | 923.621 | 5.16E-08 | 923.621 | 5.15E-08 | 923.621 | 2.87E+13 | 923.621 | 0.0055 |
| 769.5925 | 1.57E+11 | 923.895 | 1.45E+21 | 923.895 | 0.38551 | 923.895 | 5.16E-08 | 923.895 | 5.10E-08 | 923.895 | 2.88E+13 | 923.895 | 0.00546 |
| 769.8759 | 1.57E+11 | 924.169 | 1.45E+21 | 924.169 | 0.38546 | 924.169 | 5.17E-08 | 924.169 | 5.08E-08 | 924.169 | 2.88E+13 | 924.169 | 0.00544 |
| 770.1593 | 1.57E+11 | 924.444 | 1.45E+21 | 924.444 | 0.38541 | 924.444 | 5.17E-08 | 924.444 | 5.21E-08 | 924.444 | 2.89E+13 | 924.444 | 0.0054 |
| 770.4427 | 1.57E+11 | 924.718 | 1.45E+21 | 924.718 | 0.38535 | 924.718 | 5.17E-08 | 924.718 | 5.12E-08 | 924.718 | 2.89E+13 | 924.718 | 0.00538 |
| 770.7261 | 1.57E+11 | 924.992 | 1.45E+21 | 924.992 | 0.3853 | 924.992 | 5.18E-08 | 924.992 | 5.01E-08 | 924.992 | 2.89E+13 | 924.992 | 0.00539 |
| 771.0095 | 1.57E+11 | 925.266 | 1.45E+21 | 925.266 | 0.38525 | 925.266 | 5.18E-08 | 925.266 | 5.11E-08 | 925.266 | 2.89E+13 | 925.266 | 0.00538 |
| 771.2929 | 1.57E+11 | 925.541 | 1.45E+21 | 925.541 | 0.38519 | 925.541 | 5.18E-08 | 925.541 | 5.20E-08 | 925.541 | 2.90E+13 | 925.541 | 0.00538 |
| 771.5762 | 1.58E+11 | 925.815 | 1.45E+21 | 925.815 | 0.38514 | 925.815 | 5.18E-08 | 925.815 | 5.14E-08 | 925.815 | 2.90E+13 | 925.815 | 0.00537 |
| 771.8595 | 1.59E+11 | 926.089 | 1.45E+21 | 926.089 | 0.38509 | 926.089 | 5.18E-08 | 926.089 | 5.19E-08 | 926.089 | 2.90E+13 | 926.089 | 0.00534 |
| 772.1429 | 1.59E+11 | 926.363 | 1.46E+21 | 926.363 | 0.38503 | 926.363 | 5.18E-08 | 926.363 | 5.28E-08 | 926.363 | 2.90E+13 | 926.363 | 0.00536 |
| 772.4262 | 1.59E+11 | 926.637 | 1.46E+21 | 926.637 | 0.38498 | 926.637 | 5.19E-08 | 926.637 | 5.24E-08 | 926.637 | 2.91E+13 | 926.637 | 0.0053 |
| 772.7095 | 1.59E+11 | 926.911 | 1.46E+21 | 926.911 | 0.38493 | 926.911 | 5.20E-08 | 926.911 | 5.16E-08 | 926.911 | 2.92E+13 | 926.911 | 0.00527 |
| 772.9927 | 1.59E+11 | 927.185 | 1.46E+21 | 927.185 | 0.38487 | 927.185 | 5.20E-08 | 927.185 | 5.25E-08 | 927.185 | 2.92E+13 | 927.185 | 0.00524 |
| 773.276 | 1.60E+11 | 927.459 | 1.46E+21 | 927.459 | 0.38482 | 927.459 | 5.20E-08 | 927.459 | 5.23E-08 | 927.459 | 2.92E+13 | 927.459 | 0.00523 |
| 773.5592 | 1.60E+11 | 927.733 | 1.46E+21 | 927.733 | 0.38477 | 927.733 | 5.20E-08 | 927.733 | 5.27E-08 | 927.733 | 2.92E+13 | 927.733 | 0.00519 |
| 773.8425 | 1.60E+11 | 928.008 | 1.46E+21 | 928.008 | 0.38471 | 928.008 | 5.21E-08 | 928.008 | 5.04E-08 | 928.008 | 2.93E+13 | 928.008 | 0.00522 |
| 774.1257 | 1.61E+11 | 928.282 | 1.46E+21 | 928.282 | 0.38466 | 928.282 | 5.21E-08 | 928.282 | 5.18E-08 | 928.282 | 2.93E+13 | 928.282 | 0.0052 |
| 774.4089 | 1.61E+11 | 928.556 | 1.46E+21 | 928.556 | 0.38461 | 928.556 | 5.21E-08 | 928.556 | 5.32E-08 | 928.556 | 2.93E+13 | 928.556 | 0.00515 |
| 774.692 | 1.61E+11 | 928.83 | 1.46E+21 | 928.83 | 0.38455 | 928.83 | 5.21E-08 | 928.83 | 5.30E-08 | 928.83 | 2.93E+13 | 928.83 | 0.00513 |
| 774.9752 | 1.61E+11 | 929.103 | 1.47E+21 | 929.103 | 0.3845 | 929.103 | 5.22E-08 | 929.103 | 5.26E-08 | 929.103 | 2.94E+13 | 929.103 | 0.00513 |
| 775.2584 | 1.62E+11 | 929.377 | 1.47E+21 | 929.377 | 0.38445 | 929.377 | 5.22E-08 | 929.377 | 5.24E-08 | 929.377 | 2.94E+13 | 929.377 | 0.00511 |
| 775.5415 | 1.62E+11 | 929.651 | 1.47E+21 | 929.651 | 0.38439 | 929.651 | 5.22E-08 | 929.651 | 5.37E-08 | 929.651 | 2.95E+13 | 929.651 | 0.00511 |
| 775.8246 | 1.62E+11 | 929.925 | 1.47E+21 | 929.925 | 0.38434 | 929.925 | 5.23E-08 | 929.925 | 5.15E-08 | 929.925 | 2.95E+13 | 929.925 | 0.00507 |
| 776.1077 | 1.62E+11 | 930.199 | 1.47E+21 | 930.199 | 0.38429 | 930.199 | 5.23E-08 | 930.199 | 5.24E-08 | 930.199 | 2.95E+13 | 930.199 | 0.00506 |
| 776.3908 | 1.62E+11 | 930.473 | 1.47E+21 | 930.473 | 0.38423 | 930.473 | 5.24E-08 | 930.473 | 5.20E-08 | 930.473 | 2.96E+13 | 930.473 | 0.00505 |
| 776.6739 | 1.62E+11 | 930.747 | 1.47E+21 | 930.747 | 0.38418 | 930.747 | 5.24E-08 | 930.747 | 5.04E-08 | 930.747 | 2.96E+13 | 930.747 | 0.00499 |
| 776.957 | 1.63E+11 | 931.021 | 1.47E+21 | 931.021 | 0.38413 | 931.021 | 5.25E-08 | 931.021 | 5.20E-08 | 931.021 | 2.97E+13 | 931.021 | 0.00496 |
| 777.24 | 1.64E+11 | 931.295 | 1.47E+21 | 931.295 | 0.38407 | 931.295 | 5.25E-08 | 931.295 | 5.12E-08 | 931.295 | 2.97E+13 | 931.295 | 0.00495 |
| 777.5231 | 1.62E+11 | 931.568 | 1.47E+21 | 931.568 | 0.38402 | 931.568 | 5.26E-08 | 931.568 | 5.18E-08 | 931.568 | 2.98E+13 | 931.568 | 0.00495 |
| 777.8061 | 1.63E+11 | 931.842 | 1.48E+21 | 931.842 | 0.38397 | 931.842 | 5.26E-08 | 931.842 | 5.39E-08 | 931.842 | 2.98E+13 | 931.842 | 0.00493 |
| 778.0891 | 1.64E+11 | 932.116 | 1.48E+21 | 932.116 | 0.38391 | 932.116 | 5.26E-08 | 932.116 | 5.42E-08 | 932.116 | 2.98E+13 | 932.116 | 0.00494 |
| 778.3721 | 1.65E+11 | 932.39 | 1.48E+21 | 932.39 | 0.38386 | 932.39 | 5.26E-08 | 932.39 | 5.36E-08 | 932.39 | 2.98E+13 | 932.39 | 0.00492 |
| 778.655 | 1.64E+11 | 932.663 | 1.48E+21 | 932.663 | 0.38381 | 932.663 | 5.26E-08 | 932.663 | 5.25E-08 | 932.663 | 2.98E+13 | 932.663 | 0.00487 |
| 778.938 | 1.64E+11 | 932.937 | 1.48E+21 | 932.937 | 0.38375 | 932.937 | 5.25E-08 | 932.937 | 5.24E-08 | 932.937 | 2.98E+13 | 932.937 | 0.00492 |
| 779.2209 | 1.64E+11 | 933.211 | 1.48E+21 | 933.211 | 0.3837 | 933.211 | 5.25E-08 | 933.211 | 5.23E-08 | 933.211 | 2.98E+13 | 933.211 | 0.00486 |
| 779.5039 | 1.65E+11 | 933.484 | 1.48E+21 | 933.484 | 0.38365 | 933.484 | 5.25E-08 | 933.484 | 5.37E-08 | 933.484 | 2.98E+13 | 933.484 | 0.00486 |
| 779.7868 | 1.65E+11 | 933.758 | 1.48E+21 | 933.758 | 0.38359 | 933.758 | 5.26E-08 | 933.758 | 5.33E-08 | 933.758 | 2.99E+13 | 933.758 | 0.0048 |
| 780.0697 | 1.66E+11 | 934.032 | 1.48E+21 | 934.032 | 0.38354 | 934.032 | 5.26E-08 | 934.032 | 5.22E-08 | 934.032 | 2.99E+13 | 934.032 | 0.0048 |
| 780.3526 | 1.66E+11 | 934.305 | 1.48E+21 | 934.305 | 0.38349 | 934.305 | 5.25E-08 | 934.305 | 5.25E-08 | 934.305 | 2.99E+13 | 934.305 | 0.00482 |
| 780.6354 | 1.65E+11 | 934.579 | 1.49E+21 | 934.579 | 0.38343 | 934.579 | 5.25E-08 | 934.579 | 5.23E-08 | 934.579 | 2.99E+13 | 934.579 | 0.00476 |
| 780.9183 | 1.66E+11 | 934.853 | 1.49E+21 | 934.853 | 0.38338 | 934.853 | 5.25E-08 | 934.853 | 5.43E-08 | 934.853 | 2.99E+13 | 934.853 | 0.00478 |
| 781.2011 | 1.66E+11 | 935.126 | 1.49E+21 | 935.126 | 0.38333 | 935.126 | 5.25E-08 | 935.126 | 5.17E-08 | 935.126 | 3.00E+13 | 935.126 | 0.00478 |
| 781.484 | 1.67E+11 | 935.4 | 1.49E+21 | 935.4 | 0.38327 | 935.4 | 5.25E-08 | 935.4 | 5.28E-08 | 935.4 | 3.00E+13 | 935.4 | 0.00473 |
| 781.7668 | 1.67E+11 | 935.673 | 1.49E+21 | 935.673 | 0.38322 | 935.673 | 5.26E-08 | 935.673 | 5.23E-08 | 935.673 | 3.00E+13 | 935.673 | 0.00472 |
| 782.0496 | 1.67E+11 | 935.947 | 1.49E+21 | 935.947 | 0.38316 | 935.947 | 5.26E-08 | 935.947 | 5.33E-08 | 935.947 | 3.00E+13 | 935.947 | 0.00471 |
| 782.3323 | 1.67E+11 | 936.22 | 1.49E+21 | 936.22 | 0.38311 | 936.22 | 5.26E-08 | 936.22 | 5.30E-08 | 936.22 | 3.01E+13 | 936.22 | 0.00468 |
| 782.6151 | 1.67E+11 | 936.494 | 1.49E+21 | 936.494 | 0.38306 | 936.494 | 5.27E-08 | 936.494 | 5.54E-08 | 936.494 | 3.01E+13 | 936.494 | 0.00466 |
| 782.8978 | 1.67E+11 | 936.767 | 1.49E+21 | 936.767 | 0.383 | 936.767 | 5.27E-08 | 936.767 | 5.28E-08 | 936.767 | 3.01E+13 | 936.767 | 0.00464 |
| 783.1806 | 1.67E+11 | 937.04 | 1.49E+21 | 937.04 | 0.38295 | 937.04 | 5.27E-08 | 937.04 | 5.22E-08 | 937.04 | 3.02E+13 | 937.04 | 0.00462 |
| 783.4633 | 1.68E+11 | 937.314 | 1.50E+21 | 937.314 | 0.3829 | 937.314 | 5.28E-08 | 937.314 | 5.20E-08 | 937.314 | 3.02E+13 | 937.314 | 0.00461 |
| 783.746 | 1.68E+11 | 937.587 | 1.50E+21 | 937.587 | 0.38284 | 937.587 | 5.27E-08 | 937.587 | 5.29E-08 | 937.587 | 3.02E+13 | 937.587 | 0.00459 |
| 784.0287 | 1.68E+11 | 937.861 | 1.50E+21 | 937.861 | 0.38279 | 937.861 | 5.27E-08 | 937.861 | 5.18E-08 | 937.861 | 3.02E+13 | 937.861 | 0.00458 |
| 784.3114 | 1.68E+11 | 938.134 | 1.50E+21 | 938.134 | 0.38274 | 938.134 | 5.27E-08 | 938.134 | 5.18E-08 | 938.134 | 3.02E+13 | 938.134 | 0.00456 |
| 784.594 | 1.68E+11 | 938.407 | 1.50E+21 | 938.407 | 0.38268 | 938.407 | 5.27E-08 | 938.407 | 5.09E-08 | 938.407 | 3.02E+13 | 938.407 | 0.00454 |
| 784.8767 | 1.69E+11 | 938.681 | 1.50E+21 | 938.681 | 0.38263 | 938.681 | 5.27E-08 | 938.681 | 5.33E-08 | 938.681 | 3.03E+13 | 938.681 | 0.00455 |
| 785.1593 | 1.70E+11 | 938.954 | 1.50E+21 | 938.954 | 0.38258 | 938.954 | 5.28E-08 | 938.954 | 5.15E-08 | 938.954 | 3.03E+13 | 938.954 | 0.00449 |
| 785.4419 | 1.70E+11 | 939.227 | 1.50E+21 | 939.227 | 0.38252 | 939.227 | 5.28E-08 | 939.227 | 5.30E-08 | 939.227 | 3.03E+13 | 939.227 | 0.00449 |
| 785.7245 | 1.70E+11 | 939.501 | 1.50E+21 | 939.501 | 0.38247 | 939.501 | 5.27E-08 | 939.501 | 5.28E-08 | 939.501 | 3.03E+13 | 939.501 | 0.00448 |
| 786.0071 | 1.70E+11 | 939.774 | 1.50E+21 | 939.774 | 0.38241 | 939.774 | 5.28E-08 | 939.774 | 5.08E-08 | 939.774 | 3.04E+13 | 939.774 | 0.00445 |
| 786.2896 | 1.69E+11 | 940.047 | 1.51E+21 | 940.047 | 0.38325 | 940.047 | 5.28E-08 | 940.047 | 5.21E-08 | 940.047 | 3.04E+13 | 940.047 | 0.00444 |
| 786.5722 | 1.71E+11 | 940.32 | 1.51E+21 | 940.32 | 0.3832 | 940.32 | 5.28E-08 | 940.32 | 5.16E-08 | 940.32 | 3.05E+13 | 940.32 | 0.00439 |
| 786.8547 | 1.71E+11 | 940.593 | 1.51E+21 | 940.593 | 0.38315 | 940.593 | 5.27E-08 | 940.593 | 5.38E-08 | 940.593 | 3.05E+13 | 940.593 | 0.00439 |
| 787.1372 | 1.70E+11 | 940.867 | 1.51E+21 | 940.867 | 0.3831 | 940.867 | 5.28E-08 | 940.867 | 5.37E-08 | 940.867 | 3.05E+13 | 940.867 | 0.00435 |
| 787.4198 | 1.70E+11 | 941.14 | 1.51E+21 | 941.14 | 0.38305 | 941.14 | 5.28E-08 | 941.14 | 5.28E-08 | 941.14 | 3.06E+13 | 941.14 | 0.00437 |
| 787.7022 | 1.71E+11 | 941.413 | 1.51E+21 | 941.413 | 0.383 | 941.413 | 5.28E-08 | 941.413 | 5.42E-08 | 941.413 | 3.05E+13 | 941.413 | 0.00432 |
| 787.9847 | 1.71E+11 | 941.686 | 1.51E+21 | 941.686 | 0.38295 | 941.686 | 5.28E-08 | 941.686 | 5.23E-08 | 941.686 | 3.06E+13 | 941.686 | 0.00434 |
| 788.2672 | 1.71E+11 | 941.959 | 1.51E+21 | 941.959 | 0.3829 | 941.959 | 5.28E-08 | 941.959 | 5.33E-08 | 941.959 | 3.06E+13 | 941.959 | 0.0043 |
| 788.5496 | 1.72E+11 | 942.232 | 1.51E+21 | 942.232 | 0.38285 | 942.232 | 5.28E-08 | 942.232 | 5.25E-08 | 942.232 | 3.06E+13 | 942.232 | 0.00427 |
| 788.8321 | 1.72E+11 | 942.505 | 1.51E+21 | 942.505 | 0.3828 | 942.505 | 5.28E-08 | 942.505 | 5.30E-08 | 942.505 | 3.06E+13 | 942.505 | 0.00426 |
| 789.1145 | 1.73E+11 | 942.778 | 1.52E+21 | 942.778 | 0.38275 | 942.778 | 5.28E-08 | 942.778 | 5.31E-08 | 942.778 | 3.06E+13 | 942.778 | 0.00422 |
| 789.3969 | 1.72E+11 | 943.051 | 1.52E+21 | 943.051 | 0.3827 | 943.051 | 5.29E-08 | 943.051 | 5.17E-08 | 943.051 | 3.07E+13 | 943.051 | 0.00422 |
| 789.6792 | 1.71E+11 | 943.324 | 1.52E+21 | 943.324 | 0.38265 | 943.324 | 5.28E-08 | 943.324 | 5.33E-08 | 943.324 | 3.07E+13 | 943.324 | 0.00424 |
| 789.9616 | 1.73E+11 | 943.597 | 1.52E+21 | 943.597 | 0.3826 | 943.597 | 5.29E-08 | 943.597 | 5.41E-08 | 943.597 | 3.07E+13 | 943.597 | 0.00418 |
| 790.244 | 1.72E+11 | 943.87 | 1.52E+21 | 943.87 | 0.38255 | 943.87 | 5.29E-08 | 943.87 | 5.23E-08 | 943.87 | 3.07E+13 | 943.87 | 0.00419 |
| 790.5263 | 1.74E+11 | 944.143 | 1.52E+21 | 944.143 | 0.3825 | 944.143 | 5.29E-08 | 944.143 | 5.25E-08 | 944.143 | 3.08E+13 | 944.143 | 0.00415 |
| 790.8086 | 1.73E+11 | 944.416 | 1.52E+21 | 944.416 | 0.38245 | 944.416 | 5.30E-08 | 944.416 | 5.48E-08 | 944.416 | 3.08E+13 | 944.416 | 0.00413 |
| 791.0909 | 1.72E+11 | 944.689 | 1.52E+21 | 944.689 | 0.3824 | 944.689 | 5.30E-08 | 944.689 | 5.35E-08 | 944.689 | 3.09E+13 | 944.689 | 0.00412 |
| 791.3732 | 1.73E+11 | 944.962 | 1.52E+21 | 944.962 | 0.38235 | 944.962 | 5.30E-08 | 944.962 | 5.24E-08 | 944.962 | 3.09E+13 | 944.962 | 0.00414 |
| 791.6555 | 1.74E+11 | 945.235 | 1.52E+21 | 945.235 | 0.3823 | 945.235 | 5.31E-08 | 945.235 | 5.35E-08 | 945.235 | 3.09E+13 | 945.235 | 0.00408 |
| 791.9378 | 1.74E+11 | 945.508 | 1.52E+21 | 945.508 | 0.38225 | 945.508 | 5.32E-08 | 945.508 | 5.27E-08 | 945.508 | 3.10E+13 | 945.508 | 0.00406 |
| 792.22 | 1.74E+11 | 945.781 | 1.53E+21 | 945.781 | 0.3822 | 945.781 | 5.32E-08 | 945.781 | 5.30E-08 | 945.781 | 3.10E+13 | 945.781 | 0.00404 |
| 792.5022 | 1.73E+11 | 946.053 | 1.53E+21 | 946.053 | 0.38215 | 946.053 | 5.32E-08 | 946.053 | 5.21E-08 | 946.053 | 3.11E+13 | 946.053 | 0.00401 |
| 792.7845 | 1.74E+11 | 946.326 | 1.53E+21 | 946.326 | 0.3821 | 946.326 | 5.32E-08 | 946.326 | 5.51E-08 | 946.326 | 3.11E+13 | 946.326 | 0.004 |
| 793.0667 | 1.75E+11 | 946.599 | 1.53E+21 | 946.599 | 0.38205 | 946.599 | 5.32E-08 | 946.599 | 5.38E-08 | 946.599 | 3.11E+13 | 946.599 | 0.004 |
| 793.3488 | 1.74E+11 | 946.872 | 1.53E+21 | 946.872 | 0.382 | 946.872 | 5.32E-08 | 946.872 | 5.09E-08 | 946.872 | 3.11E+13 | 946.872 | 0.00396 |
| 793.631 | 1.75E+11 | 947.144 | 1.53E+21 | 947.144 | 0.38195 | 947.144 | 5.32E-08 | 947.144 | 5.28E-08 | 947.144 | 3.11E+13 | 947.144 | 0.00397 |
| 793.9132 | 1.75E+11 | 947.417 | 1.53E+21 | 947.417 | 0.3819 | 947.417 | 5.32E-08 | 947.417 | 5.44E-08 | 947.417 | 3.11E+13 | 947.417 | 0.00396 |
| 794.1953 | 1.75E+11 | 947.69 | 1.53E+21 | 947.69 | 0.38186 | 947.69 | 5.32E-08 | 947.69 | 5.43E-08 | 947.69 | 3.11E+13 | 947.69 | 0.00395 |
| 794.4774 | 1.76E+11 | 947.963 | 1.53E+21 | 947.963 | 0.38181 | 947.963 | 5.33E-08 | 947.963 | 5.24E-08 | 947.963 | 3.12E+13 | 947.963 | 0.00389 |
| 794.7595 | 1.75E+11 | 948.235 | 1.53E+21 | 948.235 | 0.38176 | 948.235 | 5.33E-08 | 948.235 | 5.29E-08 | 948.235 | 3.12E+13 | 948.235 | 0.00389 |
| 795.0416 | 1.74E+11 | 948.508 | 1.54E+21 | 948.508 | 0.38171 | 948.508 | 5.33E-08 | 948.508 | 5.33E-08 | 948.508 | 3.12E+13 | 948.508 | 0.0039 |
| 795.3237 | 1.76E+11 | 948.781 | 1.54E+21 | 948.781 | 0.38166 | 948.781 | 5.34E-08 | 948.781 | 5.22E-08 | 948.781 | 3.13E+13 | 948.781 | 0.00386 |
| 795.6058 | 1.76E+11 | 949.053 | 1.54E+21 | 949.053 | 0.38161 | 949.053 | 5.33E-08 | 949.053 | 5.33E-08 | 949.053 | 3.13E+13 | 949.053 | 0.00385 |
| 795.8878 | 1.76E+11 | 949.326 | 1.54E+21 | 949.326 | 0.38156 | 949.326 | 5.33E-08 | 949.326 | 5.21E-08 | 949.326 | 3.13E+13 | 949.326 | 0.00384 |
| 796.1698 | 1.76E+11 | 949.598 | 1.54E+21 | 949.598 | 0.38151 | 949.598 | 5.33E-08 | 949.598 | 5.32E-08 | 949.598 | 3.13E+13 | 949.598 | 0.00382 |
| 796.4519 | 1.76E+11 | 949.871 | 1.54E+21 | 949.871 | 0.38146 | 949.871 | 5.33E-08 | 949.871 | 5.40E-08 | 949.871 | 3.13E+13 | 949.871 | 0.00381 |
| 796.7339 | 1.77E+11 | 950.143 | 1.54E+21 | 950.143 | 0.38141 | 950.143 | 5.33E-08 | 950.143 | 5.41E-08 | 950.143 | 3.13E+13 | 950.143 | 0.0038 |
| 797.0158 | 1.77E+11 | 950.416 | 1.54E+21 | 950.416 | 0.38136 | 950.416 | 5.33E-08 | 950.416 | 5.24E-08 | 950.416 | 3.14E+13 | 950.416 | 0.00377 |
| 797.2978 | 1.78E+11 | 950.688 | 1.54E+21 | 950.688 | 0.38131 | 950.688 | 5.33E-08 | 950.688 | 5.52E-08 | 950.688 | 3.14E+13 | 950.688 | 0.00375 |
| 797.5798 | 1.76E+11 | 950.961 | 1.54E+21 | 950.961 | 0.38126 | 950.961 | 5.34E-08 | 950.961 | 5.47E-08 | 950.961 | 3.14E+13 | 950.961 | 0.00372 |
| 797.8617 | 1.76E+11 | 951.233 | 1.55E+21 | 951.233 | 0.38121 | 951.233 | 5.34E-08 | 951.233 | 5.44E-08 | 951.233 | 3.14E+13 | 951.233 | 0.00373 |
| 798.1436 | 1.77E+11 | 951.506 | 1.55E+21 | 951.506 | 0.38116 | 951.506 | 5.34E-08 | 951.506 | 5.25E-08 | 951.506 | 3.15E+13 | 951.506 | 0.0037 |
| 798.4255 | 1.77E+11 | 951.778 | 1.55E+21 | 951.778 | 0.38111 | 951.778 | 5.35E-08 | 951.778 | 5.32E-08 | 951.778 | 3.15E+13 | 951.778 | 0.00371 |
| 798.7074 | 1.77E+11 | 952.051 | 1.55E+21 | 952.051 | 0.38106 | 952.051 | 5.35E-08 | 952.051 | 5.44E-08 | 952.051 | 3.15E+13 | 952.051 | 0.00369 |
| 798.9893 | 1.78E+11 | 952.323 | 1.55E+21 | 952.323 | 0.38101 | 952.323 | 5.34E-08 | 952.323 | 5.31E-08 | 952.323 | 3.15E+13 | 952.323 | 0.00364 |
| 799.2712 | 1.77E+11 | 952.595 | 1.55E+21 | 952.595 | 0.38096 | 952.595 | 5.35E-08 | 952.595 | 5.28E-08 | 952.595 | 3.16E+13 | 952.595 | 0.00364 |
| 799.553 | 1.77E+11 | 952.868 | 1.55E+21 | 952.868 | 0.38091 | 952.868 | 5.35E-08 | 952.868 | 5.24E-08 | 952.868 | 3.16E+13 | 952.868 | 0.00363 |
| 799.8348 | 1.78E+11 | 953.14 | 1.55E+21 | 953.14 | 0.38086 | 953.14 | 5.35E-08 | 953.14 | 5.28E-08 | 953.14 | 3.16E+13 | 953.14 | 0.00366 |
| 800.1167 | 1.78E+11 | 953.412 | 1.55E+21 | 953.412 | 0.38082 | 953.412 | 5.35E-08 | 953.412 | 5.51E-08 | 953.412 | 3.16E+13 | 953.412 | 0.0036 |
| 800.3985 | 1.79E+11 | 953.685 | 1.55E+21 | 953.685 | 0.38077 | 953.685 | 5.35E-08 | 953.685 | 5.26E-08 | 953.685 | 3.17E+13 | 953.685 | 0.00358 |
| 800.6802 | 1.78E+11 | 953.957 | 1.55E+21 | 953.957 | 0.38072 | 953.957 | 5.36E-08 | 953.957 | 5.38E-08 | 953.957 | 3.17E+13 | 953.957 | 0.00359 |
| 800.962 | 1.78E+11 | 954.229 | 1.56E+21 | 954.229 | 0.38067 | 954.229 | 5.36E-08 | 954.229 | 5.59E-08 | 954.229 | 3.17E+13 | 954.229 | 0.00359 |
| 801.2438 | 1.79E+11 | 954.502 | 1.56E+21 | 954.502 | 0.38062 | 954.502 | 5.37E-08 | 954.502 | 5.17E-08 | 954.502 | 3.18E+13 | 954.502 | 0.00355 |
| 801.5255 | 1.79E+11 | 954.774 | 1.56E+21 | 954.774 | 0.38057 | 954.774 | 5.37E-08 | 954.774 | 5.30E-08 | 954.774 | 3.18E+13 | 954.774 | 0.00354 |
| 801.8072 | 1.78E+11 | 955.046 | 1.56E+21 | 955.046 | 0.38052 | 955.046 | 5.38E-08 | 955.046 | 5.35E-08 | 955.046 | 3.19E+13 | 955.046 | 0.00352 |
| 802.0889 | 1.79E+11 | 955.318 | 1.56E+21 | 955.318 | 0.38047 | 955.318 | 5.38E-08 | 955.318 | 5.29E-08 | 955.318 | 3.19E+13 | 955.318 | 0.00347 |
| 802.3706 | 1.80E+11 | 955.59 | 1.56E+21 | 955.59 | 0.38042 | 955.59 | 5.38E-08 | 955.59 | 5.30E-08 | 955.59 | 3.19E+13 | 955.59 | 0.00349 |
| 802.6523 | 1.79E+11 | 955.862 | 1.56E+21 | 955.862 | 0.38037 | 955.862 | 5.38E-08 | 955.862 | 5.34E-08 | 955.862 | 3.20E+13 | 955.862 | 0.00346 |
| 802.9339 | 1.79E+11 | 956.135 | 1.56E+21 | 956.135 | 0.38032 | 956.135 | 5.39E-08 | 956.135 | 5.33E-08 | 956.135 | 3.20E+13 | 956.135 | 0.00345 |
| 803.2156 | 1.79E+11 | 956.407 | 1.56E+21 | 956.407 | 0.38027 | 956.407 | 5.39E-08 | 956.407 | 5.55E-08 | 956.407 | 3.20E+13 | 956.407 | 0.00346 |
| 803.4972 | 1.80E+11 | 956.679 | 1.56E+21 | 956.679 | 0.38022 | 956.679 | 5.39E-08 | 956.679 | 5.34E-08 | 956.679 | 3.21E+13 | 956.679 | 0.00341 |
| 803.7788 | 1.80E+11 | 956.951 | 1.57E+21 | 956.951 | 0.38017 | 956.951 | 5.39E-08 | 956.951 | 5.40E-08 | 956.951 | 3.21E+13 | 956.951 | 0.00341 |
| 804.0604 | 1.81E+11 | 957.223 | 1.57E+21 | 957.223 | 0.38012 | 957.223 | 5.39E-08 | 957.223 | 5.52E-08 | 957.223 | 3.21E+13 | 957.223 | 0.00339 |
| 804.342 | 1.81E+11 | 957.495 | 1.57E+21 | 957.495 | 0.38007 | 957.495 | 5.39E-08 | 957.495 | 5.39E-08 | 957.495 | 3.21E+13 | 957.495 | 0.00338 |
| 804.6236 | 1.81E+11 | 957.767 | 1.57E+21 | 957.767 | 0.38003 | 957.767 | 5.39E-08 | 957.767 | 5.29E-08 | 957.767 | 3.21E+13 | 957.767 | 0.00337 |
| 804.9051 | 1.80E+11 | 958.039 | 1.57E+21 | 958.039 | 0.37998 | 958.039 | 5.40E-08 | 958.039 | 5.41E-08 | 958.039 | 3.22E+13 | 958.039 | 0.00335 |
| 805.1867 | 1.80E+11 | 958.311 | 1.57E+21 | 958.311 | 0.37993 | | | | | 958.311 | 3.23E+13 | 958.311 | 0.00334 |
| 805.4682 | 1.82E+11 | 958.583 | 1.57E+21 | 958.583 | 0.37988 | | | | | 958.583 | 3.23E+13 | 958.583 | 0.00332 |
| 805.7497 | 1.81E+11 | 958.855 | 1.57E+21 | 958.855 | 0.37983 | | | | | 958.855 | 3.23E+13 | 958.855 | 0.00334 |
| 806.0312 | 1.80E+11 | 959.127 | 1.57E+21 | 959.127 | 0.37978 | | | | | 959.127 | 3.23E+13 | 959.127 | 0.00331 |
| 806.3126 | 1.82E+11 | 959.399 | 1.57E+21 | 959.399 | 0.37973 | | | | | 959.399 | 3.24E+13 | 959.399 | 0.00327 |
| 806.5941 | 1.81E+11 | 959.671 | 1.57E+21 | 959.671 | 0.37968 | | | | | 959.671 | 3.24E+13 | 959.671 | 0.00327 |
| 806.8755 | 1.81E+11 | 959.942 | 1.58E+21 | 959.942 | 0.37963 | | | | | 959.942 | 3.24E+13 | 959.942 | 0.00324 |
| 807.157 | 1.81E+11 | 960.214 | 1.58E+21 | 960.214 | 0.37958 | | | | | | | | |
| 807.4384 | 1.82E+11 | 960.486 | 1.58E+21 | 960.486 | 0.37953 | | | | | 1152.85 | 4.11E+13 | | |
| 807.7198 | 1.83E+11 | 960.758 | 1.58E+21 | 960.758 | 0.37948 | | | | | 1154.57 | 4.11E+13 | | |
| 808.0011 | 1.82E+11 | 961.03 | 1.58E+21 | 961.03 | 0.37944 | | | | | 1156.29 | 4.11E+13 | | |
| 808.2825 | 1.81E+11 | 961.301 | 1.58E+21 | 961.301 | 0.37938 | | | | | 1158.02 | 4.12E+13 | | |
| 808.5639 | 1.82E+11 | 961.573 | 1.58E+21 | 961.573 | 0.37934 | | | | | 1159.74 | 4.11E+13 | | |
| 808.8452 | 1.83E+11 | 961.845 | 1.58E+21 | 961.845 | 0.37929 | | | | | 1161.46 | 4.11E+13 | | |
| 809.1265 | 1.82E+11 | 962.117 | 1.58E+21 | 962.117 | 0.37924 | | | | | 1163.18 | 4.11E+13 | | |
| 809.4078 | 1.83E+11 | 962.388 | 1.58E+21 | 962.388 | 0.37919 | | | | | 1164.9 | 4.11E+13 | | |
| 809.6891 | 1.83E+11 | 962.66 | 1.59E+21 | 962.66 | 0.37914 | | | | | 1166.62 | 4.11E+13 | | |
| 809.9704 | 1.84E+11 | 962.932 | 1.59E+21 | 962.932 | 0.37909 | | | | | 1168.34 | 4.11E+13 | | |
| 810.2516 | 1.84E+11 | 963.203 | 1.59E+21 | 963.203 | 0.37904 | | | | | 1170.06 | 4.11E+13 | | |
| 810.5328 | 1.84E+11 | 963.475 | 1.59E+21 | 963.475 | 0.37899 | | | | | 1171.78 | 4.11E+13 | | |
| 810.8141 | 1.84E+11 | 963.746 | 1.59E+21 | 963.746 | 0.37894 | | | | | 1173.5 | 4.11E+13 | | |
| 811.0953 | 1.83E+11 | 964.018 | 1.59E+21 | 964.018 | 0.37889 | | | | | 1175.22 | 4.11E+13 | | |
| 811.3765 | 1.84E+11 | 964.29 | 1.59E+21 | 964.29 | 0.37885 | | | | | 1176.94 | 4.11E+13 | | |
| 811.6576 | 1.83E+11 | 964.561 | 1.59E+21 | 964.561 | 0.3788 | | | | | 1178.66 | 4.10E+13 | | |
| 811.9388 | 1.84E+11 | 964.833 | 1.59E+21 | 964.833 | 0.37875 | | | | | 1180.38 | 4.10E+13 | | |
| 812.2199 | 1.85E+11 | 965.104 | 1.59E+21 | 965.104 | 0.3787 | | | | | 1182.1 | 4.11E+13 | | |
| 812.5011 | 1.85E+11 | 965.376 | 1.59E+21 | 965.376 | 0.37865 | | | | | 1183.82 | 4.10E+13 | | |
| 812.7822 | 1.84E+11 | 965.647 | 1.60E+21 | 965.647 | 0.3786 | | | | | 1185.54 | 4.10E+13 | | |
| 813.0633 | 1.86E+11 | 965.919 | 1.60E+21 | 965.919 | 0.37855 | | | | | 1187.26 | 4.09E+13 | | |
| 813.3443 | 1.85E+11 | 966.19 | 1.60E+21 | 966.19 | 0.3785 | | | | | 1188.98 | 4.09E+13 | | |
| 813.6254 | 1.86E+11 | 966.461 | 1.60E+21 | 966.461 | 0.37845 | | | | | 1190.7 | 4.10E+13 | | |
| 813.9065 | 1.85E+11 | 966.733 | 1.60E+21 | 966.733 | 0.3784 | | | | | 1192.42 | 4.10E+13 | | |
| 814.1875 | 1.84E+11 | 967.004 | 1.60E+21 | 967.004 | 0.37836 | | | | | 1194.14 | 4.10E+13 | | |
| 814.4685 | 1.85E+11 | 967.276 | 1.60E+21 | 967.276 | 0.37831 | | | | | 1195.86 | 4.10E+13 | | |
| 814.7495 | 1.85E+11 | 967.547 | 1.60E+21 | 967.547 | 0.37826 | | | | | 1197.58 | 4.10E+13 | | |
| 815.0305 | 1.85E+11 | 967.818 | 1.60E+21 | 967.818 | 0.37821 | | | | | 1199.3 | 4.10E+13 | | |
| 815.3115 | 1.85E+11 | 968.089 | 1.60E+21 | 968.089 | 0.37816 | | | | | 1201.02 | 4.10E+13 | | |
| 815.5924 | 1.85E+11 | 968.361 | 1.60E+21 | 968.361 | 0.37811 | | | | | 1202.73 | 4.10E+13 | | |
| 815.8734 | 1.86E+11 | 968.632 | 1.61E+21 | 968.632 | 0.37806 | | | | | 1204.45 | 4.09E+13 | | |
| 816.1543 | 1.85E+11 | 968.903 | 1.61E+21 | 968.903 | 0.37801 | | | | | 1206.17 | 4.10E+13 | | |
| 816.4352 | 1.86E+11 | 969.174 | 1.61E+21 | 969.174 | 0.37797 | | | | | 1207.89 | 4.09E+13 | | |

| | | | | | | | |
|---|---|---|---|---|---|---|---|
| 816.7161 | 1.86E+11 | 969.446 | 1.61E+21 | 969.446 | 0.37792 | 1209.61 | 4.09E+13 |
| 816.9969 | 1.86E+11 | 969.717 | 1.61E+21 | 969.717 | 0.37787 | 1211.32 | 4.09E+13 |
| 817.2778 | 1.87E+11 | 969.988 | 1.61E+21 | 969.988 | 0.37782 | 1213.04 | 4.09E+13 |
| 817.5586 | 1.86E+11 | 970.259 | 1.61E+21 | 970.259 | 0.37777 | 1214.76 | 4.09E+13 |
| 817.8395 | 1.87E+11 | 970.53 | 1.61E+21 | 970.53 | 0.37772 | 1216.48 | 4.09E+13 |
| 818.1203 | 1.88E+11 | 970.801 | 1.61E+21 | 970.801 | 0.37767 | 1218.19 | 4.08E+13 |
| 818.4011 | 1.89E+11 | 971.072 | 1.61E+21 | 971.072 | 0.37762 | 1219.91 | 4.08E+13 |
| 818.6818 | 1.88E+11 | 971.344 | 1.62E+21 | 971.344 | 0.37757 | 1221.63 | 4.08E+13 |
| 818.9626 | 1.87E+11 | 971.615 | 1.62E+21 | 971.615 | 0.37753 | 1223.35 | 4.08E+13 |
| 819.2434 | 1.87E+11 | 971.886 | 1.62E+21 | 971.886 | 0.37748 | 1225.06 | 4.08E+13 |
| 819.5241 | 1.87E+11 | 972.157 | 1.62E+21 | 972.157 | 0.37743 | 1226.78 | 4.07E+13 |
| 819.8048 | 1.88E+11 | 972.428 | 1.62E+21 | 972.428 | 0.37738 | 1228.5 | 4.07E+13 |
| 820.0855 | 1.88E+11 | 972.699 | 1.62E+21 | 972.699 | 0.37733 | 1230.21 | 4.06E+13 |
| 820.3662 | 1.89E+11 | 972.969 | 1.62E+21 | 972.969 | 0.37728 | 1231.93 | 4.06E+13 |
| 820.6468 | 1.88E+11 | 973.24 | 1.62E+21 | 973.24 | 0.37723 | 1233.65 | 4.06E+13 |
| 820.9275 | 1.88E+11 | 973.511 | 1.62E+21 | 973.511 | 0.37719 | 1235.36 | 4.06E+13 |
| 821.2081 | 1.89E+11 | 973.782 | 1.62E+21 | 973.782 | 0.37714 | 1237.08 | 4.06E+13 |
| 821.4888 | 1.89E+11 | 974.053 | 1.62E+21 | 974.053 | 0.37709 | 1238.79 | 4.05E+13 |
| 821.7694 | 1.89E+11 | 974.324 | 1.63E+21 | 974.324 | 0.37704 | 1240.51 | 4.05E+13 |
| 822.0499 | 1.89E+11 | 974.595 | 1.63E+21 | 974.595 | 0.37699 | 1242.23 | 4.04E+13 |
| 822.3305 | 1.89E+11 | 974.866 | 1.63E+21 | 974.866 | 0.37694 | 1243.94 | 4.04E+13 |
| 822.6111 | 1.89E+11 | 975.136 | 1.63E+21 | 975.136 | 0.37689 | 1245.66 | 4.03E+13 |
| 822.8916 | 1.90E+11 | 975.407 | 1.63E+21 | 975.407 | 0.37684 | 1247.37 | 4.03E+13 |
| 823.1721 | 1.89E+11 | 975.678 | 1.63E+21 | 975.678 | 0.3768 | 1249.09 | 4.03E+13 |
| 823.4526 | 1.90E+11 | 975.949 | 1.63E+21 | 975.949 | 0.37675 | 1250.8 | 4.02E+13 |
| 823.7331 | 1.88E+11 | 976.219 | 1.63E+21 | 976.219 | 0.3767 | 1252.52 | 4.01E+13 |
| 824.0136 | 1.90E+11 | 976.49 | 1.63E+21 | 976.49 | 0.37665 | 1254.23 | 4.01E+13 |
| 824.2941 | 1.90E+11 | 976.761 | 1.63E+21 | 976.761 | 0.3766 | 1255.95 | 4.01E+13 |
| 824.5745 | 1.90E+11 | 977.031 | 1.63E+21 | 977.031 | 0.37655 | 1257.66 | 4.01E+13 |
| 824.8549 | 1.90E+11 | 977.302 | 1.64E+21 | 977.302 | 0.37651 | 1259.38 | 4.00E+13 |
| 825.1353 | 1.90E+11 | 977.573 | 1.64E+21 | 977.573 | 0.37646 | 1261.09 | 4.00E+13 |
| 825.4157 | 1.90E+11 | 977.843 | 1.64E+21 | 977.843 | 0.37641 | 1262.8 | 4.00E+13 |
| 825.6961 | 1.90E+11 | 978.114 | 1.64E+21 | 978.114 | 0.37636 | 1264.52 | 4.00E+13 |
| 825.9765 | 1.90E+11 | 978.384 | 1.64E+21 | 978.384 | 0.37631 | 1266.23 | 4.00E+13 |
| 826.2568 | 1.90E+11 | 978.655 | 1.64E+21 | 978.655 | 0.37626 | 1267.95 | 4.00E+13 |
| 826.5371 | 1.91E+11 | 978.926 | 1.64E+21 | 978.926 | 0.37622 | 1269.66 | 3.99E+13 |
| 826.8175 | 1.90E+11 | 979.196 | 1.64E+21 | 979.196 | 0.37617 | 1271.37 | 3.98E+13 |
| 827.0978 | 1.90E+11 | 979.467 | 1.64E+21 | 979.467 | 0.37612 | 1273.09 | 3.97E+13 |
| 827.378 | 1.91E+11 | 979.737 | 1.64E+21 | 979.737 | 0.37607 | 1274.8 | 3.97E+13 |
| 827.6583 | 1.91E+11 | 980.007 | 1.64E+21 | 980.007 | 0.37602 | 1276.51 | 3.96E+13 |
| 827.9386 | 1.91E+11 | 980.278 | 1.65E+21 | 980.278 | 0.37597 | 1278.23 | 3.96E+13 |
| 828.2188 | 1.91E+11 | 980.548 | 1.65E+21 | 980.548 | 0.37593 | 1279.94 | 3.96E+13 |
| 828.499 | 1.91E+11 | 980.819 | 1.65E+21 | 980.819 | 0.37588 | 1281.65 | 3.95E+13 |
| 828.7792 | 1.90E+11 | 981.089 | 1.65E+21 | 981.089 | 0.37583 | 1283.37 | 3.94E+13 |
| 829.0594 | 1.91E+11 | 981.359 | 1.65E+21 | 981.359 | 0.37578 | 1285.08 | 3.94E+13 |
| 829.3396 | 1.91E+11 | 981.63 | 1.65E+21 | 981.63 | 0.37573 | 1286.79 | 3.94E+13 |
| 829.6197 | 1.92E+11 | 981.9 | 1.65E+21 | 981.9 | 0.37568 | 1288.5 | 3.93E+13 |
| 829.8999 | 1.92E+11 | 982.17 | 1.65E+21 | 982.17 | 0.37564 | 1290.22 | 3.92E+13 |
| 830.18 | 1.92E+11 | 982.441 | 1.65E+21 | 982.441 | 0.37559 | 1291.93 | 3.92E+13 |
| 830.4601 | 1.91E+11 | 982.711 | 1.65E+21 | 982.711 | 0.37554 | 1293.64 | 3.92E+13 |
| 830.7402 | 1.91E+11 | 982.981 | 1.65E+21 | 982.981 | 0.37549 | 1295.35 | 3.92E+13 |
| 831.0202 | 1.92E+11 | 983.251 | 1.66E+21 | 983.251 | 0.37544 | 1297.06 | 3.91E+13 |
| 831.3003 | 1.90E+11 | 983.522 | 1.66E+21 | 983.522 | 0.37539 | 1298.78 | 3.90E+13 |
| 831.5803 | 1.91E+11 | 983.792 | 1.66E+21 | 983.792 | 0.37534 | 1300.49 | 3.89E+13 |
| 831.8604 | 1.92E+11 | 984.062 | 1.66E+21 | 984.062 | 0.3753 | 1302.2 | 3.88E+13 |
| 832.1404 | 1.92E+11 | 984.332 | 1.66E+21 | 984.332 | 0.37525 | 1303.91 | 3.87E+13 |
| 832.4204 | 1.92E+11 | 984.602 | 1.66E+21 | 984.602 | 0.3752 | 1305.62 | 3.87E+13 |
| 832.7003 | 1.93E+11 | 984.872 | 1.66E+21 | 984.872 | 0.37515 | 1307.33 | 3.86E+13 |
| 832.9803 | 1.93E+11 | 985.142 | 1.66E+21 | 985.142 | 0.3751 | 1309.04 | 3.86E+13 |
| 833.2602 | 1.93E+11 | 985.412 | 1.66E+21 | 985.412 | 0.37506 | 1310.75 | 3.85E+13 |
| 833.5402 | 1.92E+11 | 985.682 | 1.66E+21 | 985.682 | 0.37501 | 1312.47 | 3.85E+13 |
| 833.8201 | 1.93E+11 | 985.952 | 1.66E+21 | 985.952 | 0.37496 | 1314.18 | 3.84E+13 |
| 834.1 | 1.93E+11 | 986.222 | 1.66E+21 | 986.222 | 0.37491 | 1315.89 | 3.82E+13 |
| 834.3798 | 1.93E+11 | 986.492 | 1.67E+21 | 986.492 | 0.37486 | 1317.6 | 3.82E+13 |
| 834.6597 | 1.94E+11 | 986.762 | 1.67E+21 | 986.762 | 0.37482 | 1319.31 | 3.81E+13 |
| 834.9395 | 1.94E+11 | 987.032 | 1.67E+21 | 987.032 | 0.37477 | 1321.02 | 3.81E+13 |
| 835.2194 | 1.93E+11 | 987.302 | 1.67E+21 | 987.302 | 0.37472 | 1322.73 | 3.80E+13 |
| 835.4992 | 1.94E+11 | 987.572 | 1.67E+21 | 987.572 | 0.37467 | 1324.44 | 3.79E+13 |
| 835.779 | 1.94E+11 | 987.842 | 1.67E+21 | 987.842 | 0.37462 | 1326.15 | 3.79E+13 |
| 836.0588 | 1.94E+11 | 988.112 | 1.67E+21 | 988.112 | 0.37457 | 1327.86 | 3.77E+13 |
| 836.3385 | 1.94E+11 | 988.382 | 1.67E+21 | 988.382 | 0.37453 | 1329.57 | 3.76E+13 |
| 836.6183 | 1.95E+11 | 988.652 | 1.67E+21 | 988.652 | 0.37448 | 1331.28 | 3.75E+13 |
| 836.898 | 1.93E+11 | 988.921 | 1.67E+21 | 988.921 | 0.37443 | 1332.98 | 3.74E+13 |
| 837.1777 | 1.94E+11 | 989.191 | 1.67E+21 | 989.191 | 0.37438 | 1334.69 | 3.73E+13 |
| 837.4574 | 1.94E+11 | 989.461 | 1.68E+21 | 989.461 | 0.37433 | 1336.4 | 3.72E+13 |
| 837.7371 | 1.93E+11 | 989.731 | 1.68E+21 | 989.731 | 0.37429 | 1338.11 | 3.71E+13 |
| 838.0168 | 1.94E+11 | 990 | 1.68E+21 | 990 | 0.37424 | 1339.82 | 3.70E+13 |
| 838.2964 | 1.93E+11 | 990.27 | 1.68E+21 | 990.27 | 0.37419 | 1341.53 | 3.68E+13 |
| 838.576 | 1.94E+11 | 990.54 | 1.68E+21 | 990.54 | 0.37414 | 1343.24 | 3.68E+13 |
| 838.8556 | 1.93E+11 | 990.81 | 1.68E+21 | 990.81 | 0.3741 | 1344.95 | 3.67E+13 |
| 839.1352 | 1.95E+11 | 991.079 | 1.68E+21 | 991.079 | 0.37405 | 1346.65 | 3.66E+13 |
| 839.4148 | 1.95E+11 | 991.349 | 1.68E+21 | 991.349 | 0.374 | 1348.36 | 3.65E+13 |
| 839.6944 | 1.94E+11 | 991.618 | 1.68E+21 | 991.618 | 0.37395 | 1350.07 | 3.64E+13 |
| 839.9739 | 1.94E+11 | 991.888 | 1.68E+21 | 991.888 | 0.3739 | 1351.78 | 3.63E+13 |
| 840.2535 | 1.94E+11 | 992.158 | 1.68E+21 | 992.158 | 0.37385 | 1353.49 | 3.62E+13 |
| 840.533 | 1.95E+11 | 992.427 | 1.69E+21 | 992.427 | 0.37381 | 1355.19 | 3.60E+13 |
| 840.8125 | 1.95E+11 | 992.697 | 1.69E+21 | 992.697 | 0.37376 | 1356.9 | 3.58E+13 |
| 841.092 | 1.94E+11 | 992.966 | 1.69E+21 | 992.966 | 0.37371 | 1358.61 | 3.57E+13 |
| 841.3715 | 1.95E+11 | 993.236 | 1.69E+21 | 993.236 | 0.37366 | 1360.32 | 3.56E+13 |
| 841.6509 | 1.95E+11 | 993.505 | 1.69E+21 | 993.505 | 0.37362 | 1362.02 | 3.55E+13 |
| 841.9303 | 1.96E+11 | 993.775 | 1.69E+21 | 993.775 | 0.37357 | 1363.73 | 3.54E+13 |
| 842.2098 | 1.94E+11 | 994.044 | 1.69E+21 | 994.044 | 0.37352 | 1365.44 | 3.53E+13 |
| 842.4892 | 1.95E+11 | 994.314 | 1.69E+21 | 994.314 | 0.37347 | 1367.14 | 3.52E+13 |
| 842.7686 | 1.96E+11 | 994.583 | 1.69E+21 | 994.583 | 0.37342 | 1368.85 | 3.51E+13 |
| 843.0479 | 1.95E+11 | 994.852 | 1.69E+21 | 994.852 | 0.37338 | 1370.56 | 3.50E+13 |
| 843.3273 | 1.95E+11 | 995.122 | 1.69E+21 | 995.122 | 0.37333 | 1372.26 | 3.50E+13 |
| 843.6066 | 1.95E+11 | 995.391 | 1.70E+21 | 995.391 | 0.37328 | 1373.97 | 3.48E+13 |
| 843.8859 | 1.95E+11 | 995.66 | 1.70E+21 | 995.66 | 0.37323 | 1375.67 | 3.48E+13 |
| 844.1652 | 1.95E+11 | 995.93 | 1.70E+21 | 995.93 | 0.37318 | 1377.38 | 3.47E+13 |
| 844.4445 | 1.95E+11 | 996.199 | 1.70E+21 | 996.199 | 0.37314 | 1379.09 | 3.46E+13 |
| 844.7238 | 1.96E+11 | 996.468 | 1.70E+21 | 996.468 | 0.37309 | 1380.79 | 3.45E+13 |
| 845.0031 | 1.96E+11 | 996.737 | 1.70E+21 | 996.737 | 0.37304 | 1382.5 | 3.44E+13 |
| 845.2823 | 1.95E+11 | 997.007 | 1.70E+21 | 997.007 | 0.37299 | 1384.2 | 3.43E+13 |
| 845.5615 | 1.96E+11 | 997.276 | 1.70E+21 | 997.276 | 0.37295 | 1385.91 | 3.42E+13 |
| 845.8407 | 1.94E+11 | 997.545 | 1.70E+21 | 997.545 | 0.3729 | 1387.61 | 3.41E+13 |
| 846.1199 | 1.96E+11 | 997.814 | 1.70E+21 | 997.814 | 0.37285 | 1389.32 | 3.40E+13 |
| 846.3991 | 1.96E+11 | 998.083 | 1.70E+21 | 998.083 | 0.3728 | 1391.02 | 3.40E+13 |
| 846.6782 | 1.95E+11 | 998.353 | 1.70E+21 | 998.353 | 0.37276 | 1392.73 | 3.39E+13 |
| 846.9574 | 1.96E+11 | 998.622 | 1.71E+21 | 998.622 | 0.37271 | 1394.43 | 3.38E+13 |
| 847.2365 | 1.95E+11 | 998.891 | 1.71E+21 | 998.891 | 0.37266 | 1396.14 | 3.37E+13 |
| 847.5156 | 1.95E+11 | 999.16 | 1.71E+21 | 999.16 | 0.37261 | 1397.84 | 3.37E+13 |
| 847.7947 | 1.96E+11 | 999.429 | 1.71E+21 | 999.429 | 0.37257 | 1399.55 | 3.36E+13 |
| 848.0738 | 1.95E+11 | 999.698 | 1.71E+21 | 999.698 | 0.37252 | 1401.25 | 3.35E+13 |
| 848.3528 | 1.96E+11 | 999.967 | 1.71E+21 | 999.967 | 0.37247 | 1402.95 | 3.34E+13 |
| 848.6319 | 1.97E+11 | 1000.91 | 1.71E+21 | 1000.91 | 0.3723 | 1404.66 | 3.34E+13 |
| 848.9109 | 1.95E+11 | 1002.64 | 1.72E+21 | 1002.64 | 0.372 | 1406.36 | 3.34E+13 |
| 849.1899 | 1.97E+11 | 1004.37 | 1.72E+21 | 1004.37 | 0.37169 | 1408.06 | 3.32E+13 |
| 849.4689 | 1.95E+11 | 1006.11 | 1.73E+21 | 1006.11 | 0.37139 | 1409.77 | 3.32E+13 |
| 849.7478 | 1.96E+11 | 1007.84 | 1.73E+21 | 1007.84 | 0.37108 | 1411.47 | 3.32E+13 |
| 850.0268 | 1.97E+11 | 1009.57 | 1.74E+21 | 1009.57 | 0.37078 | 1413.17 | 3.31E+13 |
| 850.3057 | 1.96E+11 | 1011.3 | 1.75E+21 | 1011.3 | 0.37047 | 1414.88 | 3.31E+13 |
| 850.5847 | 1.97E+11 | 1013.03 | 1.75E+21 | 1013.03 | 0.37017 | 1416.58 | 3.30E+13 |
| 850.8636 | 1.97E+11 | 1014.76 | 1.76E+21 | 1014.76 | 0.36986 | 1418.28 | 3.29E+13 |
| 851.1425 | 1.97E+11 | 1016.49 | 1.76E+21 | 1016.49 | 0.36956 | 1419.99 | 3.29E+13 |
| 851.4213 | 1.96E+11 | 1018.22 | 1.77E+21 | 1018.22 | 0.36926 | 1421.69 | 3.29E+13 |
| 851.7002 | 1.96E+11 | 1019.95 | 1.77E+21 | 1019.95 | 0.36895 | 1423.39 | 3.29E+13 |
| 851.979 | 1.97E+11 | 1021.68 | 1.78E+21 | 1021.68 | 0.36865 | 1425.09 | 3.28E+13 |
| 852.2579 | 1.97E+11 | 1023.41 | 1.78E+21 | 1023.41 | 0.36835 | 1426.8 | 3.28E+13 |
| 852.5367 | 1.96E+11 | 1025.14 | 1.79E+21 | 1025.14 | 0.36805 | 1428.5 | 3.27E+13 |
| 852.8155 | 1.96E+11 | 1026.87 | 1.79E+21 | 1026.87 | 0.36775 | 1430.2 | 3.26E+13 |
| 853.0942 | 1.96E+11 | 1028.6 | 1.80E+21 | 1028.6 | 0.36745 | 1431.9 | 3.26E+13 |
| 853.373 | 1.97E+11 | 1030.33 | 1.80E+21 | 1030.33 | 0.36715 | 1433.6 | 3.26E+13 |
| 853.6517 | 1.96E+11 | 1032.06 | 1.81E+21 | 1032.06 | 0.36684 | 1435.3 | 3.25E+13 |
| 853.9305 | 1.96E+11 | 1033.79 | 1.81E+21 | 1033.79 | 0.36654 | 1437.01 | 3.24E+13 |
| 854.2092 | 1.96E+11 | 1035.52 | 1.82E+21 | 1035.52 | 0.36624 | 1438.71 | 3.24E+13 |
| 854.4879 | 1.96E+11 | 1037.25 | 1.82E+21 | 1037.25 | 0.36595 | 1440.41 | 3.23E+13 |
| 854.7665 | 1.97E+11 | 1038.98 | 1.83E+21 | 1038.98 | 0.36565 | 1442.11 | 3.23E+13 |
| 855.0452 | 1.97E+11 | 1040.7 | 1.83E+21 | 1040.7 | 0.36535 | 1443.81 | 3.22E+13 |
| 855.3238 | 1.96E+11 | 1042.43 | 1.84E+21 | 1042.43 | 0.36505 | 1445.51 | 3.21E+13 |
| 855.6024 | 1.96E+11 | 1044.16 | 1.84E+21 | 1044.16 | 0.36475 | 1447.21 | 3.20E+13 |
| 855.8811 | 1.97E+11 | 1045.89 | 1.85E+21 | 1045.89 | 0.36445 | 1448.91 | 3.19E+13 |
| 856.1596 | 1.98E+11 | 1047.62 | 1.85E+21 | 1047.62 | 0.36415 | 1450.61 | 3.18E+13 |
| 856.4382 | 1.97E+11 | 1049.35 | 1.86E+21 | 1049.35 | 0.36386 | 1452.31 | 3.18E+13 |
| 856.7168 | 1.98E+11 | 1051.08 | 1.86E+21 | 1051.08 | 0.36356 | 1454.01 | 3.17E+13 |
| 856.9953 | 1.97E+11 | 1052.8 | 1.87E+21 | 1052.8 | 0.36326 | 1455.71 | 3.16E+13 |
| 857.2738 | 1.98E+11 | 1054.53 | 1.87E+21 | 1054.53 | 0.36297 | 1457.41 | 3.15E+13 |
| 857.5523 | 1.98E+11 | 1056.26 | 1.88E+21 | 1056.26 | 0.36267 | 1459.11 | 3.14E+13 |
| 857.8308 | 1.98E+11 | 1057.99 | 1.88E+21 | 1057.99 | 0.36238 | 1460.81 | 3.13E+13 |
| 858.1093 | 1.98E+11 | 1059.72 | 1.89E+21 | 1059.72 | 0.36208 | 1462.51 | 3.12E+13 |
| 858.3878 | 1.97E+11 | 1061.44 | 1.89E+21 | 1061.44 | 0.36178 | 1464.21 | 3.12E+13 |
| 858.6662 | 1.97E+11 | 1063.17 | 1.90E+21 | 1063.17 | 0.36149 | 1465.91 | 3.12E+13 |
| 858.9446 | 1.97E+11 | 1064.9 | 1.90E+21 | 1064.9 | 0.36119 | 1467.61 | 3.11E+13 |
| 859.223 | 1.97E+11 | 1066.63 | 1.91E+21 | 1066.63 | 0.3609 | 1469.31 | 3.10E+13 |
| 859.5014 | 1.98E+11 | 1068.35 | 1.91E+21 | 1068.35 | 0.36061 | 1471 | 3.09E+13 |
| 859.7798 | 1.98E+11 | 1070.08 | 1.92E+21 | 1070.08 | 0.36031 | 1472.7 | 3.08E+13 |
| 860.0581 | 1.98E+11 | 1071.81 | 1.92E+21 | 1071.81 | 0.36002 | 1474.4 | 3.07E+13 |
| 860.3365 | 1.98E+11 | 1073.53 | 1.93E+21 | 1073.53 | 0.35973 | 1476.1 | 3.07E+13 |
| 860.6148 | 1.99E+11 | 1075.26 | 1.93E+21 | 1075.26 | 0.35943 | 1477.8 | 3.06E+13 |
| 860.8931 | 1.98E+11 | 1076.99 | 1.94E+21 | 1076.99 | 0.35914 | 1479.5 | 3.05E+13 |
| 861.1714 | 1.98E+11 | 1078.71 | 1.94E+21 | 1078.71 | 0.35885 | 1481.19 | 3.04E+13 |
| 861.4497 | 1.98E+11 | 1080.44 | 1.94E+21 | 1080.44 | 0.35856 | 1482.89 | 3.04E+13 |
| 861.7279 | 1.99E+11 | 1082.17 | 1.95E+21 | 1082.17 | 0.35827 | 1484.59 | 3.03E+13 |
| 862.0062 | 1.98E+11 | 1083.89 | 1.95E+21 | 1083.89 | 0.35798 | 1486.29 | 3.02E+13 |
| 862.2844 | 1.98E+11 | 1085.62 | 1.96E+21 | 1085.62 | 0.35768 | 1487.98 | 3.01E+13 |
| 862.5626 | 1.98E+11 | 1087.34 | 1.96E+21 | 1087.34 | 0.3574 | 1489.68 | 3.01E+13 |
| 862.8408 | 1.98E+11 | 1089.07 | 1.97E+21 | 1089.07 | 0.35711 | 1491.38 | 3.00E+13 |
| 863.1189 | 1.97E+11 | 1090.8 | 1.97E+21 | 1090.8 | 0.35681 | 1493.08 | 2.99E+13 |
| 863.3971 | 1.98E+11 | 1092.52 | 1.98E+21 | 1092.52 | 0.35652 | 1494.77 | 2.98E+13 |
| 863.6752 | 1.98E+11 | 1094.25 | 1.98E+21 | 1094.25 | 0.35624 | 1496.47 | 2.97E+13 |
| 863.9533 | 1.99E+11 | 1095.97 | 1.98E+21 | 1095.97 | 0.35595 | 1498.17 | 2.97E+13 |
| 864.2314 | 1.98E+11 | 1097.7 | 1.99E+21 | 1097.7 | 0.35566 | 1499.86 | 2.95E+13 |
| 864.5095 | 1.97E+11 | 1099.42 | 1.99E+21 | 1099.42 | 0.35537 | 1501.56 | 2.94E+13 |
| 864.7876 | 1.98E+11 | 1101.15 | 2.00E+21 | 1101.15 | 0.35508 | 1503.25 | 2.93E+13 |
| 865.0656 | 1.99E+11 | 1102.87 | 2.00E+21 | 1102.87 | 0.3548 | 1504.95 | 2.92E+13 |
| 865.3437 | 1.99E+11 | 1104.6 | 2.01E+21 | 1104.6 | 0.35451 | 1506.65 | 2.91E+13 |
| 865.6217 | 1.98E+11 | 1106.32 | 2.01E+21 | 1106.32 | 0.35422 | 1508.34 | 2.90E+13 |
| 865.8997 | 1.98E+11 | 1108.05 | 2.01E+21 | 1108.05 | 0.35394 | 1510.04 | 2.90E+13 |
| 866.1777 | 1.99E+11 | 1109.77 | 2.02E+21 | 1109.77 | 0.35365 | 1511.73 | 2.89E+13 |
| 866.4556 | 1.97E+11 | 1111.5 | 2.02E+21 | 1111.5 | 0.35336 | 1513.43 | 2.88E+13 |
| 866.7336 | 1.98E+11 | 1113.22 | 2.03E+21 | 1113.22 | 0.35308 | 1515.12 | 2.87E+13 |
| 867.0115 | 1.98E+11 | 1114.95 | 2.03E+21 | 1114.95 | 0.35279 | 1516.82 | 2.86E+13 |
| 867.2894 | 1.98E+11 | 1116.67 | 2.03E+21 | 1116.67 | 0.35251 | 1518.51 | 2.85E+13 |
| 867.5673 | 1.98E+11 | 1118.39 | 2.04E+21 | 1118.39 | 0.35222 | 1520.21 | 2.84E+13 |
| 867.8452 | 1.98E+11 | 1120.12 | 2.04E+21 | 1120.12 | 0.35194 | 1521.9 | 2.83E+13 |
| 868.1231 | 1.98E+11 | 1121.84 | 2.05E+21 | 1121.84 | 0.35165 | 1523.6 | 2.83E+13 |
| 868.4009 | 1.97E+11 | 1123.57 | 2.05E+21 | 1123.57 | 0.35137 | 1525.29 | 2.82E+13 |
| 868.6788 | 1.97E+11 | 1125.29 | 2.05E+21 | 1125.29 | 0.35109 | 1526.99 | 2.81E+13 |
| 868.9566 | 1.96E+11 | 1127.01 | 2.06E+21 | 1127.01 | 0.3508 | 1528.68 | 2.80E+13 |
| 869.2344 | 1.96E+11 | 1128.74 | 2.06E+21 | 1128.74 | 0.35052 | 1530.37 | 2.80E+13 |
| 869.5122 | 1.97E+11 | 1130.46 | 2.07E+21 | 1130.46 | 0.35024 | 1532.07 | 2.79E+13 |
| 869.7899 | 1.97E+11 | 1132.18 | 2.07E+21 | 1132.18 | 0.34996 | 1533.76 | 2.78E+13 |
| 870.0677 | 1.98E+11 | 1133.91 | 2.07E+21 | 1133.91 | 0.34967 | 1535.45 | 2.78E+13 |
| 870.3454 | 1.97E+11 | 1135.63 | 2.08E+21 | 1135.63 | 0.34939 | 1537.15 | 2.77E+13 |
| 870.6231 | 1.96E+11 | 1137.35 | 2.08E+21 | 1137.35 | 0.34911 | 1538.84 | 2.76E+13 |
| 870.9008 | 1.97E+11 | 1139.07 | 2.08E+21 | 1139.07 | 0.34883 | 1540.53 | 2.76E+13 |
| 871.1785 | 1.97E+11 | 1140.8 | 2.09E+21 | 1140.8 | 0.34855 | 1542.23 | 2.75E+13 |
| 871.4561 | 1.97E+11 | 1142.52 | 2.09E+21 | 1142.52 | 0.34827 | 1543.92 | 2.74E+13 |
| 871.7338 | 1.96E+11 | 1144.24 | 2.10E+21 | 1144.24 | 0.34799 | 1545.61 | 2.73E+13 |
| 872.0114 | 1.96E+11 | 1145.96 | 2.10E+21 | 1145.96 | 0.34771 | 1547.31 | 2.73E+13 |
| 872.289 | 1.97E+11 | 1147.69 | 2.10E+21 | 1147.69 | 0.34743 | 1549 | 2.72E+13 |
| 872.5666 | 1.97E+11 | 1149.41 | 2.11E+21 | 1149.41 | 0.34715 | 1550.69 | 2.71E+13 |
| 872.8442 | 1.96E+11 | 1151.13 | 2.11E+21 | 1151.13 | 0.34687 | 1552.38 | 2.71E+13 |
| 873.1217 | 1.96E+11 | 1152.85 | 2.11E+21 | 1152.85 | 0.34659 | 1554.07 | 2.70E+13 |
| 873.3993 | 1.97E+11 | 1154.57 | 2.12E+21 | 1154.57 | 0.34632 | 1555.77 | 2.69E+13 |
| 873.6768 | 1.96E+11 | 1156.29 | 2.12E+21 | 1156.29 | 0.34604 | 1557.46 | 2.68E+13 |
| 873.9543 | 1.97E+11 | 1158.02 | 2.12E+21 | 1158.02 | 0.34576 | 1559.15 | 2.67E+13 |
| 874.2318 | 1.96E+11 | 1159.74 | 2.13E+21 | 1159.74 | 0.34548 | 1560.84 | 2.66E+13 |
| 874.5093 | 1.97E+11 | 1161.46 | 2.13E+21 | 1161.46 | 0.34521 | 1562.53 | 2.65E+13 |
| 874.7867 | 1.97E+11 | 1163.18 | 2.13E+21 | 1163.18 | 0.34493 | 1564.22 | 2.65E+13 |
| 875.0642 | 1.96E+11 | 1164.9 | 2.14E+21 | 1164.9 | 0.34465 | 1565.91 | 2.64E+13 |
| 875.3416 | 1.97E+11 | 1166.62 | 2.14E+21 | 1166.62 | 0.34438 | 1567.6 | 2.63E+13 |
| 875.619 | 1.97E+11 | 1168.34 | 2.14E+21 | 1168.34 | 0.3441 | 1569.3 | 2.62E+13 |
| 875.8964 | 1.96E+11 | 1170.06 | 2.15E+21 | 1170.06 | 0.34383 | 1570.99 | 2.61E+13 |
| 876.1737 | 1.97E+11 | 1171.78 | 2.15E+21 | 1171.78 | 0.34355 | 1572.68 | 2.60E+13 |
| 876.4511 | 1.96E+11 | 1173.5 | 2.15E+21 | 1173.5 | 0.34328 | 1574.37 | 2.59E+13 |
| 876.7284 | 1.96E+11 | 1175.22 | 2.16E+21 | 1175.22 | 0.343 | 1576.06 | 2.58E+13 |
| 877.0057 | 1.95E+11 | 1176.94 | 2.16E+21 | 1176.94 | 0.34273 | 1577.75 | 2.57E+13 |
| 877.283 | 1.95E+11 | 1178.66 | 2.16E+21 | 1178.66 | 0.34246 | 1579.44 | 2.56E+13 |
| 877.5603 | 1.95E+11 | 1180.38 | 2.17E+21 | 1180.38 | 0.34218 | 1581.13 | 2.55E+13 |
| 877.8376 | 1.96E+11 | 1182.1 | 2.17E+21 | 1182.1 | 0.34191 | | |
| 878.1148 | 1.96E+11 | 1183.82 | 2.17E+21 | 1183.82 | 0.34164 | | |
| 878.3921 | 1.96E+11 | 1185.54 | 2.18E+21 | 1185.54 | 0.34136 | | |
| 878.6693 | 1.95E+11 | 1187.26 | 2.18E+21 | 1187.26 | 0.34109 | | |
| 878.9465 | 1.95E+11 | 1188.98 | 2.18E+21 | 1188.98 | 0.34082 | | |
| 879.2237 | 1.95E+11 | 1190.7 | 2.19E+21 | 1190.7 | 0.34055 | | |
| 879.5008 | 1.93E+11 | 1192.42 | 2.19E+21 | 1192.42 | 0.34028 | | |
| 879.778 | 1.94E+11 | 1194.14 | 2.19E+21 | 1194.14 | 0.34001 | | |
| 880.0551 | 1.94E+11 | 1195.86 | 2.19E+21 | 1195.86 | 0.33974 | | |
| 880.3322 | 1.94E+11 | 1197.58 | 2.20E+21 | 1197.58 | 0.33947 | | |
| 880.6093 | 1.94E+11 | 1199.3 | 2.20E+21 | 1199.3 | 0.3392 | | |
| 880.8864 | 1.92E+11 | 1201.02 | 2.20E+21 | 1201.02 | 0.33893 | | |
| 881.1634 | 1.93E+11 | 1202.73 | 2.21E+21 | 1202.73 | 0.33866 | | |
| 881.4405 | 1.93E+11 | 1204.45 | 2.21E+21 | 1204.45 | 0.33839 | | |
| 881.7175 | 1.94E+11 | 1206.17 | 2.21E+21 | 1206.17 | 0.33812 | | |
| 881.9945 | 1.93E+11 | 1207.89 | 2.21E+21 | 1207.89 | 0.33785 | | |
| 882.2715 | 1.92E+11 | 1209.61 | 2.22E+21 | 1209.61 | 0.33758 | | |
| 882.5485 | 1.92E+11 | 1211.32 | 2.22E+21 | 1211.32 | 0.33732 | | |
| 882.8254 | 1.92E+11 | 1213.04 | 2.22E+21 | 1213.04 | 0.33705 | | |
| 883.1024 | 1.91E+11 | 1214.76 | 2.23E+21 | 1214.76 | 0.33678 | | |
| 883.3793 | 1.92E+11 | 1216.48 | 2.23E+21 | 1216.48 | 0.33652 | | |
| 883.6562 | 1.91E+11 | 1218.19 | 2.23E+21 | 1218.19 | 0.33625 | | |
| 883.9331 | 1.92E+11 | 1219.91 | 2.23E+21 | 1219.91 | 0.33598 | | |
| 884.2099 | 1.91E+11 | 1221.63 | 2.24E+21 | 1221.63 | 0.33572 | | |
| 884.4868 | 1.91E+11 | 1223.35 | 2.24E+21 | 1223.35 | 0.33545 | | |
| 884.7636 | 1.91E+11 | 1225.06 | 2.24E+21 | 1225.06 | 0.33519 | | |
| 885.0405 | 1.90E+11 | 1226.78 | 2.24E+21 | 1226.78 | 0.33492 | | |
| 885.3173 | 1.89E+11 | 1228.5 | 2.25E+21 | 1228.5 | 0.33466 | | |
| 885.594 | 1.90E+11 | 1230.21 | 2.25E+21 | 1230.21 | 0.33439 | | |
| 885.8708 | 1.88E+11 | 1231.93 | 2.25E+21 | 1231.93 | 0.33413 | | |
| 886.1475 | 1.89E+11 | 1233.65 | 2.25E+21 | 1233.65 | 0.33387 | | |
| 886.4243 | 1.89E+11 | 1235.36 | 2.26E+21 | 1235.36 | 0.3336 | | |
| 886.701 | 1.89E+11 | 1237.08 | 2.26E+21 | 1237.08 | 0.33334 | | |
| 886.9777 | 1.88E+11 | 1238.79 | 2.26E+21 | 1238.79 | 0.33308 | | |
| 887.2544 | 1.89E+11 | 1240.51 | 2.26E+21 | 1240.51 | 0.33281 | | |
| 887.531 | 1.89E+11 | 1242.23 | 2.27E+21 | 1242.23 | 0.33255 | | |
| 887.8077 | 1.87E+11 | 1243.94 | 2.27E+21 | 1243.94 | 0.33229 | | |
| 888.0843 | 1.88E+11 | 1245.66 | 2.27E+21 | 1245.66 | 0.33203 | | |
| 888.3609 | 1.87E+11 | 1247.37 | 2.27E+21 | 1247.37 | 0.33177 | | |
| 888.6375 | 1.88E+11 | 1249.09 | 2.28E+21 | 1249.09 | 0.33151 | | |
| 888.9141 | 1.87E+11 | 1250.8 | 2.28E+21 | 1250.8 | 0.33125 | | |
| 889.1906 | 1.87E+11 | 1252.52 | 2.28E+21 | 1252.52 | 0.33098 | | |
| 889.4672 | 1.87E+11 | 1254.23 | 2.28E+21 | 1254.23 | 0.33073 | | |
| 889.7437 | 1.86E+11 | 1255.95 | 2.28E+21 | 1255.95 | 0.33047 | | |
| 890.0202 | 1.87E+11 | 1257.66 | 2.29E+21 | 1257.66 | 0.33021 | | |
| 890.2967 | 1.86E+11 | 1259.38 | 2.29E+21 | 1259.38 | 0.32995 | | |
| 890.5731 | 1.87E+11 | 1261.09 | 2.29E+21 | 1261.09 | 0.32969 | | |
| 890.8496 | 1.86E+11 | 1262.8 | 2.29E+21 | 1262.8 | 0.32943 | | |
| 891.126 | 1.85E+11 | 1264.52 | 2.30E+21 | 1264.52 | 0.32917 | | |
| 891.4024 | 1.85E+11 | 1266.23 | 2.30E+21 | 1266.23 | 0.32892 | | |
| 891.6788 | 1.87E+11 | 1267.95 | 2.30E+21 | 1267.95 | 0.32866 | | |
| 891.9552 | 1.85E+11 | 1269.66 | 2.30E+21 | 1269.66 | 0.3284 | | |
| 892.2316 | 1.86E+11 | 1271.37 | 2.30E+21 | 1271.37 | 0.32814 | | |
| 892.5079 | 1.86E+11 | 1273.09 | 2.31E+21 | 1273.09 | 0.32789 | | |
| 892.7842 | 1.85E+11 | 1274.8 | 2.31E+21 | 1274.8 | 0.32763 | | |
| 893.0606 | 1.85E+11 | 1276.51 | 2.31E+21 | 1276.51 | 0.32738 | | |
| 893.3368 | 1.85E+11 | 1278.23 | 2.31E+21 | 1278.23 | 0.32712 | | |
| 893.6131 | 1.84E+11 | 1279.94 | 2.31E+21 | 1279.94 | 0.32687 | | |
| 893.8894 | 1.84E+11 | 1281.65 | 2.32E+21 | 1281.65 | 0.32661 | | |
| 894.1656 | 1.85E+11 | 1283.37 | 2.32E+21 | 1283.37 | 0.32636 | | |
| 894.4418 | 1.84E+11 | 1285.08 | 2.32E+21 | 1285.08 | 0.3261 | | |
| 894.718 | 1.83E+11 | 1286.79 | 2.32E+21 | 1286.79 | 0.32585 | | |
| 894.9942 | 1.83E+11 | 1288.5 | 2.32E+21 | 1288.5 | 0.32559 | | |
| 895.2704 | 1.82E+11 | 1290.22 | 2.33E+21 | 1290.22 | 0.32534 | | |
| 895.5466 | 1.82E+11 | 1291.93 | 2.33E+21 | 1291.93 | 0.32509 | | |
| 895.8227 | 1.82E+11 | 1293.64 | 2.33E+21 | 1293.64 | 0.32483 | | |
| 896.0988 | 1.82E+11 | 1295.35 | 2.33E+21 | 1295.35 | 0.32458 | | |
| 896.3749 | 1.81E+11 | 1297.06 | 2.33E+21 | 1297.06 | 0.32433 | | |
| 896.651 | 1.81E+11 | 1298.78 | 2.33E+21 | 1298.78 | 0.32408 | | |
| 896.9271 | 1.81E+11 | 1300.49 | 2.34E+21 | 1300.49 | 0.32383 | | |
| 897.2031 | 1.81E+11 | 1302.2 | 2.34E+21 | 1302.2 | 0.32358 | | |
| 897.4791 | 1.81E+11 | 1303.91 | 2.34E+21 | 1303.91 | 0.32332 | | |
| 897.7551 | 1.80E+11 | 1305.62 | 2.34E+21 | 1305.62 | 0.32308 | | |
| 898.0311 | 1.80E+11 | 1307.33 | 2.34E+21 | 1307.33 | 0.32282 | | |
| 898.3071 | 1.80E+11 | 1309.04 | 2.34E+21 | 1309.04 | 0.32258 | | |
| 898.5831 | 1.81E+11 | 1310.75 | 2.35E+21 | 1310.75 | 0.32232 | | |
| 898.859 | 1.79E+11 | 1312.47 | 2.35E+21 | 1312.47 | 0.32208 | | |
| 899.1349 | 1.79E+11 | 1314.18 | 2.35E+21 | 1314.18 | 0.32183 | | |
| 899.4108 | 1.79E+11 | 1315.89 | 2.35E+21 | 1315.89 | 0.32158 | | |
| 899.6867 | 1.79E+11 | 1317.6 | 2.35E+21 | 1317.6 | 0.32133 | | |
| 899.9626 | 1.79E+11 | 1319.31 | 2.35E+21 | 1319.31 | 0.32108 | | |
| 900.2385 | 1.79E+11 | 1321.02 | 2.36E+21 | 1321.02 | 0.32083 | | |
| 900.5143 | 1.79E+11 | 1322.73 | 2.36E+21 | 1322.73 | 0.32059 | | |
| 900.7901 | 1.79E+11 | 1324.44 | 2.36E+21 | 1324.44 | 0.32034 | | |

| | | | | | |
|---|---|---|---|---|---|
| 901.0659 | 1.80E+11 | 1326.15 | 2.36E+21 | 1326.15 | 0.32009 |
| 901.3417 | 1.78E+11 | 1327.86 | 2.36E+21 | 1327.86 | 0.31985 |
| 901.6175 | 1.79E+11 | 1329.57 | 2.36E+21 | 1329.57 | 0.3196 |
| 901.8932 | 1.77E+11 | 1331.28 | 2.36E+21 | 1331.28 | 0.31936 |
| 902.1689 | 1.78E+11 | 1332.98 | 2.37E+21 | 1332.98 | 0.31911 |
| 902.4446 | 1.77E+11 | 1334.69 | 2.37E+21 | 1334.69 | 0.31886 |
| 902.7203 | 1.78E+11 | 1336.4 | 2.37E+21 | 1336.4 | 0.31862 |
| 902.996 | 1.77E+11 | 1338.11 | 2.37E+21 | 1338.11 | 0.31837 |
| 903.2717 | 1.78E+11 | 1339.82 | 2.37E+21 | 1339.82 | 0.31813 |
| 903.5473 | 1.77E+11 | 1341.53 | 2.37E+21 | 1341.53 | 0.31789 |
| 903.8229 | 1.77E+11 | 1343.24 | 2.37E+21 | 1343.24 | 0.31764 |
| 904.0985 | 1.76E+11 | 1344.95 | 2.37E+21 | 1344.95 | 0.3174 |
| 904.3741 | 1.76E+11 | 1346.65 | 2.38E+21 | 1346.65 | 0.31716 |
| 904.6497 | 1.76E+11 | 1348.36 | 2.38E+21 | 1348.36 | 0.31691 |
| 904.9252 | 1.76E+11 | 1350.07 | 2.38E+21 | 1350.07 | 0.31667 |
| 905.2008 | 1.75E+11 | 1351.78 | 2.38E+21 | 1351.78 | 0.31643 |
| 905.4763 | 1.74E+11 | 1353.49 | 2.38E+21 | 1353.49 | 0.31619 |
| 905.7518 | 1.75E+11 | 1355.19 | 2.38E+21 | 1355.19 | 0.31595 |
| 906.0273 | 1.74E+11 | 1356.9 | 2.38E+21 | 1356.9 | 0.3157 |
| 906.3027 | 1.75E+11 | 1358.61 | 2.38E+21 | 1358.61 | 0.31546 |
| 906.5782 | 1.75E+11 | 1360.32 | 2.38E+21 | 1360.32 | 0.31522 |
| 906.8536 | 1.74E+11 | 1362.02 | 2.39E+21 | 1362.02 | 0.31498 |
| 907.129 | 1.73E+11 | 1363.73 | 2.39E+21 | 1363.73 | 0.31474 |
| 907.4044 | 1.74E+11 | 1365.44 | 2.39E+21 | 1365.44 | 0.3145 |
| 907.6798 | 1.74E+11 | 1367.14 | 2.39E+21 | 1367.14 | 0.31426 |
| 907.9551 | 1.73E+11 | 1368.85 | 2.39E+21 | 1368.85 | 0.31403 |
| 908.2305 | 1.73E+11 | 1370.56 | 2.39E+21 | 1370.56 | 0.31379 |
| 908.5058 | 1.73E+11 | 1372.26 | 2.39E+21 | 1372.26 | 0.31355 |
| 908.7811 | 1.73E+11 | 1373.97 | 2.39E+21 | 1373.97 | 0.31331 |
| 909.0564 | 1.72E+11 | 1375.67 | 2.39E+21 | 1375.67 | 0.31307 |
| 909.3316 | 1.73E+11 | 1377.38 | 2.40E+21 | 1377.38 | 0.31283 |
| 909.6069 | 1.72E+11 | 1379.09 | 2.40E+21 | 1379.09 | 0.3126 |
| 909.8821 | 1.72E+11 | 1380.79 | 2.40E+21 | 1380.79 | 0.31236 |
| 910.1573 | 1.71E+11 | 1382.5 | 2.40E+21 | 1382.5 | 0.31212 |
| 910.4325 | 1.72E+11 | 1384.2 | 2.40E+21 | 1384.2 | 0.31189 |
| 910.7077 | 1.71E+11 | 1385.91 | 2.40E+21 | 1385.91 | 0.31165 |
| 910.9829 | 1.70E+11 | 1387.61 | 2.40E+21 | 1387.61 | 0.31142 |
| 911.258 | 1.71E+11 | 1389.32 | 2.40E+21 | 1389.32 | 0.31118 |
| 911.5331 | 1.70E+11 | 1391.02 | 2.40E+21 | 1391.02 | 0.31095 |
| 911.8082 | 1.70E+11 | 1392.73 | 2.40E+21 | 1392.73 | 0.31071 |
| 912.0833 | 1.70E+11 | 1394.43 | 2.40E+21 | 1394.43 | 0.31048 |
| 912.3584 | 1.69E+11 | 1396.14 | 2.40E+21 | 1396.14 | 0.31024 |
| 912.6334 | 1.69E+11 | 1397.84 | 2.41E+21 | 1397.84 | 0.31001 |
| 912.9085 | 1.69E+11 | 1399.55 | 2.41E+21 | 1399.55 | 0.30978 |
| 913.1835 | 1.69E+11 | 1401.25 | 2.41E+21 | 1401.25 | 0.30954 |
| 913.4585 | 1.69E+11 | 1402.95 | 2.41E+21 | 1402.95 | 0.30931 |
| 913.7335 | 1.68E+11 | 1404.66 | 2.41E+21 | 1404.66 | 0.30908 |
| 914.0084 | 1.69E+11 | 1406.36 | 2.41E+21 | 1406.36 | 0.30884 |
| 914.2834 | 1.69E+11 | 1408.06 | 2.41E+21 | 1408.06 | 0.30861 |
| 914.5583 | 1.68E+11 | 1409.77 | 2.41E+21 | 1409.77 | 0.30838 |
| 914.8332 | 1.68E+11 | 1411.47 | 2.41E+21 | 1411.47 | 0.30815 |
| 915.1081 | 1.67E+11 | 1413.17 | 2.41E+21 | 1413.17 | 0.30792 |
| 915.383 | 1.67E+11 | 1414.88 | 2.41E+21 | 1414.88 | 0.30769 |
| 915.6578 | 1.66E+11 | 1416.58 | 2.41E+21 | 1416.58 | 0.30746 |
| 915.9327 | 1.66E+11 | 1418.28 | 2.41E+21 | 1418.28 | 0.30723 |
| 916.2075 | 1.66E+11 | 1419.99 | 2.41E+21 | 1419.99 | 0.307 |
| 916.4823 | 1.66E+11 | 1421.69 | 2.41E+21 | 1421.69 | 0.30677 |
| 916.7571 | 1.65E+11 | 1423.39 | 2.41E+21 | 1423.39 | 0.30654 |
| 917.0318 | 1.66E+11 | 1425.09 | 2.42E+21 | 1425.09 | 0.30631 |
| 917.3066 | 1.65E+11 | 1426.8 | 2.42E+21 | 1426.8 | 0.30608 |
| 917.5813 | 1.65E+11 | 1428.5 | 2.42E+21 | 1428.5 | 0.30585 |
| 917.856 | 1.65E+11 | 1430.2 | 2.42E+21 | 1430.2 | 0.30563 |
| 918.1307 | 1.64E+11 | 1431.9 | 2.42E+21 | 1431.9 | 0.3054 |
| 918.4054 | 1.64E+11 | 1433.6 | 2.42E+21 | 1433.6 | 0.30517 |
| 918.68 | 1.64E+11 | 1435.3 | 2.42E+21 | 1435.3 | 0.30494 |
| 918.9547 | 1.64E+11 | 1437.01 | 2.42E+21 | 1437.01 | 0.30472 |
| 919.2293 | 1.63E+11 | 1438.71 | 2.42E+21 | 1438.71 | 0.30449 |
| 919.5039 | 1.63E+11 | 1440.41 | 2.42E+21 | 1440.41 | 0.30427 |
| 919.7785 | 1.63E+11 | 1442.11 | 2.42E+21 | 1442.11 | 0.30404 |
| 920.053 | 1.62E+11 | 1443.81 | 2.42E+21 | 1443.81 | 0.30381 |
| 920.3276 | 1.62E+11 | 1445.51 | 2.42E+21 | 1445.51 | 0.30359 |
| 920.6021 | 1.62E+11 | 1447.21 | 2.42E+21 | 1447.21 | 0.30336 |
| 920.8766 | 1.62E+11 | 1448.91 | 2.42E+21 | 1448.91 | 0.30314 |
| 921.1511 | 1.61E+11 | 1450.61 | 2.42E+21 | 1450.61 | 0.30291 |
| 921.4256 | 1.61E+11 | 1452.31 | 2.42E+21 | 1452.31 | 0.30269 |
| 921.7 | 1.60E+11 | 1454.01 | 2.42E+21 | 1454.01 | 0.30247 |
| 921.9745 | 1.61E+11 | 1455.71 | 2.42E+21 | 1455.71 | 0.30224 |
| 922.2489 | 1.60E+11 | 1457.41 | 2.42E+21 | 1457.41 | 0.30202 |
| 922.5233 | 1.59E+11 | 1459.11 | 2.42E+21 | 1459.11 | 0.3018 |
| 922.7977 | 1.59E+11 | 1460.81 | 2.42E+21 | 1460.81 | 0.30158 |
| 923.072 | 1.59E+11 | 1462.51 | 2.42E+21 | 1462.51 | 0.30135 |
| 923.3464 | 1.59E+11 | 1464.21 | 2.42E+21 | 1464.21 | 0.30113 |
| 923.6207 | 1.58E+11 | 1465.91 | 2.42E+21 | 1465.91 | 0.30091 |
| 923.895 | 1.58E+11 | 1467.61 | 2.42E+21 | 1467.61 | 0.30069 |
| 924.1693 | 1.57E+11 | 1469.31 | 2.42E+21 | 1469.31 | 0.30047 |
| 924.4436 | 1.56E+11 | 1471 | 2.42E+21 | 1471 | 0.30025 |
| 924.7179 | 1.56E+11 | 1472.7 | 2.42E+21 | 1472.7 | 0.30003 |
| 924.9921 | 1.56E+11 | 1474.4 | 2.43E+21 | 1474.4 | 0.29981 |
| 925.2663 | 1.56E+11 | 1476.1 | 2.43E+21 | 1476.1 | 0.29959 |
| 925.5405 | 1.56E+11 | 1477.8 | 2.43E+21 | 1477.8 | 0.29937 |
| 925.8147 | 1.56E+11 | 1479.5 | 2.43E+21 | 1479.5 | 0.29915 |
| 926.0889 | 1.55E+11 | 1481.19 | 2.43E+21 | 1481.19 | 0.29893 |
| 926.363 | 1.56E+11 | 1482.89 | 2.43E+21 | 1482.89 | 0.29871 |
| 926.6371 | 1.54E+11 | 1484.59 | 2.43E+21 | 1484.59 | 0.2985 |
| 926.9113 | 1.54E+11 | 1486.29 | 2.43E+21 | 1486.29 | 0.29828 |
| 927.1853 | 1.53E+11 | 1487.98 | 2.43E+21 | 1487.98 | 0.29806 |
| 927.4594 | 1.53E+11 | 1489.68 | 2.43E+21 | 1489.68 | 0.29784 |
| 927.7335 | 1.52E+11 | 1491.38 | 2.43E+21 | 1491.38 | 0.29763 |
| 928.0075 | 1.53E+11 | 1493.08 | 2.43E+21 | 1493.08 | 0.29741 |
| 928.2815 | 1.53E+11 | 1494.77 | 2.43E+21 | 1494.77 | 0.29719 |
| 928.5555 | 1.51E+11 | 1496.47 | 2.43E+21 | 1496.47 | 0.29698 |
| 928.8295 | 1.51E+11 | 1498.17 | 2.43E+21 | 1498.17 | 0.29676 |
| 929.1035 | 1.51E+11 | 1499.86 | 2.43E+21 | 1499.86 | 0.29655 |
| 929.3774 | 1.51E+11 | 1501.56 | 2.43E+21 | 1501.56 | 0.29633 |
| 929.6514 | 1.51E+11 | 1503.25 | 2.43E+21 | 1503.25 | 0.29612 |
| 929.9253 | 1.50E+11 | 1504.95 | 2.43E+21 | 1504.95 | 0.2959 |
| 930.1992 | 1.50E+11 | 1506.65 | 2.43E+21 | 1506.65 | 0.29569 |
| 930.473 | 1.50E+11 | 1508.34 | 2.43E+21 | 1508.34 | 0.29547 |
| 930.7469 | 1.48E+11 | 1510.04 | 2.43E+21 | 1510.04 | 0.29526 |
| 931.0207 | 1.48E+11 | 1511.73 | 2.43E+21 | 1511.73 | 0.29505 |
| 931.2945 | 1.48E+11 | 1513.43 | 2.43E+21 | 1513.43 | 0.29483 |
| 931.5683 | 1.48E+11 | 1515.12 | 2.43E+21 | 1515.12 | 0.29462 |
| 931.8421 | 1.47E+11 | 1516.82 | 2.43E+21 | 1516.82 | 0.29441 |
| 932.1159 | 1.47E+11 | 1518.51 | 2.42E+21 | 1518.51 | 0.2942 |
| 932.3896 | 1.47E+11 | 1520.21 | 2.42E+21 | 1520.21 | 0.29399 |
| 932.6634 | 1.46E+11 | 1521.9 | 2.42E+21 | 1521.9 | 0.29378 |
| 932.9371 | 1.47E+11 | 1523.6 | 2.42E+21 | 1523.6 | 0.29356 |
| 933.2108 | 1.45E+11 | 1525.29 | 2.42E+21 | 1525.29 | 0.29335 |
| 933.4844 | 1.45E+11 | 1526.99 | 2.42E+21 | 1526.99 | 0.29314 |
| 933.7581 | 1.44E+11 | 1528.68 | 2.42E+21 | 1528.68 | 0.29293 |
| 934.0317 | 1.44E+11 | 1530.37 | 2.42E+21 | 1530.37 | 0.29272 |
| 934.3053 | 1.44E+11 | 1532.07 | 2.42E+21 | 1532.07 | 0.29251 |
| 934.5789 | 1.43E+11 | 1533.76 | 2.42E+21 | 1533.76 | 0.2923 |
| 934.8525 | 1.43E+11 | 1535.45 | 2.42E+21 | 1535.45 | 0.29209 |
| 935.1261 | 1.44E+11 | 1537.15 | 2.42E+21 | 1537.15 | 0.29189 |
| 935.3996 | 1.42E+11 | 1538.84 | 2.42E+21 | 1538.84 | 0.29168 |
| 935.6731 | 1.42E+11 | 1540.53 | 2.42E+21 | 1540.53 | 0.29147 |
| 935.9466 | 1.42E+11 | 1542.23 | 2.42E+21 | 1542.23 | 0.29126 |
| 936.2201 | 1.41E+11 | 1543.92 | 2.42E+21 | 1543.92 | 0.29105 |
| 936.4936 | 1.41E+11 | 1545.61 | 2.42E+21 | 1545.61 | 0.29085 |
| 936.767 | 1.40E+11 | 1547.31 | 2.42E+21 | 1547.31 | 0.29064 |
| 937.0405 | 1.40E+11 | 1549 | 2.42E+21 | 1549 | 0.29043 |
| 937.3139 | 1.40E+11 | 1550.69 | 2.42E+21 | 1550.69 | 0.29023 |
| 937.5873 | 1.39E+11 | 1552.38 | 2.42E+21 | 1552.38 | 0.29002 |
| 937.8607 | 1.38E+11 | 1554.07 | 2.42E+21 | 1554.07 | 0.28982 |
| 938.134 | 1.38E+11 | 1555.77 | 2.42E+21 | 1555.77 | 0.28961 |
| 938.4074 | 1.37E+11 | 1557.46 | 2.42E+21 | 1557.46 | 0.28941 |
| 938.6807 | 1.38E+11 | 1559.15 | 2.42E+21 | 1559.15 | 0.2892 |
| 938.954 | 1.36E+11 | 1560.84 | 2.42E+21 | 1560.84 | 0.289 |
| 939.2273 | 1.36E+11 | 1562.53 | 2.42E+21 | 1562.53 | 0.28879 |
| 939.5005 | 1.36E+11 | 1564.22 | 2.42E+21 | 1564.22 | 0.28859 |
| 939.7738 | 1.35E+11 | 1565.91 | 2.42E+21 | 1565.91 | 0.28839 |
| 940.047 | 1.35E+11 | 1567.6 | 2.42E+21 | 1567.6 | 0.28818 |
| 940.3202 | 1.34E+11 | 1569.3 | 2.42E+21 | 1569.3 | 0.28798 |
| 940.5934 | 1.34E+11 | 1570.99 | 2.42E+21 | 1570.99 | 0.28778 |
| 940.8666 | 1.33E+11 | 1572.68 | 2.41E+21 | 1572.68 | 0.28758 |
| 941.1397 | 1.34E+11 | 1574.37 | 2.41E+21 | 1574.37 | 0.28737 |
| 941.4129 | 1.32E+11 | 1576.06 | 2.41E+21 | 1576.06 | 0.28717 |
| 941.686 | 1.33E+11 | 1577.75 | 2.41E+21 | 1577.75 | 0.28697 |
| 941.9591 | 1.32E+11 | 1579.44 | 2.41E+21 | 1579.44 | 0.28677 |
| 942.2322 | 1.31E+11 | 1581.13 | 2.41E+21 | 1581.13 | 0.28657 |
| 942.5052 | 1.31E+11 | | | | |
| 942.7783 | 1.29E+11 | | | | |
| 943.0513 | 1.30E+11 | | | | |
| 943.3243 | 1.30E+11 | | | | |
| 943.5973 | 1.29E+11 | | | | |
| 943.8703 | 1.29E+11 | | | | |
| 944.1432 | 1.28E+11 | | | | |
| 944.4162 | 1.28E+11 | | | | |
| 944.6891 | 1.27E+11 | | | | |
| 944.962 | 1.28E+11 | | | | |
| 945.2348 | 1.26E+11 | | | | |
| 945.5077 | 1.26E+11 | | | | |
| 945.7805 | 1.26E+11 | | | | |
| 946.0534 | 1.25E+11 | | | | |
| 946.3262 | 1.25E+11 | | | | |
| 946.5989 | 1.24E+11 | | | | |
| 946.8717 | 1.23E+11 | | | | |
| 947.1445 | 1.24E+11 | | | | |
| 947.4172 | 1.23E+11 | | | | |
| 947.6899 | 1.23E+11 | | | | |
| 947.9626 | 1.22E+11 | | | | |
| 948.2353 | 1.22E+11 | | | | |
| 948.5079 | 1.22E+11 | | | | |
| 948.7805 | 1.21E+11 | | | | |
| 949.0532 | 1.21E+11 | | | | |
| 949.3258 | 1.20E+11 | | | | |
| 949.5983 | 1.20E+11 | | | | |
| 949.8709 | 1.20E+11 | | | | |
| 950.1434 | 1.19E+11 | | | | |
| 950.4159 | 1.18E+11 | | | | |
| 950.6885 | 1.18E+11 | | | | |
| 950.9609 | 1.17E+11 | | | | |
| 951.2334 | 1.18E+11 | | | | |
| 951.5059 | 1.17E+11 | | | | |
| 951.7783 | 1.17E+11 | | | | |
| 952.0507 | 1.17E+11 | | | | |
| 952.3231 | 1.15E+11 | | | | |
| 952.5955 | 1.15E+11 | | | | |
| 952.8678 | 1.15E+11 | | | | |
| 953.1401 | 1.16E+11 | | | | |
| 953.4125 | 1.14E+11 | | | | |
| 953.6848 | 1.14E+11 | | | | |
| 953.957 | 1.14E+11 | | | | |
| 954.2293 | 1.14E+11 | | | | |
| 954.5015 | 1.13E+11 | | | | |
| 954.7737 | 1.13E+11 | | | | |
| 955.046 | 1.12E+11 | | | | |
| 955.3181 | 1.11E+11 | | | | |
| 955.5903 | 1.12E+11 | | | | |
| 955.8624 | 1.11E+11 | | | | |
| 956.1346 | 1.11E+11 | | | | |

| | |
|---|---|
| 956.4067 | 1.11E+11 |
| 956.6788 | 1.09E+11 |
| 956.9508 | 1.10E+11 |
| 957.2229 | 1.09E+11 |
| 957.4949 | 1.09E+11 |
| 957.7669 | 1.09E+11 |
| 958.0389 | 1.08E+11 |
| 958.3109 | 1.08E+11 |
| 958.5829 | 1.07E+11 |
| 958.8548 | 1.08E+11 |
| 959.1267 | 1.07E+11 |
| 959.3986 | 1.06E+11 |
| 959.6705 | 1.06E+11 |
| 959.9424 | 1.05E+11 |
| 960.2142 | 1.06E+11 |
| 960.4861 | 1.05E+11 |
| 960.7579 | 1.05E+11 |
| 961.0296 | 1.05E+11 |
| 961.3014 | 1.04E+11 |
| 961.5732 | 1.04E+11 |
| 961.8449 | 1.03E+11 |
| 962.1166 | 1.03E+11 |
| 962.3883 | 1.02E+11 |
| 962.66 | 1.02E+11 |
| 962.9316 | 1.02E+11 |
| 963.2033 | 1.02E+11 |
| 963.4749 | 1.02E+11 |
| 963.7465 | 1.02E+11 |
| 964.0181 | 1.02E+11 |
| 964.2896 | 1.01E+11 |
| 964.5612 | 1.00E+11 |
| 964.8327 | 9.97E+10 |
| 965.1042 | 9.93E+10 |
| 965.3757 | 9.91E+10 |
| 965.6471 | 9.83E+10 |
| 965.9186 | 9.77E+10 |
| 966.19 | 9.71E+10 |
| 966.4614 | 9.77E+10 |
| 966.7328 | 9.69E+10 |
| 967.0042 | 9.65E+10 |
| 967.2755 | 9.61E+10 |
| 967.5469 | 9.54E+10 |
| 967.8182 | 9.54E+10 |
| 968.0895 | 9.48E+10 |
| 968.3607 | 9.50E+10 |
| 968.632 | 9.45E+10 |
| 968.9032 | 9.41E+10 |
| 969.1745 | 9.36E+10 |
| 969.4457 | 9.35E+10 |
| 969.7168 | 9.26E+10 |
| 969.988 | 9.25E+10 |
| 970.2591 | 9.20E+10 |
| 970.5303 | 9.08E+10 |
| 970.8014 | 9.12E+10 |
| 971.0725 | 9.08E+10 |
| 971.3435 | 9.01E+10 |
| 971.6146 | 8.96E+10 |
| 971.8856 | 9.01E+10 |
| 972.1566 | 8.98E+10 |
| 972.4276 | 8.94E+10 |
| 972.6986 | 8.92E+10 |
| 972.9695 | 8.83E+10 |
| 973.2404 | 8.88E+10 |
| 973.5113 | 8.79E+10 |
| 973.7822 | 8.78E+10 |
| 974.0531 | 8.66E+10 |
| 974.324 | 8.68E+10 |
| 974.5948 | 8.67E+10 |
| 974.8656 | 8.59E+10 |
| 975.1364 | 8.57E+10 |
| 975.4072 | 8.58E+10 |
| 975.6779 | 8.50E+10 |
| 975.9487 | 8.52E+10 |
| 976.2194 | 8.46E+10 |
| 976.4901 | 8.41E+10 |
| 976.7608 | 8.32E+10 |
| 977.0314 | 8.32E+10 |
| 977.3021 | 8.33E+10 |
| 977.5727 | 8.25E+10 |
| 977.8433 | 8.21E+10 |
| 978.1139 | 8.25E+10 |
| 978.3845 | 8.15E+10 |
| 978.655 | 8.16E+10 |
| 978.9255 | 8.09E+10 |
| 979.1961 | 8.04E+10 |
| 979.4665 | 8.04E+10 |
| 979.737 | 7.92E+10 |
| 980.0075 | 7.93E+10 |
| 980.2779 | 7.89E+10 |
| 980.5483 | 7.79E+10 |
| 980.8187 | 7.83E+10 |
| 981.0891 | 7.74E+10 |
| 981.3594 | 7.76E+10 |
| 981.6298 | 7.69E+10 |
| 981.9001 | 7.66E+10 |
| 982.1704 | 7.58E+10 |
| 982.4406 | 7.54E+10 |
| 982.7109 | 7.48E+10 |
| 982.9811 | 7.46E+10 |
| 983.2514 | 7.37E+10 |
| 983.5216 | 7.38E+10 |
| 983.7917 | 7.36E+10 |
| 984.0619 | 7.35E+10 |
| 984.332 | 7.24E+10 |
| 984.6022 | 7.18E+10 |
| 984.8723 | 7.09E+10 |
| 985.1423 | 7.07E+10 |
| 985.4124 | 7.03E+10 |
| 985.6824 | 6.97E+10 |
| 985.9525 | 6.87E+10 |
| 986.2225 | 6.85E+10 |
| 986.4925 | 6.77E+10 |
| 986.7624 | 6.73E+10 |
| 987.0324 | 6.62E+10 |
| 987.3023 | 6.57E+10 |
| 987.5722 | 6.49E+10 |
| 987.8421 | 6.45E+10 |
| 988.112 | 6.35E+10 |
| 988.3818 | 6.24E+10 |
| 988.6516 | 6.16E+10 |
| 988.9215 | 6.09E+10 |
| 989.1912 | 5.97E+10 |
| 989.461 | 5.91E+10 |
| 989.7308 | 5.81E+10 |
| 990.0005 | 5.72E+10 |
| 990.2702 | 5.61E+10 |
| 990.5399 | 5.54E+10 |
| 990.8096 | 5.43E+10 |
| 991.0792 | 5.35E+10 |
| 991.3489 | 5.22E+10 |
| 991.6185 | 5.12E+10 |
| 991.8881 | 5.03E+10 |
| 992.1577 | 4.89E+10 |
| 992.4272 | 4.81E+10 |
| 992.6968 | 4.73E+10 |
| 992.9663 | 4.67E+10 |
| 993.2358 | 4.58E+10 |
| 993.5052 | 4.45E+10 |
| 993.7747 | 4.31E+10 |
| 994.0441 | 4.25E+10 |
| 994.3136 | 4.10E+10 |
| 994.583 | 4.04E+10 |
| 994.8523 | 3.90E+10 |
| 995.1217 | 3.79E+10 |
| 995.391 | 3.68E+10 |
| 995.6604 | 3.62E+10 |
| 995.9297 | 3.58E+10 |
| 996.199 | 3.37E+10 |
| 996.4682 | 3.32E+10 |
| 996.7375 | 3.20E+10 |
| 997.0067 | 3.15E+10 |
| 997.2759 | 3.03E+10 |
| 997.5451 | 2.93E+10 |
| 997.8142 | 2.79E+10 |
| 998.0834 | 2.75E+10 |
| 998.3525 | 2.64E+10 |
| 998.6216 | 2.54E+10 |
| 998.8907 | 2.43E+10 |
| 999.1598 | 2.33E+10 |
| 999.4288 | 2.25E+10 |
| 999.6978 | 2.18E+10 |
| 999.9669 | 2.09E+10 |
| 1000.236 | 1.98E+10 |
| 1000.505 | 1.90E+10 |
| 1000.774 | 1.79E+10 |
| 1001.043 | 1.73E+10 |
| 1001.312 | 1.64E+10 |
| 1001.58 | 1.57E+10 |
| 1001.849 | 1.49E+10 |
| 1002.118 | 1.40E+10 |
| 1002.387 | 1.32E+10 |
| 1002.656 | 1.26E+10 |
| 1002.925 | 1.19E+10 |
| 1003.193 | 1.11E+10 |
| 1003.462 | 1.08E+10 |
| 1003.731 | 9.93E+09 |
| 1004 | 9.26E+09 |
| 1004.268 | 8.79E+09 |
| 1004.537 | 8.29E+09 |
| 1004.806 | 7.59E+09 |
| 1005.074 | 7.23E+09 |
| 1005.343 | 6.67E+09 |
| 1005.611 | 6.24E+09 |
| 1005.88 | 5.86E+09 |
| 1006.149 | 5.45E+09 |

on (nIR)

Data for Figure 4
Temperature 1577 K

| Figure 4 | Figure 4 | Figure 4 | Figure 4 | Figure 4 | Figure 4 | Figure 4 | Figure 4 | Figure 4 | Figure 4 |
|---|---|---|---|---|---|---|---|---|---|
| Wavelengt | P/s/sr/nm | Fit | Wavelengt | P/s/sr/nm | Interpolated | Wavelengt | Counts/s | Wavelengt | Counts/s |
| 600.0147 | 415.3916 | 375.9186 | 960.214 | 6398.898 | 6473.285 | 437.661 | 0.05 | 879.476 | 0.14087 |
| 600.3057 | 388.8458 | 377.5787 | 960.486 | 6580.162 | 6479.601 | 437.9568 | 0.01667 | 881.2142 | 0.51534 |
| 600.5967 | 389.32 | 379.244 | 960.758 | 6538.009 | 6485.916 | 438.2526 | 0.08333 | 882.9523 | -0.09165 |
| 600.8877 | 420.0086 | 380.9144 | 961.03 | 6457.604 | 6492.23 | 438.5484 | 0.05 | 884.6903 | 0.0825 |
| 601.1786 | 399.1233 | 382.5899 | 961.301 | 6449.635 | 6498.544 | 438.8442 | -0.01667 | 886.4282 | 0.04951 |
| 601.4696 | 403.5408 | 384.2706 | 961.573 | 6862.715 | 6504.856 | 439.1399 | 0 | 888.166 | 0.077 |
| 601.7605 | 410.102 | 385.9564 | 961.845 | 6508.672 | 6511.168 | 439.4357 | 0.05 | 889.9037 | 0.15551 |
| 602.0515 | 422.7719 | 387.6473 | 962.117 | 6490.414 | 6517.478 | 439.7314 | 0.08333 | 891.6413 | -0.13821 |
| 602.3424 | 429.2657 | 389.3435 | 962.388 | 6504.135 | 6523.787 | 440.0272 | 0 | 893.3788 | -0.53091 |
| 602.6333 | 409.6535 | 391.0448 | 962.66 | 6503.403 | 6530.096 | 440.3229 | 0.13333 | 895.1162 | 0.17976 |
| 602.9242 | 411.1076 | 392.7512 | 962.932 | 6510.271 | 6536.403 | 440.6187 | -0.05 | 896.8535 | 0.28453 |
| 603.215 | 433.5581 | 394.4629 | 963.203 | 6658.39 | 6542.71 | 440.9144 | 0.05 | 898.5908 | 0.27981 |
| 603.5059 | 431.2202 | 396.1797 | 963.475 | 6686.11 | 6549.015 | 441.2101 | 0.08333 | 900.3279 | -0.06004 |
| 603.7968 | 432.2792 | 397.9017 | 963.746 | 6690.679 | 6555.319 | 441.5058 | 0.01667 | 902.0649 | -0.1652 |
| 604.0876 | 416.5326 | 399.629 | 964.018 | 6668.503 | 6561.623 | 441.8015 | -0.03333 | 903.8019 | -0.19171 |
| 604.3785 | 422.1509 | 401.3614 | 964.29 | 6740.766 | 6567.925 | 442.0972 | -0.05 | 905.5387 | 0.24178 |
| 604.6693 | 428.893 | 403.099 | 964.561 | 6621.863 | 6574.226 | 442.3929 | -0.01667 | 907.2754 | -0.04212 |
| 604.9601 | 452.2577 | 404.8419 | 964.833 | 6555.586 | 6580.527 | 442.6886 | -0.03333 | 909.0121 | -0.19171 |
| 605.2509 | 420.9528 | 406.59 | 965.104 | 6833.686 | 6586.826 | 442.9843 | -0.01667 | 910.7486 | 0.18401 |
| 605.5417 | 453.0158 | 408.3433 | 965.376 | 6718.86 | 6593.124 | 443.28 | 0.06667 | 912.4851 | -0.24918 |
| 605.8325 | 449.5076 | 410.1019 | 965.647 | 6669.678 | 6599.421 | 443.5757 | 0 | 914.2214 | 0.14083 |
| 606.1233 | 428.5218 | 411.8657 | 965.919 | 6920.779 | 6605.717 | 443.8713 | 0.05 | 915.9577 | 0.12586 |
| 606.414 | 412.4106 | 413.6347 | 966.19 | 6695.398 | 6612.012 | 444.167 | 0.05 | 917.6939 | 0.05652 |
| 606.7048 | 459.5492 | 415.409 | 966.461 | 6782.565 | 6618.305 | 444.4626 | 0.01667 | 919.4299 | -0.27586 |
| 606.9955 | 441.8993 | 417.1886 | 966.733 | 6681.486 | 6624.598 | 444.7583 | 0.05 | 921.1659 | -0.24605 |
| 607.2863 | 429.4063 | 418.9734 | 967.004 | 6658.303 | 6630.89 | 445.0539 | 0.05 | 922.9017 | -0.06929 |
| 607.577 | 434.4886 | 420.7635 | 967.276 | 6851.251 | 6637.18 | 445.3495 | 0.01667 | 924.6375 | 0.26182 |
| 607.8677 | 412.1338 | 422.5589 | 967.547 | 6810.299 | 6643.469 | 445.6451 | -0.05 | 926.3732 | 0.18846 |
| 608.1584 | 466.8479 | 424.3596 | 967.818 | 6538.095 | 6649.757 | 445.9408 | -0.03333 | 928.1088 | 0.00449 |
| 608.4491 | 437.5152 | 426.1656 | 968.089 | 6769.54 | 6656.045 | 446.2364 | 0.03333 | 929.8442 | -0.0798 |
| 608.7398 | 438.1257 | 427.9769 | 968.361 | 6729.912 | 6662.33 | 446.532 | 0.05 | 931.5796 | -0.36496 |
| 609.0304 | 448.1587 | 429.7935 | 968.632 | 6878.492 | 6668.615 | 446.8276 | 0.03333 | 933.3149 | 0.19271 |
| 609.3211 | 435.9929 | 431.6154 | 968.903 | 6817.259 | 6674.899 | 447.1232 | 0 | 935.05 | -0.12837 |
| 609.6117 | 446.0794 | 433.4426 | 969.174 | 6825.039 | 6681.181 | 447.4187 | 0.03333 | 936.7851 | -0.96371 |
| 609.9023 | 461.4177 | 435.2751 | 969.446 | 7156.948 | 6687.463 | 447.7143 | -0.01667 | 938.5201 | -0.04568 |
| 610.193 | 470.4864 | 437.113 | 969.717 | 6960.628 | 6693.743 | 448.0099 | 0.01667 | 940.255 | -0.03158 |
| 610.4836 | 456.4652 | 438.9562 | 969.988 | 6940.492 | 6700.022 | 448.3054 | 0 | 941.9898 | -0.02442 |
| 610.7742 | 465.8425 | 440.8047 | 970.259 | 6844.973 | 6706.3 | 448.601 | 0.05 | 943.7244 | 0.08921 |
| 611.0648 | 463.6762 | 442.6586 | 970.53 | 6793 | 6712.576 | 448.8965 | 0.03333 | 945.459 | 0.13226 |
| 611.3553 | 470.4098 | 444.5179 | 970.801 | 6860.735 | 6718.851 | 449.1921 | 0 | 947.1935 | 0.28996 |
| 611.6459 | 494.6231 | 446.3825 | 971.072 | 6757.306 | 6725.126 | 449.4876 | -0.08333 | 948.9279 | 0.00952 |
| 611.9365 | 481.4401 | 448.2525 | 971.344 | 6729.712 | 6731.399 | 449.7831 | 0.05 | 950.6622 | 0.20819 |
| 612.227 | 473.5165 | 450.1278 | 971.615 | 6882.448 | 6737.67 | 450.0786 | -0.06667 | 952.3964 | 0.00706 |
| 612.5175 | 484.3274 | 452.0086 | 971.886 | 6969.258 | 6743.941 | 450.3742 | -0.01667 | 954.1304 | -0.26796 |
| 612.8081 | 476.411 | 453.8947 | 972.157 | 6697.995 | 6750.21 | 450.6697 | 0.01667 | 955.8644 | -0.1478 |
| 613.0986 | 470.5611 | 455.7862 | 972.428 | 7046.858 | 6756.478 | 450.9652 | 0.08333 | 957.5983 | 0.10281 |
| 613.3891 | 502.353 | 457.6831 | 972.699 | 7051.314 | 6762.745 | 451.2606 | 0.05 | 959.3321 | 0.29399 |
| 613.6796 | 480.3306 | 459.5854 | 972.969 | 6951.765 | 6769.011 | 451.5561 | -0.01667 | 961.0658 | 0.60729 |
| 613.97 | 489.081 | 461.4931 | 973.24 | 6929.411 | 6775.275 | 451.8516 | 0.03333 | 962.7993 | 0.33344 |
| 614.2605 | 499.8333 | 463.4062 | 973.511 | 6970.972 | 6781.538 | 452.1471 | 0.05 | 964.5328 | 0.32235 |
| 614.551 | 486.0349 | 465.3247 | 973.782 | 6812.773 | 6787.8 | 452.4425 | 0.01667 | 966.2662 | 0.2845 |
| 614.8414 | 492.2017 | 467.2487 | 974.053 | 6967.261 | 6794.06 | 452.738 | 0 | 967.9995 | 0.28327 |
| 615.1318 | 491.0168 | 469.1781 | 974.324 | 6858.017 | 6800.319 | 453.0334 | 0.05 | 969.7327 | 0.62243 |
| 615.4223 | 517.6026 | 471.1129 | 974.595 | 6885.652 | 6806.577 | 453.3289 | 0.01667 | 971.4657 | 1.16219 |
| 615.7127 | 488.6888 | 473.0532 | 974.866 | 6864.542 | 6812.834 | 453.6243 | -0.03333 | 973.1987 | 1.1114 |
| 616.0031 | 496.2503 | 474.9989 | 975.136 | 6824.347 | 6819.089 | 453.9197 | 0.01667 | 974.9316 | 1.15659 |
| 616.2935 | 515.6955 | 476.9501 | 975.407 | 6840.422 | 6825.343 | 454.2152 | -0.08333 | 976.6643 | 1.2091 |
| 616.5838 | 487.373 | 478.9067 | 975.678 | 6937.888 | 6831.596 | 454.5106 | -0.06667 | 978.397 | 1.20972 |
| 616.8742 | 496.0942 | 480.8688 | 975.949 | 7112.564 | 6837.847 | 454.806 | 0 | 980.1296 | 2.13259 |
| 617.1646 | 490.0849 | 482.8363 | 976.219 | 6975.428 | 6844.097 | 455.1014 | 0.01667 | 981.862 | 1.74028 |
| 617.4549 | 520.3669 | 484.8094 | 976.49 | 6810.835 | 6850.346 | 455.3968 | 0.06667 | 983.5944 | 2.06897 |
| 617.7452 | 520.3386 | 486.7879 | 976.761 | 6996.231 | 6856.593 | 455.6922 | 0.01667 | 985.3267 | 2.12689 |
| 618.0356 | 491.2555 | 488.7719 | 977.031 | 6961.453 | 6862.839 | 455.9875 | -0.01667 | 987.0588 | 2.25523 |
| 618.3259 | 506.4959 | 490.7614 | 977.302 | 7042.542 | 6869.084 | 456.2829 | -0.06667 | 988.7909 | 2.99269 |
| 618.6162 | 516.0075 | 492.7563 | 977.573 | 7036.952 | 6875.327 | 456.5783 | 0 | 990.5228 | 2.5217 |
| 618.9065 | 532.5556 | 494.7568 | 977.843 | 7024.684 | 6881.569 | 456.8736 | -0.01667 | 992.2547 | 2.73488 |
| 619.1967 | 513.9973 | 496.7628 | 978.114 | 7087.675 | 6887.81 | 457.169 | 0.01667 | 993.9864 | 3.27489 |
| 619.487 | 525.3569 | 498.7743 | 978.384 | 7009.572 | 6894.049 | 457.4643 | 0.01667 | 995.7181 | 3.20443 |
| 619.7773 | 529.1914 | 500.7913 | 978.655 | 7026.081 | 6900.287 | 457.7597 | -0.05 | 997.4496 | 3.49295 |
| 620.0675 | 518.3409 | 502.8138 | 978.926 | 7048.687 | 6906.523 | 458.055 | 0.01667 | 999.1811 | 3.79532 |
| 620.3577 | 493.6294 | 504.8419 | 979.196 | 6946.592 | 6912.758 | 458.3503 | -0.05 | 1000.912 | 3.64718 |
| 620.648 | 550.2727 | 506.8755 | 979.467 | 7103.992 | 6918.991 | 458.6456 | 0.01667 | 1002.644 | 3.67509 |
| 620.9382 | 517.3316 | 508.9146 | 979.737 | 7054.034 | 6925.224 | 458.9409 | 0.05 | 1004.375 | 4.19089 |
| 621.2284 | 508.1821 | 510.9593 | 980.13 | 8906.659 | 6934.268 | 459.2362 | 0.1 | 1006.106 | 4.6543 |
| 621.5186 | 517.4924 | 513.0095 | 981.862 | 6748.366 | 6974.154 | 459.5315 | 0.08333 | 1007.837 | 4.92991 |
| 621.8087 | 510.7533 | 515.0653 | 983.594 | 7400.501 | 7013.997 | 459.8268 | -0.01667 | 1009.568 | 4.7526 |
| 622.0989 | 516.1273 | 517.1267 | 985.327 | 7061.327 | 7053.793 | 460.1221 | -0.06667 | 1011.298 | 5.31902 |
| 622.389 | 504.3997 | 519.1936 | 987.059 | 6937.12 | 7093.542 | 460.4174 | 0.01667 | 1013.029 | 4.98413 |
| 622.6792 | 550.3912 | 521.266 | 988.791 | 8678.653 | 7133.241 | 460.7126 | 0.01667 | 1014.759 | 5.36907 |
| 622.9693 | 520.5362 | 523.3441 | 990.523 | 6742.035 | 7172.89 | 461.0079 | 0 | 1016.49 | 6.07089 |
| 623.2594 | 540.3371 | 525.4277 | 992.255 | 6929.319 | 7212.486 | 461.3031 | -0.01667 | 1018.22 | 5.72975 |
| 623.5495 | 571.6961 | 527.5169 | 993.986 | 7835.588 | 7252.028 | 461.5984 | 0.13333 | 1019.95 | 5.68797 |
| 623.8396 | 536.4991 | 529.6117 | 995.718 | 7259.576 | 7291.514 | 461.8936 | 0.06667 | 1021.68 | 5.94931 |
| 624.1297 | 573.1808 | 531.7121 | 997.45 | 7501.995 | 7330.943 | 462.1888 | 0.03333 | 1023.41 | 6.79775 |
| 624.4198 | 517.5413 | 533.818 | 999.181 | 7783.697 | 7370.313 | 462.4841 | 0.03333 | 1025.14 | 6.67053 |
| 624.7099 | 544.4182 | 535.9296 | 1000.91 | 7159.482 | 7409.623 | 462.7793 | 0.03333 | 1026.87 | 6.46145 |
| 624.9999 | 557.8056 | 538.0468 | 1002.64 | 6892.557 | 7448.87 | 463.0745 | 0.05 | 1028.6 | 7.01888 |
| 625.29 | 557.5934 | 540.1696 | 1004.37 | 7492.08 | 7488.055 | 463.3697 | 0.03333 | 1030.329 | 6.77752 |
| 625.58 | 547.5822 | 542.298 | 1006.11 | 7993.624 | 7527.174 | 463.6649 | 0.06667 | 1032.059 | 6.62519 |
| 625.87 | 561.2814 | 544.4321 | 1007.84 | 8112.961 | 7566.227 | 463.9601 | -0.01667 | 1033.788 | 7.62611 |
| 626.16 | 535.8973 | 546.5717 | 1009.57 | 7607.078 | 7605.212 | 464.2552 | 0.03333 | 1035.517 | 7.31182 |
| 626.45 | 546.9237 | 548.717 | 1011.3 | 8237.773 | 7644.128 | 464.5504 | -0.01667 | 1037.246 | 7.2819 |
| 626.74 | 575.9234 | 550.868 | 1013.03 | 7486.201 | 7682.972 | 464.8456 | 0.01667 | 1038.975 | 7.16595 |
| 627.03 | 581.1677 | 553.0245 | 1014.76 | 7819.21 | 7721.745 | 465.1407 | 0.06667 | 1040.704 | 7.55247 |
| 627.3199 | 562.2405 | 555.1868 | 1016.49 | 8571.928 | 7760.444 | 465.4359 | 0.01667 | 1042.433 | 8.35575 |
| 627.6099 | 595.1694 | 557.3546 | 1018.22 | 7859.266 | 7799.067 | 465.731 | -0.06667 | 1044.162 | 8.32416 |
| 627.8998 | 576.6848 | 559.5282 | 1019.95 | 7652.629 | 7837.615 | 466.0261 | -0.08333 | 1045.89 | 8.58875 |
| 628.1897 | 631.9645 | 561.7073 | 1021.68 | 7791.988 | 7876.084 | 466.3213 | 0.06667 | 1047.619 | 8.53586 |
| 628.4797 | 601.333 | 563.8922 | 1023.41 | 8701.469 | 7914.474 | 466.6164 | 0.08333 | 1049.347 | 8.02921 |
| 628.7696 | 601.18 | 566.0827 | 1025.14 | 8386.434 | 7952.783 | 466.9115 | 0.01667 | 1051.076 | 7.91738 |
| 629.0595 | 570.3103 | 568.2789 | 1026.87 | 7895.927 | 7991.01 | 467.2066 | -0.01667 | 1052.804 | 8.1265 |
| 629.3493 | 587.1011 | 570.4808 | 1028.6 | 8411.792 | 8029.154 | 467.5017 | 0.05 | 1054.532 | 8.45103 |
| 629.6392 | 606.7724 | 572.6883 | 1030.33 | 8075.436 | 8067.213 | 467.7968 | 0.03333 | 1056.26 | 8.36446 |
| 629.9291 | 590.5423 | 574.9015 | 1032.06 | 7706.679 | 8105.186 | 468.0919 | -0.05 | 1057.988 | 8.23406 |
| 630.2189 | 585.7839 | 577.1205 | 1033.79 | 8801.668 | 8143.072 | 468.3869 | -0.01667 | 1059.716 | 8.54982 |
| 630.5087 | 587.4855 | 579.3451 | 1035.52 | 8384.263 | 8180.869 | 468.682 | 0.03333 | 1061.443 | 8.59088 |
| 630.7986 | 575.9522 | 581.5754 | 1037.25 | 8222.108 | 8218.577 | 468.9771 | 0.06667 | 1063.171 | 8.96139 |
| 631.0884 | 605.1295 | 583.8114 | 1038.98 | 8028.396 | 8256.193 | 469.2721 | -0.05 | 1064.898 | 9.61512 |
| 631.3782 | 600.8846 | 586.0532 | 1040.7 | 8371.801 | 8293.717 | 469.5672 | 0.06667 | 1066.625 | 9.88303 |
| 631.668 | 648.1905 | 588.3006 | 1042.43 | 9061.523 | 8331.147 | 469.8622 | 0.01667 | 1068.353 | 9.53313 |
| 631.9577 | 630.1285 | 590.5538 | 1044.16 | 8962.253 | 8368.483 | 470.1572 | 0.01667 | 1070.08 | 9.33179 |
| 632.2475 | 629.0469 | 592.8126 | 1045.89 | 9190.819 | 8405.723 | 470.4523 | -0.03333 | 1071.807 | 9.18645 |
| 632.5373 | 640.5244 | 595.0772 | 1047.62 | 8977.373 | 8442.865 | 470.7473 | 0.01667 | 1073.534 | 9.34518 |
| 632.827 | 623.3277 | 597.3476 | 1049.35 | 8381.62 | 8479.909 | 471.0423 | -0.01667 | 1075.26 | 9.84844 |
| 633.1167 | 642.4573 | 599.6236 | 1051.08 | 8243.933 | 8516.853 | 471.3373 | -0.05 | 1076.987 | 10.27631 |
| 633.4065 | 650.0158 | 601.9054 | 1052.8 | 8361.131 | 8553.697 | 471.6323 | 0 | 1078.714 | 10.96043 |
| 633.6962 | 646.7672 | 604.193 | 1054.53 | 8571.848 | 8590.439 | 471.9273 | 0.1 | 1080.44 | 9.89129 |
| 633.9859 | 636.7661 | 606.4862 | 1056.26 | 8490.979 | 8627.078 | 472.2222 | -0.05 | 1082.166 | 9.85962 |
| 634.2756 | 645.207 | 608.7853 | 1057.99 | 8332.017 | 8663.612 | 472.5172 | 0.08333 | 1083.893 | 10.46017 |
| 634.5652 | 629.3854 | 611.0901 | 1059.72 | 8454.479 | 8700.042 | 472.8122 | 0.06667 | 1085.619 | 10.2376 |
| 634.8549 | 666.0639 | 613.4006 | 1061.44 | 8461.488 | 8736.365 | 473.1071 | 0.03333 | 1087.345 | 10.96347 |
| 635.1445 | 628.6813 | 615.7169 | 1063.17 | 8708.539 | 8772.582 | 473.4021 | 0.03333 | 1089.071 | 10.88655 |
| 635.4342 | 670.0726 | 618.039 | 1064.9 | 9261.152 | 8808.689 | 473.697 | 0.03333 | 1090.797 | 10.68287 |
| 635.7238 | 664.3596 | 620.3668 | 1066.63 | 9524.366 | 8844.687 | 473.9919 | 0.08333 | 1092.522 | 10.90003 |
| 636.0134 | 655.2198 | 622.7004 | 1068.35 | 9082.418 | 8880.575 | 474.2869 | -0.03333 | 1094.248 | 11.33229 |
| 636.303 | 661.0508 | 625.0398 | 1070.08 | 8800.291 | 8916.351 | 474.5818 | -0.06667 | 1095.973 | 11.31459 |
| 636.5926 | 682.6327 | 627.385 | 1071.81 | 8607.997 | 8952.015 | 474.8767 | -0.03333 | 1097.699 | 11.81998 |
| 636.8822 | 669.4516 | 629.7359 | 1073.53 | 8681.66 | 8987.565 | 475.1716 | 0.05 | 1099.424 | 11.50132 |
| 637.1718 | 648.3546 | 632.0927 | 1075.26 | 9100.463 | 9023.001 | 475.4665 | 0.08333 | 1101.149 | 11.96248 |
| 637.4613 | 674.676 | 634.4552 | 1076.99 | 9415.884 | 9058.321 | 475.7614 | 0.01667 | 1102.874 | 11.90616 |
| 637.7509 | 682.3357 | 636.8235 | 1078.71 | 9917.249 | 9093.525 | 476.0563 | 0.01667 | 1104.599 | 11.61794 |
| 638.0404 | 669.2777 | 639.1976 | 1080.44 | 8823.578 | 9128.612 | 476.3511 | 0.06667 | 1106.324 | 12.09667 |
| 638.3299 | 700.7594 | 641.5776 | 1082.17 | 8709.906 | 9163.58 | 476.646 | | 1108.049 | 11.99723 |
| 638.6194 | 662.8902 | 643.9633 | 1083.89 | 9158.917 | 9198.429 | 476.9408 | | 1109.773 | 12.15563 |
| 638.9089 | 680.9518 | 646.3548 | 1085.62 | 8906.908 | 9233.158 | 477.2357 | 0.01667 | 1111.498 | 12.95312 |
| 639.1984 | 674.767 | 648.7522 | 1087.34 | 9461.907 | 9267.766 | 477.5305 | 0.01667 | 1113.222 | 12.8633 |
| 639.4879 | 678.2065 | 651.1554 | 1089.07 | 9315.904 | 9302.252 | 477.8254 | 0.06667 | 1114.946 | 12.36085 |
| 639.7774 | 678.8145 | 653.5643 | 1090.8 | 9089.424 | 9336.616 | 478.1202 | -0.08333 | 1116.67 | 12.82014 |
| 640.0668 | 694.8796 | 655.9792 | 1092.52 | 9179.159 | 9370.856 | 478.415 | 0 | 1118.394 | 12.77078 |
| 640.3563 | 680.5812 | 658.3998 | 1094.25 | 9473.677 | 9404.971 | 478.7098 | -0.01667 | 1120.118 | 12.65256 |
| 640.6457 | 706.9795 | 660.8263 | 1095.97 | 9407.898 | 9438.961 | 479.0046 | -0.01667 | 1121.842 | 12.8759 |
| 640.9351 | 672.8847 | 663.2586 | 1097.7 | 9744.778 | 9472.826 | 479.2994 | -0.01667 | 1123.566 | 13.33893 |
| 641.2245 | 670.3677 | 665.6967 | 1099.42 | 9422.974 | 9506.563 | 479.5942 | -0.08333 | 1125.29 | 13.30024 |
| 641.5139 | 723.6095 | 668.1407 | | | | 479.889 | -0.03333 | 1127.013 | 13.446 |
| 641.8033 | 675.9482 | 670.5905 | | | | 480.1838 | -0.03333 | 1128.736 | 13.40935 |
| 642.0927 | 653.9851 | 673.0462 | | | | 480.4785 | 0.03333 | 1130.46 | 13.2039 |
| 642.382 | 715.9667 | 675.5078 | | | | 480.7733 | -0.01667 | 1132.183 | 13.33416 |
| 642.6714 | 713.9127 | 677.9751 | | | | 481.068 | 0.06667 | 1133.906 | 13.67824 |
| 642.9607 | 673.6238 | 680.4484 | | | | 481.3628 | -0.03333 | 1135.629 | 13.86002 |
| 643.25 | 699.251 | 682.9275 | | | | 481.6575 | 0.06667 | 1137.352 | 14.24741 |
| 643.5393 | 714.3407 | 685.4125 | | | | 481.9522 | -0.03333 | 1139.074 | 14.34059 |
| 643.8286 | 703.4165 | 687.9033 | | | | 482.247 | 0.01667 | 1140.797 | 14.10724 |
| 644.1179 | 750.1063 | 690.4 | | | | 482.5417 | 0.01667 | 1142.519 | 14.11876 |
| 644.4072 | 730.6788 | 692.9026 | | | | 482.8364 | -0.03333 | 1144.242 | 14.11221 |

| | | |
|---|---|---|
| 644.6965 | 716.6711 | 695.411 |
| 644.9857 | 736.0685 | 697.9254 |
| 645.2749 | 726.6648 | 700.4456 |
| 645.5642 | 703.7279 | 702.9717 |
| 645.8534 | 705.6605 | 705.5037 |
| 646.1426 | 782.9442 | 708.0415 |
| 646.4318 | 731.5769 | 710.5853 |
| 646.721 | 724.4231 | 713.135 |
| 647.0101 | 710.8116 | 715.6906 |
| 647.2993 | 687.1416 | 718.252 |
| 647.5884 | 744.1692 | 720.8194 |
| 647.8776 | 759.6415 | 723.3927 |
| 648.1667 | 759.4426 | 725.9719 |
| 648.4558 | 731.6861 | 728.557 |
| 648.7449 | 758.631 | 731.148 |
| 649.034 | 768.7656 | 733.745 |
| 649.3231 | 728.3692 | 736.3479 |
| 649.6121 | 750.077 | 738.9567 |
| 649.9012 | 743.3736 | 741.5714 |
| 650.1902 | 762.4773 | 744.192 |
| 650.4792 | 777.2697 | 746.8186 |
| 650.7683 | 761.6362 | 749.4512 |
| 651.0573 | 743.6214 | 752.0896 |
| 651.3462 | 758.8013 | 754.734 |
| 651.6352 | 765.6686 | 757.3844 |
| 651.9242 | 786.1536 | 760.0406 |
| 652.2131 | 822.2207 | 762.7029 |
| 652.5021 | 786.295 | 765.371 |
| 652.791 | 757.5353 | 768.0452 |
| 653.0799 | 806.2827 | 770.7253 |
| 653.3688 | 815.7174 | 773.4113 |
| 653.6577 | 772.9494 | 776.1033 |
| 653.9466 | 819.5775 | 778.8013 |
| 654.2355 | 792.2858 | 781.5052 |
| 654.5243 | 827.213 | 784.2151 |
| 654.8132 | 812.9563 | 786.9309 |
| 655.102 | 823.9759 | 789.6528 |
| 655.3908 | 804.3778 | 792.3806 |
| 655.6797 | 812.4204 | 795.1143 |
| 655.9685 | 833.8986 | 797.8541 |
| 656.2572 | 847.3604 | 800.5998 |
| 656.546 | 810.0916 | 803.3515 |
| 656.8348 | 841.6067 | 806.1092 |
| 657.1235 | 820.9885 | 808.8729 |
| 657.4123 | 831.445 | 811.6425 |
| 657.701 | 819.6354 | 814.4182 |
| 657.9897 | 858.4261 | 817.1998 |
| 658.2784 | 852.5465 | 819.9874 |
| 658.5671 | 861.1453 | 822.781 |
| 658.8558 | 818.566 | 825.5806 |
| 659.1444 | 884.8844 | 828.3862 |
| 659.4331 | 872.5768 | 831.1978 |
| 659.7217 | 868.9432 | 834.0154 |
| 660.0103 | 859.912 | 836.839 |
| 660.299 | 855.9194 | 839.6687 |
| 660.5876 | 883.5132 | 842.5043 |
| 660.8762 | 883.2085 | 845.3459 |
| 661.1647 | 840.397 | 848.1935 |
| 661.4533 | 899.0617 | 851.0471 |
| 661.7419 | 886.7957 | 853.9068 |
| 662.0304 | 860.705 | 856.7724 |
| 662.3189 | 917.1579 | 859.6441 |
| 662.6074 | 897.5112 | 862.5218 |
| 662.8959 | 904.2725 | 865.4055 |
| 663.1844 | 890.4138 | 868.2952 |
| 663.4729 | 895.7607 | 871.1909 |
| 663.7614 | 889.7329 | 874.0927 |
| 664.0498 | 916.6211 | 877.0004 |
| 664.3383 | 910.2114 | 879.9142 |
| 664.6267 | 902.9571 | 882.834 |
| 664.9151 | 902.886 | 885.7599 |
| 665.2035 | 930.0568 | 888.6917 |
| 665.4919 | 900.3478 | 891.6296 |
| 665.7803 | 899.94 | 894.5736 |
| 666.0687 | 927.9601 | 897.5235 |
| 666.357 | 892.7859 | 900.4795 |
| 666.6454 | 918.7909 | 903.4415 |
| 666.9337 | 913.7684 | 906.4096 |
| 667.222 | 926.4651 | 909.3836 |
| 667.5103 | 927.2197 | 912.3638 |
| 667.7986 | 922.8442 | 915.3499 |
| 668.0869 | 948.6193 | 918.3421 |
| 668.3752 | 915.1015 | 921.3403 |
| 668.6635 | 928.2792 | 924.3446 |
| 668.9517 | 944.4124 | 927.3549 |
| 669.2399 | 964.2656 | 930.3712 |
| 669.5282 | 975.2157 | 933.3936 |
| 669.8164 | 978.8774 | 936.422 |
| 670.1046 | 961.1613 | 939.4565 |
| 670.3927 | 994.2304 | 942.497 |
| 670.6809 | 977.987 | 945.5435 |
| 670.9691 | 958.4652 | 948.5961 |
| 671.2572 | 988.3246 | 951.6547 |
| 671.5454 | 979.2276 | 954.7194 |
| 671.8335 | 983.4325 | 957.7901 |
| 672.1216 | 990.5519 | 960.8669 |
| 672.4097 | 1007.808 | 963.9497 |
| 672.6978 | 968.9097 | 967.0386 |
| 672.9858 | 990.1936 | 970.1335 |
| 673.2739 | 992.32 | 973.2344 |
| 673.5619 | 983.8248 | 976.3414 |
| 673.85 | 1001.996 | 979.4545 |
| 674.138 | 983.4174 | 982.5736 |
| 674.426 | 974.7976 | 985.6987 |
| 674.714 | 1014.707 | 988.8299 |
| 675.002 | 1011.733 | 991.9672 |
| 675.29 | 1003.342 | 995.1105 |
| 675.5779 | 1021.09 | 998.2598 |
| 675.8659 | 1002.884 | 1001.415 |
| 676.1538 | 1028.512 | 1004.577 |
| 676.4417 | 1029.375 | 1007.744 |
| 676.7296 | 1020.647 | 1010.918 |
| 677.0175 | 1026.27 | 1014.097 |
| 677.3054 | 1041.754 | 1017.283 |
| 677.5933 | 1018.5 | 1020.475 |
| 677.8811 | 1014.322 | 1023.672 |
| 678.169 | 1043.635 | 1026.876 |
| 678.4568 | 1020.285 | 1030.086 |
| 678.7446 | 1073.285 | 1033.302 |
| 679.0324 | 1051.499 | 1036.524 |
| 679.3202 | 1050.545 | 1039.752 |
| 679.608 | 1079.784 | 1042.986 |
| 679.8958 | 1051.478 | 1046.226 |
| 680.1835 | 1057.443 | 1049.472 |
| 680.4713 | 1041.639 | 1052.724 |
| 680.759 | 1080.363 | 1055.982 |
| 681.0467 | 1086.05 | 1059.247 |
| 681.3344 | 1107.236 | 1062.517 |
| 681.6221 | 1100.214 | 1065.793 |
| 681.9098 | 1115.401 | 1069.076 |
| 682.1975 | 1057.084 | 1072.364 |
| 682.4851 | 1120.704 | 1075.659 |
| 682.7728 | 1068.321 | 1078.959 |
| 683.0604 | 1110.008 | 1082.266 |
| 683.348 | 1111.42 | 1085.579 |
| 683.6356 | 1119.97 | 1088.897 |
| 683.9232 | 1118.221 | 1092.222 |
| 684.2108 | 1111.375 | 1095.553 |
| 684.4983 | 1098.2 | 1098.89 |
| 684.7859 | 1146.221 | 1102.232 |
| 685.0734 | 1153.87 | 1105.581 |
| 685.361 | 1125.649 | 1108.936 |
| 685.6485 | 1134.435 | 1112.297 |
| 685.936 | 1127.451 | 1115.664 |
| 686.2235 | 1129.713 | 1119.037 |
| 686.5109 | 1130.257 | 1122.416 |
| 686.7984 | 1122.019 | 1125.801 |
| 687.0858 | 1134.465 | 1129.193 |
| 687.3733 | 1157.129 | 1132.59 |
| 687.6607 | 1163.737 | 1135.993 |
| 687.9481 | 1162.612 | 1139.402 |
| 688.2355 | 1183.593 | 1142.818 |
| 688.5229 | 1159.829 | 1146.239 |
| 688.8103 | 1173.235 | 1149.666 |
| 689.0976 | 1180.406 | 1153.1 |
| 689.385 | 1198.141 | 1156.539 |
| 689.6723 | 1164.409 | 1159.984 |
| 689.9596 | 1205.023 | 1163.436 |

| | | | |
|---|---|---|---|
| 483.1311 | 0 | 1145.964 | 14.5871 |
| 483.4258 | 0.05 | 1147.686 | 14.45227 |
| 483.7204 | 0.01667 | 1149.408 | 14.79942 |
| 484.0151 | -0.03333 | 1151.13 | 14.88791 |
| 484.3098 | 0 | 1152.852 | 14.93099 |
| 484.6044 | 0.13333 | 1154.573 | 14.82 |
| 484.8991 | -0.1 | 1156.295 | 14.79785 |
| 485.1937 | 0.05 | 1158.016 | 15.59495 |
| 485.4883 | -0.03333 | 1159.738 | 15.55294 |
| 485.783 | 0.05 | 1161.459 | 15.01798 |
| 486.0776 | 0.01667 | 1163.18 | 15.4959 |
| 486.3722 | 0.03333 | 1164.901 | 15.99186 |
| 486.6668 | -0.05 | 1166.622 | 15.62658 |
| 486.9614 | 0.08333 | 1168.343 | 15.53709 |
| 487.256 | 0.03333 | 1170.063 | 15.84435 |
| 487.5506 | 0.06667 | 1171.784 | 16.08297 |
| 487.8451 | -0.01667 | 1173.504 | 15.97154 |
| 488.1397 | 0.03333 | 1175.225 | 16.5977 |
| 488.4342 | 0.03333 | 1176.945 | 16.83885 |
| 488.7288 | 0 | 1178.665 | 16.32014 |
| 489.0233 | -0.05 | 1180.385 | 16.10004 |
| 489.3179 | 0.03333 | 1182.105 | 16.56715 |
| 489.6124 | -0.01667 | 1183.824 | 16.79466 |
| 489.9069 | -0.03333 | 1185.544 | 16.9828 |
| 490.2014 | -0.08333 | 1187.263 | 17.53277 |
| 490.4959 | -0.01667 | 1188.983 | 16.83286 |
| 490.7904 | 0.01667 | 1190.702 | 16.27352 |
| 491.0849 | 0.03333 | 1192.421 | 17.02129 |
| 491.3793 | 0.05 | 1194.14 | 17.13705 |
| 491.6738 | 0.08333 | 1195.859 | 17.35175 |
| 491.9683 | 0.06667 | 1197.578 | 17.71338 |
| 492.2627 | -0.05 | 1199.297 | 17.2377 |
| 492.5572 | 0.1 | 1201.015 | 17.15259 |
| 492.8516 | 0.01667 | 1202.734 | 17.17886 |
| 493.146 | -0.01667 | 1204.452 | 17.16651 |
| 493.4404 | 0.1 | 1206.17 | 17.81614 |
| 493.7349 | 0.01667 | 1207.888 | 17.73866 |
| 494.0293 | 0 | 1209.606 | 17.7165 |
| 494.3237 | 0.11667 | 1211.324 | 18.34867 |
| 494.618 | 0.05 | 1213.042 | 17.45773 |
| 494.9124 | -0.03333 | 1214.759 | 17.64369 |
| 495.2068 | 0.05 | 1216.477 | 18.6239 |
| 495.5011 | -0.01667 | 1218.194 | 18.26037 |
| 495.7955 | 0 | 1219.912 | 18.85106 |
| 496.0898 | -0.05 | 1221.629 | 18.65387 |
| 496.3842 | 0.03333 | 1223.346 | 18.49454 |
| 496.6785 | 0.06667 | 1225.063 | 18.80955 |
| 496.9728 | -0.01667 | 1226.779 | 18.38037 |
| 497.2671 | -0.01667 | 1228.496 | 19.17441 |
| 497.5615 | 0.03333 | 1230.213 | 18.80857 |
| 497.8558 | -0.01667 | 1231.929 | 19.47354 |
| 498.15 | 0 | 1233.645 | 19.92952 |
| 498.4443 | 0.13333 | 1235.362 | 19.69752 |
| 498.7386 | 0.05 | 1237.078 | 19.99449 |
| 499.0329 | 0.03333 | 1238.794 | 19.28742 |
| 499.3271 | 0.03333 | 1240.51 | 19.8355 |
| 499.6214 | 0 | 1242.225 | 20.41849 |
| 499.9156 | 0 | 1243.941 | 19.52853 |
| 500.2098 | -0.03333 | 1245.656 | 19.70582 |
| 500.5041 | 0 | 1247.372 | 20.50304 |
| 500.7983 | 0.03333 | 1249.087 | 20.32483 |
| 501.0925 | 0.13333 | 1250.802 | 20.20346 |
| 501.3867 | 0.06667 | 1252.517 | 20.83723 |
| 501.6809 | 0 | 1254.232 | 20.5985 |
| 501.9751 | 0.01667 | 1255.947 | 20.44452 |
| 502.2692 | -0.01667 | 1257.661 | 21.47087 |
| 502.5634 | -0.05 | 1259.376 | 20.97642 |
| 502.8575 | -0.01667 | 1261.09 | 21.09634 |
| 503.1517 | 0.01667 | 1262.805 | 21.25374 |
| 503.4458 | 0.01667 | 1264.519 | 20.71782 |
| 503.74 | 0.01667 | 1266.233 | 21.74283 |
| 504.0341 | 0.06667 | 1267.947 | 21.7364 |
| 504.3282 | 0.11667 | 1269.661 | 21.58899 |
| 504.6223 | 0.03333 | 1271.374 | 22.48888 |
| 504.9164 | 0 | 1273.088 | 22.23724 |
| 505.2105 | -0.01667 | 1274.801 | 21.44772 |
| 505.5046 | -0.06667 | 1276.515 | 22.1284 |
| 505.7987 | 0.01667 | 1278.228 | 22.24701 |
| 506.0927 | -0.01667 | 1279.941 | 21.70628 |
| 506.3868 | 0.06667 | 1281.654 | 21.81479 |
| 506.6808 | 0.03333 | 1283.367 | 22.33423 |
| 506.9749 | -0.03333 | 1285.079 | 22.47621 |
| 507.2689 | -0.06667 | 1286.792 | 22.24366 |
| 507.5629 | 0.05 | 1288.504 | 22.52053 |
| 507.857 | 0.01667 | 1290.217 | 23.36659 |
| 508.151 | 0 | 1291.929 | 23.05301 |
| 508.445 | 0.05 | 1293.641 | 22.69104 |
| 508.7389 | 0 | 1295.353 | 23.29841 |
| 509.0329 | -0.03333 | 1297.065 | 23.0281 |
| 509.3269 | 0.01667 | 1298.776 | 22.67464 |
| 509.6209 | 0 | 1300.488 | 23.00029 |
| 509.9148 | 0 | 1302.199 | 22.79823 |
| 510.2088 | 0.05 | 1303.911 | 22.86101 |
| 510.5027 | -0.01667 | 1305.622 | 23.03256 |
| 510.7966 | 0.05 | 1307.333 | 22.64722 |
| 511.0906 | -0.03333 | 1309.044 | 23.2887 |
| 511.3845 | -0.01667 | 1310.755 | 23.1688 |
| 511.6784 | 0.03333 | 1312.466 | 23.6269 |
| 511.9723 | 0 | 1314.176 | 23.46172 |
| 512.2662 | -0.01667 | 1315.887 | 23.21227 |
| 512.5601 | 0 | 1317.597 | 24.0175 |
| 512.8539 | -0.11667 | 1319.307 | 23.81305 |
| 513.1478 | -0.03333 | 1321.017 | 23.73836 |
| 513.4416 | 0 | 1322.727 | 23.63441 |
| 513.7355 | 0 | 1324.437 | 24.18537 |
| 514.0293 | 0 | 1326.147 | 24.28313 |
| 514.3232 | -0.06667 | 1327.857 | 23.93532 |
| 514.617 | 0.05 | 1329.566 | 24.31612 |
| 514.9108 | -0.01667 | 1331.275 | 24.21344 |
| 515.2046 | -0.01667 | 1332.985 | 24.23032 |
| 515.4984 | 0.11667 | 1334.694 | 23.77357 |
| 515.7922 | -0.03333 | 1336.403 | 23.50462 |
| 516.086 | 0.03333 | 1338.112 | 23.91773 |
| 516.3797 | 0.01667 | 1339.82 | 23.52317 |
| 516.6735 | -0.06667 | 1341.529 | 23.21741 |
| 516.9672 | -0.05 | 1343.237 | 23.53826 |
| 517.261 | -0.03333 | 1344.946 | 24.50019 |
| 517.5547 | -0.01667 | 1346.654 | 24.47795 |
| 517.8484 | 0.11667 | 1348.362 | 23.68066 |
| 518.1422 | 0.1 | 1350.07 | 24.05022 |
| 518.4359 | 0 | 1351.778 | 24.09216 |
| 518.7296 | -0.05 | 1353.486 | 24.19205 |
| 519.0232 | -0.05 | 1355.193 | 24.25095 |
| 519.3169 | 0.05 | 1356.901 | 24.31665 |
| 519.6106 | 0.08333 | 1358.608 | 23.88107 |
| 519.9043 | 0.01667 | 1360.315 | 23.67662 |
| 520.1979 | 0.05 | 1362.022 | 23.62778 |
| 520.4916 | 0 | 1363.729 | 23.26495 |
| 520.7852 | -0.06667 | 1365.436 | 23.43048 |
| 521.0788 | 0.01667 | 1367.143 | 23.80209 |
| 521.3725 | -0.03333 | 1368.849 | 23.37255 |
| 521.6661 | 0.08333 | 1370.556 | 23.06934 |
| 521.9597 | 0.05 | 1372.262 | 23.03159 |
| 522.2533 | 0 | 1373.968 | 23.14452 |
| 522.5469 | 0 | 1375.674 | 23.32801 |
| 522.8404 | 0.06667 | 1377.38 | 23.49637 |
| 523.134 | 0.01667 | 1379.086 | 23.38416 |
| 523.4276 | -0.05 | 1380.792 | 23.37547 |
| 523.7211 | -0.05 | 1382.497 | 23.23903 |
| 524.0147 | 0.08333 | 1384.203 | 22.65909 |
| 524.3082 | 0.1 | 1385.908 | 22.39127 |
| 524.6017 | 0.08333 | 1387.613 | 22.57763 |
| 524.8952 | 0.06667 | 1389.318 | 22.72861 |
| 525.1887 | -0.03333 | 1391.023 | 22.13685 |
| 525.4822 | 0.06667 | 1392.728 | 22.32574 |
| 525.7757 | 0.01667 | 1394.432 | 22.46413 |
| 526.0692 | -0.03333 | 1396.137 | 22.76713 |
| 526.3627 | -0.08333 | 1397.841 | 22.42572 |
| 526.6561 | -0.01667 | 1399.545 | 22.63037 |
| 526.9496 | 0.05 | 1401.25 | 23.02494 |
| 527.243 | 0.06667 | 1402.954 | 21.62551 |
| 527.5365 | 0.01667 | 1404.657 | 21.9824 |
| 527.8299 | -0.03333 | 1406.361 | 22.29499 |
| 528.1233 | -0.01667 | 1408.065 | 22.05258 |
| 528.4167 | 0.06667 | 1409.768 | 22.76579 |
| 528.7101 | 0.01667 | 1411.472 | 22.68085 |
| 529.0035 | -0.01667 | 1413.175 | 22.47677 |
| 529.2969 | 0.03333 | 1414.878 | 22.60701 |

| | | | | | | |
|---|---|---|---|---|---|---|
| 690.2469 | 1187.702 | 1166.893 | 529.5903 | 0.05 | 1416.581 | 22.88033 |
| 690.5342 | 1184.105 | 1170.357 | 529.8836 | 0.01667 | 1418.284 | 22.85624 |
| 690.8215 | 1206.358 | 1173.826 | 530.177 | 0.03333 | 1419.986 | 23.06427 |
| 691.1088 | 1177.281 | 1177.302 | 530.4703 | 0.05 | 1421.689 | 23.52593 |
| 691.396 | 1233.201 | 1180.783 | 530.7637 | -0.06667 | 1423.391 | 22.94943 |
| 691.6833 | 1223.351 | 1184.271 | 531.057 | 0.11667 | 1425.093 | 23.22924 |
| 691.9705 | 1201.151 | 1187.764 | 531.3503 | 0 | 1426.796 | 23.50918 |
| 692.2577 | 1212.545 | 1191.264 | 531.6436 | 0.1 | 1428.498 | 22.98017 |
| 692.5449 | 1257.879 | 1194.77 | 531.9369 | 0.01667 | 1430.2 | 22.86774 |
| 692.8321 | 1212.789 | 1198.281 | 532.2302 | 0.01667 | 1431.901 | 22.99198 |
| 693.1193 | 1200.054 | 1201.799 | 532.5235 | 0.01667 | 1433.603 | 22.71395 |
| 693.4065 | 1211.943 | 1205.323 | 532.8168 | 0.01667 | 1435.304 | 23.29183 |
| 693.6936 | 1227.425 | 1208.852 | 533.1101 | 0 | 1437.006 | 23.79274 |
| 693.9808 | 1207.89 | 1212.388 | 533.4033 | 0.01667 | 1438.707 | 23.06473 |
| 694.2679 | 1249.785 | 1215.929 | 533.6966 | 0.05 | 1440.408 | 23.6752 |
| 694.555 | 1247.364 | 1219.477 | 533.9898 | 0.05 | 1442.109 | 23.93371 |
| 694.8421 | 1240.093 | 1223.031 | 534.283 | 0.01667 | 1443.81 | 23.66412 |
| 695.1292 | 1252.043 | 1226.59 | 534.5762 | -0.01667 | 1445.511 | 23.76267 |
| 695.4163 | 1226.517 | 1230.156 | 534.8695 | 0 | 1447.211 | 23.76726 |
| 695.7033 | 1237.825 | 1233.728 | 535.1627 | -0.06667 | 1448.912 | 23.80234 |
| 695.9904 | 1230.183 | 1237.305 | 535.4559 | -0.05 | 1450.612 | 24.21283 |
| 696.2774 | 1257.31 | 1240.889 | 535.749 | 0.03333 | 1452.312 | 23.98885 |
| 696.5644 | 1244.893 | 1244.478 | 536.0422 | 0.11667 | 1454.012 | 24.04111 |
| 696.8514 | 1290.681 | 1248.074 | 536.3354 | 0.11667 | 1455.712 | 24.38015 |
| 697.1384 | 1222.693 | 1251.676 | 536.6285 | 0.05 | 1457.412 | 24.09517 |
| 697.4254 | 1269.614 | 1255.283 | 536.9217 | 0.03333 | 1459.111 | 24.38359 |
| 697.7124 | 1286.221 | 1258.897 | 537.2148 | 0.03333 | 1460.811 | 24.59866 |
| 697.9993 | 1298.821 | 1262.516 | 537.5079 | 0.01667 | 1462.51 | 24.41695 |
| 698.2863 | 1301.364 | 1266.142 | 537.8011 | -0.05 | 1464.209 | 24.60343 |
| 698.5732 | 1318.635 | 1269.773 | 538.0942 | 0.01667 | 1465.909 | 24.56101 |
| 698.8601 | 1277.514 | 1273.411 | 538.3873 | -0.01667 | 1467.607 | 24.75702 |
| 699.147 | 1326.535 | 1277.054 | 538.6804 | 0.03333 | 1469.306 | 24.19889 |
| 699.4339 | 1312.876 | 1280.704 | 538.9735 | 0.01667 | 1471.005 | 23.88404 |
| 699.7208 | 1331.558 | 1284.359 | 539.2665 | 0.03333 | 1472.704 | 24.83288 |
| 700.0077 | 1323.116 | 1288.02 | 539.5596 | 0.08333 | 1474.402 | 25.28227 |
| 700.2945 | 1303.229 | 1291.688 | 539.8526 | 0.06667 | 1476.1 | 24.5508 |
| 700.5814 | 1321.999 | 1295.361 | 540.1457 | -0.05 | 1477.798 | 24.17315 |
| 700.8682 | 1351.227 | 1299.04 | 540.4387 | 0 | 1479.496 | 24.72012 |
| 701.155 | 1352.43 | 1302.725 | 540.7318 | -0.03333 | 1481.194 | 25.11148 |
| 701.4418 | 1351.743 | 1306.417 | 541.0248 | 0 | 1482.892 | 24.72758 |
| 701.7286 | 1346.812 | 1310.114 | 541.3178 | 0.05 | 1484.59 | 24.55045 |
| 702.0153 | 1357.964 | 1313.817 | 541.6108 | 0.03333 | 1486.287 | 24.63986 |
| 702.3021 | 1322.883 | 1317.526 | 541.9038 | 0.03333 | 1487.984 | 24.99577 |
| 702.5888 | 1320.698 | 1321.241 | 542.1968 | 0.05 | 1489.682 | 24.79827 |
| 702.8756 | 1374.603 | 1324.962 | 542.4897 | 0.01667 | 1491.379 | 24.55253 |
| 703.1623 | 1359.135 | 1328.688 | 542.7827 | 0.01667 | 1493.076 | 24.69976 |
| 703.449 | 1356.882 | 1332.421 | 543.0756 | -0.05 | 1494.772 | 24.32357 |
| 703.7357 | 1375.408 | 1336.16 | 543.3686 | 0.06667 | 1496.469 | 23.63694 |
| 704.0224 | 1389.476 | 1339.905 | 543.6615 | 0.03333 | 1498.166 | 23.5095 |
| 704.309 | 1361.049 | 1343.655 | 543.9545 | -0.05 | 1499.862 | 24.17928 |
| 704.5957 | 1363.72 | 1347.412 | 544.2474 | 0.01667 | 1501.558 | 24.31036 |
| 704.8823 | 1393.993 | 1351.174 | 544.5403 | 0.03333 | 1503.254 | 24.11826 |
| 705.169 | 1350.79 | 1354.942 | 544.8332 | 0 | 1504.95 | 24.05995 |
| 705.4556 | 1339.726 | 1358.717 | 545.1261 | 0.06667 | 1506.646 | 24.87313 |
| 705.7422 | 1388.206 | 1362.497 | 545.4189 | 0.05 | 1508.342 | 24.37534 |
| 706.0288 | 1417.849 | 1366.283 | 545.7118 | 0 | 1510.038 | 23.56784 |
| 706.3153 | 1422.977 | 1370.075 | 546.0047 | 0 | 1511.733 | 23.75426 |
| 706.6019 | 1393.612 | 1373.873 | 546.2975 | 0.05 | 1513.428 | 23.9013 |
| 706.8884 | 1425.953 | 1377.677 | 546.5904 | 0.08333 | 1515.123 | 23.48006 |
| 707.175 | 1421.195 | 1381.487 | 546.8832 | 0.08333 | 1516.818 | 23.57043 |
| 707.4615 | 1416.537 | 1385.302 | 547.176 | 0.03333 | 1518.513 | 24.21899 |
| 707.748 | 1451.775 | 1389.124 | 547.4688 | 0.06667 | 1520.208 | 23.66459 |
| 708.0345 | 1431.75 | 1392.952 | 547.7616 | -0.01667 | 1521.903 | 23.45187 |
| 708.321 | 1408.484 | 1396.785 | 548.0544 | 0.1 | 1523.597 | 23.58632 |
| 708.6074 | 1457.145 | 1400.624 | 548.3472 | -0.03333 | 1525.291 | 23.03491 |
| 708.8939 | 1454.553 | 1404.469 | 548.64 | -0.05 | 1526.986 | 22.93879 |
| 709.1803 | 1434.43 | 1408.32 | 548.9328 | -0.03333 | 1528.68 | 23.39201 |
| 709.4667 | 1447.127 | 1412.177 | 549.2255 | 0.1 | 1530.374 | 23.80666 |
| 709.7532 | 1456.212 | 1416.04 | 549.5183 | 0.05 | 1532.067 | 23.68172 |
| 710.0396 | 1418.728 | 1419.909 | 549.811 | 0.03333 | 1533.761 | 23.04459 |
| 710.3259 | 1448.349 | 1423.783 | 550.1037 | 0.01667 | 1535.455 | 23.35578 |
| 710.6123 | 1489.154 | 1427.664 | 550.3964 | 0 | 1537.148 | 22.6927 |
| 710.8987 | 1455.511 | 1431.55 | 550.6892 | -0.01667 | 1538.841 | 23.00027 |
| 711.185 | 1506.03 | 1435.442 | 550.9819 | 0.01667 | 1540.534 | 23.64089 |
| 711.4714 | 1452.246 | 1439.34 | 551.2745 | 0.03333 | 1542.227 | 23.08403 |
| 711.7577 | 1478.44 | 1443.244 | 551.5672 | -0.01667 | 1543.92 | 23.09805 |
| 712.044 | 1477.179 | 1447.154 | 551.8599 | 0.06667 | 1545.613 | 22.98359 |
| 712.3303 | 1504.233 | 1451.07 | 552.1526 | 0.01667 | 1547.305 | 23.28806 |
| 712.6165 | 1465.994 | 1454.991 | 552.4452 | 0.11667 | 1548.998 | 22.77274 |
| 712.9028 | 1478.631 | 1458.919 | 552.7379 | -0.05 | 1550.69 | 22.5891 |
| 713.1891 | 1468.484 | 1462.852 | 553.0305 | 0 | 1552.382 | 23.05674 |
| 713.4753 | 1525.918 | 1466.791 | 553.3231 | 0.01667 | 1554.074 | 23.28012 |
| 713.7615 | 1505.322 | 1470.736 | 553.6157 | 0 | 1555.766 | 22.86694 |
| 714.0477 | 1470.571 | 1474.686 | 553.9083 | 0.03333 | 1557.457 | 22.64026 |
| 714.3339 | 1523.814 | 1478.643 | 554.2009 | 0.08333 | 1559.149 | 22.50408 |
| 714.6201 | 1503.793 | 1482.605 | 554.4935 | 0.08333 | 1560.84 | 22.62613 |
| 714.9063 | 1548.237 | 1486.573 | 554.7861 | -0.05 | 1562.532 | 22.44963 |
| 715.1924 | 1501.82 | 1490.547 | 555.0787 | 0.01667 | 1564.223 | 20.98225 |
| 715.4786 | 1534.89 | 1494.527 | 555.3712 | 0.1 | 1565.914 | 21.58616 |
| 715.7647 | 1533.872 | 1498.513 | 555.6638 | -0.03333 | 1567.605 | 22.44681 |
| 716.0508 | 1502.908 | 1502.504 | 555.9563 | -0.06667 | 1569.295 | 21.62214 |
| 716.3369 | 1517.146 | 1506.502 | 556.2488 | 0.01667 | 1570.986 | 21.36981 |
| 716.623 | 1536.57 | 1510.505 | 556.5413 | -0.01667 | 1572.676 | 20.91124 |
| 716.9091 | 1557.163 | 1514.514 | 556.8338 | 0 | 1574.367 | 20.65166 |
| 717.1952 | 1571.664 | 1518.528 | 557.1263 | -0.01667 | 1576.057 | 19.9437 |
| 717.4812 | 1534.255 | 1522.549 | 557.4188 | 0.08333 | 1577.747 | 19.2307 |
| 717.7672 | 1571.201 | 1526.575 | 557.7113 | 0 | 1579.437 | 19.21777 |
| 718.0533 | 1583.467 | 1530.607 | 558.0038 | 0.03333 | 1581.126 | 18.61261 |
| 718.3393 | 1564.263 | 1534.645 | 558.2962 | 0.01667 | 1582.816 | 17.94538 |
| 718.6253 | 1573.32 | 1538.688 | 558.5887 | 0.01667 | 1584.505 | 17.42745 |
| 718.9112 | 1598.054 | 1542.738 | 558.8811 | 0 | 1586.195 | 17.61459 |
| 719.1972 | 1547.067 | 1546.793 | 559.1736 | 0.1 | 1587.884 | 16.91792 |
| 719.4832 | 1605.162 | 1550.854 | 559.466 | 0.03333 | 1589.573 | 15.83804 |
| 719.7691 | 1577.226 | 1554.921 | 559.7584 | -0.01667 | 1591.262 | 15.60409 |
| 720.055 | 1569.969 | 1558.993 | 560.0508 | 0 | 1592.95 | 14.96263 |
| 720.3409 | 1569.851 | 1563.071 | 560.3432 | 0 | 1594.639 | 14.63927 |
| 720.6268 | 1654.416 | 1567.155 | 560.6356 | -0.05 | 1596.327 | 13.44969 |
| 720.9127 | 1613.88 | 1571.245 | 560.9279 | -0.01667 | 1598.016 | 12.81549 |
| 721.1986 | 1592.819 | 1575.34 | 561.2203 | 0.03333 | 1599.704 | 12.95632 |
| 721.4845 | 1607.883 | 1579.442 | 561.5127 | 0 | 1601.392 | 12.37909 |
| 721.7703 | 1605.814 | 1583.548 | 561.805 | -0.03333 | 1603.08 | 11.5364 |
| 722.0561 | 1598.933 | 1587.661 | 562.0973 | 0.01667 | 1604.767 | 11.27621 |
| 722.3419 | 1599.585 | 1591.78 | 562.3897 | -0.01667 | 1606.455 | 10.43889 |
| 722.6278 | 1653.257 | 1595.904 | 562.682 | 0.03333 | 1608.142 | 9.70578 |
| 722.9135 | 1612.077 | 1600.034 | 562.9743 | 0.11667 | 1609.83 | 9.34183 |
| 723.1993 | 1652.925 | 1604.169 | 563.2666 | 0.03333 | 1611.517 | 8.76331 |
| 723.4851 | 1635.264 | 1608.31 | 563.5589 | -0.03333 | 1613.204 | 7.89758 |
| 723.7708 | 1666.414 | 1612.457 | 563.8511 | 0.01667 | 1614.891 | 7.98805 |
| 724.0566 | 1656.114 | 1616.61 | 564.1434 | -0.03333 | 1616.577 | 7.01282 |
| 724.3423 | 1610.895 | 1620.769 | 564.4357 | 0.01667 | 1618.264 | 6.51091 |
| 724.628 | 1678.876 | 1624.933 | 564.7279 | 0 | 1619.95 | 6.53277 |
| 724.9137 | 1696.616 | 1629.102 | 565.0201 | 0.03333 | 1621.637 | 5.5747 |
| 725.1994 | 1651.02 | 1633.278 | 565.3124 | -0.11667 | 1623.323 | 4.8675 |
| 725.485 | 1643.484 | 1637.459 | 565.6046 | 0 | 1625.009 | 4.07564 |
| 725.7707 | 1656.697 | 1641.646 | 565.8968 | 0.01667 | 1626.695 | 4.58891 |
| 726.0563 | 1641.965 | 1645.839 | 566.189 | -0.06667 | 1628.38 | 4.09708 |
| 726.3419 | 1674.739 | 1650.037 | 566.4812 | 0.01667 | 1630.066 | 2.79944 |
| 726.6275 | 1654.019 | 1654.241 | 566.7733 | 0.05 | 1631.751 | 2.97122 |
| 726.9131 | 1685.526 | 1658.45 | 567.0655 | 0 | 1633.437 | 2.61956 |
| 727.1987 | 1689.72 | 1662.665 | 567.3577 | -0.01667 | 1635.122 | 2.0506 |
| 727.4843 | 1718.363 | 1666.886 | 567.6498 | 0.03333 | 1636.807 | 2.20554 |
| 727.7698 | 1720.552 | 1671.113 | 567.9419 | 0 | 1638.492 | 2.23761 |
| 728.0554 | 1641.896 | 1675.345 | 568.2341 | 0.1 | 1640.176 | 1.78343 |
| 728.3409 | 1697.327 | 1679.583 | 568.5262 | -0.01667 | 1641.861 | 0.6723 |
| 728.6264 | 1684.402 | 1683.826 | 568.8183 | 0 | 1643.545 | 0.44522 |
| 728.9119 | 1700.347 | 1688.075 | 569.1104 | 0 | 1645.23 | 0.18297 |
| 729.1974 | 1740.746 | 1692.33 | 569.4025 | -0.05 | 1646.914 | 0.30944 |
| 729.4829 | 1672.748 | 1696.59 | 569.6946 | 0.03333 | 1648.598 | 0.52812 |
| 729.7683 | 1719.248 | 1700.856 | 569.9866 | 0 | 1650.282 | 0.3765 |
| 730.0538 | 1712.228 | 1705.128 | 570.2787 | 0.03333 | 1651.965 | 0.43034 |
| 730.3392 | 1758.517 | 1709.405 | 570.5707 | -0.03333 | 1653.649 | -0.34532 |
| 730.6246 | 1710.347 | 1713.688 | 570.8628 | 0.01667 | 1655.332 | 0.28261 |
| 730.91 | 1769.94 | 1717.976 | 571.1548 | 0 | 1657.015 | 0.30331 |
| 731.1954 | 1724.499 | 1722.27 | 571.4468 | -0.01667 | 1658.698 | 0.64099 |
| 731.4808 | 1717.989 | 1726.57 | 571.7388 | 0.01667 | 1660.381 | 0.28152 |
| 731.7662 | 1710.002 | 1730.875 | 572.0308 | 0.01667 | 1662.064 | -0.09958 |
| 732.0515 | 1766.664 | 1735.186 | 572.3228 | 0.06667 | 1663.747 | 0.5861 |
| 732.3368 | 1739.658 | 1739.502 | 572.6148 | 0 | 1665.429 | -0.29588 |
| 732.6222 | 1705.162 | 1743.824 | 572.9067 | -0.1 | 1667.112 | -0.52693 |
| 732.9075 | 1756.252 | 1748.152 | 573.1987 | 0 | 1668.794 | -0.34865 |
| 733.1928 | 1762.5 | 1752.485 | 573.4906 | 0.06667 | 1670.476 | -0.11981 |
| 733.478 | 1737.207 | 1756.823 | 573.7826 | -0.01667 | 1672.158 | 0.02399 |
| 733.7633 | 1799.941 | 1761.168 | 574.0745 | 0.01667 | 1673.84 | 0.09099 |
| 734.0485 | 1780.35 | 1765.517 | 574.3664 | 0 | 1675.521 | 0.32944 |
| 734.3338 | 1764.596 | 1769.873 | 574.6583 | 0.03333 | 1677.203 | -0.04638 |
| 734.619 | 1762.836 | 1774.233 | 574.9502 | -0.03333 | 1678.884 | 0.12001 |
| 734.9042 | 1755.165 | 1778.6 | 575.2421 | 0.05 | 1680.565 | 0.39887 |
| 735.1894 | 1712.321 | 1782.972 | 575.534 | 0 | 1682.246 | 0.06191 |

| | | |
|---|---|---|
| 735.4746 | 1786.437 | 1787.349 |
| 735.7597 | 1771.5 | 1791.732 |
| 736.0449 | 1769.96 | 1796.121 |
| 736.33 | 1821.588 | 1800.515 |
| 736.6152 | 1770.707 | 1804.914 |
| 736.9003 | 1802.416 | 1809.319 |
| 737.1854 | 1804.497 | 1813.729 |
| 737.4704 | 1850.02 | 1818.145 |
| 737.7555 | 1776.513 | 1822.567 |
| 738.0406 | 1804.079 | 1826.994 |
| 738.3256 | 1813.076 | 1831.426 |
| 738.6106 | 1831.569 | 1835.864 |
| 738.8956 | 1852.905 | 1840.307 |
| 739.1806 | 1808.71 | 1844.756 |
| 739.4656 | 1833.437 | 1849.211 |
| 739.7506 | 1867.549 | 1853.67 |
| 740.0355 | 1850.278 | 1858.135 |
| 740.3205 | 1877.188 | 1862.606 |
| 740.6054 | 1842.422 | 1867.082 |
| 740.8903 | 1895.407 | 1871.563 |
| 741.1752 | 1874.692 | 1876.05 |
| 741.4601 | 1875.043 | 1880.543 |
| 741.745 | 1862.147 | 1885.04 |
| 742.0298 | 1859.337 | 1889.544 |
| 742.3147 | 1870.639 | 1894.052 |
| 742.5995 | 1932.053 | 1898.566 |
| 742.8843 | 1894.065 | 1903.085 |
| 743.1691 | 1902.906 | 1907.61 |
| 743.4539 | 1934.071 | 1912.14 |
| 743.7387 | 1934.92 | 1916.676 |
| 744.0234 | 1896.298 | 1921.216 |
| 744.3082 | 1913.219 | 1925.763 |
| 744.5929 | 1920.6 | 1930.314 |
| 744.8776 | 1973.719 | 1934.871 |
| 745.1623 | 1934.536 | 1939.433 |
| 745.447 | 1970.352 | 1944.001 |
| 745.7317 | 1915.781 | 1948.574 |
| 746.0163 | 1967.035 | 1953.152 |
| 746.301 | 1981.398 | 1957.736 |
| 746.5856 | 1978.987 | 1962.325 |
| 746.8702 | 1952.16 | 1966.919 |
| 747.1548 | 1967.196 | 1971.519 |
| 747.4394 | 1975.638 | 1976.123 |
| 747.7239 | 1982.149 | 1980.734 |
| 748.0085 | 1950.693 | 1985.349 |
| 748.293 | 1985.657 | 1989.97 |
| 748.5776 | 1985.136 | 1994.596 |
| 748.8621 | 1989.291 | 1999.227 |
| 749.1466 | 2029.456 | 2003.863 |
| 749.4311 | 1981.383 | 2008.505 |
| 749.7155 | 2007.206 | 2013.152 |
| 750 | 2011.885 | 2017.805 |
| 750.2844 | 2048.736 | 2022.462 |
| 750.5689 | 2039.975 | 2027.125 |
| 750.8533 | 2026.225 | 2031.793 |
| 751.1377 | 2025.085 | 2036.466 |
| 751.4221 | 2050.789 | 2041.145 |
| 751.7064 | 2014.746 | 2045.828 |
| 751.9908 | 2065.658 | 2050.517 |
| 752.2751 | 2016.307 | 2055.211 |
| 752.5595 | 2079.654 | 2059.911 |
| 752.8438 | 2090.928 | 2064.615 |
| 753.1281 | 2094.04 | 2069.325 |
| 753.4124 | 2132.129 | 2074.04 |
| 753.6966 | 2053.824 | 2078.76 |
| 753.9809 | 2079.363 | 2083.485 |
| 754.2651 | 2106.258 | 2088.215 |
| 754.5493 | 2081.732 | 2092.951 |
| 754.8336 | 2086.458 | 2097.691 |
| 755.1178 | 2110.945 | 2102.437 |
| 755.4019 | 2074.453 | 2107.188 |
| 755.6861 | 2080.049 | 2111.944 |
| 755.9703 | 2103.996 | 2116.705 |
| 756.2544 | 2093.069 | 2121.471 |
| 756.5385 | 2081.061 | 2126.243 |
| 756.8226 | 2121.543 | 2131.019 |
| 757.1067 | 2164.772 | 2135.801 |
| 757.3908 | 2139.413 | 2140.588 |
| 757.6749 | 2125.176 | 2145.379 |
| 757.9589 | 2162.959 | 2150.176 |
| 758.243 | 2134.671 | 2154.978 |
| 758.527 | 2123.046 | 2159.785 |
| 758.811 | 2105.303 | 2164.597 |
| 759.095 | 2113.314 | 2169.414 |
| 759.379 | 2180.17 | 2174.236 |
| 759.6629 | 2173.275 | 2179.064 |
| 759.9469 | 2212.876 | 2183.896 |
| 760.2308 | 2187.328 | 2188.733 |
| 760.5147 | 2194.291 | 2193.575 |
| 760.7986 | 2202.715 | 2198.423 |
| 761.0825 | 2200.567 | 2203.275 |
| 761.3664 | 2129.28 | 2208.132 |
| 761.6503 | 2242.915 | 2212.994 |
| 761.9341 | 2194.21 | 2217.862 |
| 762.218 | 2232.969 | 2222.734 |
| 762.5018 | 2213.92 | 2227.611 |
| 762.7856 | 2237.955 | 2232.493 |
| 763.0694 | 2206.154 | 2237.38 |
| 763.3531 | 2214.894 | 2242.272 |
| 763.6369 | 2273.302 | 2247.169 |
| 763.9206 | 2260.103 | 2252.071 |
| 764.2044 | 2258.532 | 2256.978 |
| 764.4881 | 2245.754 | 2261.89 |
| 764.7718 | 2304.392 | 2266.807 |
| 765.0555 | 2311.679 | 2271.729 |
| 765.3392 | 2228.777 | 2276.655 |
| 765.6228 | 2327.503 | 2281.587 |
| 765.9065 | 2253.439 | 2286.523 |
| 766.1901 | 2273.485 | 2291.464 |
| 766.4737 | 2278.58 | 2296.41 |
| 766.7573 | 2319.637 | 2301.361 |
| 767.0409 | 2325.527 | 2306.317 |
| 767.3244 | 2327.324 | 2311.278 |
| 767.608 | 2252.506 | 2316.244 |
| 767.8915 | 2262.117 | 2321.214 |
| 768.1751 | 2314.063 | 2326.189 |
| 768.4586 | 2296.4 | 2331.169 |
| 768.7421 | 2296.349 | 2336.154 |
| 769.0256 | 2311.021 | 2341.144 |
| 769.309 | 2318.976 | 2346.139 |
| 769.5925 | 2272.578 | 2351.138 |
| 769.8759 | 2313.606 | 2356.142 |
| 770.1593 | 2390.031 | 2361.151 |
| 770.4427 | 2330.053 | 2366.165 |
| 770.7261 | 2410.104 | 2371.183 |
| 771.0095 | 2352.776 | 2376.207 |
| 771.2929 | 2344.216 | 2381.235 |
| 771.5762 | 2321.014 | 2386.267 |
| 771.8595 | 2384.424 | 2391.305 |
| 772.1429 | 2352.483 | 2396.347 |
| 772.4262 | 2422.033 | 2401.394 |
| 772.7094 | 2403.874 | 2406.446 |
| 772.9927 | 2410.843 | 2411.502 |
| 773.276 | 2385.204 | 2416.563 |
| 773.5592 | 2428.031 | 2421.629 |
| 773.8424 | 2391.221 | 2426.7 |
| 774.1257 | 2428.726 | 2431.775 |
| 774.4089 | 2440.082 | 2436.855 |
| 774.692 | 2407.314 | 2441.939 |
| 774.9752 | 2404.078 | 2447.029 |
| 775.2584 | 2378.392 | 2452.123 |
| 775.5415 | 2393.79 | 2457.221 |
| 775.8246 | 2447.29 | 2462.324 |
| 776.1077 | 2418.386 | 2467.432 |
| 776.3908 | 2426.21 | 2472.545 |
| 776.6739 | 2459.167 | 2477.662 |
| 776.957 | 2422.432 | 2482.783 |
| 777.24 | 2449.64 | 2487.91 |
| 777.5231 | 2451.556 | 2493.041 |
| 777.8061 | 2479.22 | 2498.176 |
| 778.0891 | 2472.217 | 2503.316 |
| 778.3721 | 2478.956 | 2508.461 |
| 778.655 | 2473.989 | 2513.61 |
| 778.938 | 2468.885 | 2518.764 |
| 779.2209 | 2509.388 | 2523.922 |
| 779.5039 | 2454.123 | 2529.085 |
| 779.7868 | 2530.256 | 2534.253 |
| 780.0697 | 2527.037 | 2539.425 |

| | | | |
|---|---|---|---|
| 575.8259 | 0.05 | 1683.927 | 0.33248 |
| 576.1177 | 0 | 1685.608 | -0.1056 |
| 576.4096 | 0.06667 | 1687.289 | -0.04264 |
| 576.7014 | 0.06667 | 1688.969 | 0.42105 |
| 576.9932 | 0.03333 | 1690.649 | -0.48381 |
| 577.285 | 0.01667 | 1692.329 | -0.17083 |
| 577.5768 | 0.06667 | 1694.009 | 0.15011 |
| 577.8686 | -0.03333 | 1695.689 | -0.00469 |
| 578.1604 | 0.01667 | 1697.369 | 0.19438 |
| 578.4522 | 0 | 1699.049 | -0.01014 |
| 578.744 | 0.06667 | 1700.728 | 0.27439 |
| 579.0357 | 0.08333 | 1702.407 | 0.52688 |
| 579.3275 | 0.01667 | 1704.086 | 0.43256 |
| 579.6192 | 0.01667 | 1705.765 | -0.31083 |
| 579.9109 | 0.03333 | 1707.444 | -0.33955 |
| 580.2026 | 0.05 | 1709.123 | -0.16448 |
| 580.4943 | 0.01667 | 1710.801 | -0.43714 |
| 580.786 | -0.03333 | 1712.48 | 0.03887 |
| 581.0777 | 0.06667 | 1714.158 | 0.1599 |
| 581.3694 | 0.1 | 1715.836 | 0.04163 |
| 581.661 | -0.01667 | 1717.514 | 0.02254 |
| 581.9527 | -0.01667 | 1719.191 | 0.08646 |
| 582.2443 | 0.1 | 1720.869 | -0.22077 |
| 582.536 | -0.03333 | 1722.547 | -0.40542 |
| 582.8276 | -0.05 | 1724.224 | -0.29338 |
| 583.1192 | 0.05 | 1725.901 | -0.26171 |
| 583.4108 | -0.01667 | 1727.578 | 0.14007 |
| 583.7024 | 0.03333 | 1729.255 | -0.16121 |
| 583.994 | 0.03333 | 1730.932 | -0.06421 |
| 584.2855 | -0.01667 | 1732.608 | -0.0477 |
| 584.5771 | 0.01667 | 1734.285 | 0.07645 |
| 584.8686 | 0.08333 | 1735.961 | -0.17035 |
| 585.1602 | 0.08333 | 1737.637 | -0.56045 |
| 585.4517 | 0.03333 | 1739.313 | -0.08008 |
| 585.7432 | 0 | 1740.989 | 0.06819 |
| 586.0347 | 0.08333 | 1742.664 | -0.05061 |
| 586.3262 | 0.08333 | 1744.34 | -0.09124 |
| 586.6177 | 0.03333 | 1746.015 | -0.01366 |
| 586.9092 | 0.01667 | 1747.69 | -0.06367 |
| 587.2007 | 0.05 | 1749.365 | 0.53908 |
| 587.4921 | 0.05 | 1751.04 | 0.36885 |
| 587.7836 | 0.03333 | 1752.715 | -0.19391 |
| 588.075 | 0.11667 | | |
| 588.3664 | 0.03333 | | |
| 588.6578 | 0.1 | | |
| 588.9492 | 0.03333 | | |
| 589.2406 | 0.06667 | | |
| 589.532 | 0.11667 | | |
| 589.8234 | 0.03333 | | |
| 590.1147 | 0.16667 | | |
| 590.4061 | 0.21667 | | |
| 590.6974 | 0.2 | | |
| 590.9888 | 0.28333 | | |
| 591.2801 | 0.26667 | | |
| 591.5714 | 0.26667 | | |
| 591.8627 | 0.33333 | | |
| 592.154 | 0.35 | | |
| 592.4453 | 0.51667 | | |
| 592.7365 | 0.51667 | | |
| 593.0278 | 0.68333 | | |
| 593.319 | 0.5 | | |
| 593.6103 | 0.61667 | | |
| 593.9015 | 0.78333 | | |
| 594.1927 | 0.91667 | | |
| 594.4839 | 0.81667 | | |
| 594.7751 | 0.93333 | | |
| 595.0663 | 0.66667 | | |
| 595.3575 | 0.98333 | | |
| 595.6486 | 0.93333 | | |
| 595.9398 | 1.18333 | | |
| 596.2309 | 1.3 | | |
| 596.522 | 1.23333 | | |
| 596.8132 | 1.28333 | | |
| 597.1043 | 1.38333 | | |
| 597.3954 | 1.51667 | | |
| 597.6864 | 1.51667 | | |
| 597.9775 | 1.6 | | |
| 598.2686 | 1.65 | | |
| 598.5596 | 1.68333 | | |
| 598.8507 | 1.83333 | | |
| 599.1417 | 1.95 | | |
| 599.4327 | 1.93333 | | |
| 599.7238 | 1.96667 | | |
| 600.0148 | 2.15 | | |
| 600.3057 | 2.05 | | |
| 600.5967 | 2.15 | | |
| 600.8877 | 2.36667 | | |
| 601.1786 | 2.3 | | |
| 601.4696 | 2.36667 | | |
| 601.7605 | 2.5 | | |
| 602.0515 | 2.63333 | | |
| 602.3424 | 2.73333 | | |
| 602.6333 | 2.66667 | | |
| 602.9242 | 2.76667 | | |
| 603.2151 | 2.98333 | | |
| 603.5059 | 3.03333 | | |
| 603.7968 | 3.1 | | |
| 604.0876 | 3.06667 | | |
| 604.3785 | 3.15 | | |
| 604.6693 | 3.25 | | |
| 604.9601 | 3.51667 | | |
| 605.2509 | 3.31667 | | |
| 605.5417 | 3.63333 | | |
| 605.8325 | 3.65 | | |
| 606.1233 | 3.51667 | | |
| 606.414 | 3.46667 | | |
| 606.7048 | 3.9 | | |
| 606.9955 | 3.86667 | | |
| 607.2863 | 3.75 | | |
| 607.577 | 3.85 | | |
| 607.8677 | 3.71667 | | |
| 608.1584 | 4.28333 | | |
| 608.4491 | 4.05 | | |
| 608.7398 | 4.1 | | |
| 609.0304 | 4.21667 | | |
| 609.3211 | 4.15 | | |
| 609.6117 | 4.21667 | | |
| 609.9023 | 4.4 | | |
| 610.193 | 4.46667 | | |
| 610.4836 | 4.36667 | | |
| 610.7742 | 4.48333 | | |
| 611.0648 | 4.5 | | |
| 611.3553 | 4.51667 | | |
| 611.6459 | 4.76667 | | |
| 611.9365 | 4.66667 | | |
| 612.227 | 4.58333 | | |
| 612.5175 | 4.71667 | | |
| 612.8081 | 4.65 | | |
| 613.0986 | 4.56667 | | |
| 613.3891 | 4.83333 | | |
| 613.6796 | 4.66667 | | |
| 613.97 | 4.73333 | | |
| 614.2605 | 4.88333 | | |
| 614.551 | 4.7 | | |
| 614.8414 | 4.8 | | |
| 615.1318 | 4.76667 | | |
| 615.4223 | 5.01667 | | |
| 615.7127 | 4.81667 | | |
| 616.0031 | 4.85 | | |
| 616.2935 | 5.08333 | | |
| 616.5838 | 4.81667 | | |
| 616.8742 | 4.93333 | | |
| 617.1646 | 4.86667 | | |
| 617.4549 | 5.2 | | |
| 617.7452 | 5.16667 | | |
| 618.0356 | 4.95 | | |
| 618.3259 | 5.08333 | | |
| 618.6162 | 5.18333 | | |
| 618.9065 | 5.33333 | | |
| 619.1967 | 5.16667 | | |
| 619.487 | 5.28333 | | |
| 619.7773 | 5.35 | | |
| 620.0675 | 5.26667 | | |
| 620.3577 | 5.01667 | | |
| 620.648 | 5.6 | | |
| 620.9382 | 5.28333 | | |
| 621.2284 | 5.16667 | | |
| 621.5186 | 5.31667 | | |

| | | | | |
|---|---|---|---|---|
| 780.3526 | 2597.824 | 2544.601 | 621.8087 | 5.23333 |
| 780.6354 | 2528.296 | 2549.782 | 622.0989 | 5.28333 |
| 780.9183 | 2547.917 | 2554.968 | 622.3891 | 5.11667 |
| 781.2011 | 2554.847 | 2560.158 | 622.6792 | 5.6 |
| 781.484 | 2509.184 | 2565.352 | 622.9693 | 5.35 |
| 781.7668 | 2557.747 | 2570.552 | 623.2594 | 5.51667 |
| 782.0496 | 2571.14 | 2575.755 | 623.5496 | 5.85 |
| 782.3323 | 2583.711 | 2580.963 | 623.8397 | 5.5 |
| 782.6151 | 2538.751 | 2586.175 | 624.1297 | 5.88333 |
| 782.8978 | 2586.648 | 2591.392 | 624.4198 | 5.33333 |
| 783.1806 | 2564.86 | 2596.614 | 624.7099 | 5.63333 |
| 783.4633 | 2580.181 | 2601.84 | 624.9999 | 5.7 |
| 783.746 | 2541.123 | 2607.07 | 625.29 | 5.68333 |
| 784.0287 | 2604.488 | 2612.304 | 625.58 | 5.65 |
| 784.3114 | 2545.424 | 2617.544 | 625.87 | 5.76667 |
| 784.594 | 2636.706 | 2622.787 | 626.16 | 5.6 |
| 784.8767 | 2606.323 | 2628.035 | 626.45 | 5.65 |
| 785.1593 | 2604.179 | 2633.287 | 626.74 | 5.91667 |
| 785.4419 | 2561.696 | 2638.544 | 627.03 | 5.98333 |
| 785.7245 | 2638.274 | 2643.805 | 627.3199 | 5.73333 |
| 786.0071 | 2674.804 | 2649.07 | 627.6099 | 6.11667 |
| 786.2896 | 2586.781 | 2654.34 | 627.8998 | 5.91667 |
| 786.5722 | 2617.871 | 2659.614 | 628.1897 | 6.46667 |
| 786.8547 | 2668.157 | 2664.893 | 628.4797 | 6.16667 |
| 787.1372 | 2654.431 | 2670.176 | 628.7696 | 6.1 |
| 787.4197 | 2700.568 | 2675.463 | 629.0595 | 5.86667 |
| 787.7022 | 2643.441 | 2680.754 | 629.3493 | 6.01667 |
| 787.9847 | 2625.634 | 2686.05 | 629.6392 | 6.26667 |
| 788.2672 | 2671.503 | 2691.35 | 629.9291 | 6.08333 |
| 788.5496 | 2649.197 | 2696.655 | 630.2189 | 6.08333 |
| 788.832 | 2679.063 | 2701.964 | 630.5087 | 6.06667 |
| 789.1145 | 2720.568 | 2707.277 | 630.7986 | 5.95 |
| 789.3969 | 2669.625 | 2712.594 | 631.0884 | 6.16667 |
| 789.6792 | 2684.177 | 2717.915 | 631.3782 | 6.18333 |
| 789.9616 | 2780.67 | 2723.241 | 631.668 | 6.63333 |
| 790.244 | 2760.856 | 2728.571 | 631.9578 | 6.45 |
| 790.5263 | 2675.806 | 2733.906 | 632.2475 | 6.43333 |
| 790.8086 | 2702.21 | 2739.244 | 632.5373 | 6.46667 |
| 791.0909 | 2699.038 | 2744.587 | 632.827 | 6.35 |
| 791.3732 | 2738.781 | 2749.934 | 633.1167 | 6.53333 |
| 791.6555 | 2665.54 | 2755.285 | 633.4065 | 6.55 |
| 791.9378 | 2766.404 | 2760.641 | 633.6962 | 6.53333 |
| 792.22 | 2755.099 | 2766 | 633.9859 | 6.46667 |
| 792.5022 | 2707.77 | 2771.364 | 634.2756 | 6.6 |
| 792.7845 | 2730.029 | 2776.732 | 634.5652 | 6.41667 |
| 793.0667 | 2748.477 | 2782.105 | 634.8549 | 6.76667 |
| 793.3488 | 2804.776 | 2787.481 | 635.1445 | 6.41667 |
| 793.631 | 2737.325 | 2792.861 | 635.4342 | 6.8 |
| 793.9132 | 2768.296 | 2798.246 | 635.7238 | 6.8 |
| 794.1953 | 2757.741 | 2803.635 | 636.0134 | 6.71667 |
| 794.4774 | 2779.762 | 2809.028 | 636.303 | 6.8 |
| 794.7595 | 2810.186 | 2814.425 | 636.5926 | 7.05 |
| 795.0416 | 2770.869 | 2819.826 | 636.8822 | 6.93333 |
| 795.3237 | 2818.749 | 2825.232 | 637.1718 | 6.71667 |
| 795.6058 | 2768.727 | 2830.641 | 637.4613 | 6.9 |
| 795.8878 | 2791.942 | 2836.054 | 637.7509 | 7.1 |
| 796.1698 | 2819.274 | 2841.472 | 638.0404 | 6.98333 |
| 796.4519 | 2858.19 | 2846.894 | 638.3299 | |
| 796.7339 | 2811.879 | 2852.319 | 638.6194 | 6.76667 |
| 797.0158 | 2815.287 | 2857.749 | 638.9089 | 7.08333 |
| 797.2978 | 2777.4 | 2863.183 | 639.1984 | 7.05 |
| 797.5798 | 2799.12 | 2868.621 | 639.4879 | 7.1 |
| 797.8617 | 2870.011 | 2874.063 | 639.7774 | 7.06667 |
| 798.1436 | 2808.292 | 2879.509 | 640.0668 | 7.35 |
| 798.4255 | 2848.863 | 2884.959 | 640.3563 | 7.06667 |
| 798.7074 | 2887.819 | 2890.413 | 640.6457 | 7.4 |
| 798.9893 | 2861.725 | 2895.871 | 640.9351 | 7.01667 |
| 799.2712 | 2873.564 | 2901.333 | 641.2245 | 6.96667 |
| 799.553 | 2885.169 | 2906.799 | 641.5139 | 7.46667 |
| 799.8348 | 2849.825 | 2912.269 | 641.8033 | 7.13333 |
| 800.1167 | 2926.801 | 2917.743 | 642.0927 | 6.86667 |
| 800.3985 | 2905.446 | 2923.221 | 642.382 | 7.43333 |
| 800.6802 | 2880.789 | 2928.703 | 642.6714 | 7.48333 |
| 800.962 | 2906.954 | 2934.189 | 642.9607 | 7.01667 |
| 801.2438 | 2940.787 | 2939.678 | 643.25 | 7.36667 |
| 801.5255 | 2920.973 | 2945.172 | 643.5393 | 7.53333 |
| 801.8072 | 2887.09 | 2950.67 | 643.8286 | 7.43333 |
| 802.0889 | 2908.771 | 2956.171 | 644.1179 | 7.91667 |
| 802.3706 | 2947.117 | 2961.677 | 644.4072 | 7.65 |
| 802.6523 | 2943.752 | 2967.186 | 644.6965 | 7.53333 |
| 802.9339 | 2948.979 | 2972.699 | 644.9857 | 7.75 |
| 803.2156 | 2905.574 | 2978.216 | 645.275 | 7.65 |
| 803.4972 | 2989.458 | 2983.737 | 645.5642 | 7.41667 |
| 803.7788 | 2998.545 | 2989.262 | 645.8534 | 7.43333 |
| 804.0604 | 2982.451 | 2994.791 | 646.1426 | 8.21667 |
| 804.342 | 3027.029 | 3000.323 | 646.4318 | 7.75 |
| 804.6236 | 3004.934 | 3005.859 | 646.721 | 7.73333 |
| 804.9051 | 3040.714 | 3011.4 | 647.0101 | 7.55 |
| 805.1867 | 3019.404 | 3016.944 | 647.2993 | 7.26667 |
| 805.4682 | 3032.247 | 3022.492 | 647.5884 | 7.9 |
| 805.7497 | 2980.772 | 3028.043 | 647.8776 | 7.98333 |
| 806.0312 | 3006.339 | 3033.599 | 648.1667 | 8.01667 |
| 806.3126 | 3007.978 | 3039.158 | 648.4558 | 7.78333 |
| 806.5941 | 2970.152 | 3044.721 | 648.7449 | 8.05 |
| 806.8755 | 3021.271 | 3050.288 | 649.034 | 8.11667 |
| 807.157 | 3002.236 | 3055.858 | 649.3231 | 7.71667 |
| 807.4384 | 3014.675 | 3061.432 | 649.6121 | 7.98333 |
| 807.7198 | 3075.341 | 3067.01 | 649.9012 | 7.86667 |
| 808.0011 | 3045.513 | 3072.592 | 650.1902 | 8.11667 |
| 808.2825 | 3052.692 | 3078.178 | 650.4792 | 8.25 |
| 808.5639 | 3040.886 | 3083.767 | 650.7683 | 8.11667 |
| 808.8452 | 3070.256 | 3089.36 | 651.0573 | 7.95 |
| 809.1265 | 3089.427 | 3094.956 | 651.3463 | 8.08333 |
| 809.4078 | 3108.506 | 3100.557 | 651.6352 | 8.06667 |
| 809.6891 | 3031.815 | 3106.161 | 651.9242 | 8.28333 |
| 809.9704 | 3023.199 | 3111.768 | 652.2132 | 8.68333 |
| 810.2516 | 3059.749 | 3117.38 | 652.5021 | 8.35 |
| 810.5328 | 3150.648 | 3122.995 | 652.791 | 8 |
| 810.8141 | 3094.02 | 3128.613 | 653.0799 | 8.55 |
| 811.0953 | 3055.541 | 3134.236 | 653.3688 | 8.61667 |
| 811.3765 | 3129.008 | 3139.862 | 653.6577 | 8.13333 |
| 811.6576 | 3143.194 | 3145.491 | 653.9466 | 8.66667 |
| 811.9388 | 3099.615 | 3151.124 | 654.2355 | 8.33333 |
| 812.2199 | 3107.32 | 3156.761 | 654.5244 | 8.7 |
| 812.5011 | 3188.862 | 3162.401 | 654.8132 | 8.66667 |
| 812.7822 | 3116.188 | 3168.045 | 655.102 | 8.71667 |
| 813.0633 | 3101.956 | 3173.693 | 655.3909 | 8.53333 |
| 813.3443 | 3122.545 | 3179.344 | 655.6797 | 8.56667 |
| 813.6254 | 3223.97 | 3184.999 | 655.9685 | 8.81667 |
| 813.9065 | 3164.011 | 3190.657 | 656.2572 | 8.98333 |
| 814.1875 | 3144.639 | 3196.319 | 656.546 | 8.63333 |
| 814.4685 | 3219.136 | 3201.984 | 656.8348 | 8.95 |
| 814.7495 | 3212.503 | 3207.653 | 657.1235 | 8.8 |
| 815.0305 | 3146.614 | 3213.326 | 657.4123 | 8.83333 |
| 815.3115 | 3141.367 | 3219.001 | 657.701 | 8.81667 |
| 815.5924 | 3153.27 | 3224.681 | 657.9897 | 9.18333 |
| 815.8733 | 3250.2 | 3230.364 | 658.2784 | 9.13333 |
| 816.1543 | 3171.162 | 3236.05 | 658.5671 | 9.25 |
| 816.4352 | 3207.57 | 3241.74 | 658.8558 | 8.78333 |
| 816.7161 | 3201.752 | 3247.433 | 659.1444 | 9.45 |
| 816.9969 | 3311.156 | 3253.13 | 659.4331 | 9.4 |
| 817.2778 | 3241.279 | 3258.83 | 659.7217 | 9.31667 |
| 817.5586 | 3238.562 | 3264.534 | 660.0103 | 9.23333 |
| 817.8395 | 3245.121 | 3270.241 | 660.299 | 9.16667 |
| 818.1203 | 3220.775 | 3275.952 | 660.5876 | 9.41667 |
| 818.4011 | 3313.394 | 3281.666 | 660.8762 | 9.5 |
| 818.6818 | 3301.797 | 3287.383 | 661.1647 | 9.01667 |
| 818.9626 | 3233.89 | 3293.104 | 661.4533 | 9.75 |
| 819.2434 | 3251.755 | 3298.828 | 661.7419 | 9.58333 |
| 819.5241 | 3273.282 | 3304.556 | 662.0304 | 9.36667 |
| 819.8048 | 3316.952 | 3310.287 | 662.3189 | 9.96667 |
| 820.0855 | 3268.551 | 3316.021 | 662.6074 | 9.78333 |
| 820.3662 | 3358.23 | 3321.759 | 662.8959 | 9.75 |
| 820.6468 | 3272.778 | 3327.5 | 663.1844 | 9.61667 |
| 820.9275 | 3299.183 | 3333.244 | 663.4729 | 9.76667 |
| 821.2081 | 3245.384 | 3338.992 | 663.7614 | 9.61667 |
| 821.4887 | 3275.805 | 3344.743 | 664.0498 | 10.01667 |
| 821.7693 | 3380.181 | 3350.497 | 664.3383 | 9.93333 |
| 822.0499 | 3341.591 | 3356.255 | 664.6267 | 9.91667 |
| 822.3305 | 3357.612 | 3362.016 | 664.9151 | 9.86667 |
| 822.6111 | 3329.322 | 3367.78 | 665.2035 | 10.23333 |
| 822.8916 | 3363.49 | 3373.547 | 665.4919 | 9.93333 |
| 823.1721 | 3322.061 | 3379.318 | 665.7803 | 9.86667 |
| 823.4526 | 3423.645 | 3385.092 | 666.0687 | 10.35 |
| 823.7331 | 3410.62 | 3390.869 | 666.357 | 9.9 |
| 824.0136 | 3354.296 | 3396.65 | 666.6454 | 10.25 |
| 824.2941 | 3370.2 | 3402.433 | 666.9337 | 10.2 |
| 824.5745 | 3380.279 | 3408.22 | 667.222 | 10.33333 |

| | | |
|---|---|---|
| 824.8549 | 3372.45 | 3414.011 |
| 825.1353 | 3368.741 | 3419.804 |
| 825.4157 | 3413.378 | 3425.6 |
| 825.6961 | 3450.109 | 3431.4 |
| 825.9765 | 3424.903 | 3437.203 |
| 826.2568 | 3340.436 | 3443.009 |
| 826.5371 | 3423.929 | 3448.819 |
| 826.8175 | 3415.685 | 3454.631 |
| 827.0978 | 3377.501 | 3460.447 |
| 827.378 | 3464.512 | 3466.265 |
| 827.6583 | 3482.139 | 3472.087 |
| 827.9386 | 3406.335 | 3477.912 |
| 828.2188 | 3482.856 | 3483.74 |
| 828.499 | 3460.074 | 3489.571 |
| 828.7792 | 3410.472 | 3495.406 |
| 829.0594 | 3484.876 | 3501.243 |
| 829.3396 | 3484.88 | 3507.083 |
| 829.6197 | 3467.082 | 3512.927 |
| 829.8998 | 3530.437 | 3518.773 |
| 830.18 | 3587.31 | 3524.623 |
| 830.4601 | 3523.347 | 3530.475 |
| 830.7402 | 3563.477 | 3536.331 |
| 831.0202 | 3524.023 | 3542.19 |
| 831.3003 | 3572.993 | 3548.051 |
| 831.5803 | 3565.739 | 3553.916 |
| 831.8603 | 3590.613 | 3559.784 |
| 832.1404 | 3544.154 | 3565.654 |
| 832.4203 | 3537.483 | 3571.528 |
| 832.7003 | 3556.85 | 3577.405 |
| 832.9803 | 3564.474 | 3583.284 |
| 833.2602 | 3615.659 | 3589.167 |
| 833.5401 | 3550.976 | 3595.052 |
| 833.8201 | 3560.591 | 3600.94 |
| 834.0999 | 3543.038 | 3606.832 |
| 834.3798 | 3653.595 | 3612.726 |
| 834.6597 | 3572.811 | 3618.623 |
| 834.9395 | 3577.891 | 3624.523 |
| 835.2194 | 3651.227 | 3630.426 |
| 835.4992 | 3592.34 | 3636.332 |
| 835.779 | 3694.778 | 3642.24 |
| 836.0587 | 3642.728 | 3648.152 |
| 836.3385 | 3596.088 | 3654.066 |
| 836.6183 | 3612.424 | 3659.983 |
| 836.898 | 3640.818 | 3665.903 |
| 837.1777 | 3603.124 | 3671.826 |
| 837.4574 | 3663.584 | 3677.752 |
| 837.7371 | 3701.239 | 3683.68 |
| 838.0167 | 3723.187 | 3689.611 |
| 838.2964 | 3673.171 | 3695.545 |
| 838.576 | 3644.756 | 3701.482 |
| 838.8556 | 3698.584 | 3707.422 |
| 839.1352 | 3777.288 | 3713.364 |
| 839.4148 | 3695.992 | 3719.309 |
| 839.6944 | 3707.282 | 3725.257 |
| 839.9739 | 3747.993 | 3731.208 |
| 840.2535 | 3738.513 | 3737.161 |
| 840.533 | 3694.292 | 3743.117 |
| 840.8125 | 3766.741 | 3749.076 |
| 841.092 | 3758.812 | 3755.037 |
| 841.3715 | 3737.576 | 3761.001 |
| 841.6509 | 3741.678 | 3766.968 |
| 841.9303 | 3719.646 | 3772.937 |
| 842.2098 | 3792.723 | 3778.909 |
| 842.4892 | 3728.671 | 3784.884 |
| 842.7685 | 3855.541 | 3790.861 |
| 843.0479 | 3786.015 | 3796.842 |
| 843.3273 | 3808.802 | 3802.824 |
| 843.6066 | 3795.472 | 3808.809 |
| 843.8859 | 3889.491 | 3814.797 |
| 844.1652 | 3827.476 | 3820.788 |
| 844.4445 | 3833.841 | 3826.781 |
| 844.7238 | 3796.059 | 3832.776 |
| 845.003 | 3819.373 | 3838.775 |
| 845.2823 | 3879.764 | 3844.775 |
| 845.5615 | 3842.542 | 3850.779 |
| 845.8407 | 3854.418 | 3856.784 |
| 846.1199 | 3870.623 | 3862.793 |
| 846.3991 | 3837.927 | 3868.804 |
| 846.6782 | 3890.369 | 3874.817 |
| 846.9574 | 3908.378 | 3880.833 |
| 847.2365 | 3929.955 | 3886.852 |
| 847.5156 | 3956.059 | 3892.872 |
| 847.7947 | 3865.914 | 3898.896 |
| 848.0738 | 3898.423 | 3904.922 |
| 848.3528 | 3918.26 | 3910.95 |
| 848.6319 | 3844.664 | 3916.981 |
| 848.9109 | 3924.872 | 3923.014 |
| 849.1899 | 3880.53 | 3929.05 |
| 849.4689 | 3934.416 | 3935.088 |
| 849.7478 | 3938.803 | 3941.128 |
| 850.0268 | 3993.665 | 3947.171 |
| 850.3057 | 3922.525 | 3953.216 |
| 850.5847 | 3847.088 | 3959.264 |
| 850.8636 | 3949.695 | 3965.314 |
| 851.1425 | 3897.39 | 3971.367 |
| 851.4213 | 3964.824 | 3977.421 |
| 851.7002 | 3979.752 | 3983.478 |
| 851.979 | 3932.746 | 3989.538 |
| 852.2579 | 3944.576 | 3995.6 |
| 852.5367 | 4016.577 | 4001.664 |
| 852.8155 | 4100.341 | 4007.73 |
| 853.0942 | 3992.06 | 4013.799 |
| 853.373 | 4015.598 | 4019.87 |
| 853.6517 | 4001.254 | 4025.944 |
| 853.9304 | 4027.399 | 4032.019 |
| 854.2092 | 4026.239 | 4038.097 |
| 854.4878 | 4024.256 | 4044.177 |
| 854.7665 | 3980.349 | 4050.26 |
| 855.0452 | 4137.246 | 4056.344 |
| 855.3238 | 4075.094 | 4062.431 |
| 855.6024 | 4075.806 | 4068.52 |
| 855.881 | 4072.98 | 4074.612 |
| 856.1596 | 4047.514 | 4080.705 |
| 856.4382 | 4078.236 | 4086.801 |
| 856.7168 | 4076.346 | 4092.899 |
| 856.9953 | 4076.962 | 4098.999 |
| 857.2738 | 4062.993 | 4105.101 |
| 857.5523 | 4044.682 | 4111.206 |
| 857.8308 | 4067.653 | 4117.312 |
| 858.1093 | 4108.104 | 4123.421 |
| 858.3878 | 4113.646 | 4129.532 |
| 858.6662 | 4107.651 | 4135.645 |
| 858.9446 | 4218.083 | 4141.76 |
| 859.223 | 4117.094 | 4147.877 |
| 859.5014 | 4081.877 | 4153.997 |
| 859.7798 | 4199.573 | 4160.118 |
| 860.0581 | 4203.25 | 4166.241 |
| 860.3365 | 4178.127 | 4172.367 |
| 860.6148 | 4144.932 | 4178.495 |
| 860.8931 | 4146.028 | 4184.624 |
| 861.1714 | 4204.816 | 4190.756 |
| 861.4497 | 4195.179 | 4196.89 |
| 861.7279 | 4223.989 | 4203.026 |
| 862.0061 | 4193.54 | 4209.163 |
| 862.2844 | 4204.341 | 4215.303 |
| 862.5626 | 4248.78 | 4221.445 |
| 862.8408 | 4124.678 | 4227.589 |
| 863.1189 | 4301.685 | 4233.734 |
| 863.3971 | 4297.427 | 4239.882 |
| 863.6752 | 4150.952 | 4246.032 |
| 863.9533 | 4213.502 | 4252.184 |
| 864.2314 | 4217.915 | 4258.337 |
| 864.5095 | 4211.506 | 4264.493 |
| 864.7876 | 4205.314 | 4270.65 |
| 865.0656 | 4210.815 | 4276.809 |
| 865.3437 | 4245.314 | 4282.971 |
| 865.6217 | 4238.775 | 4289.134 |
| 865.8997 | 4344.402 | 4295.299 |
| 866.1777 | 4260.549 | 4301.466 |
| 866.4556 | 4275.698 | 4307.635 |
| 866.7336 | 4216.524 | 4313.805 |
| 867.0115 | 4264.766 | 4319.978 |
| 867.2894 | 4338.614 | 4326.152 |
| 867.5673 | 4353.862 | 4332.329 |
| 867.8452 | 4271.135 | 4338.507 |
| 868.1231 | 4287.732 | 4344.686 |
| 868.4009 | 4308.057 | 4350.868 |
| 868.6788 | 4194.628 | 4357.052 |

| | |
|---|---|
| 667.5104 | 10.26667 |
| 667.7986 | 10.25 |
| 668.0869 | 10.61667 |
| 668.3752 | 10.26667 |
| 668.6635 | 10.43333 |
| 668.9517 | 10.63333 |
| 669.2399 | 10.86667 |
| 669.5282 | 10.91667 |
| 669.8164 | 10.93333 |
| 670.1046 | 10.85 |
| 670.3927 | 11.18333 |
| 670.6809 | 11.06667 |
| 670.9691 | 10.91667 |
| 671.2572 | 11.16667 |
| 671.5454 | 11.05 |
| 671.8335 | 11.15 |
| 672.1216 | 11.26667 |
| 672.4097 | 11.48333 |
| 672.6978 | 11.03333 |
| 672.9858 | 11.3 |
| 673.2739 | 11.3 |
| 673.5619 | 11.23333 |
| 673.85 | 11.43333 |
| 674.138 | 11.28333 |
| 674.426 | 11.23333 |
| 674.714 | 11.66667 |
| 675.002 | 11.65 |
| 675.29 | 11.71667 |
| 675.5779 | 11.76667 |
| 675.8659 | |
| 676.1538 | 11.5 |
| 676.4417 | 11.78333 |
| 676.7296 | 11.88333 |
| 677.0175 | 11.95 |
| 677.3054 | 12.03333 |
| 677.5933 | 11.75 |
| 677.8811 | 11.78333 |
| 678.169 | 12.06667 |
| 678.4568 | 11.85 |
| 678.7446 | 12.43333 |
| 679.0324 | 12.15 |
| 679.3202 | 12.26667 |
| 679.608 | 12.46667 |
| 679.8958 | 12.16667 |
| 680.1835 | 12.31667 |
| 680.4713 | 12.2 |
| 680.759 | 12.7 |
| 681.0467 | 12.66667 |
| 681.3344 | 12.91667 |
| 681.6221 | 12.93333 |
| 681.9098 | 13.08333 |
| 682.1975 | 12.41667 |
| 682.4851 | 13.08333 |
| 682.7728 | 12.55 |
| 683.0604 | 12.85 |
| 683.348 | 12.98333 |
| 683.6356 | 13.01667 |
| 683.9232 | 13.1 |
| 684.2108 | 12.95 |
| 684.4983 | 12.88333 |
| 684.7859 | 13.45 |
| 685.0734 | 13.53333 |
| 685.361 | 13.16667 |
| 685.6485 | 13.25 |
| 685.936 | 13.18333 |
| 686.2235 | 13.18333 |
| 686.5109 | 13.2 |
| 686.7984 | 13.08333 |
| 687.0859 | 13.26667 |
| 687.3733 | 13.55 |
| 687.6607 | 13.63333 |
| 687.9481 | 13.6 |
| 688.2355 | 13.88333 |
| 688.5229 | 13.58333 |
| 688.8103 | 13.75 |
| 689.0976 | 13.9 |
| 689.385 | 13.93333 |
| 689.6723 | 13.65 |
| 689.9596 | 14.13333 |
| 690.2469 | 13.93333 |
| 690.5342 | 13.86667 |
| 690.8215 | 14.13333 |
| 691.1088 | 13.85 |
| 691.396 | 14.2 |
| 691.6833 | 14.38333 |
| 691.9705 | 14.11667 |
| 692.2577 | 14.2 |
| 692.5449 | 14.75 |
| 692.8321 | 14.2 |
| 693.1193 | 14.13333 |
| 693.4065 | 14.26667 |
| 693.6936 | 14.31667 |
| 693.9808 | 14.18333 |
| 694.2679 | 14.75 |
| 694.555 | 14.66667 |
| 694.8421 | 14.86667 |
| 695.1292 | 14.86667 |
| 695.4163 | 14.45 |
| 695.7033 | 14.68333 |
| 695.9904 | 14.6 |
| 696.2774 | 14.9 |
| 696.5644 | 14.85 |
| 696.8514 | 15.36667 |
| 697.1384 | 14.58333 |
| 697.4254 | 15.1 |
| 697.7124 | 15.28333 |
| 697.9993 | 15.45 |
| 698.2863 | 15.45 |
| 698.5732 | 15.76667 |
| 698.8601 | 15.25 |
| 699.147 | 15.73333 |
| 699.4339 | 15.68333 |
| 699.7208 | 15.88333 |
| 700.0077 | 15.76667 |
| 700.2945 | 15.53333 |
| 700.5814 | 15.71667 |
| 700.8682 | 16.1 |
| 701.155 | 16.03333 |
| 701.4418 | 16.15 |
| 701.7286 | 16 |
| 702.0153 | 16.18333 |
| 702.3021 | 15.85 |
| 702.5889 | 15.68333 |
| 702.8756 | 16.33333 |
| 703.1623 | 16.2 |
| 703.449 | 16.15 |
| 703.7357 | 16.38333 |
| 704.0224 | 16.5 |
| 704.309 | 16.35 |
| 704.5957 | 16.33333 |
| 704.8823 | 16.5 |
| 705.169 | 16.1 |
| 705.4556 | 15.95 |
| 705.7422 | 16.53333 |
| 706.0288 | 16.86667 |
| 706.3153 | 16.93333 |
| 706.6019 | 16.53333 |
| 706.8884 | 17.01667 |
| 707.175 | 16.98333 |
| 707.4615 | 16.9 |
| 707.748 | 17.31667 |
| 708.0345 | 17.01667 |
| 708.321 | 16.71667 |
| 708.6074 | 17.43333 |
| 708.8939 | 17.36667 |
| 709.1803 | 17.11667 |
| 709.4668 | 17.25 |
| 709.7532 | 17.41667 |
| 710.0396 | 16.88333 |
| 710.326 | 17.33333 |
| 710.6123 | 17.63333 |
| 710.8987 | 17.31667 |
| 711.185 | 17.85 |
| 711.4714 | 17.16667 |
| 711.7577 | 17.53333 |
| 712.044 | 17.46667 |
| 712.3303 | 17.85 |
| 712.6166 | 17.46667 |

| | | | | |
|---|---|---|---|---|
| 868.9566 | 4302.455 | 4363.237 | 712.9028 | 17.53333 |
| 869.2344 | 4412.387 | 4369.424 | 713.1891 | 17.4 |
| 869.5122 | 4312.773 | 4375.613 | 713.4753 | 18.05 |
| 869.7899 | 4327.555 | 4381.803 | 713.7615 | 17.75 |
| 870.0677 | 4416.973 | 4387.996 | 714.0477 | 17.4 |
| 870.3454 | 4341.973 | 4394.19 | 714.3339 | 17.86667 |
| 870.6231 | 4369.919 | 4400.385 | 714.6201 | 17.93333 |
| 870.9008 | 4313.478 | 4406.583 | 714.9063 | 18.38333 |
| 871.1785 | 4415.84 | 4412.782 | 715.1925 | 17.8 |
| 871.4561 | 4350.518 | 4418.983 | 715.4786 | 18.13333 |
| 871.7338 | 4404.032 | 4425.186 | 715.7647 | 18.21667 |
| 872.0114 | 4355.931 | 4431.39 | 716.0508 | 17.86667 |
| 872.289 | 4352.208 | 4437.596 | 716.3369 | 17.98333 |
| 872.5666 | 4401.559 | 4443.804 | 716.623 | 18.26667 |
| 872.8442 | 4422.501 | 4450.013 | 716.9091 | 18.31667 |
| 873.1217 | 4480.842 | 4456.224 | 717.1952 | 18.61667 |
| 873.3993 | 4479.007 | 4462.437 | 717.4812 | 18.15 |
| 873.6768 | 4448.87 | 4468.651 | 717.7673 | 18.63333 |
| 873.9543 | 4468.924 | 4474.867 | 718.0533 | 18.81667 |
| 874.2318 | 4450.753 | 4481.085 | 718.3393 | 18.46667 |
| 874.5093 | 4442.079 | 4487.304 | 718.6253 | 18.61667 |
| 874.7867 | 4431.484 | 4493.525 | 718.9113 | 18.86667 |
| 875.0642 | 4442.218 | 4499.747 | 719.1972 | 18.31667 |
| 875.3416 | 4469.498 | 4505.971 | 719.4832 | 19.11667 |
| 875.619 | 4494.415 | 4512.197 | 719.7691 | 18.61667 |
| 875.8964 | 4449.635 | 4518.424 | 720.055 | 18.7 |
| 876.1737 | 4526.26 | 4524.652 | 720.341 | 18.55 |
| 876.4511 | 4463.373 | 4530.883 | 720.6268 | 19.53333 |
| 876.7284 | 4509.013 | 4537.114 | 720.9127 | 18.98333 |
| 877.0057 | 4602.772 | 4543.348 | 721.1986 | 18.8 |
| 877.283 | 4516.404 | 4549.582 | 721.4845 | 19.05 |
| 877.5603 | 4532.207 | 4555.819 | 721.7703 | 19.03333 |
| 877.8376 | 4545.372 | 4562.057 | 722.0561 | 18.96667 |
| 878.1148 | 4623.335 | 4568.296 | 722.342 | 18.98333 |
| 878.3921 | 4498.15 | 4574.537 | 722.6278 | 19.53333 |
| 878.6693 | 4613.085 | 4580.779 | 722.9135 | 19.15 |
| 878.9465 | 4530.06 | 4587.023 | 723.1993 | 19.65 |
| 879.2237 | 4571.22 | 4593.268 | 723.4851 | 19.41667 |
| 879.5008 | 4579.83 | 4599.514 | 723.7708 | 19.78333 |
| 879.778 | 4564.014 | 4605.762 | 724.0566 | 19.68333 |
| 880.0551 | 4614.109 | 4612.012 | 724.3423 | 19.08333 |
| 880.3322 | 4593.454 | 4618.263 | 724.628 | 20.08333 |
| 880.6093 | 4620.932 | 4624.515 | 724.9137 | 20.11667 |
| 880.8864 | 4565.206 | 4630.769 | 725.1994 | 19.65 |
| 881.1634 | 4597.137 | 4637.024 | 725.485 | 19.61667 |
| 881.4405 | 4570.512 | 4643.281 | 725.7707 | 19.86667 |
| 881.7175 | 4639.935 | 4649.539 | 726.0563 | 19.53333 |
| 881.9945 | 4684.993 | 4655.798 | 726.3419 | 19.9 |
| 882.2715 | 4592.976 | 4662.058 | 726.6276 | 19.83333 |
| 882.5485 | 4693.529 | 4668.32 | 726.9131 | 20.11667 |
| 882.8254 | 4637.22 | 4674.584 | 727.1987 | 20.2 |
| 883.1024 | 4646.63 | 4680.848 | 727.4843 | 20.61667 |
| 883.3793 | 4622.318 | 4687.114 | 727.7699 | 20.48333 |
| 883.6562 | 4615.809 | 4693.382 | 728.0554 | 19.7 |
| 883.9331 | 4644.038 | 4699.65 | 728.3409 | 20.4 |
| 884.2099 | 4720.161 | 4705.92 | 728.6264 | 20.15 |
| 884.4868 | 4643.672 | 4712.191 | 728.9119 | 20.48333 |
| 884.7636 | 4727.33 | 4718.464 | 729.1974 | 20.9 |
| 885.0404 | 4670.218 | 4724.738 | 729.4829 | 20.3 |
| 885.3172 | 4770.016 | 4731.013 | 729.7683 | 20.86667 |
| 885.594 | 4738.974 | 4737.289 | 730.0538 | 20.61667 |
| 885.8708 | 4736.835 | 4743.566 | 730.3392 | 21.23333 |
| 886.1475 | 4697.996 | 4749.845 | 730.6246 | 20.71667 |
| 886.4243 | 4766.219 | 4756.125 | 730.91 | 21.48333 |
| 886.701 | 4746.909 | 4762.406 | 731.1954 | 20.95 |
| 886.9777 | 4715.565 | 4768.689 | 731.4808 | 20.83333 |
| 887.2544 | 4779.683 | 4774.972 | 731.7662 | 20.7 |
| 887.531 | 4841.264 | 4781.257 | 732.0515 | 21.46667 |
| 887.8077 | 4829.398 | 4787.543 | 732.3368 | 21.18333 |
| 888.0843 | 4755.254 | 4793.83 | 732.6222 | 20.78333 |
| 888.3609 | 4794.041 | 4800.118 | 732.9075 | 21.5 |
| 888.6375 | 4765.924 | 4806.408 | 733.1928 | 21.5 |
| 888.9141 | 4838.878 | 4812.699 | 733.478 | 21.15 |
| 889.1906 | 4823.362 | 4818.99 | 733.7633 | 21.96667 |
| 889.4671 | 4768.85 | 4825.283 | 734.0486 | 21.7 |
| 889.7437 | 4795.576 | 4831.577 | 734.3338 | 21.56667 |
| 890.0202 | 4849.854 | 4837.872 | 734.619 | 21.51667 |
| 890.2967 | 4865.99 | 4844.169 | 734.9042 | 21.48333 |
| 890.5731 | 4792.736 | 4850.466 | 735.1894 | 21.13333 |
| 890.8496 | 4853.848 | 4856.764 | 735.4746 | 21.9 |
| 891.126 | 4830.734 | 4863.064 | 735.7598 | 21.76667 |
| 891.4024 | 4904.748 | 4869.364 | 736.0449 | 21.81667 |
| 891.6788 | 4830.556 | 4875.666 | 736.33 | 22.46667 |
| 891.9552 | 4807.574 | 4881.969 | 736.6152 | 21.81667 |
| 892.2316 | 4805.326 | 4888.272 | 736.9003 | 22.23333 |
| 892.5079 | 4889.147 | 4894.577 | 737.1854 | 22.16667 |
| 892.7842 | 4956 | 4900.883 | 737.4705 | 22.73333 |
| 893.0605 | 4918.06 | 4907.19 | 737.7555 | 21.93333 |
| 893.3368 | 4879.842 | 4913.497 | 738.0406 | 22.3 |
| 893.6131 | 4922.908 | 4919.806 | 738.3256 | 22.45 |
| 893.8894 | 4879.761 | 4926.116 | 738.6106 | 22.53333 |
| 894.1656 | 4927.97 | 4932.427 | 738.8956 | 22.88333 |
| 894.4418 | 4858.465 | 4938.739 | 739.1806 | 22.18333 |
| 894.718 | 4933.056 | 4945.051 | 739.4656 | 22.4 |
| 894.9942 | 4979.722 | 4951.365 | 739.7506 | 22.88333 |
| 895.2704 | 5017.322 | 4957.679 | 740.0355 | 22.8 |
| 895.5466 | 4978.802 | 4963.995 | 740.3205 | 23.05 |
| 895.8227 | 5001.701 | 4970.311 | 740.6054 | 22.56667 |
| 896.0988 | 4982.342 | 4976.629 | 740.8903 | 23.16667 |
| 896.3749 | 5015.364 | 4982.947 | 741.1752 | 23.1 |
| 896.651 | 4985.978 | 4989.266 | 741.4601 | 22.91667 |
| 896.9271 | 5055.301 | 4995.587 | 741.745 | 23 |
| 897.2031 | 4926.183 | 5001.907 | 742.0298 | 22.95 |
| 897.4791 | 5041.081 | 5008.229 | 742.3147 | 22.98333 |
| 897.7551 | 5037.791 | 5014.552 | 742.5995 | 23.7 |
| 898.0311 | 5068.519 | 5020.876 | 742.8843 | 23.23333 |
| 898.3071 | 5032.413 | 5027.2 | 743.1691 | 23.23333 |
| 898.5831 | 5026.547 | 5033.525 | 743.4539 | 23.78333 |
| 898.859 | 4976.93 | 5039.852 | 743.7387 | 23.68333 |
| 899.1349 | 5033.36 | 5046.179 | 744.0234 | 23.26667 |
| 899.4108 | 5036.139 | 5052.506 | 744.3082 | 23.46667 |
| 899.6867 | 5082.359 | 5058.835 | 744.5929 | 23.65 |
| 899.9626 | 5057.554 | 5065.164 | 744.8776 | 24.16667 |
| 900.2385 | 5059.762 | 5071.494 | 745.1623 | 23.85 |
| 900.5143 | 5117.617 | 5077.825 | 745.447 | 24.3 |
| 900.7901 | 5189.423 | 5084.157 | 745.7317 | 23.73333 |
| 901.0659 | 5156.772 | 5090.49 | 746.0163 | 24.2 |
| 901.3417 | 5077.929 | 5096.823 | 746.301 | 24.46667 |
| 901.6175 | 5009.161 | 5103.157 | 746.5856 | 24.5 |
| 901.8932 | 5067.483 | 5109.492 | 746.8702 | 24.01667 |
| 902.1689 | 5118.294 | 5115.827 | 747.1548 | 24.33333 |
| 902.4446 | 5079.548 | 5122.163 | 747.4394 | 24.65 |
| 902.7203 | 5190.028 | 5128.5 | 747.724 | 24.5 |
| 902.996 | 5119.386 | 5134.838 | 748.0085 | 24.31667 |
| 903.2717 | 5200.526 | 5141.176 | 748.2931 | 24.56667 |
| 903.5473 | 5253.948 | 5147.515 | 748.5776 | 24.48333 |
| 903.8229 | 5089.422 | 5153.855 | 748.8621 | 24.48333 |
| 904.0985 | 5187.818 | 5160.195 | 749.1466 | 25.03333 |
| 904.3741 | 5201.734 | 5166.536 | 749.4311 | 24.58333 |
| 904.6497 | 5154.57 | 5172.878 | 749.7155 | 24.83333 |
| 904.9252 | 5204.397 | 5179.22 | 750 | 24.88333 |
| 905.2008 | 5172.458 | 5185.563 | 750.2844 | 25.16667 |
| 905.4763 | 5142.909 | 5191.907 | 750.5689 | 25.26667 |
| 905.7518 | 5202.411 | 5198.251 | 750.8533 | 25.03333 |
| 906.0273 | 5184.323 | 5204.596 | 751.1377 | 25.01667 |
| 906.3027 | 5217.54 | 5210.941 | 751.4221 | 25.21667 |
| 906.5782 | 5179.381 | 5217.287 | 751.7064 | 24.95 |
| 906.8536 | 5221.455 | 5223.634 | 751.9908 | 25.43333 |
| 907.129 | 5300.47 | 5229.981 | 752.2751 | 24.81667 |
| 907.4044 | 5204.377 | 5236.328 | 752.5595 | 25.75 |
| 907.6798 | 5319.676 | 5242.677 | 752.8438 | 25.71667 |
| 907.9551 | 5220.677 | 5249.026 | 753.1281 | 25.78333 |
| 908.2305 | 5299.106 | 5255.375 | 753.4124 | 26.35 |
| 908.5058 | 5180.154 | 5261.725 | 753.6966 | 25.41667 |
| 908.7811 | 5250.692 | 5268.075 | 753.9809 | 25.68333 |
| 909.0564 | 5276.426 | 5274.426 | 754.2651 | 26 |
| 909.3316 | 5246.778 | 5280.778 | 754.5493 | 25.68333 |
| 909.6069 | 5211.28 | 5287.13 | 754.8336 | 25.88333 |
| 909.8821 | 5224.805 | 5293.482 | 755.1178 | 26.11667 |
| 910.1573 | 5329.484 | 5299.835 | 755.4019 | 25.66667 |
| 910.4325 | 5246.185 | 5306.189 | 755.6861 | 25.7 |
| 910.7077 | 5296.122 | 5312.543 | 755.9703 | 26.25 |
| 910.9829 | 5306.473 | 5318.897 | 756.2544 | 25.85 |
| 911.258 | 5355.805 | 5325.252 | 756.5385 | 25.65 |
| 911.5331 | 5307.04 | 5331.607 | 756.8226 | 26.03333 |
| 911.8082 | 5297.555 | 5337.963 | 757.1067 | 26.63333 |
| 912.0833 | 5441.33 | 5344.319 | 757.3908 | 26.33333 |
| 912.3584 | 5352.966 | 5350.676 | 757.6749 | 26.15 |
| 912.6334 | 5364.726 | 5357.033 | 757.9589 | 26.51667 |
| 912.9085 | 5467.96 | 5363.39 | 758.243 | 26.13333 |
| 913.1835 | 5353.673 | 5369.748 | 758.527 | 26.08333 |
| 913.4585 | 5407.072 | 5376.106 | 758.811 | 25.9 |
| 913.7335 | 5412.323 | 5382.464 | 759.095 | 25.95 |
| 914.0084 | 5422.554 | 5388.823 | 759.379 | 26.73333 |
| 914.2834 | 5482.329 | 5395.182 | 759.6629 | 26.75 |
| 914.5583 | 5422.249 | 5401.542 | 759.9469 | 27.1 |
| 914.8332 | 5490.421 | 5407.902 | 760.2308 | 26.86667 |
| 915.1081 | 5423.355 | 5414.262 | 760.5147 | 26.93333 |
| 915.383 | 5526.499 | 5420.623 | 760.7986 | 27.16667 |
| 915.6578 | 5474.455 | 5426.984 | 761.0825 | 26.86667 |
| 915.9327 | 5449.871 | 5433.345 | 761.3664 | 26.21667 |
| 916.2075 | 5478.974 | 5439.707 | 761.6503 | 27.5 |
| 916.4823 | 5509.692 | 5446.068 | 761.9341 | 26.96667 |
| 916.7571 | 5448.528 | 5452.43 | 762.218 | 27.43333 |
| 917.0318 | 5494.893 | 5458.793 | 762.5018 | 27.16667 |
| 917.3066 | 5540.37 | 5465.156 | 762.7856 | 27.5 |
| 917.5813 | 5423.809 | 5471.518 | 763.0694 | 27.16667 |
| 917.856 | 5424.794 | 5477.882 | 763.3531 | 27.2 |
| 918.1307 | 5602.143 | 5484.245 | 763.6369 | 27.78333 |
| 918.4054 | 5474.107 | 5490.609 | 763.9207 | 27.85 |
| 918.68 | 5573.971 | 5496.973 | 764.2044 | 27.51667 |
| 918.9547 | 5446.023 | 5503.337 | 764.4881 | 27.65 |
| 919.2293 | 5404.261 | 5509.701 | 764.7718 | 28.18333 |
| 919.5039 | 5517.969 | 5516.066 | 765.0555 | 28.23333 |
| 919.7785 | 5432.322 | 5522.43 | 765.3392 | 27.46667 |
| 920.053 | 5527.583 | 5528.795 | 765.6228 | 28.51667 |
| 920.3276 | 5574.995 | 5535.16 | 765.9065 | 27.53333 |
| 920.6021 | 5628.925 | 5541.526 | 766.1901 | 27.93333 |
| 920.8766 | 5549.382 | 5547.891 | 766.4737 | 28.05 |
| 921.1511 | 5503.339 | 5554.257 | 766.7573 | 28.31667 |
| 921.4256 | 5721.461 | 5560.622 | 767.0409 | 28.55 |
| 921.7 | 5533.111 | 5566.988 | 767.3245 | 28.5 |
| 921.9745 | 5649.125 | 5573.354 | 767.608 | 27.81667 |
| 922.2489 | 5655.381 | 5579.72 | 767.8915 | 27.68333 |
| 922.5233 | 5568.087 | 5586.086 | 768.1751 | 28.35 |
| 922.7977 | 5684.318 | 5592.453 | 768.4586 | 28.13333 |
| 923.072 | 5649.862 | 5598.819 | 768.7421 | 28.11667 |
| 923.3464 | 5603.409 | 5605.186 | 769.0256 | 28.23333 |
| 923.6207 | 5593.588 | 5611.552 | 769.309 | 28.33333 |
| 923.895 | 5646.057 | 5617.919 | 769.5925 | 27.76667 |
| 924.1693 | 5540.69 | 5624.286 | 769.8759 | 28.4 |
| 924.4436 | 5667.011 | 5630.653 | 770.1593 | 29.2 |
| 924.7179 | 5704.708 | 5637.019 | 770.4427 | 28.3 |
| 924.9921 | 5566.527 | 5643.386 | 770.7261 | 29.31667 |
| 925.2663 | 5668.407 | 5649.753 | 771.0095 | 28.66667 |
| 925.5405 | 5721.971 | 5656.12 | 771.2929 | 28.6 |
| 925.8147 | 5601.102 | 5662.487 | 771.5762 | 28.35 |
| 926.0889 | 5653.285 | 5668.854 | 771.8595 | 29.11667 |
| 926.363 | 5566.027 | 5675.221 | 772.1429 | 28.75 |
| 926.6371 | 5674.181 | 5681.588 | 772.4262 | 29.65 |
| 926.9112 | 5589.778 | 5687.955 | 772.7095 | 29.35 |
| 927.1853 | 5845.699 | 5694.322 | 772.9927 | 29.21667 |
| 927.4594 | 5742.412 | 5700.689 | 773.276 | 29.03333 |
| 927.7335 | 5803.142 | 5707.056 | 773.5592 | 29.48333 |
| 928.0075 | 5715.068 | 5713.423 | 773.8425 | 29.1 |
| 928.2815 | 5758.189 | 5719.789 | 774.1257 | 29.66667 |
| 928.5555 | 5745.958 | 5726.156 | 774.4089 | 29.7 |
| 928.8295 | 5724.396 | 5732.523 | 774.692 | 29.35 |
| 929.1035 | 5771.845 | 5738.889 | 774.9752 | 29.33333 |
| 929.3774 | 5708.768 | 5745.256 | 775.2584 | 28.93333 |
| 929.6514 | 5817.986 | 5751.622 | 775.5415 | 29.16667 |
| 929.9253 | 5769.354 | 5757.988 | 775.8246 | 29.86667 |
| 930.1992 | 5803.278 | 5764.354 | 776.1077 | 29.25 |
| 930.473 | 5700.966 | 5770.721 | 776.3908 | 29.61667 |
| 930.7469 | 5794.957 | 5777.086 | 776.6739 | 29.86667 |
| 931.0207 | 5877.089 | 5783.452 | 776.957 | 29.35 |
| 931.2945 | 5826.007 | 5789.818 | 777.24 | 29.85 |
| 931.5683 | 5949.077 | 5796.183 | 777.5231 | 29.76667 |
| 931.8421 | 5804.574 | 5802.549 | 777.8061 | 30.21667 |
| 932.1159 | 5854.047 | 5808.914 | 778.0891 | 30.1 |
| 932.3896 | 5822.497 | 5815.279 | 778.3721 | 30.11667 |
| 932.6634 | 5904.201 | 5821.644 | 778.655 | 29.98333 |
| 932.9371 | 5855.715 | 5828.008 | 778.938 | 29.91667 |
| 933.2107 | 5922.125 | 5834.373 | 779.2209 | 30.36667 |
| 933.4844 | 5862.475 | 5840.737 | 779.5039 | 29.81667 |
| 933.7581 | 5979.101 | 5847.101 | 779.7868 | 30.61667 |
| 934.0317 | 5916.723 | 5853.465 | 780.0697 | 30.43333 |
| 934.3053 | 6075.325 | 5859.829 | 780.3526 | 31.41667 |
| 934.5789 | 5828.431 | 5866.192 | 780.6354 | 30.55 |
| 934.8525 | 6058.838 | 5872.555 | 780.9183 | 30.76667 |
| 935.1261 | 5892.512 | 5878.918 | 781.2011 | 31 |
| 935.3996 | 5838.773 | 5885.28 | 781.484 | 30.23333 |
| 935.6731 | 5979.383 | 5891.643 | 781.7668 | 30.95 |
| 935.9466 | 5797.917 | 5898.005 | 782.0496 | 31.08333 |
| 936.2201 | 5831.084 | 5904.367 | 782.3323 | 31.03333 |
| 936.4936 | 5972.491 | 5910.728 | 782.6151 | 30.78333 |
| 936.767 | 5879.459 | 5917.089 | 782.8978 | 31.23333 |
| 937.0405 | 5971.689 | 5923.45 | 783.1806 | 31.08333 |
| 937.3139 | 5856.278 | 5929.811 | 783.4633 | 31.26667 |
| 937.5873 | 6055.577 | 5936.171 | 783.746 | 30.8 |
| 937.8607 | 5944.542 | 5942.531 | 784.0287 | 31.46667 |
| 938.134 | 6032.963 | 5948.89 | 784.3114 | 30.73333 |
| 938.4074 | 6003.396 | 5955.25 | 784.594 | 31.85 |
| 938.6807 | 5991.158 | 5961.609 | 784.8767 | 31.55 |
| 938.954 | 6090.16 | 5967.967 | 785.1593 | 31.33333 |
| 939.2273 | 6094.872 | 5974.325 | 785.4419 | 31 |
| 939.5005 | 6009.912 | 5980.683 | 785.7245 | 31.65 |
| 939.7738 | 6089.965 | 5987.041 | 786.0071 | 32.25 |
| 940.047 | 6018.889 | 5993.398 | 786.2896 | 31.33333 |
| 940.3202 | 6101.291 | 5999.754 | 786.5722 | 31.55 |
| 940.5934 | 6027.883 | 6006.11 | 786.8547 | 32.05 |
| 940.8666 | 6147.034 | 6012.466 | 787.1372 | 32.08333 |
| 941.1397 | 6077.914 | 6018.822 | 787.4198 | 32.6 |
| 941.4129 | 6134.011 | 6025.177 | 787.7022 | 31.88333 |
| 941.686 | 6129.722 | 6031.531 | 787.9847 | 31.38333 |
| 941.9591 | 6094.327 | 6037.885 | 788.2672 | 31.96667 |
| 942.2322 | 6045.807 | 6044.239 | 788.5496 | 31.66667 |
| 942.5052 | 6018.412 | 6050.592 | 788.8321 | 32.06667 |
| 942.7783 | 6012.632 | 6056.945 | 789.1145 | 32.48333 |
| 943.0513 | 6100.613 | 6063.297 | 789.3969 | 31.93333 |
| 943.3243 | 6054.701 | 6069.649 | 789.6792 | 32.13333 |
| 943.5973 | 6064.388 | 6076 | 789.9616 | 33.38333 |
| 943.8703 | 6209.273 | 6082.351 | 790.244 | 32.95 |
| 944.1432 | 6021.795 | 6088.701 | 790.5263 | 32.05 |
| 944.4162 | 5955.095 | 6095.051 | 790.8086 | 32.6 |
| 944.6891 | 6015.205 | 6101.4 | 791.0909 | 32.4 |
| 944.962 | 6197.225 | 6107.749 | 791.3732 | 32.76667 |
| 945.2348 | 6097.511 | 6114.097 | 791.6555 | 31.76667 |
| 945.5077 | 6217.821 | 6120.445 | 791.9378 | 33.1 |
| 945.7805 | 6087.471 | 6126.792 | 792.22 | 32.91667 |
| 946.0534 | 6112.676 | 6133.139 | 792.5022 | 32.4 |
| 946.3262 | 6128.889 | 6139.485 | 792.7845 | 32.55 |
| 946.5989 | 6192.569 | 6145.83 | 793.0667 | 32.76667 |
| 946.8717 | 6149.992 | 6152.175 | 793.3488 | 33.18333 |
| 947.1444 | 6093.586 | 6158.519 | 793.631 | 32.56667 |
| 947.4172 | 6201.611 | 6164.863 | 793.9132 | 32.78333 |
| 947.6899 | 6274.211 | 6171.206 | 794.1953 | 32.66667 |
| 947.9626 | 6308.534 | 6177.549 | 794.4774 | 32.95 |
| 948.2352 | 6228.608 | 6183.891 | 794.7595 | 33.31667 |
| 948.5079 | 6276.372 | 6190.232 | 795.0416 | 32.88333 |
| 948.7805 | 6414.098 | 6196.573 | 795.3237 | 33.46667 |
| 949.0531 | 6147.81 | 6202.913 | 795.6058 | 32.66667 |
| 949.3257 | 6193.157 | 6209.252 | 795.8878 | 32.83333 |
| 949.5983 | 6405.543 | 6215.591 | 796.1698 | 33.2 |
| 949.8709 | 6206.404 | 6221.929 | 796.4519 | 33.73333 |
| 950.1434 | 6299.804 | 6228.266 | 796.7339 | 33.16667 |
| 950.4159 | 6372.701 | 6234.603 | 797.0158 | 33.1 |
| 950.6884 | 6328.713 | 6240.939 | 797.2978 | 33 |
| 950.9609 | 6292.74 | 6247.274 | 797.5798 | 32.91667 |
| 951.2334 | 6314.334 | 6253.609 | 797.8617 | 33.65 |
| 951.5058 | 6193.003 | 6259.943 | 798.1436 | 33.03333 |
| 951.7783 | 6300.819 | 6266.276 | 798.4255 | 33.41667 |
| 952.0507 | 6392.135 | 6272.609 | 798.7074 | 33.8 |
| 952.3231 | 6365.28 | 6278.94 | 798.9893 | 33.46667 |
| 952.5954 | 6367.631 | 6285.271 | 799.2712 | 33.66667 |
| 952.8678 | 6313.984 | 6291.602 | 799.553 | 33.75 |
| 953.1401 | 6335.99 | 6297.931 | 799.8348 | 33.1 |
| 953.4124 | 6378.485 | 6304.26 | 800.1167 | 33.85 |
| 953.6847 | 6393.722 | 6310.588 | 800.3985 | 33.88333 |
| 953.957 | 6375.75 | 6316.916 | 800.6802 | 33.45 |
| 954.2293 | 6375.559 | 6323.242 | 800.962 | 34 |
| 954.5015 | 6325.851 | 6329.568 | 801.2438 | 33.98333 |
| 954.7737 | 6358.476 | 6335.893 | 801.5255 | 34.01667 |
| 955.0459 | 6424.462 | 6342.217 | 801.8072 | 33.66667 |
| 955.3181 | 6444.335 | 6348.541 | 802.0889 | 33.66667 |
| 955.5903 | 6497.082 | 6354.863 | 802.3706 | 34.08333 |
| 955.8624 | 6381.224 | 6361.185 | 802.6523 | 34.06667 |
| 956.1346 | 6492.059 | 6367.506 | 802.9339 | 33.91667 |
| 956.4067 | 6519.811 | 6373.826 | 803.2156 | 33.48333 |
| 956.6788 | 6407.279 | 6380.145 | 803.4972 | 34.18333 |
| 956.9508 | 6431.302 | 6386.464 | 803.7788 | 34.51667 |
| 957.2229 | 6547.144 | 6392.781 | 804.0604 | 34.16667 |
| 957.4949 | 6518.824 | 6399.098 | 804.342 | 34.63333 |
| 957.7669 | 6542.514 | 6405.414 | 804.6236 | 34.4 |
| 958.0389 | 6418.011 | 6411.729 | 804.9051 | 34.71667 |
| 958.3109 | 6322.253 | 6418.043 | 805.1867 | 34.68333 |
| 958.5829 | 6545.993 | 6424.356 | 805.4682 | 34.88333 |
| 958.8548 | 6439.119 | 6430.669 | 805.7497 | 34.13333 |
| 959.1267 | 6572.524 | 6436.98 | 806.0312 | 34.35 |
| 959.3986 | 6572.058 | 6443.29 | 806.3126 | 34.38333 |
| 959.6705 | 6626.291 | 6449.6 | 806.5941 | 33.78333 |
| 959.9424 | 6548.554 | 6455.909 | 806.8755 | 34.38333 |
| 960.2142 | 6398.898 | | 807.157 | 34.15 |
| 960.486 | 6580.162 | | 807.4384 | 34.3 |
| 960.7579 | 6538.009 | | 807.7198 | 34.9 |
| 961.0296 | 6457.604 | | 808.0011 | 34.8 |
| 961.3014 | 6449.635 | | 808.2825 | 34.7 |
| 961.5732 | 6862.715 | | 808.5639 | 34.2 |
| 961.8449 | 6508.672 | | 808.8452 | 34.61667 |
| 962.1166 | 6490.414 | | 809.1265 | 35.01667 |
| 962.3883 | 6504.135 | | 809.4078 | 35.13333 |
| 962.66 | 6503.403 | | 809.6891 | 34.41667 |
| 962.9316 | 6510.271 | | 809.9704 | 34.35 |
| 963.2033 | 6658.39 | | 810.2516 | 34.83333 |
| 963.4749 | 6686.11 | | 810.5328 | 35.63333 |
| 963.7465 | 6690.679 | | 810.8141 | 35 |
| 964.0181 | 6668.503 | | 811.0953 | 34.65 |
| 964.2896 | 6740.766 | | 811.3765 | 35.38333 |
| 964.5612 | 6621.863 | | 811.6576 | 35.25 |
| 964.8327 | 6555.586 | | 811.9388 | 35.08333 |
| 965.1042 | 6833.686 | | 812.2199 | 35.08333 |
| 965.3757 | 6718.86 | | 812.5011 | 35.95 |
| 965.6471 | 6669.678 | | 812.7822 | 35.16667 |
| 965.9186 | 6920.779 | | 813.0633 | 34.93333 |
| 966.19 | 6695.399 | | 813.3443 | 35 |
| 966.4614 | 6782.565 | | 813.6254 | 35.96667 |
| 966.7328 | 6681.486 | | 813.9065 | 35.36667 |
| 967.0042 | 6658.303 | | 814.1875 | 35.31667 |
| 967.2755 | 6851.251 | | 814.4685 | 35.96667 |
| 967.5469 | 6810.299 | | 814.7495 | 35.8 |
| 967.8182 | 6538.095 | | 815.0305 | 35.16667 |
| 968.0895 | 6769.54 | | 815.3115 | 35.05 |
| 968.3607 | 6729.912 | | 815.5924 | 35.25 |
| 968.632 | 6878.492 | | 815.8734 | 36.08333 |
| 968.9032 | 6817.259 | | 816.1543 | 35.06667 |
| 969.1745 | 6825.039 | | 816.4352 | 35.53333 |
| 969.4457 | 7156.948 | | 816.7161 | 35.41667 |
| 969.7168 | 6960.628 | | 816.9969 | 36.51667 |
| 969.988 | 6940.492 | | 817.2778 | 35.76667 |
| 970.2591 | 6844.973 | | 817.5586 | 35.83333 |
| 970.5303 | 6793 | | 817.8395 | 35.8 |
| 970.8014 | 6860.735 | | 818.1203 | 35.65 |
| 971.0724 | 6757.306 | | 818.4011 | 36.45 |
| 971.3435 | 6729.713 | | 818.6818 | 36.23333 |
| 971.6146 | 6882.448 | | 818.9626 | 35.53333 |
| 971.8856 | 6969.258 | | 819.2434 | 35.78333 |
| 972.1566 | 6697.995 | | 819.5241 | 35.88333 |
| 972.4276 | 7046.858 | | 819.8048 | 36.26667 |
| 972.6985 | 7051.314 | | 820.0855 | 35.76667 |
| 972.9695 | 6951.765 | | 820.3662 | 36.7 |
| 973.2404 | 6929.412 | | 820.6468 | 35.83333 |
| 973.5113 | 6970.972 | | 820.9275 | 36.05 |
| 973.7822 | 6812.773 | | 821.2081 | 35.45 |
| 974.0531 | 6967.261 | | 821.4888 | 35.7 |
| 974.324 | 6858.017 | | 821.7694 | 36.6 |
| 974.5948 | 6885.652 | | 822.0499 | 36.45 |
| 974.8656 | 6864.542 | | 822.3305 | 36.41667 |
| 975.1364 | 6824.347 | | 822.6111 | 36.16667 |
| 975.4072 | 6840.422 | | 822.8916 | 36.15 |
| 975.6779 | 6937.888 | | 823.1721 | 35.88333 |
| 975.9487 | 7112.564 | | 823.4526 | 36.66667 |
| 976.2194 | 6975.428 | | 823.7331 | 36.53333 |
| 976.4901 | 6810.835 | | 824.0136 | 36.01667 |
| 976.7608 | 6996.231 | | 824.2941 | 36.21667 |
| 977.0314 | 6961.453 | | 824.5745 | 36.21667 |
| 977.3021 | 7042.542 | | 824.8549 | 36.18333 |
| 977.5727 | 7036.952 | | 825.1353 | 35.91667 |
| 977.8433 | 7024.684 | | 825.4157 | 36.41667 |
| 978.1139 | 7087.675 | | 825.6961 | 36.73333 |
| 978.3845 | 7009.572 | | 825.9765 | 36.58333 |
| 978.655 | 7026.081 | | 826.2568 | 35.68333 |
| 978.9255 | 7048.687 | | 826.5371 | 36.58333 |
| 979.196 | 6946.592 | | 826.8175 | 36.35 |
| 979.4665 | 7103.992 | | 827.0978 | 36.11667 |
| 979.737 | 7054.034 | | 827.378 | 36.98333 |
| 980.1296 | 8906.659 | | 827.6583 | 37.01667 |
| 981.862 | 6748.366 | | 827.9386 | 36.25 |
| 983.5944 | 7400.501 | | 828.2188 | 36.91667 |
| 985.3267 | 7061.327 | | 828.499 | 36.7 |
| 987.0588 | 6937.12 | | 828.7792 | 36.11667 |
| 988.7909 | 8678.653 | | 829.0594 | 36.93333 |
| 990.5228 | 6742.035 | | 829.3396 | 36.76667 |
| 992.2547 | 6929.319 | | 829.6197 | 36.58333 |
| 993.9864 | 7835.588 | | 829.8999 | 37.3 |
| 995.7181 | 7259.576 | | 830.18 | 37.68333 |
| 997.4496 | 7501.995 | | 830.4601 | 37.26667 |
| 999.1811 | 7783.697 | | 830.7402 | 37.38333 |
| 1000.912 | 7159.482 | | 831.0202 | 36.88333 |
| 1002.644 | 6892.557 | | 831.3003 | 37.18333 |
| 1004.375 | 7492.08 | | 831.5803 | 37.41667 |
| 1006.106 | 7993.624 | | 831.8604 | 37.43333 |
| 1007.837 | 8112.961 | | 832.1404 | 37.06667 |
| 1009.568 | 7607.078 | | 832.4204 | 36.91667 |
| 1011.298 | 8237.773 | | 832.7003 | 37.15 |
| 1013.029 | 7486.201 | | 832.9803 | 37.08333 |
| 1014.759 | 7819.21 | | 833.2602 | 37.55 |
| 1016.49 | 8571.928 | | 833.5402 | 37.03333 |
| 1018.22 | 7859.266 | | 833.8201 | 37 |
| 1019.95 | 7652.629 | | 834.1 | 36.75 |
| 1021.68 | 7791.988 | | 834.3798 | 38.06667 |
| 1023.41 | 8701.469 | | 834.6597 | 37.16667 |
| 1025.14 | 8386.434 | | 834.9395 | 36.95 |
| 1026.87 | 7895.927 | | 835.2194 | 38.05 |
| 1028.6 | 8411.792 | | 835.4992 | 37.15 |
| 1030.329 | 8075.436 | | 835.779 | 38.18333 |
| 1032.059 | 7706.679 | | 836.0588 | 37.51667 |
| 1033.788 | 8801.668 | | 836.3385 | 37.15 |
| 1035.517 | 8384.263 | | 836.6183 | 37.31667 |
| 1037.246 | 8222.108 | | 836.898 | 37.53333 |
| 1038.975 | 8028.396 | | 837.1777 | 37.1 |
| 1040.704 | 8371.801 | | 837.4574 | 37.43333 |
| 1042.433 | 9061.523 | | 837.7371 | 37.98333 |
| 1044.162 | 8962.253 | | 838.0168 | 38.26667 |
| 1045.89 | 9190.819 | | 838.2964 | 37.55 |
| 1047.619 | 8977.373 | | 838.576 | 37.05 |
| 1049.347 | 8381.62 | | 838.8556 | 37.68333 |
| 1051.076 | 8243.933 | | 839.1352 | 38.23333 |
| 1052.804 | 8361.131 | | 839.4148 | 37.51667 |
| 1054.532 | 8571.848 | | 839.6944 | 37.51667 |
| 1056.26 | 8490.979 | | 839.9739 | 37.98333 |
| 1057.988 | 8332.017 | | 840.2535 | 37.85 |
| 1059.716 | 8454.479 | | 840.533 | 37.26667 |
| 1061.443 | 8461.488 | | 840.8125 | 38 |
| 1063.171 | 8708.539 | | 841.092 | 37.8 |
| 1064.898 | 9261.152 | | 841.3715 | 37.4 |
| 1066.625 | 9524.366 | | 841.6509 | 37.58333 |
| 1068.353 | 9082.418 | | 841.9303 | 37.28333 |
| 1070.08 | 8800.291 | | 842.2098 | 37.98333 |
| 1071.807 | 8607.997 | | 842.4892 | 37.45 |
| 1073.534 | 8681.66 | | 842.7686 | 38.75 |
| 1075.26 | 9100.463 | | 843.0479 | 37.68333 |
| 1076.987 | 9415.884 | | 843.3273 | 38.11667 |
| 1078.714 | 9917.249 | | 843.6066 | 37.7 |
| 1080.44 | 8823.578 | | 843.8859 | 38.63333 |
| 1082.166 | 8709.906 | | 844.1652 | 38.06667 |
| 1083.893 | 9158.917 | | 844.4445 | 38.08333 |
| 1085.619 | 8906.908 | | 844.7238 | 37.51667 |
| 1087.345 | 9461.907 | | 845.0031 | 37.85 |
| 1089.071 | 9315.904 | | 845.2823 | 38.28333 |
| 1090.797 | 9089.424 | | 845.5615 | 37.95 |
| 1092.522 | 9179.159 | | 845.8407 | 37.93333 |
| 1094.248 | 9473.677 | | 846.1199 | 37.75 |
| 1095.973 | 9407.898 | | 846.3991 | 37.6 |
| 1097.699 | 9744.778 | | 846.6782 | 38.21667 |
| 1099.424 | 9422.974 | | 846.9574 | 38.23333 |
| 1101.149 | 9714.747 | 9525.105 | 847.2365 | 38.2 |
| 1102.874 | 9624.889 | 9558.55 | 847.5156 | 38.6 |
| 1104.599 | 9283.255 | 9591.866 | 847.7947 | 37.65 |
| 1106.324 | 9580.896 | 9625.053 | 848.0738 | 37.93333 |
| 1108.049 | 9523.041 | 9658.109 | 848.3528 | 38.08333 |
| 1109.773 | 9581.2 | 9691.035 | 848.6319 | 37.55 |
| 1111.498 | 10139.42 | 9723.828 | 848.9109 | 38.21667 |
| 1113.222 | 10071.19 | 9756.49 | 849.1899 | 37.83333 |
| 1114.946 | 9600.502 | 9789.018 | 849.4689 | 38.26667 |
| 1116.67 | 9871.508 | 9821.412 | 849.7478 | 38.11667 |
| 1118.394 | 9785.413 | 9853.671 | 850.0268 | 38.66667 |
| 1120.118 | 9635.95 | 9885.796 | 850.3057 | 38.28333 |
| 1121.842 | 9745.148 | 9917.785 | 850.5847 | 37.5 |
| 1123.566 | 10072.61 | 9949.637 | 850.8636 | 38.28333 |
| 1125.29 | 9965.349 | 9981.353 | 851.1425 | 37.93333 |
| 1127.013 | 10040.66 | 10012.93 | 851.4213 | 38.43333 |
| 1128.736 | 10024.8 | 10044.37 | 851.7002 | 38.51667 |
| 1130.46 | 9809.164 | 10075.67 | 851.979 | 37.63333 |
| 1132.183 | 9840.652 | 10106.83 | 852.2579 | 37.95 |
| 1133.906 | 10034.47 | 10137.85 | 852.5367 | 38.53333 |
| 1135.629 | 10092.64 | 10168.73 | 852.8155 | 39.05 |
| 1137.351 | 10350.53 | 10199.47 | 853.0942 | 38.05 |
| 1139.074 | 10373.79 | 10230.07 | 853.373 | 38.6 |
| 1140.797 | 10177.23 | 10260.53 | 853.6517 | 38.23333 |
| 1142.519 | 10130.33 | 10290.84 | 853.9305 | 38.5 |
| 1144.242 | 10073.36 | 10321.01 | 854.2092 | 38.06667 |
| 1145.964 | 10346.24 | 10351.04 | 854.4879 | 38.4 |
| 1147.686 | 10196.76 | 10380.92 | 854.7665 | 37.88333 |
| 1149.408 | 10420.32 | 10410.66 | 855.0452 | 39.2 |
| 1151.13 | 10415.02 | 10440.25 | 855.3238 | 38.68333 |
| 1152.852 | 10387.9 | 10469.7 | 855.6024 | 38.53333 |
| 1154.573 | 10252.79 | 10499 | 855.8811 | 38.61667 |
| 1156.295 | 10192.95 | 10528.16 | 856.1596 | 38.25 |
| 1158.016 | 10718.5 | 10557.17 | 856.4382 | 38.46667 |
| 1159.738 | 10640.95 | 10586.03 | 856.7168 | 38.53333 |
| 1161.459 | 10224.16 | 10614.75 | 856.9953 | 38.56667 |
| 1163.18 | 10577.82 | 10643.32 | 857.2738 | 38.23333 |
| 1164.901 | 10903.3 | 10671.74 | 857.5523 | 37.9 |
| 1166.622 | 10591.4 | 10700.01 | 857.8308 | 38.25 |
| 1168.343 | 10556.81 | 10728.13 | 858.1093 | 38.58333 |
| 1170.063 | 10766.7 | 10756.1 | 858.3878 | 38.53333 |
| 1171.784 | 10854.65 | 10783.92 | 858.6662 | 38.46667 |
| 1173.504 | 10746.34 | 10811.6 | 858.9446 | 39.31667 |
| 1175.225 | 11168.28 | 10839.12 | 859.223 | 38.65 |
| 1176.945 | 11247.65 | 10866.49 | 859.5014 | 38.16667 |
| 1178.665 | 10846.75 | 10893.72 | 859.7798 | 39.01667 |
| 1180.385 | 10703.58 | 10920.79 | 860.0581 | 39.06667 |
| 1182.105 | 11002.45 | 10947.71 | 860.3365 | 38.9 |
| 1183.824 | 11109.31 | 10974.48 | 860.6148 | 38.45 |
| 1185.544 | 11237.68 | 11001.09 | 860.8931 | 38.26667 |
| 1187.263 | 11576.84 | 11027.56 | 861.1714 | 38.76667 |
| 1188.983 | 11052.84 | 11053.87 | 861.4497 | 38.7 |
| 1190.702 | 10673.48 | 11080.03 | 861.7279 | 39.1 |
| 1192.421 | 11122.42 | 11106.04 | 862.0062 | 38.78333 |
| 1194.14 | 11184.67 | 11131.9 | 862.2844 | 38.71667 |
| 1195.859 | 11312.51 | 11157.6 | 862.5626 | 38.96667 |
| 1197.578 | 11488.61 | 11183.15 | 862.8408 | 38.05 |
| 1199.297 | 11144.55 | 11208.54 | 863.1189 | 39.55 |
| 1201.015 | 11066.34 | 11233.79 | 863.3971 | 39.21667 |
| 1202.734 | 11012.11 | 11258.87 | 863.6752 | 38.06667 |
| 1204.452 | 11014.1 | 11283.81 | 863.9533 | 38.56667 |
| 1206.17 | 11441.35 | 11308.59 | 864.2314 | 38.45 |
| 1207.888 | 11283.6 | 11333.22 | 864.5095 | 38.55 |
| 1209.606 | 11259.55 | 11357.69 | 864.7876 | 38.3 |
| 1211.324 | 11601.16 | 11382.01 | 865.0656 | 38.41667 |
| 1213.042 | 11005.02 | 11406.17 | 865.3437 | 38.66667 |
| 1214.759 | 11115.14 | 11430.18 | 865.6217 | 38.5 |
| 1216.477 | 11716.93 | 11454.04 | 865.8997 | 39.46667 |
| 1218.194 | 11451.5 | 11477.74 | 866.1777 | 38.73333 |
| 1219.912 | 11757.32 | 11501.28 | 866.4556 | 38.93333 |
| 1221.629 | 11583.9 | 11524.67 | 866.7336 | 38.26667 |
| 1223.346 | 11499.69 | 11547.91 | 867.0115 | 38.86667 |
| 1225.063 | 11700.36 | 11570.99 | 867.2894 | 39.21667 |
| 1226.779 | 11321.19 | 11593.92 | 867.5673 | 39.45 |
| 1228.496 | 11723.27 | 11616.69 | 867.8452 | 38.66667 |
| 1230.213 | 11478.27 | 11639.3 | 868.1231 | 38.53333 |
| 1231.929 | 11858.26 | 11661.76 | 868.4009 | 38.61667 |
| 1233.645 | 12065.42 | 11684.07 | 868.6788 | 37.81667 |
| 1235.362 | 11823.05 | 11706.22 | 868.9566 | 38.7 |
| 1237.078 | 12002.06 | 11728.22 | 869.2344 | 39.33333 |
| 1238.794 | 11536.29 | 11750.06 | 869.5122 | 38.65 |
| 1240.51 | 11829.1 | 11771.75 | 869.7899 | 38.63333 |
| 1242.225 | 12186.55 | 11793.28 | 870.0677 | 39.23333 |
| 1243.941 | 11616.51 | 11814.65 | 870.3454 | 38.78333 |
| 1245.656 | 11638 | 11835.88 | 870.6231 | 39.08333 |
| 1247.372 | 12021.06 | 11856.94 | 870.9008 | 38.16667 |
| 1249.087 | 11869.06 | 11877.85 | 871.1785 | 39.01667 |
| 1250.802 | 11745.05 | 11898.61 | 871.4561 | 38.46667 |
| 1252.517 | 12110.59 | 11919.21 | 871.7338 | 38.95 |
| 1254.232 | 11905.75 | 11939.66 | 872.0114 | 38.48333 |
| 1255.947 | 11708.88 | 11959.96 | 872.289 | 38.15 |
| 1257.661 | 12304.83 | 11980.1 | 872.5666 | 38.51667 |
| 1259.376 | 11958.15 | 12000.08 | 872.8442 | 38.7 |
| 1261.09 | 11970.09 | 12019.91 | 873.1217 | 38.9 |
| 1262.805 | 12035.56 | 12039.59 | 873.3993 | 38.86667 |
| 1264.519 | 11652.83 | 12059.11 | 873.6768 | 38.78333 |
| 1266.233 | 12201.75 | 12078.48 | 873.9543 | 38.88333 |
| 1267.947 | 12190.15 | 12097.7 | 874.2318 | 38.73333 |
| 1269.661 | 12049.12 | 12116.76 | 874.5093 | 38.51667 |
| 1271.374 | 12533.68 | 12135.67 | 874.7867 | 38.36667 |
| 1273.088 | 12358.04 | 12154.42 | 875.0642 | 38.61667 |
| 1274.801 | 11899.05 | 12173.02 | 875.3416 | 38.46667 |
| 1276.514 | 12259.28 | 12191.47 | 875.619 | 38.76667 |
| 1278.228 | 12357.35 | 12209.77 | 875.8964 | 38.33333 |
| 1279.941 | 12080.6 | 12227.91 | 876.1737 | 38.85 |
| 1281.654 | 12032.59 | 12245.9 | 876.4511 | 38.18333 |
| 1283.367 | 12302.2 | 12263.74 | 876.7284 | 38.55 |
| 1285.079 | 12372.76 | 12281.43 | 877.0057 | 39.01667 |
| 1286.792 | 12197.93 | 12298.96 | 877.283 | 38.38333 |
| 1288.504 | 12341.83 | 12316.35 | 877.5603 | 38.33333 |
| 1290.217 | 12754.01 | 12333.58 | 877.8376 | 38.36667 |
| 1291.929 | 12549.02 | 12350.66 | 878.1148 | 38.78333 |
| 1293.641 | 12384.31 | 12367.59 | 878.3921 | 37.9 |
| 1295.353 | 12674.82 | 12384.36 | 878.6693 | 38.76667 |
| 1297.065 | 12480.06 | 12400.99 | 878.9465 | 37.86667 |
| 1298.776 | 12312.38 | 12417.47 | 879.2237 | 38.11667 |
| 1300.488 | 12436.87 | 12433.8 | 879.5008 | 38.3 |
| 1302.199 | 12265.88 | 12449.97 | 879.778 | 37.98333 |
| 1303.911 | 12302.79 | 12466 | 880.0551 | 38.26667 |
| 1305.622 | 12381.06 | 12481.88 | 880.3322 | 38 |
| 1307.333 | 12138.56 | 12497.6 | 880.6093 | 38.25 |
| 1309.044 | 12522.63 | 12513.18 | 880.8864 | 37.66667 |
| 1310.755 | 12405.63 | 12528.61 | 881.1634 | 37.58333 |
| 1312.466 | 12611.79 | 12543.9 | 881.4405 | 37.65 |
| 1314.176 | 12548.3 | 12559.03 | 881.7175 | 37.93333 |
| 1315.887 | 12367.88 | 12574.02 | 881.9945 | 38.2 |
| 1317.597 | 12760.77 | 12588.86 | 882.2715 | 37.48333 |
| 1319.307 | 12626.62 | 12603.55 | 882.5485 | 38.06667 |
| 1321.017 | 12552.45 | 12618.09 | 882.8254 | 37.61667 |
| 1322.727 | 12470.36 | 12632.49 | 883.1024 | 37.46667 |
| 1324.437 | 12784 | 12646.74 | 883.3793 | 37.38333 |
| 1326.147 | 12836.04 | 12660.84 | 883.6562 | 37.18333 |
| 1327.857 | 12667.21 | 12674.8 | 883.9331 | 37.46667 |
| 1329.566 | 12867.13 | 12688.61 | 884.2099 | 37.91667 |
| 1331.275 | 12828.27 | 12702.28 | 884.4868 | 37.35 |
| 1332.985 | 12827.1 | 12715.8 | 884.7636 | 37.7 |
| 1334.694 | 12641.12 | 12729.18 | 885.0405 | 37.15 |
| 1336.403 | 12504.03 | 12742.41 | 885.3173 | 37.93333 |
| 1338.112 | 12654.08 | 12755.5 | 885.594 | 37.65 |
| 1339.82 | 12447.95 | 12768.45 | 885.8708 | 37.46667 |
| 1341.529 | 12300.98 | 12781.25 | 886.1475 | 37.08333 |
| 1343.237 | 12473.85 | 12793.91 | 886.4243 | 37.4 |
| 1344.946 | 12980.64 | 12806.43 | 886.701 | 37.2 |
| 1346.654 | 13012.84 | 12818.8 | 886.9777 | 36.85 |
| 1348.362 | 12663.73 | 12831.03 | 887.2544 | 37.31667 |
| 1350.07 | 12849.31 | 12843.12 | 887.531 | 37.56667 |
| 1351.778 | 12929.96 | 12855.07 | 887.8077 | 37.36667 |
| 1353.486 | 13030.76 | 12866.88 | 888.0843 | 36.91667 |
| 1355.193 | 13095.14 | 12878.55 | 888.3609 | 36.96667 |
| 1356.901 | 13125.39 | 12890.08 | 888.6375 | 36.73333 |
| 1358.608 | 12915.3 | 12901.46 | 888.9141 | 37.43333 |
| 1360.315 | 12903.5 | 12912.71 | 889.1906 | 37 |
| 1362.022 | 12849.32 | 12923.82 | 889.4672 | 36.5 |
| 1363.729 | 12655.19 | 12934.79 | 889.7437 | 36.71667 |
| 1365.436 | 12854.93 | 12945.62 | 890.0202 | 36.95 |
| 1367.143 | 13111.37 | 12956.31 | 890.2967 | 37.16667 |
| 1368.849 | 12919.41 | 12966.87 | 890.5731 | 36.53333 |

| | | |
|---|---|---|
| 1370.556 | 12811.84 | 12977.29 |
| 1372.262 | 12847.59 | 12987.57 |
| 1373.968 | 12948.59 | 12997.71 |
| 1375.674 | 13087.36 | 13007.72 |
| 1377.38 | 13164.06 | 13017.6 |
| 1379.086 | 13142.1 | 13027.33 |
| 1380.792 | 13184.85 | 13036.94 |
| 1382.497 | 13173.45 | 13046.4 |
| 1384.203 | 12885.74 | 13055.74 |
| 1385.908 | 12809.79 | 13064.94 |
| 1387.613 | 12985.81 | 13074 |
| 1389.318 | 13212.65 | 13082.94 |
| 1391.023 | 12933.26 | 13091.74 |
| 1392.728 | 13005.14 | 13100.41 |
| 1394.432 | 13102.27 | 13108.95 |
| 1396.137 | 13271.54 | 13117.35 |
| 1397.841 | 13077.75 | 13125.63 |
| 1399.545 | 13252.8 | 13133.77 |
| 1401.25 | 13542.39 | 13141.79 |
| 1402.954 | 12724.92 | 13149.67 |
| 1404.657 | 12915.72 | 13157.43 |
| 1406.361 | 13151.46 | 13165.06 |
| 1408.065 | 13051.06 | 13172.56 |
| 1409.768 | 13493.08 | 13179.93 |
| 1411.471 | 13416.66 | 13187.17 |
| 1413.175 | 13153.17 | 13194.29 |
| 1414.878 | 13248.03 | 13201.28 |
| 1416.581 | 13510.01 | 13208.14 |
| 1418.284 | 13355.23 | 13214.88 |
| 1419.986 | 13371.68 | 13221.49 |
| 1421.689 | 13717.35 | 13227.98 |
| 1423.391 | 13294.54 | 13234.34 |
| 1425.093 | 13353.84 | 13240.58 |
| 1426.796 | 13555.41 | 13246.7 |
| 1428.498 | 13246.76 | 13252.69 |
| 1430.2 | 13120.94 | 13258.56 |
| 1431.901 | 13133.92 | 13264.31 |
| 1433.603 | 12915.78 | 13269.94 |
| 1435.304 | 13267.85 | 13275.44 |
| 1437.006 | 13528.94 | 13280.83 |
| 1438.707 | 13061.09 | 13286.09 |
| 1440.408 | 13408.25 | 13291.23 |
| 1442.109 | 13487.37 | 13296.26 |
| 1443.81 | 13291.3 | 13301.16 |
| 1445.511 | 13386.42 | 13305.95 |
| 1447.211 | 13366.75 | 13310.62 |
| 1448.912 | 13288.05 | 13315.17 |
| 1450.612 | 13479.29 | 13319.6 |
| 1452.312 | 13298.82 | 13323.92 |
| 1454.012 | 13289.9 | 13328.12 |
| 1455.712 | 13467.83 | 13332.21 |
| 1457.412 | 13264.17 | 13336.18 |
| 1459.111 | 13378.04 | 13340.03 |
| 1460.811 | 13491.92 | 13343.77 |
| 1462.51 | 13417.79 | 13347.4 |
| 1464.209 | 13575.02 | 13350.91 |
| 1465.909 | 13502.28 | 13354.31 |
| 1467.607 | 13518 | 13357.6 |
| 1469.306 | 13194.66 | 13360.77 |
| 1471.005 | 12979.32 | 13363.83 |
| 1472.704 | 13477.92 | 13366.79 |
| 1474.402 | 13715.2 | 13369.63 |
| 1476.1 | 13265.58 | 13372.36 |
| 1477.798 | 13158.68 | 13374.98 |
| 1479.496 | 13424.22 | 13377.49 |
| 1481.194 | 13703.24 | 13379.9 |
| 1482.892 | 13569.3 | 13382.19 |
| 1484.59 | 13425.72 | 13384.38 |
| 1486.287 | 13464.76 | 13386.46 |
| 1487.984 | 13680.5 | 13388.43 |
| 1489.682 | 13614.21 | 13390.29 |
| 1491.379 | 13436.35 | 13392.05 |
| 1493.076 | 13546.51 | 13393.71 |
| 1494.772 | 13393.37 | 13395.26 |
| 1496.469 | 13034.95 | 13396.7 |
| 1498.166 | 12978.72 | 13398.04 |
| 1499.862 | 13402.98 | 13399.28 |
| 1501.558 | 13478.53 | 13400.41 |
| 1503.254 | 13343.93 | 13401.44 |
| 1504.95 | 13333.76 | 13402.37 |
| 1506.646 | 13789.88 | 13403.2 |
| 1508.342 | 13618.16 | 13403.92 |
| 1510.038 | 13217.56 | 13404.55 |
| 1511.733 | 13295.18 | 13405.07 |
| 1513.428 | 13401 | 13405.49 |
| 1515.123 | 13234.12 | 13405.82 |
| 1516.818 | 13302.36 | 13406.04 |
| 1518.513 | 13713.58 | 13406.17 |
| 1520.208 | 13404.43 | 13406.2 |
| 1521.903 | 13265.62 | 13406.13 |
| 1523.597 | 13337.53 | 13405.96 |
| 1525.291 | 13079.34 | 13405.7 |
| 1526.986 | 13058.77 | 13405.34 |
| 1528.68 | 13327.18 | 13404.88 |
| 1530.374 | 13620.85 | 13404.33 |
| 1532.067 | 13516.47 | 13403.69 |
| 1533.761 | 13197.32 | 13402.95 |
| 1535.455 | 13402.2 | 13402.11 |
| 1537.148 | 13086.77 | 13401.19 |
| 1538.841 | 13352.48 | 13400.17 |
| 1540.534 | 13645.72 | 13399.06 |
| 1542.227 | 13301.83 | 13397.85 |
| 1543.92 | 13408.88 | 13396.55 |
| 1545.613 | 13358.32 | 13395.17 |
| 1547.305 | 13562.25 | 13393.69 |
| 1548.998 | 13277.79 | 13392.12 |
| 1550.69 | 13255.39 | 13390.46 |
| 1552.382 | 13517.76 | 13388.72 |
| 1554.074 | 13656.3 | 13386.88 |
| 1555.766 | 13569.48 | 13384.96 |
| 1557.457 | 13372.15 | 13382.94 |
| 1559.149 | 13322.66 | 13380.84 |
| 1560.84 | 13418.49 | 13378.66 |
| 1562.532 | 13426.11 | 13376.38 |
| 1564.223 | 12644.52 | 13374.02 |
| 1565.914 | 13027.47 | 13371.58 |
| 1567.605 | 13668.82 | 13369.05 |
| 1569.295 | 13350.62 | 13366.43 |
| 1570.986 | 13371.47 | 13363.73 |
| 1572.676 | 13353.44 | 13360.95 |
| 1574.367 | 13406.26 | 13358.08 |
| 1576.057 | 13122.66 | 13355.13 |
| 1577.747 | 12870.03 | 13352.1 |
| 1579.437 | 13186.28 | 13348.99 |
| 1581.126 | 13037.1 | 13345.79 |

| | |
|---|---|
| 890.8496 | 37.05 |
| 891.126 | 36.66667 |
| 891.4024 | 37.13333 |
| 891.6788 | 36.63333 |
| 891.9552 | 36.18333 |
| 892.2316 | 36.28333 |
| 892.5079 | 36.73333 |
| 892.7842 | 37.1 |
| 893.0606 | 36.7 |
| 893.3368 | 36.25 |
| 893.6131 | 36.61667 |
| 893.8894 | 36.08333 |
| 894.1656 | 36.33333 |
| 894.4418 | 35.83333 |
| 894.718 | 36.36667 |
| 894.9942 | 36.53333 |
| 895.2704 | 36.8 |
| 895.5466 | 36.35 |
| 895.8227 | 36.38333 |
| 896.0988 | 35.96667 |
| 896.3749 | 36.06667 |
| 896.651 | 36.2 |
| 896.9271 | 36.56667 |
| 897.2031 | 35.21667 |
| 897.4791 | 36.18333 |
| 897.7551 | 35.85 |
| 898.0311 | 36.21667 |
| 898.3071 | 35.85 |
| 898.5831 | 35.61667 |
| 898.859 | 35.6 |
| 899.1349 | 35.71667 |
| 899.4108 | 35.71667 |
| 899.6867 | 35.78333 |
| 899.9626 | 35.66667 |
| 900.2385 | 35.61667 |
| 900.5143 | 35.9 |
| 900.7901 | 36.35 |
| 901.0659 | 35.86667 |
| 901.3417 | 35.45 |
| 901.6175 | 34.93333 |
| 901.8932 | 35.16667 |
| 902.1689 | 35.46667 |
| 902.4446 | 35.11667 |
| 902.7203 | 35.83333 |
| 902.996 | 34.98333 |
| 903.2717 | 35.76667 |
| 903.5473 | 35.76667 |
| 903.8229 | 34.86667 |
| 904.0985 | 35.33333 |
| 904.3741 | 35.5 |
| 904.6497 | 34.91667 |
| 904.9252 | 35.35 |
| 905.2008 | 34.95 |
| 905.4763 | 34.7 |
| 905.7518 | 35.03333 |
| 906.0273 | 34.71667 |
| 906.3027 | 34.96667 |
| 906.5782 | 34.73333 |
| 906.8536 | 34.9 |
| 907.129 | 35.06667 |
| 907.4044 | 34.46667 |
| 907.6798 | 35.23333 |
| 907.9551 | 34.51667 |
| 908.2305 | 34.91667 |
| 908.5058 | 34.23333 |
| 908.7811 | 34.75 |
| 909.0564 | 34.8 |
| 909.3316 | 34.35 |
| 909.6069 | 34.11667 |
| 909.8821 | 34.01667 |
| 910.1573 | 34.6 |
| 910.4325 | 34.23333 |
| 910.7077 | 34.41667 |
| 910.9829 | 34.26667 |
| 911.258 | 34.4 |
| 911.5331 | 34.06667 |
| 911.8082 | 33.9 |
| 912.0833 | 34.6 |
| 912.3584 | 34.01667 |
| 912.6334 | 33.98333 |
| 912.9085 | 34.46667 |
| 913.1835 | 33.83333 |
| 913.4585 | 34.05 |
| 913.7335 | 34 |
| 914.0084 | 34.11667 |
| 914.2834 | 34.08333 |
| 914.5583 | 33.88333 |
| 914.8332 | 33.8 |
| 915.1081 | 33.56667 |
| 915.383 | 34.11667 |
| 915.6578 | 33.8 |
| 915.9327 | 33.43333 |
| 916.2075 | 33.61667 |
| 916.4823 | 33.71667 |
| 916.7571 | 33.43333 |
| 917.0318 | 33.4 |
| 917.3066 | 33.6 |
| 917.5813 | 32.91667 |
| 917.856 | 32.86667 |
| 918.1307 | 33.66667 |
| 918.4054 | 32.88333 |
| 918.68 | 33.5 |
| 918.9547 | 32.63333 |
| 919.2293 | 32.35 |
| 919.5039 | 32.96667 |
| 919.7785 | 32.31667 |
| 920.053 | 32.88333 |
| 920.3276 | 32.76667 |
| 920.6021 | 33.16667 |
| 920.8766 | 32.68333 |
| 921.1511 | 32.41667 |
| 921.4256 | 33.4 |
| 921.7 | 32.16667 |
| 921.9745 | 32.75 |
| 922.2489 | 32.68333 |
| 922.5233 | 32.03333 |
| 922.7977 | 32.73333 |
| 923.072 | 32.35 |
| 923.3464 | 32.15 |
| 923.6207 | 31.93333 |
| 923.895 | 31.95 |
| 924.1693 | 31.28333 |
| 924.4436 | 31.96667 |
| 924.7179 | 32 |
| 924.9921 | 31.33333 |
| 925.2663 | 31.46667 |
| 925.5405 | 31.83333 |
| 925.8147 | 31.06667 |
| 926.0889 | 31.26667 |
| 926.363 | 30.88333 |
| 926.6371 | 31.25 |
| 926.9113 | 30.48333 |
| 927.1853 | 31.75 |
| 927.4594 | 31.11667 |
| 927.7335 | 31.28333 |
| 928.0075 | 30.81667 |
| 928.2815 | 30.91667 |
| 928.5555 | 30.55 |
| 928.8295 | 30.51667 |
| 929.1035 | 30.63333 |
| 929.3774 | 30.26667 |
| 929.6514 | 30.71667 |
| 929.9253 | 30.3 |
| 930.1992 | 30.31667 |
| 930.473 | 29.71667 |
| 930.7469 | 30.23333 |
| 931.0207 | 30.3 |
| 931.2945 | 29.81667 |
| 931.5683 | 30.46667 |
| 931.8421 | 29.83333 |
| 932.1159 | 29.96667 |
| 932.3896 | 29.66667 |
| 932.6634 | 29.86667 |
| 932.9371 | 29.6 |
| 933.2108 | 29.83333 |
| 933.4844 | 29.55 |
| 933.7581 | 29.71667 |
| 934.0317 | 29.63333 |
| 934.3053 | 30.1 |
| 934.5789 | 28.96667 |
| 934.8525 | 29.85 |
| 935.1261 | 29.08333 |
| 935.3996 | 28.61667 |
| 935.6731 | 29.28333 |
| 935.9466 | 28.35 |
| 936.2201 | 28.36667 |
| 936.4936 | 28.78333 |
| 936.767 | 28.36667 |
| 937.0405 | 28.8 |
| 937.3139 | 27.95 |
| 937.5873 | 28.8 |
| 937.8607 | 28.28333 |
| 938.134 | 28.41667 |
| 938.4074 | 28.23333 |
| 938.6807 | 27.98333 |
| 938.954 | 28.46667 |
| 939.2273 | 28.45 |
| 939.5005 | 27.86667 |
| 939.7738 | 28.1 |
| 940.047 | 27.55 |
| 940.3202 | 27.81667 |
| 940.5934 | 27.33333 |
| 940.8666 | 27.78333 |
| 941.1397 | 27.3 |
| 941.4129 | 27.48333 |
| 941.686 | 27.58333 |
| 941.9591 | 27.1 |
| 942.2322 | 26.65 |
| 942.5052 | 26.66667 |
| 942.7783 | 26.5 |
| 943.0513 | 26.56667 |
| 943.3243 | 26.36667 |
| 943.5973 | 26.43333 |
| 943.8703 | 26.83333 |
| 944.1432 | 25.96667 |
| 944.4162 | 25.7 |
| 944.6891 | 25.68333 |
| 944.962 | 26.41667 |
| 945.2348 | 25.8 |
| 945.5077 | 26.11667 |
| 945.7805 | 25.48333 |
| 946.0534 | 25.4 |
| 946.3262 | 25.51667 |
| 946.5989 | 25.65 |
| 946.8717 | 25.46667 |
| 947.1445 | 24.98333 |
| 947.4172 | 25.45 |
| 947.6899 | 25.46667 |
| 947.9626 | 25.55 |
| 948.2353 | 25.33333 |
| 948.5079 | 25.26667 |
| 948.7805 | 25.63333 |
| 949.0532 | 24.46667 |
| 949.3258 | 24.65 |
| 949.5983 | 25.3 |
| 949.8709 | 24.43333 |
| 950.1434 | 24.71667 |
| 950.4159 | 24.9 |
| 950.6885 | 24.61667 |
| 950.9609 | 24.53333 |
| 951.2334 | 24.38333 |
| 951.5059 | 23.75 |
| 951.7783 | 24.23333 |
| 952.0507 | 24.4 |
| 952.3231 | 24.18333 |
| 952.5955 | 24.08333 |
| 952.8678 | 23.68333 |
| 953.1401 | 23.78333 |
| 953.4125 | 23.83333 |
| 953.6848 | 23.85 |
| 953.957 | 23.5 |
| 954.2293 | 23.68333 |
| 954.5015 | 23.18333 |
| 954.7737 | 23.3 |
| 955.046 | 23.36667 |
| 955.3181 | 23.31667 |
| 955.5903 | 23.48333 |
| 955.8624 | 22.93333 |
| 956.1346 | 23.16667 |
| 956.4067 | 23.18333 |
| 956.6788 | 22.56667 |
| 956.9508 | 22.8 |
| 957.2229 | 23.06667 |
| 957.4949 | 22.71667 |
| 957.7669 | 22.83333 |
| 958.0389 | 22.35 |
| 958.3109 | 21.91667 |
| 958.5829 | 22.41667 |
| 958.8548 | 21.98333 |
| 959.1267 | 22.46667 |
| 959.3986 | 22.41667 |
| 959.6705 | 22.45 |
| 959.9424 | 22.1 |
| 960.2142 | 21.4 |
| 960.4861 | 22 |
| 960.7579 | 21.9 |
| 961.0296 | 21.55 |
| 961.3014 | 21.45 |
| 961.5732 | 22.66667 |
| 961.8449 | 21.5 |
| 962.1166 | 21.21667 |
| 962.3883 | 21.23333 |
| 962.66 | 21.2 |
| 962.9316 | 21.03333 |
| 963.2033 | 21.48333 |
| 963.4749 | 21.51667 |
| 963.7465 | 21.43333 |
| 964.0181 | 21.31667 |
| 964.2896 | 21.46667 |
| 964.5612 | 21 |
| 964.8327 | 20.7 |
| 965.1042 | 21.31667 |
| 965.3757 | 21.06667 |
| 965.6471 | 20.71667 |
| 965.9186 | 21.28333 |
| 966.19 | 20.51667 |
| 966.4614 | 20.71667 |
| 966.7328 | 20.33333 |
| 967.0042 | 20.2 |
| 967.2755 | 20.51667 |
| 967.5469 | 20.36667 |
| 967.8182 | 19.35 |
| 968.0895 | 20.11667 |
| 968.3607 | 19.81667 |
| 968.632 | 20.16667 |
| 968.9032 | 20.1 |
| 969.1745 | 19.88333 |
| 969.4457 | 20.68333 |
| 969.7168 | 20.13333 |
| 969.988 | 19.85 |
| 970.2591 | 19.48333 |
| 970.5303 | 19.25 |
| 970.8014 | 19.35 |
| 971.0725 | 19.01667 |
| 971.3435 | 18.98333 |
| 971.6146 | 19.15 |
| 971.8856 | 19.36667 |
| 972.1566 | 18.58333 |
| 972.4276 | 19.26667 |
| 972.6986 | 19.31667 |
| 972.9695 | 18.86667 |
| 973.2404 | 18.86667 |
| 973.5113 | 18.88333 |
| 973.7822 | 18.33333 |
| 974.0531 | 18.65 |
| 974.324 | 18.28333 |
| 974.5948 | 18.36667 |
| 974.8656 | 18.23333 |
| 975.1364 | 17.91667 |
| 975.4072 | 17.95 |
| 975.6779 | 18.11667 |
| 975.9487 | 18.46667 |
| 976.2194 | 18.06667 |
| 976.4901 | 17.48333 |
| 976.7608 | 17.83333 |
| 977.0314 | 17.61667 |
| 977.3021 | 17.75 |
| 977.5727 | 17.68333 |
| 977.8433 | 17.65 |
| 978.1139 | 17.66667 |
| 978.3845 | 17.38333 |
| 978.655 | 17.4 |
| 978.9255 | 17.28333 |
| 979.1961 | 16.88333 |
| 979.4665 | 17.33333 |
| 979.737 | 17.03333 |
| 980.0075 | 16.75 |
| 980.2779 | 16.76667 |
| 980.5483 | 16.53333 |
| 980.8187 | 16.73333 |
| 981.0891 | 16.86667 |
| 981.3594 | 16.01667 |
| 981.6298 | 16.58333 |
| 981.9001 | 16.3 |
| 982.1704 | 16.05 |
| 982.4406 | 16.26667 |
| 982.7109 | 16.2 |
| 982.9811 | 15.6 |
| 983.2514 | 15.93333 |
| 983.5216 | 16.01667 |
| 983.7917 | 15.6 |
| 984.0619 | 15.98333 |
| 984.332 | 15.58333 |
| 984.6022 | 15.25 |
| 984.8723 | 15.56667 |
| 985.1423 | 15.46667 |
| 985.4124 | 15.53333 |
| 985.6824 | 14.46667 |
| 985.9525 | 15.2 |
| 986.2225 | 14.48333 |
| 986.4925 | 14.61667 |
| 986.7624 | 14.05 |
| 987.0324 | 14.33333 |
| 987.3023 | 13.61667 |
| 987.5722 | 14 |
| 987.8421 | 13.76667 |
| 988.112 | 13.93333 |
| 988.3818 | 13.01667 |
| 988.6516 | 13 |
| 988.9215 | 13.28333 |
| 989.1912 | 12.8 |
| 989.461 | 12.7 |
| 989.7308 | 12.53333 |
| 990.0005 | 11.91667 |
| 990.2702 | 12.1 |
| 990.5399 | 12.16667 |
| 990.8096 | 11.23333 |
| 991.0792 | 11.11667 |
| 991.3489 | 11.11667 |
| 991.6185 | 10.96667 |
| 991.8881 | 10.46667 |
| 992.1577 | 10.95 |
| 992.4272 | 10.31667 |
| 992.6968 | 10.26667 |
| 992.9663 | 9.68333 |
| 993.2358 | 9.65 |
| 993.5052 | 9.38333 |
| 993.7747 | 9.08333 |
| 994.0441 | 8.8 |
| 994.3136 | 8.68333 |
| 994.583 | 8.68333 |
| 994.8523 | 8.18333 |
| 995.1217 | 7.85 |
| 995.391 | 7.81667 |
| 995.6604 | 7.45 |
| 995.9297 | 7.33333 |
| 996.199 | 7 |
| 996.4682 | 6.95 |
| 996.7375 | 6.98333 |
| 997.0067 | 6.51667 |
| 997.2759 | 6.25 |
| 997.5451 | 6.1 |
| 997.8142 | 5.98333 |
| 998.0834 | 5.6 |
| 998.3525 | 5.13333 |
| 998.6216 | 5.18333 |
| 998.8907 | 5.01667 |
| 999.1598 | 4.81667 |
| 999.4288 | 4.53333 |
| 999.6978 | 4.45 |
| 999.9669 | 4.38333 |
| 1000.236 | 4.06667 |
| 1000.505 | 3.88333 |
| 1000.774 | 3.85 |
| 1001.043 | 3.6 |
| 1001.312 | 3.25 |
| 1001.58 | 3.35 |
| 1001.849 | 3.08333 |
| 1002.118 | 3.05 |
| 1002.387 | 2.93333 |
| 1002.656 | 2.61667 |
| 1002.925 | 2.61667 |
| 1003.193 | 2.31667 |
| 1003.462 | 2.3 |
| 1003.731 | 2.15 |
| 1004 | 1.91667 |
| 1004.268 | 2.05 |
| 1004.537 | 1.76667 |
| 1004.806 | 1.7 |
| 1005.074 | 1.66667 |
| 1005.343 | 1.56667 |
| 1005.611 | 1.35 |
| 1005.88 | 1.38333 |
| 1006.149 | 1.18333 |

Thermal blackbody emission from a single carbon black nanoparticle being heated to four different temperatures

| Figure 5 | Figure 5 | Figure 5 | Figure 5 | Figure 5 | Figure 5 | Figure 5 | Figure 5 | Figure 5 | Figure 5 | Figure 5 | Figure 5 |
|---|---|---|---|---|---|---|---|---|---|---|---|
| Wavelengt | P/s/sr/m^2 | Fit T = 1616 K | Wavelengt | P/s/sr/m^2 | Fit T = 1667 K | Wavelengt | P/s/sr/m^2 | Fit T = 1756 K | Wavelengt | P/s/sr/m^2 | Fit T = 1924 K |
| 600.015 | 253.2791 | 250.3973 | 600.015 | 455.2083 | 419.7902 | 600.015 | 914.8093 | 859.3435 | 600.015 | 2989.581 | 2835.901 |
| 600.306 | 249.8462 | 251.4554 | 600.306 | 469.085 | 421.4658 | 600.306 | 848.7592 | 862.4745 | 600.306 | 2915.652 | 2844.455 |
| 600.597 | 260.01 | 252.5162 | 600.597 | 447.2003 | 423.1442 | 600.597 | 855.6902 | 865.6114 | 600.597 | 2874.004 | 2853.024 |
| 600.888 | 280.3201 | 253.5802 | 600.888 | 457.9754 | 424.8256 | 600.888 | 928.2312 | 868.7541 | 600.888 | 2924.731 | 2861.604 |
| 601.179 | 287.0944 | 254.6474 | 601.179 | 447.1857 | 426.5129 | 601.179 | 920.9158 | 871.9027 | 601.179 | 2914.707 | 2870.199 |
| 601.47 | 283.8947 | 255.7173 | 601.47 | 454.8385 | 428.2061 | 601.47 | 910.0233 | 875.0601 | 601.47 | 2926.709 | 2878.805 |
| 601.761 | 270.0908 | 256.7904 | 601.761 | 491.7296 | 429.9021 | 601.761 | 939.7017 | 878.2234 | 601.761 | 3007.628 | 2887.426 |
| 602.051 | 284.2316 | 257.8667 | 602.051 | 475.2472 | 431.6011 | 602.051 | 934.2731 | 881.3955 | 602.051 | 3009.33 | 2896.059 |
| 602.342 | 270.84 | 258.946 | 602.342 | 452.001 | 433.306 | 602.342 | 914.1403 | 884.5734 | 602.342 | 2946.094 | 2904.704 |
| 602.633 | 267.7976 | 260.0282 | 602.633 | 441.0763 | 435.0168 | 602.633 | 901.9244 | 887.7572 | 602.633 | 2943.071 | 2913.363 |
| 602.924 | 287.9571 | 261.1136 | 602.924 | 468.8678 | 436.7304 | 602.924 | 949.7167 | 890.9498 | 602.924 | 3096.158 | 2922.034 |
| 603.215 | 284.8267 | 262.202 | 603.215 | 460.719 | 438.4471 | 603.215 | 947.5688 | 894.1483 | 603.215 | 2981.13 | 2930.717 |
| 603.506 | 279.5756 | 263.2936 | 603.506 | 470.6021 | 440.1695 | 603.506 | 957.3109 | 897.3556 | 603.506 | 3063.909 | 2939.403 |
| 603.797 | 271.303 | 264.3881 | 603.797 | 475.4144 | 441.895 | 603.797 | 928.9853 | 900.5687 | 604.088 | 3028.286 | 2956.833 |
| 604.088 | 265.4354 | 265.4859 | 604.088 | 483.8743 | 443.6262 | 604.088 | 950.8113 | 903.7877 | 604.378 | 3120.249 | 2965.578 |
| 604.378 | 271.7264 | 266.5866 | 604.378 | 485.7112 | 445.3634 | 604.378 | 936.6969 | 907.0126 | 604.669 | 3120.866 | 2974.322 |
| 604.669 | 266.6335 | 267.6902 | 604.669 | 478.8154 | 447.1035 | 604.669 | 936.691 | 910.2493 | 604.96 | 3101.234 | 2983.096 |
| 604.96 | 289.9657 | 268.797 | 604.96 | 463.6974 | 448.8465 | 604.96 | 969.3801 | 913.4888 | 605.251 | 3074.532 | 2991.87 |
| 605.251 | 283.7524 | 269.9071 | 605.251 | 465.7192 | 450.5954 | 605.251 | 1002.465 | 916.7372 | 605.542 | 3098.476 | 3000.644 |
| 605.542 | 301.7106 | 271.0201 | 605.542 | 467.8555 | 452.3472 | 605.542 | 967.7984 | 919.9914 | 605.833 | 3126.588 | 3009.447 |
| 605.833 | 287.6839 | 272.1361 | 605.833 | 469.1378 | 454.1049 | 605.833 | 996.2209 | 923.2515 | 606.123 | 3154.053 | 3018.25 |
| 606.123 | 283.9432 | 273.2552 | 606.123 | 496.052 | 455.8685 | 606.123 | 946.2219 | 926.5204 | 606.414 | 3059.214 | 3027.083 |
| 606.414 | 281.1027 | 274.3777 | 606.414 | 499.5234 | 457.635 | 606.414 | 961.692 | 929.7981 | 606.705 | 3193.697 | 3035.915 |
| 606.705 | 293.776 | 275.503 | 606.705 | 494.9164 | 459.4044 | 606.705 | 986.0386 | 933.0788 | 606.996 | 3099.444 | 3044.777 |
| 606.996 | 279.8717 | 276.6316 | 606.996 | 479.8718 | 461.1797 | 606.996 | 954.1477 | 936.3712 | 607.286 | 3109.157 | 3053.639 |
| 607.286 | 288.3626 | 277.7631 | 607.286 | 464.2608 | 462.9609 | 607.286 | 971.889 | 939.6665 | 607.577 | 3139.293 | 3062.501 |
| 607.577 | 268.3768 | 278.8978 | 607.577 | 516.9419 | 464.745 | 607.577 | 1000.024 | 942.9706 | 607.868 | 3107.191 | 3071.392 |
| 607.868 | 277.0253 | 280.0357 | 607.868 | 493.7632 | 466.5321 | 607.868 | 987.2827 | 946.2806 | 608.158 | 3167.522 | 3080.283 |
| 608.158 | 283.8173 | 281.1766 | 608.158 | 499.1654 | 468.325 | 608.158 | 974.3949 | 949.5994 | 608.449 | 3202.031 | 3089.203 |
| 608.449 | 264.7408 | 282.3207 | 608.449 | 490.2419 | 470.1237 | 608.449 | 960.3246 | 952.924 | 608.74 | 3155.08 | 3098.124 |
| 608.74 | 276.0852 | 283.4678 | 608.74 | 481.1013 | 471.9255 | 608.74 | 951.2368 | 956.2546 | 609.03 | 3140.702 | 3107.074 |
| 609.03 | 308.5594 | 284.6181 | 609.03 | 507.2144 | 473.7301 | 609.03 | 994.3077 | 959.5939 | 609.321 | 3215.94 | 3116.024 |
| 609.321 | 292.1292 | 285.7713 | 609.321 | 484.5932 | 475.5406 | 609.321 | 1007.333 | 962.9391 | 609.612 | 3175.386 | 3124.974 |
| 609.612 | 306.2442 | 286.928 | 609.612 | 492.3987 | 477.357 | 609.612 | 982.5378 | 966.2902 | 609.902 | 3131.693 | 3133.953 |
| 609.902 | 303.8615 | 288.0874 | 609.902 | 476.3945 | 479.1763 | 609.902 | 957.6073 | 969.65 | 610.193 | 3188.239 | 3142.932 |
| 610.193 | 293.411 | 289.2503 | 610.193 | 483.2962 | 481.0015 | 610.193 | 997.2509 | 973.0158 | 610.484 | 3179.993 | 3151.941 |
| 610.484 | 292.6427 | 290.4161 | 610.484 | 482.6418 | 482.8296 | 610.484 | 993.9291 | 976.3903 | 610.774 | 3228.822 | 3160.949 |
| 610.774 | 294.9175 | 291.5852 | 610.774 | 488.1291 | 484.6607 | 610.774 | 1058.644 | 979.7707 | 611.065 | 3266.499 | 3169.987 |
| 611.065 | 300.1055 | 292.7575 | 611.065 | 518.8991 | 486.4976 | 611.065 | 1012.639 | 983.157 | 611.355 | 3154.347 | 3179.025 |
| 611.355 | 311.8841 | 293.9315 | 611.355 | 484.065 | 488.3404 | 611.355 | 1022.859 | 986.5521 | 611.646 | 3270.343 | 3188.063 |
| 611.646 | 308.2719 | 295.1111 | 611.646 | 475.9573 | 490.1861 | 611.646 | 1023.514 | 989.953 | 611.936 | 3300.978 | 3197.13 |
| 611.936 | 304.3545 | 296.2937 | 611.936 | 484.8456 | 492.0377 | 611.936 | 1029.453 | 993.3628 | 612.227 | 3259.251 | 3206.197 |
| 612.227 | 307.82 | 297.4763 | 612.227 | 499.7845 | 493.8923 | 612.227 | 1021.439 | 996.7784 | 612.518 | 3279.176 | 3215.265 |
| 612.518 | 329.1383 | 298.6647 | 612.518 | 519.3833 | 495.7527 | 612.518 | 1060.158 | 1000.2 | 612.808 | 3239.151 | 3224.391 |
| 612.808 | 310.1176 | 299.856 | 612.808 | 521.0031 | 497.616 | 612.808 | 1037.643 | 1003.63 | 613.099 | 3315.797 | 3233.487 |
| 613.099 | 306.7988 | 301.0503 | 613.099 | 519.4918 | 499.4852 | 613.099 | 1015.767 | 1007.066 | 613.389 | 3328.444 | 3242.613 |
| 613.389 | 281.4211 | 302.2476 | 613.389 | 529.2281 | 501.3573 | 613.389 | 1072.782 | 1010.508 | 613.68 | 3325.275 | 3251.739 |
| 613.68 | 322.8968 | 303.4477 | 613.68 | 501.419 | 503.2353 | 613.68 | 1031.369 | 1013.959 | 613.97 | 3270.343 | 3260.894 |
| 613.97 | 309.5512 | 304.6508 | 613.97 | 505.8235 | 505.1192 | 613.97 | 1054.85 | 1017.416 | 614.261 | 3321.46 | 3270.05 |
| 614.261 | 304.9648 | 305.8598 | 614.261 | 529.6566 | 507.006 | 614.261 | 1054.853 | 1020.879 | 614.551 | 3353.357 | 3279.205 |
| 614.551 | 307.2595 | 307.0688 | 614.551 | 501.2605 | 508.8958 | 614.551 | 1061.975 | 1024.35 | 614.841 | 3370.464 | 3288.39 |
| 614.841 | 332.6419 | 308.2807 | 614.841 | 548.8152 | 510.7914 | 614.841 | 1065.493 | 1027.827 | 615.132 | 3446.465 | 3297.604 |
| 615.132 | 316.4118 | 309.4984 | 615.132 | 521.1292 | 512.6929 | 615.132 | 1049.952 | 1031.313 | 615.422 | 3393.851 | 3306.788 |
| 615.422 | 302.7229 | 310.7191 | 615.422 | 530.8978 | 514.5973 | 615.422 | 1033.473 | 1034.805 | 615.713 | 3430.179 | 3316.031 |
| 615.713 | 317.703 | 311.9398 | 615.713 | 538.7649 | 516.5046 | 615.713 | 1115.457 | 1038.303 | 616.003 | 3362.365 | 3325.245 |
| 616.003 | 327.8471 | 313.1664 | 616.003 | 515.3896 | 518.4179 | 616.003 | 1075.499 | 1041.809 | 616.293 | 3375.453 | 3334.489 |
| 616.293 | 327.9674 | 314.3959 | 616.293 | 513.8784 | 520.3369 | 616.293 | 1038.681 | 1045.322 | 616.584 | 3402.478 | 3343.732 |
| 616.584 | 316.5204 | 315.6284 | 616.584 | 544.0556 | 522.259 | 616.584 | 1056.493 | 1048.843 | 616.874 | 3331.114 | 3353.005 |
| 616.874 | 324.6897 | 316.8637 | 616.874 | 544.663 | 524.1869 | 616.874 | 1034.658 | 1052.37 | 617.165 | 3367.119 | 3362.277 |
| 617.165 | 312.3037 | 318.102 | 617.165 | 531.7928 | 526.1177 | 617.165 | 1052.397 | 1055.903 | 617.455 | 3345.258 | 3371.579 |
| 617.455 | 308.1838 | 319.3433 | 617.455 | 528.2979 | 528.0514 | 617.455 | 1045.436 | 1059.442 | 617.745 | 3422.755 | 3380.881 |
| 617.745 | 337.5746 | 320.5875 | 617.745 | 514.6032 | 529.994 | 617.745 | 1071.69 | 1062.99 | 618.036 | 3374.866 | 3390.183 |
| 618.036 | 327.1223 | 321.8375 | 618.036 | 529.7739 | 531.9395 | 618.036 | 1081.614 | 1066.546 | 618.326 | 3372.02 | 3399.515 |
| 618.326 | 328.4076 | 323.0876 | 618.326 | 558.0409 | 533.8879 | 618.326 | 1076.553 | 1070.106 | 618.616 | 3478.538 | 3408.846 |
| 618.616 | 342.3547 | 324.3435 | 618.616 | 564.6315 | 535.8422 | 618.616 | 1104.303 | 1073.674 | 618.906 | 3501.866 | 3418.207 |
| 618.906 | 340.0307 | 325.5994 | 618.906 | 548.466 | 537.7995 | 618.906 | 1100.541 | 1077.251 | 619.197 | 3295.168 | 3427.567 |
| 619.197 | 329.822 | 326.8612 | 619.197 | 523.1892 | 539.7626 | 619.197 | 1050.029 | 1080.831 | 619.487 | 3524.52 | 3436.957 |
| 619.487 | 328.0643 | 328.1259 | 619.487 | 556.9522 | 541.7316 | 619.487 | 1109.066 | 1084.423 | 619.777 | 3434.522 | 3446.318 |
| 619.777 | 332.8972 | 329.3936 | 619.777 | 548.9325 | 543.7035 | 619.777 | 1101.357 | 1088.017 | 620.068 | 3515.364 | 3455.738 |
| 620.068 | 345.6823 | 330.6641 | 620.068 | 539.9181 | 545.6783 | 620.068 | 1104.825 | 1091.621 | 620.358 | 3499.225 | 3465.128 |
| 620.358 | 359.6529 | 331.9377 | 620.358 | 560.2064 | 547.659 | 620.358 | 1097.302 | 1095.23 | 620.648 | 3513.017 | 3474.547 |
| 620.648 | 342.9974 | 333.2141 | 620.648 | 563.0704 | 549.6456 | 620.648 | 1093.308 | 1098.848 | 620.938 | 3476.249 | 3483.996 |
| 620.938 | 335.2242 | 334.4935 | 620.938 | 533.125 | 551.6351 | 620.938 | 1099.294 | 1102.472 | 621.228 | 3457.616 | 3493.415 |
| 621.228 | 357.1059 | 335.7788 | 621.228 | 546.987 | 553.6305 | 621.228 | 1116.815 | 1106.102 | 621.519 | 3446.348 | 3502.893 |
| 621.519 | 332.3426 | 337.064 | 621.519 | 539.9621 | 555.6288 | 621.519 | 1106.061 | 1109.741 | 621.809 | 3503.979 | 3512.342 |
| 621.809 | 347.5603 | 338.3552 | 621.809 | 559.7956 | 557.633 | 621.809 | 1116.325 | 1113.385 | 622.099 | 3548.376 | 3521.82 |
| 622.099 | 327.4774 | 339.6463 | 622.099 | 550.8956 | 559.6401 | 622.099 | 1131.226 | 1117.035 | 622.389 | 3576.018 | 3531.327 |
| 622.389 | 353.1298 | 340.9433 | 622.389 | 603.1042 | 561.6531 | 622.389 | 1148.248 | 1120.695 | 622.679 | 3652.341 | 3540.805 |
| 622.679 | 347.2405 | 342.2432 | 622.679 | 559.6342 | 563.669 | 622.679 | 1150.605 | 1124.36 | 622.969 | 3544.063 | 3550.342 |
| 622.969 | 343.4317 | 343.5461 | 622.969 | 579.6996 | 565.6908 | 622.969 | 1149.384 | 1128.031 | 623.259 | 3524.343 | 3559.85 |
| 623.259 | 329.9511 | 344.8519 | 623.259 | 550.6491 | 567.7155 | 623.259 | 1127.083 | 1131.71 | 623.55 | 3573.318 | 3569.386 |
| 623.55 | 356.8535 | 346.1606 | 623.55 | 580.8674 | 569.7461 | 623.55 | 1153.337 | 1135.396 | 623.84 | 3595.238 | 3578.923 |
| 623.84 | 348.9512 | 347.4723 | 623.84 | 573.106 | 571.7826 | 623.84 | 1146.024 | 1139.09 | 624.13 | 3553.687 | 3588.489 |
| 624.13 | 363.7934 | 348.7898 | 624.13 | 563.4167 | 573.822 | 624.13 | 1174.99 | 1142.788 | 624.42 | 3610.702 | 3598.055 |
| 624.42 | 362.0914 | 350.1074 | 624.42 | 583.6962 | 575.8673 | 624.42 | 1181.601 | 1146.497 | 624.71 | 3708.359 | 3607.651 |
| 624.71 | 354.1979 | 351.4308 | 624.71 | 619.0321 | 577.9155 | 624.71 | 1157.386 | 1150.209 | 625 | 3654.748 | 3617.246 |
| 625 | 363.8902 | 352.7542 | 625 | 569.5407 | 579.9695 | 625 | 1179.761 | 1153.929 | 625.29 | 3685.265 | 3626.842 |
| 625.29 | 369.1017 | 354.0835 | 625.29 | 585.4598 | 582.0265 | 625.29 | 1167.345 | 1157.656 | 625.58 | 3696.973 | 3636.466 |
| 625.58 | 351.6567 | 355.4157 | 625.58 | 616.6259 | 584.0894 | 625.58 | 1200.249 | 1161.392 | 625.87 | 3695.976 | 3646.091 |
| 625.87 | 363.6584 | 356.7508 | 625.87 | 586.2198 | 586.1552 | 625.87 | 1194.826 | 1165.133 | 626.16 | 3698.47 | 3655.716 |
| 626.16 | 342.968 | 358.0889 | 626.16 | 633.1406 | 588.2269 | 626.16 | 1207.118 | 1168.88 | 626.45 | 3671.826 | 3665.37 |
| 626.45 | 357.9393 | 359.4299 | 626.45 | 608.8145 | 590.3015 | 626.45 | 1189.38 | 1172.636 | 626.74 | 3702.842 | 3675.024 |
| 626.74 | 352.763 | 360.7768 | 626.74 | 601.147 | 592.382 | 626.74 | 1200.114 | 1176.398 | 627.03 | 3711.616 | 3684.708 |
| 627.03 | 372.4116 | 362.1237 | 627.03 | 602.5173 | 594.4683 | 627.03 | 1219.052 | 1180.166 | 627.32 | 3679.074 | 3694.391 |
| 627.32 | 384.786 | 363.4764 | 627.32 | 620.3936 | 596.5576 | 627.32 | 1222.192 | 1183.942 | 627.61 | 3754.634 | 3704.075 |
| 627.61 | 375.7745 | 364.8292 | 627.61 | 590.3661 | 598.6498 | 627.61 | 1212.544 | 1187.725 | 627.9 | 3782.657 | 3713.787 |
| 627.9 | 373.0807 | 366.1878 | 627.9 | 620.464 | 600.7479 | 627.9 | 1182.02 | 1191.513 | 628.19 | 3789.495 | 3723.5 |
| 628.19 | 390.2058 | 367.5494 | 628.19 | 627.7765 | 602.8519 | 628.19 | 1204.351 | 1195.307 | 628.48 | 3796.537 | 3733.213 |
| 628.48 | 382.0981 | 368.9139 | 628.48 | 612.5001 | 604.9588 | 628.48 | 1255.406 | 1199.11 | 628.77 | 3831.427 | 3742.955 |
| 628.77 | 376.2087 | 370.2813 | 628.77 | 638.1349 | 607.0715 | 628.77 | 1209.099 | 1202.922 | 629.059 | 3747.386 | 3752.697 |
| 629.059 | 398.3663 | 371.6516 | 629.059 | 634.0209 | 609.1872 | 629.059 | 1225.179 | 1206.737 | 629.349 | 3812.441 | 3762.469 |
| 629.349 | 384.3546 | 373.0249 | 629.349 | 627.8558 | 611.3088 | 629.349 | 1254.872 | 1210.56 | 629.639 | 3871.452 | 3772.24 |
| 629.639 | 390.1324 | 374.4041 | 629.639 | 641.3217 | 613.4362 | 629.639 | 1249.271 | 1214.392 | 629.929 | 3945.692 | 3782.012 |
| 629.929 | 389.2668 | 375.7833 | 629.929 | 643.6369 | 615.5666 | 629.929 | 1222.779 | 1218.228 | 630.219 | 3877.35 | 3791.813 |
| 630.219 | 398.7683 | 377.1683 | 630.219 | 653.9982 | 617.6999 | 630.219 | 1264.427 | 1222.072 | 630.509 | 3869.603 | 3801.614 |
| 630.509 | 372.3001 | 378.5563 | 630.509 | 627.2689 | 619.839 | 630.509 | 1234.332 | 1225.925 | 630.799 | 3905.139 | 3811.414 |
| 630.799 | 375.17 | 379.9472 | 630.799 | 619.6131 | 621.9841 | 630.799 | 1285.358 | 1229.78 | 631.088 | 3987.8 | 3821.245 |
| 631.088 | 411.7324 | 381.341 | 631.088 | 659.6528 | 624.132 | 631.088 | 1255.867 | 1233.645 | 631.378 | 3865.759 | 3831.075 |
| 631.378 | 387.6176 | 382.7378 | 631.378 | 622.6326 | 626.2859 | 631.378 | 1265.336 | 1237.518 | 631.668 | 3953.409 | 3840.934 |
| 631.668 | 412.7418 | 384.1375 | 631.668 | 661.5953 | 628.4426 | 631.668 | 1279.744 | 1241.395 | 631.958 | 4009.485 | 3850.794 |
| 631.958 | 407.4688 | 385.5401 | 631.958 | 648.8865 | 630.6053 | 631.958 | 1301.611 | 1245.28 | 632.248 | 4073.015 | 3860.653 |
| 632.248 | 403.8125 | 386.9486 | 632.248 | 654.5821 | 632.7709 | 632.248 | 1304.827 | 1249.171 | 632.537 | 3988.622 | 3870.542 |
| 632.537 | 417.4838 | 388.3571 | 632.537 | 672.1444 | 634.9423 | 632.537 | 1294.073 | 1253.071 | 632.827 | 4056.436 | 3880.431 |
| 632.827 | 413.3727 | 389.7715 | 632.827 | 662.9393 | 637.1167 | 632.827 | 1303.392 | 1256.976 | 633.117 | 4051.066 | 3890.32 |
| 633.117 | 419.4498 | 391.1888 | 633.117 | 684.5833 | 639.2969 | 633.117 | 1339.814 | 1260.888 | 633.406 | 4104.383 | 3900.238 |
| 633.406 | 414.4144 | 392.609 | 633.406 | 666.2639 | 641.483 | 633.406 | 1293.53 | 1264.805 | 633.696 | 4019.902 | 3910.156 |
| 633.696 | 408.1965 | 394.0322 | 633.696 | 661.5366 | 643.6721 | 633.696 | 1313.724 | 1268.731 | 633.986 | 4115.211 | 3920.104 |
| 633.986 | 410.409 | 395.4583 | 633.986 | 680.5426 | 645.867 | 633.986 | 1315.561 | 1272.663 | 634.276 | 4135.165 | 3930.052 |
| 634.276 | 408.9682 | 396.8874 | 634.276 | 689.0993 | 648.0649 | 634.276 | 1316.207 | 1276.604 | 634.565 | 4063.977 | 3939.999 |
| 634.565 | 416.2367 | 398.3223 | 634.565 | 667.8514 | 650.2657 | 634.565 | 1345.859 | 1280.551 | 634.855 | 4144.232 | 3949.947 |
| 634.855 | 427.1673 | 399.7572 | 634.855 | 685.3697 | 652.4753 | 634.855 | 1341.404 | 1284.504 | 635.145 | 4149.045 | 3959.924 |
| 635.145 | 388.765 | 401.198 | 635.145 | 679.5978 | 654.6848 | 635.145 | 1372.585 | 1288.462 | 635.434 | 4178.799 | 3969.93 |
| 635.434 | 460.7601 | 402.6417 | 635.434 | 692.8847 | 656.9032 | 635.434 | 1379.32 | 1292.429 | 635.724 | 4141.973 | 3979.907 |
| 635.724 | 401.0689 | 404.0883 | 635.724 | 706.8846 | 659.1246 | 635.724 | 1332.493 | 1296.403 | 636.013 | 4134.314 | 3989.913 |
| 636.013 | 419.0684 | 405.5379 | 636.013 | 712.5421 | 661.3488 | 636.013 | 1349.697 | 1300.382 | 636.303 | 4197.198 | 3999.949 |
| 636.303 | 408.9213 | 406.9904 | 636.303 | 688.4244 | 663.579 | 636.303 | 1381.793 | 1304.369 | 636.593 | 4099.424 | 4009.955 |
| 636.593 | 429.7525 | 408.4459 | 636.593 | 679.0344 | 665.815 | 636.593 | 1377.802 | 1308.363 | 636.882 | 4152.478 | 4019.991 |
| 636.882 | 412.7594 | 409.9072 | 636.882 | 698.5744 | 668.0539 | 636.882 | 1377.723 | 1312.363 | 637.172 | 4174.985 | 4030.055 |
| 637.172 | 431.8506 | 411.3685 | 637.172 | 722.3312 | 670.2958 | 637.172 | 1358.952 | 1316.368 | 637.461 | 4167.854 | 4040.12 |
| 637.461 | 416.225 | 412.8357 | 637.461 | 686.4584 | 672.5464 | 637.461 | 1376.288 | 1320.382 | 637.751 | 4152.243 | 4050.185 |
| 637.751 | 412.5775 | 414.3059 | 637.751 | 699.8861 | 674.7971 | 637.751 | 1369.058 | 1324.403 | 638.04 | 4225.896 | 4060.25 |
| 638.04 | 438.4881 | 415.776 | 638.04 | 730.1602 | 677.0566 | 638.04 | 1369.354 | 1328.431 | 638.33 | 4179.738 | 4070.345 |
| 638.33 | 432.8453 | 417.252 | 638.33 | 753.2127 | 679.3161 | 638.33 | 1384.672 | 1332.463 | 638.619 | 4286.051 | 4080.439 |
| 638.619 | 415.8552 | 418.7339 | 638.619 | 704.8159 | 681.5844 | 638.619 | 1385.306 | 1336.504 | 638.909 | 4261.725 | 4090.562 |
| 638.909 | 460.7366 | 420.2157 | 638.909 | 680.6542 | 683.8556 | 638.909 | 1390.517 | 1340.553 | 639.198 | 4227.686 | 4100.686 |
| 639.198 | 444.2542 | 421.7005 | 639.198 | 689.1081 | 686.1297 | 639.198 | 1383.043 | 1344.606 | 639.488 | 4226.542 | 4110.81 |
| 639.488 | 436.1817 | 423.1912 | 639.488 | 722.029 | 688.4097 | 639.488 | 1411.604 | 1348.667 | 639.777 | 4293.593 | 4120.933 |
| 639.777 | 423.7722 | 424.6848 | 639.777 | 726.0491 | 690.6956 | 639.777 | 1401.929 | 1352.734 | 640.067 | 4197.756 | 4131.086 |
| 640.067 | 459.1374 | 426.1784 | 640.067 | 714.4553 | 692.9844 | 640.067 | 1388.724 | 1356.807 | 640.356 | 4309.409 | 4141.269 |
| 640.356 | 439.7235 | 427.6779 | 640.356 | 736.5483 | 695.2762 | 640.356 | 1388.868 | 1360.889 | 640.646 | 4318.506 | 4151.422 |
| 640.646 | 437.5638 | 429.1803 | 640.646 | 737.2526 | 697.5768 | 640.646 | 1371.074 | 1364.976 | 640.935 | 4285.171 | 4161.604 |
| 640.935 | 435.7709 | 430.6885 | 640.935 | 733.6492 | 699.8773 | 640.935 | 1426.962 | 1369.07 | 641.225 | 4312.813 | 4171.786 |
| 641.225 | 463.3013 | 432.1968 | 641.225 | 712.8532 | 702.1867 | 641.225 | 1419.001 | 1373.172 | 641.514 | 4334.028 | 4181.998 |
| 641.514 | 425.973 | 433.708 | 641.514 | 742.4347 | 704.496 | 641.514 | 1415.636 | 1377.28 | 641.803 | 4274.46 | 4192.21 |
| 641.803 | 433.8812 | 435.2251 | 641.803 | 716.1426 | 706.8142 | 641.803 | 1428.682 | 1381.394 | 642.093 | 4356.476 | 4202.421 |
| 642.093 | 442.0065 | 436.7451 | 642.093 | 744.0721 | 709.1353 | 642.093 | 1445.989 | 1385.514 | 642.382 | 4300.166 | 4212.662 |
| 642.382 | 441.7746 | 438.2651 | 642.382 | 718.0412 | 711.4593 | 642.382 | 1462.765 | 1389.643 | 642.671 | 4253.626 | 4222.903 |
| 642.671 | 463.0137 | 439.791 | 642.671 | 741.4928 | 713.7892 | 642.671 | 1450.657 | 1393.777 | 642.961 | 4294.678 | 4233.144 |
| 642.961 | 452.7522 | 441.3228 | 642.961 | 747.6403 | 716.125 | 642.961 | 1454.933 | 1397.918 | 643.25 | 4348.466 | 4243.415 |
| 643.25 | 441.8216 | 442.8545 | 643.25 | 729.899 | 718.4637 | 643.25 | 1444.935 | 1402.064 | 643.539 | 4418.568 | 4253.685 |
| 643.539 | 478.7215 | 444.3892 | 643.539 | 707.8559 | 720.8053 | 643.539 | 1478.942 | 1406.219 | 643.829 | 4335.055 | 4263.955 |
| 643.829 | 465.9628 | 445.9297 | 643.829 | 734.7906 | 723.1558 | 643.829 | 1418.414 | 1410.38 | 644.118 | 4341.364 | 4274.255 |
| 644.118 | 445.3634 | 447.4703 | 644.118 | 737.9715 | 725.5062 | 644.118 | 1410.536 | 1414.547 | 644.407 | 4365.632 | 4284.555 |
| 644.407 | 462.7819 | 449.0167 | 644.407 | 762.7788 | 727.8655 | 644.407 | 1467.586 | 1418.72 | 644.696 | 4373.76 | 4294.854 |
| 644.696 | 445.7742 | 450.5661 | 644.696 | 760.8949 | 730.2247 | 644.696 | 1467.116 | 1422.901 | 644.986 | 4374.523 | 4305.183 |
| 644.986 | 478.1405 | 452.1184 | 644.986 | 724.5115 | 732.5928 | 644.986 | 1464.214 | 1427.088 | 645.275 | 4341.57 | 4315.512 |
| 645.275 | 445.9004 | 453.6736 | 645.275 | 752.661 | 734.9638 | 645.275 | 1468.008 | 1431.282 | 645.564 | 4431.538 | 4325.841 |
| 645.564 | 461.6933 | 455.2317 | 645.564 | 729.8785 | 737.3377 | 645.564 | 1465.414 | 1435.484 | 645.853 | 4376.841 | 4336.2 |
| 645.853 | 484.0709 | 456.7958 | 645.853 | 742.8132 | 739.7175 | 645.853 | 1448.803 | 1439.689 | 646.143 | 4466.692 | 4346.558 |
| 646.143 | 472.8234 | 458.3598 | 646.143 | 784.3084 | 742.1031 | 646.143 | 1472.307 | 1443.902 | 646.432 | 4373.965 | 4356.917 |
| 646.432 | 450.5543 | 459.9297 | 646.432 | 743.4559 | 744.4917 | 646.432 | 1459.287 | 1448.122 | 646.721 | 4371.97 | 4367.275 |
| 646.721 | 477.6504 | 461.5025 | 646.721 | 777.0194 | 746.8832 | 646.721 | 1475.471 | 1452.351 | 647.01 | 4471.387 | 4377.663 |
| 647.01 | 486.4595 | 463.0783 | 647.01 | 780.0066 | 749.2836 | 647.01 | 1460.714 | 1456.585 | 647.299 | 4573.562 | 4388.08 |
| 647.299 | 473.3222 | 464.657 | 647.299 | 765.6163 | 751.6839 | 647.299 | 1482.906 | 1460.825 | 647.588 | 4472.179 | 4398.468 |
| 647.588 | 472.4008 | 466.2386 | 647.588 | 750.158 | 754.0901 | 647.588 | 1531.156 | 1465.071 | 647.878 | 4453.399 | 4408.885 |
| 647.878 | 506.1551 | 467.8261 | 647.878 | 802.349 | 756.5021 | 647.878 | 1478.455 | 1469.323 | 648.167 | 4540.961 | 4419.302 |
| 648.167 | 481.3654 | 469.4136 | 648.167 | 765.1645 | 758.9201 | 648.167 | 1511.919 | 1473.584 | 648.456 | 4466.222 | 4429.748 |
| 648.456 | 498.2528 | 471.007 | 648.456 | 763.4244 | 761.338 | 648.456 | 1502.197 | 1477.85 | 648.745 | 4554.812 | 4440.194 |
| 648.745 | 481.3272 | 472.6033 | 648.745 | 800.1012 | 763.7648 | 648.745 | 1527.615 | 1482.123 | 649.034 | 4566.256 | 4450.641 |
| 649.034 | 495.157 | 474.2025 | 649.034 | 797.7567 | 766.1944 | 649.034 | 1513.532 | 1486.401 | 649.323 | 4607.161 | 4461.087 |
| 649.323 | 477.3893 | 475.8047 | 649.323 | 754.7767 | 768.627 | 649.323 | 1527.327 | 1490.688 | 649.612 | 4587.706 | 4471.563 |
| 649.612 | 482.231 | 477.4098 | 649.612 | 792.6098 | 771.0655 | 649.612 | 1541.236 | 1494.978 | 649.901 | 4539.083 | 4482.039 |
| 649.901 | 462.4122 | 479.0179 | 649.901 | 799.2004 | 773.5098 | 649.901 | 1519.718 | 1499.277 | 650.19 | 4508.536 | 4492.544 |
| 650.19 | 476.0189 | 480.6318 | 650.19 | 812.8365 | 775.9571 | 650.19 | 1554.998 | 1503.585 | 650.479 | 4555.046 | 4503.02 |
| 650.479 | 458.9496 | 482.2457 | 650.479 | 794.3645 | 778.4073 | 650.479 | 1546.494 | 1507.896 | 653.369 | 4733.164 | 4608.511 |
| 650.768 | 491.8558 | 483.8655 | 650.768 | 808.79 | 780.8663 | 650.768 | 1534.622 | 1512.215 | 653.658 | 4774.597 | 4619.104 |
| 651.057 | 484.9864 | 485.4882 | 651.057 | 786.7909 | 783.3254 | 651.057 | 1526.784 | 1516.54 | 653.947 | 4755.846 | 4629.726 |
| 651.346 | 505.5946 | 487.1138 | 651.346 | 795.7906 | 785.7902 | 651.346 | 1498.943 | 1520.871 | 654.235 | 4730.904 | 4640.32 |
| 651.635 | 508.08 | 488.7424 | 651.635 | 788.575 | 788.261 | 651.635 | 1547.404 | 1525.208 | 654.524 | 4779.204 | 4650.942 |
| 651.924 | 475.5876 | 490.3739 | 651.924 | 819.8027 | 790.7376 | 651.924 | 1577.619 | 1529.551 | 654.813 | 4814.681 | 4661.594 |
| 652.213 | 519.3598 | 492.0113 | 652.213 | 816.5455 | 793.2142 | 652.213 | 1590.469 | 1533.903 | 655.102 | 4777.179 | 4672.216 |
| 652.502 | 506.9532 | 493.6487 | 652.502 | 809.0306 | 795.6997 | 652.502 | 1550.277 | 1538.261 | 655.391 | 4835.133 | 4682.868 |
| 652.791 | 501.1167 | 495.292 | 652.791 | 844.2725 | 798.188 | 652.791 | 1591.596 | 1542.624 | 655.68 | 4820.08 | 4693.549 |
| 653.08 | 495.9346 | 496.9382 | 653.08 | 823.4296 | 800.6793 | 653.08 | 1626.216 | 1546.996 | 655.968 | 4801.241 | 4704.201 |
| 653.369 | 516.801 | 498.5873 | 653.369 | 832.7639 | 803.1765 | 653.369 | 1576.666 | 1551.371 | 656.257 | 4804.557 | 4714.882 |
| 653.658 | 485.3415 | 500.2393 | 653.658 | 827.4526 | 805.6766 | 653.658 | 1619.992 | 1555.755 | 656.546 | 4942.561 | 4725.563 |
| 653.947 | 530.1202 | 501.8943 | 653.947 | 805.7734 | 808.1825 | 653.947 | 1617.953 | 1560.145 | 656.835 | 4842.675 | 4736.274 |
| 654.235 | 486.9701 | 503.5552 | 654.235 | 851.3503 | 810.6944 | 654.235 | 1626.451 | 1564.541 | 657.124 | 4830.526 | 4746.955 |
| 654.524 | 558.9124 | 505.2161 | 654.524 | 835.4987 | 813.2091 | 654.524 | 1620.039 | 1568.942 | 657.412 | 4829.852 | 4757.666 |
| 654.813 | 500.7822 | 506.8828 | 654.813 | 856.908 | 815.7298 | 654.813 | 1648.608 | 1573.353 | 657.701 | 4812.685 | 4768.405 |
| 655.102 | 525.701 | 508.5525 | 655.102 | 849.4928 | 818.2534 | 655.102 | 1602.902 | 1577.766 | 657.99 | 4890.711 | 4779.116 |
| 655.391 | 507.948 | 510.2221 | 655.391 | 850.3555 | 820.7799 | 655.391 | 1594.407 | 1582.188 | 658.278 | 4884.46 | 4789.856 |
| 655.68 | 517.7371 | 511.9006 | 655.68 | 844.6129 | 823.3152 | 655.68 | 1619.845 | 1586.616 | 658.567 | 4827.974 | 4800.596 |
| 655.968 | 495.738 | 513.5791 | 655.968 | 829.9674 | 825.8505 | 655.968 | 1650.257 | 1591.05 | 658.856 | 5025.78 | 4811.365 |
| 656.257 | 517.5229 | 515.2605 | 656.257 | 865.7963 | 828.3946 | 656.257 | 1657.094 | 1595.493 | 659.144 | 5007.705 | 4822.105 |
| 656.546 | 531.1032 | 516.9477 | 656.546 | 839.7888 | 830.9387 | 656.546 | 1633.481 | 1599.938 | 659.433 | 4839.271 | 4832.874 |
| 656.835 | 523.2126 | 518.635 | 656.835 | 847.4094 | 833.4916 | 656.835 | 1650.424 | 1604.393 | 659.722 | 4902.096 | 4843.672 |
| 657.124 | 538.5008 | 520.3281 | 657.124 | 882.4988 | 836.0475 | 657.124 | 1623.804 | 1608.853 | 660.01 | 4916.563 | 4854.442 |
| 657.412 | 543.6389 | 522.0242 | 657.412 | 857.2161 | 838.6062 | 657.412 | 1656.226 | 1613.319 | 660.299 | 4946.259 | 4865.24 |
| 657.701 | 545.4025 | 523.7232 | 657.701 | 876.6418 | 841.1709 | 657.701 | 1667.957 | 1617.791 | 660.588 | 5009.025 | 4876.039 |
| 657.99 | 571.6858 | 525.4252 | 657.99 | 893.9576 | 843.7385 | 657.99 | 1665.231 | 1622.269 | 660.876 | 5014.659 | 4886.867 |
| 658.278 | 545.9277 | 527.13 | 658.278 | 860.5936 | 846.3119 | 658.278 | 1697.175 | 1626.756 | 661.165 | 5056.914 | 4897.695 |
| 658.567 | 535.7777 | 528.8408 | 658.567 | 869.0534 | 848.8883 | 658.567 | 1709.256 | 1631.248 | 661.453 | 4917.648 | 4908.522 |
| 658.856 | 536.9602 | 530.5515 | 658.856 | 872.5453 | 851.4706 | 658.856 | 1661.458 | 1635.747 | 661.742 | 5081.798 | 4919.35 |
| 659.144 | 562.3192 | 532.2682 | 659.144 | 855.5523 | 854.0587 | 659.144 | 1694.904 | 1640.251 | 662.03 | 5045.47 | 4930.178 |
| 659.433 | 543.457 | 533.9877 | 659.433 | 857.357 | 856.6498 | 659.433 | 1676.969 | 1644.761 | 662.319 | 5039.543 | 4941.035 |
| 659.722 | 529.0579 | 535.7102 | 659.722 | 861.6441 | 859.2438 | 659.722 | 1694.654 | 1649.277 | 662.607 | 5079.274 | 4951.893 |
| 660.01 | 563.7072 | 537.4356 | 660.01 | 893.4675 | 861.8436 | 660.01 | 1644.749 | 1653.802 | 662.896 | 5018.738 | 4962.779 |
| 660.299 | 530.2757 | 539.164 | 660.299 | 886.5248 | 864.4494 | 660.299 | 1663.926 | 1658.33 | 663.184 | 5109.997 | 4973.636 |
| 660.588 | 536.4027 | 540.8982 | 660.588 | 901.9831 | 867.058 | 660.588 | 1728.062 | 1662.866 | 663.473 | 5125.315 | 4984.523 |
| 660.876 | 534.5951 | 542.6324 | 660.876 | 872.9855 | 869.6726 | 660.876 | 1697.113 | 1667.409 | 663.761 | 5111.993 | 4995.439 |
| 661.165 | 585.1605 | 544.3725 | 661.165 | 898.4442 | 872.2901 | 661.165 | 1725.826 | 1671.957 | 664.05 | 5085.29 | 5006.325 |
| 661.453 | 566.1897 | 546.1155 | 661.453 | 904.0049 | 874.9105 | 661.453 | 1687.826 | 1676.511 | 664.338 | 5125.638 | 5017.241 |
| 661.742 | 574.0567 | 547.8615 | 661.742 | 906.6312 | 877.5397 | 661.742 | 1729.902 | 1681.071 | 664.627 | 5129.188 | 5028.157 |
| 662.03 | 549.2494 | 549.6104 | 662.03 | 905.2109 | 880.1689 | 662.03 | 1716.222 | 1685.64 | 664.915 | 5172.999 | 5039.073 |
| 662.319 | 573.2146 | 551.3622 | 662.319 | 888.7432 | 882.804 | 662.319 | 1744.111 | 1690.212 | 665.204 | 5175.551 | 5050.018 |
| 662.607 | 538.9879 | 553.1199 | 662.607 | 927.4272 | 885.4449 | 662.607 | 1743.82 | 1694.792 | 665.492 | 5121.471 | 5060.934 |
| 662.896 | 575.3156 | 554.8776 | 662.896 | 902.6316 | 888.0888 | 662.896 | 1747.268 | 1699.379 | 665.78 | 5123.73 | 5071.909 |
| 663.184 | 563.6162 | 556.6412 | 663.184 | 893.7111 | 890.7385 | 663.184 | 1749.328 | 1703.971 | 666.069 | 5198.528 | 5082.854 |
| 663.473 | 566.0635 | 558.4077 | 663.473 | 910.9036 | 893.3912 | 663.473 | 1727.194 | 1708.569 | 666.357 | 5163.902 | 5093.799 |
| 663.761 | 546.7493 | 560.1771 | 663.761 | 909.8942 | 896.0498 | 663.761 | 1720.251 | 1713.173 | 666.645 | 5183.122 | 5104.774 |
| 664.05 | 587.7192 | 561.9495 | 664.05 | 935.1358 | 898.7113 | 664.05 | 1735.305 | 1717.783 | 666.934 | 5128.22 | 5115.749 |
| 664.338 | 593.1332 | 563.7248 | 664.338 | 933.7478 | 901.3786 | 664.338 | 1797.255 | 1722.402 | 667.222 | 5249.41 | 5126.753 |
| 664.627 | 551.9139 | 565.503 | 664.627 | 926.183 | 904.0489 | 664.627 | 1780.691 | 1727.024 | 667.51 | 5219.949 | 5137.757 |
| 664.915 | 552.6533 | 567.2871 | 664.915 | 899.0693 | 906.7251 | 664.915 | 1722.839 | 1731.654 | 667.799 | 5157.593 | 5148.731 |
| 665.204 | 566.2924 | 569.0712 | 665.204 | 956.8004 | 909.4042 | 665.204 | 1745.93 | 1736.29 | 668.087 | 5201.286 | 5159.764 |
| 665.492 | 560.673 | 570.8612 | 665.492 | 893.7199 | 912.0891 | 665.492 | 1760.423 | 1740.93 | 668.375 | 5176.373 | 5170.768 |
| 665.78 | 559.8983 | 572.6541 | 665.78 | 957.1994 | 914.777 | 665.78 | 1786.926 | 1745.578 | 668.663 | 5252.931 | 5181.802 |
| 666.069 | 593.7318 | 574.4499 | 666.069 | 921.4351 | 917.4708 | 666.069 | 1779.215 | 1750.232 | 668.952 | 5247.532 | 5192.835 |
| 666.357 | 575.1806 | 576.2487 | 666.357 | 942.1842 | 920.1675 | 666.357 | 1757.048 | 1754.892 | 669.24 | 5256.511 | 5203.868 |
| 666.645 | 578.0094 | 578.0504 | 666.645 | 939.3818 | 922.8701 | 666.645 | 1800.028 | 1759.56 | 669.528 | 5222.443 | 5214.902 |
| 666.934 | 590.1108 | 579.858 | 666.934 | 942.1196 | 925.5756 | 666.934 | 1813.306 | 1764.232 | 669.816 | 5389.938 | 5225.964 |
| 667.222 | 585.7855 | 581.6656 | 667.222 | 925.3702 | 928.2869 | 667.222 | 1792.469 | 1768.909 | 670.105 | 5278.431 | 5237.027 |
| 667.51 | 587.1617 | 583.4791 | 667.51 | 957.2933 | 931.0012 | 667.51 | 1841.972 | 1773.595 | 670.393 | 5290.081 | 5248.089 |
| 667.799 | 596.5077 | 585.2954 | 667.799 | 919.0436 | 933.7214 | 667.799 | 1811.082 | 1778.284 | 670.681 | 5326.672 | 5259.181 |
| 668.087 | 596.6046 | 587.1148 | 668.087 | 948.989 | 936.4445 | 668.087 | 1781.753 | 1782.982 | 670.969 | 5390.29 | 5270.244 |
| 668.375 | 555.127 | 588.937 | 668.375 | 976.1233 | 939.1735 | 668.375 | 1811.965 | 1787.683 | 671.257 | 5365.7 | 5281.336 |
| 668.663 | 595.0171 | 590.7622 | 668.663 | 979.4773 | 941.9054 | 668.663 | 1895.176 | 1792.393 | 671.545 | 5431.547 | 5292.428 |
| 668.952 | 564.3997 | 592.5903 | 668.952 | 946.8909 | 944.6432 | 668.952 | 1787.859 | 1797.109 | 671.833 | 5345.687 | 5303.549 |
| 669.24 | 582.0852 | 594.4243 | 669.24 | 957.4518 | 947.3839 | 669.24 | 1830.748 | 1801.83 | 672.122 | 5289.464 | 5314.641 |
| 669.528 | 592.6549 | 596.2583 | 669.528 | 965.8324 | 950.1305 | 669.528 | 1834.854 | 1806.557 | 672.41 | 5425.708 | 5325.763 |
| 669.816 | 632.4745 | 598.0982 | 669.816 | 1011.506 | 952.88 | 669.816 | 1894.838 | 1811.29 | 672.698 | 5433.249 | 5336.913 |
| 670.105 | 601.5813 | 599.941 | 670.105 | 935.5525 | 955.6354 | 670.105 | 1846.887 | 1816.029 | 672.986 | 5316.402 | 5348.035 |
| 670.393 | 606.5639 | 601.7867 | 670.393 | 999.8537 | 958.3937 | 670.393 | 1858.971 | 1820.774 | 673.274 | 5366.052 | 5359.185 |
| 670.681 | 601.4756 | 603.6353 | 670.681 | 967.4727 | 961.1579 | 670.681 | 1840.218 | 1825.525 | 673.562 | 5484.131 | 5370.307 |
| 670.969 | 611.3205 | 605.4869 | 670.969 | 985.3901 | 963.925 | 670.969 | 1859.843 | 1830.282 | 673.85 | 5481.901 | 5381.487 |
| 671.257 | 608.7852 | 607.3444 | 671.257 | 977.7108 | 966.6951 | 671.257 | 1835.69 | 1835.047 | 674.138 | 5418.753 | 5392.637 |
| 671.545 | 603.5444 | 609.2019 | 671.545 | 958.3673 | 969.474 | 671.545 | 1886.417 | 1839.816 | 674.426 | 5439.675 | 5403.817 |
| 671.833 | 608.524 | 611.0652 | 671.833 | 972.2822 | 972.2528 | 671.833 | 1843.751 | 1844.59 | 674.714 | 5386.446 | 5414.968 |
| 672.122 | 653.6109 | 612.9315 | 672.122 | 961.8651 | 975.0376 | 672.122 | 1861.339 | 1849.373 | 675.002 | 5431.195 | 5426.148 |
| 672.41 | 603.7204 | 614.8007 | 672.41 | 993.2542 | 977.8282 | 672.41 | 1901.455 | 1854.159 | 675.29 | 5420.338 | 5437.357 |
| 672.698 | 639.2647 | 616.6728 | 672.698 | 964.6029 | 980.6217 | 672.698 | 1910.625 | 1858.954 | 675.578 | 5456.519 | 5448.537 |
| 672.986 | 610.4989 | 618.5479 | 672.986 | 1005.567 | 983.4211 | 672.986 | 1883.438 | 1863.751 | 675.866 | 5444.84 | 5459.747 |
| 673.274 | 617.4358 | 620.4259 | 673.274 | 976.584 | 986.2234 | 673.274 | 1895.058 | 1868.558 | 676.154 | 5489.971 | 5470.956 |
| 673.562 | 630.2238 | 622.3098 | 673.562 | 983.4563 | 989.0287 | 673.562 | 1933.455 | 1873.367 | 676.442 | 5579.088 | 5482.165 |
| 673.85 | 628.2578 | 624.1936 | 673.85 | 1016.23 | 991.8398 | 673.85 | 1914.185 | 1878.186 | 676.73 | 5450.503 | 5493.404 |
| 674.138 | 634.5872 | 626.0834 | 674.138 | 1026.92 | 994.6569 | 674.138 | 1924.235 | 1883.007 | 677.018 | 5514.091 | 5504.613 |
| 674.426 | 641.9584 | 627.9761 | 674.426 | 1025.582 | 997.4768 | 674.426 | 1912.844 | 1887.837 | 677.305 | 5528.851 | 5515.852 |
| 674.714 | 629.6927 | 629.8717 | 674.714 | 977.7607 | 1000.3 | 674.714 | 1882.009 | 1892.673 | 677.593 | 5525.037 | 5527.091 |
| 675.002 | 618.9118 | 631.7702 | 675.002 | 1003.8 | 1003.128 | 675.002 | 1888.87 | 1897.511 | 677.881 | 5575.273 | 5538.359 |
| 675.29 | 639.787 | 633.6717 | 675.29 | 1001.99 | 1005.96 | 675.29 | 1899.266 | 1902.359 | 678.169 | 5609.283 | 5549.598 |
| 675.578 | 648.9217 | 635.5791 | 675.578 | 1025.673 | 1008.798 | 675.578 | 1907.579 | 1907.213 | 678.457 | 5597.311 | 5560.866 |
| 675.866 | 642.8475 | 637.4864 | 675.866 | 1009.678 | 1011.638 | 675.866 | 1881.029 | 1912.069 | 678.745 | 5489.736 | 5572.134 |
| 676.154 | 657.1615 | 639.3996 | 676.154 | 998.7034 | 1014.485 | 676.154 | 1923.525 | 1916.934 | 679.032 | 5654.12 | 5583.402 |
| 676.442 | 669.9172 | 641.3129 | 676.442 | 1012.023 | 1017.334 | 676.442 | 1948.731 | 1921.805 | 679.32 | 5561.482 | 5594.699 |
| 676.73 | 644.2913 | 643.2319 | 676.73 | 1015.247 | 1020.189 | 676.73 | 1939.913 | 1926.679 | 679.608 | 5663.129 | 5605.996 |
| 677.018 | 657.7014 | 645.154 | 677.018 | 1029.3 | 1023.047 | 677.018 | 1956.989 | 1931.562 | 679.896 | 5555.085 | 5617.294 |
| 677.305 | 661.6217 | 647.0789 | 677.305 | 1029.118 | 1025.911 | 677.305 | 1921.394 | 1936.451 | 680.184 | 5587.891 | 5628.591 |
| 677.593 | 633.7979 | 649.0068 | 677.593 | 995.232 | 1028.778 | 677.593 | 1964.521 | 1941.342 | 680.471 | 5639.419 | 5639.889 |
| 677.881 | 663.7433 | 650.9406 | 677.881 | 1013.592 | 1031.651 | 677.881 | 1949.145 | 1946.243 | 680.759 | 5676.216 | 5651.215 |
| 678.169 | 653.0621 | 652.8743 | 678.169 | 1048.896 | 1034.526 | 678.169 | 1925.908 | 1951.149 | 681.047 | 5628.797 | 5662.513 |
| 678.457 | 643.9245 | 654.814 | 678.457 | 1014.669 | 1037.405 | 678.457 | 1957.411 | 1956.058 | 681.334 | 5624.16 | 5673.839 |
| 678.745 | 646.6476 | 656.7536 | 678.745 | 1059.956 | 1040.289 | 678.745 | 1971.349 | 1960.976 | 681.622 | 5666.151 | 5685.195 |
| 679.032 | 661.2989 | 658.6991 | 679.032 | 1059.489 | 1043.18 | 679.032 | 1961.936 | 1965.897 | 681.91 | 5741.272 | 5696.522 |
| 679.32 | 677.1417 | 660.6475 | 679.32 | 1068.172 | 1046.073 | 679.32 | 1959.301 | 1970.827 | 682.197 | 5808.469 | 5707.878 |
| 679.608 | 649.1095 | 662.5989 | 679.608 | 1065.957 | 1048.969 | 679.608 | 2021.838 | 1975.76 | 682.485 | 5824.637 | 5719.205 |
| 679.896 | 685.1262 | 664.5561 | 679.896 | 1066.532 | 1051.872 | 679.896 | 1942.95 | 1980.701 | 682.773 | 5821.146 | 5730.561 |
| 680.184 | 644.0242 | 666.5134 | 680.184 | 1050.974 | 1054.777 | 680.184 | 1997.289 | 1985.646 | 683.06 | 5809.555 | 5741.947 |
| 680.471 | 658.2765 | 668.4735 | 680.471 | 1047.147 | 1057.687 | 680.471 | 2003.22 | 1990.596 | 683.636 | 5813.663 | 5764.688 |
| 680.759 | 701.7436 | 670.4396 | 680.759 | 1069.73 | 1060.601 | 680.759 | 1992.098 | 1995.555 | 683.923 | 5887.991 | 5776.073 |
| 681.047 | 650.204 | 672.4056 | 681.047 | 1086.072 | 1063.518 | 681.047 | 2021.489 | 2000.517 | 684.211 | 5853.864 | 5787.459 |
| 681.334 | 655.8146 | 674.3775 | 681.334 | 1083.536 | 1066.441 | 681.334 | 2028.121 | 2005.485 | 684.498 | 5777.599 | 5798.844 |
| 681.622 | 667.0151 | 676.3523 | 681.622 | 1068.647 | 1069.369 | 681.622 | 2000.505 | 2010.459 | 684.786 | 5858.207 | 5810.23 |
| 681.91 | 686.9924 | 678.3301 | 681.91 | 1094.919 | 1072.301 | 681.91 | 1983.301 | 2015.439 | 685.073 | 5847.731 | 5821.644 |
| 682.197 | 687.1186 | 680.3108 | 682.197 | 1074.425 | 1075.235 | 682.197 | 1986.379 | 2020.424 | 685.361 | 5852.954 | 5833.059 |
| 682.485 | 690.8922 | 682.2974 | 682.485 | 1062.89 | 1078.175 | 682.485 | 2014.723 | 2025.415 | 685.648 | 5867.861 | 5844.474 |
| 682.773 | 730.4947 | 684.284 | 682.773 | 1088.598 | 1081.119 | 682.773 | 2031.895 | 2030.413 | 685.936 | 5867.744 | 5855.889 |
| 683.06 | 691.4879 | 686.2735 | 683.06 | 1114.503 | 1084.068 | 683.348 | 2076.13 | 2040.425 | 686.223 | 5859.263 | 5867.333 |
| 683.348 | 688.9673 | 688.2689 | 683.348 | 1071.406 | 1087.02 | 683.636 | 2111.032 | 2045.437 | 686.511 | 5880.978 | 5878.777 |
| 683.636 | 714.561 | 690.2672 | 683.636 | 1105.794 | 1089.977 | 683.923 | 2048.131 | 2050.458 | 686.798 | 5860.642 | 5890.192 |
| 683.923 | 681.021 | 692.2685 | 683.923 | 1131.285 | 1092.938 | 684.211 | 2022.942 | 2055.481 | 687.086 | 5969.684 | 5901.636 |
| 684.211 | 706.2508 | 694.2697 | 684.211 | 1143.947 | 1095.902 | 684.498 | 2050.273 | 2060.514 | 687.661 | 5995.624 | 5924.553 |
| 684.498 | 679.8501 | 696.2798 | 684.498 | 1106.947 | 1098.872 | 684.786 | 2015.092 | 2065.549 | 688.236 | 6051.671 | 5947.5 |
| 684.786 | 705.1915 | 698.2898 | 684.786 | 1074.921 | 1101.847 | 685.073 | 2085.996 | 2070.59 | 688.523 | 6025.408 | 5958.974 |
| 685.073 | 703.2724 | 700.3028 | 685.073 | 1081.518 | 1104.822 | 685.361 | 2037.854 | 2075.637 | 688.81 | 6032.979 | 5970.447 |
| 685.361 | 701.365 | 702.3187 | 685.361 | 1107.695 | 1107.807 | 685.648 | 2066.541 | 2080.69 | 689.098 | 6039.082 | 5981.921 |
| 685.648 | 722.0495 | 704.3405 | 685.648 | 1101.733 | 1110.791 | 685.936 | 2070.015 | 2085.749 | 689.385 | 5995.213 | 5993.423 |
| 685.936 | 708.9123 | 706.3652 | 685.936 | 1112.205 | 1113.781 | 686.223 | 2093.74 | 2090.814 | 689.672 | 6145.982 | 6004.926 |
| 686.223 | 674.9644 | 708.39 | 686.223 | 1080.813 | 1116.777 | 686.511 | 2139.909 | 2095.885 | 689.96 | 6072.006 | 6016.4 |
| 686.511 | 717.9649 | 710.4206 | 686.511 | 1129.325 | 1119.776 | 686.798 | 2109.183 | 2100.958 | 690.247 | 6075.997 | 6027.932 |
| 686.798 | 748.6351 | 712.4541 | 686.798 | 1086.826 | 1122.778 | 687.086 | 2153.71 | 2106.041 | 690.534 | 6097.858 | 6039.434 |
| 687.086 | 720.1979 | 714.4906 | 687.086 | 1108.259 | 1125.786 | 687.373 | 2090.855 | 2111.126 | 690.822 | 6131.662 | 6050.937 |
| 687.373 | 710.086 | 716.5299 | 687.373 | 1115.096 | 1128.796 | 687.661 | 2165.518 | 2116.22 | 691.109 | 6192.903 | 6062.469 |
| 687.661 | 717.9561 | 718.5723 | 687.661 | 1147.11 | 1131.813 | 687.948 | 2094.098 | 2121.317 | 691.396 | 6172.832 | 6074.002 |
| 687.948 | 720.0923 | 720.6205 | 687.948 | 1112.804 | 1134.832 | 688.236 | 2166.768 | 2126.42 | 691.683 | 6079.958 | 6085.534 |
| 688.236 | 734.5588 | 722.6687 | 688.236 | 1111.161 | 1137.855 | 688.523 | 2164.165 | 2131.529 | 691.971 | 6236.889 | 6097.066 |
| 688.523 | 713.2405 | 724.7228 | 688.523 | 1132.708 | 1140.883 | 688.81 | 2149.886 | 2136.64 | 692.258 | 6271.691 | 6108.627 |
| 688.81 | 732.7776 | 726.7798 | 688.81 | 1158.786 | 1143.914 | 689.098 | 2163.285 | 2141.761 | 692.545 | 6080.105 | 6120.159 |
| 689.098 | 705.7842 | 728.8368 | 689.098 | 1160.232 | 1146.951 | 689.385 | 2138.081 | 2146.884 | 692.832 | 6154.521 | 6131.721 |
| 689.385 | 744.6707 | 730.8996 | 689.385 | 1168.877 | 1149.991 | 689.672 | 2145.948 | 2152.017 | 693.406 | 6268.346 | 6154.844 |
| 689.672 | 744.7617 | 732.9654 | 689.672 | 1196.639 | 1153.037 | 689.96 | 2125.59 | 2157.152 | 693.694 | 6296.604 | 6166.405 |
| 689.96 | 730.4272 | 735.0342 | 689.96 | 1150.373 | 1156.086 | 690.247 | 2185.319 | 2162.293 | 693.981 | 6234.982 | 6177.996 |
| 690.247 | 728.8837 | 737.1088 | 690.247 | 1161.198 | 1159.138 | 690.534 | 2201.273 | 2167.44 | 694.268 | 6177.116 | 6189.558 |
| 690.534 | 749.4508 | 739.1834 | 690.534 | 1152.119 | 1162.196 | 690.822 | 2215.065 | 2172.59 | 694.555 | 6249.977 | 6201.148 |
| 690.822 | 752.2033 | 741.2609 | 690.822 | 1169.265 | 1165.256 | 691.109 | 2205.293 | 2177.748 | 694.842 | 6255.699 | 6212.739 |
| 691.109 | 728.8602 | 743.3444 | 691.109 | 1178.074 | 1168.323 | 691.396 | 2195.847 | 2182.91 | 695.129 | 6328.207 | 6224.33 |
| 691.396 | 721.4773 | 745.4278 | 691.396 | 1170.752 | 1171.392 | 691.683 | 2182.842 | 2188.08 | 695.416 | 6275.564 | 6235.921 |
| 691.683 | 742.7193 | 747.5171 | 691.683 | 1155.303 | 1174.464 | 691.971 | 2220.934 | 2193.253 | 695.703 | 6207.898 | 6247.541 |
| 691.971 | 736.0084 | 749.6093 | 691.971 | 1189.529 | 1177.542 | 692.258 | 2183.814 | 2198.43 | 695.99 | 6321.928 | 6259.132 |
| 692.258 | 738.4293 | 751.7044 | 692.258 | 1168.936 | 1180.624 | 692.545 | 2207.271 | 2203.615 | 696.277 | 6369.788 | 6270.752 |
| 692.545 | 726.8091 | 753.8025 | 692.545 | 1205.868 | 1183.711 | 692.832 | 2249.955 | 2208.806 | 696.564 | 6287.654 | 6282.372 |
| 692.832 | 757.6583 | 755.9035 | 692.832 | 1200.155 | 1186.8 | 693.119 | 2260.988 | 2214 | 696.851 | 6277.354 | 6293.992 |
| 693.119 | 753.2479 | 758.0075 | 693.119 | 1182.378 | 1189.893 | 693.406 | 2248.291 | 2219.199 | 697.138 | 6342.116 | 6305.613 |
| 693.406 | 750.918 | 760.1144 | 693.406 | 1189.347 | 1192.992 | 693.694 | 2221.383 | 2224.405 | 697.425 | 6339.299 | 6317.262 |
| 693.694 | 751.3728 | 762.2271 | 693.694 | 1191.023 | 1196.094 | 693.981 | 2240.967 | 2229.616 | 697.712 | 6417.823 | 6328.882 |
| 693.981 | 743.7464 | 764.3399 | 694.268 | 1189.133 | 1202.309 | 694.268 | 2240.177 | 2234.831 | 697.999 | 6331.787 | 6340.532 |
| 694.268 | 764.1257 | 766.4585 | 694.555 | 1269.075 | 1205.425 | 694.555 | 2277.514 | 2240.054 | 698.286 | 6392.529 | 6352.181 |
| 694.555 | 779.1644 | 768.5771 | 694.842 | 1205.994 | 1208.541 | 694.842 | 2261.584 | 2245.28 | 698.573 | 6463.717 | 6363.831 |
| 694.842 | 776.5704 | 770.7016 | 695.129 | 1197.259 | 1211.663 | 695.129 | 2258.01 | 2250.512 | 698.86 | 6439.567 | 6375.48 |
| 695.129 | 773.9999 | 772.8291 | 695.416 | 1201.461 | 1214.792 | 695.416 | 2222.894 | 2255.75 | 699.147 | 6418.381 | 6387.13 |
| 695.416 | 782.9468 | 774.9594 | 695.703 | 1210.989 | 1217.923 | 695.703 | 2221.884 | 2260.991 | 699.434 | 6490.42 | 6398.809 |
| 695.703 | 787.991 | 777.0927 | 695.99 | 1189.262 | 1221.056 | 695.99 | 2284.425 | 2266.24 | 699.721 | 6525.281 | 6410.488 |
| 695.99 | 781.8875 | 779.229 | 696.277 | 1217.136 | 1224.196 | 696.277 | 2233.161 | 2271.493 | 700.008 | 6508.848 | 6422.137 |
| 696.277 | 780.6962 | 781.3681 | 696.564 | 1209.298 | 1227.339 | 696.564 | 2257.54 | 2276.751 | 700.295 | 6495.79 | 6433.816 |
| 696.564 | 769.0408 | 783.5102 | 696.851 | 1200.146 | 1230.485 | 696.851 | 2258.86 | 2282.013 | 700.581 | 6588.575 | 6445.495 |
| 696.851 | 748.3886 | 785.6582 | 697.138 | 1190.336 | 1233.636 | 697.138 | 2270.73 | 2287.283 | 700.868 | 6571.145 | 6457.203 |
| 697.138 | 797.1639 | 787.8062 | 697.425 | 1226.843 | 1236.791 | 697.425 | 2289.889 | 2292.556 | 701.155 | 6513.396 | 6468.882 |
| 697.425 | 770.4933 | 789.96 | 697.712 | 1258.44 | 1239.948 | 697.712 | 2371.327 | 2297.835 | 701.442 | 6578.892 | 6480.59 |
| 697.712 | 808.6256 | 792.1138 | 697.999 | 1199.606 | 1243.111 | 697.999 | 2318.035 | 2303.117 | 701.729 | 6522.317 | 6492.269 |
| 697.999 | 800.0778 | 794.2736 | 698.286 | 1233.739 | 1246.28 | 698.286 | 2325.53 | 2308.408 | 702.015 | 6563.398 | 6503.977 |
| 698.286 | 770.2351 | 796.4362 | 698.573 | 1247.542 | 1249.45 | 698.573 | 2313.766 | 2313.701 | 702.302 | 6531.384 | 6515.685 |
| 698.573 | 798.2086 | 798.6018 | 698.86 | 1258.153 | 1252.625 | 698.86 | 2330.559 | 2319.001 | 702.589 | 6611.845 | 6527.393 |
| 698.86 | 774.7335 | 800.7703 | 699.147 | 1259.403 | 1255.802 | 699.147 | 2361.749 | 2324.306 | 702.876 | 6747.091 | 6539.102 |
| 699.147 | 832.8754 | 802.9417 | 699.434 | 1284.997 | 1258.986 | 699.434 | 2328.743 | 2329.614 | 703.162 | 6598.875 | 6550.839 |
| 699.434 | 790.585 | 805.1161 | 699.721 | 1263.708 | 1262.173 | 699.721 | 2343.394 | 2334.929 | 704.596 | 6664.605 | 6609.498 |
| 699.721 | 799.6582 | 807.2934 | 700.008 | 1282.602 | 1265.366 | 700.008 | 2389.089 | 2340.249 | 704.882 | 6721.063 | 6621.235 |
| 700.008 | 822.2148 | 809.4737 | 700.295 | 1234.654 | 1268.558 | 700.295 | 2312.149 | 2345.575 | 705.456 | 6723.058 | 6644.739 |
| 700.295 | 791.5446 | 811.6598 | 700.581 | 1263.74 | 1271.757 | 700.581 | 2298.005 | 2350.903 | 705.742 | 6735.06 | 6656.477 |
| 700.581 | 808.2207 | 813.8459 | 700.868 | 1279.327 | 1274.961 | 700.868 | 2372.659 | 2356.241 | 706.029 | 6809.124 | 6668.244 |
| 700.868 | 822.6373 | 816.035 | 701.155 | 1261.891 | 1278.168 | 701.155 | 2378.266 | 2361.579 | 706.315 | 6765.196 | 6680.011 |
| 701.155 | 834.9588 | 818.2299 | 701.442 | 1280.756 | 1281.379 | 701.442 | 2413.796 | 2366.925 | 706.602 | 6788.26 | 6691.778 |
| 701.442 | 818.5351 | 820.4277 | 701.729 | 1269.21 | 1284.592 | 701.729 | 2355.378 | 2372.274 | 706.888 | 6805.28 | 6703.545 |
| 701.729 | 814.0278 | 822.6256 | 702.015 | 1263.335 | 1287.811 | 702.015 | 2396.865 | 2377.63 | 707.175 | 6849.912 | 6715.311 |
| 702.015 | 817.6078 | 824.8293 | 702.302 | 1303.422 | 1291.033 | 702.302 | 2442.506 | 2382.991 | 707.461 | 6808.42 | 6727.108 |
| 702.302 | 825.3898 | 827.036 | 702.589 | 1276.05 | 1294.261 | 702.589 | 2363.592 | 2388.355 | 707.748 | 6847.564 | 6738.875 |
| 702.589 | 831.282 | 829.2456 | 702.876 | 1307.744 | 1297.491 | 702.876 | 2418.333 | 2393.725 | 708.034 | 6774.058 | 6750.671 |
| 702.876 | 842.8112 | 831.4581 | 703.162 | 1315.112 | 1300.725 | 703.162 | 2396.278 | 2399.101 | 708.321 | 6800.849 | 6762.467 |
| 703.162 | 810.8206 | 833.6735 | 703.449 | 1296.79 | 1303.965 | 703.449 | 2391.063 | 2404.482 | 708.607 | 6829.048 | 6774.234 |
| 703.449 | 832.2386 | 835.8919 | 703.736 | 1298.31 | 1307.207 | 703.736 | 2431.361 | 2409.867 | 708.894 | 6885.829 | 6786.03 |
| 703.736 | 845.596 | 838.1133 | 704.022 | 1292.403 | 1310.452 | 704.022 | 2406.331 | 2415.257 | 709.18 | 6880.899 | 6797.856 |
| 704.022 | 859.9862 | 840.3405 | 704.309 | 1289.489 | 1313.704 | 704.309 | 2440.514 | 2420.651 | 709.467 | 6949.505 | 6809.652 |
| 704.309 | 839.7301 | 842.5677 | 704.596 | 1325.879 | 1316.955 | 704.596 | 2524.466 | 2426.05 | 709.753 | 6941.905 | 6821.448 |
| 704.596 | 846.7022 | 844.7978 | 704.882 | 1321.926 | 1320.215 | 705.456 | 2415.072 | 2442.28 | 710.04 | 6911.534 | 6833.274 |
| 704.882 | 854.7219 | 847.0338 | 705.169 | 1366.669 | 1323.475 | 705.742 | 2484.57 | 2447.7 | 710.326 | 6915.495 | 6845.07 |
| 705.169 | 842.2918 | 849.2698 | 705.456 | 1343.027 | 1326.741 | 706.029 | 2456.456 | 2453.123 | 710.612 | 7028.968 | 6856.896 |
| 705.456 | 860.5232 | 851.5117 | 705.742 | 1305.796 | 1330.01 | 706.315 | 2456.547 | 2458.554 | 710.899 | 6943.607 | 6868.721 |
| 705.742 | 884.3445 | 853.7565 | 706.029 | 1365.959 | 1333.285 | 706.602 | 2492.417 | 2463.989 | 711.185 | 7065.032 | 6880.547 |
| 706.029 | 880.7734 | 856.0013 | 706.315 | 1344.471 | 1336.563 | 706.888 | 2500.082 | 2469.426 | 711.471 | 6958.22 | 6892.372 |
| 706.315 | 865.1566 | 858.252 | 706.602 | 1352.758 | 1339.843 | 708.034 | 2531.485 | 2491.229 | 711.758 | 7095.637 | 6904.198 |
| 706.602 | 868.7717 | 860.5056 | 706.888 | 1328.2 | 1343.13 | 708.321 | 2481.46 | 2496.692 | 712.044 | 6936.212 | 6916.023 |
| 706.888 | 872.6158 | 862.7621 | 707.175 | 1372.84 | 1346.416 | 708.607 | 2513.418 | 2502.162 | 712.33 | 6944.81 | 6927.849 |
| 707.175 | 887.455 | 865.0216 | 707.461 | 1365.848 | 1349.712 | 708.894 | 2592.588 | 2507.632 | 712.617 | 7085.367 | 6939.704 |
| 707.461 | 905.1434 | 867.284 | 707.748 | 1374.041 | 1353.007 | 709.18 | 2534.549 | 2513.11 | 712.903 | 7029.32 | 6951.529 |
| 707.748 | 891.029 | 869.5493 | 708.034 | 1377.5 | 1356.308 | 709.467 | 2562.455 | 2518.592 | 713.189 | 7118.085 | 6963.384 |
| 708.034 | 894.1131 | 871.8176 | 708.321 | 1368.556 | 1359.612 | 709.753 | 2545.934 | 2524.079 | 713.475 | 7068.347 | 6975.239 |
| 708.321 | 853.1637 | 874.0888 | 708.607 | 1373.389 | 1362.919 | 710.04 | 2536.465 | 2529.569 | 713.762 | 7191.826 | 6987.094 |
| 708.607 | 843.9292 | 876.363 | 708.894 | 1367.124 | 1366.232 | 710.326 | 2619.561 | 2535.065 | 714.048 | 7102.445 | 6998.949 |
| 708.894 | 886.8153 | 878.643 | 709.18 | 1364.26 | 1369.548 | 710.612 | 2576.038 | 2540.567 | 714.334 | 7122.81 | 7010.804 |
| 709.18 | 877.5514 | 880.923 | 709.467 | 1394.332 | 1372.867 | 710.899 | 2602.706 | 2546.072 | 714.62 | 7123.925 | 7022.659 |
| 709.467 | 900.8475 | 883.206 | 709.753 | 1402.616 | 1376.192 | 711.185 | 2566.17 | 2551.583 | 714.906 | 7211.369 | 7034.514 |
| 709.753 | 932.4215 | 885.4948 | 710.04 | 1405.21 | 1379.519 | 711.471 | 2559.832 | 2557.1 | 715.192 | 7162.248 | 7046.398 |
| 710.04 | 876.0314 | 887.7836 | 710.326 | 1414.242 | 1382.85 | 711.758 | 2599.566 | 2562.619 | 715.479 | 7149.982 | 7058.253 |
| 710.326 | 917.1157 | 890.0783 | 710.612 | 1376.28 | 1386.186 | 712.044 | 2571.968 | 2568.142 | 715.765 | 7183.023 | 7070.137 |
| 710.612 | 910.6425 | 892.373 | 710.899 | 1418.303 | 1389.522 | 712.33 | 2605.67 | 2573.67 | 716.051 | 7222.256 | 7081.992 |
| 710.899 | 905.9944 | 894.6735 | 711.185 | 1395.215 | 1392.868 | 712.617 | 2653.301 | 2579.204 | 716.337 | 7143.497 | 7093.877 |
| 711.185 | 933.1404 | 896.977 | 711.471 | 1398.578 | 1396.213 | 712.903 | 2631.475 | 2584.742 | 716.623 | 7162.6 | 7105.761 |
| 711.471 | 910.5339 | 899.2805 | 711.758 | 1438.996 | 1399.564 | 713.189 | 2629.394 | 2590.285 | 716.909 | 7226.129 | 7117.645 |
| 711.758 | 926.9665 | 901.5899 | 712.044 | 1399.206 | 1402.918 | 713.475 | 2717.027 | 2595.834 | 717.195 | 7223.84 | 7129.529 |
| 712.044 | 932.9702 | 903.9022 | 712.33 | 1413.878 | 1406.275 | 713.762 | 2657.007 | 2601.385 | 717.481 | 7181.703 | 7141.414 |
| 712.33 | 934.3054 | 906.2174 | 712.617 | 1432.141 | 1409.635 | 714.048 | 2696.465 | 2606.94 | 717.767 | 7229.915 | 7153.327 |
| 712.617 | 918.8265 | 908.5356 | 712.903 | 1424.175 | 1413 | 714.334 | 2683.272 | 2612.501 | 718.053 | 7275.31 | 7165.211 |
| 712.903 | 917.236 | 910.8567 | 713.189 | 1422.279 | 1416.369 | 714.62 | 2645.181 | 2618.067 | 718.339 | 7324.343 | 7177.096 |
| 713.189 | 929.4754 | 913.1807 | 713.475 | 1463.073 | 1419.744 | 714.906 | 2665.188 | 2623.637 | 718.625 | 7303.597 | 7189.009 |
| 713.475 | 933.4162 | 915.5077 | 713.762 | 1515.226 | 1423.118 | 715.192 | 2684.731 | 2629.212 | 718.911 | 7332.178 | 7200.894 |
| 713.762 | 954.4411 | 917.8376 | 714.048 | 1424.547 | 1426.499 | 715.479 | 2615.027 | 2634.79 | 719.197 | 7386.171 | 7212.807 |
| 714.048 | 926.8344 | 920.1704 | 714.334 | 1451.247 | 1429.882 | 715.765 | 2665.61 | 2640.375 | 719.483 | 7268.737 | 7224.721 |
| 714.334 | 922.9317 | 922.5062 | 714.62 | 1461.967 | 1433.271 | 716.051 | 2677.14 | 2645.962 | 719.769 | 7405.244 | 7236.634 |
| 714.62 | 932.8294 | 924.8449 | 714.906 | 1475.444 | 1436.663 | 716.337 | 2651.666 | 2651.555 | 720.055 | 7355.301 | 7248.548 |
| 715.192 | 939.9511 | 929.5311 | 715.192 | 1469.126 | 1440.058 | 716.623 | 2727.417 | 2657.153 | 720.341 | 7354.479 | 7260.462 |
| 715.479 | 992.4972 | 931.8786 | 715.479 | 1415.697 | 1443.456 | 716.909 | 2643.79 | 2662.755 | 720.627 | 7406.653 | 7272.375 |
| 716.337 | 985.3461 | 938.9417 | 715.765 | 1484.385 | 1446.857 | 717.195 | 2719.318 | 2668.36 | 720.913 | 7363.4 | 7284.289 |
| 716.623 | 942.4541 | 941.3009 | 716.051 | 1508.785 | 1450.264 | 717.481 | 2713.538 | 2673.97 | 721.199 | 7331.885 | 7296.202 |
| 716.909 | 970.859 | 943.666 | 716.337 | 1465.209 | 1453.674 | 717.767 | 2648.873 | 2679.584 | 721.484 | 7416.6 | 7308.145 |
| 717.195 | 956.7035 | 946.0312 | 716.623 | 1482.771 | 1457.087 | 718.053 | 2720.533 | 2685.203 | 721.77 | 7341.744 | 7320.059 |
| 717.481 | 976.138 | 948.4021 | 716.909 | 1470.949 | 1460.505 | 718.339 | 2750.757 | 2690.828 | 722.056 | 7412.962 | 7332.002 |
| 717.767 | 945.77 | 950.7731 | 717.195 | 1525.411 | 1463.927 | 718.625 | 2700.799 | 2696.457 | 722.342 | 7366.452 | 7343.915 |
| 718.053 | 962.3698 | 953.147 | 717.481 | 1503.752 | 1467.351 | 718.911 | 2728.814 | 2702.088 | 722.628 | 7543.307 | 7355.858 |
| 718.339 | 950.3388 | 955.5268 | 717.767 | 1484.153 | 1470.778 | 719.197 | 2770.969 | 2707.725 | 722.914 | 7403.894 | 7367.801 |
| 718.625 | 944.0299 | 957.9066 | 718.053 | 1509.248 | 1474.212 | 719.483 | 2760.215 | 2713.367 | 723.199 | 7378.365 | 7379.715 |
| 718.911 | 916.0858 | 960.2923 | 718.339 | 1452.227 | 1477.648 | 719.769 | 2748.152 | 2719.013 | 723.485 | 7445.269 | 7391.658 |
| 719.197 | 974.8703 | 962.6779 | 718.625 | 1501.675 | 1481.087 | 720.055 | 2736.179 | 2724.662 | 723.771 | 7352.102 | 7403.601 |
| 719.483 | 955.7176 | 965.0695 | 718.911 | 1502.338 | 1484.529 | 720.341 | 2722.197 | 2730.316 | 724.057 | 7488.815 | 7415.544 |
| 719.769 | 977.4144 | 967.461 | 719.197 | 1502.787 | 1487.974 | 720.627 | 2782.581 | 2735.974 | 724.342 | 7467.6 | 7427.487 |
| 720.055 | 969.9141 | 969.8584 | 719.483 | 1521.587 | 1491.425 | 720.913 | 2793.432 | 2741.637 | 724.628 | 7494.743 | 7439.43 |
| 720.341 | 981.4756 | 972.2587 | 719.769 | 1529.865 | 1494.879 | 721.199 | 2732.241 | 2747.304 | 724.914 | 7570.098 | 7451.402 |
| 720.627 | 977.4614 | 974.659 | 720.055 | 1504.946 | 1498.335 | 721.484 | 2795.125 | 2752.976 | 725.199 | 7485.852 | 7463.345 |
| 720.913 | 989.9589 | 977.0652 | 720.341 | 1528.146 | 1501.798 | 721.77 | 2734.463 | 2758.651 | 725.485 | 7465.37 | 7475.288 |
| 721.199 | 985.2434 | 979.4714 | 720.627 | 1495.768 | 1505.263 | 722.056 | 2835.59 | 2764.332 | 725.771 | 7471.268 | 7487.26 |
| 721.484 | 1001.057 | 981.8835 | 720.913 | 1521.863 | 1508.732 | 722.342 | 2786.715 | 2770.016 | 726.056 | 7618.222 | 7499.203 |
| 721.77 | 977.3088 | 984.2985 | 721.199 | 1529.164 | 1512.203 | 722.628 | 2806.939 | 2775.703 | 726.342 | 7649.297 | 7511.175 |
| 723.199 | 1006.521 | 996.4116 | 721.484 | 1539.742 | 1515.678 | 722.914 | 2864.201 | 2781.395 | 726.628 | 7551.171 | 7523.118 |
| 723.485 | 989.6713 | 998.8413 | 721.77 | 1489.679 | 1519.158 | 723.199 | 2837.556 | 2787.094 | 726.913 | 7698.125 | 7535.091 |
| 723.771 | 978.5265 | 1001.277 | 722.056 | 1535.379 | 1522.641 | 723.485 | 2801.475 | 2792.792 | 727.199 | 7631.691 | 7547.063 |
| 724.057 | 1000.212 | 1003.712 | 722.342 | 1532.395 | 1526.127 | 723.771 | 2799.718 | 2798.5 | 727.484 | 7557.128 | 7559.006 |
| 724.342 | 1003.363 | 1006.154 | 722.628 | 1533.841 | 1529.616 | 724.057 | 2792.458 | 2804.207 | 727.77 | 7550.584 | 7570.978 |
| 724.628 | 1021.099 | 1008.598 | 722.914 | 1571.971 | 1533.108 | 724.342 | 2846.078 | 2809.92 | 728.055 | 7572.445 | 7582.951 |
| 724.914 | 1029.523 | 1011.042 | 723.199 | 1547.081 | 1536.606 | 724.628 | 2806.936 | 2815.64 | 728.341 | 7647.742 | 7594.923 |
| 725.199 | 1028.197 | 1013.493 | 723.485 | 1526.661 | 1540.106 | 724.914 | 2886.784 | 2821.362 | 728.626 | 7619.953 | 7606.895 |
| 725.485 | 1010.602 | 1015.943 | 723.771 | 1571.375 | 1543.61 | 725.199 | 2858.15 | 2827.087 | 728.912 | 7684.862 | 7618.867 |
| 725.771 | 1024.608 | 1018.399 | 724.057 | 1522.744 | 1547.117 | 725.485 | 2786.087 | 2832.817 | 729.197 | 7682.162 | 7630.84 |
| 726.056 | 1057.476 | 1020.858 | 724.342 | 1544.185 | 1550.629 | 725.771 | 2864.039 | 2838.551 | 729.483 | 7698.477 | 7642.812 |
| 726.342 | 1039.882 | 1023.317 | 724.628 | 1597.858 | 1554.144 | 726.056 | 2914.983 | 2844.291 | 729.768 | 7640.993 | 7654.814 |
| 726.628 | 1009.182 | 1025.782 | 724.914 | 1554.1 | 1557.663 | 726.342 | 2885.361 | 2850.031 | 730.054 | 7633.51 | 7666.786 |
| 726.913 | 1030.477 | 1028.247 | 725.199 | 1593.58 | 1561.184 | 726.628 | 2840.218 | 2855.779 | 730.339 | 7673.183 | 7678.758 |
| 727.199 | 1036.616 | 1030.718 | 725.485 | 1596.194 | 1564.708 | 726.913 | 2844.109 | 2861.527 | 730.625 | 7632.131 | 7690.76 |
| 727.484 | 1051.07 | 1033.188 | 725.771 | 1550.145 | 1568.238 | 727.77 | 2942.191 | 2878.805 | 730.91 | 7654.197 | 7702.732 |
| 727.77 | 1060.129 | 1035.665 | 726.056 | 1589.034 | 1571.768 | 728.055 | 2882.209 | 2884.571 | 731.195 | 7728.144 | 7714.734 |
| 728.055 | 1052.945 | 1038.142 | 726.342 | 1567.543 | 1575.304 | 728.341 | 2891.975 | 2890.34 | 731.481 | 7793.287 | 7726.706 |
| 728.341 | 1062.35 | 1040.624 | 726.628 | 1584.979 | 1578.843 | 728.626 | 2920.074 | 2896.115 | 731.766 | 7791.351 | 7738.708 |
| 728.626 | 1018.695 | 1043.106 | 726.913 | 1570.272 | 1582.388 | 728.912 | 2908.844 | 2901.896 | 732.052 | 7797.161 | 7750.68 |
| 729.483 | 1017.648 | 1050.575 | 727.199 | 1594.219 | 1585.933 | 729.197 | 2930.532 | 2907.677 | 732.337 | 7763.591 | 7762.682 |
| 729.768 | 1054.779 | 1053.072 | 727.484 | 1577.308 | 1589.483 | 729.483 | 2946.328 | 2913.463 | 732.622 | 7691.083 | 7774.683 |
| 730.054 | 1035.049 | 1055.569 | 727.77 | 1605.599 | 1593.034 | 729.768 | 2918.584 | 2919.256 | 732.907 | 7779.144 | 7786.656 |
| 730.339 | 1056.942 | 1058.069 | 728.055 | 1576.663 | 1596.59 | 730.054 | 2871.839 | 2925.048 | 733.193 | 7666.375 | 7798.657 |
| 730.625 | 1035.395 | 1060.572 | 729.197 | 1648.203 | 1610.851 | 730.339 | 2949.409 | 2930.846 | 733.478 | 7762.564 | 7810.659 |
| 730.91 | 1066.482 | 1063.081 | 729.483 | 1614.235 | 1614.425 | 730.625 | 2895.402 | 2936.645 | 733.763 | 7809.573 | 7822.66 |
| 731.195 | 1040.923 | 1065.59 | 729.768 | 1588.439 | 1618.002 | 730.91 | 2948.793 | 2942.455 | 734.049 | 7760.451 | 7834.662 |
| 731.481 | 1061.617 | 1068.102 | 730.054 | 1595.572 | 1621.582 | 731.195 | 2910.006 | 2948.265 | 734.334 | 7761.596 | 7846.664 |
| 731.766 | 1060.61 | 1070.616 | 730.339 | 1641.695 | 1625.168 | 731.481 | 2917.63 | 2954.075 | 734.619 | 7797.571 | 7858.636 |
| 732.052 | 1066.406 | 1073.134 | 730.625 | 1594.612 | 1628.754 | 731.766 | 2957.42 | 2959.885 | 734.904 | 7745.662 | 7870.638 |
| 732.337 | 1092.827 | 1075.655 | 730.91 | 1608.075 | 1632.346 | 732.052 | 2924.57 | 2965.725 | 735.189 | 7872.604 | 7882.639 |
| 732.622 | 1065.604 | 1078.178 | 731.195 | 1629.391 | 1635.94 | 732.337 | 2947.385 | 2971.535 | 735.475 | 7876.8 | 7894.67 |
| 732.907 | 1106.348 | 1080.705 | 731.481 | 1629.799 | 1639.538 | 732.622 | 3003.549 | 2977.374 | 735.76 | 7884.752 | 7906.672 |
| 733.193 | 1073.237 | 1083.234 | 731.766 | 1624.109 | 1643.138 | 732.907 | 2941.985 | 2983.213 | 736.045 | 7865.649 | 7918.674 |
| 733.478 | 1090.306 | 1085.767 | 732.052 | 1655.064 | 1646.742 | 733.193 | 2975.085 | 2989.053 | 736.33 | 7842.409 | 7930.675 |
| 733.763 | 1073.392 | 1088.302 | 732.337 | 1658.717 | 1650.348 | 733.478 | 3003.607 | 2994.892 | 736.615 | 7813.711 | 7942.677 |
| 734.049 | 1118.808 | 1090.84 | 732.622 | 1651.052 | 1653.96 | 733.763 | 3006.248 | 3000.732 | 736.9 | 7906.114 | 7954.678 |
| 734.334 | 1050.299 | 1093.378 | 732.907 | 1621.037 | 1657.576 | 734.049 | 3006.072 | 3006.571 | 737.185 | 7818.963 | 7966.709 |
| 734.619 | 1091.55 | 1095.922 | 733.193 | 1647.825 | 1661.194 | 734.334 | 2960.853 | 3012.44 | 737.47 | 7955.324 | 7978.711 |
| 734.904 | 1080.793 | 1098.47 | 733.478 | 1647.939 | 1664.815 | 734.619 | 2990.285 | 3018.279 | 737.756 | 7895.081 | 7990.713 |
| 735.189 | 1070 | 1101.017 | 733.763 | 1661.34 | 1668.439 | 735.189 | 3027.053 | 3030.017 | 738.041 | 7932.847 | 8002.714 |
| 735.475 | 1114.946 | 1103.57 | 734.049 | 1683.202 | 1672.066 | 735.475 | 3085.095 | 3035.886 | 738.326 | 7882.317 | 8014.745 |
| 735.76 | 1065.731 | 1106.122 | 734.334 | 1701.641 | 1675.695 | 735.76 | 3008.479 | 3041.754 | 738.611 | 7881.964 | 8026.747 |
| 736.045 | 1110.815 | 1108.681 | 734.619 | 1656.026 | 1679.331 | 736.045 | 3049.853 | 3047.623 | 738.896 | 8054.653 | 8038.778 |
| 736.33 | 1076.092 | 1111.24 | 734.904 | 1624.719 | 1682.967 | 736.33 | 3025.967 | 3053.492 | 739.181 | 7971.61 | 8050.78 |
| 736.615 | 1080.584 | 1113.805 | 735.189 | 1656.337 | 1686.608 | 736.615 | 3039.759 | 3059.39 | 739.466 | 8108.147 | 8062.781 |
| 736.9 | 1104.972 | 1116.369 | 735.475 | 1673.272 | 1690.253 | 736.9 | 3052.201 | 3065.259 | 739.751 | 8126.047 | 8074.812 |
| 737.185 | 1114.383 | 1118.937 | 736.33 | 1739.263 | 1701.207 | 737.185 | 3004.752 | 3071.157 | 740.036 | 8071.086 | 8086.814 |
| 737.47 | 1078.847 | 1121.507 | 736.615 | 1711.172 | 1704.863 | 738.041 | 3113.001 | 3088.851 | 740.32 | 8063.339 | 8098.845 |
| 737.756 | 1088.023 | 1124.081 | 736.9 | 1748.677 | 1708.522 | 738.326 | 3079.373 | 3094.749 | 740.605 | 7987.808 | 8110.846 |
| 738.041 | 1117.012 | 1126.657 | 737.185 | 1691.494 | 1712.187 | 738.611 | 3041.52 | 3100.648 | 740.89 | 8050.868 | 8122.877 |
| 738.326 | 1095.016 | 1129.237 | 737.47 | 1712.851 | 1715.855 | 738.896 | 3127.996 | 3106.575 | 741.175 | 8075.164 | 8134.879 |
| 738.611 | 1118.415 | 1131.819 | 737.756 | 1680.514 | 1719.526 | 739.181 | 3122.949 | 3112.473 | 741.46 | 8163.02 | 8146.91 |
| 738.896 | 1119.753 | 1134.404 | 738.041 | 1740.947 | 1723.197 | 739.466 | 3132.632 | 3118.401 | 741.745 | 8175.109 | 8158.912 |
| 739.181 | 1137.297 | 1136.992 | 738.326 | 1679.19 | 1726.874 | 739.751 | 3161.36 | 3124.328 | 742.03 | 8074.988 | 8170.943 |
| 739.466 | 1115.061 | 1139.583 | 738.611 | 1670.346 | 1730.557 | 740.036 | 3138.883 | 3130.226 | 742.315 | 8125.078 | 8182.974 |
| 739.751 | 1144.944 | 1142.174 | 738.896 | 1711.953 | 1734.239 | 740.32 | 3098.828 | 3136.154 | 742.599 | 8098.434 | 8194.975 |
| 740.036 | 1116.689 | 1144.771 | 739.181 | 1720.483 | 1737.925 | 740.605 | 3139.293 | 3142.11 | 742.884 | 8238.14 | 8207.006 |
| 740.32 | 1151.646 | 1147.368 | 739.466 | 1745.378 | 1741.613 | 740.89 | 3129.551 | 3148.038 | 743.169 | 8194.095 | 8219.008 |
| 740.605 | 1169.426 | 1149.971 | 739.751 | 1782.178 | 1745.308 | 741.175 | 3192.875 | 3153.965 | 743.454 | 8343.807 | 8231.039 |
| 740.89 | 1152.964 | 1152.574 | 740.036 | 1741.203 | 1749.005 | 741.46 | 3126.734 | 3159.922 | 743.739 | 8242.307 | 8243.07 |
| 741.175 | 1130.393 | 1155.182 | 740.32 | 1737.279 | 1752.702 | 741.745 | 3163.15 | 3165.85 | 744.023 | 8215.868 | 8255.071 |
| 741.46 | 1138.178 | 1157.791 | 740.605 | 1758.527 | 1756.406 | 742.03 | 3138.619 | 3171.806 | 744.308 | 8238.492 | 8267.102 |
| 741.745 | 1184.881 | 1160.403 | 740.89 | 1761.127 | 1760.112 | 742.315 | 3114.791 | 3177.763 | 744.593 | 8273.793 | 8279.133 |
| 742.03 | 1149.214 | 1163.017 | 741.175 | 1781.04 | 1763.821 | 742.599 | 3145.221 | 3183.72 | 744.878 | 8230.716 | 8291.135 |
| 742.315 | 1151.459 | 1165.635 | 741.46 | 1784.523 | 1767.533 | 742.884 | 3204.613 | 3189.677 | 745.162 | 8358.949 | 8303.166 |
| 742.599 | 1131.76 | 1168.255 | 741.745 | 1776.271 | 1771.248 | 743.169 | 3165.82 | 3195.634 | 745.447 | 8187.405 | 8315.197 |
| 742.884 | 1199.82 | 1170.878 | 742.03 | 1780.785 | 1774.966 | 743.454 | 3235.864 | 3201.62 | 745.732 | 8363.79 | 8327.199 |
| 743.169 | 1147.139 | 1173.502 | 743.454 | 1798.332 | 1793.605 | 743.739 | 3188.092 | 3207.577 | 746.016 | 8227.723 | 8339.23 |
| 743.454 | 1183.127 | 1176.131 | 743.739 | 1815.856 | 1797.34 | 744.023 | 3214.443 | 3213.533 | 746.301 | 8315.931 | 8351.261 |
| 743.739 | 1200.771 | 1178.76 | 744.023 | 1789.45 | 1801.082 | 744.308 | 3250.096 | 3219.52 | 746.586 | 8300.613 | 8363.262 |
| 744.023 | 1193.652 | 1181.395 | 744.308 | 1813.004 | 1804.823 | 744.593 | 3227.912 | 3225.506 | 746.87 | 8246.298 | 8375.293 |
| 744.308 | 1216.62 | 1184.03 | 744.593 | 1786.645 | 1808.57 | 744.878 | 3244.286 | 3231.492 | 747.155 | 8308.507 | 8387.324 |
| 744.593 | 1194.967 | 1186.668 | 744.878 | 1773.07 | 1812.32 | 745.162 | 3269.023 | 3237.478 | 747.439 | 8404.138 | 8399.326 |
| 744.878 | 1185.454 | 1189.312 | 745.162 | 1814.322 | 1816.071 | 745.447 | 3214.472 | 3243.464 | 747.724 | 8290.137 | 8411.357 |
| 745.162 | 1200.199 | 1191.956 | 745.447 | 1805.627 | 1819.827 | 745.732 | 3181.49 | 3249.45 | 748.009 | 8367.84 | 8423.358 |
| 745.447 | 1182.308 | 1194.603 | 745.732 | 1806.287 | 1823.585 | 746.016 | 3254.996 | 3255.436 | 748.293 | 8509.483 | 8435.389 |
| 745.732 | 1194.885 | 1197.25 | 746.87 | 1810.654 | 1838.651 | 746.301 | 3261.129 | 3261.452 | 748.578 | 8352.2 | 8447.42 |
| 746.016 | 1173.188 | 1199.902 | 747.155 | 1801.472 | 1842.424 | 746.586 | 3253.06 | 3267.438 | 748.862 | 8387.53 | 8459.422 |
| 746.301 | 1182.962 | 1202.558 | 747.439 | 1854.276 | 1846.201 | 746.87 | 3303.795 | 3273.454 | 749.147 | 8442.315 | 8471.453 |
| 746.586 | 1186.187 | 1205.214 | 747.724 | 1832.274 | 1849.98 | 747.155 | 3289.915 | 3279.469 | 749.431 | 8443.283 | 8483.455 |
| 746.87 | 1180.116 | 1207.875 | 748.009 | 1856.427 | 1853.763 | 747.439 | 3319.318 | 3285.485 | 749.716 | 8541.996 | 8495.486 |
| 747.155 | 1208.353 | 1210.537 | 748.293 | 1860.415 | 1857.551 | 747.724 | 3209.865 | 3291.5 | 750 | 8460.361 | 8507.517 |
| 748.578 | 1234.514 | 1223.888 | 748.578 | 1878.69 | 1861.339 | 748.009 | 3320.726 | 3297.516 | 750.284 | 8547.278 | 8519.518 |
| 748.862 | 1211.3 | 1226.567 | 748.862 | 1870.304 | 1865.131 | 748.293 | 3291.881 | 3303.531 | 750.569 | 8499.623 | 8531.549 |
| 749.147 | 1234.094 | 1229.249 | 749.147 | 1865.832 | 1868.925 | 748.578 | 3261.452 | 3309.546 | 750.853 | 8620.109 | 8543.551 |
| 749.431 | 1216.221 | 1231.931 | 749.431 | 1822.679 | 1872.722 | 748.862 | 3309.664 | 3315.591 | 751.138 | 8530.992 | 8555.582 |
| 749.716 | 1241.635 | 1234.616 | 749.716 | 1871.037 | 1876.525 | 749.147 | 3287.333 | 3321.607 | 751.422 | 8558.839 | 8567.584 |
| 750 | 1221.168 | 1237.307 | 750 | 1849.038 | 1880.328 | 749.431 | 3370.846 | 3327.652 | 751.706 | 8548.305 | 8579.614 |
| 750.284 | 1188.347 | 1239.998 | 750.284 | 1875.909 | 1884.134 | 749.716 | 3284.868 | 3333.696 | 751.991 | 8584.896 | 8591.616 |
| 750.569 | 1190.31 | 1242.692 | 750.569 | 1874.526 | 1887.942 | 750 | 3302.152 | 3339.741 | 752.275 | 8472.128 | 8603.647 |
| 751.422 | 1262.082 | 1250.788 | 750.853 | 1877.164 | 1891.757 | 750.284 | 3343.703 | 3345.786 | 752.559 | 8470.807 | 8615.649 |
| 751.706 | 1273.658 | 1253.49 | 751.138 | 1893.744 | 1895.572 | 750.569 | 3324.571 | 3351.831 | 752.844 | 8669.906 | 8627.65 |
| 751.991 | 1246.847 | 1256.199 | 751.422 | 1843.387 | 1899.389 | 750.853 | 3361.896 | 3357.876 | 753.128 | 8447.596 | 8639.681 |
| 752.275 | 1278.268 | 1258.907 | 751.706 | 1898.744 | 1903.21 | 751.138 | 3340.24 | 3363.921 | 753.412 | 8616.793 | 8651.683 |
| 752.559 | 1264.761 | 1261.618 | 751.991 | 1856.48 | 1907.036 | 751.422 | 3323.866 | 3369.965 | 753.697 | 8612.626 | 8663.714 |
| 752.844 | 1250.573 | 1264.333 | 752.275 | 1915.209 | 1910.863 | 751.706 | 3338.274 | 3376.04 | 753.981 | 8633.372 | 8675.716 |
| 753.128 | 1243.81 | 1267.05 | 752.559 | 1917.26 | 1914.692 | 751.991 | 3410.695 | 3382.084 | 754.265 | 8697.195 | 8687.717 |
| 753.412 | 1238.267 | 1269.767 | 752.844 | 1874.271 | 1918.525 | 752.275 | 3350.628 | 3388.159 | 754.549 | 8626.829 | 8699.719 |
| 753.697 | 1275.551 | 1272.49 | 753.128 | 1876.305 | 1922.36 | 752.559 | 3361.544 | 3394.233 | 754.834 | 8672.693 | 8711.75 |
| 753.981 | 1277.379 | 1275.213 | 753.412 | 1908.929 | 1926.201 | 752.844 | 3420.202 | 3400.307 | 755.118 | 8520.985 | 8723.751 |
| 754.265 | 1239.364 | 1277.942 | 753.697 | 1907.708 | 1930.042 | 753.128 | 3381.556 | 3406.381 | 755.402 | 8625.948 | 8735.753 |
| 754.549 | 1262.064 | 1280.671 | 753.981 | 1863.455 | 1933.886 | 753.412 | 3413.6 | 3412.455 | 755.686 | 8633.783 | 8747.755 |
| 754.834 | 1242.624 | 1283.403 | 754.834 | 1964.228 | 1945.436 | 753.697 | 3378.915 | 3418.53 | 755.97 | 8663.655 | 8759.756 |
| 755.118 | 1258.678 | 1286.138 | 755.118 | 1914.226 | 1949.292 | 753.981 | 3405.882 | 3424.604 | 756.254 | 8695.259 | 8771.758 |
| 755.686 | 1283.503 | 1291.614 | 755.402 | 1892.203 | 1953.15 | 754.265 | 3429.71 | 3430.707 | 756.539 | 8630.614 | 8783.76 |
| 755.97 | 1319.52 | 1294.354 | 755.686 | 1961.073 | 1957.012 | 754.549 | 3379.884 | 3436.781 | 756.823 | 8748.84 | 8795.761 |
| 756.254 | 1273.309 | 1297.098 | 755.97 | 1977.849 | 1960.877 | 754.834 | 3377.507 | 3442.885 | 757.107 | 8771.024 | 8807.763 |
| 756.539 | 1286.294 | 1299.845 | 756.254 | 1958.532 | 1964.744 | 755.118 | 3345.845 | 3448.959 | 757.391 | 8679.266 | 8819.765 |
| 756.823 | 1270.463 | 1302.594 | 756.539 | 1911.676 | 1968.615 | 755.402 | 3457.351 | 3455.063 | 757.675 | 8785.051 | 8831.766 |
| 757.107 | 1285.666 | 1305.347 | 756.823 | 1961.977 | 1972.485 | 755.686 | 3440.537 | 3461.166 | 757.959 | 8770.408 | 8843.768 |
| 757.391 | 1312.75 | 1308.102 | 757.107 | 1977.969 | 1976.361 | 755.97 | 3406.675 | 3467.27 | 758.243 | 8755.912 | 8855.769 |
| 757.675 | 1306.946 | 1310.857 | 757.391 | 1969.02 | 1980.241 | 756.254 | 3416.534 | 3473.373 | 758.527 | 8791.653 | 8867.771 |
| 757.959 | 1278.846 | 1313.616 | 757.675 | 2010.212 | 1984.12 | 756.539 | 3464.071 | 3479.477 | 758.811 | 8850.194 | 8879.773 |
| 758.243 | 1301.159 | 1316.377 | 757.959 | 1954.236 | 1988.005 | 756.823 | 3464.159 | 3485.61 | 759.095 | 8745.202 | 8891.745 |
| 758.527 | 1304.992 | 1319.141 | 758.243 | 1947.54 | 1991.89 | 757.107 | 3409.844 | 3491.713 | 759.379 | 8741.505 | 8903.747 |
| 758.811 | 1323.12 | 1321.908 | 758.527 | 1999.091 | 1995.781 | 757.391 | 3499.401 | 3497.817 | 759.663 | 8852.043 | 8915.748 |
| 759.095 | 1289.874 | 1324.678 | 758.811 | 1959.415 | 1999.672 | 757.675 | 3454.534 | 3503.95 | 759.947 | 8800.222 | 8927.721 |
| 759.379 | 1310.062 | 1327.448 | 759.095 | 1947.578 | 2003.566 | 757.959 | 3476.249 | 3510.082 | 760.231 | 8946.647 | 8939.722 |
| 759.663 | 1304.725 | 1330.224 | 759.379 | 2034.057 | 2007.466 | 758.243 | 3485.668 | 3516.186 | 760.515 | 8915.484 | 8951.695 |
| 759.947 | 1311.896 | 1333 | 759.663 | 2002.152 | 2011.366 | 758.527 | 3496.027 | 3522.319 | 760.799 | 8830.064 | 8963.696 |
| 760.231 | 1345.603 | 1335.779 | 759.947 | 2008.466 | 2015.268 | 758.811 | 3484.465 | 3528.452 | 761.083 | 8905.361 | 8975.668 |
| 760.515 | 1319.587 | 1338.558 | 760.231 | 1992.28 | 2019.174 | 759.095 | 3471.437 | 3534.584 | 761.366 | 8891.921 | 8987.67 |
| 760.799 | 1357.253 | 1341.343 | 760.515 | 2007.407 | 2023.083 | 759.379 | 3503.157 | 3540.717 | 761.65 | 8847.054 | 8999.642 |
| 761.083 | 1339.741 | 1344.127 | 760.799 | 1979.413 | 2026.994 | 759.663 | 3441.799 | 3546.85 | 761.934 | 8947.234 | 9011.615 |
| 761.366 | 1306.905 | 1346.918 | 761.083 | 1995.232 | 2030.906 | 759.947 | 3478.127 | 3553.012 | 762.218 | 8829.008 | 9023.616 |
| 761.65 | 1318.725 | 1349.709 | 761.366 | 2041.599 | 2034.823 | 760.515 | 3550.782 | 3565.278 | 762.502 | 8859.232 | 9035.589 |
| 761.934 | 1318.751 | 1352.502 | 761.65 | 2022.375 | 2038.743 | 760.799 | 3535.406 | 3571.44 | 762.786 | 9025.054 | 9047.561 |
| 762.218 | 1324.775 | 1355.296 | 761.934 | 2014.94 | 2042.664 | 761.083 | 3539.808 | 3577.603 | 763.069 | 8892.948 | 9059.533 |
| 762.502 | 1294.114 | 1358.095 | 762.218 | 1995.925 | 2046.587 | 761.366 | 3553.658 | 3583.735 | 763.353 | 8885.73 | 9071.506 |
| 762.786 | 1335.732 | 1360.895 | 762.502 | 2005.01 | 2050.516 | 761.65 | 3552.807 | 3589.898 | 763.637 | 9031.128 | 9083.478 |
| 763.069 | 1348.423 | 1363.697 | 762.786 | 2028.904 | 2054.445 | 761.934 | 3555.184 | 3596.06 | 763.921 | 9028.898 | 9095.45 |
| 763.353 | 1322.633 | 1366.502 | 763.069 | 2006.236 | 2058.377 | 762.218 | 3562.256 | 3602.222 | 764.204 | 8929.804 | 9107.422 |
| 763.637 | 1360.272 | 1369.31 | 763.353 | 2043.82 | 2062.312 | 762.502 | 3531.944 | 3608.384 | 764.488 | 9030.307 | 9119.395 |
| 763.921 | 1378.037 | 1372.119 | 763.637 | 2050.009 | 2066.25 | 762.786 | 3609.998 | 3614.546 | 764.772 | 9040.694 | 9131.367 |
| 764.204 | 1371.629 | 1374.933 | 763.921 | 2041.12 | 2070.191 | 763.069 | 3616.19 | 3620.738 | 765.055 | 9002.313 | 9143.31 |
| 764.488 | 1339.57 | 1377.747 | 764.204 | 2084.053 | 2074.132 | 763.353 | 3562.608 | 3626.9 | 765.339 | 9038.288 | 9155.282 |
| 764.772 | 1377.638 | 1380.564 | 764.488 | 2055.191 | 2078.079 | 763.637 | 3678.076 | 3633.062 | 765.623 | 8997.119 | 9167.254 |
| 765.055 | 1373.612 | 1383.384 | 764.772 | 2058.222 | 2082.026 | 763.921 | 3578.982 | 3639.254 | 765.906 | 9078.255 | 9179.197 |
| 765.339 | 1369.017 | 1386.204 | 765.055 | 2080.239 | 2085.978 | 764.204 | 3596.001 | 3645.446 | 766.19 | 9065.607 | 9191.17 |
| 765.623 | 1385.329 | 1389.029 | 765.339 | 2078.108 | 2089.931 | 764.488 | 3657.33 | 3651.608 | 766.474 | 9213.383 | 9203.113 |
| 765.906 | 1368.709 | 1391.855 | 765.623 | 2084.843 | 2093.886 | 764.772 | 3653.955 | 3657.799 | 766.757 | 9142.958 | 9215.056 |
| 766.19 | 1345.424 | 1394.684 | 765.906 | 2105.774 | 2097.845 | 765.055 | 3582.18 | 3663.991 | 767.041 | 9142.899 | 9227.028 |
| 766.474 | 1403.235 | 1397.513 | 766.19 | 2072.644 | 2101.806 | 765.339 | 3598.525 | 3670.182 | 767.324 | 9150.059 | 9238.971 |
| 766.757 | 1415.01 | 1400.347 | 766.474 | 2036.921 | 2105.768 | 765.623 | 3620.767 | 3676.374 | 767.608 | 9123.004 | 9250.914 |
| 767.041 | 1392.281 | 1403.182 | 768.459 | 2134.786 | 2133.586 | 765.906 | 3637.083 | 3682.566 | 767.892 | 9225.678 | 9262.857 |
| 767.324 | 1409.699 | 1406.02 | 769.026 | 2079.147 | 2141.555 | 766.19 | 3629.013 | 3688.757 | 768.175 | 9247.305 | 9274.8 |
| 767.608 | 1396.732 | 1408.86 | 769.592 | 2076.535 | 2149.534 | 769.592 | 3729.017 | 3763.32 | 768.459 | 9183.57 | 9286.743 |
| 767.892 | 1390.682 | 1411.703 | 769.876 | 2127.925 | 2153.528 | 769.876 | 3684.473 | 3769.541 | 768.742 | 9273.89 | 9298.686 |
| 768.175 | 1399.156 | 1414.547 | 770.159 | 2122.423 | 2157.524 | 770.159 | 3763.349 | 3775.762 | 769.026 | 9186.416 | 9310.629 |
| 768.459 | 1432.315 | 1417.396 | 771.293 | 2175.794 | 2173.531 | 770.443 | 3760.884 | 3782.012 | 769.309 | 9303.41 | 9322.572 |
| 768.742 | 1366.816 | 1420.245 | 771.576 | 2195.801 | 2177.54 | 770.726 | 3791.02 | 3788.233 | 769.592 | 9178.699 | 9334.485 |
| 769.309 | 1437.723 | 1425.95 | 771.86 | 2177.769 | 2181.551 | 771.009 | 3767.281 | 3794.454 | 769.876 | 9273.831 | 9346.428 |
| 769.592 | 1372.949 | 1428.805 | 772.143 | 2167.701 | 2185.565 | 771.293 | 3812.236 | 3800.704 | 770.159 | 9312.771 | 9358.371 |
| 769.876 | 1414.908 | 1431.666 | 772.426 | 2185.533 | 2189.58 | 771.576 | 3778.315 | 3806.954 | 770.443 | 9282.429 | 9370.285 |
| 770.159 | 1410.776 | 1434.524 | 772.709 | 2169.051 | 2193.597 | 771.86 | 3787.176 | 3813.175 | 770.726 | 9364.445 | 9382.228 |
| 770.443 | 1447.215 | 1437.388 | 772.993 | 2192.59 | 2197.617 | 772.143 | 3806.191 | 3819.425 | 771.009 | 9397.604 | 9394.141 |
| 770.726 | 1396.198 | 1440.255 | 773.276 | 2201.482 | 2201.64 | 772.426 | 3866.522 | 3825.676 | 771.293 | 9335.6 | 9406.055 |
| 771.009 | 1427.399 | 1443.122 | 773.559 | 2209.055 | 2205.666 | 772.709 | 3803.169 | 3831.926 | 771.576 | 9304.73 | 9417.968 |
| 771.293 | 1460.561 | 1445.992 | 773.842 | 2216.218 | 2209.692 | 772.993 | 3805.047 | 3838.176 | 771.86 | 9355.994 | 9429.911 |
| 771.576 | 1443.656 | 1448.862 | 774.126 | 2186.408 | 2213.721 | 773.276 | 3857.161 | 3844.426 | 772.143 | 9364.504 | 9441.825 |
| 771.86 | 1447.171 | 1451.737 | 774.409 | 2209.401 | 2217.756 | 773.559 | 3807.717 | 3850.676 | 772.426 | 9407.522 | 9453.739 |
| 772.143 | 1454.185 | 1454.613 | 774.692 | 2211.752 | 2221.787 | 774.975 | 3841.873 | 3881.986 | 772.709 | 9447.459 | 9465.623 |
| 772.426 | 1417.607 | 1457.492 | 774.975 | 2196.214 | 2225.825 | 775.258 | 3870.572 | 3888.266 | 772.993 | 9354.879 | 9477.536 |
| 772.709 | 1450.531 | 1460.373 | 775.258 | 2204.255 | 2229.866 | 775.542 | 3909.482 | 3894.546 | 773.276 | 9396.459 | 9489.45 |

| | | |
|---|---|---|
| 772.993 | 1465.755 | 1463.258 |
| 773.276 | 1435.451 | 1466.142 |
| 773.559 | 1479.91 | 1469.03 |
| 773.842 | 1445.229 | 1471.92 |
| 774.126 | 1442.887 | 1474.81 |
| 774.409 | 1486.096 | 1477.707 |
| 774.692 | 1451.934 | 1480.603 |
| 774.975 | 1482.845 | 1483.499 |
| 775.258 | 1437.737 | 1486.401 |
| 775.542 | 1487.144 | 1489.303 |
| 775.825 | 1494.641 | 1492.208 |
| 776.108 | 1509.595 | 1495.116 |
| 776.391 | 1477.266 | 1498.024 |
| 776.674 | 1481.26 | 1500.938 |
| 776.957 | 1488.983 | 1503.852 |
| 777.24 | 1492.282 | 1506.766 |
| 777.523 | 1503.717 | 1509.685 |
| 777.806 | 1500.803 | 1512.605 |
| 778.089 | 1504.327 | 1515.528 |
| 778.372 | 1485.248 | 1518.451 |
| 778.655 | 1517.35 | 1521.379 |
| 778.938 | 1505.187 | 1524.308 |
| 779.221 | 1513.934 | 1527.236 |
| 779.504 | 1492.091 | 1530.17 |
| 781.201 | 1507.896 | 1547.812 |
| 781.484 | 1499.01 | 1550.758 |
| 781.767 | 1525.86 | 1553.707 |
| 782.05 | 1528.243 | 1556.659 |
| 782.332 | 1545.667 | 1559.614 |
| 782.615 | 1567.185 | 1562.569 |
| 782.898 | 1540.881 | 1565.527 |
| 783.181 | 1556.354 | 1568.488 |
| 783.463 | 1572.581 | 1571.448 |
| 783.746 | 1550.151 | 1574.412 |
| 784.029 | 1529.675 | 1577.379 |
| 784.311 | 1564.963 | 1580.345 |
| 784.594 | 1557.208 | 1583.315 |
| 784.877 | 1513.817 | 1586.288 |
| 785.159 | 1564.905 | 1589.263 |
| 785.442 | 1566.178 | 1592.239 |
| 785.724 | 1552.363 | 1595.217 |
| 786.007 | 1545.15 | 1598.195 |
| 786.29 | 1569.861 | 1601.177 |
| 786.572 | 1592.203 | 1604.161 |
| 786.855 | 1582.672 | 1607.148 |
| 787.137 | 1582.781 | 1610.135 |
| 787.42 | 1643.091 | 1613.126 |
| 787.702 | 1598.404 | 1616.116 |
| 787.985 | 1579.445 | 1619.112 |
| 788.267 | 1597.688 | 1622.108 |
| 788.55 | 1608.862 | 1625.104 |
| 788.832 | 1587.3 | 1628.103 |
| 789.114 | 1579.627 | 1631.104 |
| 789.397 | 1640.782 | 1634.109 |
| 789.679 | 1632.346 | 1637.114 |
| 791.091 | 1622.096 | 1652.17 |
| 791.373 | 1652.848 | 1655.187 |
| 791.656 | 1621.756 | 1658.206 |
| 791.938 | 1621.94 | 1661.226 |
| 792.22 | 1655.862 | 1664.248 |
| 792.502 | 1671.449 | 1667.274 |
| 792.784 | 1666.898 | 1670.299 |
| 793.067 | 1633.22 | 1673.327 |
| 793.349 | 1682.535 | 1676.359 |
| 793.631 | 1664.85 | 1679.39 |
| 793.913 | 1648.127 | 1682.424 |
| 794.195 | 1619.801 | 1685.458 |
| 794.477 | 1666.886 | 1688.495 |
| 794.76 | 1696.198 | 1691.535 |
| 795.042 | 1658.385 | 1694.578 |
| 795.324 | 1657.573 | 1697.621 |
| 795.606 | 1666.634 | 1700.664 |
| 795.888 | 1686.462 | 1703.71 |
| 796.17 | 1696.844 | 1706.759 |
| 796.452 | 1656.625 | 1709.811 |
| 796.734 | 1662.588 | 1712.862 |
| 797.298 | 1680.728 | 1718.972 |
| 797.58 | 1747.993 | 1722.029 |
| 797.862 | 1700.987 | 1725.087 |
| 798.144 | 1702.419 | 1728.148 |
| 798.426 | 1748.612 | 1731.211 |
| 798.707 | 1732.561 | 1734.275 |
| 799.271 | 1744.02 | 1740.41 |
| 799.553 | 1682.77 | 1743.48 |
| 799.835 | 1716.783 | 1746.552 |
| 800.117 | 1715.565 | 1749.624 |
| 800.398 | 1739.862 | 1752.7 |
| 800.68 | 1702.745 | 1755.775 |
| 800.962 | 1745.751 | 1758.853 |
| 801.244 | 1729.902 | 1761.934 |
| 801.525 | 1742.209 | 1765.015 |
| 801.807 | 1785.931 | 1768.099 |
| 802.089 | 1725.284 | 1771.186 |
| 803.779 | 1747.485 | 1789.731 |
| 804.06 | 1771.535 | 1792.827 |
| 804.342 | 1801.589 | 1795.926 |
| 804.624 | 1776.559 | 1799.028 |
| 805.75 | 1792.657 | 1811.446 |
| 806.031 | 1800.656 | 1814.553 |
| 806.313 | 1814.14 | 1817.664 |
| 807.157 | 1748.157 | 1827.004 |
| 809.689 | 1833.7 | 1855.113 |
| 809.97 | 1812.778 | 1858.241 |
| 810.252 | 1855.394 | 1861.375 |
| 810.533 | 1818.022 | 1864.508 |
| 810.814 | 1856.313 | 1867.642 |
| 811.095 | 1864.318 | 1870.782 |
| 811.376 | 1810.011 | 1873.919 |
| 811.658 | 1848.434 | 1877.059 |
| 811.939 | 1854.863 | 1880.202 |
| 812.22 | 1877.329 | 1883.344 |
| 812.501 | 1879.647 | 1886.49 |
| 812.782 | 1848.484 | 1889.636 |
| 813.063 | 1915.881 | 1892.781 |
| 813.625 | 1888.268 | 1899.081 |
| 813.906 | 1853.22 | 1902.233 |
| 814.187 | 1893.289 | 1905.387 |
| 814.469 | 1905.2 | 1908.542 |
| 775.542 | 2213.008 | 2233.906 |
| 775.825 | 2234.622 | 2237.95 |
| 776.108 | 2249.418 | 2241.997 |
| 776.391 | 2231.274 | 2246.043 |
| 776.674 | 2256.616 | 2250.095 |
| 776.957 | 2253.722 | 2254.148 |
| 777.24 | 2220.097 | 2258.203 |
| 777.523 | 2243.96 | 2262.258 |
| 777.806 | 2228.63 | 2266.32 |
| 778.089 | 2257.429 | 2270.381 |
| 778.372 | 2263.174 | 2274.445 |
| 778.655 | 2225.194 | 2278.512 |
| 778.938 | 2276.795 | 2282.582 |
| 779.221 | 2306.585 | 2286.652 |
| 779.504 | 2301.353 | 2290.725 |
| 779.787 | 2283.313 | 2294.801 |
| 780.07 | 2282.608 | 2298.877 |
| 780.353 | 2322.369 | 2302.958 |
| 780.635 | 2298.73 | 2307.04 |
| 780.918 | 2291.01 | 2311.122 |
| 781.201 | 2304.596 | 2315.209 |
| 781.484 | 2327.669 | 2319.297 |
| 781.767 | 2306.359 | 2323.388 |
| 782.05 | 2332.913 | 2327.481 |
| 782.332 | 2342.358 | 2331.577 |
| 782.615 | 2347.655 | 2335.674 |
| 782.898 | 2346.376 | 2339.773 |
| 783.181 | 2332.079 | 2343.873 |
| 783.463 | 2355.771 | 2347.978 |
| 783.746 | 2320.025 | 2352.083 |
| 784.029 | 2306.565 | 2356.191 |
| 785.159 | 2376.119 | 2372.641 |
| 785.442 | 2344.09 | 2376.758 |
| 785.724 | 2349.046 | 2380.878 |
| 786.007 | 2381.926 | 2385.001 |
| 786.29 | 2377.474 | 2389.127 |
| 786.572 | 2345.742 | 2393.252 |
| 786.855 | 2379.96 | 2397.381 |
| 787.137 | 2448.739 | 2401.51 |
| 787.42 | 2422.338 | 2405.641 |
| 787.702 | 2433.536 | 2409.776 |
| 787.985 | 2414.315 | 2413.913 |
| 788.267 | 2438.219 | 2418.051 |
| 788.55 | 2423.488 | 2422.191 |
| 788.832 | 2402.311 | 2426.332 |
| 789.114 | 2427.549 | 2430.478 |
| 789.397 | 2460.94 | 2434.624 |
| 790.244 | 2373.363 | 2447.072 |
| 790.526 | 2373.771 | 2451.227 |
| 790.809 | 2413.723 | 2455.382 |
| 791.091 | 2411.052 | 2459.54 |
| 791.373 | 2405.304 | 2463.701 |
| 791.656 | 2420.451 | 2467.862 |
| 792.22 | 2439.249 | 2476.19 |
| 792.502 | 2495.712 | 2480.357 |
| 792.784 | 2429.219 | 2484.526 |
| 793.067 | 2428.53 | 2488.696 |
| 793.349 | 2465.632 | 2492.869 |
| 793.631 | 2458.363 | 2497.045 |
| 793.913 | 2471.257 | 2501.22 |
| 794.195 | 2484.468 | 2505.399 |
| 794.477 | 2451.503 | 2509.58 |
| 794.76 | 2488.902 | 2513.762 |
| 795.042 | 2519.378 | 2517.943 |
| 795.324 | 2517.902 | 2522.131 |
| 795.606 | 2506.828 | 2526.318 |
| 795.888 | 2486.827 | 2530.505 |
| 796.17 | 2528.768 | 2534.699 |
| 796.452 | 2482.552 | 2538.889 |
| 796.734 | 2501.467 | 2543.085 |
| 797.016 | 2527.521 | 2547.281 |
| 797.298 | 2440.091 | 2551.477 |
| 797.58 | 2559.491 | 2555.676 |
| 798.144 | 2500.592 | 2564.084 |
| 798.426 | 2565.469 | 2568.286 |
| 798.707 | 2544.766 | 2572.493 |
| 798.989 | 2507.039 | 2576.701 |
| 799.271 | 2561.977 | 2580.912 |
| 799.553 | 2554.186 | 2585.123 |
| 799.835 | 2605.265 | 2589.337 |
| 800.117 | 2589.748 | 2593.551 |
| 800.398 | 2530.687 | 2597.767 |
| 800.68 | 2540.215 | 2601.987 |
| 800.962 | 2595.37 | 2606.207 |
| 801.244 | 2537.788 | 2610.429 |
| 801.525 | 2575.607 | 2614.652 |
| 801.807 | 2617.821 | 2618.877 |
| 802.089 | 2618.12 | 2623.103 |
| 802.371 | 2529.511 | 2627.331 |
| 802.652 | 2588.97 | 2631.56 |
| 802.934 | 2594.986 | 2635.791 |
| 803.216 | 2622.839 | 2640.025 |
| 803.497 | 2627.786 | 2644.26 |
| 803.779 | 2562.024 | 2648.494 |
| 804.06 | 2612.096 | 2652.731 |
| 804.342 | 2627.962 | 2656.971 |
| 804.624 | 2628.259 | 2661.212 |
| 804.905 | 2610.101 | 2665.455 |
| 805.187 | 2627.443 | 2669.698 |
| 805.75 | 2710.172 | 2678.19 |
| 806.031 | 2641.927 | 2682.439 |
| 808.845 | 2732.004 | 2725.005 |
| 809.127 | 2691.154 | 2729.269 |
| 809.408 | 2708.493 | 2733.535 |
| 809.689 | 2689.065 | 2737.802 |
| 809.97 | 2743.486 | 2742.072 |
| 810.252 | 2721.719 | 2746.341 |
| 810.533 | 2712.91 | 2750.614 |
| 810.814 | 2707.845 | 2754.886 |
| 811.095 | 2724.401 | 2759.161 |
| 811.376 | 2701.621 | 2763.437 |
| 811.658 | 2765.18 | 2767.712 |
| 811.939 | 2751.776 | 2771.991 |
| 812.22 | 2749.273 | 2776.272 |
| 812.501 | 2763.763 | 2780.553 |
| 812.782 | 2756.984 | 2784.834 |
| 813.063 | 2746.142 | 2789.119 |
| 813.344 | 2784.201 | 2793.406 |
| 775.825 | 3847.742 | 3900.796 |
| 776.108 | 3902.85 | 3907.075 |
| 776.391 | 3879.551 | 3913.355 |
| 776.674 | 3917.14 | 3919.635 |
| 776.957 | 3894.575 | 3925.914 |
| 777.24 | 3875.854 | 3932.194 |
| 777.523 | 3963.24 | 3938.473 |
| 777.806 | 3899.358 | 3944.753 |
| 778.089 | 3943.579 | 3951.062 |
| 778.372 | 3939.178 | 3957.341 |
| 778.655 | 3910.861 | 3963.621 |
| 778.938 | 3940.41 | 3969.93 |
| 779.221 | 3942.523 | 3976.209 |
| 779.504 | 3985.306 | 3982.518 |
| 779.787 | 3974.214 | 3988.827 |
| 780.07 | 3977.119 | 3995.107 |
| 781.201 | 4022.338 | 4020.343 |
| 781.484 | 3974.302 | 4026.652 |
| 781.767 | 3942.493 | 4032.961 |
| 782.05 | 3970.429 | 4039.269 |
| 782.332 | 3977.501 | 4045.578 |
| 782.615 | 4017.085 | 4051.887 |
| 782.898 | 4033.841 | 4058.226 |
| 783.181 | 4028.295 | 4064.534 |
| 783.463 | 4031.669 | 4070.843 |
| 783.746 | 4006.463 | 4077.182 |
| 784.029 | 4048.366 | 4083.491 |
| 784.594 | 4052.65 | 4096.167 |
| 784.877 | 4133.903 | 4102.476 |
| 785.442 | 4023.13 | 4115.153 |
| 785.724 | 4095.169 | 4121.491 |
| 786.007 | 4056.729 | 4127.829 |
| 786.29 | 4044.375 | 4134.167 |
| 786.572 | 4054.91 | 4140.506 |
| 786.855 | 4080.057 | 4146.844 |
| 787.137 | 4157.643 | 4153.182 |
| 787.42 | 4095.492 | 4159.521 |
| 788.55 | 4119.231 | 4184.903 |
| 788.832 | 4167.355 | 4191.271 |
| 789.114 | 4167.091 | 4197.609 |
| 789.397 | 4180.589 | 4203.976 |
| 789.679 | 4195.555 | 4210.315 |
| 789.962 | 4153.065 | 4216.682 |
| 790.244 | 4202.157 | 4223.021 |
| 790.526 | 4110.81 | 4229.388 |
| 790.809 | 4128.005 | 4235.756 |
| 791.091 | 4169.908 | 4242.123 |
| 791.656 | 4175.542 | 4254.829 |
| 791.938 | 4253.714 | 4261.197 |
| 792.22 | 4216.154 | 4267.565 |
| 792.502 | 4206.764 | 4273.932 |
| 792.784 | 4251.836 | 4280.3 |
| 793.067 | 4214.687 | 4286.667 |
| 793.349 | 4244.354 | 4293.064 |
| 793.631 | 4193.765 | 4299.432 |
| 793.913 | 4226.571 | 4305.8 |
| 794.195 | 4196.171 | 4312.167 |
| 794.477 | 4226.923 | 4318.564 |
| 794.76 | 4224.899 | 4324.932 |
| 795.042 | 4291.539 | 4331.299 |
| 795.324 | 4234.699 | 4337.696 |
| 795.606 | 4259.877 | 4344.064 |
| 795.888 | 4340.719 | 4350.461 |
| 796.734 | 4379.218 | 4369.623 |
| 797.016 | 4324.198 | 4375.99 |
| 797.298 | 4309.057 | 4382.387 |
| 797.58 | 4348.876 | 4388.784 |
| 797.862 | 4386.084 | 4395.152 |
| 799.271 | 4357.562 | 4427.136 |
| 799.553 | 4406.508 | 4433.533 |
| 799.835 | 4459.561 | 4439.93 |
| 800.117 | 4437.231 | 4446.327 |
| 800.398 | 4412.171 | 4452.724 |
| 800.68 | 4388.256 | 4459.121 |
| 800.962 | 4412.993 | 4465.518 |
| 801.244 | 4359.704 | 4471.945 |
| 801.525 | 4415.663 | 4478.342 |
| 801.807 | 4474.321 | 4484.738 |
| 802.089 | 4425.288 | 4491.135 |
| 802.371 | 4444.567 | 4497.562 |
| 802.652 | 4436.732 | 4503.959 |
| 802.934 | 4470.008 | 4510.356 |
| 803.216 | 4482.479 | 4516.782 |
| 803.497 | 4510.062 | 4523.179 |
| 803.779 | 4498.354 | 4529.605 |
| 804.06 | 4418.304 | 4536.002 |
| 804.342 | 4440.899 | 4542.399 |
| 804.624 | 4493.189 | 4548.825 |
| 804.905 | 4557.834 | 4555.252 |
| 805.187 | 4558.802 | 4561.649 |
| 805.468 | 4537.968 | 4568.075 |
| 805.75 | 4550.381 | 4574.472 |
| 806.031 | 4542.194 | 4580.898 |
| 806.313 | 4517.105 | 4587.325 |
| 806.594 | 4576.849 | 4593.722 |
| 806.876 | 4527.082 | 4600.148 |
| 807.157 | 4504.78 | 4606.574 |
| 807.438 | 4578.052 | 4613 |
| 807.72 | 4542.399 | 4619.427 |
| 808.001 | 4619.603 | 4625.824 |
| 808.283 | 4596.539 | 4632.25 |
| 808.564 | 4686.125 | 4638.676 |
| 808.845 | 4603.552 | 4645.103 |
| 809.127 | 4677.645 | 4651.529 |
| 809.408 | 4567.283 | 4657.955 |
| 809.689 | 4578.316 | 4664.382 |
| 809.97 | 4646.717 | 4670.808 |
| 810.252 | 4635.243 | 4677.234 |
| 810.533 | 4598.152 | 4683.66 |
| 810.814 | 4556.132 | 4690.087 |
| 811.095 | 4630.93 | 4696.513 |
| 811.376 | 4615.7 | 4702.939 |
| 811.658 | 4688.355 | 4709.366 |
| 811.939 | 4651.764 | 4715.792 |
| 812.22 | 4678.819 | 4722.218 |
| 812.501 | 4626.205 | 4728.645 |
| 812.782 | 4647.949 | 4735.071 |
| 813.063 | 4678.731 | 4741.497 |
| 773.559 | 9447.195 | 9501.364 |
| 773.842 | 9422.81 | 9513.248 |
| 774.126 | 9412.628 | 9525.161 |
| 774.409 | 9423.69 | 9537.046 |
| 774.692 | 9454.296 | 9548.959 |
| 774.975 | 9495.436 | 9560.844 |
| 775.258 | 9515.83 | 9572.728 |
| 775.542 | 9532.057 | 9584.612 |
| 775.825 | 9406.436 | 9596.496 |
| 776.108 | 9551.806 | 9608.381 |
| 776.391 | 9611.961 | 9620.265 |
| 776.674 | 9592.594 | 9632.12 |
| 776.957 | 9548.578 | 9644.004 |
| 777.24 | 9497.93 | 9655.888 |
| 777.523 | 9626.633 | 9667.743 |
| 777.806 | 9505.56 | 9679.598 |
| 778.089 | 9540.89 | 9691.482 |
| 778.655 | 9703.66 | 9715.192 |
| 778.938 | 9723.144 | 9727.047 |
| 779.221 | 9685.878 | 9738.902 |
| 779.504 | 9628.951 | 9750.757 |
| 779.787 | 9604.038 | 9762.583 |
| 780.07 | 9715.046 | 9774.438 |
| 780.353 | 9681.594 | 9786.263 |
| 780.635 | 9783.27 | 9798.118 |
| 780.918 | 9802.725 | 9809.944 |
| 781.201 | 9751.696 | 9821.769 |
| 781.484 | 9731.977 | 9833.595 |
| 781.767 | 9812.673 | 9845.42 |
| 782.05 | 9777.607 | 9857.246 |
| 782.332 | 9698.143 | 9869.071 |
| 782.615 | 9902.788 | 9880.897 |
| 782.898 | 9783.299 | 9892.693 |
| 783.181 | 9867.839 | 9904.519 |
| 783.463 | 9790.782 | 9916.315 |
| 783.746 | 9806.862 | 9928.141 |
| 784.029 | 9738.462 | 9939.937 |
| 784.311 | 9944.338 | 9951.733 |
| 784.594 | 9888.438 | 9963.529 |
| 784.877 | 9872.974 | 9975.326 |
| 785.159 | 9781.568 | 9987.092 |
| 785.442 | 9908.686 | 9998.889 |
| 785.724 | 9973.858 | 10010.66 |
| 786.007 | 9925.324 | 10022.45 |
| 786.29 | 9951.645 | 10034.22 |
| 786.572 | 9966.522 | 10045.99 |
| 786.855 | 9990.614 | 10057.75 |
| 787.137 | 9945.659 | 10069.52 |
| 787.42 | 10061.13 | 10081.29 |
| 787.702 | 9928.375 | 10093.05 |
| 789.679 | 10093.38 | 10175.25 |
| 789.962 | 10026.21 | 10186.98 |
| 790.244 | 10131.08 | 10198.69 |
| 790.526 | 10001.82 | 10210.43 |
| 790.809 | 10083.37 | 10222.14 |
| 791.091 | 10158.93 | 10233.84 |
| 791.373 | 10055.64 | 10245.58 |
| 791.656 | 10188.48 | 10257.26 |
| 791.938 | 10184.58 | 10268.97 |
| 792.22 | 10129.79 | 10280.68 |
| 792.502 | 10192.38 | 10292.39 |
| 792.784 | 10133.78 | 10304.06 |
| 793.067 | 10154.29 | 10315.74 |
| 793.349 | 10204.47 | 10327.45 |
| 793.631 | 10281.21 | 10339.13 |
| 793.913 | 10185.02 | 10350.81 |
| 794.195 | 10183.11 | 10362.46 |
| 794.477 | 10172.13 | 10374.14 |
| 794.76 | 10266.48 | 10385.82 |
| 795.042 | 10262.37 | 10397.47 |
| 795.324 | 10196.72 | 10409.12 |
| 795.606 | 10319.68 | 10420.77 |
| 795.888 | 10362.96 | 10432.41 |
| 796.17 | 10328.57 | 10444.06 |
| 796.452 | 10331.3 | 10455.71 |
| 796.734 | 10328.68 | 10467.33 |
| 797.016 | 10333.82 | 10478.98 |
| 797.298 | 10389.57 | 10490.6 |
| 797.58 | 10430.6 | 10502.22 |
| 797.862 | 10466.72 | 10513.84 |
| 798.144 | 10382.88 | 10525.46 |
| 798.426 | 10495.45 | 10537.08 |
| 798.707 | 10512.76 | 10548.68 |
| 798.989 | 10442.92 | 10560.3 |
| 799.271 | 10558.18 | 10571.89 |
| 799.553 | 10496.62 | 10583.48 |
| 799.835 | 10559.09 | 10595.07 |
| 800.117 | 10514.72 | 10606.66 |
| 800.398 | 10580.16 | 10618.25 |
| 800.68 | 10501.55 | 10629.81 |
| 800.962 | 10590.87 | 10641.4 |
| 801.244 | 10522.97 | 10652.96 |
| 801.525 | 10475.08 | 10664.52 |
| 801.807 | 10532.92 | 10676.09 |
| 802.089 | 10573.53 | 10687.65 |
| 802.371 | 10566.25 | 10699.18 |
| 802.652 | 10613.2 | 10710.74 |
| 802.934 | 10479.6 | 10722.27 |
| 803.216 | 10661.53 | 10733.81 |
| 803.497 | 10612.79 | 10745.34 |
| 803.779 | 10709.83 | 10756.87 |
| 804.06 | 10588.41 | 10768.4 |
| 804.342 | 10653.29 | 10779.9 |
| 804.624 | 10722.63 | 10791.44 |
| 804.905 | 10733.89 | 10802.94 |
| 805.187 | 10738.59 | 10814.44 |
| 805.468 | 10701.85 | 10825.95 |
| 805.75 | 10678.17 | 10837.45 |
| 806.031 | 10810.6 | 10848.92 |
| 806.313 | 10826.5 | 10860.42 |
| 806.594 | 10769.08 | 10871.9 |
| 806.876 | 10759.04 | 10883.37 |
| 807.157 | 10918 | 10894.84 |
| 807.438 | 10913.77 | 10906.32 |
| 807.72 | 10730.61 | 10917.76 |
| 808.001 | 10924.54 | 10929.24 |
| 808.283 | 10876.97 | 10940.68 |
| 808.564 | 10945.55 | 10952.12 |

| | | | | | | | | | | | |
|---|---|---|---|---|---|---|---|---|---|---|---|
| 814.75 | 1900.152 | 1911.699 | 813.625 | 2812.359 | 2797.69 | 813.344 | 4739.355 | 4747.953 | 808.845 | 10717.52 | 10963.57 |
| 815.03 | 1910.426 | 1914.857 | 816.716 | 2827.99 | 2844.928 | 813.625 | 4659.452 | 4754.379 | 809.127 | 10824.21 | 10975.01 |
| 815.311 | 1877.775 | 1918.017 | 816.997 | 2866.008 | 2849.229 | 813.906 | 4714.677 | 4760.805 | 809.408 | 10854.32 | 10986.46 |
| 815.592 | 1879.568 | 1921.177 | 817.278 | 2823.692 | 2853.531 | 814.187 | 4724.654 | 4767.232 | 809.689 | 10877.15 | 10997.87 |
| 815.873 | 1860.682 | 1924.341 | 817.559 | 2870.982 | 2857.836 | 814.469 | 4731.814 | 4773.658 | 809.97 | 10916.76 | 11009.29 |
| 816.716 | 1936.741 | 1933.836 | 817.839 | 2873.96 | 2862.141 | 814.75 | 4705.815 | 4780.114 | 810.252 | 10871.52 | 11020.7 |
| 816.997 | 1932.929 | 1937.002 | 818.12 | 2846.911 | 2866.448 | 815.03 | 4708.779 | 4786.54 | 810.533 | 10857.17 | 11032.12 |
| 817.278 | 1921.788 | 1940.172 | 818.401 | 2794.626 | 2870.756 | 815.311 | 4737.008 | 4792.966 | 810.814 | 10810.36 | 11043.53 |
| 817.559 | 1938.789 | 1943.344 | 818.682 | 2875.354 | 2875.064 | 815.592 | 4780.877 | 4799.393 | 811.095 | 10954.74 | 11054.94 |
| 818.963 | 1898.902 | 1959.219 | 818.963 | 2795.689 | 2879.374 | 815.873 | 4812.392 | 4805.848 | 811.376 | 10976.68 | 11066.33 |
| 820.927 | 1985.872 | 1981.502 | 819.243 | 2879.521 | 2883.685 | 816.154 | 4844.289 | 4812.275 | 811.658 | 11007.52 | 11077.72 |
| 821.208 | 1990.954 | 1984.692 | 819.524 | 2861.742 | 2887.999 | 816.435 | 4761.334 | 4818.701 | 811.939 | 10921.84 | 11089.1 |
| 821.489 | 1982.661 | 1987.882 | 819.805 | 2876.499 | 2892.312 | 816.716 | 4778.646 | 4825.157 | 812.22 | 10853.03 | 11100.49 |
| 821.769 | 1949.653 | 1991.074 | 820.085 | 2881.302 | 2896.626 | 816.997 | 4829.441 | 4831.583 | 812.501 | 10975.28 | 11111.87 |
| 822.05 | 2011.732 | 1994.267 | 820.366 | 2879.835 | 2900.942 | 817.278 | 4783.576 | 4838.009 | 812.782 | 11092.3 | 11123.23 |
| 822.611 | 2015.735 | 2000.655 | 820.647 | 2912.198 | 2905.259 | 817.559 | 4875.041 | 4844.465 | 813.063 | 10993.79 | 11134.61 |
| 823.453 | 1935.377 | 2010.251 | 820.927 | 2851.084 | 2909.578 | 817.839 | 4841.53 | 4850.891 | 813.344 | 11072.79 | 11145.97 |
| 824.855 | 2015.324 | 2026.266 | 821.208 | 2873.007 | 2913.897 | 818.12 | 4756.375 | 4857.317 | 813.625 | 11011.57 | 11157.33 |
| 825.135 | 2034.967 | 2029.474 | 821.489 | 2860.316 | 2918.217 | 818.401 | 4774.89 | 4863.773 | 813.906 | 11104.15 | 11168.68 |
| 825.416 | 2046.455 | 2032.681 | 821.769 | 2880.498 | 2922.539 | 819.524 | 4893.146 | 4889.508 | 814.187 | 11156.53 | 11180.01 |
| 825.696 | 2027.49 | 2035.891 | 822.05 | 2932.589 | 2926.862 | 819.805 | 4854.999 | 4895.963 | 814.469 | 11076.86 | 11191.36 |
| 825.976 | 2021.753 | 2039.104 | 822.331 | 2880.677 | 2931.184 | 820.085 | 4896.433 | 4902.39 | 814.75 | 10999.34 | 11202.69 |
| 826.257 | 1990.942 | 2042.315 | 822.611 | 2914.291 | 2935.5 | 820.366 | 4819.875 | 4908.816 | 815.03 | 11102.78 | 11214.02 |
| 826.537 | 2047.4 | 2045.528 | 822.892 | 2958.858 | 2939.843 | 820.647 | 4818.231 | 4915.271 | 815.311 | 11089.54 | 11225.34 |
| 827.098 | 2075.608 | 2051.96 | 823.172 | 2934.019 | 2944.157 | 820.927 | 4884.901 | 4921.698 | 815.592 | 11060.75 | 11236.67 |
| 827.378 | 1997.063 | 2055.176 | 823.453 | 2956.716 | 2948.5 | 821.208 | 4874.865 | 4928.153 | 815.873 | 11125.31 | 11248 |
| 827.658 | 2041.153 | 2058.395 | 823.733 | 3002.786 | 2952.813 | 821.489 | 4944.263 | 4934.58 | 816.435 | 11102.42 | 11270.59 |
| 827.939 | 2035.809 | 2061.614 | 824.014 | 2981.922 | 2957.156 | 821.769 | 4942.767 | 4941.035 | 816.716 | 11257.33 | 11281.89 |
| 828.219 | 2068.102 | 2064.836 | 824.294 | 3006.835 | 2961.47 | 822.05 | 4928.946 | 4947.462 | 816.997 | 11254.48 | 11293.19 |
| 828.499 | 2060.599 | 2068.058 | 824.574 | 2924.376 | 2965.813 | 822.331 | 4950.484 | 4953.888 | 817.278 | 11184.7 | 11304.48 |
| 828.779 | 2057.614 | 2071.28 | 824.855 | 2930.885 | 2970.155 | 822.611 | 4911.662 | 4960.344 | 817.559 | 11212.29 | 11315.75 |
| 829.059 | 2052.23 | 2074.505 | 825.696 | 3069.69 | 2983.155 | 822.892 | 4901.656 | 4966.77 | 817.839 | 11114.4 | 11327.02 |
| 829.34 | 2078.325 | 2077.73 | 825.976 | 2963.524 | 2987.498 | 823.172 | 4952.391 | 4973.226 | 818.12 | 11183.59 | 11338.29 |
| 829.62 | 2090.744 | 2080.958 | 826.537 | 3057.894 | 2996.154 | 823.453 | 4898.604 | 4979.652 | 818.401 | 11198.41 | 11349.56 |
| 829.9 | 2053.756 | 2084.185 | 826.817 | 2999.118 | 3000.497 | 823.733 | 5036.872 | 4986.108 | 818.682 | 11174.29 | 11360.82 |
| 830.18 | 2072.286 | 2087.413 | 827.098 | 2976.993 | 3004.84 | 824.014 | 4982.293 | 4992.534 | 818.963 | 11190.01 | 11372.09 |
| 830.46 | 2091.642 | 2090.644 | 827.378 | 3012.968 | 3009.183 | 824.294 | 4967.68 | 4998.96 | 819.243 | 11232.8 | 11383.33 |
| 830.74 | 2063.932 | 2093.875 | 827.658 | 3007.54 | 3013.526 | 824.574 | 4971.993 | 5005.416 | 819.524 | 11300.46 | 11394.57 |
| 831.02 | 2115.281 | 2097.108 | 827.939 | 2993.836 | 3017.869 | 824.855 | 4935.577 | 5011.842 | 819.805 | 11435.01 | 11405.81 |
| 831.3 | 2066.115 | 2100.342 | 828.219 | 2933.98 | 3022.211 | 825.135 | 5048.903 | 5018.298 | 820.085 | 11341.69 | 11417.05 |
| 831.58 | 2093.373 | 2103.579 | 828.499 | 2973.706 | 3026.554 | 825.416 | 5072.789 | 5024.724 | 820.366 | 11396.65 | 11428.26 |
| 831.86 | 2088.91 | 2106.812 | 829.059 | 3011.736 | 3035.269 | 826.817 | 5092.802 | 5056.914 | 820.647 | 11320.54 | 11439.5 |
| 832.14 | 2080.972 | 2110.052 | 829.34 | 3084.949 | 3039.612 | 827.098 | 5021.203 | 5063.37 | 820.927 | 11364.87 | 11450.71 |
| 832.42 | 2099.227 | 2113.289 | 829.62 | 3023.444 | 3043.955 | 827.378 | 4997.552 | 5069.796 | 821.208 | 11334.44 | 11461.91 |
| 834.1 | 2108.233 | 2132.744 | 829.9 | 3030.633 | 3048.298 | 827.658 | 5039.191 | 5076.252 | 821.489 | 11495.98 | 11473.12 |
| 834.38 | 2124.143 | 2135.989 | 830.18 | 3039.994 | 3052.641 | 827.939 | 5065.805 | 5082.678 | 821.769 | 11541.47 | 11484.3 |
| 834.66 | 2119.075 | 2139.237 | 830.46 | 3031.161 | 3057.013 | 828.219 | 5120.238 | 5089.104 | 822.05 | 11527.26 | 11495.51 |
| 834.94 | 2141.931 | 2142.486 | 830.74 | 3061.532 | 3061.356 | 829.9 | 5109.675 | 5127.75 | 822.331 | 11499.59 | 11506.69 |
| 835.219 | 2168.053 | 2145.734 | 831.02 | 3046.86 | 3065.699 | 830.18 | 5124.875 | 5134.177 | 822.611 | 11463.62 | 11517.87 |
| 835.499 | 2097.396 | 2148.985 | 831.3 | 3043.398 | 3070.071 | 830.46 | 5117.979 | 5140.603 | 822.892 | 11536.36 | 11529.05 |
| 835.779 | 2143.184 | 2152.237 | 831.58 | 3102.379 | 3074.414 | 830.74 | 5162.112 | 5147.059 | 823.172 | 11556.99 | 11540.2 |
| 836.059 | 2125.689 | 2155.488 | 831.86 | 3035.446 | 3078.786 | 831.02 | 5152.34 | 5153.485 | 823.733 | 11670.26 | 11562.53 |
| 836.339 | 2131.678 | 2158.742 | 832.14 | 3047.829 | 3083.129 | 831.3 | 5167.775 | 5159.911 | 824.014 | 11443.4 | 11573.69 |
| 836.618 | 2150.602 | 2161.996 | 832.42 | 3047.682 | 3087.502 | 831.58 | 5189.05 | 5166.367 | 824.294 | 11552.85 | 11584.84 |
| 836.898 | 2179.383 | 2165.254 | 832.7 | 3054.724 | 3091.844 | 831.86 | 5194.742 | 5172.793 | 824.574 | 11551 | 11595.96 |
| 837.178 | 2176.918 | 2168.511 | 832.98 | 3104.58 | 3096.217 | 832.14 | 5136.73 | 5179.219 | 824.855 | 11567.55 | 11607.11 |
| 837.457 | 2180.873 | 2171.768 | 833.26 | 3134.745 | 3100.56 | 832.42 | 5220.418 | 5185.675 | 825.135 | 11517.02 | 11618.23 |
| 837.737 | 2228.331 | 2175.028 | 833.54 | 3126.177 | 3104.932 | 832.7 | 5093.33 | 5192.101 | 825.416 | 11576.12 | 11629.35 |
| 838.017 | 2129.833 | 2178.288 | 833.82 | 3113.764 | 3109.304 | 833.26 | 5177.547 | 5204.954 | 825.696 | 11688.21 | 11640.47 |
| 838.296 | 2173.957 | 2181.548 | 834.1 | 3095.659 | 3113.647 | 833.54 | 5098.348 | 5211.41 | 825.976 | 11564.06 | 11651.59 |
| 838.576 | 2174.476 | 2184.811 | 834.38 | 3105.665 | 3118.019 | 833.82 | 5187.729 | 5217.836 | 826.257 | 11626.89 | 11662.69 |
| 838.856 | 2177.111 | 2188.074 | 834.66 | 3092.695 | 3122.391 | 834.38 | 5195.711 | 5230.689 | 826.537 | 11597.07 | 11673.78 |
| 839.135 | 2180.586 | 2191.337 | 834.94 | 3063.528 | 3126.764 | 834.66 | 5233.799 | 5237.115 | 826.817 | 11538.85 | 11684.87 |
| 839.415 | 2178.08 | 2194.603 | 835.219 | 3063.616 | 3131.107 | 834.94 | 5153.954 | 5243.571 | 827.098 | 11735.78 | 11695.96 |
| 839.694 | 2206.285 | 2197.869 | 835.499 | 3126.852 | 3135.479 | 835.219 | 5220.154 | 5249.997 | 827.378 | 11644.67 | 11707.05 |
| 839.974 | 2154.05 | 2201.135 | 836.059 | 3161.067 | 3144.223 | 835.499 | 5198.176 | 5256.423 | 827.658 | 11580.76 | 11718.12 |
| 840.253 | 2190.727 | 2204.404 | 836.339 | 3191.203 | 3148.595 | 835.779 | 5241.37 | 5262.849 | 827.939 | 11566.09 | 11729.18 |
| 840.533 | 2218.122 | 2207.673 | 836.618 | 3099.063 | 3152.968 | 836.059 | 5273.296 | 5269.276 | 828.219 | 11670.08 | 11740.24 |
| 840.812 | 2176.63 | 2210.942 | 836.898 | 3132.75 | 3157.311 | 836.339 | 5249.351 | 5275.702 | 828.499 | 11786.84 | 11751.3 |
| 841.092 | 2198.747 | 2214.214 | 837.178 | 3149.006 | 3161.683 | 836.618 | 5207.419 | 5282.128 | 828.779 | 11708.11 | 11762.37 |
| 841.371 | 2177.801 | 2217.486 | 837.457 | 3088.939 | 3166.055 | 836.898 | 5220.624 | 5288.555 | 829.059 | 11773.75 | 11773.4 |
| 841.651 | 2202.673 | 2220.757 | 837.737 | 3170.897 | 3170.427 | 838.576 | 5329.959 | 5327.112 | 829.34 | 11735.84 | 11784.43 |
| 841.93 | 2220.056 | 2224.032 | 838.017 | 3184.248 | 3174.8 | 838.856 | 5298.033 | 5333.539 | 829.62 | 11788.75 | 11795.47 |
| 842.21 | 2208.462 | 2227.307 | 838.296 | 3161.536 | 3179.172 | 839.135 | 5300.116 | 5339.965 | 829.9 | 11744.26 | 11806.5 |
| 842.489 | 2202.453 | 2230.582 | 838.576 | 3113.119 | 3183.544 | 839.415 | 5322.095 | 5346.391 | 830.18 | 11676.51 | 11817.5 |
| 842.769 | 2247.543 | 2233.86 | 838.856 | 3177.264 | 3187.946 | 839.694 | 5327.171 | 5352.818 | 830.46 | 11735.19 | 11828.54 |
| 843.048 | 2219.405 | 2237.137 | 839.415 | 3159.13 | 3196.69 | 839.974 | 5333.656 | 5359.244 | 830.74 | 11947.35 | 11839.54 |
| 844.724 | 2224.346 | 2256.821 | 839.694 | 3236.891 | 3201.062 | 840.253 | 5392.579 | 5365.67 | 831.02 | 11679.09 | 11850.54 |
| 845.003 | 2299.854 | 2260.108 | 839.974 | 3214.648 | 3205.434 | 840.533 | 5339.32 | 5372.097 | 831.3 | 11901.02 | 11861.52 |
| 845.282 | 2281.05 | 2263.391 | 841.092 | 3254.145 | 3222.953 | 841.651 | 5356.544 | 5397.772 | 831.58 | 11939.54 | 11872.52 |
| 845.562 | 2290.643 | 2266.678 | 841.371 | 3174.359 | 3227.325 | 841.93 | 5314.729 | 5404.169 | 831.86 | 11936.02 | 11883.5 |
| 845.841 | 2295.922 | 2269.964 | 841.651 | 3217.407 | 3231.727 | 842.21 | 5480.346 | 5410.596 | 832.14 | 11834.2 | 11894.47 |
| 846.12 | 2282.462 | 2273.251 | 841.93 | 3199.536 | 3236.099 | 842.489 | 5343.046 | 5417.022 | 832.42 | 11887.25 | 11905.45 |
| 846.399 | 2269.524 | 2276.54 | 842.21 | 3272.397 | 3240.471 | 842.769 | 5399.738 | 5423.419 | 832.7 | 11695.32 | 11916.42 |
| 846.678 | 2303.005 | 2279.83 | 842.489 | 3229.32 | 3244.873 | 843.048 | 5408.542 | 5429.845 | 832.98 | 11943.98 | 11927.37 |
| 846.957 | 2298.548 | 2283.119 | 842.769 | 3221.016 | 3249.245 | 843.327 | 5445.309 | 5436.272 | 833.26 | 11828.01 | 11938.31 |
| 847.236 | 2323.986 | 2286.411 | 843.048 | 3236.334 | 3253.617 | 843.607 | 5416.347 | 5442.669 | 833.54 | 11965.84 | 11949.26 |
| 847.516 | 2311.128 | 2289.701 | 843.327 | 3184.424 | 3258.019 | 843.886 | 5502.002 | 5449.095 | 833.82 | 12008.56 | 11960.2 |
| 847.795 | 2266.686 | 2292.996 | 843.607 | 3289.739 | 3262.391 | 844.165 | 5455.727 | 5455.492 | 834.1 | 11922.11 | 11971.12 |
| 848.074 | 2327.552 | 2296.289 | 845.562 | 3227.853 | 3293.114 | 844.445 | 5452.704 | 5461.918 | 834.38 | 11917.8 | 11982.06 |
| 848.353 | 2286.922 | 2299.584 | 845.841 | 3319.054 | 3297.486 | 844.724 | 5469.283 | 5468.315 | 834.66 | 11974.55 | 11992.98 |
| 848.632 | 2305.843 | 2302.879 | 846.12 | 3326.713 | 3301.888 | 845.003 | 5423.624 | 5474.741 | 834.94 | 11833.47 | 12003.9 |
| 848.911 | 2328.223 | 2306.175 | 846.399 | 3302.739 | 3306.26 | 845.282 | 5471.514 | 5481.138 | 835.219 | 11894.33 | 12014.78 |
| 850.027 | 2306.06 | 2319.368 | 846.678 | 3329.676 | 3310.662 | 845.562 | 5399.709 | 5487.535 | 835.499 | 11868.27 | 12025.7 |
| 850.585 | 2285.332 | 2325.967 | 846.957 | 3319.993 | 3315.034 | 845.841 | 5527.326 | 5493.962 | 835.779 | 11990.16 | 12036.58 |
| 850.864 | 2342.658 | 2329.268 | 847.236 | 3362.101 | 3319.435 | 846.12 | 5547.25 | 5500.359 | 836.059 | 11915.48 | 12047.47 |
| 852.815 | 2326.897 | 2352.403 | 847.516 | 3304.03 | 3323.837 | 846.399 | 5501.943 | 5506.756 | 836.339 | 11901.75 | 12058.36 |
| 853.094 | 2327.352 | 2355.71 | 847.795 | 3342.558 | 3328.209 | 846.678 | 5575.596 | 5513.182 | 836.618 | 11933.85 | 12069.21 |
| 853.373 | 2351.464 | 2359.02 | 848.074 | 3375.893 | 3332.611 | 846.957 | 5536.158 | 5519.579 | 836.898 | 12052.43 | 12080.1 |
| 853.93 | 2358.806 | 2365.64 | 848.353 | 3328.356 | 3337.012 | 847.236 | 5594.288 | 5525.976 | 837.178 | 12043.42 | 12090.96 |
| 854.209 | 2419.653 | 2368.95 | 848.632 | 3326.155 | 3341.414 | 847.516 | 5614.624 | 5532.373 | 837.457 | 11908.88 | 12101.79 |
| 854.488 | 2406.011 | 2372.26 | 848.911 | 3302.269 | 3345.786 | 847.795 | 5593.261 | 5538.77 | 837.737 | 12033.42 | 12112.64 |
| 854.767 | 2352.778 | 2375.573 | 849.19 | 3387.132 | 3350.188 | 848.074 | 5558.195 | 5545.196 | 838.017 | 12169.45 | 12123.5 |
| 855.045 | 2364.384 | 2378.886 | 849.469 | 3376.891 | 3354.589 | 848.353 | 5609.518 | 5551.593 | 838.296 | 12146.24 | 12134.33 |
| 855.602 | 2449.672 | 2385.514 | 849.748 | 3388.188 | 3358.962 | 849.19 | 5479.466 | 5570.784 | 838.576 | 12151.99 | 12145.16 |
| 855.881 | 2354.231 | 2388.83 | 850.027 | 3392.149 | 3363.363 | 850.027 | 5620.023 | 5589.975 | 838.856 | 12114.2 | 12155.98 |
| 856.16 | 2384.077 | 2392.146 | 851.979 | 3391.973 | 3394.145 | 850.306 | 5658.639 | 5596.342 | 839.135 | 12041.1 | 12166.78 |
| 856.438 | 2421.763 | 2395.462 | 852.258 | 3348.368 | 3398.546 | 850.585 | 5557.403 | 5602.739 | 839.415 | 12127.08 | 12177.58 |
| 856.717 | 2395.447 | 2398.778 | 852.537 | 3445.056 | 3402.948 | 850.864 | 5616.678 | 5609.136 | 839.694 | 12070.62 | 12188.38 |
| 856.995 | 2399.658 | 2402.097 | 853.094 | 3366.415 | 3411.751 | 851.142 | 5561.746 | 5615.533 | 839.974 | 12128.99 | 12199.18 |
| 857.274 | 2343.471 | 2405.415 | 853.373 | 3448.167 | 3416.123 | 851.421 | 5667.266 | 5621.93 | 840.253 | 12316.73 | 12209.98 |
| 857.552 | 2382.196 | 2408.734 | 853.652 | 3395.671 | 3420.525 | 851.7 | 5550.889 | 5628.298 | 840.533 | 12145.95 | 12220.75 |
| 857.831 | 2357.919 | 2412.053 | 853.93 | 3444.205 | 3424.927 | 851.979 | 5584.018 | 5634.695 | 840.812 | 12226.97 | 12231.52 |
| 858.109 | 2411.498 | 2415.375 | 854.209 | 3418.882 | 3429.328 | 852.258 | 5567.116 | 5641.092 | 841.092 | 12199.15 | 12242.28 |
| 858.388 | 2434.143 | 2418.696 | 854.488 | 3456.412 | 3433.73 | 852.537 | 5646.784 | 5647.459 | 841.371 | 12203.7 | 12253.05 |
| 858.666 | 2418.667 | 2422.018 | 854.767 | 3460.638 | 3438.131 | 852.815 | 5691.769 | 5653.856 | 841.651 | 12181.87 | 12263.82 |
| 858.945 | 2405.691 | 2425.34 | 855.045 | 3382.437 | 3442.533 | 853.094 | 5611.19 | 5660.224 | 841.93 | 12126.93 | 12274.56 |
| 859.223 | 2389.112 | 2428.662 | 855.324 | 3461.166 | 3446.934 | 853.373 | 5654.502 | 5666.621 | 842.21 | 12380.67 | 12285.3 |
| 859.501 | 2431.675 | 2431.986 | 856.438 | 3535.641 | 3464.541 | 853.652 | 5713.278 | 5672.988 | 842.489 | 12193.69 | 12296.04 |
| 860.893 | 2444.859 | 2448.613 | 857.274 | 3429.093 | 3477.745 | 853.93 | 5651.391 | 5679.385 | 844.165 | 12497.08 | 12360.28 |
| 861.171 | 2445.25 | 2451.94 | 857.552 | 3446.318 | 3482.147 | 854.209 | 5711.135 | 5685.753 | 844.445 | 12307.13 | 12370.99 |
| 861.45 | 2422.224 | 2455.268 | 857.831 | 3494.53 | 3486.549 | 854.488 | 5692.502 | 5692.121 | 844.724 | 12351.12 | 12381.64 |
| 861.728 | 2437.265 | 2458.595 | 858.109 | 3470.703 | 3490.95 | 854.767 | 5658.082 | 5698.518 | 845.003 | 12278.58 | 12392.32 |
| 862.006 | 2478.725 | 2461.926 | 858.388 | 3486.343 | 3495.352 | 855.045 | 5611.542 | 5704.885 | 845.282 | 12398.57 | 12403 |
| 862.284 | 2439.956 | 2465.256 | 858.666 | 3435.197 | 3499.753 | 855.324 | 5674.866 | 5711.253 | 845.562 | 12429.82 | 12413.65 |
| 862.563 | 2500.985 | 2468.587 | 858.945 | 3489.307 | 3504.155 | 855.602 | 5693.089 | 5717.62 | 845.841 | 12503.12 | 12424.3 |
| 862.841 | 2435.326 | 2471.917 | 859.223 | 3451.747 | 3508.557 | 855.881 | 5676.539 | 5723.988 | 846.12 | 12417.91 | 12434.96 |
| 863.119 | 2499.674 | 2475.248 | 859.501 | 3476.396 | 3512.958 | 856.16 | 5733.231 | 5730.385 | 846.399 | 12427.71 | 12445.58 |
| 863.397 | 2444.088 | 2478.578 | 859.78 | 3511.872 | 3517.36 | 856.438 | 5672.284 | 5736.753 | 846.678 | 12512.89 | 12456.2 |
| 863.675 | 2525.176 | 2481.912 | 860.058 | 3538.751 | 3521.791 | 856.717 | 5672.988 | 5743.12 | 846.957 | 12587.08 | 12466.82 |
| 865.066 | 2463.422 | 2498.582 | 860.336 | 3567.479 | 3526.192 | 856.995 | 5699.603 | 5749.488 | 847.236 | 12500.45 | 12477.45 |
| 865.344 | 2476.918 | 2501.919 | 860.615 | 3513.809 | 3530.594 | 857.274 | 5770.322 | 5755.855 | 847.516 | 12487.6 | 12488.07 |
| 865.622 | 2493.955 | 2505.255 | 861.171 | 3506.943 | 3539.397 | 857.552 | 5726.922 | 5762.194 | 847.795 | 12586.64 | 12498.66 |
| 865.9 | 2522.788 | 2508.591 | 861.45 | 3431.529 | 3543.798 | 857.831 | 5775.516 | 5768.561 | 848.074 | 12641.3 | 12509.26 |
| 867.845 | 2518.545 | 2531.961 | 861.728 | 3524.197 | 3548.2 | 858.109 | 5684.697 | 5774.929 | 848.353 | 12463.21 | 12519.85 |
| 868.123 | 2469.875 | 2535.3 | 862.006 | 3538.751 | 3552.602 | 858.666 | 5803.305 | 5787.664 | 848.632 | 12578.54 | 12530.41 |
| 868.401 | 2515.754 | 2538.642 | 862.284 | 3546.469 | 3557.003 | 859.78 | 5829.156 | 5813.076 | 848.911 | 12595.06 | 12541.01 |
| 868.679 | 2507.215 | 2541.985 | 862.563 | 3504.214 | 3561.405 | 860.058 | 5888.636 | 5819.444 | 849.19 | 12465.56 | 12551.57 |
| 868.957 | 2498.019 | 2545.327 | 862.841 | 3526.632 | 3565.806 | 860.336 | 5844.65 | 5825.782 | 849.469 | 12564.48 | 12562.13 |
| 869.79 | 2558.353 | 2555.354 | 863.119 | 3531.386 | 3570.237 | 860.615 | 5788.192 | 5832.15 | 849.748 | 12589.72 | 12572.67 |
| 870.068 | 2533.096 | 2558.699 | 865.066 | 3557.326 | 3601.048 | 860.893 | 5886.876 | 5838.488 | 850.027 | 12652.34 | 12583.23 |
| 870.345 | 2574.201 | 2562.041 | 865.344 | 3590.074 | 3605.45 | 861.171 | 5821.615 | 5844.826 | 850.306 | 12559.02 | 12593.77 |
| 870.623 | 2553.038 | 2565.386 | 865.622 | 3603.308 | 3609.851 | 861.45 | 5779.565 | 5851.194 | 850.585 | 12485.22 | 12604.3 |
| 870.901 | 2539.37 | 2568.732 | 865.9 | 3575.959 | 3614.282 | 861.728 | 5863.371 | 5857.532 | 850.864 | 12717.63 | 12614.81 |
| 871.178 | 2594.765 | 2572.077 | 866.178 | 3546.439 | 3618.684 | 862.006 | 5880.978 | 5863.87 | 851.142 | 12579.27 | 12625.34 |
| 871.456 | 2539.147 | 2575.422 | 866.456 | 3610.233 | 3623.086 | 862.284 | 5828.071 | 5870.208 | 851.421 | 12518.44 | 12635.84 |
| 871.734 | 2582.045 | 2578.77 | 866.734 | 3565.689 | 3627.487 | 862.563 | 5796.057 | 5876.547 | 851.7 | 12634.99 | 12646.35 |
| 872.011 | 2573.676 | 2582.115 | 867.012 | 3618.097 | 3631.889 | 862.841 | 5910.498 | 5882.885 | 851.979 | 12606.76 | 12656.83 |
| 873.122 | 2613.17 | 2595.508 | 867.289 | 3603.425 | 3636.29 | 863.675 | 5770.263 | 5901.9 | 852.258 | 12658.12 | 12667.33 |
| 873.399 | 2537.739 | 2598.856 | 867.567 | 3609.265 | 3640.692 | 863.953 | 5833.235 | 5908.238 | 852.537 | 12674.31 | 12677.81 |
| 873.677 | 2593.483 | 2602.204 | 867.845 | 3581.212 | 3645.093 | 864.51 | 5902.34 | 5920.915 | 852.815 | 12748.79 | 12688.28 |
| 873.954 | 2602.765 | 2605.552 | 868.123 | 3647.412 | 3649.495 | 865.344 | 5821.527 | 5939.9 | 853.094 | 12682.53 | 12698.73 |
| 874.232 | 2540.048 | 2608.903 | 868.401 | 3592.069 | 3653.897 | 865.622 | 5922.294 | 5946.209 | 853.373 | 12581.79 | 12709.2 |
| 874.509 | 2626.683 | 2612.251 | 868.679 | 3591.864 | 3658.328 | 865.9 | 5931.273 | 5952.547 | 853.652 | 12664.45 | 12719.65 |
| 874.787 | 2626.427 | 2615.603 | 868.957 | 3571.47 | 3662.729 | 866.178 | 5948.116 | 5958.856 | 853.93 | 12699.55 | 12730.1 |
| 875.064 | 2634.597 | 2618.954 | 869.79 | 3570.648 | 3675.934 | 866.456 | 5913.755 | 5965.195 | 854.209 | 12683 | 12740.51 |
| 875.342 | 2571.692 | 2622.305 | 870.345 | 3665.634 | 3684.737 | 866.734 | 5870.062 | 5971.503 | 854.488 | 12674.29 | 12750.96 |
| 875.619 | 2601.027 | 2625.656 | 870.623 | 3622.264 | 3689.139 | 867.012 | 5941.485 | 5977.812 | 854.767 | 12763.55 | 12761.38 |
| 875.896 | 2628.628 | 2629.007 | 870.901 | 3642.541 | 3693.54 | 867.289 | 5957.8 | 5984.151 | 855.045 | 12688.55 | 12771.79 |
| 876.174 | 2612.425 | 2632.361 | 871.178 | 3656.479 | 3697.942 | 867.567 | 5942.248 | 5990.46 | 855.324 | 12658.03 | 12782.21 |
| 876.451 | 2628.898 | 2635.712 | 871.456 | 3657.741 | 3702.343 | 867.845 | 5958.533 | 5996.769 | 855.602 | 12825.08 | 12792.6 |
| 876.728 | 2623.54 | 2639.066 | 871.734 | 3702.813 | 3706.745 | 868.401 | 5906.712 | 6009.386 | 855.881 | 12714.72 | 12802.99 |
| 877.006 | 2643.541 | 2642.417 | 872.011 | 3762.41 | 3711.146 | 868.679 | 5862.256 | 6015.695 | 856.16 | 12756.71 | 12813.37 |
| 877.283 | 2639.952 | 2645.771 | 872.289 | 3656.332 | 3715.548 | 868.957 | 5983.182 | 6022.004 | 856.438 | 12825.55 | 12823.76 |
| 877.56 | 2614.508 | 2649.125 | 872.567 | 3646.531 | 3719.95 | 870.623 | 6020.126 | 6059.799 | 856.717 | 12807.54 | 12834.12 |
| 877.838 | 2610.109 | 2652.479 | 872.844 | 3710.002 | 3724.351 | 870.901 | 6115.288 | 6066.108 | 856.995 | 12749.64 | 12844.48 |
| 878.115 | 2613.945 | 2655.833 | 873.122 | 3717.69 | 3728.753 | 871.178 | 6016.488 | 6072.388 | 857.274 | 12813.46 | 12854.84 |
| 878.392 | 2633.065 | 2659.187 | 873.399 | 3673.498 | 3733.154 | 871.456 | 6031.688 | 6078.697 | 857.552 | 12777.81 | 12865.2 |
| 878.669 | 2704.553 | 2662.541 | 873.677 | 3733.33 | 3737.556 | 871.734 | 6084.829 | 6084.976 | 857.831 | 12799.85 | 12875.53 |
| 878.946 | 2670.217 | 2665.898 | 873.954 | 3708.065 | 3741.958 | 872.011 | 6015.578 | 6091.285 | 858.109 | 12888.35 | 12885.85 |
| 879.224 | 2635.574 | 2669.252 | 874.232 | 3719.421 | 3746.359 | 873.122 | 6034.769 | 6116.403 | 858.388 | 12908.92 | 12896.18 |
| 879.501 | 2754.056 | 2672.609 | 874.509 | 3715.46 | 3750.761 | 873.399 | 5993.306 | 6122.683 | 858.666 | 12887.53 | 12906.51 |
| 879.778 | 2713.629 | 2675.963 | 874.787 | 3678.047 | 3755.162 | 873.677 | 6114.995 | 6128.963 | 858.945 | 12903.84 | 12916.81 |
| 880.055 | 2684.379 | 2679.32 | 875.064 | 3736.998 | 3759.564 | 873.954 | 5994.04 | 6135.242 | 859.223 | 12995.1 | 12927.11 |
| 880.332 | 2703.144 | 2682.677 | 875.342 | 3717.896 | 3763.965 | 874.232 | 6120.57 | 6141.522 | 859.501 | 12855.45 | 12937.41 |
| 880.609 | 2675.1 | 2686.034 | 875.619 | 3728.753 | 3768.367 | 874.509 | 6128.2 | 6147.801 | 859.78 | 12978.38 | 12947.71 |
| 880.886 | 2694.091 | 2689.391 | 875.896 | 3749.851 | 3772.769 | 874.787 | 6159.686 | 6154.081 | 860.058 | 12926.94 | 12957.98 |
| 881.163 | 2686.312 | 2692.747 | 876.174 | 3750.379 | 3777.17 | 875.064 | 6136.475 | 6160.331 | 860.336 | 13018.96 | 12968.25 |
| 881.44 | 2750.1 | 2696.104 | 876.451 | 3766.46 | 3781.572 | 875.342 | 6053.461 | 6166.611 | 860.615 | 13032.81 | 12978.52 |
| 881.718 | 2722.411 | 2699.461 | 876.728 | 3728.518 | 3785.973 | 875.619 | 6141.639 | 6172.861 | 860.893 | 12822.32 | 12988.76 |
| 881.995 | 2660.686 | 2702.818 | 877.006 | 3809.243 | 3790.375 | 875.896 | 6166.053 | 6179.141 | 861.171 | 12913.97 | 12999.03 |
| 882.272 | 2703.531 | 2706.178 | 877.283 | 3767.604 | 3794.776 | 877.006 | 6237.564 | 6204.171 | 861.45 | 13013.73 | 13009.27 |
| 882.548 | 2747.142 | 2709.535 | 877.56 | 3824.766 | 3799.149 | 879.501 | 6261.157 | 6260.423 | 861.728 | 13002.11 | 13019.49 |
| 882.825 | 2726.924 | 2712.895 | 877.838 | 3802.318 | 3803.55 | 879.778 | 6341.471 | 6266.644 | 862.006 | 13004.17 | 13029.73 |
| 883.102 | 2724.354 | 2716.252 | 878.115 | 3777.434 | 3807.952 | 880.055 | 6303.118 | 6272.894 | 862.284 | 12941.72 | 13039.94 |
| 883.379 | 2763.531 | 2719.612 | 878.392 | 3791.813 | 3812.353 | 880.332 | 6221.865 | 6279.115 | 862.563 | 12938.03 | 13050.15 |
| 884.21 | 2671.456 | 2729.688 | 878.669 | 3805.017 | 3816.755 | 880.886 | 6350.831 | 6291.586 | 862.841 | 13132.93 | 13060.36 |
| 884.487 | 2737.368 | 2733.048 | 878.946 | 3787.352 | 3821.157 | 881.718 | 6224.8 | 6310.249 | 863.675 | 12888.38 | 13090.91 |
| 884.764 | 2753.219 | 2736.408 | 879.224 | 3823.152 | 3825.529 | 881.995 | 6309.046 | 6316.499 | 864.231 | 13001.82 | 13111.24 |
| 885.04 | 2730.211 | 2739.768 | 879.501 | 3782.775 | 3829.93 | 882.272 | 6373.72 | 6322.72 | 864.51 | 12980.81 | 13121.43 |
| 885.317 | 2716.994 | 2743.128 | 879.778 | 3877.732 | 3834.332 | 882.548 | 6411.28 | 6328.912 | 864.788 | 13054.52 | 13131.55 |
| 885.594 | 2745.197 | 2746.488 | 880.055 | 3838.499 | 3838.734 | 882.825 | 6337.597 | 6335.133 | 865.066 | 13040.7 | 13141.7 |
| 885.871 | 2712.907 | 2749.848 | 880.332 | 3924.388 | 3843.135 | 883.102 | 6362.011 | 6341.353 | 865.344 | 13035.13 | 13151.83 |
| 886.148 | 2765.303 | 2753.208 | 880.609 | 3822.624 | 3847.507 | 883.379 | 6421.932 | 6347.574 | 865.622 | 13056.37 | 13161.98 |
| 886.424 | 2731.54 | 2756.57 | 881.995 | 3801.056 | 3869.486 | 883.656 | 6414.273 | 6353.766 | 865.9 | 13000.12 | 13172.07 |
| 886.701 | 2784.262 | 2759.93 | 882.272 | 3879.756 | 3873.887 | 884.487 | 6260.716 | 6372.399 | 866.178 | 13064.09 | 13182.2 |
| 886.978 | 2792.176 | 2763.29 | 882.548 | 3818.046 | 3878.26 | 884.764 | 6262.037 | 6378.591 | 866.456 | 13129.58 | 13192.29 |
| 887.254 | 2745.405 | 2766.653 | 882.825 | 3816.168 | 3882.661 | 885.04 | 6332.697 | 6384.812 | 866.734 | 13115.26 | 13202.39 |
| 887.531 | 2725.976 | 2770.013 | 883.102 | 3892.081 | 3887.034 | 885.317 | 6323.776 | 6391.003 | 867.012 | 13029.38 | 13212.48 |
| 888.914 | 2781.912 | 2786.821 | 883.379 | 3887.386 | 3891.435 | 885.594 | 6373.778 | 6397.195 | 867.289 | 13119.67 | 13222.55 |
| 889.191 | 2817.145 | 2790.184 | 883.656 | 3919.547 | 3895.837 | 885.871 | 6368.32 | 6403.386 | 867.567 | 13091.79 | 13232.64 |
| 889.467 | 2808.183 | 2793.544 | 883.933 | 3829.167 | 3900.209 | 886.148 | 6427.536 | 6409.578 | 867.845 | 13048.68 | 13242.7 |
| 889.744 | 2830.118 | 2796.906 | 884.21 | 3892.462 | 3904.61 | 886.424 | 6429.209 | 6415.769 | 868.123 | 13106.73 | 13252.74 |
| 890.02 | 2793.479 | 2800.269 | 884.487 | 3798.943 | 3908.983 | 886.701 | 6493.619 | 6421.961 | 868.401 | 13085.16 | 13262.81 |
| 890.297 | 2843.135 | 2803.629 | 884.764 | 3945.897 | 3913.384 | 886.978 | 6421.521 | 6428.153 | 868.679 | 13003.44 | 13272.84 |
| 890.573 | 2768.877 | 2806.992 | 885.04 | 3935.979 | 3917.757 | 887.254 | 6502.95 | 6434.344 | 868.957 | 13300.22 | 13282.88 |
| 890.85 | 2864.879 | 2810.355 | 885.317 | 3905.696 | 3922.158 | 887.531 | 6459.756 | 6440.506 | 869.234 | 13136.3 | 13292.88 |
| 893.061 | 2871.942 | 2837.254 | 885.594 | 3889.088 | 3926.53 | 887.808 | 6531.824 | 6446.698 | 872.011 | 13264.07 | 13392.71 |
| 893.337 | 2836.022 | 2840.617 | 885.871 | 3916.26 | 3930.932 | 888.084 | 6518.179 | 6452.86 | 872.289 | 13337.78 | 13402.63 |
| 893.613 | 2846.691 | 2843.98 | 886.148 | 3973.363 | 3935.304 | 889.467 | 6419.936 | 6483.73 | 872.567 | 13406.5 | 13412.58 |
| 893.889 | 2881.602 | 2847.343 | 886.424 | 3977.677 | 3939.706 | 889.744 | 6522.816 | 6489.892 | 872.844 | 13461.87 | 13422.49 |
| 894.166 | 2848.323 | 2850.705 | 886.701 | 3966.585 | 3944.078 | 890.02 | 6569.443 | 6496.054 | 873.122 | 13387.84 | 13432.38 |
| 894.442 | 2879.685 | 2854.068 | 887.254 | 4040.971 | 3952.852 | 890.297 | 6555.945 | 6502.216 | 873.399 | 13392.09 | 13442.3 |
| 894.718 | 2935.324 | 2857.431 | 887.531 | 3929.611 | 3957.224 | 890.573 | 6554.126 | 6508.349 | 873.677 | 13397.49 | 13452.19 |
| 894.994 | 2947.355 | 2860.794 | 887.808 | 4027.767 | 3961.626 | 890.85 | 6589.25 | 6514.511 | 873.954 | 13429.36 | 13462.08 |
| 895.27 | 2886.878 | 2864.157 | 888.084 | 3974.713 | 3965.998 | 891.126 | 6606.769 | 6520.674 | 874.232 | 13411.61 | 13471.97 |
| 895.547 | 2922.442 | 2867.519 | 888.361 | 4035.161 | 3970.37 | 891.402 | 6529.242 | 6526.807 | 874.509 | 13497.7 | 13481.83 |
| 895.823 | 2944.597 | 2870.882 | 888.637 | 3960.158 | 3974.772 | 891.679 | 6581.914 | 6532.969 | 874.787 | 13439.51 | 13491.69 |
| 896.099 | 2898.201 | 2874.245 | 888.914 | 3966.996 | 3979.144 | 891.955 | 6605.712 | 6539.102 | 875.064 | 13386.46 | 13501.55 |
| 896.375 | 2915.247 | 2877.608 | 889.191 | 3950.563 | 3983.516 | 892.232 | 6478.448 | 6545.264 | 875.342 | 13387.55 | 13511.41 |
| 896.651 | 2905.353 | 2880.971 | 889.467 | 3976.092 | 3987.918 | 892.508 | 6481.793 | 6551.397 | 875.619 | 13491.28 | 13521.24 |
| 896.927 | 2935.559 | 2884.334 | 889.744 | 4006.61 | 3992.29 | 892.784 | 6572.7 | 6557.53 | 875.896 | 13470.12 | 13531.07 |
| 897.203 | 2894.134 | 2887.696 | 890.02 | 3940.205 | 3996.662 | 893.061 | 6466.27 | 6563.662 | 876.174 | 13390.57 | 13540.9 |
| 897.479 | 2921.536 | 2891.059 | 890.297 | 4011.686 | 4001.034 | 894.166 | 6553.275 | 6588.194 | 876.451 | 13536.55 | 13550.73 |
| 897.755 | 2898.733 | 2894.419 | 890.573 | 4026.828 | 4005.407 | 894.442 | 6651.401 | 6594.327 | 876.728 | 13524.73 | 13560.53 |
| 898.031 | 2923.819 | 2897.782 | 890.85 | 4038.33 | 4009.779 | 894.718 | 6672.939 | 6600.46 | 877.006 | 13526.11 | 13570.33 |
| 898.307 | 2924.934 | 2901.145 | 891.126 | 4068.173 | 4014.18 | 895.27 | 6580.917 | 6612.696 | 877.283 | 13521.35 | 13580.1 |
| 898.583 | 2826.726 | 2904.507 | 891.402 | 4124.161 | 4018.553 | 895.547 | 6554.448 | 6618.8 | 877.56 | 13467.6 | 13589.9 |
| 898.859 | 2942.044 | 2907.87 | 891.679 | 4022.749 | 4022.925 | 895.823 | 6629.334 | 6624.903 | 877.838 | 13585.41 | 13599.67 |
| 899.135 | 2951.61 | 2911.233 | 891.955 | 4037.949 | 4027.297 | 896.099 | 6656.8 | 6631.036 | 878.115 | 13550.84 | 13609.44 |
| 899.411 | 2944.333 | 2914.593 | 892.232 | 4021.252 | 4031.669 | 896.375 | 6783.125 | 6637.139 | 878.392 | 13483.68 | 13619.19 |
| 899.687 | 2935.148 | 2917.956 | 892.508 | 4087.1 | 4036.042 | 896.651 | 6686.907 | 6643.243 | 878.669 | 13668.66 | 13628.93 |
| 899.963 | 2978.02 | 2921.319 | 892.784 | 4023.365 | 4040.414 | 896.927 | 6776.317 | 6649.346 | 878.946 | 13611.56 | 13638.67 |
| 900.238 | 2967.045 | 2924.678 | 893.337 | 4133.903 | 4049.158 | 897.203 | 6747.267 | 6655.45 | 879.224 | 13482.03 | 13648.41 |
| 900.79 | 2970.009 | 2931.404 | 894.166 | 4130.265 | 4062.275 | 897.479 | 6687.64 | 6661.554 | 879.501 | 13533.88 | 13658.13 |
| 901.066 | 2858.109 | 2934.767 | 894.442 | 4122.225 | 4066.618 | 897.755 | 6797.768 | 6667.628 | 879.778 | 13541.69 | 13667.87 |
| 901.342 | 2974 | 2938.112 | 894.718 | 4103.973 | 4070.99 | 898.031 | 6656.917 | 6673.731 | 880.055 | 13613.55 | 13677.55 |
| 901.617 | 2929.579 | 2941.487 | 894.994 | 4128.621 | 4075.362 | 898.307 | 6760.442 | 6679.835 | 880.332 | 13761.06 | 13687.26 |
| 901.893 | 2948.999 | 2944.861 | 895.27 | 4059.429 | 4079.735 | 898.583 | 6712.436 | 6685.909 | 880.609 | 13567.34 | 13696.95 |
| 902.169 | 2971.388 | 2948.206 | 895.547 | 4091.824 | 4084.107 | 898.859 | 6767.837 | 6692.012 | 880.886 | 13812.94 | 13706.63 |
| 902.445 | 2976.23 | 2951.581 | 895.823 | 4114.536 | 4088.45 | 899.135 | 6719.83 | 6698.087 | 881.163 | 13617.63 | 13716.31 |
| 902.72 | 2962.82 | 2954.926 | 896.099 | 4113.392 | 4092.822 | 900.238 | 6653.484 | 6722.413 | 881.44 | 13693.16 | 13725.97 |
| 902.996 | 2985.766 | 2958.301 | 896.375 | 4147.636 | 4097.194 | 900.514 | 6754.368 | 6728.458 | 881.718 | 13553.57 | 13735.65 |
| 903.272 | 2912.008 | 2961.646 | 896.651 | 4063.595 | 4101.537 | 900.79 | 6726.345 | 6734.532 | 881.995 | 13863.03 | 13745.28 |
| 903.547 | 2973.442 | 2965.02 | 896.927 | 4139.978 | 4105.909 | 901.066 | 6707.623 | 6740.606 | 882.825 | 13623.73 | 13774.18 |
| 903.823 | 2875.029 | 2968.366 | 897.203 | 4163.981 | 4110.282 | 901.342 | 6781.805 | 6746.68 | 883.102 | 13760.13 | 13783.81 |
| 904.099 | 2984.915 | 2971.74 | 897.479 | 4096.549 | 4114.624 | 901.617 | 6678.309 | 6752.725 | 883.379 | 13721.27 | 13793.4 |
| 904.374 | 3020.598 | 2975.085 | 897.755 | 4134.49 | 4118.997 | 901.893 | 6816.137 | 6758.799 | 883.656 | 13909.69 | 13803.03 |
| 904.65 | 3009.065 | 2978.46 | 898.031 | 4078.15 | 4123.34 | 902.169 | 6760.824 | 6764.844 | 883.933 | 13725.91 | 13812.59 |
| 905.201 | 2933.98 | 2985.18 | 898.307 | 4119.261 | 4127.712 | 902.445 | 6822.769 | 6770.889 | 884.21 | 13703.11 | 13822.19 |
| 905.476 | 2975.496 | 2988.525 | 898.583 | 4090.592 | 4132.084 | 902.72 | 6760.296 | 6776.963 | 884.487 | 13751 | 13831.75 |
| 906.027 | 2987.762 | 2995.244 | 898.859 | 4217.357 | 4136.427 | 902.996 | 6825.439 | 6783.008 | 884.764 | 13993.94 | 13841.32 |
| 906.303 | 3055.517 | 2998.619 | 901.066 | 4161.721 | 4171.258 | 903.272 | 6744.685 | 6789.053 | 885.594 | 13785.74 | 13869.99 |
| 906.578 | 2987.439 | 3001.964 | 901.342 | 4132.964 | 4175.601 | 903.547 | 6772.444 | 6795.097 | 885.871 | 13837.42 | 13879.53 |
| 906.854 | 3061.004 | 3005.309 | 901.617 | 4163.57 | 4179.944 | 903.823 | 6788.612 | 6801.142 | 886.148 | 13873.77 | 13889.03 |
| 907.129 | 3068.751 | 3008.684 | 901.893 | 4202.128 | 4184.287 | 904.099 | 6850.851 | 6807.187 | 886.424 | 13992.94 | 13898.54 |
| 907.404 | 2997.944 | 3012.029 | 902.169 | 4147.108 | 4188.659 | 904.374 | 6887.12 | 6813.203 | 886.701 | 13875.15 | 13908.05 |
| 907.68 | 3061.268 | 3015.404 | 902.445 | 4131.321 | 4193.002 | 904.65 | 6787.204 | 6819.247 | 886.978 | 13891.17 | 13917.55 |
| 907.955 | 3060.975 | 3018.749 | 902.72 | 4212.838 | 4197.345 | 904.925 | 6866.99 | 6825.263 | 887.254 | 14024.95 | 13927.06 |
| 908.23 | 3021.184 | 3022.123 | 902.996 | 4232.029 | 4201.688 | 905.201 | 6901.821 | 6831.308 | 887.531 | 13912.92 | 13936.54 |
| 908.506 | 3023.62 | 3025.469 | 903.272 | 4231.325 | 4206.031 | 905.476 | 6902.437 | 6837.323 | 887.808 | 14096.67 | 13946.02 |
| 908.781 | 2995.655 | 3028.814 | 904.374 | 4253.215 | 4223.402 | 905.752 | 6926.822 | 6843.339 | 888.084 | 14006.11 | 13955.47 |
| 909.056 | 3041.549 | 3032.188 | 904.65 | 4224.224 | 4227.745 | 906.027 | 6858.568 | 6849.384 | 888.361 | 14117.06 | 13964.92 |
| 909.332 | 2998.414 | 3035.534 | 904.925 | 4197.433 | 4232.059 | 906.303 | 6873.181 | 6855.399 | 888.637 | 14053.36 | 13974.36 |
| 909.607 | 3041.637 | 3038.879 | 905.201 | 4263.544 | 4236.401 | 906.578 | 6958.396 | 6861.415 | 888.914 | 13933.52 | 13983.81 |
| 909.882 | 3030.868 | 3042.253 | 905.476 | 4308.969 | 4240.744 | 906.854 | 6847.447 | 6867.43 | 889.191 | 14082.76 | 13993.23 |
| 910.157 | 3070.394 | 3045.599 | 905.752 | 4257.881 | 4245.087 | 907.129 | 6881.134 | 6873.416 | 889.467 | 14090.48 | 14002.68 |
| 910.433 | 3069.22 | 3048.973 | 906.027 | 4278.216 | 4249.401 | 907.404 | 6809.388 | 6879.432 | 889.744 | 14178.39 | 14012.07 |
| 910.708 | 3037.764 | 3052.318 | 906.303 | 4297.437 | 4253.744 | 907.68 | 6909.626 | 6885.447 | 890.02 | 13967.79 | 14021.49 |
| 910.983 | 3062.53 | 3055.663 | 906.578 | 4323.993 | 4258.087 | 907.955 | 6933.219 | 6891.433 | 890.297 | 14085.87 | 14030.88 |
| 911.258 | 3028.315 | 3059.038 | 906.854 | 4375.462 | 4262.4 | 908.23 | 6911.592 | 6897.449 | 890.573 | 13941.94 | 14040.27 |
| 911.533 | 3070.599 | 3062.383 | 907.129 | 4284.731 | 4266.743 | 908.506 | 6976.032 | 6903.435 | 890.85 | 14235.38 | 14049.66 |
| 911.808 | 3124.269 | 3065.728 | 907.404 | 4330.801 | 4271.086 | 908.781 | 6858.979 | 6909.421 | 891.126 | 13995.87 | 14059.02 |
| 912.083 | 3084.743 | 3069.074 | 907.68 | 4218.648 | 4275.399 | 909.056 | 6923.33 | 6915.407 | 891.402 | 14154.86 | 14068.38 |
| 912.358 | 3103.171 | 3072.448 | 907.955 | 4287.518 | 4279.742 | 909.332 | 6956.313 | 6921.393 | 891.679 | 14072.46 | 14077.74 |
| 912.633 | 3083.217 | 3075.793 | 908.23 | 4231.736 | 4284.056 | 909.607 | 7026.826 | 6927.379 | 891.955 | 14191.27 | 14087.07 |
| 912.908 | 3037.412 | 3079.139 | 908.506 | 4254.536 | 4288.399 | 909.882 | 6959.247 | 6933.366 | 892.232 | 14029.33 | 14096.41 |
| 913.183 | 3077.525 | 3082.484 | 908.781 | 4222.404 | 4292.712 | 910.157 | 6914.468 | 6939.352 | 892.508 | 14005.23 | 14105.74 |
| 913.458 | 3049.237 | 3085.858 | 909.056 | 4338.049 | 4297.026 | 910.433 | 6895.806 | 6945.338 | 892.784 | 14253.31 | 14115.07 |
| 913.733 | 3088.851 | 3089.203 | 909.332 | 4342.509 | 4301.369 | 910.708 | 7007.928 | 6951.295 | 893.061 | 14087.87 | 14124.37 |
| 914.008 | 3073.006 | 3092.549 | 909.607 | 4325.636 | 4305.682 | 910.983 | 7006.373 | 6957.281 | 893.337 | 14152.39 | 14133.67 |
| 914.283 | 3116.493 | 3095.894 | 909.882 | 4330.067 | 4309.996 | 911.258 | 7022.982 | 6963.238 | 893.613 | 14185.52 | 14142.97 |
| 914.558 | 3070.893 | 3099.239 | 910.157 | 4244.852 | 4314.339 | 911.533 | 6993.609 | 6969.224 | 893.889 | 14017.85 | 14152.25 |
| 914.833 | 3156.518 | 3102.614 | 910.433 | 4317.097 | 4318.652 | 911.808 | 7012.917 | 6975.181 | 894.166 | 14088.39 | 14161.52 |
| 915.108 | 3053.962 | 3105.959 | 910.708 | 4353.278 | 4322.966 | 912.083 | 6980.169 | 6981.137 | 894.442 | 14007.61 | 14170.79 |
| 915.383 | 3111.505 | 3109.304 | 910.983 | 4302.132 | 4327.279 | 912.358 | 7063.271 | 6987.094 | 894.718 | 14315.99 | 14180.06 |
| 915.658 | 3127.262 | 3112.649 | 911.258 | 4287.548 | 4331.593 | 912.633 | 6947.48 | 6993.051 | 896.099 | 14194.53 | 14226.22 |
| 915.933 | 3157.105 | 3115.994 | 911.533 | 4389.459 | 4335.906 | 912.908 | 7022.747 | 6999.008 | 896.375 | 14287.84 | 14235.44 |
| 916.207 | 3182.81 | 3119.34 | 911.808 | 4350.432 | 4340.22 | 913.183 | 7131.818 | 7004.935 | 896.651 | 14374.5 | 14244.62 |
| 917.856 | 3185.07 | 3139.44 | 912.083 | 4380.274 | 4344.534 | 913.458 | 6982.164 | 7010.892 | 896.927 | 14445.86 | 14253.81 |
| 918.131 | 3166.495 | 3142.785 | 912.358 | 4389.43 | 4348.847 | 913.733 | 7080.085 | 7016.849 | 897.203 | 14337.2 | 14262.99 |
| 918.405 | 3196.191 | 3146.131 | 912.633 | 4372.616 | 4353.161 | 914.008 | 7076.123 | 7022.776 | 897.479 | 14336.03 | 14272.18 |
| 918.68 | 3284.692 | 3149.476 | 912.908 | 4394.682 | 4357.474 | 914.283 | 7009.718 | 7028.704 | 897.755 | 14294.65 | 14281.33 |
| 919.504 | 3129.903 | 3159.511 | 913.183 | 4379.159 | 4361.788 | 914.558 | 6943.929 | 7034.661 | 898.031 | 14312.05 | 14290.49 |
| 919.778 | 3129.669 | 3162.857 | 913.458 | 4372.117 | 4366.101 | 914.833 | 7034.074 | 7040.588 | 898.307 | 14470.95 | 14299.64 |
| 920.053 | 3163.414 | 3166.202 | 913.733 | 4346.734 | 4370.415 | 915.108 | 7054.38 | 7046.516 | 898.583 | 14337.93 | 14308.77 |
| 920.328 | 3188.415 | 3169.547 | 915.383 | 4448.235 | 4396.208 | 915.383 | 6963.942 | 7052.443 | 898.859 | 14452.2 | 14317.89 |
| 920.602 | 3272.456 | 3172.892 | 915.658 | 4417.336 | 4400.522 | 915.658 | 7195.348 | 7058.37 | 899.135 | 14393.31 | 14327.02 |
| 920.877 | 3214.678 | 3176.237 | 915.933 | 4410.557 | 4404.806 | 915.933 | 7161.661 | 7064.298 | 899.411 | 14187.96 | 14336.12 |
| 921.151 | 3217.935 | 3179.553 | 916.207 | 4431.773 | 4409.119 | 916.207 | 7098.425 | 7070.196 | 899.687 | 14459.51 | 14345.24 |
| 921.426 | 3147.862 | 3182.898 | 916.482 | 4410.763 | 4413.404 | 916.482 | 7155.264 | 7076.123 | 899.963 | 14424.06 | 14354.31 |
| 921.7 | 3249.362 | 3186.244 | 916.757 | 4486.47 | 4417.717 | 916.757 | 7129.94 | 7082.022 | 900.238 | 14344.07 | 14363.41 |
| 921.974 | 3260.777 | 3189.589 | 917.032 | 4348.524 | 4422.001 | 917.032 | 7157.787 | 7087.949 | 900.514 | 14406.92 | 14372.47 |
| 922.249 | 3281.582 | 3192.934 | 917.307 | 4436.116 | 4426.286 | 917.307 | 7181.086 | 7093.847 | 900.79 | 14420.45 | 14381.54 |
| 922.523 | 3223.07 | 3196.279 | 917.581 | 4555.134 | 4430.599 | 917.581 | 7177.829 | 7099.745 | 901.066 | 14297.06 | 14390.61 |
| 922.798 | 3218.052 | 3199.595 | 917.856 | 4421.297 | 4434.883 | 917.856 | 7054.321 | 7105.643 | 901.342 | 14555.58 | 14399.64 |
| 923.072 | 3267.849 | 3202.94 | 918.131 | 4432.859 | 4439.167 | 918.131 | 7165.916 | 7111.542 | 901.617 | 14400.17 | 14408.68 |
| 923.346 | 3263.858 | 3206.285 | 918.405 | 4560.093 | 4443.452 | 918.405 | 7128.473 | 7117.44 | 901.893 | 14514.09 | 14417.72 |
| 923.621 | 3242.202 | 3209.631 | 918.68 | 4443.305 | 4447.736 | 918.68 | 7110.749 | 7123.338 | 902.169 | 14402.58 | 14426.76 |
| 923.895 | 3255.554 | 3212.947 | 918.955 | 4531.131 | 4452.02 | 918.955 | 7248.284 | 7129.236 | 902.445 | 14661.1 | 14435.77 |
| 924.169 | 3234.103 | 3216.292 | 919.229 | 4424.055 | 4456.304 | 919.229 | 7093.7 | 7135.105 | 902.72 | 14616.35 | 14444.78 |
| 924.444 | 3345.933 | 3219.637 | 919.504 | 4455.453 | 4460.588 | 919.778 | 7221.816 | 7146.872 | 902.996 | 14594.58 | 14453.76 |
| 924.718 | 3287.832 | 3222.953 | 919.778 | 4470.037 | 4464.873 | 920.602 | 7022.512 | 7164.478 | 903.272 | 14390.58 | 14462.76 |
| 924.992 | 3226.415 | 3226.298 | 921.426 | 4542.076 | 4490.549 | 920.877 | 7180.588 | 7170.347 | 903.547 | 14502.58 | 14471.74 |
| 925.266 | 3258.371 | 3229.643 | 921.7 | 4558.656 | 4494.833 | 921.151 | 7271.436 | 7176.215 | 903.823 | 14505.43 | 14480.69 |
| 925.541 | 3278.119 | 3232.959 | 921.974 | 4526.055 | 4499.088 | 921.426 | 7202.009 | 7182.084 | 904.099 | 14607.34 | 14489.67 |
| 925.815 | 3297.222 | 3236.304 | 922.249 | 4498.589 | 4503.372 | 921.7 | 7363.194 | 7187.953 | 904.374 | 14588.8 | 14498.62 |
| 926.089 | 3221.163 | 3239.649 | 922.523 | 4634.744 | 4507.627 | 921.974 | 7297.229 | 7193.792 | 904.65 | 14457.19 | 14507.54 |
| 926.363 | 3208.545 | 3242.965 | 922.798 | 4663.325 | 4511.911 | 923.072 | 7146.373 | 7217.179 | 904.925 | 14528.23 | 14516.49 |
| 926.637 | 3290.854 | 3246.31 | 923.072 | 4665.761 | 4516.166 | 923.346 | 7263.836 | 7223.048 | 905.201 | 14493.13 | 14525.41 |
| 926.911 | 3248.1 | 3249.626 | 924.444 | 4496.183 | 4537.499 | 923.621 | 7284.025 | 7228.858 | 905.476 | 14562.39 | 14534.33 |
| 927.185 | 3323.397 | 3252.972 | 924.718 | 4669.986 | 4541.754 | 923.895 | 7216.064 | 7234.698 | 905.752 | 14589.18 | 14543.22 |
| 927.459 | 3231.462 | 3256.287 | 924.992 | 4544.013 | 4546.008 | 924.169 | 7404.657 | 7240.537 | 906.027 | 14672.63 | 14552.14 |
| 928.556 | 3253.5 | 3269.61 | 925.266 | 4589.525 | 4550.263 | 924.444 | 7338.604 | 7246.377 | 906.303 | 14589.18 | 14561.01 |
| 928.83 | 3311.424 | 3272.955 | 925.541 | 4598.505 | 4554.518 | 924.718 | 7280.357 | 7252.187 | 906.578 | 14593.75 | 14569.9 |
| 929.103 | 3332.171 | 3276.271 | 925.815 | 4509.886 | 4558.773 | 924.992 | 7212.514 | 7258.026 | 906.854 | 14737.28 | 14578.76 |
| 929.377 | 3296.195 | 3279.586 | 926.089 | 4530.221 | 4563.028 | 925.266 | 7335.787 | 7263.836 | 907.129 | 14720.99 | 14587.62 |
| 929.651 | 3327.476 | 3282.932 | 926.363 | 4624.268 | 4567.283 | 925.541 | 7308.116 | 7269.646 | 907.404 | 14651.47 | 14596.48 |
| 929.925 | 3356.027 | 3286.247 | 926.637 | 4618.869 | 4571.538 | 925.815 | 7183.346 | 7275.456 | 907.68 | 14589.76 | 14605.35 |
| 930.199 | 3360.487 | 3289.563 | 926.911 | 4667.873 | 4575.792 | 926.089 | 7408.824 | 7281.266 | 907.955 | 14683.61 | 14614.18 |
| 930.473 | 3334.577 | 3292.909 | 927.185 | 4602.055 | 4580.047 | 926.911 | 7225.601 | 7298.697 | 908.23 | 14660.16 | 14622.98 |
| 930.747 | 3438.278 | 3296.224 | 927.459 | 4593.311 | 4584.273 | 927.185 | 7397.849 | 7304.477 | 908.506 | 14767.03 | 14631.81 |
| 931.021 | 3347.195 | 3299.54 | 927.733 | 4697.628 | 4588.528 | 927.459 | 7426.02 | 7310.287 | 908.781 | 14497.83 | 14640.62 |
| 931.295 | 3330.586 | 3302.856 | 928.008 | 4660.097 | 4592.783 | 927.733 | 7448.145 | 7316.068 | 909.056 | 14591.14 | 14649.42 |
| 931.568 | 3419.146 | 3306.201 | 928.282 | 4679.934 | 4597.008 | 928.008 | 7324.402 | 7321.849 | 909.332 | 14790.8 | 14658.22 |
| 931.842 | 3368.674 | 3309.517 | 928.556 | 4639.674 | 4601.263 | 928.282 | 7358.675 | 7327.63 | 909.607 | 14691.32 | 14667 |
| 932.116 | 3388.188 | 3312.833 | 928.83 | 4642.198 | 4605.518 | 928.556 | 7447.793 | 7333.41 | 909.882 | 14609.92 | 14675.77 |
| 932.39 | 3345.434 | 3316.149 | 929.103 | 4737.682 | 4609.743 | 928.83 | 7497.237 | 7339.191 | 910.157 | 14673.1 | 14684.54 |
| 932.663 | 3367.647 | 3319.465 | 929.377 | 4644.927 | 4613.969 | 929.103 | 7407.797 | 7344.972 | 910.433 | 14519.6 | 14693.29 |
| 932.937 | 3344.29 | 3322.781 | 929.651 | 4663.413 | 4618.224 | 929.377 | 7360.172 | 7350.753 | 910.708 | 14762.57 | 14702.03 |
| 933.211 | 3307.463 | 3326.126 | 929.925 | 4612.678 | 4622.449 | 929.651 | 7460.176 | 7356.504 | 910.983 | 14788.42 | 14710.78 |
| 933.484 | 3344.436 | 3329.442 | 930.199 | 4714.149 | 4626.675 | 929.925 | 7393.154 | 7362.285 | 911.258 | 14807.79 | 14719.49 |
| 933.758 | 3348.31 | 3332.757 | 930.473 | 4726.004 | 4630.93 | 930.199 | 7337.46 | 7368.036 | 911.533 | 14635.22 | 14728.21 |
| 934.032 | 3327.505 | 3336.073 | 930.747 | 4719.343 | 4635.155 | 930.473 | 7382.825 | 7373.817 | 911.808 | 14799.4 | 14736.92 |
| 934.305 | 3241.938 | 3339.389 | 931.021 | 4723.685 | 4639.381 | 930.747 | 7594.747 | 7379.568 | 912.083 | 14806.12 | 14745.64 |
| 934.579 | 3429.768 | 3342.705 | 931.295 | 4667.081 | 4643.606 | 931.021 | 7557.333 | 7385.32 | 912.358 | 14947.49 | 14754.32 |
| 934.853 | 3276.359 | 3346.021 | 931.568 | 4664.235 | 4647.832 | 931.568 | 7393.213 | 7396.822 | 912.633 | 14680 | 14763.01 |
| 935.126 | 3319.787 | 3349.337 | 931.842 | 4742.172 | 4652.057 | 933.484 | 7358.47 | 7436.936 | 912.908 | 14906.68 | 14771.67 |
| 935.4 | 3352.535 | 3352.653 | 932.116 | 4588.176 | 4656.283 | 933.758 | 7492.219 | 7442.658 | 913.183 | 14810.55 | 14780.32 |
| 935.673 | 3357.172 | 3355.968 | 932.39 | 4672.04 | 4660.508 | 934.032 | 7621.831 | 7448.38 | 913.458 | 14782.64 | 14788.98 |
| 935.947 | 3374.69 | 3359.255 | 932.663 | 4685.744 | 4664.734 | 934.305 | 7387.726 | 7454.102 | 913.733 | 15055.16 | 14797.64 |
| 936.22 | 3428.859 | 3362.571 | 932.937 | 4677.029 | 4668.959 | 934.579 | 7550.115 | 7459.794 | 914.008 | 14982.15 | 14806.26 |
| 936.494 | 3409.756 | 3365.887 | 933.211 | 4734.103 | 4673.155 | 934.853 | 7496.944 | 7465.516 | 914.833 | 14686.89 | 14832.14 |
| 936.767 | 3398.634 | 3369.203 | 933.484 | 4753.176 | 4677.381 | 935.126 | 7573.15 | 7471.209 | 915.108 | 14924.05 | 14840.74 |
| 937.04 | 3314.535 | 3372.518 | 933.758 | 4683.103 | 4681.606 | 935.4 | 7451.314 | 7476.931 | 915.383 | 15101.4 | 14849.34 |
| 937.314 | 3384.461 | 3375.834 | 934.032 | 4702.529 | 4685.803 | 935.673 | 7547.767 | 7482.624 | 915.658 | 14914.92 | 14857.91 |
| 937.587 | 3470.409 | 3379.121 | 934.305 | 4692.786 | 4690.028 | 935.947 | 7455.217 | 7488.317 | 915.933 | 14879.65 | 14866.48 |
| 937.861 | 3458.965 | 3382.437 | 934.579 | 4783.547 | 4694.254 | 936.22 | 7591.46 | 7494.009 | 916.207 | 14823.11 | 14875.04 |
| 938.134 | 3418.735 | 3385.752 | 934.853 | 4792.291 | 4698.45 | 936.494 | 7490.459 | 7499.702 | 916.482 | 14844 | 14883.61 |
| 938.407 | 3527.982 | 3389.039 | 935.126 | 4727.089 | 4702.646 | 936.767 | 7527.373 | 7505.365 | 916.757 | 14912.25 | 14892.15 |
| 938.681 | 3385.43 | 3392.355 | 935.4 | 4775.155 | 4706.871 | 937.04 | 7659.509 | 7511.058 | 917.032 | 15039.78 | 14900.69 |
| 938.954 | 3427.186 | 3395.671 | 935.673 | 4869.055 | 4711.068 | 937.314 | 7699.534 | 7516.751 | 917.307 | 15051.96 | 14909.2 |
| 939.227 | 3461.87 | 3398.957 | 935.947 | 4722.57 | 4715.264 | 938.681 | 7704.258 | 7545.068 | 917.581 | 14900.37 | 14917.74 |
| 939.501 | 3512.811 | 3402.273 | 936.22 | 4746.515 | 4719.489 | 938.954 | 7587.616 | 7550.731 | 917.856 | 14984.64 | 14926.25 |
| 939.774 | 3386.985 | 3405.56 | 936.494 | 4725.094 | 4723.685 | 939.227 | 7566.753 | 7556.394 | 918.131 | 15135.74 | 14934.76 |
| 940.047 | 3451.189 | 3408.875 | 936.767 | 4741.38 | 4727.882 | 939.501 | 7687.62 | 7562.028 | 918.405 | 15114.4 | 14943.24 |
| 940.32 | 3487.546 | 3412.162 | 937.04 | 4757.372 | 4732.078 | 939.774 | 7626.203 | 7567.692 | 918.68 | 14946.91 | 14951.72 |
| 940.593 | 3437.838 | 3415.478 | 937.314 | 4717.817 | 4736.274 | 940.047 | 7651.292 | 7573.326 | 918.955 | 15181.13 | 14960.2 |
| 940.867 | 3465.597 | 3418.764 | 937.587 | 4656.899 | 4740.47 | 940.32 | 7560.209 | 7578.96 | 919.229 | 14960.46 | 14968.65 |
| 941.14 | 3471.642 | 3422.08 | 937.861 | 4869.524 | 4744.666 | 940.593 | 7692.961 | 7584.623 | 919.504 | 15038.55 | 14977.1 |
| 941.413 | 3410.988 | 3425.367 | 938.134 | 4833.079 | 4748.862 | 940.867 | 7581.073 | 7590.257 | 919.778 | 15008 | 14985.55 |
| 941.959 | 3486.637 | 3431.969 | 938.407 | 4907.906 | 4753.029 | 941.14 | 7723.42 | 7595.862 | 920.053 | 14985.79 | 14994 |
| 942.232 | 3377.008 | 3435.256 | 938.681 | 4851.419 | 4757.225 | 941.413 | 7655.987 | 7601.496 | 920.328 | 15063.49 | 15002.43 |
| 942.505 | 3436.517 | 3438.571 | 939.227 | 4779.497 | 4765.618 | 941.686 | 7790.089 | 7607.13 | 920.602 | 14942.07 | 15010.85 |
| 942.778 | 3535.993 | 3441.858 | 939.501 | 4724.947 | 4769.785 | 941.959 | 7735.656 | 7612.735 | 920.877 | 15159.86 | 15019.24 |
| 943.051 | 3472.757 | 3445.144 | 939.774 | 4901.685 | 4773.981 | 942.505 | 7605.692 | 7623.973 | 921.151 | 15160.15 | 15027.66 |
| 943.324 | 3452.392 | 3448.431 | 940.047 | 4853.444 | 4778.148 | 942.778 | 7696.805 | 7629.578 | 921.426 | 15188.53 | 15036.05 |
| 943.597 | 3418.295 | 3451.747 | 940.32 | 4739.913 | 4782.344 | 943.051 | 7772.747 | 7635.183 | 921.7 | 15189.64 | 15044.42 |
| 943.87 | 3497.259 | 3455.033 | 940.593 | 4864.448 | 4786.511 | 943.324 | 7664.497 | 7640.787 | 921.974 | 15050.9 | 15052.81 |
| 944.143 | 3492.183 | 3458.32 | 941.959 | 4847.135 | 4807.374 | 943.597 | 7692.139 | 7646.392 | 922.249 | 15202.84 | 15061.17 |
| 944.416 | 3501.661 | 3461.606 | 942.232 | 5015.275 | 4811.541 | 943.87 | 7900.627 | 7651.997 | 922.523 | 15317.46 | 15069.51 |
| 944.689 | 3456.882 | 3464.893 | 942.505 | 4847.957 | 4815.708 | 944.143 | 7806.492 | 7657.572 | 922.798 | 15182.89 | 15077.87 |
| 944.962 | 3535.259 | 3468.179 | 943.051 | 4885.986 | 4824.041 | 944.689 | 7755.698 | 7668.752 | 923.072 | 15201.32 | 15086.2 |
| 945.235 | 3479.389 | 3471.466 | 943.324 | 4783.987 | 4828.208 | 944.962 | 7579.195 | 7674.357 | 923.346 | 15210.89 | 15094.54 |
| 945.508 | 3494.706 | 3474.752 | 943.597 | 4888.099 | 4832.375 | 945.235 | 7780.493 | 7679.932 | 923.621 | 15168.16 | 15102.84 |
| 945.781 | 3448.255 | 3478.039 | 943.87 | 4754.262 | 4836.542 | 945.508 | 7719.546 | 7685.507 | 923.895 | 15213.44 | 15111.15 |
| 946.053 | 3613.314 | 3481.325 | 944.143 | 4968.178 | 4840.679 | 946.872 | 7750.24 | 7713.325 | 924.169 | 15226.85 | 15119.45 |
| 946.326 | 3483.027 | 3484.612 | 944.416 | 4945.73 | 4844.846 | 947.144 | 7720.045 | 7718.871 | 924.444 | 15098.18 | 15127.75 |
| 946.599 | 3499.665 | 3487.898 | 944.689 | 4894.936 | 4849.013 | 947.417 | 7852.562 | 7724.417 | 924.718 | 15396.93 | 15136.03 |
| 946.872 | 3571.851 | 3491.185 | 944.962 | 4842.528 | 4853.151 | 947.69 | 7754.143 | 7729.963 | 924.992 | 15333.31 | 15144.3 |
| 947.144 | 3534.643 | 3494.471 | 945.235 | 4949.927 | 4857.317 | 947.963 | 7856.23 | 7735.509 | 925.266 | 15211.59 | 15152.55 |
| 947.417 | 3629.013 | 3497.758 | 945.508 | 4911.868 | 4861.455 | 948.235 | 7917.148 | 7741.055 | 925.541 | 15176.52 | 15160.8 |
| 947.69 | 3544.297 | 3501.044 | 945.781 | 4924.632 | 4865.592 | 948.508 | 7757.84 | 7746.601 | 925.815 | 15202.02 | 15169.04 |
| 947.963 | 3575.842 | 3504.331 | 946.053 | 4927.919 | 4869.759 | 948.781 | 7747.834 | 7752.147 | 926.089 | 15150.23 | 15177.29 |
| 948.235 | 3575.314 | 3507.588 | 946.326 | 4840.621 | 4873.897 | 949.053 | 7877.24 | 7757.664 | 926.363 | 15243.66 | 15185.5 |
| 948.508 | 3512.841 | 3510.875 | 946.599 | 4931.997 | 4878.034 | 949.326 | 7680.284 | 7763.18 | 926.637 | 15347.54 | 15193.72 |
| 948.781 | 3589.839 | 3514.161 | 946.872 | 4857.464 | 4882.172 | 949.598 | 7817.32 | 7768.726 | 926.911 | 15397.19 | 15201.94 |
| 949.053 | 3571.528 | 3517.418 | 947.144 | 4977.979 | 4886.309 | 949.871 | 7819.638 | 7774.243 | 927.185 | 15353.26 | 15210.12 |
| 949.326 | 3534.849 | 3520.705 | 947.417 | 4916.152 | 4890.447 | 950.143 | 7858.137 | 7779.76 | 927.459 | 15358.46 | 15218.31 |
| 949.598 | 3644.242 | 3523.991 | 947.69 | 4870.141 | 4894.584 | 950.416 | 7934.167 | 7785.247 | 927.733 | 15382.37 | 15226.5 |
| 949.871 | 3564.28 | 3527.249 | 947.963 | 5025.34 | 4898.722 | 950.688 | 7967.032 | 7790.764 | 928.008 | 15532.29 | 15234.68 |
| 950.143 | 3620.973 | 3530.535 | 948.235 | 4999.606 | 4902.859 | 950.961 | 8165.984 | 7796.28 | 928.282 | 15375.77 | 15242.84 |
| 950.416 | 3605.215 | 3533.792 | 948.508 | 4900.453 | 4906.997 | 951.233 | 7901.185 | 7801.768 | 928.83 | 15344.75 | 15259.13 |
| 950.688 | 3548.787 | 3537.079 | 948.781 | 4915.888 | 4911.134 | 951.506 | 7927.33 | 7807.284 | 929.103 | 15301.79 | 15267.25 |
| 950.961 | 3567.802 | 3540.336 | 949.053 | 4848.426 | 4915.242 | 951.778 | 7938.95 | 7812.772 | 929.377 | 15270.98 | 15275.38 |
| 951.233 | 3619.153 | 3543.622 | 949.326 | 4996.759 | 4919.38 | 952.051 | 7908.256 | 7818.259 | 929.651 | 15281.72 | 15283.48 |
| 951.506 | 3517.301 | 3546.88 | 949.598 | 5046.703 | 4923.488 | 952.323 | 7962.777 | 7823.746 | 929.925 | 15496.4 | 15291.61 |
| 951.778 | 3594.358 | 3550.137 | 949.871 | 5011.109 | 4927.625 | 952.595 | 7919.348 | 7829.233 | 930.199 | 15336.8 | 15299.71 |
| 952.051 | 3695.536 | 3553.423 | 950.143 | 5071.85 | 4931.733 | 952.868 | 7908.198 | 7834.721 | 930.473 | 15374.51 | 15307.78 |
| 952.323 | 3683.886 | 3556.68 | 950.416 | 5011.05 | 4935.871 | 953.14 | 7872.134 | 7840.179 | 930.747 | 15588.98 | 15315.85 |
| 952.595 | 3611.143 | 3559.938 | 950.688 | 4943.852 | 4939.979 | 953.412 | 7945.494 | 7845.666 | 931.021 | 15581.67 | 15323.92 |
| 952.868 | 3588.166 | 3563.224 | 950.961 | 5108.031 | 4944.087 | 953.685 | 8036.166 | 7851.124 | 931.295 | 15597.84 | 15331.99 |
| 953.14 | 3564.251 | 3566.481 | 951.233 | 5062.255 | 4948.225 | 953.957 | 7789.883 | 7856.582 | 931.568 | 15417.47 | 15340.03 |
| 953.412 | 3615.691 | 3569.738 | 951.506 | 5131.653 | 4952.333 | 954.229 | 8017.416 | 7862.04 | 931.842 | 15693.62 | 15348.07 |
| 953.685 | 3525.517 | 3572.996 | 951.778 | 4994.206 | 4956.441 | 954.502 | 7873.631 | 7867.498 | 932.116 | 15616.68 | 15356.11 |
| 953.957 | 3653.544 | 3576.253 | 952.051 | 5084.82 | 4960.549 | 954.774 | 8112.666 | 7872.956 | 932.39 | 15478.21 | 15364.12 |
| 954.229 | 3560.378 | 3579.539 | 952.323 | 5126.078 | 4964.657 | 955.046 | 7844.287 | 7878.414 | 932.663 | 15542.44 | 15372.13 |
| 954.502 | 3651.49 | 3582.796 | 952.595 | 5021.438 | 4968.765 | 955.318 | 8113.663 | 7883.872 | 932.937 | 15329.05 | 15380.14 |
| 954.774 | 3626.02 | 3586.054 | 953.14 | 5076.399 | 4976.982 | 955.59 | 7927.066 | 7889.3 | 933.211 | 15560.17 | 15388.15 |
| 955.046 | 3622.323 | 3589.311 | 953.412 | 4870.816 | 4981.06 | 955.862 | 7997.403 | 7894.758 | 933.484 | 15443.52 | 15396.13 |
| 955.318 | 3643.626 | 3592.568 | 953.685 | 4973.989 | 4985.169 | 956.135 | 8039.423 | 7900.187 | 933.758 | 15505.44 | 15404.09 |
| 955.59 | 3678.047 | 3595.825 | 953.957 | 5020.704 | 4989.277 | 956.407 | 8043.884 | 7905.616 | 934.032 | 15663.48 | 15412.07 |
| 955.862 | 3703.488 | 3599.082 | 954.229 | 5117.568 | 4993.355 | 956.679 | 8124.873 | 7911.044 | 934.305 | 15358.28 | 15420.02 |
| 956.135 | 3655.892 | 3602.31 | 954.502 | 5121.617 | 4997.464 | 956.951 | 8029.329 | 7916.473 | 934.579 | 15723.99 | 15427.97 |
| 956.407 | 3691.163 | 3605.567 | 954.774 | 4969.499 | 5001.542 | 957.223 | 8215.516 | 7921.901 | 934.853 | 15227.85 | 15435.89 |
| 956.679 | 3750.467 | 3608.824 | 955.046 | 5014.395 | 5005.621 | 957.495 | 8001.306 | 7927.301 | 935.126 | 15578.18 | 15443.85 |
| 956.951 | 3646.385 | 3612.082 | 955.318 | 5106.036 | 5009.729 | 957.767 | 8044.823 | 7932.729 | 935.4 | 15544.03 | 15451.74 |
| 957.223 | 3743.836 | 3615.339 | 955.59 | 5060.582 | 5013.808 | 958.039 | 7987.191 | 7938.129 | 935.673 | 15656.32 | 15459.66 |
| 957.495 | 3632.945 | 3618.567 | 955.862 | 5061.639 | 5017.887 | 958.311 | 8000.719 | 7943.557 | 935.947 | 15531.64 | 15467.56 |
| 957.767 | 3668.481 | 3621.824 | 956.135 | 5059.379 | 5021.966 | 958.583 | 8080.534 | 7948.956 | 936.22 | 15776.4 | 15475.45 |
| 958.039 | 3784.77 | 3625.081 | 956.407 | 5107.914 | 5026.045 | 958.855 | 8054.917 | 7954.356 | 936.494 | 15580.53 | 15483.34 |
| 958.311 | 3684.385 | 3628.309 | 956.679 | 5069.151 | 5030.153 | 959.127 | 8124.521 | 7959.755 | 936.767 | 15764.72 | 15491.21 |
| 958.583 | 3678.516 | 3631.566 | 956.951 | 5180.1 | 5034.202 | 959.399 | 8238.257 | 7965.125 | 937.04 | 15652.16 | 15499.07 |
| 958.855 | 3681.861 | 3634.823 | 957.223 | 5189.959 | 5038.281 | 959.671 | 8119.121 | 7970.524 | 937.314 | 15628.15 | 15506.94 |
| 959.127 | 3625.609 | 3638.051 | 957.495 | 5202.313 | 5042.36 | 959.942 | 8193.244 | 7975.923 | 937.587 | 15557.49 | 15514.77 |
| 959.399 | 3704.955 | 3641.308 | 957.767 | 5183.122 | 5046.439 | | | | 937.861 | 15799.79 | 15522.6 |
| 959.671 | 3726.552 | 3644.536 | 958.039 | 5273.853 | 5050.517 | 1101.15 | 10154.29 | 10245.58 | 938.134 | 15895.62 | 15530.44 |
| 959.942 | 3633.532 | 3647.793 | 958.311 | 5100.079 | 5054.567 | 1102.87 | 10219.64 | 10266.21 | 938.954 | 15504.94 | 15553.86 |
| | | | 958.583 | 5083.617 | 5058.646 | 1104.6 | 10615.84 | 10286.69 | 939.227 | 15588.34 | 15561.63 |
| 1101.15 | 5205.042 | 5161.29 | 958.855 | 5207.155 | 5062.695 | 1106.32 | 10456.01 | 10306.97 | 939.501 | 15536.6 | 15569.41 |
| 1102.87 | 5230.689 | 5176.989 | 959.127 | 5163.961 | 5066.774 | 1108.05 | 10221.64 | 10327.07 | 939.774 | 15714.92 | 15577.18 |
| 1104.6 | 4894.555 | 5192.63 | 959.399 | 5210.705 | 5070.823 | 1109.77 | 10369.35 | 10347.02 | 940.047 | 15393.58 | 15584.96 |
| 1106.32 | 4863.978 | 5208.182 | 959.671 | 5198.381 | 5074.902 | 1111.5 | 10423.23 | 10366.77 | 940.32 | 15916.31 | 15592.71 |
| 1108.05 | 5112.374 | 5223.675 | 959.942 | 5302.61 | 5078.951 | 1113.22 | 9958.453 | 10386.34 | 940.593 | 15764.02 | 15600.45 |
| 1109.77 | 4877.271 | 5239.081 | | | | 1114.95 | 10182.23 | 10405.74 | 940.867 | 15828.46 | 15608.17 |
| 1111.5 | 5199.467 | 5254.369 | 1101.15 | 6987.065 | 6898.828 | 1116.67 | 10397.11 | 10424.96 | 941.14 | 15786.14 | 15615.89 |
| 1113.22 | 4943.06 | 5269.628 | 1102.87 | 7048.423 | 6916.816 | 1118.39 | 10572.44 | 10444.01 | 941.413 | 15793.16 | 15623.61 |
| 1114.95 | 5030.475 | 5284.769 | 1104.6 | 7074.803 | 6934.686 | 1120.12 | 10406.5 | 10462.87 | 941.686 | 15610.52 | 15631.32 |
| 1116.67 | 5368.311 | 5299.852 | 1106.32 | 6329.675 | 6952.439 | 1121.84 | 10682.6 | 10481.57 | 941.959 | 15498.72 | 15639.01 |
| 1118.39 | 5202.342 | 5314.817 | 1108.05 | 6684.706 | 6970.104 | 1123.57 | 10505.48 | 10500.08 | 942.232 | 15737.26 | 15646.7 |
| 1120.12 | 5208.974 | 5329.724 | 1109.77 | 6947.685 | 6987.622 | 1125.29 | 9987.239 | 10518.42 | 942.505 | 15722.73 | 15654.36 |
| 1121.84 | 5093.301 | 5344.543 | 1111.5 | 6877.231 | 7005.023 | 1127.01 | 10182.17 | 10536.59 | 942.778 | 15761.43 | 15662.05 |
| 1123.57 | 5096.294 | 5359.303 | 1113.22 | 7330.828 | 7022.336 | 1128.74 | 10215.45 | 10554.57 | 943.051 | 15897.56 | 15669.71 |
| 1125.29 | 5394.398 | 5373.945 | 1114.95 | 6954.875 | 7039.502 | 1130.46 | 9875.292 | 10572.38 | 943.324 | 15721.82 | 15677.34 |
| 1127.01 | 5592.88 | 5388.529 | 1116.67 | 6661.025 | 7056.58 | 1132.18 | 9939.057 | 10590.02 | 943.597 | 15802.08 | 15684.99 |
| 1128.74 | 5545.255 | 5403.025 | 1118.39 | 6604.48 | 7073.512 | 1133.91 | 10330.5 | 10607.48 | 943.87 | 15779.75 | 15692.62 |
| 1130.46 | 5350.5 | 5417.433 | 1120.12 | 6738.552 | 7090.355 | 1135.63 | 10408.15 | 10624.76 | 944.143 | 16103.03 | 15700.22 |
| 1132.18 | 5843.74 | 5431.753 | 1121.84 | 7224.339 | 7107.052 | 1137.35 | 10720.22 | 10641.87 | 944.416 | 15782.21 | 15707.85 |
| 1133.91 | 5661.192 | 5445.984 | 1123.57 | 7062.772 | 7123.661 | 1139.07 | 10803.85 | 10658.8 | 944.689 | 15744.71 | 15715.45 |
| 1135.63 | 5004.565 | 5460.157 | 1125.29 | 7119.288 | 7140.152 | 1140.8 | 10467.92 | 10675.56 | 944.962 | 15561.43 | 15723.02 |
| 1137.35 | 5534.075 | 5474.213 | 1127.01 | 7422.234 | 7156.496 | 1142.52 | 10547.3 | 10692.14 | 945.235 | 15867.51 | 15730.62 |
| 1139.07 | 5721.846 | 5488.21 | 1128.74 | 7499.849 | 7172.753 | 1144.24 | 10910.57 | 10708.54 | 945.508 | 15881.98 | 15738.19 |
| 1140.8 | 5487.271 | 5502.119 | 1130.46 | 7185.283 | 7188.892 | 1145.96 | 10829.61 | 10724.8 | 945.781 | 15842.75 | 15745.76 |
| 1142.52 | 5307.393 | 5515.94 | 1132.18 | 7323.287 | 7204.914 | 1147.69 | 10570.51 | 10740.85 | 946.053 | 15725.87 | 15753.31 |
| 1144.24 | 5533.723 | 5529.673 | 1133.91 | 7257.527 | 7220.789 | 1149.41 | 10624.76 | 10756.75 | 946.326 | 15906.19 | 15760.85 |
| 1145.96 | 5892.07 | 5543.318 | 1135.63 | 7196.668 | 7236.576 | 1151.13 | 10622.68 | 10772.45 | 946.599 | 15854.16 | 15768.39 |
| 1147.69 | 5435.068 | 5556.875 | 1137.35 | 7664.702 | 7252.245 | 1152.85 | 10562.44 | 10788 | 946.872 | 15983.83 | 15775.9 |
| 1149.41 | 5400.149 | 5570.373 | 1139.07 | 7295.41 | 7267.798 | 1154.57 | 10734.39 | 10803.38 | 947.144 | 15866.22 | 15783.41 |
| 1151.13 | 5379.168 | 5583.754 | 1140.8 | 7124.6 | 7283.232 | 1156.29 | 10639.2 | 10818.58 | 947.417 | 15770.06 | 15790.93 |
| 1152.85 | 5101.194 | 5597.076 | 1142.52 | 7308.732 | 7298.55 | 1158.02 | 10963.54 | 10833.63 | 947.69 | 15964.52 | 15798.44 |
| 1154.57 | 5273.384 | 5610.281 | 1144.24 | 7230.854 | 7313.75 | 1159.74 | 10944.99 | 10848.48 | 947.963 | 16084.63 | 15805.92 |
| 1156.29 | 5587.04 | 5623.427 | 1145.96 | 7368.74 | 7328.833 | 1161.46 | 10629.9 | 10863.18 | 948.235 | 15809.76 | 15813.4 |
| 1158.02 | 5630.557 | 5636.485 | 1147.69 | 7067.35 | 7343.798 | 1163.18 | 10542.63 | 10877.68 | 948.508 | 15825.46 | 15820.86 |
| 1159.74 | 5910.762 | 5649.455 | 1149.41 | 6708.093 | 7358.617 | 1164.9 | 10362.43 | 10892.03 | 948.781 | 15771.53 | 15828.34 |
| 1161.46 | 5845.442 | 5662.337 | 1151.13 | 7164.859 | 7373.347 | 1166.62 | 10565.17 | 10906.23 | 949.053 | 15914.11 | 15835.79 |
| 1163.18 | 5821.468 | 5675.131 | 1152.85 | 7606.748 | 7387.961 | 1168.34 | 10733.81 | 10920.23 | 949.326 | 15953.17 | 15843.22 |
| 1164.9 | 5576.124 | 5687.836 | 1154.57 | 7406.389 | 7402.456 | 1170.06 | 10793.49 | 10934.08 | 949.598 | 15905.78 | 15850.64 |
| 1166.62 | 5529.292 | 5700.454 | 1156.29 | 7445.827 | 7416.835 | 1171.78 | 10696.71 | 10947.75 | 949.871 | 15919.25 | 15858.06 |
| 1168.34 | 5998.617 | 5713.013 | 1158.02 | 7690.173 | 7431.096 | 1173.5 | 10835.42 | 10961.25 | 950.143 | 15984.24 | 15865.49 |
| 1170.06 | 5862.491 | 5725.455 | 1159.74 | 7456.332 | 7445.24 | 1175.22 | 11021.49 | 10974.57 | 950.416 | 16061.53 | 15872.88 |
| 1171.78 | 5771.965 | 5737.809 | 1161.46 | 7394.563 | 7459.296 | 1176.94 | 10739.35 | 10987.75 | 950.688 | 15935.47 | 15880.28 |
| 1173.5 | 5531.58 | 5750.104 | 1163.18 | 6861.092 | 7473.205 | 1178.66 | 10793.7 | 11000.75 | 950.961 | 16016.96 | 15887.67 |
| 1175.22 | 5820.119 | 5762.282 | 1164.9 | 6941.699 | 7486.996 | 1180.38 | 11097.96 | 11013.6 | 951.233 | 16170.66 | 15895.04 |
| 1176.94 | 6029.546 | 5774.401 | 1166.62 | 7602.992 | 7500.67 | 1182.1 | 10788.36 | 11026.28 | 951.506 | 16056.93 | 15902.4 |
| 1178.66 | 5909.001 | 5786.432 | 1168.34 | 7234.023 | 7514.227 | 1183.82 | 10831.46 | 11038.78 | 951.778 | 16131.46 | 15909.77 |
| 1180.38 | 5796.057 | 5798.345 | 1170.06 | 7658.834 | 7527.667 | 1185.54 | 11208.03 | 11051.1 | 952.051 | 16049.91 | 15917.1 |
| 1182.1 | 5861.053 | 5810.2 | 1171.78 | 7803.088 | 7541.018 | 1187.26 | 11031.59 | 11063.28 | 952.323 | 16148.8 | 15924.44 |
| 1183.82 | 5948.527 | 5821.967 | 1173.5 | 7174.044 | 7554.223 | 1188.98 | 11263.96 | 11075.31 | 952.595 | 16269.35 | 15931.78 |
| 1185.54 | 5684.374 | 5833.646 | 1175.22 | 7217.884 | 7567.31 | 1190.7 | 11019.23 | 11087.13 | 952.868 | 16156.26 | 15939.08 |
| 1187.26 | 6282.724 | 5845.237 | 1176.94 | 7310.111 | 7580.28 | 1192.42 | 10707.04 | 11098.84 | 953.14 | 16164.88 | 15946.42 |
| 1188.98 | 5807.178 | 5856.74 | 1178.66 | 7467.16 | 7593.162 | 1194.14 | 10880.17 | 11110.35 | 953.412 | 15867.34 | 15953.7 |
| 1190.7 | 5623.163 | 5868.154 | 1180.38 | 7727.674 | 7605.897 | 1195.86 | 10763.68 | 11121.7 | 953.685 | 16236.86 | 15961 |
| 1192.42 | 6002.021 | 5879.481 | 1182.1 | 7803 | 7618.545 | 1197.58 | 10849.68 | 11132.91 | 953.957 | 16071.39 | 15968.28 |
| 1194.14 | 5537.302 | 5890.72 | 1183.82 | 7633.363 | 7631.045 | 1199.3 | 11039.48 | 11143.94 | 954.229 | 16069.05 | 15975.56 |
| 1195.86 | 5662.571 | 5901.9 | 1185.54 | 7499.115 | 7643.458 | 1201.02 | 11573.33 | 11154.83 | 954.502 | 15887.11 | 15982.8 |
| 1197.58 | 5580.233 | 5912.962 | 1187.26 | 7393.859 | 7655.753 | 1202.73 | 11310.97 | 11165.54 | 954.774 | 16146.54 | 15990.05 |
| 1199.3 | 5577.768 | 5923.937 | 1188.98 | 7878.296 | 7667.93 | 1204.45 | 11051.72 | 11176.08 | 955.046 | 15917.13 | 15997.3 |
| 1201.02 | 5940.957 | 5934.853 | 1190.7 | 7819.227 | 7679.961 | 1206.17 | 11167.42 | 11186.49 | 955.318 | 16393.33 | 16004.55 |
| 1202.73 | 6027.462 | 5945.652 | 1192.42 | 7559.358 | 7691.904 | 1207.89 | 10954.76 | 11196.73 | 955.59 | 16364.63 | 16011.77 |
| 1204.45 | 5770.733 | 5956.391 | 1194.14 | 7781.168 | 7703.73 | 1209.61 | 11280.04 | 11206.8 | 955.862 | 16217.58 | 16018.99 |
| 1206.17 | 5931.889 | 5967.043 | 1195.86 | 7294.148 | 7715.467 | 1211.32 | 11544.25 | 11216.72 | 956.135 | 16260.1 | 16026.17 |
| 1207.89 | 5767.564 | 5977.578 | 1197.58 | 7214.773 | 7727.058 | 1213.04 | 11402.79 | 11226.49 | 956.407 | 16054.93 | 16033.39 |
| 1209.61 | 5726.306 | 5988.053 | 1199.3 | 7549.352 | 7738.532 | 1214.76 | 11143.56 | 11236.08 | 956.679 | 16329.88 | 16040.55 |
| 1211.32 | 6018.923 | 5998.441 | 1201.02 | 7611.561 | 7749.917 | 1216.48 | 11274.26 | 11245.53 | 956.951 | 16416.8 | 16047.74 |
| 1213.04 | 6091.197 | 6008.741 | 1202.73 | 7464.665 | 7761.156 | 1218.19 | 11451.26 | 11254.83 | 957.223 | 16335.75 | 16054.9 |
| 1214.76 | 6030.455 | 6018.952 | 1204.45 | 7792.055 | 7772.306 | 1219.91 | 10960.08 | 11263.96 | 957.495 | 16260.43 | 16062.06 |
| 1216.48 | 5822.583 | 6029.076 | 1206.17 | 7854.176 | 7783.34 | 1221.63 | 10944.99 | 11272.94 | 957.767 | 16204.7 | 16069.22 |
| 1218.19 | 5885.027 | 6039.112 | 1207.89 | 8031.237 | 7794.256 | 1223.35 | 11222.23 | 11281.77 | 958.039 | 16042.61 | 16076.35 |
| 1219.91 | 5949.055 | 6049.089 | 1209.61 | 8248.088 | 7805.054 | 1225.06 | 11260.56 | 11290.46 | 958.311 | 16247.81 | 16083.48 |
| 1221.63 | 5826.163 | 6058.948 | 1211.32 | 7624.208 | 7815.735 | 1226.78 | 11143.94 | 11298.97 | 958.583 | 16220.14 | 16090.61 |
| 1223.35 | 6301.534 | 6068.72 | 1213.04 | 7627.671 | 7826.299 | 1228.5 | 11306.92 | 11307.33 | 958.855 | 16313.19 | 16097.72 |
| 1225.06 | 6287.097 | 6078.432 | 1214.76 | 8000.191 | 7836.775 | 1230.21 | 11550.24 | 11315.55 | 959.127 | 16123.68 | 16104.82 |
| 1226.78 | 6035.268 | 6088.057 | 1216.48 | 7500.318 | 7847.133 | 1231.93 | 11240.98 | 11323.62 | 959.399 | 16438.1 | 16111.92 |
| 1228.5 | 6495.203 | 6097.565 | 1218.19 | 7727.616 | 7857.374 | 1233.65 | 11007.38 | 11331.54 | 959.671 | 16426.51 | 16118.99 |
| 1230.21 | 6203.672 | 6107.013 | 1219.91 | 8274.703 | 7867.498 | 1235.36 | 11489.91 | 11339.29 | 959.942 | 16506.92 | 16126.06 |
| 1231.93 | 6041.635 | 6116.374 | 1221.63 | 7823.423 | 7877.504 | 1237.08 | 11306.77 | 11346.89 | | | |
| 1233.65 | 6183.924 | 6125.647 | 1223.35 | 8133.852 | 7887.422 | 1238.79 | 11180.42 | 11354.37 | 1101.15 | 18229.02 | 18597.72 |
| 1235.36 | 6124.297 | 6134.831 | 1225.06 | 8240.106 | 7897.194 | 1240.51 | 11473.65 | 11361.68 | 1102.87 | 18705.47 | 18613.69 |
| 1237.08 | 6238.18 | 6143.928 | 1226.78 | 8040.363 | 7906.877 | 1242.23 | 11225.52 | 11368.84 | 1104.6 | 18478.12 | 18629.36 |
| 1238.79 | 6343.73 | 6152.966 | 1228.5 | 8177.516 | 7916.443 | 1243.94 | 11127.22 | 11375.85 | 1106.32 | 18185.15 | 18644.67 |
| 1240.51 | 6488.836 | 6161.886 | 1230.21 | 8078.304 | 7925.921 | 1245.66 | 11111.87 | 11382.74 | 1108.05 | 18709.93 | 18659.7 |
| 1242.23 | 6109.508 | 6170.748 | 1231.93 | 8282.508 | 7935.253 | 1247.37 | 11188.96 | 11389.46 | 1109.77 | 18348.42 | 18674.4 |
| 1243.94 | 6059.388 | 6179.522 | 1233.65 | 7942.119 | 7944.496 | 1249.09 | 11432.48 | 11396.04 | 1111.5 | 18771.35 | 18688.81 |
| 1245.66 | 6363.156 | 6188.208 | 1235.36 | 7945.905 | 7953.622 | 1250.8 | 11632.08 | 11402.46 | 1113.22 | 18727.95 | 18702.86 |
| 1247.37 | 6099.413 | 6196.806 | 1237.08 | 8374.501 | 7962.66 | 1252.52 | 11374.12 | 11408.74 | 1114.95 | 18486.01 | 18716.65 |
| 1249.09 | 5966.368 | 6205.315 | 1238.79 | 7700.59 | 7971.581 | 1254.23 | 11144.18 | 11414.91 | 1116.67 | 18431.73 | 18730.09 |
| 1250.8 | 5912.2 | 6213.737 | 1240.51 | 7857.961 | 7980.384 | 1255.95 | 11271.03 | 11420.89 | 1118.39 | 18458.87 | 18743.24 |
| 1252.52 | 6268.111 | 6222.1 | 1242.23 | 7831.405 | 7989.069 | 1257.66 | 11377.49 | 11426.76 | 1120.12 | 18538.42 | 18756.09 |
| 1254.23 | 6255.933 | 6230.346 | 1243.94 | 7624.149 | 7997.638 | 1259.38 | 11382.07 | 11432.45 | 1121.84 | 18367.81 | 18768.62 |
| 1255.95 | 6168.225 | 6238.533 | 1245.66 | 8499.242 | 8006.118 | 1261.09 | 11627.74 | 11438.03 | 1123.57 | 18655.3 | 18780.86 |
| 1257.66 | 6175.502 | 6246.631 | 1247.37 | 7807.9 | 8014.481 | 1262.8 | 11814.16 | 11443.46 | 1125.29 | 18645.44 | 18792.77 |
| 1259.38 | 6014.052 | 6254.642 | 1249.09 | 7764.413 | 8022.756 | 1264.52 | 11724.63 | 11448.77 | 1127.01 | 18473.63 | 18804.39 |
| 1261.09 | 6349.276 | 6262.594 | 1250.8 | 8298.735 | 8030.914 | 1266.23 | 11665.41 | 11453.9 | 1128.74 | 18604.59 | 18815.72 |
| 1262.8 | 6159.656 | 6270.429 | 1252.52 | 7752.499 | 8038.954 | 1267.95 | 11592.67 | 11458.92 | 1130.46 | 18730.12 | 18826.75 |
| 1264.52 | 6047.563 | 6278.205 | 1254.23 | 8182.005 | 8046.906 | 1269.66 | 11394.98 | 11463.79 | 1132.18 | 19170.19 | 18837.49 |
| 1266.23 | 6356.055 | 6285.894 | 1255.95 | 8366.695 | 8054.741 | 1271.37 | 11641.7 | 11468.52 | 1133.91 | 18937.38 | 18847.91 |
| 1267.95 | 6112.501 | 6293.494 | 1257.66 | 7870.96 | 8062.458 | 1273.09 | 11394.22 | 11473.12 | 1135.63 | 18856.71 | 18858.03 |
| 1269.66 | 6083.45 | 6301.006 | 1259.38 | 8012.809 | 8070.088 | 1274.8 | 11219.24 | 11477.58 | 1137.35 | 18872.15 | 18867.89 |
| 1271.37 | 6391.913 | 6308.459 | 1261.09 | 7965.682 | 8077.6 | 1276.51 | 11456.57 | 11481.9 | 1139.07 | 18891.9 | 18877.43 |
| 1273.09 | 6332.697 | 6315.824 | 1262.8 | 7794.637 | 8085.024 | 1278.23 | 11664.5 | 11486.09 | 1140.8 | 18940.22 | 18886.7 |
| 1274.8 | 5995.771 | 6323.072 | 1264.52 | 7885.603 | 8092.301 | 1279.94 | 11945.47 | 11490.14 | 1142.52 | 18418.17 | 18895.65 |
| 1276.51 | 6594.209 | 6330.291 | 1266.23 | 7891.707 | 8099.52 | 1281.65 | 11870.35 | 11494.05 | 1144.24 | 18697.4 | 18904.34 |
| 1278.23 | 6368.966 | 6337.392 | 1267.95 | 8063.309 | 8106.621 | 1283.37 | 11493.72 | 11497.83 | 1145.96 | 18975.09 | 18912.73 |
| 1279.94 | 6487.985 | 6344.435 | 1269.66 | 8184.382 | 8113.605 | 1285.08 | 11261.41 | 11501.47 | 1147.69 | 18786.05 | 18920.83 |
| 1281.65 | 6551.808 | 6351.389 | 1271.37 | 7851.417 | 8120.501 | 1286.79 | 11320.68 | 11504.99 | 1149.41 | 18625.66 | 18928.66 |
| 1283.37 | 6121.715 | 6358.255 | 1273.09 | 7998.313 | 8127.279 | 1288.5 | 11497.95 | 11508.37 | 1151.13 | 18752.37 | 18936.2 |
| 1285.08 | 6595.882 | 6365.034 | 1274.8 | 8374.149 | 8133.969 | 1290.22 | 11761.96 | 11511.62 | 1152.85 | 18892.01 | 18943.45 |
| 1286.79 | 6477.744 | 6371.754 | 1276.51 | 8386.737 | 8140.542 | 1291.93 | 11856.18 | 11514.73 | 1154.57 | 18882.65 | 18950.44 |
| 1288.5 | 6552.923 | 6378.385 | 1278.23 | 8232.183 | 8147.027 | 1293.64 | 11569.22 | 11517.73 | 1156.29 | 18898.47 | 18957.13 |
| 1290.22 | 6267.436 | 6384.929 | 1279.94 | 8089.337 | 8153.395 | 1295.35 | 11533.69 | 11520.6 | 1158.02 | 18783.29 | 18963.55 |
| 1291.93 | 6278.029 | 6391.414 | 1281.65 | 8228.457 | 8159.675 | 1297.06 | 11545.87 | 11523.33 | 1159.74 | 18955.51 | 18969.69 |
| 1293.64 | 6629.921 | 6397.811 | 1283.37 | 8357.364 | 8165.837 | 1298.78 | 11518.46 | 11525.91 | 1161.46 | 18902.14 | 18975.55 |
| 1295.35 | 6267.964 | 6404.12 | 1285.08 | 8084.995 | 8171.911 | 1300.49 | 11567.7 | 11528.41 | 1163.18 | 18842.16 | 18981.16 |
| 1297.06 | 6213.414 | 6410.341 | 1286.79 | 8034.993 | 8177.897 | 1302.2 | 11532.31 | 11530.76 | 1164.9 | 18996.45 | 18986.47 |
| 1298.78 | 6410.81 | 6416.503 | 1288.5 | 8384.008 | 8183.766 | 1303.91 | 11657.34 | 11532.96 | 1166.62 | 19123.01 | 18991.55 |
| 1300.49 | 6644.153 | 6422.577 | 1290.22 | 8526.884 | 8189.517 | 1305.62 | 11761.22 | 11535.07 | 1168.34 | 19250.62 | 18996.33 |
| 1302.2 | 6737.642 | 6428.593 | 1291.93 | 8721.991 | 8195.181 | 1307.33 | 11832.85 | 11537.03 | 1170.06 | 18954.25 | 19000.85 |
| 1303.91 | 6688.286 | 6434.491 | 1293.64 | 8267.337 | 8200.756 | 1309.04 | 11768.26 | 11538.88 | 1171.78 | 19183.95 | 19005.1 |
| 1305.62 | 6646.236 | 6440.33 | 1295.35 | 8270.624 | 8206.243 | 1310.75 | 11702.04 | 11540.59 | 1173.5 | 19012.88 | 19009.09 |
| 1307.33 | 6592.214 | 6446.111 | 1297.06 | 8410.946 | 8211.613 | 1312.47 | 11562.24 | 11542.17 | 1175.22 | 19037 | 19012.82 |
| 1309.04 | 6463.629 | 6451.804 | 1298.78 | 8375.469 | 8216.866 | 1314.18 | 11423.83 | 11543.67 | 1176.94 | 19018.16 | 19016.28 |
| 1310.75 | 6483.495 | 6457.408 | 1300.49 | 8423.828 | 8222.06 | 1315.89 | 11508.01 | 11545.02 | 1178.66 | 18671.17 | 19019.48 |
| 1312.47 | 6502.304 | 6462.925 | 1302.2 | 8339.376 | 8227.136 | 1317.6 | 11709.28 | 11546.22 | 1180.38 | 18835.35 | 19022.45 |
| 1314.18 | 6413.95 | 6468.383 | 1303.91 | 8420.894 | 8232.125 | 1319.31 | 12080.22 | 11547.33 | 1182.1 | 18950.38 | 19025.12 |
| 1315.89 | 6197.95 | 6473.782 | 1305.62 | 8425.002 | 8236.996 | 1321.02 | 11908.15 | 11548.33 | 1183.82 | 18950.64 | 19027.55 |
| 1317.6 | 6332.051 | 6479.064 | 1307.33 | 8261.85 | 8241.779 | 1322.73 | 11908.73 | 11549.18 | 1185.54 | 19206.58 | 19029.75 |
| 1319.31 | 6671.795 | 6484.287 | 1309.04 | 8235.352 | 8246.474 | 1324.44 | 11948.02 | 11549.95 | 1187.26 | 19276.3 | 19031.66 |
| 1321.02 | 6616.716 | 6489.452 | 1310.75 | 8315.52 | 8251.081 | 1326.15 | 11379.43 | 11550.56 | 1188.98 | 18930.19 | 19033.33 |
| 1322.73 | 6487.955 | 6494.528 | 1312.47 | 8141.364 | 8255.57 | 1327.86 | 11693.23 | 11551.09 | 1190.7 | 18882.36 | 19034.77 |
| 1324.44 | 6401.274 | 6499.517 | 1314.18 | 8413.499 | 8259.972 | 1329.57 | 11662.45 | 11551.47 | 1192.42 | 18521.66 | 19035.94 |
| 1326.15 | 6302.59 | 6504.447 | 1315.89 | 8171.97 | 8264.285 | 1331.28 | 11546.92 | 11551.74 | 1194.14 | 18553.68 | 19036.88 |
| 1327.86 | 6198.449 | 6509.288 | 1317.6 | 8381.25 | 8268.511 | 1332.98 | 11931.77 | 11551.91 | 1195.86 | 18942.72 | 19037.56 |
| 1329.57 | 6337.128 | 6514.071 | 1319.31 | 8355.31 | 8272.619 | 1334.69 | 11745.9 | 11551.94 | 1197.58 | 18702.54 | 19038 |
| 1331.28 | 6365.533 | 6518.766 | 1321.02 | 8088.105 | 8276.639 | 1336.4 | 11789.39 | 11551.88 | 1199.3 | 18811.58 | 19038.2 |
| 1332.98 | 6634.117 | 6523.403 | 1322.73 | 8342.487 | 8280.571 | 1338.11 | 11561.21 | 11551.71 | 1201.02 | 19240.09 | 19038.15 |
| 1334.69 | 6596.058 | 6527.951 | 1324.44 | 8205.539 | 8284.415 | 1339.82 | 11594.08 | 11551.38 | 1202.73 | 19152.32 | 19037.88 |
| 1336.4 | 6462.573 | 6532.411 | 1326.15 | 8544.167 | 8288.171 | 1341.53 | 11586.98 | 11550.97 | 1204.45 | 18850.9 | 19037.35 |
| 1338.11 | 6375.832 | 6536.813 | 1327.86 | 8480.843 | 8291.81 | 1343.24 | 11589.21 | 11550.47 | 1206.17 | 18783.38 | 19036.59 |
| 1339.82 | 6569.883 | 6541.156 | 1329.57 | 8241.603 | 8295.39 | 1344.95 | 11983.35 | 11549.83 | 1207.89 | 18769.21 | 19035.59 |
| 1341.53 | 6541.713 | 6545.411 | 1331.28 | 8398.592 | 8298.853 | 1346.65 | 11389.7 | 11549.07 | 1209.61 | 18869.92 | 19034.36 |
| 1343.24 | 6515.157 | 6549.607 | 1332.98 | 8682.964 | 8302.227 | 1348.36 | 11278.08 | 11548.21 | 1211.32 | 19163.74 | 19032.89 |
| 1344.95 | 6771.035 | 6553.715 | 1334.69 | 8435.477 | 8305.514 | 1350.07 | 11458.75 | 11547.25 | 1213.04 | 19173.63 | 19031.22 |
| 1346.65 | 6608.47 | 6557.735 | 1336.4 | 8112.372 | 8308.712 | 1351.78 | 11397.83 | 11546.19 | 1214.76 | 18927.72 | 19029.28 |
| 1348.36 | 6637.873 | 6561.726 | 1338.11 | 8483.337 | 8311.822 | 1353.49 | 11494.52 | 11544.99 | 1216.48 | 18967.13 | 19027.14 |
| 1350.07 | 6644.651 | 6565.599 | 1339.82 | 8450.355 | 8314.845 | 1355.19 | 11283 | 11543.7 | 1218.19 | 18706.59 | 19024.76 |
| 1351.78 | 6700.053 | 6569.443 | 1341.53 | 8262.994 | 8317.779 | 1356.9 | 11572.89 | 11542.29 | 1219.91 | 18517.88 | 19022.15 |
| 1353.49 | 6554.507 | 6573.17 | 1343.24 | 8172.498 | 8320.626 | 1358.61 | 11749.63 | 11540.79 | 1221.63 | 19063.26 | 19019.34 |
| 1355.19 | 6605.008 | 6576.867 | 1344.95 | 8212.523 | 8323.384 | 1360.32 | 11442.87 | 11539.18 | 1223.35 | 19219.9 | 19016.28 |
| 1356.9 | 6814.259 | 6580.476 | 1346.65 | 8157.503 | 8326.054 | 1362.02 | 11527.79 | 11537.45 | 1225.06 | 19143.87 | 19013 |
| 1358.61 | 6477.773 | 6583.998 | 1348.36 | 8360.152 | 8328.607 | 1363.73 | 11521.63 | 11535.63 | 1226.78 | 19015.32 | 19009.53 |
| 1360.32 | 6270.84 | 6587.49 | 1350.07 | 8483.191 | 8331.101 | 1365.44 | 11766.86 | 11533.69 | 1228.5 | 19293.23 | 19005.81 |
| 1362.02 | 6458.788 | 6590.864 | 1351.78 | 8396.92 | 8333.508 | 1367.14 | 11660.57 | 11531.66 | 1230.21 | 19387.95 | 19001.88 |
| 1363.73 | 6488.806 | 6594.209 | 1353.49 | 8423.945 | 8335.826 | 1368.85 | 11542.23 | 11529.52 | 1231.93 | 19488.75 | 18997.74 |
| 1365.44 | 6547.435 | 6597.467 | 1355.19 | 8236.585 | 8338.056 | 1370.56 | 11666.03 | 11527.29 | 1233.65 | 19439.01 | 18993.4 |
| 1367.14 | 6722.383 | 6600.665 | 1356.9 | 8147.966 | 8340.198 | 1372.26 | 11500.74 | 11524.95 | 1235.36 | 19027.32 | 18988.82 |
| 1368.85 | 6748.675 | 6603.775 | 1358.61 | 8133.999 | 8342.252 | 1373.97 | 11519.96 | 11522.51 | 1237.08 | 19404.71 | 18984.04 |
| 1370.56 | 6517.886 | 6606.827 | 1360.32 | 8462.65 | 8344.247 | 1375.67 | 11544.84 | 11519.96 | 1238.79 | 19109.33 | 18979.08 |
| 1372.26 | 6319.023 | 6609.82 | 1362.02 | 8314.728 | 8346.125 | 1377.38 | 11650.92 | 11517.32 | 1240.51 | 19071.22 | 18973.85 |
| 1373.97 | 6668.273 | 6612.725 | 1363.73 | 8145.824 | 8347.915 | 1379.09 | 11657.76 | 11514.59 | 1242.23 | 19017.08 | 18968.45 |
| 1375.67 | 6654.981 | 6615.572 | 1365.44 | 8117.684 | 8349.647 | 1380.79 | 11559.45 | 11511.74 | 1243.94 | 18764.13 | 18962.85 |
| 1377.38 | 6459.961 | 6618.359 | 1367.14 | 8039.13 | 8351.29 | 1382.5 | 11680.56 | 11508.81 | 1245.66 | 19082.6 | 18957.04 |
| 1379.09 | 6758.447 | 6621.059 | 1368.85 | 8148.817 | 8352.845 | 1384.2 | 11635.37 | 11505.75 | 1247.37 | 18741.77 | 18951.02 |
| 1380.79 | 6960.92 | 6623.7 | 1370.56 | 8335.767 | 8354.312 | 1385.91 | 11512.03 | 11502.64 | 1249.09 | 18740.66 | 18944.8 |
| 1382.5 | 6455.589 | 6626.282 | 1372.26 | 8490.057 | 8355.692 | 1387.61 | 11571.54 | 11499.39 | 1250.8 | 18793.3 | 18938.41 |
| 1384.2 | 6589.338 | 6628.806 | 1373.97 | 8289.052 | 8357.012 | 1389.32 | 11618.05 | 11496.07 | 1252.52 | 18915.99 | 18931.77 |
| 1385.91 | 6969.899 | 6631.241 | 1375.67 | 8245.271 | 8358.244 | 1391.02 | 11467.55 | 11492.64 | 1254.23 | 19001.44 | 18924.97 |
| 1387.61 | 6414.948 | 6633.618 | 1377.38 | 8106.063 | 8359.389 | 1392.73 | 11611.45 | 11489.15 | 1255.95 | 19102.82 | 18917.95 |
| 1389.32 | 6514.511 | 6635.936 | 1379.09 | 8024.487 | 8360.445 | 1394.43 | 11627.41 | 11485.54 | 1257.66 | 19135.57 | 18910.76 |
| 1391.02 | 6988.679 | 6638.166 | 1380.79 | 8133.471 | 8361.414 | 1396.14 | 11219.24 | 11481.81 | 1259.38 | 18952.7 | 18903.37 |
| 1392.73 | 6833.714 | 6640.367 | 1382.5 | 8340.99 | 8362.323 | 1397.84 | 11357.89 | 11478.02 | 1261.09 | 19042.66 | 18895.8 |
| 1394.43 | 6679.13 | 6642.48 | 1384.2 | 8347.27 | 8363.145 | 1399.55 | 11936.7 | 11474.12 | 1262.8 | 19025.97 | 18888.02 |
| 1396.14 | 6836.384 | 6644.534 | 1385.91 | 8530.317 | 8363.878 | 1401.25 | 11768.18 | 11470.16 | 1264.52 | 19093.05 | 18880.04 |
| 1397.84 | 6775.232 | 6646.5 | 1387.61 | 8381.045 | 8364.553 | 1402.95 | 11416.34 | 11466.08 | 1266.23 | 19139.76 | 18871.88 |
| 1399.55 | 6681.507 | 6648.437 | 1389.32 | 8174.024 | 8365.14 | 1404.66 | 11517.76 | 11461.91 | 1267.95 | 19191.41 | 18863.55 |
| 1401.25 | 6542.682 | 6650.285 | 1391.02 | 8379.93 | 8365.639 | 1406.36 | 11227.75 | 11457.66 | 1269.66 | 19066.17 | 18855.04 |
| 1402.95 | 6784.71 | 6652.075 | 1392.73 | 8034.699 | 8366.079 | 1408.06 | 11218.04 | 11453.32 | 1271.37 | 18892.98 | 18846.32 |
| 1404.66 | 6975.856 | 6653.807 | 1394.43 | 8162.814 | 8366.431 | 1409.77 | 11336.88 | 11448.89 | 1273.09 | 18978.93 | 18837.43 |
| 1406.36 | 6279.467 | 6655.479 | 1396.14 | 8381.397 | 8366.695 | 1411.47 | 11338.99 | 11444.37 | 1274.8 | 19079.73 | 18828.37 |
| 1408.06 | 6352.68 | 6657.093 | 1397.84 | 8562.947 | 8366.901 | 1413.17 | 11363.73 | 11439.76 | 1276.51 | 19414.89 | 18819.12 |
| 1409.77 | 6656.536 | 6658.619 | 1399.55 | 8723.253 | 8367.018 | 1414.88 | 11592.03 | 11435.06 | 1278.23 | 19059.1 | 18809.7 |
| 1411.47 | 6561.579 | 6660.116 | 1401.25 | 8438.823 | 8367.077 | 1416.58 | 11714.65 | 11430.28 | 1279.94 | 18818.62 | 18800.11 |
| 1413.17 | 6641.893 | 6661.524 | 1402.95 | 8378.844 | 8367.048 | 1418.28 | 11382.33 | 11425.41 | 1281.65 | 18839.34 | 18790.31 |
| 1414.88 | 6774.087 | 6662.874 | 1404.66 | 8545.077 | 8366.96 | 1419.99 | 11478.82 | 11420.48 | 1283.37 | 18677.89 | 18780.36 |
| 1416.58 | 6835.24 | 6664.165 | 1406.36 | 8214.284 | 8366.784 | 1421.69 | 11542.17 | 11415.43 | 1285.08 | 18927.93 | 18770.24 |
| 1418.28 | 6788.495 | 6665.398 | 1408.06 | 8296.476 | 8366.519 | 1423.39 | 11421.63 | 11410.33 | 1286.79 | 19132.16 | 18759.94 |
| 1419.99 | 6736.527 | 6666.571 | 1409.77 | 8548.891 | 8366.197 | 1425.09 | 11600.95 | 11405.1 | 1288.5 | 19114.94 | 18749.49 |
| 1421.69 | 6542.183 | 6667.686 | 1411.47 | 8616.646 | 8365.786 | 1426.8 | 11624.71 | 11399.82 | 1290.22 | 18854.95 | 18738.84 |
| 1423.39 | 6602.484 | 6668.743 | 1413.17 | 8467.404 | 8365.316 | 1428.5 | 11465.49 | 11394.45 | 1291.93 | 19074.97 | 18728.04 |
| 1425.09 | 6594.562 | 6669.74 | 1414.88 | 8493.989 | 8364.788 | 1430.2 | 11615.03 | 11388.99 | 1293.64 | 19055.16 | 18717.09 |
| 1426.8 | 6575.4 | 6670.65 | 1416.58 | 8318.718 | 8364.172 | 1431.9 | 11535.77 | 11383.48 | 1295.35 | 18802.69 | 18705.94 |
| 1428.5 | 6453.3 | 6671.53 | 1418.28 | 8260.559 | 8363.468 | 1433.6 | 11562.51 | 11377.84 | 1297.06 | 18895.97 | 18694.65 |
| 1430.2 | 6748.998 | 6672.352 | 1419.99 | 8689.859 | 8362.705 | 1435.3 | 11591.88 | 11372.15 | 1298.78 | 18727.89 | 18683.2 |
| 1431.9 | 6615.63 | 6673.086 | 1421.69 | 8413.323 | 8361.883 | 1437.01 | 11367.05 | 11366.4 | 1300.49 | 18830.95 | 18671.58 |
| 1433.6 | 6352.886 | 6673.79 | 1423.39 | 8461.183 | 8360.973 | 1438.71 | 11358.62 | 11360.53 | 1302.2 | 19242.2 | 18659.81 |
| 1435.3 | 6770.859 | 6674.406 | 1425.09 | 8444.897 | 8360.005 | 1440.41 | 11411.74 | 11354.6 | 1303.91 | 19181.87 | 18647.87 |
| 1437.01 | 6881.691 | 6674.993 | 1426.8 | 8383.539 | 8358.949 | 1442.11 | 11396.24 | 11348.59 | 1305.62 | 18951.55 | 18635.78 |
| 1438.71 | 7037.83 | 6675.521 | 1428.5 | 8210.528 | 8357.863 | 1443.81 | 11525.97 | 11342.51 | 1307.33 | 18910.56 | 18623.52 |
| 1440.41 | 7086.042 | 6675.961 | 1430.2 | 8028.508 | 8356.66 | 1445.51 | 11763.63 | 11336.35 | 1309.04 | 18616.33 | 18611.13 |
| 1442.11 | 7025.329 | 6676.372 | 1431.9 | 8002.773 | 8355.427 | 1447.21 | 11304.02 | 11330.1 | 1310.75 | 18977.23 | 18598.57 |
| 1443.81 | 6633.589 | 6676.724 | 1433.6 | 8491.407 | 8354.107 | 1448.91 | 11308.42 | 11323.79 | 1312.47 | 18989.64 | 18585.87 |
| 1445.51 | 6503.39 | 6677.018 | 1435.3 | 8742.532 | 8352.728 | 1450.61 | 11406.51 | 11317.4 | 1314.18 | 18416.94 | 18573.02 |
| 1447.21 | 6628.16 | 6677.252 | 1437.01 | 8348.972 | 8351.261 | 1452.31 | 11572.45 | 11310.94 | 1315.89 | 18414.32 | 18559.99 |
| 1448.91 | 6680.04 | 6677.429 | 1438.71 | 8354.547 | 8349.735 | 1454.01 | 11387.97 | 11304.4 | 1317.6 | 18738.46 | 18546.84 |
| 1450.61 | 6848.797 | 6677.546 | 1440.41 | 8512.74 | 8348.15 | 1455.71 | 10949.31 | 11297.79 | 1319.31 | 18827.87 | 18533.55 |
| 1452.31 | 6533.35 | 6677.605 | 1442.11 | 8444.75 | 8346.507 | 1457.41 | 11218.95 | 11291.1 | 1321.02 | 18763.55 | 18520.11 |
| 1454.01 | 6554.859 | 6677.605 | 1443.81 | 8358.068 | 8344.776 | 1459.11 | 11237.93 | 11284.35 | 1322.73 | 18800.9 | 18506.49 |
| 1455.71 | 6576.486 | 6677.575 | 1445.51 | 8280.043 | 8342.986 | 1460.81 | 11443.57 | 11277.52 | 1324.44 | 18791.69 | 18492.76 |
| 1457.41 | 6649.699 | 6677.458 | 1447.21 | 8310.209 | 8341.137 | 1462.51 | 11551.74 | 11270.62 | 1326.15 | 18556.2 | 18478.88 |
| 1459.11 | 6703.633 | 6677.311 | 1448.91 | 8193.244 | 8339.23 | 1464.21 | 11321.24 | 11263.64 | 1327.86 | 18511.16 | 18464.88 |
| 1460.81 | 6425.746 | 6677.106 | 1450.61 | 8265.694 | 8337.234 | 1465.91 | 11220.88 | 11256.6 | 1329.57 | 18679.94 | 18450.71 |
| 1462.51 | 6458.641 | 6676.842 | 1452.31 | 8556.022 | 8335.18 | 1467.61 | 11234.29 | 11249.49 | 1331.28 | 18788.75 | 18436.42 |
| 1464.21 | 6528.186 | 6676.519 | 1454.01 | 8316.547 | 8333.067 | 1469.31 | 11097.64 | 11242.31 | 1332.98 | 18830.33 | 18421.98 |
| 1465.91 | 6627.603 | 6676.137 | 1455.71 | 8362.939 | 8330.896 | 1471 | 11458.72 | 11235.06 | 1334.69 | 18521.25 | 18407.43 |
| 1467.61 | 6853.932 | 6675.727 | 1457.41 | 8327.404 | 8328.666 | 1472.7 | 11784.58 | 11227.75 | 1336.4 | 18280.4 | 18392.73 |
| 1469.31 | 6760.53 | 6675.228 | 1459.11 | 8127.925 | 8326.377 | 1474.4 | 11414.17 | 11220.36 | 1338.11 | 18246.65 | 18377.88 |
| 1471 | 6714.959 | 6674.7 | 1460.81 | 8295.214 | 8324 | 1476.1 | 11297.91 | 11212.9 | 1339.82 | 18322.27 | 18362.91 |
| 1472.7 | 6828.755 | 6674.113 | 1462.51 | 8459.099 | 8321.594 | 1477.8 | 11334.97 | 11205.39 | 1341.53 | 18197.09 | 18347.8 |
| 1474.4 | 6930.813 | 6673.496 | 1464.21 | 8684.988 | 8319.1 | 1479.5 | 11123.55 | 11197.79 | 1343.24 | 18298.86 | 18332.57 |
| 1476.1 | 6808.39 | 6672.792 | 1465.91 | 8367.37 | 8316.547 | 1481.19 | 11125.11 | 11190.13 | 1344.95 | 18516.97 | 18317.23 |
| 1477.8 | 6557.97 | 6672.059 | 1467.61 | 8063.838 | 8313.935 | 1482.89 | 11201.58 | 11182.41 | 1346.65 | 18434.54 | 18301.73 |
| 1479.5 | 6975.239 | 6671.266 | 1469.31 | 8221.531 | 8311.265 | 1484.59 | 11202.07 | 11174.64 | 1348.36 | 18317.2 | 18286.12 |
| 1481.19 | 7016.409 | 6670.415 | 1471 | 8107.765 | 8308.536 | 1486.29 | 11076.63 | 11166.8 | 1350.07 | 18413.12 | 18270.39 |
| 1482.89 | 6801.025 | 6669.535 | 1472.7 | 8299.146 | 8305.748 | 1487.98 | 11171.26 | 11158.88 | 1351.78 | 18493.17 | 18254.52 |
| 1484.59 | 6744.068 | 6668.596 | 1474.4 | 8123.934 | 8302.902 | 1489.68 | 11250.4 | 11150.93 | 1353.49 | 18286.5 | 18238.53 |
| 1486.29 | 6710.558 | 6667.598 | 1476.1 | 8247.501 | 8299.997 | 1491.38 | 11353.75 | 11142.89 | 1355.19 | 18367.2 | 18222.44 |
| 1487.98 | 6636.787 | 6666.542 | 1477.8 | 8680.293 | 8297.033 | 1493.08 | 11295.42 | 11134.79 | 1356.9 | 18580.03 | 18206.22 |
| 1489.68 | 6405.616 | 6665.456 | 1479.5 | 8378.198 | 8294.011 | 1494.77 | 11205.89 | 11126.63 | 1358.61 | 18128.75 | 18189.84 |
| 1491.38 | 6864.173 | 6664.312 | 1481.19 | 8206.478 | 8290.93 | 1496.47 | 11158.29 | 11118.42 | 1360.32 | 17971.38 | 18173.38 |
| 1493.08 | 6826.994 | 6663.138 | 1482.89 | 8191.689 | 8287.79 | 1498.17 | 11160.32 | 11110.11 | 1362.02 | 18225.35 | 18156.8 |
| 1494.77 | 6583.88 | 6661.876 | 1484.59 | 8447.596 | 8284.591 | 1499.86 | 10982.7 | 11101.78 | 1363.73 | 18095.62 | 18140.11 |
| 1496.47 | 6814.464 | 6660.585 | 1486.29 | 8526.414 | 8281.364 | 1501.56 | 11175.99 | 11093.39 | 1365.44 | 18191.9 | 18123.29 |
| 1498.17 | 6711.145 | 6659.265 | 1487.98 | 8185.849 | 8278.048 | 1503.25 | 11329.57 | 11084.93 | 1367.14 | 18274.82 | 18106.36 |
| 1499.86 | 6806.6 | 6657.886 | 1489.68 | 8161.523 | 8274.673 | 1504.95 | 11027.86 | 11076.39 | 1368.85 | 18307.6 | 18089.31 |
| 1501.56 | 6439.538 | 6656.448 | 1491.38 | 8235.558 | 8271.24 | 1506.65 | 11074.34 | 11067.83 | 1370.56 | 18344.96 | 18072.17 |
| 1503.25 | 6631.593 | 6654.951 | 1493.08 | 8306.599 | 8267.777 | 1508.34 | 11046.38 | 11059.2 | 1372.26 | 17944.94 | 18054.92 |
| 1504.95 | 6792.28 | 6653.425 | 1494.77 | 8652.622 | 8264.256 | 1510.04 | 10954.68 | 11050.51 | 1373.97 | 18212.56 | 18037.55 |
| 1506.65 | 6432.026 | 6651.841 | 1496.47 | 8339.699 | 8260.647 | 1511.73 | 10831.64 | 11041.77 | 1375.67 | 18284.6 | 18020.06 |
| 1508.34 | 6807.979 | 6650.227 | 1498.17 | 8409.714 | 8257.008 | 1513.43 | 11262.02 | 11032.97 | 1377.38 | 17992.15 | 18002.45 |
| 1510.04 | 6485.637 | 6648.554 | 1499.86 | 8357.804 | 8253.311 | 1515.12 | 11059.7 | 11024.1 | 1379.09 | 18122.65 | 17984.76 |
| 1511.73 | 6473.87 | 6646.852 | 1501.56 | 7858.636 | 8249.584 | 1516.82 | 11282.18 | 11015.18 | 1380.79 | 17974.61 | 17966.98 |
| 1513.43 | 6720.505 | 6645.092 | 1503.25 | 8101.222 | 8245.769 | 1518.51 | 11113.69 | 11006.2 | 1382.5 | 17619.96 | 17949.05 |
| 1515.12 | 6805.162 | 6643.272 | 1504.95 | 8298.295 | 8241.925 | 1520.21 | 11122.38 | 10997.2 | 1384.2 | 17804.65 | 17931.06 |
| 1516.82 | 6806.747 | 6641.424 | 1506.65 | 8160.203 | 8238.023 | 1521.9 | 11018.47 | 10988.1 | 1385.91 | 18024.43 | 17912.96 |
| 1518.51 | 6794.012 | 6639.516 | 1508.34 | 8223.85 | 8234.061 | 1523.6 | 10427.43 | 10978.97 | 1387.61 | 17572.89 | 17894.73 |
| 1520.21 | 6900.06 | 6637.58 | 1510.04 | 8382.805 | 8230.041 | 1525.29 | 10658.07 | 10969.79 | 1389.32 | 17850.89 | 17876.42 |
| 1521.9 | 6489.041 | 6635.584 | 1511.73 | 8377.054 | 8225.992 | 1526.99 | 10583.71 | 10960.55 | 1391.02 | 18232.3 | 17857.99 |
| 1523.6 | 6460.02 | 6633.559 | 1513.43 | 8266.017 | 8221.854 | 1528.68 | 10684.86 | 10951.27 | 1392.73 | 17782.38 | 17839.51 |
| 1525.29 | 6755.219 | 6631.476 | 1515.12 | 8112.959 | 8217.687 | 1530.37 | 10627.61 | 10941.94 | 1394.43 | 17840.15 | 17820.87 |
| 1526.99 | 6357.17 | 6629.363 | 1516.82 | 8339.523 | 8213.491 | 1532.07 | 10662.15 | 10932.55 | 1396.14 | 18161 | 17802.18 |
| 1528.68 | 6595.501 | 6627.192 | 1518.51 | 8200.991 | 8209.207 | 1533.76 | 10728.41 | 10923.1 | 1397.84 | 17922.87 | 17783.4 |
| 1530.37 | 6750.847 | 6624.991 | 1520.21 | 8346.8 | 8204.893 | 1535.45 | 10813.47 | 10913.6 | 1399.55 | 17887.78 | 17764.5 |
| 1532.07 | 6306.61 | 6622.732 | 1521.9 | 8289.022 | 8200.551 | 1537.15 | 10978.33 | 10904.06 | 1401.25 | 17855.03 | 17745.52 |
| 1533.76 | 6582.589 | 6620.443 | 1523.6 | 8184.265 | 8196.12 | 1538.84 | 11037.13 | 10894.46 | 1402.95 | 17797.4 | 17726.45 |
| 1535.45 | 6681.419 | 6618.095 | 1525.29 | 8428.259 | 8191.659 | 1540.53 | 10822.22 | 10884.84 | 1404.66 | 18009.35 | 17707.25 |
| 1537.15 | 6548.374 | 6615.718 | 1526.99 | 8415.201 | 8187.14 | 1542.23 | 10884.78 | 10875.15 | 1406.36 | 17659.16 | 17688.01 |
| 1538.84 | 6608.588 | 6613.312 | 1528.68 | 8590.853 | 8182.592 | 1543.92 | 11064.69 | 10865.41 | 1408.06 | 17218.18 | 17668.67 |
| 1540.53 | 6661.319 | 6610.847 | 1530.37 | 8234.707 | 8177.985 | 1545.61 | 10900.95 | 10855.61 | 1409.77 | 17769.05 | 17649.21 |
| 1542.23 | 6573.493 | 6608.324 | 1532.07 | 7923.486 | 8173.319 | 1547.31 | 10516.02 | 10845.78 | 1411.47 | 17854.71 | 17629.7 |
| 1543.92 | 6454.122 | 6605.8 | 1533.76 | 8053.127 | 8168.624 | 1549 | 10365.77 | 10835.92 | 1413.17 | 17533.63 | 17610.1 |
| 1545.61 | 6390.094 | 6603.218 | 1535.45 | 7966.328 | 8163.871 | 1550.69 | 10707.72 | 10826 | 1414.88 | 17869.44 | 17590.41 |
| 1547.31 | 6376.067 | 6600.577 | 1537.15 | 8045.556 | 8159.088 | 1552.38 | 10743.72 | 10816.03 | 1416.58 | 17906.68 | 17570.63 |
| 1549 | 6691.543 | 6597.907 | 1538.84 | 8021.172 | 8154.246 | 1554.07 | 10613.41 | 10805.99 | 1418.28 | 17791.77 | 17550.76 |
| 1550.69 | 6802.874 | 6595.207 | 1540.53 | 7888.039 | 8149.346 | 1555.77 | 10854.12 | 10795.96 | 1419.99 | 17672.86 | 17530.81 |
| 1552.38 | 6494.323 | 6592.449 | 1542.23 | 8555.611 | 8144.416 | 1557.46 | 10986.43 | 10785.83 | 1421.69 | 17619.4 | 17510.8 |
| 1554.07 | 6544.648 | 6589.661 | 1543.92 | 8090.57 | 8139.457 | 1559.15 | 10531.07 | 10775.68 | 1423.39 | 17588.65 | 17490.7 |
| 1555.77 | 6472.168 | 6586.844 | 1545.61 | 7983.905 | 8134.439 | 1560.84 | 10349.75 | 10765.5 | 1425.09 | 17620.69 | 17470.51 |
| 1557.46 | 6600.988 | 6583.968 | 1547.31 | 8633.314 | 8129.362 | 1562.53 | 10643.57 | 10755.26 | 1426.8 | 17451.17 | 17450.23 |
| 1559.15 | 6550.194 | 6581.063 | 1549 | 7987.954 | 8124.257 | 1564.22 | 10439.49 | 10744.99 | 1428.5 | 17221.26 | 17429.9 |

| | | |
|---|---|---|
| 1560.84 | 6518.297 | 6578.129 |
| 1562.53 | 6722.031 | 6575.136 |
| 1564.22 | 6662.434 | 6572.113 |
| 1565.91 | 6415.799 | 6569.062 |
| 1567.6 | 6210.01 | 6565.951 |
| 1569.3 | 6730.365 | 6562.811 |
| 1570.99 | 6587.167 | 6559.642 |
| 1572.68 | 6162.767 | 6556.444 |
| 1574.37 | 6618.565 | 6553.187 |
| 1576.06 | 6495.233 | 6549.9 |
| 1577.75 | 6671.149 | 6546.584 |
| 1579.44 | 6865.552 | 6543.239 |
| 1581.13 | 6384.665 | 6539.835 |
| -- | -- | -- |
| -- | -- | -- |
| -- | -- | -- |
| -- | -- | -- |
| -- | -- | -- |
| -- | -- | -- |
| -- | -- | -- |
| -- | -- | -- |
| -- | -- | -- |
| -- | -- | -- |
| -- | -- | -- |
| -- | -- | -- |
| -- | -- | -- |
| -- | -- | -- |
| -- | -- | -- |
| -- | -- | -- |
| 1550.69 | 8024.693 | 8119.092 |
| 1552.38 | 8246.503 | 8113.898 |
| 1554.07 | 7970.553 | 8108.646 |
| 1555.77 | 8061.402 | 8103.364 |
| 1557.46 | 7880.174 | 8098.053 |
| 1559.15 | 7767.347 | 8092.683 |
| 1560.84 | 8203.544 | 8087.283 |
| 1562.53 | 7910.897 | 8081.825 |
| 1564.22 | 8166.746 | 8076.338 |
| 1565.91 | 8166.453 | 8070.792 |
| 1567.6 | 7908.638 | 8065.217 |
| 1569.3 | 7934.989 | 8059.612 |
| 1570.99 | 7913.304 | 8053.949 |
| 1572.68 | 8086.139 | 8048.256 |
| 1574.37 | 7908.755 | 8042.505 |
| 1576.06 | 8053.068 | 8036.753 |
| 1577.75 | 8074.548 | 8030.943 |
| 1579.44 | 8073.228 | 8025.074 |
| 1581.13 | 8165.367 | 8019.206 |
| 1565.91 | 10588.7 | 10734.66 |
| 1567.6 | 10920.76 | 10724.3 |
| 1569.3 | 10760.54 | 10713.91 |
| 1570.99 | 10529.69 | 10703.46 |
| 1572.68 | 10686.8 | 10692.96 |
| 1574.37 | 10554.51 | 10682.45 |
| 1576.06 | 10655.43 | 10671.89 |
| 1577.75 | 10627.93 | 10661.27 |
| 1579.44 | 10096.87 | 10650.62 |
| 1581.13 | 10317.42 | 10639.93 |
| 1430.2 | 17634.57 | 17409.47 |
| 1431.9 | 17475.03 | 17388.99 |
| 1433.6 | 17391.93 | 17368.42 |
| 1435.3 | 17486.65 | 17347.76 |
| 1437.01 | 17429.84 | 17327.05 |
| 1438.71 | 17605.81 | 17306.27 |
| 1440.41 | 17545.57 | 17285.41 |
| 1442.11 | 17826.57 | 17264.49 |
| 1443.81 | 17585.51 | 17243.48 |
| 1445.51 | 17330.19 | 17222.41 |
| 1447.21 | 17292.57 | 17201.28 |
| 1448.91 | 16945.52 | 17180.06 |
| 1450.61 | 17426.4 | 17158.79 |
| 1452.31 | 16979.09 | 17137.46 |
| 1454.01 | 16929.88 | 17116.06 |
| 1455.71 | 17290.4 | 17094.58 |
| 1457.41 | 16798.15 | 17073.08 |
| 1459.11 | 16957.43 | 17051.48 |
| 1460.81 | 17165.54 | 17029.82 |
| 1462.51 | 16828.73 | 17008.11 |
| 1464.21 | 17185.52 | 16986.33 |
| 1465.91 | 17441.49 | 16964.5 |
| 1467.61 | 17083.52 | 16942.64 |
| 1469.31 | 16975.51 | 16920.69 |
| 1471 | 16889.12 | 16898.68 |
| 1472.7 | 17004.32 | 16876.62 |
| 1474.4 | 16910.45 | 16854.52 |
| 1476.1 | 16899.57 | 16832.34 |
| 1477.8 | 16662.14 | 16810.13 |
| 1479.5 | 16866.73 | 16787.85 |
| 1481.19 | 17043.64 | 16765.52 |
| 1482.89 | 16694.1 | 16743.16 |
| 1484.59 | 16885.33 | 16720.74 |
| 1486.29 | 16831.37 | 16698.27 |
| 1487.98 | 16734.04 | 16675.73 |
| 1489.68 | 16514.08 | 16653.16 |
| 1491.38 | 16606.42 | 16630.54 |
| 1493.08 | 16563.23 | 16607.89 |
| 1494.77 | 16386.43 | 16585.18 |
| 1496.47 | 16551.22 | 16562.4 |
| 1498.17 | 16611.35 | 16539.6 |
| 1499.86 | 16710.8 | 16516.77 |
| 1501.56 | 15983.45 | 16493.89 |
| 1503.25 | 16173.57 | 16470.94 |
| 1504.95 | 16561.52 | 16447.99 |
| 1506.65 | 16131.52 | 16424.96 |
| 1508.34 | 16574.08 | 16401.92 |
| 1510.04 | 16413.66 | 16378.83 |
| 1511.73 | 16077.97 | 16355.71 |
| 1513.43 | 16420.7 | 16332.52 |
| 1515.12 | 16276.86 | 16309.31 |
| 1516.82 | 16407.79 | 16286.07 |
| 1518.51 | 16504.39 | 16262.8 |
| 1520.21 | 16184.16 | 16239.48 |
| 1521.9 | 16044.1 | 16216.12 |
| 1523.6 | 16069.16 | 16192.73 |
| 1525.29 | 16075.03 | 16169.31 |
| 1526.99 | 15722.52 | 16145.87 |
| 1528.68 | 15823.29 | 16122.39 |
| 1530.37 | 15925.64 | 16098.89 |
| 1532.07 | 15953.02 | 16075.33 |
| 1533.76 | 15932.25 | 16051.76 |
| 1535.45 | 15912.64 | 16028.14 |
| 1537.15 | 15881.66 | 16004.52 |
| 1538.84 | 16217.2 | 15980.84 |
| 1542.23 | 16358.17 | 15933.45 |
| 1543.92 | 15813.43 | 15909.68 |
| 1545.61 | 15493.47 | 15885.91 |
| 1547.31 | 15326.06 | 15862.11 |
| 1549 | 15855.28 | 15838.29 |
| 1550.69 | 15628.27 | 15814.46 |
| 1552.38 | 15325.15 | 15790.57 |
| 1554.07 | 15727.75 | 15766.69 |
| 1555.77 | 15526.36 | 15742.77 |
| 1557.46 | 15573.58 | 15718.83 |
| 1559.15 | 15474.92 | 15694.85 |
| 1560.84 | 15203.9 | 15670.88 |
| 1562.53 | 15390.15 | 15646.88 |
| 1564.22 | 15289.91 | 15622.84 |
| 1565.91 | 15035.26 | 15598.81 |
| 1567.6 | 14950.87 | 15574.75 |
| 1569.3 | 15245.69 | 15550.66 |
| 1570.99 | 14898.64 | 15526.57 |
| 1572.68 | 14803.56 | 15502.45 |
| 1574.37 | 15390.41 | 15478.33 |
| 1576.06 | 14877.77 | 15454.18 |
| 1577.75 | 15026.72 | 15430 |
| 1579.44 | 15452.24 | 15405.82 |
| 1581.13 | 14710.19 | 15381.64 |